%% file: WhitePaperKevSterileNeutrinos.tex
\documentclass[a4paper,bbold,11pt]{article}
\pdfoutput=1 
\usepackage{jcappub} 

\usepackage[T1]{fontenc} 
\usepackage{tabularx}
\usepackage{subfigure, wrapfig}
\usepackage{footnote}
\usepackage[table]{xcolor}
\usepackage{multirow}
\usepackage{mciteplus}
\usepackage{bbold}
\usepackage{upgreek}
\usepackage[yyyymmdd,hhmmss]{datetime}
\usepackage{epstopdf}
\usepackage{slashed}

\usepackage{morefloats}
\mciteErrorOnUnknownfalse
\usepackage{placeins}

\renewcommand{\Im}{\mathrm{Im}}

\newcommand{\nr}[1]{(\ref{#1})}
\newcommand{\tr}{{\rm Tr\,}}

\newcommand{\fr}[2]{{\frac{#1}{#2}}}

\renewcommand{\vec}[1]{{\bf #1}}
\newcommand{\la}[1]{\label{#1}}
\newcommand{\ba}{\begin{eqnarray}}
\newcommand{\ea}{\end{eqnarray}}

\def\lsi{\raise0.3ex\hbox{$<$\kern-0.75em\raise-1.1ex\hbox{$\sim$}}}
\def\gsi{\raise0.3ex\hbox{$>$\kern-0.75em\raise-1.1ex\hbox{$\sim$}}}
\newcommand{\lsim}{\mathop{\lsi}}
\newcommand{\gsim}{\mathop{\gsi}}

\newcommand{\rmii}[1]{{\mbox{\tiny\rm{#1}}}}

\newcommand{\re}{\mathop{\mbox{Re}}}
\newcommand{\im}{\mathop{\mbox{Im}}}
\newcommand{\Tint}[1]{{\hbox{$\sum$}\!\!\!\!\!\!\!\int\,}_{\!\!\!\!\raise-0.9ex\hbox{$\scriptstyle{#1}$}}}
\newcommand{\Tinti}[1]{{{\Sigma}\!\!\!\!\raise0.3ex\hbox{$\int$}_\rmii{${#1}$}}}

\newcommand{\unit}{{\mathbbm{1}}} 
\newcommand{\bi}{\begin{itemize}}
\newcommand{\ei}{\end{itemize}}
\newcommand{\hide}[1]{ }
\newcommand{\bsl}[1]{\,\slash\!\!\!\!{#1}\,}
\newcommand{\msl}[1]{\,\slash\!\!\!{#1}\,}

\newcommand{\kmsec}{$\mathrm{km}\,\mathrm{s}^{-1}$}
\newcommand{\Vhalo}{$V_\mathrm{h,max}$}

\newcommand{\Vrot}{$V_\mathrm{rot,HI}$}
\newcommand{\vrot}{$V_\mathrm{rot}$}
\newcommand{\mwdm}{$m_\mathrm{WDM}$}

\usepackage{xspace}
\usepackage{units,bm}
\usepackage[colorlinks=true,linkcolor=blue,citecolor=blue]{hyperref}
\usepackage{etoolbox} %
\preto\tabular{\setcounter{magicrownumbers}{0}}%
\newcounter{magicrownumbers}%

 \newcommand{\eV}{\:\mathrm{eV}} 
\newcommand{\keV}{\:\mathrm{keV}} 
\newcommand{\MeV}{\:\mathrm{MeV}} \newcommand{\GeV}{\:\mathrm{GeV}}
\newcommand{\TeV}{\:\mathrm{TeV}} 
\def\be{\begin{eqnarray}} \def\ee{\end{eqnarray}} 
 \renewcommand{\S}{\mathcal{S}} %
\newcommand{\dm}{{\textsc{dm}}} 
\newcommand{\xmm}{\textit{XMM-Newton}\xspace}
\newcommand{\chan}{\textsl{Chandra}\xspace} 
\newcommand{\suza}{\textsl{Suzaku}\xspace} 
 
\renewcommand{\S}{\mathcal{S}} 
 
\newcommand{\adsurl}[1]{\href{#1}{ADS}}

\title{\boldmath 
A White Paper on keV Sterile Neutrino Dark Matter
}

\collaboration{Editors: M.~Drewes\footnote{marcodrewes@gmail.com}, T.~Lasserre\footnote{thierry.lasserre@cea.fr}, A.~Merle\footnote{\emph{Corresponding author}: amerle@mpp.mpg.de}, S.~Mertens\footnote{smertens@lbl.gov} \vspace{1cm}}

\author[61]{Authors: R.~ Adhikari}
\author[84]{M.~Agostini}
\author[39,73]{N.~Anh~Ky}
\author[57]{T.~Araki}
\author[34]{M.~Archidiacono}
\author[70]{M.~Bahr}
\author[2]{J.~Baur}
\author[69]{J.~Behrens}
\author[64]{F.~Bezrukov}
\author[31]{P.S.~Bhupal~Dev}

% ...
\affiliation[2]{Commissariat \`{a} l'\'energie atomique et aux \'energies alternatives, Centre de Saclay,DSM/IRFU, 91191 Gif-sur-Yvette, France}
% ...
\affiliation[31]{Consortium for Fundamental Physics, School of Physics and Astronomy, University of Manchester, Manchester M13 9PL, United Kingdom}
% ...
\affiliation[34]{Department of Physics and Astronomy, University of Aarhus,  DK-8000 Aarhus C, Denmark}
% ...
\affiliation[39]{Institute of physics, Vietnam academy of science and technology, 10 Dao Tan, Ba Dinh, Hanoi, Vietnam}
% ...
\affiliation[57]{Department of physics, Saitama University, Shimo-Okubo 255, 338-8570 Saitama Sakura-ku, Japan}
% ...
\affiliation[61]{Centre for Theoretical Physics, Jamia Millia Islamia (Central University), New Delhi-110025, India}	
% ...
\affiliation[64]{University of Connecticut}
% ...
\affiliation[69]{Westf\"{a}lische Wilhelms Universit\"at M\"unster, Institut f\"ur Kernphysik, Wilhelm Klemm-Str.9, D-48149 M\"unster}
\affiliation[70]{University of California, Santa Barbara}	
% ...
\affiliation[73]{Laboratory for high energy physics and cosmology, Faculty of physics, VNU university of science, 334 Nguyen Trai, Thanh Xuan, Hanoi, Vietnam}
% ...
\affiliation[84]{Gran Sasso Science Institute (INFN), L'Aquila, Italy}
% ...

\abstract{
We present a comprehensive review of keV-scale sterile neutrino Dark Matter, collecting views and insights from all disciplines involved -- cosmology, astrophysics, nuclear, and particle physics -- in each case viewed from both theoretical and experimental/observational perspectives. After reviewing the role of active neutrinos in particle physics, astrophysics, and cosmology, we focus on sterile neutrinos in the context of the Dark Matter puzzle. Here, we first review the physics motivation for sterile neutrino Dark Matter, based on challenges and tensions in purely cold Dark Matter scenarios. We then round out the discussion by critically summarizing all known constraints on sterile neutrino Dark Matter arising from astrophysical observations, laboratory experiments, and theoretical considerations. In this context, we provide a balanced discourse on the possibly positive signal from X-ray observations. Another focus of the paper concerns the construction of particle physics models, aiming to explain how sterile neutrinos of keV-scale masses could arise in concrete settings beyond the Standard Model of elementary particle physics. The paper ends with an extensive review of current and future astrophysical and laboratory searches, highlighting new ideas and their experimental challenges, as well as future perspectives for the discovery of sterile neutrinos.
}

\begin{document}
\maketitle
\flushbottom

\newpage
\section*{Executive Summary}

Despite decades of searching, the nature and origin of Dark Matter (DM) remains one of the biggest mysteries in modern physics. Astrophysical observations over a vast range of physical scales and epochs clearly show that the movement of celestial bodies, the gravitational distortion of light and the formation of structures in the Universe cannot be explained by the known laws of gravity and observed matter distribution~\cite{Ade:2015xua,Persic:1995ru,Faber:1976sn,Kaiser:1992ps,Clowe:2003tk,Percival:2007yw,Dave:1998gm}. They can, however, be brought into very good agreement if one postulates the presence of large amounts of non-luminous DM in and between the galaxies, a substance which is much more abundant in the Universe than ordinary matter~\cite{Ade:2015xua}. Generic ideas for what could be behind DM, such as Massive Compact Halo Objects (MACHOs)~\cite{Paczynski:1985jf,Griest:1990vu,Lasserre:2000xw,Bennett:2005at} are largely ruled out~\cite{Clowe:2006eq, Yoo:2003fr} or at least disfavored~\cite{Griest:2013esa,Pani:2014rca}. Alternative explanations based on a modification of the law of gravity~\cite{Milgrom:1983ca} have not been able to match the observations on various different scales. Thus, the existence of one or several new elementary particles appears to be the most attractive explanation.

As a first step, the suitability of known particles within the well-tested Standard Model (SM) has been examined. Indeed, the neutral, weakly interacting, massive neutrino could in principle be a DM candidate. However, neutrinos are so light that even with the upper limit for their mass~\cite{Kraus:2004zw,Lobashev:1999tp} they could not make up all of the DM energy density~\cite{Kolb:1990vq}. Moreover, neutrinos are produced with such large (relativistic) velocities that they would act as \emph{hot} DM (HDM), preventing the formation of structures such as galaxies or galaxy clusters~\cite{White:1984yj}. 

Consequently, explaining DM in terms of a new elementary particle clearly requires physics beyond the SM. There are multiple suggested extensions to the SM, providing a variety of suitable DM candidates, but to date there is no clear evidence telling us which of these is correct. Typically, extensions of the SM are sought at high energies, resulting in DM candidates with masses above the electroweak scale. In fact, there is a class of good DM candidates available at those scales, which are called Weakly Interacting Massive Particles (WIMPs). If these particles couple with a strength comparable to the SM weak interaction, they would have been produced in the early Universe via thermal freeze-out in suitable amounts~\cite{Gondolo:1990dk} \footnote{Note that this is true independently of the WIMP mass -- up to logarithmic corrections -- as long as they freeze out cold, since the main dependence on the mass drops out in the formula for the DM abundance~\cite{Jungman:1995df}.} WIMPs generically avoid the structure formation problem associated with SM neutrinos, as they are much more massive and therefore non-relativistic at the time of galaxy formation, because their velocities have been considerably redishifted.\footnote{NOte that, of course, also othe reasons can be responsible for DM having non-relativistic speed such as, e.g., strong interactions of condensation.} That is, WIMPs act as \emph{cold} DM (CDM). Typical examples for WIMPs are neutralinos as predicted by supersymmetry~\cite{Jungman:1995df,Gelmini:2006pw,Belanger:2005kh,Gunion:2005rw} or Kaluza-Klein bosons as predicted by models based on extra spatial dimensions~\cite{Servant:2002aq,Kong:2005hn,Bonnevier:2011km,Melbeus:2012wi}. More minimal extensions of the SM also predict WIMPs, e.g.\ an inert scalar doublet~\cite{LopezHonorez:2006gr,Dolle:2009fn}.

One of the advantages of WIMPs is that there is a variety of ways to test their existence. WIMPs could annihilate in regions of sufficiently high density, such as the center of a galaxy, thereby producing detectable (indirect) signals~\cite{Cirelli:2012tf} in e.g.\ photons, antimatter, or neutrinos. The same interactions that are responsible for the annihilation of two WIMPs in outer space can also be responsible for their production at colliders~\cite{Goodman:2010ku} or their scattering with ordinary matter in direct search experiments~\cite{Baudis:2012ig}.\footnote{At the level of amplitudes, this relation between ``break it'', ``make it'' and ``shake it'' can be visualized by rotating the Feynman diagram in steps of 90~degrees.} While a lot of experiments are currently taking data, no conclusive evidence for WIMPs has yet been found. Direct searches keep on pushing the limit on DM-matter cross sections towards smaller and smaller values~\cite{Angloher:2014myn,Aprile:2012nq,Akerib:2013tjd}, indirect searches yield some interesting but still inconclusive hints~\cite{Accardo:2014lma,Ackermann:2013uma,Adriani:2013uda}, and as of today the LHC has not discovered a hint of a DM-like particle~\cite{ATLAS:2012ky,Chatrchyan:2012me,Khachatryan:2014qwa,Aad:2014vma}.  WIMPs are certainly not yet excluded, nevertheless the current experimental results suggest the thorough exploration of alternative DM candidates.

A seemingly unrelated issue arose recently in $N$-body simulations of cosmological structure formation. Advanced simulations~\cite{Springel:2005nw} revealed some discrepancies between purely CDM scenarios and observations at small scales (a few 10~kpc or smaller). For example, there seem to be too few dwarf satellite galaxies observed compared to simulations (the missing satellite problem)~\cite{Klypin:1999uc,Moore:1999nt}; the density profile of galaxies is observed to be cored, whereas simulations predict a cusp profile (the cusp-core problem)~\cite{Dubinski:1991bm,Navarro:1995iw} and, finally, the observed dwarf satellite galaxies seem to be less dense than expected. This could possibly be explained if larger and very dense galaxies exist but are invisible due to a suppression of star formation~\cite{Maccio':2009dx,GarrisonKimmel:2013aq,Geen:2011fj}. However, no mechanism is known to suppress star formation in these types of galaxies: they are too big to fail producing enough stars (too-big-to-fail problem)~\cite{BoylanKolchin:2011de,BoylanKolchin:2011dk}.

While the discrepancy between simulation and observation is apparent, its origin is not so clear. A natural possibility would be that earlier simulations did not include baryons, although we clearly know they exist. The full inclusion of baryons and their interactions is highly non-trivial and only recently has it been attempted~\cite{Brooks:2012vi,Zhu:2015}. Another source for the discrepancy could arise from astrophysical feedback effects~\cite{Maccio':2009dx,GarrisonKimmel:2013aq}. These include, for example, relatively large supernova rates in dwarf galaxies which could wipe out all the visible material so that many dwarfs are simply invisible~\cite{Geen:2011fj}. Finally, it could also be that the DM velocity spectrum is not as cold as assumed~\cite{Herpich:2013yga}. It has been shown that a \emph{warm} DM (WDM) spectrum can significantly affect structure formation and strongly reduce the build-up of small objects~\cite{Lovell:2011rd}. Even more generally, the DM spectrum need not be thermal at all. It could have various shapes depending on the production mechanism~(see Sec.~5) and thereby influence structure formation in non-trivial ways. Thus, DM may be not simply \emph{cold}, \emph{warm}, or \emph{hot}, but the spectra could be more complicated resembling, e.g., mixed scenarios~\cite{Boyarsky:2008xj}. In any case, resolving the small-scale structure problem by modifying the DM spectrum would require a new DM candidate.

The candidate particle discussed in this White Paper is a \emph{sterile neutrino with a keV-scale mass}. A sterile neutrino is a hypothetical particle which, however, is connected to and can mix with the known active neutrinos. In SM language, sterile neutrinos are right-handed fermions with zero hypercharge and no color, i.e., they are total singlets under the SM gauge group and thus perfectly neutral. These properties allow sterile neutrinos to have a mass that does not depend on the Higgs mechanism. This so-called Majorana mass~\cite{Majorana:1937vz} can exist independently of electroweak symmetry breaking, unlike the fermion masses in the SM. In particular, the Majorana mass can have an arbitrary scale that is very different from all other fermion masses. Typically, it is assumed to be very large, but in fact it is just unrelated to the electroweak scale and could also be comparatively small. Observationally and experimentally the magnitude of the Majorana mass is almost unconstrained~\cite{Ibarra:2011xn,Ruchayskiy:2012si,Abada:2012mc,Ruchayskiy:2011aa,Abada:2013aba,Merle:2013gea,Drewes:2013gca,Hernandez:2014fha,Fernandez-Martinez:2015hxa,Drewes:2015jna,Drewes:2015iva,Antusch:2015mia,deGouvea:2015euy,Deppisch:2015qwa}. 

Depending on the choice of the Majorana mass, the implications for particle physics and cosmology are very different, , see e.g.~\cite{Drewes:2013gca}. Two reasons motivate a keV mass scale for a sterile neutrino DM candidate. First, fermionic DM can not have an arbitrarily small mass, since in dense regions (e.g.\ in galaxy cores) it cannot be packed within an infinitely small volume, due to the Pauli principle. This results in a lower bound on the mass, the so-called Tremaine-Gunn bound~\cite{Tremaine:1979we}. Second, sterile neutrinos typically have a small mixing with the active neutrinos, which would enable a DM particle to decay into an active neutrino and a mono-energetic photon. Since the decay rate scales with the fifth power of the initial state mass, a non-observation of the corresponding X-ray peak leads to an upper bound of a few tens of keV.\footnote{This only holds if active-sterile mixing is not switched off or forbidden, which may be the case in certain scenarios, see Sec.~6.} It is these two constraints, the phase space and X-ray bounds, which enforce keV-scale masses for sterile neutrinos acting as DM. 

This White Paper attempts to shed light on sterile neutrino DM from \emph{all} perspectives: astrophysics, cosmology, nuclear, and particle physics, as well as experiments, observations, and theory. Progress in the question of sterile neutrino DM requires expertise from all these different areas. The goal of this document is thus to advance the field by stimulating fruitful discussions between these communities. Furthermore, it should provide a comprehensive compendium of the current knowledge of the topic, and serve as a future reference.\footnote{The reader should be warned that the texts contributed to this work by the different authors cannot treat the various topics in full detail. They should, however, serve as possible overview and we made a great effort to ensure that they do contain all the relevant references, so that the present White Paper can guide the inclined reader to more specific information.} The list of authors indicates that there is great interest in the subject among scientists from many areas of physics. 

This White Paper is laid out as follows. First, sterile neutrinos are introduced from the particle physics (Sec.~1) and cosmology/astrophysics (Sec.~2) perspectives. Sec.~3 reviews the current tensions of CDM simulations with small-scale structure observations, and discusses attempts to tackle them. Sec.~4 gives a comprehensive summary of current constraints on keV sterile neutrino DM, arising from all accessible observables. The different sterile neutrino DM production mechanisms in the early Universe, and how they are constrained by astrophysical observations, are treated in Sec.~5. Sec.~6 turns to particle physics by reviewing attempts to explain or motivate the keV mass scale in various scenarios of physics beyond the SM. Current and future astrophysical and laboratory searches are discussed in Secs.~7 and 8, respectively, highlighting new ideas, their experimental challenges, and future perspectives for the discovery or exclusion of sterile neutrino DM. We end by giving an overall conclusion, involving all the viewpoints discussed in this paper.

Let us now start our journey into the fascinating world of keV sterile neutrino DM and address one of the biggest questions in modern science:

\begin{center}
\emph{What is Dark Matter and where did it come from?}
\end{center}

\vspace{0.1cm}

\paragraph{Note added} Several sections in the White Paper make reference to Japan's spaceborne Astro-H/Hitomi X-ray observatory. The Japanese Space Agency (JAXA) successfully launched the Astro-H satellite from Tanegashima Space Center in Japan on the 16th of February 2016, but after an apparant break off of bigger parts of the satellite occuring on March 26th, it was finally decided on April 28th to give up on the spacecraft. We give a short wrap-up of the events in the paragraph right before Sec.~\ref{sec:LymanAlphaBounds}. More detailed information can be found on the JAXA webpage, \url{http://global.jaxa.jp/projects/sat/astro_h/}.

\newpage
\section{Neutrinos in the Standard Model of Particle Physics and Beyond \\} 
\begin{flushright} Section Editors: \\Carlo Giunti, Andr\'e de Gouvea
\end{flushright}
\label{sec:neutr-stand-model}
\input{kevnuwp_section1.tex}

\newpage
\section{Neutrinos in The Standard Model of Cosmology and Beyond \\} 
\begin{flushright} Section Editors: \\Julien Lesgourgues, Alessandro Mirizzi
\end{flushright}
\label{sec:neutrino-cosmology}
\input{kevnuwp_section2.tex}

\newpage
\section{Dark Matter at Galactic Scales: Observational Constraints and Simulations \\} 
\label{sec:DMGalactic}
\begin{flushright} Section Editors: \\Aurel Schneider, Francesco Shankar, Oleg Ruchayskiy
\end{flushright}
\input{kevnuwp_section3.tex}

\newpage
\section{Observables Related to keV Neutrino Dark Matter \\} 
\begin{flushright} Section Editors: \\Marco Drewes, George Fuller
\end{flushright}
\label{sec:kev-neutrino-observables}
\input{kevnuwp_section4.tex}

\newpage
\section{Constraining keV Neutrino Production Mechanisms \\} 
\begin{flushright} Section Editors: \\Marco Drewes, Fedor Bezrukov, George Fuller
\end{flushright}
\label{sec:production-mechanisms}
\input{kevnuwp_section5.tex}

\newpage
\section{keV Neutrino Theory and Model Building (Particle Physics) \\} 
\begin{flushright} Section Editors: \\Alexander Merle, Viviana Niro
\end{flushright}
\label{sec:kev-neutrino-theory}
\input{kevnuwp_section6.tex}

\newpage
\section{Current and Future keV Neutrino Search with Astrophysical Experiments \\} 
\begin{flushright} Section Editors: \\Steen Hansen, Alexei Boyarsky
\end{flushright}
\label{sec:current-future-astro-exp}
\input{kevnuwp_section7.tex}

\newpage
\section{Current and Future keV Neutrino Search with Laboratory Experiment \\} 
\begin{flushright} Section Editors: \\Susanne Mertens, Loredana Gastaldo
\end{flushright}
\label{sec:current-future-lab-exp}
\input{kevnuwp_section8.tex}

\newpage
\section{Discussion - Pro and Cons for keV Neutrino as Dark Matter and Perspectives \\} 
\begin{flushright} Section Editors: \\Marco Drewes, Thierry Lasserre, Alexander Merle, Susanne Mertens
\end{flushright}
\label{sec:discussion}
\input{kevnuwp_section9.tex}


\newpage
\acknowledgments
\input{acknowledgments.tex}


\newpage
\bibliographystyle{JHEP}
\bibliography{all}

\end{document}

%% file: kevnuwp_section1.tex
\label{Section1}

The existence of sterile neutrinos is an exciting possible manifestation of new physics beyond the standard scenario of three-neutrino mixing, which has been established by the observation of neutrino flavor oscillations in many solar, reactor, and accelerator experiments (see the recent reviews in Refs.~\cite{Bellini:2013wra,Wang:2015rma}). Sterile neutrinos~\cite{Pontecorvo:1967fh} are observable through their mixing with the active neutrinos. In this Section we present a brief introduction to the standard theory of three-neutrino mixing in Subsection~\ref{sub:Stephen} and a summary of its current phenomenological status in Subsection~\ref{sub:Concha}. In Subsection~\ref{sub:Andre} we summarize the open questions in neutrino physics
and in Subsection~\ref{sub:Paul} we present a general introduction to sterile neutrinos.

\input{kevnuwp_section1-1.tex}
\input{kevnuwp_section1-2.tex}
\input{kevnuwp_section1-3.tex}
\input{kevnuwp_section1-4.tex}
\input{kevnuwp_section1-5.tex}

%% file: kevnuwp_section1-1.tex
\subsection{Introduction: Massive Neutrinos and Lepton Mixing (Author: S.~Parke)}
\label{sub:Stephen}

In the Standard Model (SM), as constructed around 1970, the neutrinos, $(\nu_e,~\nu_\mu,~\nu_\tau)$, are massless and interact diagonally in flavor, as follows
\begin{eqnarray}
W^+  \rightarrow e^+ +\nu_e,  \quad & \quad  W^- \rightarrow e^- +\bar{\nu}_e,  \quad & \quad Z  \rightarrow  \nu_e + \bar{\nu}_e,  \nonumber \\
W^+  \rightarrow \mu^+ +\nu_\mu,  \quad & \quad W^-  \rightarrow \mu^- +\bar{\nu}_\mu,  \quad & \quad  Z  \rightarrow  \nu_\mu + \bar{\nu}_\mu, \\
W^+  \rightarrow \tau^+ +\nu_\tau,  \quad & \quad W^-  \rightarrow \tau^- +\bar{\nu}_\tau,  \quad & \quad Z \rightarrow  \nu_\tau + \bar{\nu}_\tau. \nonumber
\end{eqnarray}
Since they travel at the speed of light, their character cannot change from production to detection. Therefore, in flavor terms,  massless neutrinos are relatively uninteresting compared to quarks.

Since then many experiments have seen neutrino flavor transitions, therefore neutrinos must have a mass and, like the quarks, there is a mixing matrix relating the neutrino flavor states, $\nu_{e}, \nu_\mu, \nu_\tau$, with the mass eigenstates, $\nu_{1}, \nu_2, \nu_3$:
\begin{eqnarray}
|\nu_\alpha \rangle = \sum_{j=1}^{3} U_{\alpha j}  
|\nu_j \rangle
\qquad
(\alpha=e,\mu,\tau),
\end{eqnarray}
where the mixing matrix $U$ is unitary and referred to as the PMNS\footnote{Pontecorvo-Maki-Nakagawa-Sakata~\cite{Pontecorvo:1957cp,Pontecorvo:1957qd,Maki:1962mu,Nakagawa01111963}.} matrix. By convention, the mass eigenstates are labeled such that $|U_{e1}|^2 > |U_{e2}|^2 > |U_{e3}|^2$, which implies that
\begin{center}
$\nu_1$ component of $\nu_e  \quad > \quad  \nu_2$ component  of $\nu_e  \quad $  $ > \quad \nu_3$ component of 
$\nu_e $.
\end{center}

\begin{figure}
\begin{center}
\includegraphics[width=\textwidth]{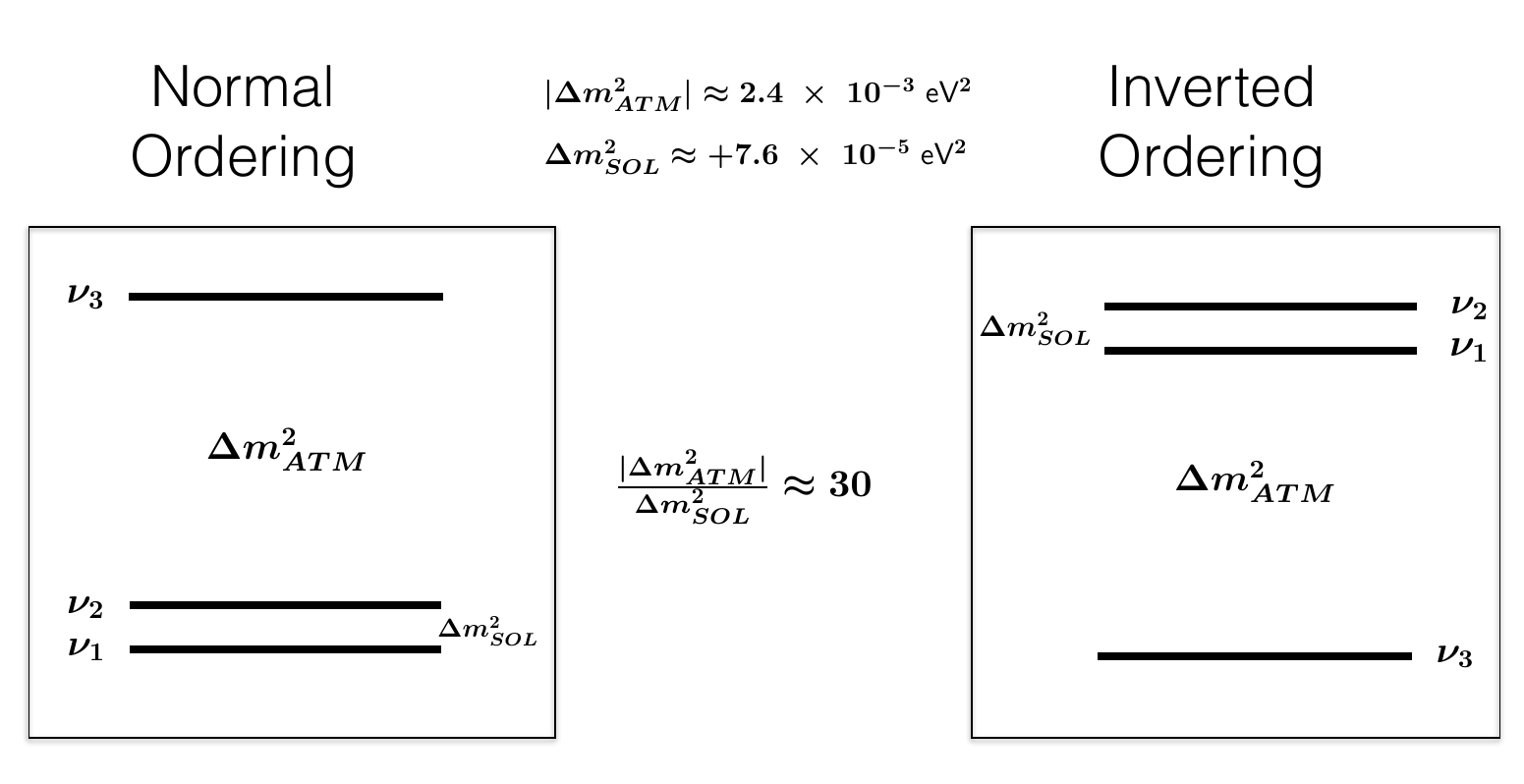}
\end{center}
\caption{\label{fig:mass0}What is known about the square of the neutrino masses for the two atmospheric mass orderings.}
\end{figure}

With this choice of labeling of the neutrino mass eigenstates, the solar neutrino oscillations/transformations are governed by $\Delta m^2_{21}\equiv m^2_2-m^2_1$, as these two are electron neutrino rich, and the atmospheric neutrino oscillations by $\Delta m^2_{31}$ and $\Delta m^2_{32}$. The SNO experiment \cite{Ahmed:2003kj} determined the mass ordering of the solar pair, $\nu_1$ and $\nu_2$, such that $m^2_2 > m^2_1$, i.e.\ $\Delta m^2_{21}>0$. The atmospheric neutrino mass ordering, 
\begin{equation}
m^2_3 > m^2_2\ \ \text{or}\ \ m^2_3 < m^2_1,
\end{equation}
is still to be determined, see Fig.~\ref{fig:mass0}.  If $m^2_3 > m^2_2$, the ordering is known as the normal ordering (NO), whereas if  $m^2_3 < m^2_1$ the ordering is known as the inverted ordering (IO). 
  
The mass splittings of the neutrinos are approximately~\cite{Gonzalez-Garcia:2014bfa}:
\begin{eqnarray}
 \Delta m^2_{32} \simeq \pm 2.5 \times 10^{-3} {\rm eV^2} \quad  &{\rm and} &\quad
\Delta m^2_{21} \simeq + 7.5 \times 10^{-5} {\rm eV^2},
\end{eqnarray}
and the sum of the masses of the neutrinos satisfies
 \begin{eqnarray}
 \sqrt{\delta m^2_A} \simeq 0.05 ~{\rm eV} < \sum_{i=1}^{3} m_{i} < 0.5 ~{\rm eV}.
 \end{eqnarray}
So the sum of neutrino masses ranges from $10^{-7} $ to  $10^{-6} $ times $m_e$, however the mass of the lightest neutrino, $m$, could be very small. If $m \ll \sqrt{\delta m^2_\odot} \sim 0.01~{\rm eV}^2$, then this is an additional scale to be explained by a theory of neutrino masses and mixings. 

The standard representation \cite{Agashe:2014kda} of the PMNS mixing matrix is given as follows:
\begin{eqnarray}
U
&=&
 \left( \begin{array}{lll}
 U_{e 1} & U_{e 2} & U_{e 3} \\
  U_{\mu 1} & U_{\mu 2} & U_{\mu 3} \\
   U_{\tau 1} & U_{\tau 2} & U_{\tau 3} 
   \end{array}  \right)
=
\left( \begin{array}{ccc}
  1  &  0  & 0\\
  0  & c_{23} &  s_{23} \\
  0  & - s_{23} & c_{23} 
\end{array} \right) \hspace*{-0.25cm}
\left( \begin{array}{ccc}
  c_{13} & 0 & {s_{13} e^{-i\delta}} \\
0& 1 & 0\\
 {- s_{13} e^{i\delta}} &0 & c_{13} 
\end{array} \right) \hspace*{-0.25cm}
\left( \begin{array}{ccc}
  c_{12}       & s_{12}  & 0\\
- s_{12} & c_{12} & 0\\
0& 0& 1 
\end{array} \right)
  \begin{pmatrix}
    e^{i\alpha_1} & 0 & 0 \\
    0& e^{i\alpha_2} & 0 \\
    0 & 0 & 1
  \end{pmatrix}
\nonumber
\\[0.3cm]
& = &
  \begin{pmatrix}
    c_{12} c_{13}
    & s_{12} c_{13}
    & s_{13} e^{-i\delta}
    \\
    - s_{12} c_{23} - c_{12} s_{13} s_{23} e^{i\delta}
    & \hphantom{+} c_{12} c_{23} - s_{12} s_{13} s_{23}
    e^{i\delta}
    & c_{13} s_{23}
    \\
    \hphantom{+} s_{12} s_{23} - c_{12} s_{13} c_{23} e^{i\delta}
    & - c_{12} s_{23} - s_{12} s_{13} c_{23} e^{i\delta}
    & c_{13} c_{23}
  \end{pmatrix}
  \begin{pmatrix}
    e^{i\alpha_1} & 0 & 0 \\
    0& e^{i\alpha_2} & 0 \\
    0 & 0 & 1
  \end{pmatrix},
\label{eq:pmns}
\end{eqnarray}
where $s_{ij} = \sin \theta_{ij}$ and $c_{ij} = \cos \theta_{ij}$. The Dirac phase, $\delta$, allows for the possibility of CP violation in the neutrino oscillation appearance channels. The Majorana phases $\alpha_1$ and $\alpha_2$ are unobservable in oscillations since oscillations depend on $U_{\alpha i}^* U_{\beta i}$ but they have observable, CP conserving effects, in neutrinoless double beta decay. If the neutrinos are Dirac, then neutrinoless double beta decay will be absent and the Majorana phases in the PMNS matrix are non-physical and can be set to zero. Note that there is some arbitrariness involved in which parameter combinations are called the physical phases, which is the reason why the ``distribution'' of the phases in eq.~\eqref{eq:pmns} looks a little asymmetric. This can be avoided when using the symmetric parametrization instead~\cite{Rodejohann:2011vc}.

\begin{figure}[t!]
\begin{center}
\includegraphics[width=0.8\textwidth]{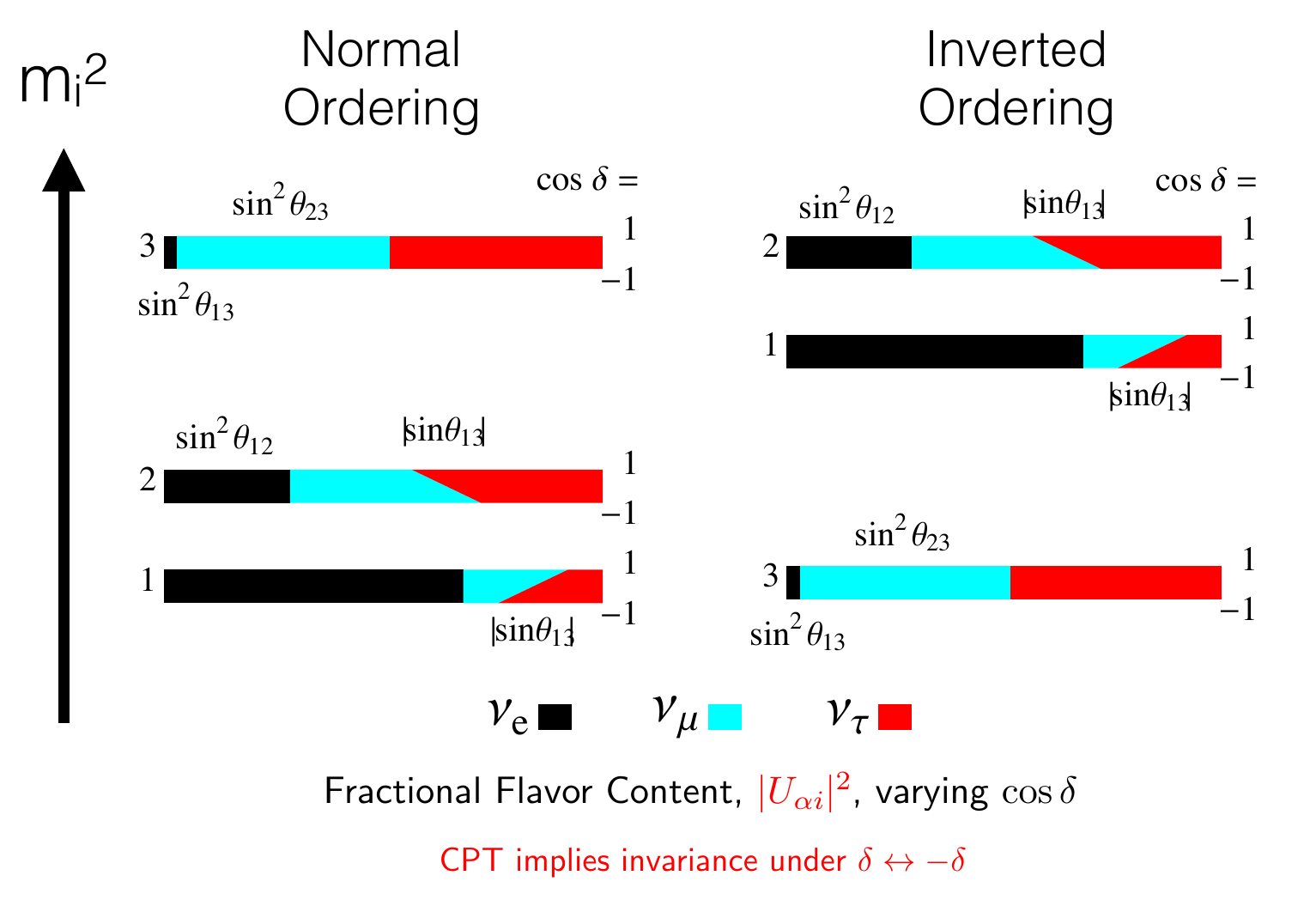}
\end{center}
\caption{\label{fig:deltaCP}The flavor content of the neutrino mass eigenstates (figure similar to Fig.~1 in Ref.~\cite{Mena:2003ug}). The width of the lines is used to show how these fractions change as $\cos \delta$ varies from $-1$ to $+1$. Of course, this figure must be the same for neutrinos and anti-neutrinos, if CPT is conserved.}
\end{figure}

The approximate values of the mixing parameters are as follows:
\begin{eqnarray}
\sin^2 \theta_{13} & \equiv &  |U_{e 3}|^2  \approx  0.02, \\
\sin^2 \theta_{12}  & \equiv & \vert U_{e2}\vert^2 /(1- \vert U_{e3}\vert^2)   \approx  1/3, \\
\sin^2\theta_{23}  & \equiv  & \vert U_{\mu 3}\vert^2 /(1- \vert U_{e3}\vert^2)  \approx 1/2, \\
0 \le & \delta&  < 2 \pi.
\end{eqnarray}
More precise values will be given in the next section.  These mixing angles and mass splittings are summarized in Fig.~\ref{fig:deltaCP}, which also shows the dependence of the flavor fractions on the CP violating Dirac phase $\delta$.

%% file: kevnuwp_section1-2.tex
\begingroup 

\newcommand{\Dmq}{\Delta m^2}
\newcommand{\eVq}{\ensuremath{\text{eV}^2}}
\newcommand{\Nuc}[2]{\ensuremath{\mbox{}^{#1}\text{#2}}}
\newcommand{\sign}{\mathop{\rm sign}}

\subsection{\label{sub:Concha}Current status of Three-Neutrino Masses and Mixings
(Authors: M.C.~Gonzalez-Garcia, M.~Maltoni, T.~Schwetz, R.~Shrock)}

\subsubsection{Neutrino oscillations}

Thanks to remarkable discoveries by a number of neutrino oscillation experiments it is now an established fact that neutrinos have mass and that leptonic flavors are not symmetries of Nature~\cite{Pontecorvo:1967fh,Gribov:1968kq}. Historically neutrino oscillations were first observed in the disappearance of solar $\nu_e$'s and atmospheric $\nu_\mu$'s which could be interpreted as flavor oscillations with two very different wavelengths. Over the last 15 years, these effects were confirmed also by terrestrial experiments using man-made beams from accelerators and nuclear reactors (see ref.~\cite{GonzalezGarcia:2007ib} for an overview).  In brief, at present we have observed neutrino oscillation effects in:
\begin{itemize}
\item atmospheric neutrinos, in particular in the high-statistics results of Super-Kamiokande~\cite{skatm:nu2014};

\item event rates of solar neutrino radiochemical experiments Chlorine~\cite{Cleveland:1998nv}, GALLEX/GNO~\cite{Kaether:2010ag}, and SAGE~\cite{Abdurashitov:2009tn}, as well as time- and energy-dependent rates from the four phases in Super-Kamiokande~\cite{Hosaka:2005um, Cravens:2008aa, Abe:2010hy,sksol:nu2014}, the three phases of SNO~\cite{Aharmim:2011vm}, and Borexino~\cite{Bellini:2011rx, Bellini:2008mr};

\item disappearance results from accelerator long-baseline (LBL) experiments in the form of the energy distribution of $\nu_\mu$ and $\bar\nu_\mu$ events in MINOS~\cite{Adamson:2013whj} and T2K~\cite{Abe:2014ugx};

\item LBL $\nu_e$ appearance results for both neutrino and antineutrino events in MINOS~\cite{Adamson:2013ue} and $\nu_e$ appearance in T2K~\cite{Abe:2013hdq};

\item reactor $\bar\nu_e$ disappearance at medium baselines in the form of
the energy distribution of the near/far ratio of events at
Daya Bay~\cite{An:2012eh}
and
RENO~\cite{Ahn:2012nd}
and
the energy distribution of events in the
near
Daya Bay~\cite{An:2015rpe}
and
RENO~\cite{reno:nu2014}
detectors
and in the far
Daya Bay~\cite{An:2015rpe},
RENO~\cite{reno:nu2014} and
Double Chooz~\cite{Abe:2012tg,Abe:2014bwa}
detectors.

\item the energy spectrum of reactor $\bar\nu_e$ disappearance at LBL in KamLAND~\cite{Gando:2010aa}.

\end{itemize}
This wealth of data can be consistently described by assuming mixing among the three known neutrinos ($\nu_e$, $\nu_\mu$, $\nu_\tau$), which can be expressed as quantum superpositions of three massive states $\nu_i$ ($i=1,2,3$) with masses $m_i$.  As explained in the previous section this implies the presence of a leptonic mixing matrix in the weak charged current interactions which can be parametrized in the standard representation, see eq.~\eqref{eq:pmns}.

In this convention, disappearance of solar $\nu_e$'s and long-baseline reactor $\bar\nu_e$'s dominantly proceed via oscillations with wavelength $\propto E / \Dmq_{21}$ ($\Dmq_{ij} \equiv m_i^2 - m_j^2$ and $\Dmq_{21} \geq 0$ by convention) and amplitudes controlled by $\theta_{12}$, while disappearance of atmospheric and LBL accelerator $\nu_\mu$'s dominantly proceed via oscillations with wavelength $\propto E / |\Dmq_{31}| \ll E/\Dmq_{21}$ and amplitudes controlled by $\theta_{23}$. Generically $\theta_{13}$ controls the amplitude of oscillations involving $\nu_e$ flavor with $E/|\Dmq_{31}|$ wavelengths. So, given the observed hierarchy between the solar and atmospheric wavelengths, there are two possible non-equivalent orderings for the mass eigenvalues, which are conventionally chosen as:
\begin{align}
  \label{eq:normal}
  m_{1} < m_{2} < m_{3}
  \quad \text{with} \quad
  \Dmq_{21} &\ll \hphantom{+} (\Dmq_{32} \simeq \Dmq_{31} > 0) \,,
  \\
  \label{eq:inverted}
  m_{3} < m_{1} < m_{2}
  \quad \text{with} \quad
  \Dmq_{21} &\ll -(\Dmq_{31} \simeq \Dmq_{32} < 0) \,.
\end{align}
As it is customary, we refer to the first option, eq.~\eqref{eq:normal}, as normal ordering (NO), and to the second one, eq.~\eqref{eq:inverted}, as inverted ordering (IO); in this form they correspond to the two possible choices of the sign of $\Dmq_{31}$. In this convention the angles $\theta_{ij}$ can be taken without loss of generality to lie in the first quadrant, $\theta_{ij} \in [0, \pi/2]$, and the CP phase $\delta \in [0, 2\pi]$. In the following we adopt the (arbitrary) convention of reporting results for $\Dmq_{31}$ for NO and $\Dmq_{32}$ for IO, i.e., we always use the one which has the larger absolute value. Sometimes we will generically denote such quantity as $\Dmq_{3\ell}$, with $\ell=1$ for NO and $\ell=2$ for IO.

In summary, in total the $3\nu$ oscillation analysis of the existing data involves six parameters: 2~mass square differences (one of which can be positive or negative), 3~mixing angles, and the Dirac CP phase $\delta$. For the sake of clarity we summarize in tab.~\ref{tab:expe} which experiment contribute dominantly to the present determination of the different parameters.

\begin{table}\centering
  \caption{\label{tab:expe}Experiments contributing to the present determination of the oscillation parameters.}
  \begin{tabular*}{\textwidth}{@{\extracolsep{\fill}} l l l}  
    \hline
    Experiment 
    & Dominant & Important
    \\
    \hline\hline
    Solar Experiments
    & $\theta_{12}$ & $\Dmq_{21}$, $\theta_{13}$
    \\
    Reactor LBL (KamLAND) & $\Dmq_{21}$
    & $\theta_{12}$, $\theta_{13}$
    \\
    Reactor MBL (Daya-Bay, Reno, D-Chooz)
    & $\theta_{13}$ & $|\Dmq_{3\ell}|$
    \\
    Atmospheric Experiments
    & $\theta_{23}$ & $|\Dmq_{3\ell}|$, $\theta_{13}$,$\delta$
    \\
    Accelerator LBL $\nu_\mu$ Disapp.\ (Minos, T2K)
    & $|\Dmq_{3\ell }|$, $\theta_{23}$ & ~
    \\
    Accelerator LBL $\nu_e$ App.\ (Minos, T2K)
    & $\delta$ & $\theta_{13}$, $\theta_{23}$, $\sign(\Dmq_{3\ell})$
    \\
    \hline
  \end{tabular*}
\end{table}

The consistent determination of these leptonic parameters requires a global analysis of the data described above which, at present, is in the hands of a few phenomenological groups~\cite{Capozzi:2013csa,Forero:2014bxa,Gonzalez-Garcia:2014bfa}. Here we summarize the results from ref.~\cite{Gonzalez-Garcia:2014bfa}.  We show in fig.~\ref{fig:chisq} the one-dimensional projections of the $\Delta\chi^2$ of the global analysis as a function of each of the six parameters. The corresponding best-fit values and the derived ranges for the six parameters at the $1\sigma$ ($3\sigma$) level are given in tab.~\ref{tab:results}.  For each parameter the curves and ranges are obtained after marginalizing with respect to the other five parameters. The ranges presented in the table are shown for three scenarios. In the first and second columns we assume that the ordering of the neutrino mass states is known ``a priori'' to be normal or inverted, respectively, so that the ranges of all parameters are defined with respect to the minimum in the given scenario.  In the third column we make no assumptions on the ordering, so in this case the ranges of the parameters are defined with respect to the global minimum (which corresponds to IO) and are obtained by marginalizing also over the ordering. For this third case we only give the $3\sigma$ ranges. Of course in this case the range of $\Dmq_{3\ell}$ is composed of two disconnected intervals, one one containing the absolute minimum (IO) and the other the secondary local minimum (NO).

As mentioned, all the data described above can be consistently interpreted as oscillations of the three known active neutrinos. However, together with this data, several anomalies at short baselines (SBL) have been observed which \emph{cannot} be explained as oscillations in this framework but could be interpreted as oscillations involving an $\mathcal{O}(\text{eV})$ mass sterile state. They will be discussed in Section~\ref{sub:Paul}. In what respect the results presented here the only SBL effect which has to be treated in some form is the so-called \emph{reactor anomaly} by which the most recent reactor flux calculations~\cite{Mueller:2011nm,Huber:2011wv, Mention:2011rk}, fall short at describing the results from reactor experiments at baselines $\lesssim 100$~m from Bugey4~\cite{Declais:1994ma}, ROVNO4~\cite{Kuvshinnikov:1990ry}, Bugey3~\cite{Declais:1994su}, Krasnoyarsk~\cite{Vidyakin:1987ue,Vidyakin:1994ut}, ILL~\cite{Kwon:1981ua}, G\"osgen~\cite{Zacek:1986cu}, SRP~\cite{Greenwood:1996pb}, and ROVNO88~\cite{Afonin:1988gx}, to which we refer as reactor short-baseline experiments (RSBL).  We notice that these RSBL do not contribute to oscillation physics in the $3\nu$ framework, but they play an important role in constraining the unoscillated reactor neutrino flux if they are to be used instead of the theoretically calculated reactor fluxes. Thus, to account for the possible effect of the reactor anomaly in the determined ranges of neutrino parameters in the framework of $3\nu$ oscillations, the results in fig.~\ref{fig:chisq} are shown for two extreme choices.  The first option is to leave the normalization of reactor fluxes free and include the RSBL data, experiments (labeled ``Free+RSBL'') The second option is not to include short-baseline reactor data but assume reactor fluxes and uncertainties as predicted in~\cite{Huber:2011wv} (labeled ``Huber'').

\begin{figure}\centering
  \includegraphics[width=\textwidth]{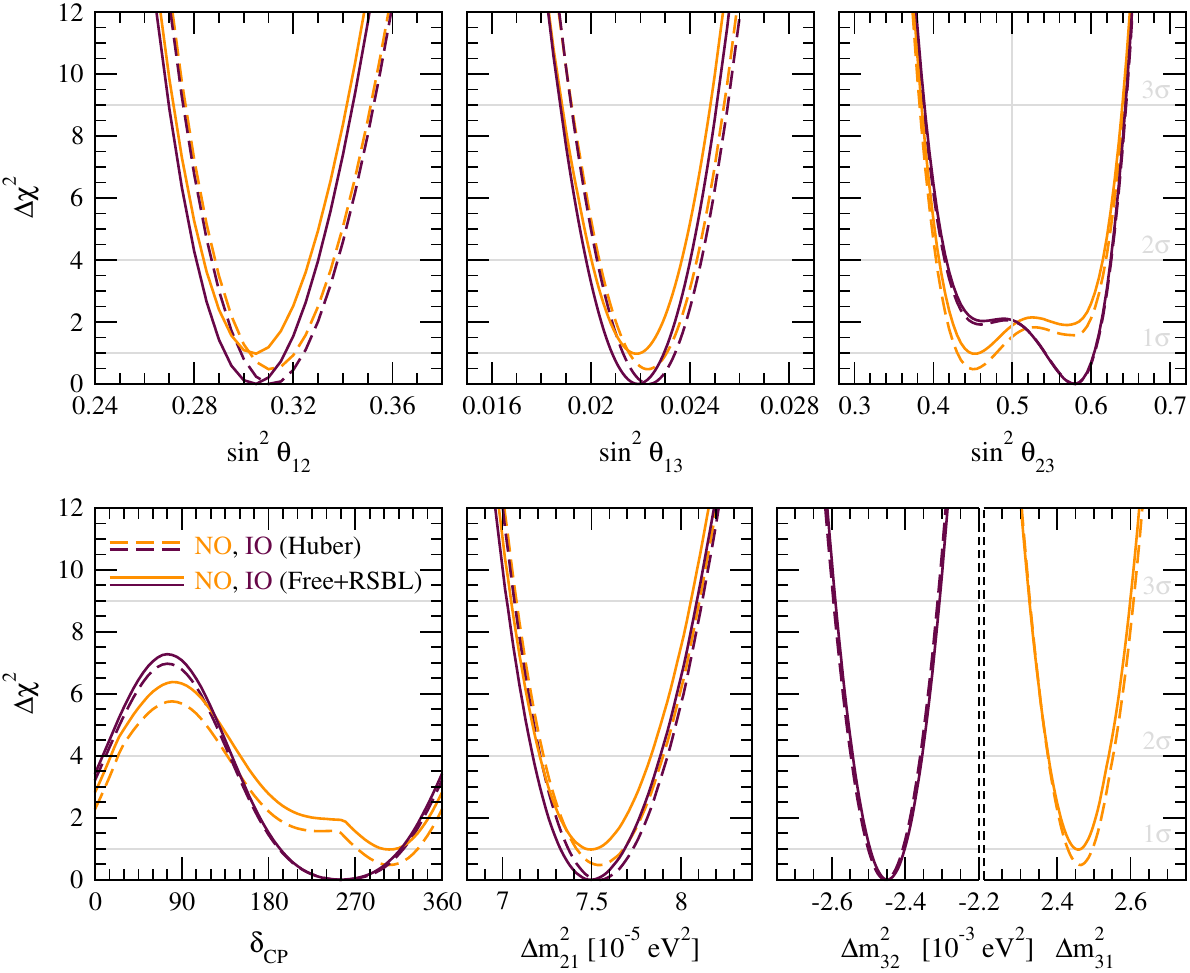}
  \caption{\label{fig:chisq}Global $3\nu$ oscillation analysis.  The orange (violet) curves are for NO (IO).  For solid curves the normalization of reactor fluxes is left free and data from short-baseline (less than 100 m) reactor experiments are included. For dashed curves, short-baseline data are not included but reactor fluxes as predicted in~\cite{Huber:2011wv} are assumed. Note that we use $\Dmq_{31}$ for NO and $\Dmq_{32}$ for IO to denote the mass square differences. (Figure similar to fig.~2 in ref.~\cite{Gonzalez-Garcia:2014bfa}.)}
\end{figure}

\begin{table}\centering
\caption{\label{tab:results}Three-flavor oscillation parameters from our fit to global data after the NOW~2014 conference. The results are presented for the ``Free Fluxes + RSBL'' in which reactor fluxes have been left free in the fit and short-baseline reactor data (RSBL) with $L \lesssim 100$~m are included. The numbers in the 1st (2nd) column are obtained assuming NO (IO), i.e., relative to the respective local minimum, whereas in the 3rd column we minimize also with respect to the ordering. Note that $\Dmq_{3\ell} \equiv \Dmq_{31} > 0$ for NO and $\Dmq_{3\ell} \equiv \Dmq_{32} < 0$ for IO.}
  \begin{footnotesize}
    \begin{tabular*}{\textwidth}{@{\extracolsep{\fill}}l|cc|cc|c}  
      \hline
      & \multicolumn{2}{c|}{Normal ordering ($\Delta\chi^2=0.97$)}
      & \multicolumn{2}{c|}{Inverted ordering (best-fit)}
      & Any ordering
      \\
      \hline\hline
      & bfp $\pm 1\sigma$ & $3\sigma$ range
      & bfp $\pm 1\sigma$ & $3\sigma$ range
      & $3\sigma$ range
      \\
      \hline\hline
      \rule{0pt}{4mm}\ignorespaces
      $\sin^2\theta_{12}$
      & $0.304_{-0.012}^{+0.013}$ & $0.270 \to 0.344$
      & $0.304_{-0.012}^{+0.013}$ & $0.270 \to 0.344$
      & $0.270 \to 0.344$
      \\[1mm]
      $\theta_{12}/^\circ$
      & $33.48_{-0.75}^{+0.78}$ & $31.29 \to 35.91$
      & $33.48_{-0.75}^{+0.78}$ & $31.29 \to 35.91$
      & $31.29 \to 35.91$
      \\[3mm]
      $\sin^2\theta_{23}$
      & $0.452_{-0.028}^{+0.052}$ & $0.382 \to 0.643$
      & $0.579_{-0.037}^{+0.025}$ & $0.389 \to 0.644$
      & $0.385 \to 0.644$
      \\[1mm]
      $\theta_{23}/^\circ$
      & $42.3_{-1.6}^{+3.0}$ & $38.2 \to 53.3$
      & $49.5_{-2.2}^{+1.5}$ & $38.6 \to 53.3$
      & $38.3 \to 53.3$
      \\[3mm]
      $\sin^2\theta_{13}$
      & $0.0218_{-0.0010}^{+0.0010}$ & $0.0186 \to 0.0250$
      & $0.0219_{-0.0010}^{+0.0011}$ & $0.0188 \to 0.0251$
      & $0.0188 \to 0.0251$
      \\[1mm]
      $\theta_{13}/^\circ$
      & $8.50_{-0.21}^{+0.20}$ & $7.85 \to 9.10$
      & $8.51_{-0.21}^{+0.20}$ & $7.87 \to 9.11$
      & $7.87 \to 9.11$
      \\[3mm]
      $\delta/^\circ$
      & $306_{-70}^{+39}$ & $\hphantom{00}0 \to 360$
      & $254_{-62}^{+63}$ & $\hphantom{00}0 \to 360$
      & $\hphantom{00}0 \to 360$
      \\[3mm]
      $\dfrac{\Dmq_{21}}{10^{-5}~\eVq}$
      & $7.50_{-0.17}^{+0.19}$ & $7.02 \to 8.09$
      & $7.50_{-0.17}^{+0.19}$ & $7.02 \to 8.09$
      & $7.02 \to 8.09$
      \\[3mm]
      $\dfrac{\Dmq_{3\ell}}{10^{-3}~\eVq}$
      & $+2.457_{-0.047}^{+0.047}$ & $+2.317 \to +2.607$
      & $-2.449_{-0.047}^{+0.048}$ & $-2.590 \to -2.307$
      & $\begin{bmatrix}
        +2.325 \to +2.599\\[-2pt]
        -2.590 \to -2.307
      \end{bmatrix}$
      \\[3mm]
      \hline
    \end{tabular*}
  \end{footnotesize}
\end{table}

From the results in the figure and table  we conclude that:
\begin{enumerate}
\item if we define the $3\sigma$ relative precision of a parameter $x$ by $2(x^\text{up} - x^\text{low}) / (x^\text{up} + x^\text{low})$, where $x^\text{up}$ ($x^\text{low}$) is the upper (lower) bound on $x$ at the $3\sigma$ level, from the numbers in the table we find $3\sigma$ relative precision of 14\% ($\theta_{12}$), 32\% ($\theta_{23}$), 15\% ($\theta_{13}$), 14\% ($\Dmq_{21}$), and 11\% ($|\Dmq_{3\ell}|$) for the various oscillation parameters;

\item for either choice of the reactor fluxes the global best-fit corresponds to IO with $\sin^2\theta_{23} > 0.5$, while the second local minimum is for NO and with $\sin^2\theta_{23} < 0.5$;

\item the statistical significance of the preference for IO versus NO is quite small, $\Delta\chi^2 \lesssim 1$;

\item the present global analysis disfavors $\theta_{13}=0$ with $\Delta\chi^2 \approx 500$. Such impressive result is mostly driven by the reactor data from Daya Bay with secondary contributions from RENO and Double Chooz;

\item the uncertainty on $\theta_{13}$ associated with the choice of reactor fluxes is at the level of $0.5\sigma$ in the global analysis. This is so because the most precise results from Daya Bay, and RENO are reactor flux normalization independent;

\item a non-maximal value of the $\theta_{23}$ mixing is slightly favored, at the level of $\sim 1.4\sigma$ for IO at of $\sim 1.0\sigma$ for NO;

\item the statistical significance of the preference of the fit for the second (first) octant of $\theta_{23}$ is $\leq 1.4\sigma$ ($\leq 1.0\sigma$) for IO (NO);

\item the best-fit for $\delta$ for all analyses and orderings occurs for $\delta \simeq 3\pi/2$, and values around $\pi/2$ are disfavored with $\Delta\chi^2 \simeq 6$.  Assigning a confidence level to this $\Delta\chi^2$ is non-trivial, due to the non-Gaussian behavior of the involved $\chi^2$ function, see ref.~\cite{Gonzalez-Garcia:2014bfa} for a discussion and a Monte Carlo study.

\end{enumerate}

From this global analysis one can also derive the $3\sigma$ ranges on the magnitude of the elements of the leptonic mixing matrix to be:
\begin{equation}
  \label{eq:umatrix}
  |U| = \begin{pmatrix}
    0.801 \to 0.845 &\qquad
    0.514 \to 0.580 &\qquad
    0.137 \to 0.158
    \\
    0.225 \to 0.517 &\qquad
    0.441 \to 0.699 &\qquad
    0.614 \to 0.793
    \\
    0.246 \to 0.529 &\qquad
    0.464 \to 0.713 &\qquad
    0.590 \to 0.776
  \end{pmatrix} .
\end{equation}

\begin{figure}\centering
  \includegraphics[width=\textwidth]{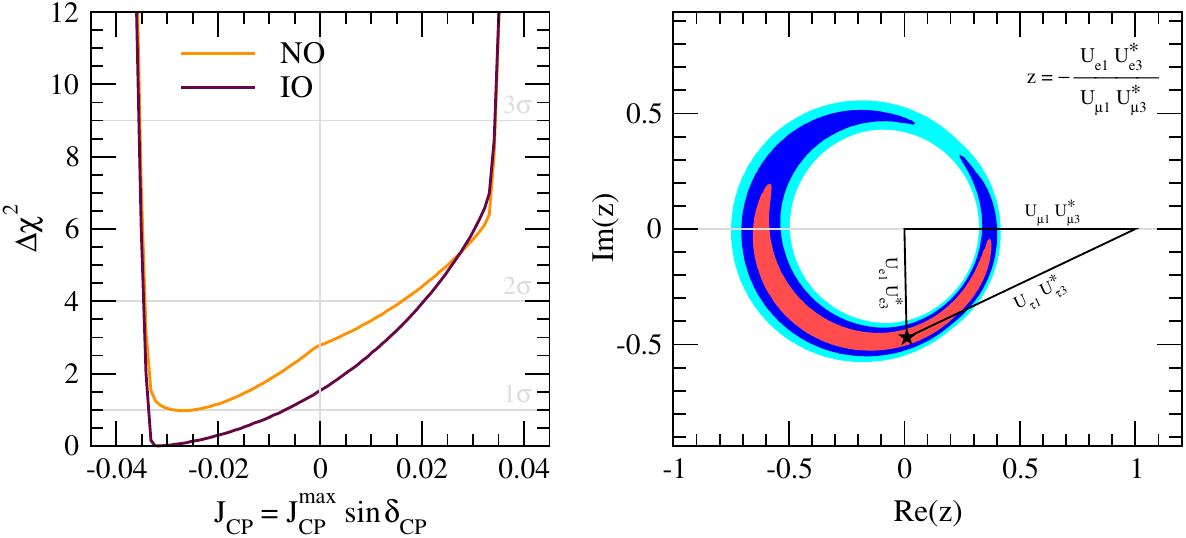}
  \caption{\label{fig:viola}Left: dependence of the global $\Delta\chi^2$ function on the Jarlskog invariant. The orange (violet) curves are for NO (IO). (Figure similar to fig.~3b in~\cite{Gonzalez-Garcia:2014bfa}.) Right: leptonic unitarity triangle. After scaling and rotating so that two of its vertices always coincide with $(0,0)$ and $(1,0)$ we plot the $1\sigma$, $2\sigma$, $3\sigma$ (2~dof) allowed regions of the third vertex. (Figure similar to fig.~4d in~\cite{Gonzalez-Garcia:2014bfa}.)}
\end{figure}

The present status of the determination of leptonic CP violation is further illustrated in fig.~\ref{fig:viola}. On the left panel we show the dependence of the $\Delta\chi^2$ of the global analysis on the Jarlskog invariant which gives a convention-independent measure of CP violation~\cite{Jarlskog:1985ht}, defined as:
\begin{equation}
  \Im\big[ U_{\alpha i} U_{\alpha j}^* U_{\beta i}^* U_{\beta j} \big]
  \equiv
  \cos\theta_{12} \sin\theta_{12} \cos\theta_{23} \sin\theta_{23}
  \cos^2\theta_{13} \sin\theta_{13} \, \sin\delta
  \equiv
  J_\text{CP}^\text{max} \, \sin\delta,
\end{equation}
where in the second equality we have used the parametrization in eq.~\eqref{eq:pmns}.  Thus the determination of the mixing angles yields at present a maximum allowed CP violation
\begin{equation}
  \label{eq:jmax}
  J_\text{CP}^\text{max} = 0.0329 \pm 0.0009 \; (\mathrel{\pm} 0.0027)
\end{equation}
at $1\sigma$ ($3\sigma$) for both orderings. The preference of the present data for non-zero $\delta$ implies a best-fit of $J_\text{CP}^\text{best} = -0.032$, which is favored over CP conservation at the $\sim 1.2\sigma$ level. These numbers can be compared to the size of the Jarlskog invariant in the quark sector, which is determined to be $J_\text{CP}^\text{quarks} = (2.96^{+0.20}_{-0.16}) \times 10^{-5}$~\cite{Agashe:2014kda}.

On the right panel of fig.~\ref{fig:viola} we recast the allowed regions for the leptonic mixing matrix in terms of one leptonic unitarity triangle. Since in the analysis $U$ is unitary by construction, any given pair of rows or columns can be used to define a triangle in the complex plane. In the figure we show the triangle corresponding to the unitarity conditions on the first and third columns which is the equivalent to the one usually shown for the quark sector.  In this figure the absence of CP violation implies a flat triangle, \textit{i.e.}, $\Im(z) = 0$.  As can be seen, the horizontal axis marginally crosses the $1\sigma$ allowed region, which for 2~dof corresponds to $\Delta\chi^2 \simeq 2.3$. This is consistent with the present preference for CP violation, $\chi^2(J_\text{CP} = 0) - \chi^2(J_\text{CP}~\text{free}) = 1.5$.

\subsubsection{Absolute Neutrino Mass Measurements}

\begin{figure}[t]\centering
  \includegraphics[width=\textwidth]{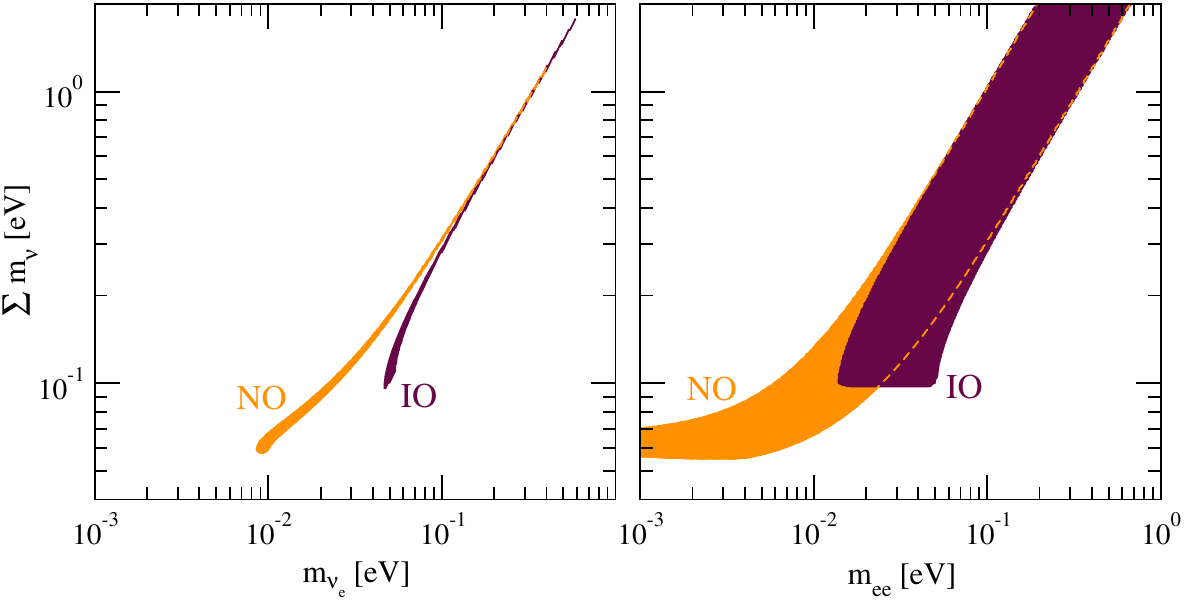}
  \caption{\label{fig:mbeta}95\% allowed regions (for 2~d.o.f.) in the planes ($m_{\nu_e}$, $\sum m_\nu$) and ($m_{ee}$, $\sum m_\nu$) obtain from projecting the results of the global analysis of oscillation data.}
\end{figure}

Oscillation experiments provide information on $\Dmq_{ij}$ and on the leptonic mixing angles $\theta_{ij}$. But they are insensitive to the absolute mass scale for the neutrinos.  Of course, the results of an oscillation experiment do provide a lower bound on the heavier mass in $\Dmq_{ij}$, $|m_i| \geq \sqrt{\Dmq_{ij}}$ for $\Dmq_{ij} > 0$, but there is no upper bound on this mass. In particular, the corresponding neutrinos could be approximately degenerate at a mass scale that is much higher than $\sqrt{\Dmq_{ij}}$.  Moreover, there is neither an upper nor a lower bound on the lighter mass $m_j$.

Information on the absolute neutrino masses, rather than mass differences, can be extracted from kinematic studies of reactions in which a neutrino or an anti-neutrino is involved. In the presence of mixing the most relevant constraint comes from the study of the end point ($E \sim E_0$) of the electron spectrum in tritium beta decay $\Nuc{3}{H} \to \Nuc{3}{He} + e^- + \bar\nu_e$. This spectrum can be effectively described by a single parameter $m_{\nu_e}$~\cite{Shro80}, if for all neutrino states $E_0 - E \gg m_i$. In this case:
\begin{equation}
  \frac{dN}{dE}
  \simeq R(E) \sum_i |U_{ei}|^2
  \sqrt{(E_0 - E)^2 - m_{\nu_e}^2} \,,
\end{equation}
where $R(E)$ contains all the $m_\nu$-independent factors, and
\begin{equation}
  \label{eq:mbeta}
  m^2_{\nu_e} = \frac{\sum_i m^2_i |U_{ei}|^2}{\sum_i |U_{ei}|^2}
  = \sum_i m^2_i |U_{ei}|^2
  = c_{13}^2 c_{12}^2 m_1^2
  + c_{13}^2 s_{12}^2 m_2^2+s_{13}^2 m_3^2 \,,
\end{equation}
where the second equality holds if unitarity is assumed.  At present we only have a bound $m_{\nu_e} \leq 2.2$~eV at 95\% CL~\cite{Kraus:2004zw,Aseev:2011dq} which is expected to be superseded soon by KATRIN~\cite{Angrik:2005ep} with about one order of magnitude improvement in sensitivity.

Direct information on neutrino masses can also be obtained from neutrinoless double beta decay $(A,Z) \to (A,Z+2) + e^- + e^-$. This process violates lepton number by two units. Hence, in order to induce $0\nu\beta\beta$ decay, $\nu$'s must be Majorana particles~\cite{Schechter:1981bd,Duerr:2011zd}. In particular, for the case in which the only effective lepton number violation at low energies is induced by the Majorana mass term for the neutrinos, the rate of $0\nu\beta\beta$ decay is proportional to the square of the \emph{effective Majorana mass of $\nu_e$}~\cite{Lindner:2005kr,Merle:2006du}:
\begin{equation}
  m_{ee}
  = \Big| \sum_i m_i U_{ei}^2 \Big|
  = \Big| m_1 c_{13}^2 c_{12}^2 e^{i 2\alpha_1} +
  m_2 c_{13}^2 s_{12}^2 e^{i 2\alpha_2} +
  m_3 s_{13}^2 e^{-i 2\delta} \Big|
\end{equation}
which, unlike eq.~\eqref{eq:mbeta}, also depends on all three CP violating phases.

Recent searches carried out with \Nuc{76}{Ge} (GERDA experiment~\cite{gerda}) and \Nuc{136}{Xe} (KamLAND-Zen~\cite{kzen} and EXO-200~\cite{exo} experiments) have established the lifetime of this decay to be longer than $10^{25}$ yr, corresponding to a limit on the neutrino mass of $m_{ee} \leq 0.2-0.4$ eV at 90\% C.L.  A series of new experiments is planned with sensitivity of up to $m_{ee} \sim 0.01$~eV~\cite{bbothers}.

Neutrino masses have also interesting cosmological effects and, as we will see in more detail in Section~2, cosmological data mostly gives information on the sum of the neutrino masses $\Sigma \equiv \sum_i m_i$. However, in particular when trying to derive conclusions on neutrino parameters, it is very important to have all systematics under control~\cite{Maneschg:2008sf}.

Correlated information on these three probes of the neutrino mass scale can be obtained by mapping the results from the global analysis of oscillations presented previously. We show in fig.~\ref{fig:mbeta} the present status of this exercise.  The relatively large width of the regions in the right panel are due to the unknown Majorana phases. Thus from a positive determination of two of these probes information can be obtained on the value of the Majorana phases and/or the mass ordering.

\endgroup 

%% file: kevnuwp_section1-3.tex
\subsection{Open questions in Neutrino Physics (Author: A.~de Gouv\^ea)}
\label{sub:Andre}

The discovery of non-zero neutrino masses and mixing in the lepton sector invites several fundamental particle physics questions. Searches for the answers to these open questions are among the key driving forces behind current and future research in neutrino physics, both experimentally and theoretically.

Some of these questions are related to the neutrino masses themselves. First and foremost, we do not know the exact values of the neutrino masses. While the oscillation data measure, sometimes very precisely, the neutrino mass-squared differences, they are not at all sensitive to the values of the masses themselves. As mentioned above, cosmological surveys and precision measurements of the $\beta$ spectrum of weak nuclear decays place complementary upper bounds on different combinations of the neutrino masses. Some information on the neutrino masses can also be extracted from searches for neutrinoless double beta decay if one assumes that neutrinos are Majorana fermions. 

The neutrino mass ordering -- inverted or normal -- is also unknown, as discussed earlier. This is a question that can, and very likely will, be addressed by neutrino oscillation experiments, including the current generation of long-baseline accelerator-based experiments, NO$\nu$A~\cite{Patterson:2012zs} and T2K~\cite{Abe:2011ks}, along with near-future measurements of the atmospheric neutrino flux (see, for example, Ref.~\cite{Aartsen:2013aaa}). Very ambitious long-baseline reactor antineutrino experiments, like JUNO~\cite{He:2014zwa}, may also be able to determine the neutrino mass ordering by measuring very precisely $\Delta m^2_{31}$ and $\Delta m^2_{32}$ (see, for example, Refs.~\cite{deGouvea:2005hk,Nunokawa:2005nx}). 

The answers to the two questions above will reveal not only the values and ordering of the masses, but they will also provide crucial information regarding the dynamical mechanism behind the neutrino masses. Current data do not allow one to distinguish scenarios where the lightest neutrino mass is zero from scenarios where all three neutrino masses are almost degenerate, $|m_i-m_j|\ll m_i,m_j$. None of the electrically charged fermions -- charged leptons and quarks -- have almost degenerate masses. Note also that, if the neutrino mass ordering is inverted, $m_1$ and $m_2$ are guaranteed to be almost degenerate even for that case where the lightest neutrino mass, $m_3$, is very small.

The fact that the neutrinos are massive and neutral invites another question: are neutrinos Majorana fermions? While all known fundamental fermions are massive Dirac fermions, i.e., the particles and antiparticles are distinct objects and all contain four degrees of freedom, the massive neutrinos might be Majorana fermions, i.e., the particle and antiparticle are related and containing only two degrees of freedom. The answer to this question is intimately tied to whether or not baryon number minus lepton number, $B-L$,  is an exact symmetry. Majorana neutrinos imply that $B-L$ is explicitly broken and one should be able to observe that experimentally (see, for example, Refs.~\cite{Avignone:2007fu,deGouvea:2013zba}, for recent overviews). The most precise probes of $B-L$ violation are searches for neutrinoless double beta decay, which are currently being pursued in earnest. In the next ten years we hope to learn much more about the nature of the neutrino, especially if the neutrino mass ordering turns out to be inverted~\cite{Lindner:2005kr}. 

Neutrino oscillations allow one to explore another important fundamental question: is there CP violation in the leptonic sector? More concretely, we hope to address whether neutrinos and antineutrinos oscillate differently. While the survival probabilities for neutrinos and antineutrinos are guaranteed to be the same, the transition probabilities need not be. If $P(\nu_{\alpha}\to\nu_{\beta})\neq P(\bar{\nu}_{\alpha}\to\bar{\nu}_{\beta})$, $\alpha,\beta=e,\mu,\tau$, $\alpha\neq\beta$, CP is violated. In the three-flavor paradigm discussed above, CP violation is governed by the CP-odd parameter $\delta$. If $\delta\neq 0,\pi$, CP-invariance is violated in neutrino oscillations. Given what is currently known about the neutrino oscillation parameters, it is clear that the current generation of neutrino experiments is not very sensitive to potential CP-invariance violation in neutrino oscillations. Next-generation experiments, especially the long-baseline proposals LBNF/DUNE (see, for example, Ref.~\cite{Adams:2013qkq}) and T2HK~\cite{Kearns:2013lea}, are designed to significantly challenge CP-invariance in the leptonic sector. We currently understand very little about the phenomenon of CP violation. We do know that it is necessary in order to dynamically generate a matter--antimatter asymmetry in the early Universe \cite{Sakharov:1967dj} -- a mechanism dubbed baryogenesis, see e.g. \cite{Canetti:2012zc} for a general discussion. Since we know that the CP violation observed in the quark sector is insufficient to explain the baryon-asymmetry, the community is eagerly awaiting information on leptonic CP violation so that it can make progress on understanding baryogenesis. 

Precision neutrino experiments, especially precision neutrino oscillation experiments, allow one to ask other fundamental physics questions. For example, are there more than three neutrinos? While we know that there are three families of SM fermions, including the three active neutrinos $\nu_e, \nu_{\mu}, \nu_{\tau}$, there could be more neutral fermionic states. Collider data, especially those from LEP, restrict the number of active neutrinos to three~\cite{Agashe:2014kda}. New fermions that do not couple to the $W$- and $Z$-bosons with SM strength -- usually referred to as sterile neutrinos -- are not severely constrained and their existence might be revealed via neutrino oscillation experiments. While these sterile neutrinos do not necessarily couple to SM particles, they mix with the active neutrinos in such a way that neutrino oscillation experiments might depend on more mixing parameters and more oscillation frequencies. If the new neutrinos are light enough, they can only be probed experimentally in neutrino oscillation experiments. There are many current neutrino oscillation experiments dedicated to the search for sterile neutrinos. 

Neutrino oscillation experiments are also sensitive to new ``weaker-than-weak'' neutrino interactions with matter. These modify neutrinos production and detection and, often more significantly, they lead to non-standard matter effects. Long-baseline experiments, especially NO$\nu$A~\cite{Patterson:2012zs} and LBNF/DUNE (see, for example, Ref.~\cite{Adams:2013qkq}), are particularly sensitive to non-standard neutrino interactions. 

Finally, neutrino oscillations are very special phenomena. They are a consequence of quantum mechanical interference, and the associated coherence lengths are of the order of several kilometers (or significantly more). This provides a unique sensitivity to more exotic physics, including the violation of Lorentz invariance and tests of the CPT-theorem, or departures from the basic laws of quantum mechanics. 

%% file: kevnuwp_section1-4.tex
\subsection{\label{sub:Paul}Sterile Neutrinos -- General Introduction (Author: P.~Langacker)}

Sterile neutrinos, also known as singlet or right-handed neutrinos, are $SU(2) \times U(1)$-singlet leptons. They therefore have no ordinary  charged or neutral current weak interactions except those induced by mixing. Most extensions of the original standard model involve one or more sterile neutrinos, with model-dependent masses which can vary from zero to extremely large. One usually defines the right-chiral component of a sterile neutrino field as $\nu_R$, i.e., $\nu_R$ annihilates a right-chiral state, where chirality coincides with helicity in the massless limit. The CP-conjugate field is then
\begin{equation}
(\nu_R)^c \equiv C\, \overline{\nu_R}^T,
\label{eqnucr}
\end{equation}
where we are following the notation in ref.~\cite{Drewes:2013gca}. In Eq.~(\ref{eqnucr}), $C$  is the charge conjugation matrix,  given by $=i \gamma_2 \gamma_0$ in the Weyl representation, and $\overline{\nu_R}\equiv(\nu_R)^\dag \gamma^0$ is the Dirac adjoint. Note that the CP conjugate in Eq.~(\ref{eqnucr}) is always well-defined, independent of whether CP is violated, and that $(\nu_R)^c$ is the field which annihilates a \emph{left}-chiral antineutrino.\footnote{Some authors use  alternative notations, such as $\nu^c_{R,L} $ for $ C\, \overline{\nu_{R,L}}^T$.}

In contrast, an active (or doublet or ordinary) neutrino is in an $SU(2)$ doublet with a charged lepton, and it has conventional weak interactions. There are three known left-chiral active neutrinos $\nu_{L,\alpha}$, where the flavor index $\alpha=e, \mu, \tau$ denotes the associated charged lepton. The CP-conjugate  $(\nu_L)^c \equiv  C\, \overline{\nu_L}^T$ (suppressing the flavor index)  is the field associated with a right-chiral antineutrino.
The number $n$ of right chiral neutrinos is unknown (and could even be zero, as there are alternative explanations of neutrino masses, see section \ref{ModelZeugByMarco}).
In the remainder of this subsections we use an illustrative toy model with only one LH and one RH neutrino flavour.

\begin{figure}\centering
\hfill
\includegraphics[height=4cm]{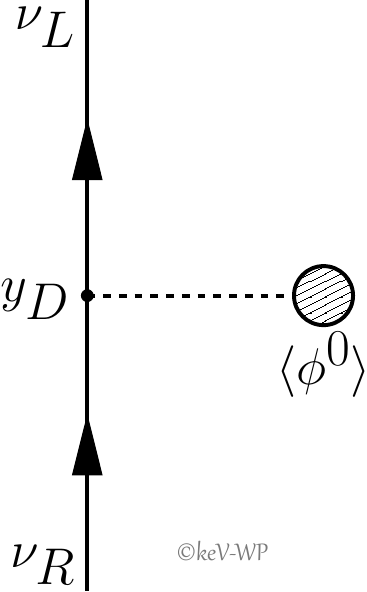}
\hfill
\includegraphics[height=4cm]{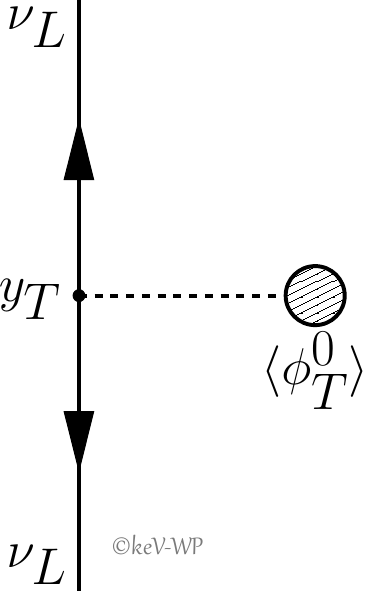}
\hfill
\includegraphics[height=4cm]{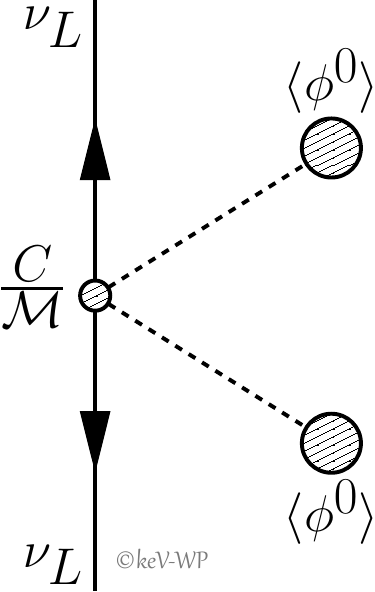}
\hfill
\null
\caption{\label{figmasses}Left: Dirac mass from the vacuum expectation value of the neutral component $\phi^0$ of a Higgs doublet. $y_D$ is a Yukawa coupling. Center: a Majorana mass  for an active neutrino due to a Higgs triplet $\phi^0_T$ with the Yukawa coupling $y_T$. An analogous diagram can generate a Majorana mass for a sterile neutrino, with $\phi^0_T$ replaced by a Higgs singlet (or bare mass). Right: Majorana mass for $\nu_L$ generated by a higher-dimensional operator involving two Higgs doublets. (Figure similar to fig.~1 in Ref.~\cite{Langacker:2011bi}.)}
\end{figure}

$\nu_L \leftrightarrow (\nu_L)^c$ and $\nu_R \leftrightarrow (\nu_R)^c$ each describe two degrees of freedom and are known as Weyl spinors. Fermion mass terms  describe transitions between left and right-chiral states. There are two possible types for neutrinos. A Dirac mass term connects the left and right components of two different Weyl spinors. These are typically active and sterile, such as
\begin{equation}
\mathcal{L}_D=- m_D \left( \overline{\nu_L} \nu_R +  \overline{\nu_R} \nu_L \right),
\label{eqnuDirac}
\end{equation}
where we have chosen the phases of the fields so that $m_D$ is real. $\mathcal{L}_D$ allows a conserved lepton number $L$, but violates weak isospin by $1/2$ unit. It can  be generated by the Higgs mechanism, as in fig.~\ref{figmasses}, and it is analogous to the quark and charged lepton masses. That is, $m_D = y_D  v$,\footnote{In the minimal seesaw model, (\ref{Lseesaw}), the number $y_D$ is to be identified with the matrix $F$. } 
where $v=174$ GeV is the expectation value of the neutral Higgs field. If eq.~\eqref{eqnuDirac} is the only neutrino mass term, then $\nu_L$ and $\nu_R$ can be combined to form a four-component Dirac spinor $\nu_D \equiv \nu_L+\nu_R$, with CP conjugate $(\nu_D)^c \equiv (\nu_L)^c + (\nu_R)^c$.

Unlike quarks and charged leptons, neutrinos are not charged under any unbroken gauge symmetries. They may therefore have Majorana mass terms, which connect a Weyl spinor with its own CP conjugate. These could be present for either active or sterile neutrinos,
\begin{equation}
\mathcal{L}_M=- \frac{1}{2} m_T \left(  \overline{\nu_L} \nu^c_L +  \overline{\nu^c_L} \nu_L \right) 
- \frac{1}{2} m_S \left(  \overline{\nu_R} \nu^c_R +  \overline{\nu^c_R} \nu_R \right)
\equiv- \frac{1}{2} m_T  \left(  \overline{\nu_a} \nu_a \right) - \frac{1}{2} m_S  \left(  \overline{\nu_s} \nu_s \right) ,
\label{eqnuMajorana}
\end{equation}
where $\nu_a\equiv \nu_L + (\nu_L)^c$ and $\nu_s\equiv \nu_R+ (\nu_R)^c$ are  active ($a$) and sterile ($s$) Majorana two-component spinors. They are  self-conjugate, i.e., $\nu_a = C\, \overline{\nu_a}^T$ and $\nu_s = C\, \overline{\nu_s}^T$. Both $m_T$ and $m_s$ violate lepton number by two units. The mass $m_T$ also violates weak isospin by one unit. It can be generated by the expectation value of a Higgs triplet $\phi_T=(\phi^0_T,\, \phi^-_T,\, \phi^{--}_T)^T$, i.e., $m_T = y_T\, \langle \phi^0_T \rangle$, where $y_T$ is a Yukawa coupling, as illustrated in fig.~\ref{figmasses}. It can also be due to a higher-dimensional operator involving two Higgs doublets with coefficient $C/\mathcal{M}$, with $C$ a dimensionless constant and $\mathcal{M}$ a new physics scale. For mass dimension 5, the only operator of this kind is given by (\ref{Leff}).
The singlet mass $m_S$ does not violate weak isospin and could in principle be due to a bare mass. However, in many extensions of the standard model a bare mass is forbidden by new broken symmetries, so that $m_S = y_S\,  \langle \phi_S \rangle$, where $y_S$ is a Yukawa coupling and $\phi_S$ is a standard model singlet.
 
Dirac and Majorana mass terms\footnote{Off-diagonal mass terms are generally defined as Majorana if the mass eigenstates are.} can be present simultaneously and can be generalized to three or more active neutrinos, to an arbitrary number of sterile neutrinos, and to non-standard assignments such as additional sterile neutrinos in which the \emph{left}-chiral component carries lepton number $L=+1$.

Sterile neutrino masses can have almost any value, but here we mention a number of important special cases. For simplicity we focus on $m_T = 0$ except when it induced by the seesaw mechanism, but generalization is straightforward. We also take $m_D$ and $m_S$ real and nonnegative.

\begin{itemize}
\item { $m_S=0$:} This is the pure Dirac case. $\nu_L$ and $\nu_R$ are the left and right-chiral components of a Dirac neutrino with conserved lepton number $L$. $\nu_R$ has no standard model interactions except for  the extremely weak Higgs-Yukawa coupling $y_D \sim 3 \times 10^{-13}\, m_D/(0.05\ \text{eV})$. This small value could occur via fine-tuning, or more likely because it is strongly suppressed by new symmetries or extra-dimensional effects, see e.g.\ ref.~\cite{Langacker:2011bi}.

\item { $m_D=0$:} This is the pure Majorana case. The sterile Majorana neutrino can have arbitrary mass, but it has no standard model interactions, and there is no active-sterile mixing.

\item { $m_S \ll m_D$:} This pseudo-Dirac limit is a perturbation on the Dirac case. The four components of the Dirac neutrino split into two Majorana neutrinos, with masses $|m_{1,2}| \sim m_D \pm m_S/2$ and left-chiral components $\sim [\nu_L \pm (\nu_R)^c]/\sqrt{2}$. The non-observation of solar oscillations into the sterile state requires $m_S < \mathcal{O}(10^{-9}\ \text{eV})$, see ref.~\cite{deGouvea:2009fp}.

\item { $m_S \gg m_D$ with $ m_T=0$:} This is the (minimal type-I) seesaw limit~\cite{Minkowski:1977sc,Yanagida:1979as,GellMann:1980vs,Glashow:1979nm,Mohapatra:1979ia,Schechter:1980gr}. There are two Majorana mass eigenstates, with one light mainly-active neutrino  and a heavier mainly-sterile state, with respective masses:
\begin{equation}
|m_1| \sim \frac{m_D^2}{m_S}, \qquad m_2 \sim m_S,
\label{eqnseesaw1}
\end{equation}
and the active-sterile mixing angle
\begin{equation}
 |\theta| \sim \frac{m_D}{m_S} \sim \sqrt{\frac{m_1}{m_2}}\sim 3 \times 10^{-3} \,\left(  \frac{m_1}{0.05\ \text{eV}} \right)^{1/2} \, \left( \frac{6\ \text{keV}}{m_S} \right)^{1/2}  .
\label{eqnseesaw2}
\end{equation}
The Higgs Yukawa coupling required is 
\begin{equation}
y_D = \frac{\sqrt{m_1 m_S}}{v} \sim 10^{-10} \, \left( \frac{m_1}{0.05\ \text{eV}} \right) \left( \frac{m_S}{6\ \text{keV}} \right).
\label{eqnseesaw3}
\end{equation}
The mass $m_S$ is often directly associated with a new physics scale, especially for $m_S \gg v$. In that limit, the sterile state is often integrated out, so that $m_1$ plays the role of $m_T$ in the low-energy effective theory. These results are generalized to three families in section \ref{sec:SeeSawwMechanism}. An important phenomenological difference in the case of several families is that there may be cancellations in the neutrino mass matrix that allow individual entries of the mixing matrix $\theta$ to be much bigger than the estimate (\ref{eqnseesaw2}).
This is in particular the case in models with an approximate $B-L$ symmetry.

We now mention several special cases:
 
\begin{itemize}
\item { $m_S = \mathcal{O}(\text{eV})$:} Oscillations of active  into eV-scale sterile neutrinos have been suggested by the LSND and MiniBooNE $\bar\nu_\mu \rightarrow \bar \nu_e$ (and to a lesser extent $\nu_\mu \rightarrow  \nu_e$) results, by possible $\bar\nu_e$ disappearance in short-baseline reactor experiments, and by  radioactive source calibrations carried out by the SAGE and GALLEX experiments, e.g., ref.~\cite{Abazajian:2012ys,Conrad:2013mka,Kopp:2013vaa}. There is, however, tension with a number of other  $\nu_\mu$, $\bar\nu_\mu$, $\nu_e$,  and $\bar \nu_e$ disappearance or appearance results in longer-baseline reactor and accelerator experiments and in atmospheric and solar neutrinos, especially for a single eV-scale  sterile state, ref.~\cite{Kopp:2013vaa,Giunti:2013aea,Conrad:2012qt}. There was some suggestion of additional radiation in the early Universe from CMB experiments, e.g., ref.~\cite{Ade:2013zuv}, but the number density of even one eV-scale neutrino with large mixing would have to be suppressed compared to the expectations of an oscillation-induced thermalized species~\cite{Mirizzi:2013kva}.

The LSND and other positive results present a challenge to theory, because they would require mixing between active and sterile states with the same chirality. This does not occur for the pure Dirac or pure Majorana cases,  or in the conventional seesaw with large $m_S$. They would require that both Dirac and Majorana mass terms are tiny.\footnote{More generally, two distinct types of mass terms, such as active-sterile Dirac, and sterile-sterile Dirac in the presence of extra sterile states, would be necessary.} A promising possibility is the minimal mini-seesaw ($m_T = 0$ and $m_D \ll m_S \sim \mathcal{O}(\text{eV})$, e.g., ref.~\cite{deGouvea:2011zz,Donini:2012tt}), which  yields a reasonable relation between mixing angles and masses, as seen in eq.~\eqref{eqnseesaw2}. The tiny values for the masses would suggest that they are suppressed by some new symmetry, ref.~\cite{Langacker:1998ut,Sayre:2005yh}.

Sterile neutrinos in the eV range (and higher,  up to kinematic limits) could lead to observable kinks or lines in $\beta$ decay and other weak decay spectra, ref.~\cite{Atre:2009rg,Kusenko:2009up,Shro80,Shrock:1980ct,Shrock:1981wq}. Both eV-scale, ref.~\cite{Barry:2011wb}, and heavier sterile neutrinos up to the TeV range, ref.~\cite{Faessler:2014kka}, could also significantly affect neutrinoless double $\beta$ decay~\cite{Escrihuela:2015wra}.

\item { $m_S = \mathcal{O}(\text{keV})$:} The keV range is especially interesting because the sterile neutrino would be a viable DM candidate. It could be warm or cold depending on the cosmological production mechanisms, e.g., produced  by oscillations between active and sterile neutrinos, ref.~\cite{Dodelson:1993je}. In particular, warm  or moderately cold dark matter could describe the low-scale structure observations better than cold dark matter candidates such as WIMPs or axions (see section~\ref{sec:DMGalactic} and e.g.~\cite{deVega:2009ku, deVega:2010yk, deVega:2011gg, deVega:2011gs, Destri:2012yn, Destri:2013pt, deVega:2013jfy, Destri:2013hha}). The keV neutrinos would not be absolutely stable, but could decay by mixing into a photon and light neutrino with a cosmologically-long lifetime. There are stringent astrophysical and cosmological constraints on sterile neutrinos in the keV and other mass ranges, as reviewed in refs.~\cite{Kusenko:2009up,Boyarsky:2009ix,Merle:2013gea}.  The observational limits on X-ray emission by the radiative decays are especially stringent, but  they are consistent with sterile DM\footnote{The relevant mixing angle, $|\theta| = \mathcal{O}( 10^{-5}-10^{-4})$ would lead to a very small $m_1$ using the seesaw formula eq.~\eqref{eqnseesaw2}, which however would still be applicable~\cite{Merle:2012xq}. The observed solar and atmospheric scales could, however, be associated with heavier steriles, as in the $\nu$MSM~\cite{Asaka:2005pn,Asaka:2005an}.} for $m_S=\mathcal{O}(\text{keV})$. There is some tentative evidence for a  3.5~keV X-ray line, which could possibly be due to a decaying sterile neutrino, see refs.~\cite{Bulbul:2014sua,Boyarsky:2014jta}. A keV-scale sterile could also explain the observed pulsar velocities, ref.~\cite{Kusenko:2009up}. However, they are very difficult to detect in a lab, see e.g.\ refs.~\cite{Bezrukov:2005mx,Asaka:2011pb,Merle:2013ibc} for investigations of neutrinoless double beta decay and Sec.~8 for direct detection attempts.

The neutrino minimal standard model ($\nu$MSM)~\cite{Asaka:2005pn,Asaka:2005an} is a minimal extension of the SM, with no new physics other than three sterile neutrinos up to the Planck scale. One is at the keV scale, to account for DM. Two heavier sterile neutrinos can account for the observed light neutrino masses by a seesaw mechanism. If they are nearly degenerate and in the range 150 MeV--100 GeV they can account for the baryon asymmetry through oscillation-induced leptogenesis. General theoretical models for keV steriles are reviewed in ref.~\cite{Merle:2013gea}.

\item { $m_S = \mathcal{O}(\text{MeV--TeV})$:} There are a variety of constraints from astrophysics, ref.~\cite{Kusenko:2009up}, from weak decays (e.g., spectral lines or from lepton violation such as $\ell_2^+ \rightarrow \ell_1^- M^+ M^+$ or  $M_1^+ \rightarrow M_2^- \ell^+\ell^+$, where $\ell$ and $M$ represent  leptons and mesons, respectively), and from same-sign dilepton production at the LHC or Tevatron, e.g., ref~\cite{Atre:2009rg,Drewes:2015iva}. These results are typically sensitive to mixings larger than those suggested by eq.~\eqref{eqnseesaw2}.

\item { $m_S \gg$ TeV:} This is the canonical minimal type-I seesaw range, in which $y_D$ is comparable to the quark and charged-lepton Yukawas (i.e., $m_D \sim m_e$ for $m_S\sim$ several TeV, while  $m_D \sim m_t$ for $m_S\sim 10^{14}$ GeV). The only direct observational consequence other than the mixing-induced active neutrino masses is the possibility of leptogenesis, usually by  heavy sterile decay. There may also be model-dependent associated signatures, such as $\mu \rightarrow e \gamma$ by sneutrino exchange for  $m_S =  \mathcal{O}(\text{TeV})$ in supersymmetric models.  A very large $m_S$ may be associated with a grand unified theory, with implications for proton decay. For example, $m_S= 10^{14}$ GeV is about two orders of magnitude below the scale of supersymmetric gauge unification. 

\item{ $m_S \sim M_P$:} For completeness we mention the possibility that $m_S$ is comparable to the Planck scale, $M_P \sim 10^{19}$ GeV. This would imply $m_1 < 10^{-5}$ eV for $y_D\sim 1$.  This is the typical \emph{maximum} value of $m_1$ expected from gravitational mechanisms,\footnote{In this case, $m_S$ might correspond, e.g.,  to a string excitation rather than a particle.} unless there are large extra dimensions or multiple contributions.

\end{itemize}
\end{itemize}

%% file: kevnuwp_section1-5.tex
\subsection{\label{sub:seesaw}The seesaw mechanism (Author: M.~Drewes)}

\subsubsection{\label{ModelZeugByMarco}Possible origins of neutrino mass}  
Any model of neutrino masses should explain the fact that they are orders of magnitude smaller than any other fermion masses in the SM (``mass puzzle''). It is convenient to classify the most popular models according to the way how this mass hierarchy is realized. 
\begin{itemize}
\item 
\textbf{Small coupling constant}: 
If the light neutrino masses $m_i$ are generated via spontaneous symmetry breaking, then their smallness could simply be due to a tiny coupling constant. For instance, in the model (\ref{Lseesaw}), Dirac masses can be generated via the standard Higgs mechanism with $F\sim 10^{-12}$ and $M_M=0$.  
\item 
\textbf{Seesaw mechanism}: If the neutrino masses $m_i$ are generated at classical level, they may be suppressed by the new heavy scale $M$. The most studied version is the type-I seesaw~\cite{Minkowski:1977sc,Glashow:1979nm,GellMann:1980vs,Mohapatra:1979ia,Yanagida:1980xy,Schechter:1980gr} discussed below in~\ref{sec:SeeSawwMechanism}, the two other possibilities~\cite{Ma:1998dn} are the type-II~\cite{Schechter:1980gr,Magg:1980ut,Cheng:1980qt,Lazarides:1980nt,Mohapatra:1980yp} and type-III~\cite{Foot:1988aq} seesaw.
\item 
\textbf{Flavor (``horizontal'') symmetry}: Individual entries of the light neutrino mass matrix $m_\nu$ may be much larger than the $m_i$ if there is a symmetry that leads to approximate lepton number conservation and cancellations in $m_\nu m_\nu^\dagger$. Prominent realizations of this idea are Froggatt-Nielsen type models~\cite{Froggatt:1978nt} and models with approximate lepton number conservation, e.g.~\cite{Chikashige:1980ui,Gelmini:1980re,Wyler:1982dd,GonzalezGarcia:1988rw,Branco:1996bq,Abada:2007ux,Gavela:2009cd}.
\item 
\textbf{Radiative masses}: Neutrinos could be massless in the classical limit, with their masses being generated by quantum corrections. The suppression due to the ``loop factor'' $(4\pi)^2$ is usually not sufficient to explain the smallness of the $m_i$, but the new particles in the loop couple to $\nu_L$ with some small coupling constants that lead to an additional suppression. Flavor symmetries or an additional seesaw-like suppression can help to make such models more ``natural'', see e.g.~\cite{Zee:1980ai,Witten:1979nr,Zee:1985id,Babu:1988ki,Ma:2006km,FileviezPerez:2009ud,Khoze:2013oga,King:2014uha,Geib:2015tvt}.
\end{itemize}
Of course, any combination of these ideas may be realized in nature. In addition to the smallness of the neutrino masses, it is desirable to find an explanation for the observed mixing pattern of neutrinos (``flavor puzzle''). While the quark mass matrix is approximately diagonal in the weak interaction basis, there is no obvious symmetry or structure in $m_\nu$. Numerous attempts have been made to identify discrete or continuous symmetries, see e.g.~\cite{King:2014nza} and Sec.~6.\footnote{See Refs.~\cite{Appelquist2002204, 
PhysRevLett.90.201801, PhysRevD.69.015002} for some seesaw implementations not covered in Sec.~6.}. The basic problem is that the reservoir of possible symmetries is practically unlimited; for any possible observed pattern of neutrino masses and mixings one could find a symmetry that ``predicts'' it. In this situation, models can only be convincing if they either predict  observables that have not been measured, such as mass~\cite{Barry:2010yk,Dorame:2011eb,King:2013psa,King:2014nza,Agostini:2015dna} or mixing~\cite{Petcov:2004rk,Marzocca:2013cr,Ballett:2013wya,King:2014nza} sum rules, or are ``simple'' and aesthetically appealing from some viewpoint. Prior to the measurement of $\theta_{13}$, models predicting $\theta_{13}=0$ seemed well-motivated, such as those leading to tri/bi-maximal mixing~\cite{Harrison:1999cf,Harrison:2002er,Ma:2004zv,Altarelli:2005yx,Xing:2002sw}. The observed $\theta_{13}\neq0$, however, requires an additional symmetry-breaking sector in these models. That makes it difficult to explain $m_\nu$ in terms of a simple symmetry and a small number of parameters, and interest in anarchic models~\cite{Hall:1999sn} with random values has grown. A more detailed overview of models that incorporate sterile neutrinos with keV masses that could act as DM candidates is given in section~6. In the following we briefly review the main features of the probably minimal and most studied mechanism for neutrino masses, the (type I) seesaw mechanism. Is particularly important in the present context because $i)$ it predicts the existence of heavy sterile neutrinos, $ii)$ these heavy particles mix with ordinary neutrinos (which is the basis for many experimental and astrophysical searches) ans $iii)$ the type I seesaw is implemented in many theories of particle physics, such as grand unified theories based on $SO(10)$ or any other theories that involve a (spontaneously broken) gauged $B-L$ symmetry.

\subsubsection{The seesaw mechanism}\label{sec:SeeSawwMechanism}
The type-I seesaw model is defined by adding $n$ RH neutrinos $\nu_R$ to the SM, i.e.,\ singlet fermions with RH chirality to the SM that couple to the SM neutrinos $\nu_L$ in the same way as the RH and LH components of the charged leptons. The Lagrangian reads  
\begin{eqnarray}
	\label{Lseesaw}
	\mathcal{L} &=&\mathcal{L}_{\rm SM}+ 
	i \overline{\nu_{R}}\slashed{\partial}\nu_{R}-
	\overline{\ell_{L}}F\nu_{R}\tilde{\Phi} -
	\tilde{\Phi}^{\dagger}\overline{\nu_{R}}F^{\dagger}\ell_L 
-{\rm \frac{1}{2}}(\overline{\nu_R^c}M_{M}\nu_{R} 
	+\overline{\nu_{R}}M_{M}^{\dagger}\nu^c_{R}). 
	\end{eqnarray}
Here, flavor and isospin indices have been suppressed. $\mathcal{L}_{\rm SM}$ is the Lagrangian of the SM, $\ell_{L}=(\nu_{L},e_{L})^{T}$ are the LH lepton doublets, $\Phi$ is the Higgs doublet and $\tilde{\Phi}=\epsilon\Phi^*$, where $\epsilon$ is the antisymmetric $SU(2)$-invariant tensor, and $F$ is a matrix of Yukawa interactions. An explicit Majorana mass term $M_{M}$ is allowed for $\nu_R$ because the $\nu_R$ are gauge singlets. This is a specific realization of the term $m_S$ in~\eqref{eqnuMajorana}. It is often assumed that the eigenvalues $M_I$ of $M_M$ are far above the electroweak scale. Then the $\nu_R$ are experimentally unobservable. The only effect they have on low energy physics is mediated by the dimension-5 operator~\cite{Weinberg:1979sa}:
\begin{eqnarray}
\label{Leff}
	\mathcal{L}_{\rm eff} &=&\mathcal{L}_{\rm SM}+ \frac{1}{2}\bar{\ell_{L}}\tilde{\Phi} F M_M^{-1}F^T \tilde{\Phi}^{T}\ell_{L}^{c},
	\end{eqnarray}
as obtained by integrating out the fields $\nu_R$ instead of (\ref{Lseesaw}). The Higgs mechanism generates a Majorana mass term $\overline{\nu_L}m_\nu  \nu_L^c$, with $m_\nu$ is given by 
\begin{eqnarray}
m_\nu= -v^2 F M_M^{-1}F^T,\label{preactivemass} 
\end{eqnarray}
where $v=174$ GeV is the Higgs field expectation value. This case is not interesting in the context of this review because superheavy $\nu_R$ are too short-lived to be DM candidates. However, experimentally the magnitude of the $M_I$ is almost unconstrained, and different choices have very different implications for particle physics, cosmology and astrophysics, see e.g.~\cite{Abazajian:2012ys,Drewes:2013gca}. There are several scenarios that predict eigenvalues of $M_M$ at or below the electroweak scale, including the inverse seesaw~\cite{Mohapatra:1986bd} and linear seesaw~\cite{Malinsky:2005bi}, the $\nu$MSM~\cite{Shaposhnikov:2006nn,Araki:2011zg}, low-scale seesaw (e.g.~\cite{Appelquist2002204,PhysRevLett.90.201801,PhysRevD.69.015002}), or Coleman-Weinberg type models~\cite{Khoze:2013oga}, see also section~6.
The full neutrino mass term after electroweak symmetry breaking reads
\begin{eqnarray}\label{neutrinomassfull}
\frac{1}{2}
(\overline{\nu_L} \  \overline{\nu_R^c})
\mathfrak{M}
\left(
\begin{tabular}{c}
$\nu_L^c$\\
$\nu_R$
\end{tabular}
\right) + h.c.
\equiv
\frac{1}{2}
(\overline{\nu_L} \ \overline{\nu_R^c})
\left(
\begin{tabular}{c c}
$0$ & $m_D$\\
$m_D^T$ & $M_M$
\end{tabular}
\right)
\left(
\begin{tabular}{c}
$\nu_L^c$\\
$\nu_R$
\end{tabular}
\right) + h.c. ,
\end{eqnarray}
where $m_D\equiv Fv$ and $v=174$ is the Higgs field vacuum expectation value. For $M_I\gg 1$ eV there is a hierarchy $m_D\ll M_M$, and one observes two distinct sets of mass eigenstates: one set of ``active'' light neutrinos that are mostly $SU(2)$ doublets and one set of ``sterile'' heavy neutrinos that are mostly gauge singlets. Mixing between the active and sterile neutrinos is suppressed by elements of the mixing matrix
\begin{equation}
\theta\equiv m_D M_M^{-1}.
\end{equation}  
This allows to rewrite (\ref{preactivemass}) as
\begin{eqnarray}
m_\nu=  -v^2 F M_M^{-1}F^T = -m_D M_M^{-1} m_D^T=-\theta M_M \theta^T. \label{activemass} 
\end{eqnarray}
The fact that heavier $M_I$ imply lighter $m_i$ motivates the name ``seesaw mechanism''. An important implication of this relation is that one RH neutrino with non-vanishing mixing is needed for each non-zero light neutrino mass. That is, if the minimal seesaw mechanism is the only source of light neutrino masses, there must be at least $n=2$ RH neutrinos because two mass splittings (``solar'' and ``atmospheric'') have been observed. If the lightest neutrino is massive, i.e.\ $m_{\rm lightest}\equiv{\rm min}(m_1,m_2,m_3)\neq 0$, then $n\geq3$ is required. A heavy neutrino that is a DM candidate would not count in this context - to ensure its longevity, the three mixing angles $\theta_{\alpha I}$ must be so tiny that the effect of the DM candidate on light neutrino masses is negligible. This has an interesting consequence for the scenario with $n=3$, in which the number of sterile flavors equals that of active flavors: If one of the heavy neutrinos composes the DM, then the lightest neutrino is effectively massless ($m_{\rm lightest}\simeq 0$). If, on the other hand, the absolute neutrino mass scale $m_{\rm lightest}$ is larger than $10^{-3}$ eV, then all $M_I$ are expected to be larger than 100 MeV due to cosmological constraints~\cite{Hernandez:2014fha}. These conclusions can of course be avoided for $n>3$ or if there is another source of neutrino mass. The matrix
\begin{equation}
\mathcal{U}=
\left[
\left(
\begin{tabular}{c c}
$1-\frac{1}{2}\theta\theta^\dagger$ & $\theta$\\
$-\theta^\dagger$ & $1-\frac{1}{2}\theta^\dagger\theta$
\end{tabular}
\right)  \ + \ \mathcal{O}[\theta^3]
\right]
\left(
\begin{tabular}{c c}
$U_\nu$ & $ $\\
$ $ & $U_N^*$
\end{tabular}
\right) 
.\end{equation}
diagonalizes the full $6\times 6$ neutrino mass matrix $\mathfrak{M}$ to second order in the small expansion parameters $|\theta_{\alpha I}|$:
\begin{eqnarray}\label{diagonalisationofmathfrakM}
\mathcal{U}^\dagger \mathfrak{M}\mathcal{U}^*=
\left(
\begin{tabular}{c c}
$U_\nu^\dagger m_\nu U_\nu$ & $ $\\
$ $ & $U_N^T M_N U_N^*$
\end{tabular}
\right).
\end{eqnarray} 
The unitary matrices $U_\nu$ and $U_N$ diagonalize the mass matrices $m_\nu$ defined in~\eqref{activemass} and $M_N\equiv M_M + \frac{1}{2}\big(\theta^{\dagger} \theta M_M + M_M^T \theta^T \theta^{*}\big)$,  
\begin{eqnarray}
\ m_\nu=U_\nu m_\nu^{\rm diag} U_\nu^T \ &,& \ M_N = U_N^* M_N^{\rm diag} U_N^\dagger, \\
 M_N^{\rm diag} = {\rm diag}(M_1,M_2,M_3) \ &,& \ m_\nu^{\rm diag}={\rm diag}(m_1,m_2,m_3).
\end{eqnarray}
The eigenvalues of $M_M$ and $M_N$ coincide in very good approximation; we do not distinguish them in what follows and use the notation $M_I$ for both. The free (quadratic) part of the neutrino Lagrangian reads $\frac{1}{2}\bar{N}(i\slashed{\partial}-M_N^{\rm diag})N+\frac{1}{2}\bar{\upnu}(i\slashed{\partial}-m_\nu^{\rm diag})\upnu$. These relations hold at tree level. Phenomenological implications of quantum corrections are e.g. \ discussed in~\cite{Drewes:2015iva,Fernandez-Martinez:2015hxa}, but are not relevant if the sterile neutrinos are DM because $\theta_{\alpha I}$ are too small. The mass terms $m_\nu$ and $M_N$ are not exactly identical to the terms $m_T$ and $m_S$ in (\ref{eqnuMajorana}) because the latter are mass terms for the chiral fields $\nu_L$ and $\nu_R$, while the former involve the mass eigenstates $\upnu$ and $N$. These can only be identified with each other if there is no Dirac mass term (\ref{eqnuDirac}), which in the minimal seesaw model (\ref{Lseesaw}) means $F=\Theta=0$ and massless active neutrinos. In other models, however, $m_T$ can be generated without a Dirac mass term and mixing with $\nu_R$.

All six mass eigenstates are Majorana fermions and can be represented by the elements of the flavor vectors
\begin{eqnarray}\label{LightMassEigenstates}
\upnu=V_\nu^{\dagger}\nu_L-U_\nu^{\dagger}\theta\nu_{R}^c +{\rm c.c.} \ {\rm and} \ 
N=V_N^\dagger\nu_R+\Theta^{T}\nu_{L}^{c} +{\rm c.c.}
\end{eqnarray}
Here, ``c.c.'' stands for the charge conjugation defined in eq.~\eqref{eqnucr}. The observed light mass eigenstates $\upnu_i$ are related to the active flavor eigenstates by the matrix
\begin{equation}\label{VnuDef}
V_\nu\equiv (1-\frac{1}{2}\theta\theta^{\dagger})U_\nu. 
\end{equation}
$V_\nu$ is the usual light neutrino neutrino mixing matrix and $U_\nu$ its unitary part, $V_N$ and $U_N$ are their equivalents in the sterile sector. Unfortunately, unitarity violation due to heavy neutrinos does not appear to be able to resolve the long-standing issues of short-baseline anomalies~\cite{Kopp:2013vaa}. The active-sterile mixing is determined by the matrix
\begin{equation}
\Theta\equiv\theta U_N^*.
\end{equation}
In summary, one can express the interaction eigenstates (singlets $\nu_R$ and doublet component $\nu_L$) as
\begin{eqnarray}\label{PMNSfull}
\left(
\begin{tabular}{c}
$\nu_L$\\
$\nu_R^c$
\end{tabular}
\right)
=
P_L\mathcal{U}
\left(
\begin{tabular}{c}
$\upnu$\\
$N$
\end{tabular}
\right).
\end{eqnarray}
The interactions of neutrinos in the SM are described by the Lagrangian term
\begin{equation}\label{WeakWW}
-\frac{g}{\sqrt{2}}\overline{\nu_L}\gamma^\mu e_L W^+_\mu
-\frac{g}{\sqrt{2}}\overline{e_L}\gamma^\mu \nu_L W^-_\mu  
- \frac{g}{2\cos\theta_W}\overline{\nu_L}\gamma^\mu\nu_L Z_\mu ,
\end{equation}
The interactions of light and heavy neutrinos can be determined by inserting $\nu_L=P_L(V_\nu \upnu + \Theta N)$ from eq.~\eqref{PMNSfull} into eq.~\eqref{WeakWW}. The unitarity violation in \eqref{VnuDef} implies a flavor-dependent suppression of the weak interactions of the light mass eigenstates $\upnu_i$ in eq.~\eqref{LightMassEigenstates}. The doublet component $\Theta^{T}_{I\alpha}\nu_{L\alpha}^{c}+\Theta^\dagger_{I\alpha}\nu_{L\alpha}$ of $N_I$ in (\ref{LightMassEigenstates}) implies that the heavy states have $\theta$-suppressed weak interactions: If kinematically allowed, they appear in any process in which $\nu_{L\alpha}$ appears, but with an amplitude suppressed by $\Theta_{\alpha I}$. The $N_I$ also interact with Higgs particles and $\nu_L$ directly via the Yukawa coupling. This makes various different direct and indirect searches for heavy neutrinos possible~\cite{Shro80,Shrock:1980ct,Shrock:1981wq}, see e.g.~\cite{Atre:2009rg,Drewes:2013gca,Drewes:2015iva,Antusch:2014woa} and references therein for a recent summary.

%% file: kevnuwp_section2.tex
\subsection{The standard model of cosmology (Author: J.~Hamann)}
\label{sec:stand-model-cosm}

Over the past fifteen years, we have seen the emergence of a theoretical description of the Universe often dubbed the ``standard model of cosmology''.\footnote{In the following we shall also refer to it as the ``base $\Lambda$CDM'' model.}

Unlike its namesake and distant relative, the SM, the cosmological standard model is of a more empirical nature.  Many of its aspects are not so much fixed by the cold hard requirements imposed by fundamental symmetries, but rather driven by the desire to describe the measurements in a maximally economical way. While this approach might imply a certain degree of arbitrariness and lack of predictivity, it is even more remarkable that it has up to now defied countless attempts to challenge its position and has been left virtually untouched by the avalanche of new information that has accompanied the evolution of cosmology into a precision science.

In this section we will briefly summarize the broad features of this model.  During the history of the Universe, all known interactions were relevant, at least  during a given range of time. The basic framework of the cosmological standard model is thus given by the theory of General Relativity for gravity, and by the SM for the weak, strong and electromagnetic interactions. The properties of the model can roughly be categorised in terms of its assumptions regarding the geometry and the energy content of the Universe.

\subsubsection{Geometry}

The cosmological standard model does have one symmetry condition, known as the {\it cosmological principle}, which stipulates that at leading order, the Universe is homogeneous and isotropic. This assumption is very constraining: the most general metric satisfying it is the Friedmann-Lema{\^i}tre-Robertson-Walker (FLRW) metric, in spherical coordinates given by $\mathrm{d}s^2 = \mathrm{d}t^2 - a^2(t) \left( \frac{\mathrm{d}r^2}{1 - \kappa r^2} + \mathrm{d}\Omega^2 \right)$, with a freedom of a one-parameter function, the scale factor $a(t)$, and one discrete {\it curvature} parameter $\kappa \in \left\{ -1,0,1 \right\}$, corresponding to a spatially open, flat, and closed geometry, respectively. In base $\Lambda$CDM however, space is taken to be flat (i.e., $\kappa = 0$).  The time evolution of the scale factor is governed by Einstein's equations and determined by the energy content up to a boundary condition, e.g., today's expansion rate, the Hubble parameter $H_0 \equiv \left. \frac{1}{a} \frac{\mathrm{d}a}{ \mathrm{d}t} \right|_{t=t_0}$.

The fact that today's Universe contains objects such as stars, galaxies, and galaxy clusters requires small deviations from exact isotropy and homogeneity at early times.  Actually, initial density fluctuations of the order of $\delta \rho/\rho \sim 10^{-5}$, which eventually start growing under the influence of gravity, are sufficient to seed the rich phenomenology of structures we observe in our neighborhood today. The FLRW metric thus ought to be interpreted as a background solution, with perturbations around it described by {\it statistically} isotropic and homogeneous random fields.

In the cosmological standard model, the state of these initial perturbations is a particularly simple one: they are taken to be Gaussian and adiabatic curvature perturbations, with an amplitude only weakly dependent on their wavenumber $k$, and described by a power-law power spectrum,
\begin{equation}
	\mathcal{P_R}(k) = A_\mathrm{s} \left( \frac{k}{k_*}
	\right)^{n_\mathrm{s}-1}, 
\end{equation} 
where $k_*$ is an arbitrary pivot scale, $A_\mathrm{s}$ the amplitude of the spectrum at $k=k_*$, and $n_\mathrm{s}$ the spectral index. Coincidentally, perturbations with these properties are a generic prediction of the simplest models of slow-roll single-field inflation -- which also explain the flat spatial geometry.

\subsubsection{Energy content}

Whereas there is no evidence for interactions not  described by the SM or by general relativity, the particle content of the SM turns out to be insufficient to explain cosmological observations.  The base $\Lambda$CDM model requires the presence of five constituents:

\begin{itemize}

\item{{\bf Baryonic matter} (also including charged leptons) with an energy density parameterized by $\omega_\mathrm{b}$. During the process of Big Bang Nucleosynthesis (discussed in section~\ref{subsec:bbn}), light elements are formed; astrophysical measurements of their abundances have traditionally been one of the observational cornerstones in support of the cosmological standard model.}

\item{{\bf Photons} making up the relic radiation of the Big Bang, the Cosmic Microwave Background (CMB). This is easily the most well-measured ingredient of the cosmic recipe. Not only has its blackbody nature been impressively confirmed~\cite{Fixsen:1996nj} and its mean temperature been determined to high precision~\cite{Fixsen:2009ug}, but the measurement of its temperature fluctuations and polarization~\cite{Adam:2015rua} makes it by far the most powerful source of information for cosmology to date.}

\item{{\bf Neutrinos}: the Cosmic Neutrino Background (CNB) is the neutrino equivalent of the CMB, but has not yet been directly detected.  In base $\Lambda$CDM, it is assumed to consist of the three standard model neutrino flavors.  The physics of the CNB is covered in more detail in section~\ref{sec:theCNB}.}

\item{{\bf Dark Matter (DM)}: this component with energy density parameter $\omega_\mathrm{c}$ cannot be identified with any standard model particle, and is characterized by its lack of electromagnetic interaction. In base $\Lambda$CDM, its velocity dispersion is zero, thereby making it {\it cold} Dark Matter (CDM).}

\item{{\bf Dark energy}: in base $\Lambda$CDM, this takes the form of a cosmological constant $\Lambda$, i.e., a non-vanishing vacuum energy, whose value is fixed by the requirement of spatial flatness.}

\end{itemize}

\subsubsection{Parameters of base $\Lambda$CDM} 

On top of the five free parameters introduced above, a sixth one needs to be taken into account: the optical depth to reionization, $\tau$, which should be derivable from first principles. In practice, however, such a calculation would prove to be extremely complicated, due to the highly non-linear nature of the problem. In any case this simple parametrization is sufficient, since current CMB data are not very sensitive to the exact details of the reionization process.

This completes the parameter space of the base $\Lambda$CDM model:
\begin{equation} 
	\Theta_{\Lambda\mathrm{CDM}} = \left\{ \omega_\mathrm{b},
	\omega_\mathrm{c}, H_0, A_\mathrm{s}, n_\mathrm{s}, \tau
	\right\}.
\end{equation}
Note that this is only one possible parametrization; there exist equivalent ones that are related by parameter transformations. Also, when inferring the values of these parameters from data, one may in addition have to vary extra ``nuisance'' parameters, depending on the data sets being considered.

\subsubsection{The cosmological standard model vs.~observations}

In terms of cosmological observations, a special role is played by the anisotropies of the CMB. The physics of CMB anisotropies is very well understood, so they can be predicted very accurately, leaving no significant theoretical ambiguity. In addition, CMB data are the only single observable able to constrain all parameters of the base $\Lambda$CDM model simultaneously. The state-of-the-art is provided by the {\it Planck} collaboration, who recently released their full mission temperature and polarization data~\cite{Adam:2015rua}, with a further re-analysis that has been announced and is to be expected by 2017.  

When faced with observational data, any model claiming the title of a ``standard model'' needs at the very least fulfill the following criteria: firstly, it ought to give a reasonably good fit to individual observations (internal consistency). Secondly, the parameter values inferred from different observables should be consistent with each other (external consistency). And finally, it should pass the Occam's razor test, i.e., not be unnecessarily complex. Let us now turn to addressing these three criteria.

\subsubsection{Internal consistency}

Does the base $\Lambda$CDM model fit current CMB data well? For the {\it Planck} data, the goodness-of-fit with respect to base $\Lambda$CDM is discussed in ref.~\cite{Ade:2015xua}: for the best fit of the $\Lambda$CDM model to the {\it Planck} temperature plus large-scale polarization data, the $\ell \geq 30$ $TT$-data have a reduced $\chi^2$ of 1.03, corresponding to a $p$-value of $p \approx 0.17$.  The analogous $p$-values for the $TE$- and $EE$-data are $p \approx 0.10$ and $p \approx 0.37$, respectively.  In other words, if the true underlying model is indeed the best-fit of base $\Lambda$CDM, the observed goodness-of-fit lies well within the expected range and the effective $\chi^2$ is not anomalously large.  In addition, the analysis of higher order correlations of the {\it Planck} temperature shows no signs of departure from Gaussianity~\cite{Ade:2015ava}.  With internal consistency thus established, it makes sense to infer estimates of the free parameters from the data; we list current constraints in table~\ref{tab:LCDMparams}.

\begin{table*}
\caption{\label{tab:LCDMparams}Constraints on the base $\Lambda$CDM parameters from {\it Planck} temperature and polarization data~\cite{Ade:2015xua}.}
\begin{center}
\begin{tabular}{|c c |}
\hline
$\Omega_{\mathrm{b}}h^2$ & $0.02225\pm0.00016$ \\
$\Omega_{\mathrm{c}}h^2$ & $0.1198\pm 0.0015$ \\
$H_0$ & $67.27\pm0.66$ \\
$\tau$ & $0.079\pm 0.017$ \\
$\ln(10^{10} A_\mathrm{s})$ & $3.094\pm0.034$ \\
$n_\mathrm{s}$ & $0.9645\pm 0.0049$ \\
\hline
\end{tabular}
\end{center}
\end{table*}

\subsubsection{External consistency}

Besides the CMB, there are a number of other observables sensitive to subsets of $\Theta_{\Lambda\mathrm{CDM}}$. If these observables happened to prefer different parameter regions within base $\Lambda$CDM, such discrepancies could be a sign for new physics, but they might as well indicate insufficient understanding of experimental or theoretical systematic uncertainties in any of the data sets.

Most cosmologically interesting observables broadly fall into one of two classes: those sensitive to properties of the homogeneous Universe (geometric observables or measurements of primordial element abundances, the latter of which will be discussed in section~\ref{subsec:bbn}), and those constraining properties of its inhomogeneities (shape measurements).

\paragraph{Geometric observables}
These observables make use of standard rulers or standard candles to relate a known distance to an observed scale.  Examples include the baryon acoustic oscillation (BAO) scale, the luminosity distances of high-redshift type~Ia supernovae (SNe~Ia), or the determination of the Hubble parameter from low-redshift SNe~Ia. Combining these measurements with CMB data can help breaking the geometric degeneracy~\cite{Bond:1997wr} which, although not relevant anymore for the base $\Lambda$CDM, can reappear in extended models (with, e.g., a free curvature parameter, additional radiation, or hot Dark Matter (HDM) components).

While all of the recent galaxy-based BAO scale measurements, as well as the latest SN~Ia data show excellent agreement with {\it Planck} data, there do remain two occurrences of slight discrepancies of $\sim 2.5\sigma$ with geometric observables. (i)~The BAO scale extracted from measurements of the Lyman-$\alpha$ forest at high redshifts~\cite{Delubac:2014aqe}.  (ii)~The Hubble parameter reported by Riess et al.~\cite{Riess:2011yx} based on SNe~Ia measurements. 

Whereas the reason for the discrepancy in (i) remains unclear, a re-analysis of the SNe~Ia data~\cite{Efstathiou:2013via} finds a value within one standard deviation of the {\it Planck} value, pointing to systematic issues as a more likely explanation for this tension. On the other hand, even in a more recent analysis, the tension seems to remain present~\cite{Riess:2016jrr}.

\paragraph{Shape measurements}
Even though the CMB power spectra are exquisitely sensitive to the structure of the Universe and the physics around the time of recombination (at a redshift $z \approx 1100$), much of the later evolution of cosmic structures only weakly affects the temperature and polarization power spectra.  It may thus be helpful to combine CMB data with lower-redshift tracers of the cosmic structure, particularly with regard to models in which the growth of perturbations at late times is affected (e.g., HDM or dark energy models).  Information about the late-time structure of the Universe can for instance be extracted from galaxy surveys (2-/3-dimensional power spectra of the galaxy distribution, redshift space distortions, weak gravitational lensing of galaxies), quasar spectra (1-dimensional power spectrum of hydrogen gas from measurements of the Lyman-$\alpha$ forest), and even from the CMB maps (weak lensing of the CMB, cluster counts). Compared to the CMB power spectra, the interpretation of most of these observations tends not to be as straightforward, due to complications introduced by, e.g., non-linear structure growth and complicated astrophysical processes.

Of particular note here is a mild tension concerning the amplitude of matter perturbations on small scales.  Assuming base $\Lambda$CDM, {\it Planck} CMB data prefer somewhat more power than has been inferred from galaxy weak lensing~\cite{Heymans:2013fya}, cluster counts~\cite{Ade:2015fva}, and redshift space distortions~\cite{Samushia:2013yga}.  It has been suggested that this could potentially be a sign of an additional HDM component~\cite{Wyman:2013lza,Hamann:2013iba,Giunti:2013aea,Battye:2013xqa}, but this interpretation is not fully compelling either, as it is not statistically required by the data~\cite{Leistedt:2014sia}. Lyman-$\alpha$ data, on the other hand, appear perfectly consistent with CMB data~\cite{Palanque-Delabrouille:2014jca} in base $\Lambda$CDM.

\subsubsection{Occam's razor}

Following the detection of the CMB's $B$-polarization on intermediate scales by the BICEP2 collaboration~\cite{Ade:2014xna}, it seemed for a while that the base $\Lambda$CDM model might have to be extended to include primordial tensor perturbations.  However, a more thorough analysis using {\it Planck} polarization data to estimate the contribution of polarized emission from galactic dust showed that the signal is consistent with foreground emission~\cite{Ade:2015tva}.

The {\it Planck} data themselves have been fitted to many classes of extended models, in a large number of combinations with other data sets.\footnote{\url{http://www.cosmos.esa.int/web/planck/pla}} None of these extended models improve the fit by a sufficient amount to warrant the introduction of extra parameters though. This in turn implies that for the time being, there does not seem to be any convincing evidence against the base $\Lambda$CDM model. With ongoing experimental efforts on many fronts of observational cosmology however, this situation may eventually change.

\subsection{Active neutrinos in Cosmology (Authors: J.~Lesgourgues, S.~Pastor)} 

\subsubsection{\label{sec:theCNB}The cosmic neutrino background}

The existence of a relic sea of neutrinos is a generic feature of the standard hot Big Bang model, in number only slightly below that of relic photons. This cosmic neutrino background (CNB) has not been detected yet, but its presence is indirectly established by the accurate agreement between the calculated and observed primordial abundances of light elements, as well as from the analysis of other cosmological observables, such as CMB anisotropies.

Produced at large temperatures by frequent weak interactions, flavor neutrinos ($\nu_{e,\mu,\tau}$) were kept in equilibrium with the rest of the primeval relativistic plasma (electrons, positrons, and photons). Neutrinos had a momentum spectrum with an equilibrium Fermi-Dirac phase-space distribution with temperature $T$: $f_{\rm eq}(p,T)=[\exp(\frac{p-\mu_\nu}{T})+1]^{-1}$, where we included a neutrino chemical potential $\mu_\nu$ that would exist in the presence of a neutrino-antineutrino asymmetry. It has been shown that, even if $\mu_\nu$ is non-zero, the contribution of the chemical potential cannot be very relevant~\cite{Mangano:2011ip}.

As the Universe cools, the weak interaction rate falls below the expansion rate of the Universe and neutrinos decouple from the rest of the plasma. An estimate of the decoupling temperature $T_{\rm dec}$ can be found by equating the thermally averaged value of the weak interaction rate $\Gamma_\nu=\langle\sigma_\nu\,n_\nu\rangle$, where $\sigma_\nu \propto G_F^2$ is the cross section of the electron-neutrino processes and $n_\nu$ is the neutrino number density, to the expansion rate given by the Hubble parameter $H =(8\pi\rho/3M_P^2)^{1/2}$. If we approximate the numerical factors to unity, with $\Gamma_\nu \approx G_F^2T^5$ and $H \approx T^2/M_P$, we obtain the rough estimate $T_{\rm dec}\approx 1$ MeV. Neutrinos were ultra-relativistic at decoupling, because they cannot possess masses much larger than $1$ eV. 

Although neutrino decoupling is not described by a unique $T_{\rm dec}$, it can be approximated as an instantaneous process. In this simple but reasonably accurate approximation, the spectrum is preserved after decoupling, because the number density of non-interacting neutrinos has remained constant in any co-moving volume since decoupling. Shortly after neutrino decoupling the temperature drops below the electron mass, favoring $e^{\pm}$ annihilations into photons over their production. If one assumes that this entropy transfer did not affect the neutrinos because they were already completely decoupled, it is easy to find that the ratio between the temperatures of relic photons and neutrinos is $T_\gamma/T_\nu=(11/4)^{1/3}\simeq 1.40102$. It turns out that neutrino decoupling and $e^{\pm}$ annihilations are sufficiently close in time so that some relic electron-neutrino interactions exist. These processes are more efficient for larger neutrino energies, leading to non-thermal distortions in the neutrino spectra at the per cent level, and to a smaller increase of the co-moving photon temperature~\cite{Mangano:2005cc}. These changes slightly enhance the contribution of relic neutrinos to the total energy density. In practice, these distortions only have small consequences on the evolution of cosmological perturbations, and for many purposes they can be safely neglected.

The neutrino number density, inferred from the Fermi-Dirac spectrum and $T_\nu$, is equal today to $113$ neutrinos and antineutrinos of each flavor per cm$^{3}$. The energy density for massive neutrinos must be calculated numerically at each time, because they contribute initially to radiation, but they behave as matter after the non-relativistic transition. In units of the critical value, the present energy density of neutrinos is $\Omega_{\nu} = \rho_\nu/\rho^0_{\rm c} = \sum m_\nu/93.14\,h^2$~eV (i.e.\ of order unity for eV masses), where $\sum m_\nu$ includes the mass of all the neutrino states that are non-relativistic today.\footnote{Alternatively, one could state $\Omega_{\nu} =\sum m_\nu/94.10\,h^2$~eV for exactly $N_\nu=3$ neutrino flavours, which may be more appropriate given that the true value of $N_{\rm eff}$, to be defined in the next subsection, is not unknown. See Ref.~\cite{Grohs:2015tfy} for more detailed explanations.} Taking into account the results of neutrino oscillation experiments, as well as the bounds from neutrinoless double beta decay and tritium beta decay experiments, the possible present values of $\Omega_\nu$ are restricted to the approximate range $0.0013\, (0.0022) \lesssim \Omega_{\nu} \lesssim 0.13$ for NO (IO), where we already included that $h\approx 0.7$.

\subsubsection{\label{subsec:neff}The effective number of neutrinos}

In the base $\Lambda$CDM model, together with photons, neutrinos fix the expansion rate while the Universe is dominated by radiation. Their contribution to the total radiation content can be parametrized in terms of the effective number of neutrinos $N_{\rm eff}$, defined as
\begin{equation}
\rho_{\rm r} = \rho_\gamma + \rho_\nu = \left[ 1 + \frac{7}{8} \left( \frac{4}{11} \right)^{4/3} \, N_{\rm eff} \right] \, \rho_\gamma \,\,.
\label{neff}
\end{equation}
Here, we have normalized $\rho_{\rm r}$ to the photon energy density because its value today is known from the measurement of the CMB temperature. This equation is valid when neutrino decoupling is complete and holds as long as all neutrinos are relativistic.

We know that the number of light neutrinos sensitive to weak interactions (i.e., flavor or active neutrinos) equals three from the analysis of the invisible $Z$-boson width at LEP, $N_\nu=2.984 \pm 0.008$~\cite{Agashe:2014kda}. The effect of the small distortions in the CNB spectra from relic $e^\pm$ annihilations leads to $N_{\rm eff}\simeq 3.046$~\cite{Mangano:2005cc} (disregarding neutrino oscillations, one may obtain the slightly higher value of $N_{\rm eff}\simeq 3.052$~\cite{Grohs:2015tfy}). Any departure of $N_{\rm eff}$ from this last value would be due to non-standard neutrino features or additional relativistic relics. A detailed discussion of cosmological scenarios where $N_{\rm eff}$ is not fixed to three can be found in~\cite{NuCosmo,Dolgov:2002wy,Sarkar:1995dd}. We will see in section~\ref{subsec:bbn} that the theory of Big Bang Nucleosynthesis (BBN) and the measurement of light element abundances provide interesting bounds on $N_{\rm eff}$.

\subsubsection{\label{sec:nuDM}Massive neutrinos as Dark Matter}

A priori, massive neutrinos are excellent DM candidates, in particular because we are certain that they exist, in contrast to other candidate particles. DM particles with a large velocity dispersion such as that of neutrinos are called HDM. 

The role of neutrinos as HDM particles has been widely discussed since the 1970s (see e.g.~\cite{Primack:2001ib}). It was realized in the mid-1980s that HDM affects the evolution of cosmological perturbations in a particular way: it erases the density contrasts on wavelengths smaller than a mass-dependent free-streaming scale. As we will see, in a Universe dominated by HDM, this suppression contradicts various observations, so the attention then turned to CDM candidates, i.e.\ particles which were non-relativistic at the epoch when the Universe became matter-dominated. Still in the mid-1990s, it appeared that a small admixture of HDM in a Universe dominated by CDM fitted the observational data on density fluctuations at small scales better than a pure CDM model. However, within the presently favored $\Lambda$CDM model, dominated at late times by a cosmological constant (or some form of dark energy), a significant contribution of HDM is not needed. Instead, one can use the available cosmological data to find how large the neutrino contribution can be to constrain the possible values of neutrino masses.

\subsubsection{\label{22-sec:neutrino_effects}Effects of standard neutrinos on cosmology}

The physical effect of standard neutrinos on cosmological observables is described briefly in~\cite{Lesgourgues:2006nd}, and in further details in sections 5.3.3 and 6.1.4 of~\cite{NuCosmo}. In few words, the most crucial point is that at least two out of the three neutrino masses are such that neutrinos become non-relativistic at some time between photon decoupling and today. This has two important consequences, explaining why standard neutrinos can be probed with cosmology. First, they provide a non-trivial background density and pressure, behaving initially as radiation and later as non-relativistic matter. Second, since neutrinos entered the non-relativistic regime only very recently, they still have considerable velocities compared to CDM particles, which prevents them from clustering on small scales: their velocity is greater than the escape velocity from gravitational potential wells on those scales. These two points have important implications for the matter and CMB power spectra in the Universe.

{\bf Matter power spectrum.} The quantity $P(k,z)=\langle | \delta \rho_m / \rho_m |^2 \rangle$, measurable through several types of astrophysical observations, characterizes the amount of clustering of total matter on different wavenumbers $k$ and redhsifts $z$. Since total matter includes non-relativistic neutrinos at late times, the absence of neutrino clustering on small scales is expected to reduce the power spectrum by $(1-f_\nu)^2$, where $f_\nu$ is the fraction of total matter in the form of neutrinos, proportional to the total neutrino mass $\sum m_\nu$. In reality, however, the effect is larger than that, because the presence of a non-clustering component contributing nevertheless to the background expansion rate also changes the growth rate of the CDM fluctuations themselves. Hence, on small scales, the matter power spectrum suppression is close to $1-8f_\nu$ according to linear perturbation theory. $N$-body simulations show that the effect is even further enhanced to about $1-10f_\nu$ through the non-linear clustering happening at a slightly later epoch~\cite{Bird:2011rb}.

While $f_\nu$ only depends on the sum $\sum m_\nu$ of all light neutrino masses, second-order effects on the growth of structure do depend on individual neutrino masses (which govern the individual value of free-streaming scales at a given time). Mass-splitting effects have been shown to be so small that it is unlikely that they can be measured one day~(see e.g.~\cite{Lesgourgues:2006nd,Jimenez:2010ev,NuCosmo}).

{\bf CMB anisotropies.} For masses smaller than 0.6 eV, neutrinos are still relativistic at the time of photon decoupling, and their mass cannot impact the evolution of acoustic oscillations. However, the presence of relativistic neutrinos is felt by the CMB perturbations: {\it (i)}~through the neutrino contribution to the background radiation density and expansion history, that depends on $N_\mathrm{eff}$; {\it (ii)}~through gravitational forces between photon/baryon clumps and neutrino overdensities, important on Hubble-crossing scales during radiation domination: the existence of such forces can be tested by measuring effective parameters with the CMB, like the sound speed and viscosity speed $(c_\mathrm{eff}^2, c_\mathrm{vis}^2)$ of non-photon radiation.

An effect of neutrino masses on the CMB can only appear at two levels: that of the background evolution around and after the non-relativistic transition; and that of secondary anisotropies (related to the behavior of photons after decoupling: Integrated Sachs Wolfe (ISW) effect, weak lensing by large scale structure, etc.~\cite{NuCosmo}).

Since the CMB is very sensitive to the redshift $z_\mathrm{eq}$ of equality between matter and radiation, the effect of neutrino masses is best discussed for a fixed $z_\mathrm{eq}$, and in particular when the density of baryons and CDM in the early Universe is kept fixed. In that case, playing with the neutrino mass still changes a bit the expansion history of the Universe at late times, when neutrinos become non-relativistic. This shows up in the angular distance to recombination (controlling the angle under which we see CMB peaks), and in the time of matter-to-cosmological-constant equality (controlling the late ISW effect). While the second effect is hard to observe due to cosmic variance, the first effect is important. It explains in particular why the CMB tends to constrain the particular combination of $\sum m_\nu$ and $H_0$ that fixes the angular peak scale.

Soon after the photon decoupling, at least one neutrino eigenstate is expected to become non-relativstic. This impacts the evolution of metric fluctuations at high redshift. Through the early ISW effect, this non-relativistic transition produces a depletion in the CMB power spectrum of the order of $(\Delta C_l/C_l) \sim -(\sum m_\nu/0.3 \, {\rm eV}) \%$  on multipoles $20<l<500$. This effect is roughly thirty times smaller than the depletion in the small-scale matter power spectrum, $\Delta P(k)/P(k) \sim -(\sum m_\nu/0.01 \, {\rm eV}) \%$, and hence difficult to observe.

Finally, since neutrino masses reduce the amplitude of matter fluctuations on small scales, they also result in a weaker lensing of CMB anisotropy patterns. This shows up directly in the temperature and polarization spectra: less lensing means less smoothing and sharper contrasts between maxima and minima, and also a larger spectrum tail at high multipoles. This effect can be seen even better in the reconstruction of the CMB lensing spectrum, that can be inferred from high-order statistics in CMB maps.

\subsubsection{\label{22-sec:neutrino bounds}Current cosmological bounds on standard neutrinos}

Recent results from the {\it Planck} satellite strongly support the existence of the standard neutrino background, through measurements of the background density and of gravitational effects of neutrinos. The previously defined parameters $(N_\mathrm{eff}, c_\mathrm{eff}^2, c_\mathrm{vis}^2)$ are expected to be equal to $(3.046, 1/3, 1/3)$ in a Universe with no relativistic relics beyond standard neutrinos. The {\it Planck} 2015 release~\cite{Ade:2015xua} gives $(2.99\pm0.20, 0.3240\pm0.0060, 0.327\pm0.037)$ at 68\%~C.L., assuming the minimal cosmological model with neutrino masses close to their smallest possible values in the normal hierarchy scenario. These results are well compatible with theoretical expectations.

The neutrino mass bound coming from CMB temperature/polarization spectra is $\sum m_\nu<0.49$~eV (95\%~C.L.) for the {\it Planck} 2015 data~\cite{Ade:2015xua}. We recall that this bound comes mainly from the angular diameter distance and lensing effects. The first effect can be better constrained by adding information on the Baryon Acoustic Oscillation (BAO) scale at redshifts $0.1<z<0.8$, in order to reduce parameter degeneracies. This gives $\sum m_\nu<0.17$~eV (95\%~C.L.) for {\it Planck} 2015 + BAO data~\cite{Ade:2015xua}.

It is important to know that the observed CMB temperature/polarization spectra present a low contrast betwen maxima and minima, already {\it slightly} smaller than what would be expected in the base $\Lambda$CDM model with negligible neutrino masses. Any parameter predicting less lensing (like $\sum m_\nu$) is hence disfavored by the data. This ``slight lensing excess'' is statistically compatible with the minimal $\Lambda$CDM model at the 2$\sigma$ level, and it could be due to a statistical fluke. However one cannot totally exclude that it may be induced by unkown systematics or by new physics; in that case, the previous upper bound on $\sum m_\nu$ would be relaxed.

Instead of looking at the smoothing of the temperature/polarization spectra, one can extract from CMB maps the lensing potential spectrum, to probe CMB lensing more directly. In that case, no ``slight lensing excess'' is observed: on the contrary, the lensing potential spectrum is a bit low on small scales. Hence lensing extraction does not strengthen the neutrino mass bound, and even makes it slightly weaker: $\sum m_\nu< 0.23$~eV (95\%~C.L.) for {\it Planck} 2015 + BAO + lensing~\cite{Ade:2015xua}.

To improve these bounds, one needs to use information from Large Scale Structure (LSS), to probe the neutrino free-streaming effect and the reduction  of the matter power spectrum on small scales. These measurements usually suffer from high systematics and modeling uncertainties. They actually give results on neutrino masses that are in moderate tension with each other. The problem is that any LSS data returning a high amplitude of fluctuations on inter-galactic scales will automatically lead to a strong neutrino mass bound, and vice-versa. However a high/low amplitude could easily be the result of unknown systematics. This problem should be solved with future LSS surveys, which will have much better sensitivity, better control on systematics, and -- even more importantly -- a wide coverage of scales and redhsifts, allowing to probe the actual scale-dependent and time-dependent effect of neutrino masses on the growth of structures, without trivial degeneracies with the global amplitude of fluctuations.

Hence the following list of neutrino mass bounds from current LSS data should be taken with a grain of salt. The measurement of the one-dimensional flux power spectrum from Lyman-$\alpha$ forests in BOSS quasars gives $\sum m_\nu< 0.15$~eV (95\%~C.L.) for {\it Planck} 2015 + Lyman-$\alpha$~\cite{Palanque-Delabrouille:2014jca}. A slightly tighter bound, $\sum m_\nu < 0.13$~eV (95\% C.L.), is found by combining {\it Planck} 2015 with a compilation of baryon acoustic oscillation data, and with the galaxy correlation spectrum of the Sloan Digital Sky Survey (SDSS)~\cite{Cuesta:2015iho}. The measurement of the structure formation rate through redshift space distorsions (RSD) in the BOSS galaxy survey gives $\sum m_\nu< 0.23$~eV (95\%~C.L.) for {\it Planck} 2015 + RSD~\cite{Gil-Marin:2014baa}. Instead, various measurements of galaxy weak lensing (GWL) and of the cluster mass function (CMF) are in tension with {\it Planck}: they would prefer a higher neutrino mass, of the order of 0.4~eV (see, e.g., Refs.~\cite{Battye:2014qga,Ade:2015xua}), but at the cost of adopting values of the Hubble rate in tension with {\it Planck}. It is not clear at the moment whether these tensions will be resolved by a better understanding of systematics, or by a change of the physical paradigm. What is clear is that the tension between CMB data and GWL/CMF data cannot be resolved by simply playing with neutrino parameters~\cite{Ade:2015xua}: extra ingredients would be needed (like for instance a departure from Einstein gravity).

In the future, we expect a sensitivity $\sigma(\sum m_\nu)$ close to 0.02~eV around the year 2025 for a LSS survey like Euclid combined with {\it Planck}~\cite{Audren:2012vy}; or 0.015~eV for a CMB satellite of next generation like Core+ combined with Euclid~\cite{Hall:2012kg}; or, under very optimistic assumptions, even 0.0075~eV for a futuristic 21cm survey like FFFT~\cite{Pritchard:2008wy}. Possibilities to detect the CMB polarization, such as CMB-S4~\cite{Wu:2014hta}, and observations aiming at dark energy, like DESI~\cite{Font-Ribera:2013rwa}, will further improve the situation. Hence there are very good prospects to detect the absolute neutrino mass scale with cosmology, and even to exclude the IO scenario if the range $\sum m_\nu \geq 0.12$~eV is found to be incompatible with the data at several $\sigma$'s. If instead the total mass is found close to $\sum m_\nu \sim 0.12$~eV, cosmological data will probably never reach the required sensitivity to discriminate between NO and IO.

\subsection{\label{subsec:bbn}Big Bang Nucleosynthesis (Author: G.~Mangano)}

\subsubsection{What it is and how it works}

Big Bang Nucleosynthesis (BBN) is the epoch in the evolution history of the Universe when light nuclei, basically $^2$H, $^3$H, $^3$He, $^4$He, and $^7$Li, were produced out of the initial plasma of protons and neutrons. This happened via weak, electromagnetic, and strong interactions. The whole phase started about one second after the Bang and lasted for approximately 15 minutes. In terms of the photon temperature this corresponds to the range from 1 MeV ($10^{10}\,\mathrm{K}$) to 20-30 keV($10^{8}\,\mathrm{K}$).

The basic picture of BBN was understood in the decade after the seminal $\alpha\beta\gamma$ paper~\cite{Alpher:1948ve}, and since then it became one of the main observational pillars of the hot Big Bang cosmological scenario. Furthermore, since the abundances produced of nuclear species are quite sensitive to the expansion rate during BBN, as well as to other key parameters such as the baryon density, BBN is a robust probe of the early Universe, and of exotic physics that could be relevant at these stages (extra particle species, new interactions, etc.).

\noindent
The dynamics of BBN can be summarized as follows: 
\begin{itemize}

\item {\it Initial conditions.} At high temperatures ($T$ > MeV) the Universe is filled by free protons and neutrons, which are maintained in mutual chemical equilibrium by charged current weak interactions with $e^\pm$ and electron neutrinos/antineutrinos. These enforce their number density ratio to follow the standard equilibrium result, usually referred to as the Saha equation:
\begin{equation}
\frac{n_n}{n_p} = \exp \left(-\frac{\Delta m}T \right) \,\, ,
\label{npratio}
\end{equation}
with $\Delta m \simeq 1.29$~MeV being the neutron-proton mass difference. In this early phase there is no room for formation of bound nuclei, since there are too many photons per baryon. As soon as, say, $^2$H is formed, it is immediately photodissociated by the inverse process,
\begin{equation}
\gamma + \, ^2\mbox{H}  \rightarrow n + p \label{2hgamma} \,\, .
\end{equation}  
This is again the effect of chemical equilibrium now applied to the electromagnetic process of eq.~\eqref{2hgamma}. Defining the baryon to photon number density ratio $\eta_\mathrm{b} = n_\mathrm{b}/n_\gamma$, the Saha equation gives
\begin{equation}
\frac{n_{^2{\rm H}}}{n_{\rm H}} \sim \eta_\mathrm{b} \left(\frac{T}{m_N} \right)^{3/2} \exp \left( \frac{B_{^2{\rm H}}}{T} \right) \,\, ,
\label{sahadeut}
\end{equation}
with $B_{^2{\rm H}}\sim 2.2$~MeV being the deuterium binding energy and $m_N$ the mean nucleon mass. 

\item {\it Weak interaction freeze-out}. As it has been already discussed, at $T_{\rm dec} \sim$~MeV weak interaction rates become slower than expansion rate of the Universe. At this scale, charged current processes freeze out, and the number of neutrons per co-moving volume is only reduced via neutron $\beta$-decays. Using eq.~\eqref{npratio}, for $T \leq T_{\rm dec}$ we have
\begin{equation}
\frac{n_n}{n_p} = \exp \left(-\frac{\Delta m}{T_{\rm dec}} \right) \exp \left( - \frac{t(T)}{\tau_n} \right) \,\, ,
\end{equation}
where $t(T) \propto (\frac{\mathrm{MeV}}{T})^2 \mathrm{s}$ is the time-temperature relation during radiation domination, and $\tau_n = (880.3 \pm 1.1)$~sec is the neutron lifetime~\cite{Agashe:2014kda}. 

Notice that the time-temperature relation is sensitive to the total radiation energy density, which means that the number of neutrons available at the onset of BBN, which eventually get almost entirely bound in $^4$He nuclei, strongly depends on $N_{\rm eff}$. 

\item {\it Deuterium formation.} The baryon to photon number density ratio is related to the current energy density of baryons through $\eta_\mathrm{b} \sim 274 \cdot 10^{-10}\,  \omega_\mathrm{b}$. The latter is accurately measured by CMB anisotropy data, leading to $\eta_\mathrm{b} \sim 6 \cdot 10^{-10}$. From eq.~\eqref{sahadeut}, we see that for $n_{^2{\rm H}}/{n_{\rm H}}$ to become of order one it is necessary to wait until the photon temperature decreases well below $B_{^2{\rm H}}$. In particular, at $T_{\rm BBN} \sim 70$ keV, the photodissociation process become ineffective and the deuterium abundance starts raising. This is the {\it deuterium bottleneck}.

\item {\it The nuclear chain}. After $^2$H formation, a complicated network of nuclear processes starts. They take place in a rapidly cooling environment with low nucleon density, resulting in a peculiar out-of-equilibrium nuclear chain. Deuterium is rapidly burned into $^4$He, which is the main BBN outcome, since it has the highest binding energy per nucleon among light nuclei. Its final density fraction, $Y_p = 4 n_{^4 {\rm He}}/n_\mathrm{b}$, is weakly sensitive to the details of the nuclear network, and can be computed in good approximation under the assumption that all neutrons are eventually bound in helium nuclei
\begin{equation}
Y_p \sim \left. 4 \frac{n_n}{2}\frac{1}{n_n+n_p} \right|_{T_{\rm BBN}} = \frac{2}{1+ \exp (\Delta m/T_{\rm dec}) \exp(t(T_{\rm BBN})/\tau_n)} \sim 0.25 \,\, ,
\end{equation}
which is in very good agreement with experimental results, see the next section. A more accurate determination of $Y_p$ and in particular of other nuclides, which are much less abundant,\footnote{for instance, $^2$H/H and  $^3$He/H are expected to be of order $10^{-5}$, while $^7{\rm Li/H}$ is of order $10^{-10}$.} can only be obtained numerically, by solving a set of coupled kinetic equations along with the Einstein equations, the constraint of conservation of baryon and electric charge, and the covariant conservation of the total stress-energy tensor, as in~\cite{Wagoner:1966pv, Kawano:1992ua, Pisanti:2007hk}. The reader interested in this issue or in the details of the nuclear reaction network can refer e.g.\ to~\cite{Serpico:2004gx,Iocco:2008va,NuCosmo}.

\end{itemize}

\subsubsection{Constraints on the baryon density and $N_{{\rm eff}}$}

In the standard (minimal) model of BBN, the only free parameter is the baryon-to-photon number density $\eta_\mathrm{b}$, or equivalently the present baryon energy density parameter $\omega_\mathrm{b}$. In fact, already in the late sixties it was understood that comparing experimental observations of {\it primordial} light nuclei abundances with theoretical BBN results was a powerful method to constrain the cosmological baryon density~\cite{Wagoner:1966pv}. Today the value of $\omega_\mathrm{b}$ is measured with an exquisite precision by CMB temperature and polarization maps, and found to be in the same range required by BBN. Let us discuss this point in more details. The most recent and precise determination by the {\it Planck} satellite, reported in Table~\ref{tab:LCDMparams}, is in agreement with previous results of several earlier CMB experiments. With this value and a standard relativistic energy density at BBN, $N_{\rm eff}=3.046$, one obtains from the numerical code of~\cite{Pisanti:2007hk}:\footnote{For the still controversial primordial origin of  $^7$Li observations see e.g.\ the BIG-BANG NUCLEOSYNTHESIS review in~\cite{Agashe:2014kda}.}
\begin{eqnarray}
Y_p &=& 0.2467 \pm 0.0006 \,\, , \\ 
^2\mbox{H/H}  &=& (2.61 \pm 0.13) \cdot 10^{-5} \,\, ,
\end{eqnarray}
where the 95\%~C.L.\ errors are mainly due to uncertainties in the neutron lifetime (for $^4$He) and in nuclear rates like $d(p,\gamma)^3$He (for deuterium). The error coming from uncertainties on $\omega_\mathrm{b}$ is strongly sub-dominant for $^4$He, and slight sub-dominant for deuterium. These values are in very good agreement with the most recent experimental determinations of these primordial nuclei ratio from~\cite{Aver:2013wba,Cooke:2013cba} (with 68\%~C.L.),
\begin{eqnarray}
Y_p &=& 0.2465 \pm 0.0097  \,\, , \\
^2\mbox{H/H}  &=& (2.53 \pm 0.04) \cdot 10^{-5}  \,\, .
\end{eqnarray} 
As {\it Planck} scientists comment in~\cite{Ade:2015xua}: {\it ``This is a remarkable success for the standard theory of BBN.''}. Indeed, it is a remarkable success of the standard cosmological model.

Because the baryon density is basically fixed by the CMB, while BBN can be exploited to constrain non-minimal cosmological models. For instance, nuclei abundances could be affected by  a non-standard value of $N_{\rm eff}$, active neutrino chemical potentials, extra sterile neutrino states, exotic properties of neutrinos and of their phase space distribution, or non-standard  neutrino interactions. For a review on these issues see e.g.~\cite{Iocco:2008va,NuCosmo}. Here we only briefly consider how $N_{\rm eff}$ can be bounded by combining the {\it Planck} results with $Y_p$ and $^2$H experimental determinations, still assuming standard BBN, but with an arbitrary number of relativistic degrees of freedom.

As we mentioned, relativistic particles, such as neutrinos, fix the expansion rate during BBN, which in turn fixes the produced abundances of light elements, and in particular that of $^4$He. This is why BBN gave the first constraints on the number of neutrino species, even before accelerators. The {\it Planck} collaboration~\cite{Ade:2015xua} found that the joint constraints on $(\omega_b, N_{\rm eff})$ obtained either from CMB data alone or from $^4$He and deuterium abundances nicely agree with each other, showing once more the remarkable consistency of the simplest cosmological models. After marginalizing over the baryon density, the CMB+BBN joint constraint on $N_{\rm eff}$ at 95\%~C.L.\ is
\begin{equation}
N_{\rm eff} = 2.99 \pm 0.39,
\end{equation}
using the $^4$He abundance estimation of~\cite{Aver:2013wba}, and 
\begin{equation}
N_{\rm eff} = 2.91 \pm 0.37
\end{equation}
if the $^2$H/H value of~\cite{Cooke:2013cba} is assumed.

\subsection{Sterile neutrinos in Cosmology}

The role of sterile neutrinos in cosmology strongly depends on the magnitude of their mass. For different choices of the mass scale, they can be responsible for various phenomena. In this White Paper we are primarily intersted in sterile neutinos with keV masses. However, in the following we briefly sketch other well-motivated mass scales and their phenomenolgical implications.  
More details are e.g.~given in the reviews~\cite{Abazajian:2012ys,Drewes:2013gca}.

\subsubsection{eV-scale (Authors: M.~Archidiacono, N.~Saviano)} 

In recent years, light sterile neutrinos ($m \sim 1$~eV) have been introduced in order to explain intriguing but controversial anomalies in laboratory oscillation experiments. Global analyses~\cite{Karagiorgi:2012usa,Giunti:2013aea, Kopp:2013vaa}, based on different scenarios with one (dubbed ``3+1'') or two (``3+2'') eV sterile neutrinos, have been performed; however there is still no consensus on a preferred model and the search for sterile neutrinos in laboratory experiments is presently open. 

Given the systematic uncertainties of any experimental measurement, it is important to use as many observations as possible to corner sterile neutrinos. In this context, cosmological observations represent a valid complementary tool to probe these scenarios, being sensitive to the number of neutrinos and to the sum of their masses.  Indeed, sterile neutrinos can be produced by oscillations with the active neutrinos in the early Universe, contributing -- if sufficiently light -- to the radiation content (parametrized in terms of the effective number of relativistic degrees of freedom $N_{\rm eff}$) in addition to photons and ordinary neutrinos and leaving possible traces on different cosmological observables. Moreover, once sterile neutrinos become non-relativistic, they contribute to the matter density and, as well as active neutrinos, they damp density perturbations on scales smaller than the characteristic mass-dependent free-streaming scale, leaving a peculiar imprint on structure formation. Therefore cosmology can reveal sterile neutrinos in two ways: constraining both $N_{\rm eff}$ (if the steriles are still relativistic) and the effective sterile neutrino mass, $m_{\rm eff}$, which is defined as the physical mass weighted by the sterile abundance.

The first cosmological observable able to constrain $N_{\rm eff}$ is Big Bang Nucleosynthesis (BBN). Indeed if these additional states are produced before the active-neutrino decoupling ($\rm T_{\rm d} \sim 1$ MeV), they would acquire quasi-thermal distributions (depending on their temperature) and behave as extra degrees of freedom at the time of primordial nucleosynthesis. This would anticipate  weak interaction decoupling leading to a larger neutron-to-proton ratio and eventually resulting into a larger $^4$He fraction. Furthermore, sterile neutrinos, can distort the $\nu_e$ phase space distribution via flavor oscillations with the active ones, leading to a possible effect on the helium and deuterium abundances. However several and independent investigations~\cite{Abazajian:2002bj,Mangano:2011ar,Cooke:2013cba} showed that one fully thermalized sterile neutrino is still marginally allowed.

Concerning Cosmic Microwave Background (CMB), the basic effect induced by a variation in $N_{\rm eff}$ is related to the impact on the expansion rate of the Universe at recombination and it is detectable at small scales, as it is explained in refs.~\cite{Hou:2011ec,Archidiacono:2013fha}. Furthermore the CMB angular power spectrum is also affected both by the background variation due to the sterile neutrino mass and by the early Integrated Sachs-Wolfe effect due to their non-relativistic transition close to the recombination epoch. The combined analysis of $N_{\rm eff}$ and $m_{\rm eff}$ performed by the {\it Planck} collaboration, using the up-to-date cosmological data~\cite{Ade:2015xua} including {\it Planck} temperature power spectrum, lensing data, and baryonic acoustic oscillations, lead to the following 95\%~C.L.\ upper bounds (for massive sterile neutrinos):
\begin{align}
N_{\rm eff}<3.7 \quad \textrm{ and }  & \quad m_{\rm eff}<0.52 \textrm{ eV},\label{eq:neff1}  \\
N_{\rm eff}<3.7 \quad \textrm{ and }  & \quad m_{\rm eff}<0.38 \textrm{ eV},
\label{eq:neff2}
\end{align}
where the former set  is computed in presence of the prior cutoff of $m_{\textrm{ster}}^{\textrm{ther}}< 10$ eV, while the latter restricting to the region where $m_{\textrm{ster}}^{\textrm{ther}}< 2$ eV.

The tension with the eV sterile neutrinos, as hinted by laboratory experiments, is twofold~\cite{Mirizzi:2013kva, Archidiacono:2014apa,Bergstrom:2014fqa}: on one hand the constraint on $N_{\rm eff}$ rules out one fully thermalized sterile neutrino at high significance (more than 99\%~C.L.), on the other hand eV sterile neutrinos are too heavy to be consistent with the cosmological mass bounds. However, it has been noticed in~\cite{Wyman:2013lza} that massive sterile neutrinos might partially alleviate the tension between {\it Planck} and local Universe observations. Compared to the base-$\Lambda$CDM model, adding sterile neutrinos increases the value of the Hubble constant, making it compatible with the measurements of the Hubble Space Telescope; at the same time additional massive neutrinos lower $\sigma_8$ (the r.m.s. of density fluctuations on a sphere of $8$ h$^{-1}$Mpc), consistently with the value preferred by lensing and clusters measurements. Finally we note here that lensing and clusters measurements may provide a detection of a non-zero effective mass $m_{\rm eff}=(0.67\pm0.18)$ eV~\cite{Battye:2014qga}.

Given this partially contradictory situation and the existence of the laboratory anomalies, it is necessary to study the physical conditions under which the sterile neutrino production can occur in the early Universe.  As already mentioned, sterile neutrinos are produced in the primordial plasma by mixing with the active species. Therefore, in order to assess their abundance, it is necessary to solve the quantum kinetic equations for the flavor evolution of the active-sterile neutrino ensemble.

\noindent{\bf Flavor evolution for the active-sterile system.}
It is commonly assumed that sterile neutrinos  $\nu_s$ are produced in the early Universe via mixing with the other neutrino species  $\nu_e, \nu_{\mu}, \nu_{\tau}$ in presence of collisions. Considering only one additional sterile state, the flavor eigenstates $\nu_\alpha$ are related to the mass eigenstates $\nu_i$ via a unitary matrix ${\mathcal U}={\mathcal U} (\theta_{12}, \theta_{13}, \theta_{23}, \theta_{14}, \theta_{24}, \theta_{34})$~\cite{Maltoni:2001bc}.

The neutrino (antineutrino) ensemble, simultaneously mixing and scattering in the universal plasma, is described in terms of  a $4\times 4$ momentum-dependent density matrix $\varrho_{\bf p}$ ($\bar \varrho_{\bf p}$), in which the diagonal components represent the flavor contents while the off-diagonal ones encode the phase information and vanish in the absence of mixing. The general evolution equation for the density matrix $\varrho_{\bf p}$ ($\bar \varrho_{\bf p}$)  is the following~\cite{Sigl:1992fn,Dolgov:1980cq,Dolgov:2002ab,Mirizzi:2012we}:
\begin{equation}
{\rm i}\,\frac{d\rho}{dt} =[{\sf\Omega},\rho]+ C[\rho]\,,
\label{drhodt}
\end{equation}
and a similar expression holds for the antineutrino matrix $\bar\rho$. The evolution in terms of the comoving observer proper time $t$ can be easily recast in function of the temperature $T$  (see~\cite{Dolgov:2002ab} for a detailed treatment). The first term on the right-hand side of eq.~\eqref{drhodt} describes the flavor oscillations Hamiltonian,
\begin{equation}
{\sf\Omega}=\frac{{\sf M}^2}{2}  \frac{1}{p}  +
\sqrt{2}\,G_{\rm F}\left[-\frac{8p}{3 }\, \bigg(\frac{{\sf E_\ell}}{m_{\rm W}^2} + \frac{{\sf E_\nu}}{m_{\rm Z}^2}\bigg)+ {\sf N_\nu}\right]\,,
\label{omega}
\end{equation}
where ${\sf M}^2= {\mathcal U}^{\dagger} {\mathcal M}^2 {\mathcal U}$ is the neutrino mass matrix, while the terms proportional to the Fermi constant $G_F$  encode the matter effects in the neutrino oscillations. At the temperatures of interest, the primordial plasma is composed almost only by $e^{\pm}$ pairs. The ${\sf E_\ell}$ term is related to the energy density of $e^{\pm}$ pairs and describes their refractive effect. Moreover, since the neutrino gas is very dense, the neutrinos themselves form a background medium; as a consequence, we must consider neutrino self-interactions, which make the flavor evolution equations non-linear. Neutrino self-interactions are described by a first order term  ${\sf N_\nu}$, proportional to the difference of the density matrices for $\nu$ and  $\bar\nu$, namely a primordial $\nu$ asymmetry, which has a negligible contribution in a standard scenario, and by a second order term ${\sf E_\nu}$, which is proportional to the energy density of $\nu$ and $\bar\nu$. Finally, the last term on the right-hand side of eq.~\eqref{drhodt} is the collisional term, which is proportional to $G_F^{2}$ and describes pair processes, annihilations, and any momentum exchange process. Concerning the mass and mixing parameters involved in the equations of motion, they can be fixed at the best-fit values obtained from the global analysis in the active sector~\cite{Capozzi:2013csa} and in the sterile sector~\cite{Giunti:2013aea}. In particular the active-sterile mixing angle turns out to be ${\mathcal O}(10^ {\circ})$, while the active-sterile mass splitting is $\Delta m^2_{\rm st} \sim {\mathcal O}$ (1) eV$^2$, much larger than the solar and atmospheric ones.

\begin{figure}[t]
\begin{center}
 \includegraphics[trim=0 2.5cm 0 5cm,angle=0,width=0.8\textwidth]{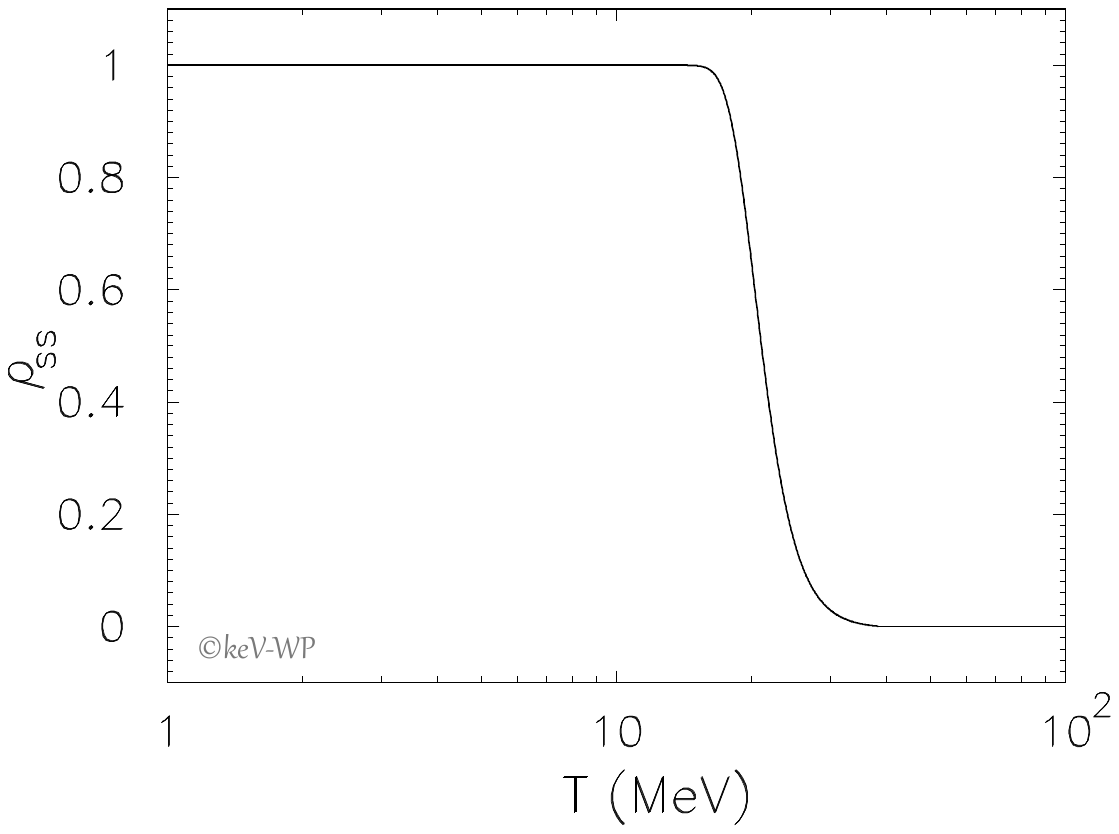} 
 \end{center}
\vspace{-5cm}
\caption{\label{rhoss}Evolution of the sterile abundance $\rho_{ss}$ in function of the temperature $T$.}
\end{figure}

\noindent{\bf Production of sterile neutrinos and cosmological implications.}
For the mass and mixing parameters preferred by laboratory anomalies, sterile neutrinos are copiously produced by oscillations with the active species~\cite{Mirizzi:2013kva} and they represent one extra degree of freedom. Fig.~\ref{rhoss} shows the behavior of the sterile component of the density matrix as a function of the temperature of the Universe $T$: at $T\sim  30$ MeV the active-sterile flavor conversions start and  the sterile neutrino production reaches $\rho_{ss}=1$ (which translates into $\Delta N_{\rm eff}=1$), in tension with the cosmological bounds, see eqs.~\eqref{eq:neff1}--\eqref{eq:neff2}. Furthermore, fixing $\Delta N_{\rm eff}$ to $1$ leads to a strong cosmological bound on $m_{\rm eff}$ (see fig.~2 of ref.~\cite{Hamann:2013iba}), which results in a severe tension with the mass range suggested by laboratory experiments. However cosmological analyses show a negative correlation between  $N_{\rm eff}$ and $m_{\rm eff}$. This anti-correlation suggests the possibility of reconciling eV sterile neutrinos with cosmology. Indeed, if the production of sterile neutrinos is suppressed in the early Universe, they do not thermalize and their physical mass is weighted by a lower abundance. Several mechanisms have been proposed in order to achieve this suppression. The first one is represented by the introduction of a primordial neutrino-antineutrino asymmetry term~\cite{Hannestad:2012ky, Saviano:2013ktj} in the equations for the flavor evolution (see eqs.~\eqref{drhodt} and~\eqref{omega}). This new term, acting as a type of matter effect, would suppress the flavor conversions and by that the sterile production. However, in order to accommodate the cosmological bounds, a very large primordial asymmetry is required, $L \sim{\mathcal O}(10^{-2})$. In the presence of such a large asymmetry, the flavor conversions occur at lower temperature (a few MeV) around the neutrino decoupling time. Therefore active neutrinos are not repopulated anymore by collisions and their abundances are depleted. This causes distortions in the electron (anti-)neutrino spectra which directly participate in the weak interactions ruling the neutron-proton interconversion. The observable consequence is an alteration to the primordial abundances of light nuclei predicted by BBN. Recently a different model has been proposed by several authors~\cite{Hannestad:2013ana, Dasgupta:2013zpn} in order to achieve an analogous suppression of the flavor evolution through a new matter term in the equations of motion. This model is based on new secret interactions among sterile neutrinos only,\footnote{The cosmological impact of secret interactions among active neutrinos has been discussed, see e.g.~\cite{Archidiacono:2013dua}.} mediated by a new gauge boson. In the presence of the new interactions, flavor conversions are shifted at lower temperature and sterile neutrinos are produced resonantly by collisions. BBN and mass bounds  strongly constrain the model, disfavoring values of the mass of the new gauge boson $M_X > 1$~MeV~\cite{Mirizzi:2014ama, Saviano:2014esa}. Interestingly, if the new light boson ($M_X < 1$~MeV) also couples to DM~\cite{Dasgupta:2013zpn, Bringmann:2013vra, Chu:2014lja, Boehm:2014vja}, secret interactions will be able to solve the small-scale problems of the CDM scenario. Finally, we mention that new interactions mediated by a very light (or even massless) pseudoscalar might represent an escape route to the cosmological mass bounds~\cite{Archidiacono:2014nda}.

\subsubsection{keV-scale (Authors: A.~Boyarsky, O.~Ruchaisky)} 

It is the keV mass scale sterile neutrinos that could play a very prominent role in cosmology. A sterile neutrino is a neutral, massive particle and its lifetime can be very long (see the discussion below). Therefore it is a viable DM candidate. However, to constitute 100\% of the DM \emph{its mass should be above} $0.4$~{keV}. Indeed, as sterile neutrinos are fermions, they satisfy the so-called Tremaine-Gunn bound~\cite{Tremaine:1979we}, i.e., their phase space distribution in a galaxy cannot exceed that of the degenerate Fermi gas. This bound is very robust as it makes no assumption about possible distribution of DM particles inside a galaxy or even in the early Universe. One can simply estimate the total mass of a galaxy, its size, and the volume in velocity space, defined by escape velocity and then define the average phase space density within the object. These quantities are very robust and close to the direct observables. The phase space density appears to be the largest for dwarf satellite galaxies of the Milky Way (for review see e.g.~\cite{Walker:13}), so that these objects provide the strongest bounds. At the same time, dwarf galaxies are the most DM dominated objects which also reduces the uncertainty in the determination of the DM mass.  Still, there are appreciable astronomical uncertainties in deducing phase-space density from astronomical observations, and under reasonable assumptions they can be changed by a factor $\sim 2$, which has already been taken into account when quoting the above bound, see e.g.~\cite{Boyarsky:2008ju,Gorbunov:2008ka,Horiuchi:2013noa}.

To be a DM candidate, keV sterile neutrinos need to be produced efficiently in the early Universe. Since they cannot thermalize easily, the simplest production mechanism is via mixing with the active neutrinos in the primordial plasma~\cite{Dodelson:1993je}.  What is important, however, is that sterile neutrino DM is practically always produced out of thermal equilibrium, and therefore its primordial momentum distribution is in general \emph{not} given by a Fermi-Dirac distribution. Indeed, sterile neutrinos in equilibrium have the same number density as ordinary neutrinos, i.e., $112$ {cm${}^{-3}$. With the lower bound on the sterile neutrino mass being $0.4$~keV, this would lead to the energy density today $\rho_\text{sterile,\,eq} \simeq 45$~{keV/cm${}^3$}, which significantly exceeds the critical density of the Universe $\rho_\text{crit} = 10.5\,h^2$~{keV/cm${}^3$}. Therefore, sterile neutrino DM cannot be a thermal relic (unless entropy dilution is exploited, see Section~5), and its primordial properties are in general different from such a particle.

Assuming the validity of Big Bang theory already below $T\sim 1$~GeV,\footnote{This hypothesis has never been rigorously proven by any observation. Strictly speaking, the hot Big Bang model is only valid for temperatures below few MeV, corresponding to Big Bang Nucleosynthesis.} one can relate the sterile neutrino mass $m_N$ and the mixing angle $\theta^2$ needed to produce the correct DM abundance.  This relation depends on one additional parameter, namely the level of \emph{lepton--anti-lepton asymmetry} present in the plasma at the epoch of production. If the lepton asymmetry is significantly larger than the baryon asymmetry, i.e.\ if $\eta_L \equiv n_L/n_\gamma \gtrsim 10^6\eta_\mathrm{b}$ ($\eta_\mathrm{b} = n_\mathrm{b}/n_\gamma= 6\times10^{-10}$ is the baryon asymmetry of the Universe),\footnote{Usually $\eta_\mathrm{b}$ and $\eta_L$ are thought to be of the same order of magnitude. However, the observational constraints on $\eta_L$ are very weak (see e.g. \cite{Canetti:2012zc} and references therein) and allow for asymmetries that are many orders of magnitude larger than $\eta_\mathrm{b}$.} 
the mixing of sterile neutrino with active neutrinos can occur \emph{resonantly}~\cite{Shi:1998km,Abazajian:2001nj,Laine:2008pg,Venumadhav:2015pla} (similar to the MSW effect~\cite{Mikheev:1986gs,Wolfenstein:1977ue}). Otherwise, \emph{non-resonant production}~\cite{Dodelson:1993je,Asaka:2006nq} takes place.  The non-resonant production defines a smallest amount of DM that can be produced for given $m_N$ and $\theta^2$, while the resonance may (significantly) enhance it depending on the lepton asymmetry present in plasma. 

The lower bound on the DM mass becomes stronger once a particular production mechanism is specified and the corresponding primordial distribution of sterile neutrinos is taken into account. As the DM particles obey dissipationless Liouville dynamics (neglecting the baryonic feedback for a moment), the maximum of their phase-space distribution remains constant. The coarse-grained distribution (i.e., the only one that can be inferred from the astronomical observations) can only decrease this maximum, see sec.~4.1. This gives a bound on the mass of the sterile neutrino DM that is stronger by a factor of few than that of a degenerate Fermi gas, bringing the lower bound to $\sim 1$~keV or above.

A keV sterile neutrino mixes with ordinary neutrinos. In the presence of sterile neutrinos, the leptonic weak neutral current is nondiagonal in mass eigenstates~\cite{PhysRevD.16.1444}, so the $N$ can decay at tree-level via $Z$-exchange, as $N \to \nu_i \bar\nu_j \nu_j$, where $\nu_i$ and $\nu_j$ are mass eigenstates.  The necessary and sufficient condition for the leptonic weak neutral current to be diagonal in mass eigenstates is that all leptons of the same chirality and charge must have the same weak $T$ and $T_3$, which is violated in the presence of sterile neutrinos~\cite{PhysRevD.16.1444}. Its keV-scale mass makes the decay $N \to \nu_\alpha\nu_\beta\bar\nu_\beta$ possible (here $\alpha,\beta$ are neutrino flavors and all possible combinations of flavors are allowed, including charge conjugated decay channels as $N$ is a Majorana particle). The total decay width for $N\to 3\nu$ is given by~\cite{Pal:82,Barger:1995ty,PhysRevD.16.1444}:
\begin{equation}
  \label{eq:1}
  \Gamma_{N\to 3\nu} = \frac{G_F^2 m_N^5}{96\pi^3} \sin^2\theta =
  \frac1{4.7\times
    10^{10}{\,\ \textrm{sec}}}\left(\frac{m_N}{50 \,\ \textrm{keV}}\right)^5 \sin^2\theta,
\end{equation}
and one requires that the corresponding lifetime should be much longer than the age of the Universe. This imposes a bound on the mixing angle $\theta^2$~\cite{Dolgov:2000ew}:
\begin{equation}
  \label{eq:2}
  \theta^2<1.1\times10^{-7}\left(\frac{50 \,\ \textrm{keV}}{m_N}\right)^5\quad\text{--- lifetime longer than the age of the Universe}.
\end{equation}
Already this bound tells us that the contribution of the DM sterile neutrino to neutrino oscillations driven by $\delta m_\nu \sim m_N \theta^2$ becomes much smaller than the solar neutrino mass difference $\sqrt{\Delta m_\odot^2} \sim 10^{-4}$~eV)~\cite{Asaka:2005an,Boyarsky:2006jm}. Therefore, at least two more sterile neutrinos are required to explain to observed mass differences if the neutrino masses are due to exactly two sterile neutrinos, and the lightest active mass is (almost) zero -- a non-trivial prediction that can be checked by the Euclid space mission~\cite{Audren:2012vy}, by CMB-S4~\cite{Wu:2014hta}, or by DESI~\cite{Font-Ribera:2013rwa}.

A further bound on the mixing angle, much stronger than~\eqref{eq:2} comes from a one loop mediated \emph{radiative decay} $N\to \nu + \gamma$ that leads to a monochromatic X-ray line signal. If sterile neutrinos are produced via mixing (whether resonantly, or not) with ordinary neutrinos, their mass should be below about 50~keV~\cite{Laine:2008pg,Boyarsky:2009ix}, both to yield the correct DM abundance and not to produce a too strong decay line. Therefore, in the minimal setting, the mass of DM sterile neutrinos should be in the range of 0.4 -- 50~keV. Note that the lower bound is fairly universal (if 100\% of DM is made of this particle) and does not depend on the production mechanism, however, taking into account more specific information about the details of the DM production may be used to derive much stronger bounds from structure formation data.

If (part of) the DM particles are born relativistic, they stream out of the smallest overdensities, erasing structure on scales below their \emph{free-streaming horizon} $\lambda_\text{FS}$ (see e.g.~\cite{Boyarsky:2008xj} for a definition). This means that the power spectrum of sterile neutrino DM is suppressed at scales below $\lambda_\text{FS}$ (as compared to a CDM power spectrum with the same cosmological parameters). For example, the suppression of the power spectrum of density perturbations due to the non-zero masses of active neutrinos (that were relativistic and cooled down during the matter-dominated epoch) is at the few per cent level (see e.g.~\cite{Lesgourgues:2006nd}). The situation is different if particles were born relativistically, but cooled down and became non-relativistic while deeply in the radiation dominated. Such DM candidates are called \emph{warm} DM (WDM), although their spectrum may not even necessarily be thermal. The free-streaming of WDM particles modifies the primordial spectrum of density perturbations in the epoch when these perturbations are still linear. Once the WDM particles became non-relativistic, their evolution is governed by the same dynamics as that of conventional CDM. The suppression of power for WDM particles however occurs on smaller scales than for the massive active neutrinos (below a comoving Mpc or similar). These scales are significantly less constrained by the current cosmological data. Currently two major methods are used: (1) the Lyman-$\alpha$ forest (see e.g.~\cite{Hansen:2001zv,Viel:2006kd,Viel:2005qj,Seljak:2006qw,Boyarsky:2008xj,Boyarsky:2008mt,Viel:2013apy,Viel:2007mv}) and (2) studying the properties of sub-structures in the Local Group~\cite{Bode:2000gq,Schneider:2013wwa,Maccio:2012qf,Maccio':2009rx,Schneider:2011yu,Goerdt:06,Moore:1999gc,Lovell:2011rd,Horiuchi:2013noa,Kennedy:2013uta,Lovell:2013ola,Shao:2012cg}. These methods are subject to large statistical and systematic uncertainties~\cite{Boyarsky:2008xj}. Based on the current observational limits, the scales of interest for WDM range from 1 Mpc (comoving scale) to tens of kpc. At the Mpc scales the suppression of the power spectrum as compared to the standard ($\Lambda$CDM) prediction cannot be more than tens of per cent, but at smaller scales (order 100 kpc or below) the suppression can be stronger. Current constraints are discussed in a recent review~\cite{Boyarsky:2012rt}.

\begin{figure}
  \centering
\includegraphics[width=0.49\linewidth]{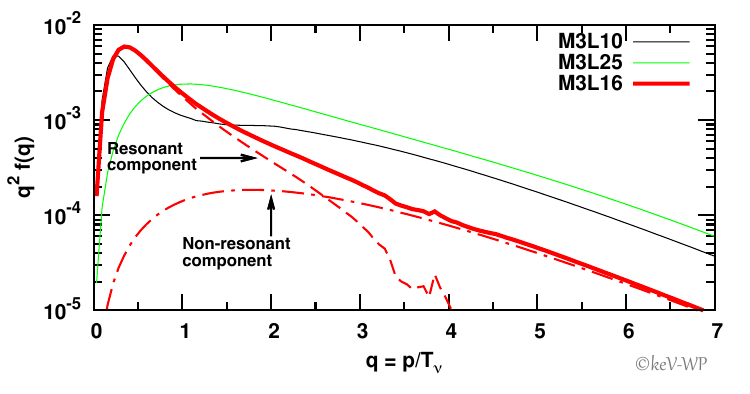}~\includegraphics[width=0.51\linewidth]{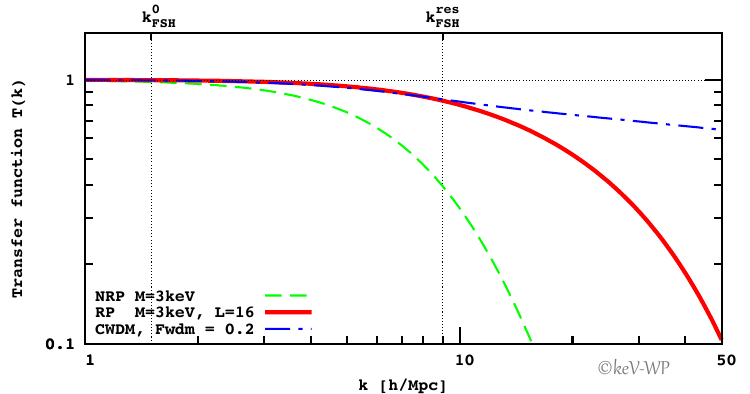}
\caption{\label{fig:rp-spectra}\textbf{Left panel:} Characteristic form of the resonantly produced (RP) sterile neutrino distribution function for $m_N =3$~keV and various values of the lepton asymmetry parameter $L_6 \equiv 10^6 (n_\nu - \bar n_\nu)/s$. The spectrum for $L_6=16$ (red solid line) is shown together with its resonant (dashed) and non-resonant (dot-dashed) components. All these spectra have the same shape for $q\gtrsim 3$. \textbf{Right panel:} Transfer functions for the spectrum with $m_N=3$~keV and $L_6=16$ together with a CWDM spectrum for $m_N=3$~keV and $F_\text{\sc wdm} \simeq 0.2$ (blue dashed-dotted line). Vertical lines mark the free-streaming horizon of non-resonant ($k_\textsc{fsh}$, left line) and resonant ($k_\textsc{FS}^{\rm res}$, right line) components. The green dashed line shows the transfer function for non-resonantly produced sterile neutrino with $m_N = 3$~keV. Reproduced from~\protect\cite{Boyarsky:2008mt}.}
\end{figure}

The production of the DM sterile neutrino via mixing becomes most efficient at temperatures $T\sim 150-500$~MeV~\cite{Dodelson:1993je,Dolgov:2000ew,Abazajian:2001nj,Asaka:2006nq} resulting in the population of ``warm'' DM particles. Resonant production, in turn, results into an efficient conversion of an excess of $\nu_e(\bar \nu_e)$ into DM neutrinos $N$~\cite{Shi:1998km,Laine:2008pg}. While non-resonant (NRP) oscillations result in (quasi-) thermal momentum distribution~\cite{Abazajian:2005gj,Asaka:2006nq}, resonantly produced (RP) sterile neutrinos are typically much colder and the dispersion of their momentum distribution is also much smaller than thermal (see fig.~\ref{fig:rp-spectra}, left panel). Therefore, resonantly produced sterile neutrinos behave in some sense like a \emph{mixture of a cold and warm DM (CWDM)} over some range of scales (see fig.~\ref{fig:rp-spectra}, right panel). As resonant production does practically not depend on the coupling constant between active and sterile neutrinos (that is why it is so efficient), for DM particles produced in that way this coupling is \emph{not} fixed by the DM abundance (as opposed to the non-resonant case) and it could even be smaller. The expected flux of X-ray photons, produced by decays of DM particles in galaxies and clusters is, on the contrary, always proportional to this coupling constant. Therefore, the bounds from X-ray observations are relaxed for resonantly produced sterile neutrinos. It has been demonstrated in~\cite{Boyarsky:2006fg,Boyarsky:2007ay,Boyarsky:2007ge,Boyarsky:2008xj,Boyarsky:2008mt,Lovell:2011rd} that resonantly produced sterile neutrino DM is fully compatible with all existing astrophysical and cosmological observations but, at the same time, it is ``warm'' enough to suppresses sub-structures in Milky-Way-size galaxies, consistent with observations~\cite{Lovell:2011rd}.  Therefore, the value of lepton asymmetry present in the plasma at these temperatures is an important parameter for such type of DM.  It defines the fraction of CDM (when the spectra are approximated as CWDM mixture) and also the lifetime of the DM particles.

\subsubsection{MeV-scale (Authors: S.~Pascoli, N.~Saviano)} 

Heavy sterile neutrinos $N_i$, $i=4, 5, ...$, with masses $M_i$ in the range ${\mathcal O}(1-100)$ MeV and mixing with active neutrinos (with $|U_{ai}| \ll 1$) emerge in low scale seesaw models in order to explain neutrino masses, as for instance in the $\nu$MSM (Neutrino Minimal Standard Model -- see~\cite{Boyarsky:2009ix} for a review). They have many possible experimental signatures depending on their masses and mixing angles.

Bounds on these heavy sterile neutrinos are obtained from several terrestrial experiments (as first obtained in~\cite{Shro80,Shrock:1980ct,Shrock:1981wq,SHROCK1982382}; for detailed reviews see ref.~\cite{Atre:2009rg} and also~\cite{Kusenko:2004qc,Ruchayskiy:2011aa,Drewes:2015iva}). For masses up to $\sim 3 \ \mathrm{MeV}$, kink searches~\cite{Shro80} pose model-independent bounds on $U_{e4}$ down to $|U_{e4}|^2 \lesssim 10^{-3}$ and for heavier masses the strongest limits are due to peak searches, looking for the spectrum of electrons in pion and kaon decays, as strong as $|U_{e4}|^2 \lesssim \mbox{few} \times 10^{-8}$ around (60--100)~MeV~\cite{Shro80,Shrock:1981wq,Shrock:1980ct,SHROCK1982382,PIENU:2011aa,Agashe:2014kda}. For heavier masses, constraints are obtained from looking for the decay products of these sterile neutrinos in accelerator experiments. If of Majorana nature, they would also induce neutrinoless double beta decay which currently provides the strongest bound, $|U_{e4}|^2 < 10^{-8}$ at 100~MeV. It should be pointed out that this latter limit could be evaded if a cancellation of the sterile neutrino contributions to the decay takes places, for instance due to the interplay among different sterile neutrinos or due to additional lepton number violating processes. For mixing with muons, peak searches~\cite{Shro80,Shrock:1980ct,Shrock:1981wq} provide the most stringent bounds on $|U_{\mu 4}|^2$ using pion decays, $|U_{\mu 4}|^2 \lesssim \mbox{few} \times 10^{-5}$, and kaon decays,  $|U_{\mu 4}|^2 \lesssim 10^{-4}$--$  10^{-9}$, depending on the heavy neutrino mass~\cite{Agashe:2014kda,Artamonov:2014urb}. Mixing with tau neutrinos is significantly less known, with bounds coming only from searches of decays of heavy neutrinos, $|U_{\tau 4}|^2 < 10^{-4}$ at 100~MeV. Furthermore,  since $N_i$  mix with active neutrinos, they can, in principle be emitted by core-collpase supernovae. In this context limits have been placed from the SN 1987A observation, requiring that the SN core should not emit too much energy in the $N_i$ channels, since this additional energy-loss would  shorten the observed neutrino burst~\cite{Dolgov:2000pj,Dolgov:2000jw}. Furthermore, in~\cite{Fuller:2009zz} the possible role of heavy sterile neutrinos in enhancing supernova explosions has been discussed. 

In this section, we focus on the constraints on $N_i$ that can be placed from cosmological arguments. The basic idea is that $N_i$ can be produced in the Early Universe by the mixing with active neutrinos, and then decay into lighter species. For sufficiently large mixing angles, they would reach equilibrium and decouple at a temperature higher than that of active neutrinos by a factor $ |U_{\alpha i}|^{2/3}$. For smaller mixing, they would be produced by decoherence in the active neutrino processes, and even a small fraction could lead to very significant effects. In particular, for masses in the range 1 MeV $< M_i  \lesssim  M_{\pi} \sim 140$ MeV, the main decay channels are $N_i \to 3 \nu$ and $N_i \to \nu e^+ e^-$ with a decay rate of~\cite{Pal:82}:
\begin{equation}
\Gamma_i \simeq \frac{1+ (g^a_L)^2 + g_R^2}{768 \pi^3} G_F^2 M_i^5 |U_{\alpha i}|^2\,\ ,
\end{equation}
where we have considered the neutral current contribution, and $g^e_L = 1/2 + \sin^2 \theta_W$, $g^{\mu, \tau} = -1/2 + \sin^2 \theta_W$, $g_R = \sin^2 \theta_W$. If $N_i$ mix with electron neutrinos, the decay $N_i \to \nu e^+ e^-$ would receive an additional contribution from the charged current interactions. The decay products of the sterile neutrinos are injected into the primordial plasma, increasing its temperature and shifting the chemical equilibrium. Depending on their mass and mixing, the $N_i$ would decay in different cosmological epochs affecting different observables. For masses heavier than that of the pion additional decay channels are kinematically allowed, e.g.\ $N_i \rightarrow \pi^0 \nu$, $N_i \rightarrow \pi^{\pm} e (\mu)$. They dominate with a much faster decay rate. Here, however, we restrict our discussion to heavy neutrino masses below the pion threshold.

The  decay of  $N_i$  into active neutrinos and other particles, occurring during or shortly before BBN, would influence the abundances of light elements  through indirect and direct effects. The decays  into electrons and positrons  will inject more energy into the electromagnetic part of the primeval plasma, thereby slowing down the cooling of the Universe and affecting the \emph{neutron-to-proton ratio} which rules the abundance of the primordial yields~\cite{Dolgov:2000pj,Dolgov:2000jw, Drewes:2015iva, Ruchayskiy:2012si}. Concerning the decay into neutrinos, the injection  can modify the spectra of the active species which directly participate in the charged current interaction governing the proton-neutron interconvention, and consequently the BBN products. Moreover, the decay products could  directly dissociate nuclei that have already formed. The most updated bounds from BBN have been discussed in~\cite{Ruchayskiy:2012si} where  it has been shown that, in the region of masses 1 MeV $< M_i  \lesssim 140$ MeV, primordial nucleosynthesis restricts the lifetime of sterile neutrinos to be well below 1~sec. For higher values of the mass of $N_i$ the bounds would relax since sterile neutrinos can decay also into heaver species, and the branching ratio into $3\nu$ becomes sub-dominant. 

If $N_i$ decay after BBN, one can make use of the Hubble constant, of Supernovae~Ia luminosity distances, of the Cosmic Microwave Background shift parameter, as well as of measurements of the Baryon Acoustic Oscillation scale to set bounds on the mixing parameters. Indeed, the decay of  additional heavy neutrino species can modify the Universe expansion rate between BBN and recombination, thereby causing entropy production diluting the density of ordinary neutrinos and producing extra radiation affecting the value of $N_{\rm eff}$. Specifically, they would produce a highly non-thermal component of active neutrinos whose distribution is dictated by the kinematics of the heavy neutrino decay and which is severely constrained by the amount of dark radiation allowed at recombination. Information on the early expansion rate, e.g.\ from Baryon Acoustic Oscillations, essentially imposes that the decay time needs to be faster than 0.1~s. A very detailed recent analysis in~\cite{Vincent:2014rja} finds that the bounds are as strong as $|U_{\alpha4}|^2 \lesssim 10^{-14}$--$10^{-17}$ for $M_i \sim 1 \ \mathrm{MeV}$--$100$ MeV.

These bounds are much stronger than those which can be obtained in terrestrial experiments, discussed briefly above. However, they can be evaded in nonstandard particle physics and/or cosmological scenarios. A possibility would be to consider a low-reheating model in which  the temperature at the end of  inflation, the so-called reheating temperature $T_{RH}$, is low enough $\ll$ 100 MeV~\cite{Gelmini:2008fq}. The abundance of sterile neutrinos becomes very suppressed with respect to that obtained within the standard framework, allowing to relax the previous bounds. For $M_i \gg 1 \ \mathrm{MeV}$, another option is to consider heavy sterile neutrinos which have additional interactions so that they decay much faster than in the standard case, $ \Gamma_i^{-1} \ll 1 \ \mathrm{s}$, and their decay products become thermalised before BBN.

\subsubsection{GeV--TeV-scale (Authors: A.~Ibarra)} 

The stringent limits on the mixing angle of sterile neutrinos with mass in the keV--MeV range reviewed in the previous section from Big Bang Nucleosynthesis or from the Cosmic Microwave Background totally disappear when the mass is larger than $\sim 1$ GeV. Nevertheless, heavy sterile neutrinos can still play an important role in cosmology. If the sterile neutrino is short lived on cosmological time scales, its out-of-equilibrium decay generates a cosmic lepton asymmetry provided $C$ and $CP$ are violated which requires the existence of more than one sterile neutrino state in Nature. Moreover, if the decays occur earlier than $\sim 10^{-12}$ s after the Big Bang, when the sphalerons are still in thermal equilibrium, the lepton asymmetry can be efficiently converted into a baryon asymmetry, thus providing an explanation for the matter-antimatter asymmetry observed in our Universe \cite{Canetti:2012zc}. This is the renowned leptogenesis mechanism~\cite{Fukugita:1986hr}, which will be reviewed in the next subsection. On the other hand, if the sterile neutrino is very long-lived on cosmological time scales, it contributes to the DM of our Universe. The longevity of the DM particle implies that the mixing angle with the active neutrinos must be tiny, which in turn implies that DM thermal production via active-sterile oscillations is highly inefficient. Thus, other mechanisms must then  be at work in order to generate the observed DM abundance  $\Omega_{\rm DM} h^2\simeq 0.12$, as measured by the {\it Planck} satellite~\cite{Ade:2013zuv}. Regardless of how the DM particle was produced, this scenario has phenomenological interest since the decay of the DM particles could lead to experimental signatures through the observation of an excess in the fluxes of cosmic antimatter, gamma-rays, or neutrinos with respect to the expected astrophysical backgrounds (for a review, see~\cite{Ibarra:2013cra}).

The most distinctive signal of sterile neutrino decay is the gamma-ray line at $E=M_N/2$ produced in the one-loop decay $N\rightarrow  \gamma \nu$. The predicted partial decay width reads~\cite{Pal:82,SHROCK1982359}:
\begin{align}
\Gamma_{\gamma \nu}=\frac{9\,\alpha_{\rm EM}\,G_F^2}{256\pi^4} \,\sin^2\theta \, m_N^5 = \frac{1}{6\times 10^{29}\,{\rm s}}\left(\frac{\sin\theta}{10^{-26}}\right)^2\left(\frac{m_N}{500\,{\rm GeV}}\right)^5\;,
\end{align}
where $\alpha_{\rm EM}=1/137$ and $G_F=1.166\times 10^{-5}\,{\rm GeV}^{-2}$. Gamma-ray lines have been searched for using data collected by the Fermi-LAT~\cite{Ackermann:2013uma}, as well as by  the H.E.S.S.~\cite{Abramowski:2013ax} and MAGIC telescopes~\cite{Aleksic:2013xea}. The strongest limits on the inverse decay width read $\Gamma^{-1}_{\gamma \nu}\gtrsim 10^{29}\,{\rm s}$ for a DM mass in the range of 10-600 GeV~\cite{Ackermann:2013uma} and $\Gamma^{-1}_{\gamma \nu}\gtrsim 4\times 10^{27}\,{\rm s}$ for $m_N=1-30$~TeV~\cite{Ibarra:2013cra}, with practically no sensitivity to the choice of the DM halo profile. These limits in turn translate, respectively, into $\sin \theta\lesssim 1.4\times 10^{-24}\left(M_N/100\,{\rm GeV}\right)^{-5/2}$ and $\sin\theta\lesssim 2.2\times 10^{-26}\left(M_N/1\,{\rm TeV}\right)^{-5/2}$.

Complementary limits on the mixing angle can be obtained from the tree-level decays $N\rightarrow W^\pm \ell^\mp$, $Z\nu$, when kinematically accessible, which produce a continuum flux of gamma-rays, antimatter particles, and neutrinos.. The decay widths of these processes read~\cite{Buchmuller:1991tu}:
\begin{eqnarray}
\Gamma_{W^+\ell^-}&=&\Gamma_{W^-\ell^+} =\frac{G_F}{8\sqrt{2}\pi} \, \sin^2\theta\, m_N^3
\left(1+2\frac{M_W^2}{m_N^2}\right)\left(1-\frac{M_W^2}{m_N^2}\right)^2\;, \\
\Gamma_{Z\nu}&=&\frac{G_F}{8\sqrt{2}\pi} \, \sin^2\theta\, m_N^3
\left(1+2\frac{M_Z^4}{m_N^4}\right) \left(1-\frac{M_Z^2}{m_N^2}\right)^2\;.
\end{eqnarray}
Numerically, for $M_N\gg M_W, M_Z$, we have:
\begin{align}
\Gamma_{W^+\ell^-}=\Gamma_{W^-\ell^+} \simeq \Gamma_{Z\nu}\simeq = \frac{1}{2\times 10^{26}\,{\rm s}}\left(\frac{\sin\theta}{10^{-26}}\right)^2\left(\frac{m_N}{500\,{\rm GeV}}\right)^3\;.
\end{align}

Limits on the DM decay width with gauge bosons in the final state were derived in~\cite{Ackermann:2012qk} from an analysis of the diffuse gamma-ray emission using Fermi-LAT data, in~\cite{Aleksic:2013xea} from observations of the dwarf galaxy Segue 1 using the MAGIC telescopes, and in~\cite{Dugger:2010ys} from Fermi-LAT observations of several nearby galaxies and clusters. Besides, limits from antiprotons were derived in~\cite{Garny:2012vt,Cholis:2010xb} from the non-observation of an excess in the antiproton-to-proton fraction measured by the PAMELA experiment~\cite{Adriani:2010rc}. Lastly, limits from positrons were derived in~\cite{Ibarra:2013zia} from the AMS-02 measurements of the positron fraction~\cite{Aguilar:2014fea} and the positron flux~\cite{Aguilar:2014mma}. All these different observations typically require $\Gamma^{-1}_{Z \nu}\gtrsim 10^{26}\,{\rm s}$ for $m_N\leq 10\,{\rm TeV}$, which translates into an upper limit on the mixing angle $\sin \theta\lesssim 1.2 \times 10^{-26}\left(M_N/500\,{\rm GeV}\right)^{-3/2}$. Tree level decays into weak gauge bosons then provide, for $M_N\lesssim 5$~TeV, the strongest limits on this scenario, due to the larger branching fraction. 

For larger masses, the strongest limits stem from the non-observation of the neutrinos produced in the two body decay $N\rightarrow Z \nu$. More specifically, limits on the high energy neutrino flux from AMANDA between $16\,{\rm TeV}$ and $2.5\times 10^3\,{\rm TeV}$~\cite{Achterberg:2007qp}, from IceCube between $340\,{\rm TeV}$ and $6.3\times 10^6\,{\rm TeV}$~\cite{Abbasi:2012cu,Abbasi:2011ji}, from Auger between $10^5\,{\rm TeV}$ and $10^8\,{\rm TeV}$ ~\cite{Abreu:2011zze} and from ANITA between $10^6\,{\rm TeV}$ and $3.2\times 10^{11}\,{\rm TeV}$~\cite{Gorham:2010kv} typically require  $\Gamma^{-1}_{W\ell, Z \nu}\gtrsim 10^{26}-10^{27}$ s for masses between $\sim 10$ TeV and the Grand Unification scale~\cite{Esmaili:2012us} , which translates into $\sin \theta\lesssim 1.4\times 10^{-28}\left(\frac{M_N}{10\,{\rm TeV}}\right)^{-3/2}$. Clearly, this scenario requires a very strong suppression of the mixing angle, assuming that the sterile neutrino constitutes to totality of the DM of the Universe. If the sterile neutrino contributes only to a fraction $f$ of the total DM density, the previous limits on the mixing angle should be scaled by a factor $f^{1/2}$.

\subsubsection{Leptogenesis (Author: P.~Di Bari)}

The same minimal extension of the SM that can account for neutrino masses and mixing, the (type~I) seesaw mechanism~\cite{Minkowski:1977sc,Yanagida:1979as,GellMann:1980vs,Glashow:1979nm,Barbieri:1979ag}, can also elegantly explain the observed matter-antimatter asymmetry of the Universe \cite{Canetti:2012zc} with leptogenesis~\cite{Fukugita:1986hr}. The out-of-equilibrium decays of heavy RH neutrinos  at temperatures above the electroweak symmetry breaking temperature $T_{\rm EW} \sim 100\,{\rm GeV}$, can produce a sizeable $B-L$ asymmetry that, first injected in the form of a lepton asymmetry, is then partly reprocessed by sphaleron processes into a baryon asymmetry able to reproduce the observed baryon-to-photon number ratio $\eta_\mathrm{b} \simeq (6.1 \pm 0.1)\times 10^{-10}$~\cite{Ade:2013zuv}. In the absence of strong mass degeneracies or fine tuned cancellations in the seesaw formula, the mechanism requires very high masses for the RH neutrinos producing the asymmetry,  $M_i \gtrsim 10^{9}\,{\rm GeV}$~\cite{Davidson:2002qv}. These can be either the lightest RH neutrinos, as in the original proposal~\cite{Fukugita:1986hr}, or the next-to-lightest RH neutrinos~\cite{DiBari:2005st,Vives:2005ra,Blanchet:2008pw}, or both~\cite{Barbieri:1999ma,Engelhard:2006yg,Bertuzzo:2009im,Antusch:2011nz}. In any case this lower bound,  even with flavor effects~\cite{Nardi:2006fx,Abada:2006fw} included~\cite{Blanchet:2006be}, also implies a similar lower bound on the reheating temperature $T_{RH} \gtrsim 10^{9}\,{\rm GeV}$. It is also interesting that there is an upper bound on the lightest neutrino mass $m_{1} \lesssim {\cal}(0.1)\,{\rm eV}$~\cite{Buchmuller:2003gz,Giudice:2003jh,Buchmuller:2004nz,Blanchet:2008pw} an upper bound that now cosmological observations, in particular the new results from the {\em Planck} satellite, indicate to be respected. 

Within a minimal extension of the SM the scale of leptogenesis can be lowered to the ${\rm TeV}$ scale thanks to a resonant enhancement of the $C\!P$ asymmetries if at least two RH neutrino masses are quasi-degenerate~\cite{Pilaftsis:1997jf}.  In recent years, in the LHC era, much attention has been devoted to scenarios of resonant leptogenesis that might produce signals at colliders  (e.g.~\cite{Pilaftsis:2005rv}) or large lepton flavor violating rates~\cite{Blanchet:2009kk}. These scenarios can be motivated and embedded within extensions of the SM beyond the minimal type I seesaw scenario such as the inverse seesaw ~\cite{Mohapatra:1986bd}. 

An interesting alternative scenario of leptogenesis  is offered by the possibility that the matter-antimatter asymmetry is generated by RH neutrino oscillations~\cite{Akhmedov:1998qx} instead of decays. This is realized within a  different region of the seesaw parameter space with RH neutrino masses of a few GeV.  The key point of leptogenesis from RH neutrino oscillations is that RH neutrinos also carry (chiral) asymmetries. These can be generated by RH neutrino oscillations and then transferred to the active sector by the Yukawa couplings in a way that the total asymmetry the sum of those in the RH and LH sector vanishes. The violation of total $B-L$ (involving both chiralities) is suppressed because the asymmetries are generated at temperatures that are much larger than the $B-L$-violating Majorana masses, but sphalerons only ``see'' the left chiral charges and therefore generate a net $B$-charge. In this regard, the scenario is similar to Dirac leptogenesis \cite{Dick:1999je}.
In the original proposal~\cite{Akhmedov:1998qx} it is essential that one of the two oscillating RH neutrinos thermalizes after sphaleron processes become inoperative. In this way, RH neutrino mixing can generate opposite asymmetries in the two RH neutrino species but only that in the thermalized RH neutrino is communicated to the active sector and then reprocessed into a baryon asymmetry. $C\!P$ violation in the RH neutrino mixing, requiring a non-vanishing value of the  Jarlskog invariant for RH neutrino mixing, is a necessary ingredient.

However, in~\cite{Asaka:2005pn} it was noticed that the same coupling between the LH and the RH neutrino sectors provides an additional  source of $C\!P$ violation that has to be taken into account. This source -- and not $C\!P$ violation in the RH neutrino mixing -- can be responsible for a generation of an asymmetry able to explain the observed one within the so called $\nu$MSM~\cite{Asaka:2005an}, where the mixing between the two heavier RH neutrinos generate the asymmetry and the lightest RH neutrino, with mass $M_1 \sim {\rm keV}$, can play the role of (warm) DM. In this way the model provided a picture where neutrino masses, mixing, DM and matter-antimatter asymmetry are all explained within just the minimal type I seesaw extension of the SM~\cite{Canetti:2012vf,Canetti:2012kh}. 

The scenario requires RH neutrino masses $M_{2,3} \sim (1$--$80)\,{\rm GeV}$ and highly degenerate ($|M_3-M_2| \sim 10^{-6}\,M_2$).  Leptogenesis occurs when the oscillating RH neutrinos are ultra-relativistic at $T_L \sim 10^4\,{\rm GeV}$. In the $\nu$MSM, the Dirac neutrino masses are below the ${\rm keV}$ scale and, therefore, neutrino Yukawa couplings are all smaller than $\sim 10^{-8}$, reintroducing somehow the problem of the smallness of neutrino Yukawa couplings that the seesaw mechanism was supposed to solve. However, despite this price to be paid, the intriguing feature is that the seesaw formula and the neutrino data enforce a mixing of the RH with LH neutrinos that is exactly what is needed in order to produce the correct amount of DM in the early Universe. This would happen at temperatures above ${\rm MeV}$, and thus prior to the onset of Big Bang Nucleosynthesis.  On the other hand a sufficient production also requires the presence of a large lepton asymmetry resonantly enhancing the active-sterile neutrino mixing. This can be generated  by the leptogenesis mechanism itself, though in this case even more fine tuned conditions  are required in order to have a much smaller baryon asymmetry reproducing the observed one~\cite{Shaposhnikov:2008pf}. 

If one does not require the lightest RH neutrino to be the DM particle, then the possibility to reproduce the correct asymmetry with leptogenesis from RH neutrino oscillations (together with correct light neutrino masses  and mixing parameters) occurs for a larger parameter space. In particular RH neutrino masses can be much heavier than a few GeV~\cite{Garbrecht:2014bfa} and there is no need for a degeneray in the heavy neutrino masses~\cite{Drewes:2012ma}. This opens up the possibility to find the heavy neutrinos that generated the BAU at existing experiments~\cite{Drewes:2012ma} or at the proposed SHiP experiment~\cite{Alekhin:2015byh}. In principle, even the relevant source of CP-violation may be probed~\cite{Canetti:2012kh,Cvetic:2014nla}, allowing to unveil the origin of matter in experiments. However, it has also been noticed that in this case a fine tuning in the seesaw formula is then necessarily required anyway~\cite{Shuve:2014zua}.
 
An alternative possibility to combine DM and Leptogenesis is to fully decouple one of  the RH neutrinos that would then play the role of DM. This  has to be induced through physics beyond a minimal type~I seesaw extension, for example from inflation or introducing the 5-dim effective operator~\cite{Anisimov:2006hv,Anisimov:2008gg}:
\begin{equation}
\hat{O} = {\lambda_{AB}\over \Lambda_{\rm eff}}\,|\Phi|^2 \, \bar{N}^{c}_A \, N_\mathrm{b}  ,
\end{equation}
where $\Phi$  is the SM Higgs field, $\lambda_{AB}$ are dimensionless couplings, and $\Lambda_{\rm eff}$ is the scale of new physics, a simple example being a  Higgs portal operator. Leptogenesis could then proceed within a 2 RH neutrino mixing scenario~\cite{Antusch:2011nz}.

Many leptogenesis models relying on extensions of the minimal type I scenarios have been also proposed (for a recent review see~\cite{Hambye:2012fh}). In some cases these models also offer new possibilities to explain simultaneously DM the and matter-antimatter asymmetry. 

As we discussed, leptogenesis scenarios at low scale, as in the inverse seesaw or in the $\nu$MSM, have the advantage that they can be probed with direct experimental tests that are already producing important constraints on the parameters of the models. In this respect the LHC14 stage will be extremely exciting.

It should be said, however, that also high energy scale leptogenesis scenarios might be testable with low energy neutrino experiments within some specific frameworks. An interesting example is provided by $SO(10)$-inspired leptogenesis~\cite{Vives:2005ra,Branco:2002kt,Akhmedov:2003dg,DiBari:2008mp,DiBari:2010ux}, which produces interesting testable constraints on low energy neutrino parameters, especially when a condition of independence of the initial conditions is imposed (strong thermal leptogenesis)~\cite{DiBari:2013qja,DiBari:2014eqa,DiBari:2014eya}. 

In conclusion we have fully entered an exciting stage where finally we are starting to test phenomenological consequences from leptogenesis scenarios. 


%% file: kevnuwp_section3.tex


Dark Matter particles were created (or have decoupled from the primordial plasma) deeply in the radiation-dominated era.  Current constraints extracted from cosmological observations of, e.g., galaxy clustering or the Cosmic Microwave Background (CMB) anisotropies, rule out the possibility that a significant fraction of Dark Matter particles remained relativistic up until the matter-radiation equality time. These observations, however, still allow for Dark Matter to possibly retain significant primordial velocities deeply in the radiation-dominated epoch.  Such Dark Matter models, often labeled as \emph{Warm Dark Matter}, WDM as opposed to its non-relativistic counterpart ``cold'' Dark Matter CDM, are still allowed by current data on CMB at least at large scales, where they are about indistinguishable.

The difference between cold and warm Dark Matter particles would appear at approximately galactic scales (co-moving Mpc or below) and it is only recently that such small-scale effects are starting to be resolvable both observationally and numerically. 

The power spectrum at (sub)-Mpc scales can be probed indirectly by the Lyman-$\alpha$ forest method (see Sec.~\ref{sec:4-2-LymanA}). A number of other astronomical observables of both the local and the high redshift Universe can be sensitive to modifications of the primordial power spectrum. Examples are the abundance and internal structure of dwarf galaxies, the formation of first stars, or the process of reionization. Some of these observations have been claimed to be in tension with the $\Lambda$CDM ``concordance'' model, and WDM scenarios have been proposed as promising alternatives.

In this section we will critically discuss some of the main issues around the nature of Dark Matter and how this can be probed via a variety of astrophysical and cosmological tests. Our aim is to go through the main ``challenger'' of small scale structure formation, providing an overview of the current state of research and discussing how sterile neutrino DM could alter this picture.

In particular, we will first provide in Sec.~\ref{sec:indentityDM} an overview of the state-of-the-art in this field, giving a brief but complete framework of the main features of warm versus cold DM and their validation via galactic observables. We will then move on to discuss the specifics of each of the topics of current interest to the community. We will start in Sec.~\ref{sec:missing-satellites} by analyzing the so-called ``missing satellite problem'', in light of the most recent discoveries and theoretical approaches. We will continue in Sec.~\ref{sec:core-cusp} by emphasizing the role of CDM and WDM in shaping the inner regions of galactic and host halo structures. Sec.~\ref{sec:TBTF} will instead be specifically devoted to a thorough examination of one of the most recent and puzzling problems, the so-called ``Too-big-to-fail'' problem, also in light of the most recent available data on the rotation curves and number densities of dwarf galaxies in different environments. We then conclude our journey in Sec.~\ref{sec:kinematicWDM} in which we consider the kinematics and formation of subhaloes in WDM cosmologies.

\subsection{\label{sec:indentityDM}Astrophysical clues to the identity of the Dark Matter (Author: C.~Frenk)}

Cold, collisionless particles are the most popular candidates for the DM. There are sound reasons for this, both from the points of view of astrophysics and particle physics. The current standard model of cosmology, $\Lambda$CDM (where $\Lambda$ stands for Einstein's cosmological constant and CDM for cold DM), is based on this idea. $\Lambda$CDM accounts for an impressive array of data on the structure of the Universe on large scales, from a few gigaparsecs down to a few megaparsecs, where the cosmic microwave background (CMB) radiation and the clustering of galaxies provide clean and well-understood diagnostics. Particle physics provides well-motivated CDM candidates, most famously the lightest supersymmetric particle, which would have the relic abundance required to provide the measured DM content of our Universe.

In recent years, an altogether different kind of elementary particle, a sterile neutrino, has emerged as alternative plausible candidate for the Dark Matter. These particles are predicted in a simple extension of the Standard Model of particle physics and, if they occur as three, could explain the observed neutrino oscillation rates as well as baryogenesis. In the neutrino Minimal Standard Model one of them would have a mass in the keV range and behave as warm WDM. These particles would be relativistic at the time of decoupling, and subsequent free streaming would damp primordial density fluctuations below some critical cutoff scale which varies inversely with the particle mass and, for keV-mass thermal relics, corresponds to a dwarf galaxy mass.

\begin{figure}[t]
\begin{center}
\includegraphics[trim=0 3cm 0 5cm,width=\textwidth]{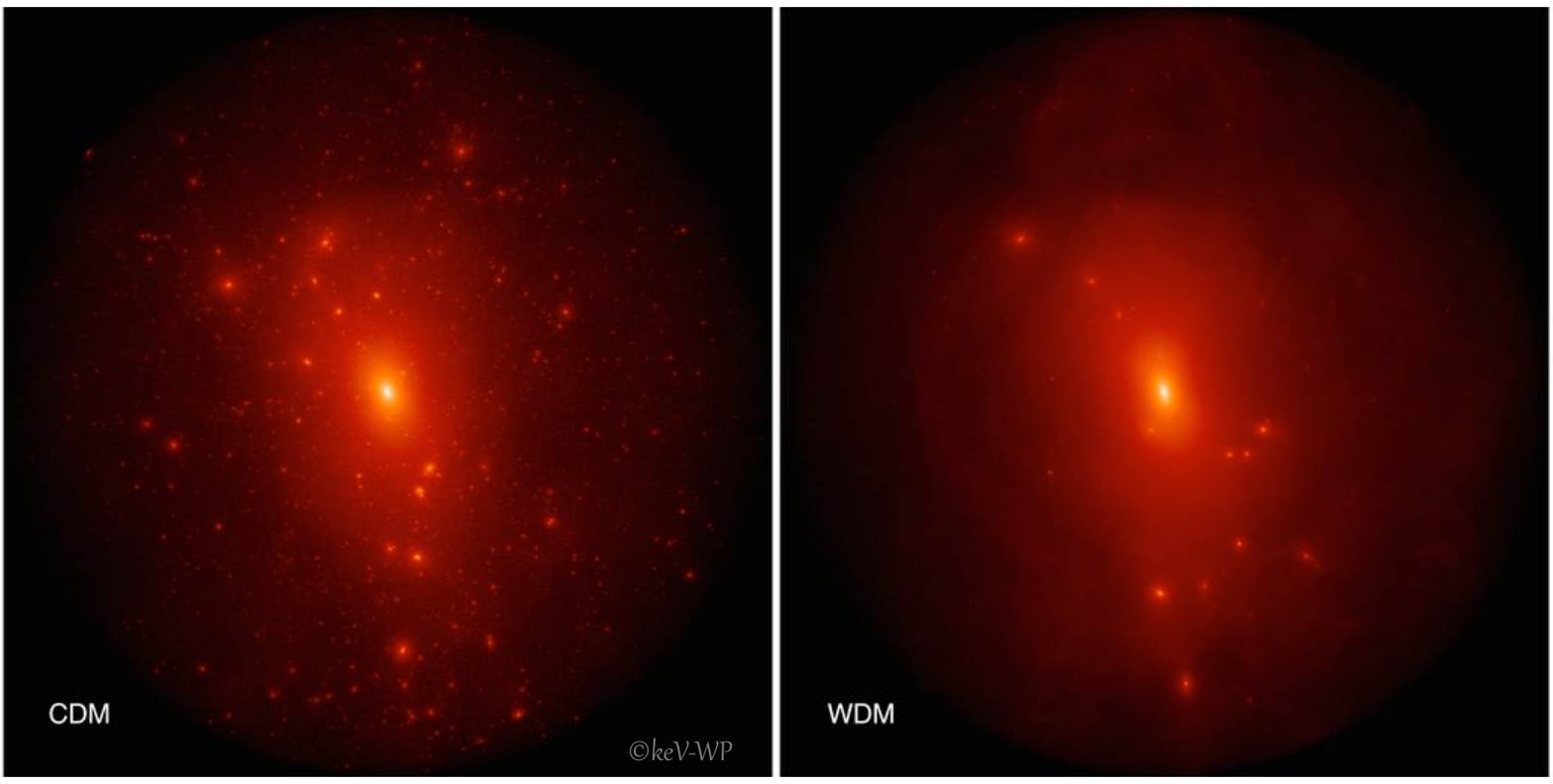}
\end{center}
\caption{\label{Frenk:fig}$N$-body simulations of galactic haloes in universes dominated by CDM (left) and WDM (right; for a particle mass of 2~keV). Intensity indicates the line-of-sight projected square of the density, and hue the projected density-weighted velocity dispersion. (Figure similar to Fig.~3 in Ref.~\cite{Lovell:2011rd}.)}
\end{figure}

CDM and WDM models produce very similar large-scale structure and are thus indistinguishable by standard CMB and galaxy clustering tests. However, as is clearly shown in Fig.~\ref{Frenk:fig}, they produce completely different structures on scales smaller than dwarf galaxies, where the WDM primordial power spectrum is cut off whereas the CDM power spectrum continues to increase (logarithmically).

The search for the identity of the Dark Matter is currently at a very exciting stage, with tentative, mutually exclusive, claims for detections of CDM~\cite{Hooper:2010mq} and WDM~\cite{Bulbul:2014sua}. While eagerly awaiting developments on the Dark Matter particle detection front, it is incumbent upon astrophysicists to develop tests that could distinguish between the two frontrunner candidates. A glance at the figure suggests that such a test should not be too difficult: the two simulated galactic Dark Matter haloes look completely different. In CDM, there is a very large number of small subhaloes whereas in WDM only a handful of the most massive ones are present. Surely, it cannot be beyond the ingenuity of observational astronomers to tell us whether the Milky Way resides in a halo like that on the left or in one like that on the right of the figure!

It turns out that distinguishing CDM from WDM merely by counting the number of small satellites orbiting the Milky Way (or similar galaxies) is not possible once galaxy formation is taken into account. This is because there are two physical processes that inhibit the formation of galaxies in small haloes: the reionization of hydrogen in the early Universe heats up gas to a temperature higher than the virial temperature of small haloes (a few kilometers per second) and, in slightly larger haloes where gas can cool, the first supernovae explosions expel the remaining gas. Models of galaxy formation going back over 10 years, see e.g.~\cite{Bullock:2000qf,Benson:2001at} and more recent simulations, confirm this conclusion. In fact, if anything, requiring that a galactic halo should contain at least as many subhaloes as there are observed satellites in the Milky Way sets a lower limit on the mass of WDM particle candidates~\cite{Kennedy:2013uta}. A similar limit can be placed from the degree of inhomogeneity in the density distribution at early times as measured by the clustering of Lyman-$\alpha$ clouds~\cite{Viel:2013apy}.

Other tests based on the observed properties of Milky Way satellites also fail to distinguish CDM from WDM. In CDM, haloes and subhaloes have cuspy ``NFW''~\cite{Navarro:1996gj} density profiles, while in WDM one would expect cores to form in their inner regions. However, detailed calculations show that such cores would be much too small to be astrophysically relevant~\cite{Shao:2012cg}. Another test is offered by the ``too-big-to-fail'' argument~\cite{BoylanKolchin:2011de} that CDM $N$-body simulations predict a population of massive, dense subhaloes whose structure is incompatible with the observed kinematics of stars in the brightest Milky Way satellites. WDM models neatly avoid this problem because the inner densities of subhaloes are slightly less dense that those of their cold Dark Matter counterparts, on account of their more recent formation epoch~\cite{Lovell:2011rd} (which has also shown in connection to the 3.5~keV hint~\cite{Horiuchi:2015qri}). However, baryon physics produce similar effects in CDM subhaloes~\cite{Sawala:2014baa}

At present, the best strategy for distinguishing CDM from WDM would seem to be the use of gravitational lensing. There are two complementary techniques that could reveal the presence of the myriad small subhaloes predicted to exist in the haloes of galaxies by CDM simulations: flux ratio anomalies and distortions of giant arcs. Whether these techniques will work in practice is still an open question.

\subsection{\label{sec:missing-satellites}Missing dwarf galaxies (Author: N.~Menci)}

The abundance and the properties of low-mass galaxies and in particular of satellites  have long been constituting a challenging issue for CDM galaxy formation models. The latter connect the physical processes involving gas and star formation to the merging histories (and hence to the power spectrum of density perturbations) of Dark Matter haloes~\cite{Kauffmann:1993gv,Cole:1994ab,Monaco:1999cu,Kauffmann:1999ce,Granato:2003ch,Menci:2006me,Croton:2005fe,Bower:2005vb}.

\begin{figure}
\centering
\hspace{-1.cm}
\includegraphics[scale=0.8]{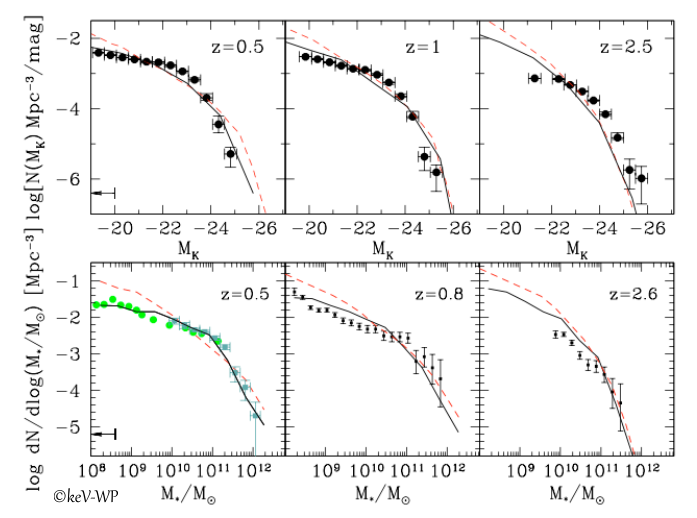}
\includegraphics[scale=0.83]{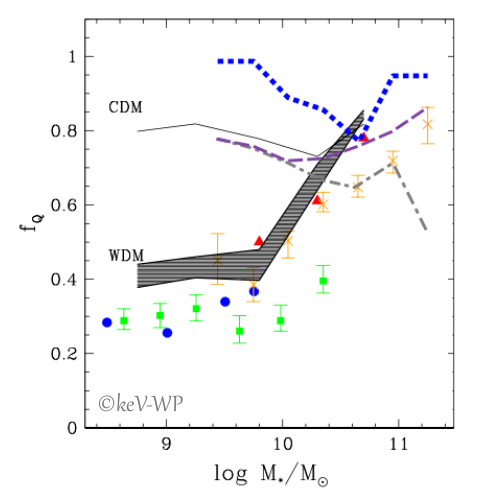}
\caption{\emph{Top Left Panel}: The evolution of the K-band luminosity functions in the WDM cosmology (solid line) is compared with data from~\cite{Cirasuolo:2008en}. Dashed line refer to the standard CDM case. \emph{Bottom Left Panel}: The evolution of the stellar mass function in the WDM cosmology (solid line) is compared with the standard CDM case (dashed); data are from~\cite{Drory:2004eh} (squares),~\cite{Fontana:2006xg}, and~\cite{Santini:2011fr}. In both the upper and the lower left panels, the arrows show the range of magnitudes and stellar masses corresponding to the free streaming mass. \emph{Right Panel}: Satellite quiescent fraction at different stellar masses. Data are from~\cite{Wetzel:2012nn}, red triangles;~\cite{Phillips:2013lca}, green squares; Ref.~\cite{Geha:2012nq} and~\cite{Wheeler:2014ega}, blue dots;~\cite{Kimm:2008rp}, yellow crosses. These are compared with predictions from a WDM model \cite{Calura:2014pla} and from different CDM models in the literature: the MORGANA model \cite[ grey dot-dashed line]{Monaco:2006df}; the Millenium-based SAM~\cite[blue short-dashed line]{DeLucia:2005yk}; the~\cite{Somerville:2006wp} code (purple dashed line); the Millenium-based model~\cite{Guo:2010ap}, petrol solid line (taken from~\cite{Phillips:2013lca}); the Rome model~\cite{Calura:2014pla}, solid line.}\label{Menci_Fig1}
\end{figure}

The large amplitude of  the CDM spectrum on small mass scales $M\lesssim 5\times10^9\,M_{\odot}$, a characteristic of dwarf galaxies, results into the following critical issues for galaxy formation models:

\begin{enumerate}

\item[(a)]The predicted abundance of local low-luminosity dwarfs exceeds the observed values unless a strong energy feedback -- heating the gas and  suppressing star formation -- is assumed , see e.g.~\cite{Somerville:1998bb,Cole:2000ex,Poli:2001fq}. However, at high redshifts the larger escape velocities of galaxies (due to their larger densities) make the feedback inefficient in suppressing star formation even in low-mass objects. This results into an over-prediction (of factors $\sim 5$) of faint galaxies increasing with redshift  when compared, e.g., to the evolution of the K-band luminosity function for $1\lesssim z\lesssim 3$~\cite[see][and Fig.~\ref{Menci_Fig1}]{Cirasuolo:2008en} or with the faint Lyman-break galaxies~\cite{Faro:2009si} at $z\gtrsim 3$; the excess of star-forming low-mass galaxies at $z\gtrsim 3$ reflects into an excess of red low-mass galaxies at $z\approx 0$~\cite{Croton:2005fe,Salimbeni:2007dq}.

\item[(b)] The number of galaxies with stellar mass $M_*\lesssim 10^{10} M_{\odot}$  in the Universe is systematically overpredicted (by a similar factor) by all theoretical CDM models~\cite{Fontana:2006xg,Fontanot:2009sy,Marchesini:2008xk,Guo:2010ap} in the whole redshift range $1\lesssim z\lesssim 3$~\cite{Santini:2011fr}.

\item[(c)]The local distribution of galaxy rotation velocities (velocity function) predicted by CDM models is characterized by an excess of galaxies in the low-velocity end~\cite{Papastergis:2011xe}. Since the rotation velocity directly probes the depth of the potential wells rather than the their gas or stellar content, the velocity distribution is less prone to the effect of energy feedback (which enters only in  selection effects or in determining the shape of the velocity curve), this is considered as a challenging issue for solutions based on feedback effects.

\end{enumerate}

\noindent As for satellite galaxies, long standing problems of CDM scenario (or, at least, points that require the improvement in the implementation of thedifferent baryonic
processes) include:

\begin{enumerate}

\item[(d)] The mass and luminosity distributions of  satellite galaxies predicted by simulations and by semi-analytic models tend to exceed the observed abundances~\cite{Klypin:1999uc,Moore:1999gc,Simon:2007dq}.  The improved implementation of processes involving baryons -- such as heating or winds caused by Supernovae explosions or by the UV background -- may be effective in  suppressing star formation (SF) in the low-mass haloes of satellite galaxies at low redshifts $z\leq 1$~\cite{Bullock:2000wn,Benson:2001at,Somerville:2001km,Governato:2006cq,Mashchenko:2007jp,Maccio':2009dx} thus reducing the discrepancy. However, recent observational results for larger samples of central galaxies are still in tension with the predictions of  CDM models~\cite{Nierenberg:2013lqa}. In addition, reducing the number of detectable satellites through feedback processes results in hosts that are too massive to be accommodated within measured galactic rotation curves~\cite{Papastergis:2014aba}, and it still does not avoid over-producing low-mass galaxies at redshifts $z\gtrsim 1$.

\item[(e)] CDM models predict most (in fact, more than 80\%~\cite{Kimm:2008rp}) of satellite galaxies to be quiescent, with low specific star formation rates, ${\rm SSFR}\leq 10^{-11}$ yr$^{-1}$, while observations indicate a much lower fraction $\sim 30$\%~\cite{Geha:2012nq,Wetzel:2012nn,Phillips:2013lca,Slater:2014yca,Wheeler:2014ega} for satellite galaxies with stellar mass in the range $10^8\,M_{\odot}\leq M_*\leq 10^{10}\,M_{\odot}$, as shown in Fig.~\ref{Menci_Fig1} (right). The discrepancy is generally ascribed to the implementation of the various baryonic processes contributing to the quenching of star formation in satellites, like the stripping of hot gas~\cite{Larson:1980mv,Cole:2000ex,Balogh:2000sf,Kawata:2007gk,McCarthy:2007ff} due to the pressure of the gas in the host halo. However the persistence of such a problem in the most recent semi-analytic models~\cite{Hirschmann:2012xp,Bower:2011aa,Weinmann:2012za} and $N$-body hydrodynamical simulations~\cite{Dave:2011pn,Weinmann:2012za,Hirschmann:2013ivp} indicates that it constitutes a major challenge for all CDM galaxy formation models, and a fundamental problem in understanding the evolution of low-mass galaxies~\cite{Weinmann:2012za}. A solution has been recently sought in the implementation of star formation~\cite{Wang:2011gma}, in reincorporation time-scales of the hot gas ejected by SNae winds~\cite{Henriques:2012ku}, or in a delayed and gradual stripping of hot gas from satellites~\cite{Simha:2008hd,Guo:2010ap,Font:2008pc}, but none of the proposals proved to be completely successful. This is because a gentle mode of strangulation should be at work in order to support cooling and star formation in satellite galaxies for several Gyr. Phenomenological estimates of  the quenching time scales required to comply with the observed quiescent fraction of satellites yield values in the range 3--6~Gyr~\cite{DeLucia:2011ac,Coe:2012kj,Wetzel:2012nn,Hirschmann:2014jta}, and is not easy to obtain such a delayed and gentle quenching time scale from a physical modelling of gas stripping.

On the other hand, Ref.~\cite{Fillingham:2015vya} has shown that the long quenching time scales may be understood in terms of a starvation scenario. However, this paper mainly deals with the satellites in the local group and it is meant to explain the small values of the quiescent fraction in such a group, while the problem is that -- when all the host haloes are considered -- the CDM quiescent fraction is instead too high. Still they show that starvation can actually be relevant for satellites with masses larger than $10^8 M_\odot$, while for smaller galaxies the authors suggest ram-pressure stripping to be more efficient in getting -- however -- short quenching times (i.e., still below 2~Gyr).

\end{enumerate}

While a solution in terms of refined treatment of baryonic processes is still possible, several authors have considered the possibility that the problem is rooted in the CDM power spectrum, and explored the effects of suppressed small-scale perturbations. Alternative DM models such as warm and mixed DM would naturally lead to a suppression of the power spectrum at small scales, resulting in a reduced dwarf galaxy abundances~\cite{Benson:2012su,Schneider:2014rda}. This could be in better agreement with observations~\cite{Maccio:2012qf,Schneider:2011yu,Menci:2012kk}, as shown in Fig.~\ref{Menci_Fig1} (left panels) for the case of a WDM spectrum corresponding to a thermal relic particle $m_X=1$ keV. Assuming a WDM spectrum also yields a quiescent fraction of satellite dwarf galaxies in agreement with observations. In fact, the delay in the average collapse time of the first structures in WDM scenarios results into a prolonged star formation history in small mass objects, thus yielding a larger fraction of star-forming low-mass satellites at low redshifts compared to the CDM case~\cite{Calura:2014pla}, as shown in the right panel of Fig.~\ref{Menci_Fig1}.\footnote{Note that the result on the right panel has not been published, but the quiescent fraction for this case was instead computed specifically for this White Paper. However, the model adopted is exactly that used to compute the results in ref.~\cite{Calura:2014pla}. All details on the model can be found therein. }

The improvement in the dwarf and satellite abundances and passive fraction compared to the CDM case illustrated in the figure requires a WDM suppression of the CDM power spectrum corresponding to thermal relics $1\lesssim m_X\lesssim 3$ keV~\cite{Menci:2012kk,Schneider:2013wwa,Calura:2014pla,Dayal:2014nva}. Indeed, a thermal relic mass $m_X=1.5$ keV, corresponds to that produced by (non-thermal) sterile neutrinos with mass $m_X\approx 6-7$ keV (depending on the production mechanism). These constitute the simplest candidates, see e.g.~\cite{Abazajian:2014gza} for a Dark Matter interpretation of the origin of the recent unidentified X-ray line reported in stacked observations of X-ray clusters with the XMM-Newton X-ray Space telescope with both CCD instruments aboard the telescope, and the Perseus cluster with the Chandra X-ray Space Telescope~\cite{Bulbul:2014sua,Boyarsky:2014jta}.  Such a value is consistent with lower limits ($m_X\geq 1$ keV for thermal relics at 2$\sigma$ level~\cite{Pacucci:2013jfa}) derived from the density of high-redshift galaxies set by the two objects already detected at $z\approx 10$ by the Cluster Lensing And Supernova survey with Hubble (CLASH). However the thermal relic equivalent mass of sterile neutrinos inferred from X-ray observations is only marginally consistent with the limits set by different authors ($m_X\geq 2.3$ keV~\cite{Polisensky:2010rw}; $m_X\geq 1.5$, ~\cite{Lovell:2013ola}; $m_X\geq 1.8$,~\cite{Horiuchi:2013noa}) from the abundance of ultra-faint Milky Way satellites measured in the Sloan Digital Sky Survey, see  e.g.~\cite{Belokurov:2010rf}, and definitely in tension with the constraint derived by comparing small scale structure in the Lyman-$\alpha$ forest of high-resolution ($z > 4$) quasar spectra with hydrodynamical $N$-body simulations which yield bounds ranging from $m_X\gtrsim 1.8$ to $m_X\gtrsim 3.3$ keV~\cite{Seljak:2006qw, Viel:2006kd, Viel:2007mv, Viel:2013apy}. Note however that various uncertainties may still affect the constraints (see discussions in~\cite{Abazajian:2011dt,Watson:2011dw,Schultz:2014eia}): the comparison of subhaloes to Milky Way dwarfs assumes a factor $\approx  4$ correction for the number of dwarfs being missed by current surveys, and lower correction factors would appreciably weaken the constraints; Lyman-$\alpha$ is also a challenging tool, as it requires disentangling the effects of pressure support and thermal broadening from those caused by DM spectrum, as well as assumptions on the thermal history of the intergalactic medium and of the ionizing background.

\subsection{\label{sec:core-cusp}Inner density profiles of small galaxies and the cusp-core problem (Author: W.~Evans)}

Cosmological $\Lambda$CDM simulations provide galaxy-scale structures that are too compact. One manifestation of this is that the inner density profiles of simulated haloes follow a power-law cusp $\rho \sim r^{-\alpha}$ with $\alpha \approx 1$.  This result was first obtained by \cite[][henceforth NFW]{Navarro:1996gj}, and it has survived the mammoth increases in computing power and resolution over the two succeeding decades, see e.g.~\cite{Moore:1999gc,Diemand:2005wv}. Cusps are a fundamental feature of dissipationless simulations, arising from the intrinsic coldness of the velocity distributions.

As this is an unambiguous prediction of $\Lambda$CDM, much effort has been expended on testing it. Large galaxies like the Milky Way are dominated by baryons in the center~\cite{Binney:2001wu,Bissantz:2001wx}. The uncertain effects of the growth of baryonic disks, bars, and bulges may have altered and homogenized the Dark Matter distribution, possibly erasing any cusps. So, the best testing ground are dwarf and low surface brightness galaxies, as here any effects of baryonic feedback are smaller.

Dwarf spirals and low surface brightness galaxies are usually endowed with abundant neutral hydrogen. Here, the gas rotation curve can be directly linked to the matter distribution, dominated by Dark Matter. Possible problems may include non-circular motions, the ambiguous position of the center and the resolution of the data~\cite{vandenBosch:2000rza}. However, high resolution H$\alpha$ rotation curves, where available, show excellent consistency with gas rotation curves~\cite{Marchesini:2002vm}, suggesting beam smearing is not a significant obstacle. Recent surveys like THINGS (``The HI Nearby Galaxies Survey'') have acquired very high spectral and spatial resolution data form the VLA (Very Large Array) on 34~galaxies. The quality of this data has overcome most of the problems and led to robust conclusions. For example,~\cite{Oh:2008ww} studied the dIrr galaxies IC~2574 and NGC~2366, correcting for non-circular motions, and concluded that both are best described by dark halo models with a kpc sized density core. Ref.~\cite{Oh:2010ea} looked at seven dwarf galaxies in the THINGS sample and showed that the rotation curves are gently rising in the inner parts, consistent with core-like models with $\alpha = -0.29 \pm 0.07$, significantly different from the $\alpha = -1$ value from dissipationless simulations. This situation seems to persist in low surface brightness (LSB) galaxies as well. This is a particularly important regime to study as the large extent and low stellar density of LSB galaxies does not readily permit the Dark Matter to be re-distributed via feedback from star formation. Ref.~\cite{deBlok:2001mf} presented 30 high resolution HI rotation curves for LSB galaxies. They fitted the data with both cored pseudo-isothermal and cusped models, and found the former to be preferred. Even for the cusped fits, the distribution of concentrations has too low a mean to match the predictions of simulations.

Dwarf spheroidal (dSph) galaxies have also been studied intensively. These galaxies are highly DM-dominated, but have little or no HI gas, and so the mass profile must be deduced from stellar dynamical modelling. This procedure is susceptible to well-known difficulties, such as the mass-anisotropy degeneracy~\cite{ma-an-de}. However, dSphs have the advantage that they are some of the closest galaxies to the Milky Way, and so samples of radial velocities of bright giant stars can be easily gathered. The size of such samples has increased from hundreds to thousands in recent years~\cite{Kleyna:2001us,Walker:2009zp}. It was soon realized that, for a single stellar population, only the mass within the half-light radius is well-constrained~\cite{Wolf:2009tu}, and the data cannot discriminate between cores and cusps~\cite{Evans:2008ik}. However, many dSphs are built from multiple populations of stars, typically at least metal-rich and metal-poor ones. If the surface brightness and line-of-sight kinematics of two or more populations can be measured, then the degeneracy in the data is broken. Ref.~\cite{Walker:2011zu} used statistical techniques to separate the metal-rich and metal-poor populations in the Fornax and Sculptor dSphs. Using a simple estimator for the mass enclosed within the half-light radius of each population, they demonstrated that the central density slopes were shallower than NFW and consistent with cores of size several hundred kpc. Ref.~\cite{Amorisco:2011hb} used full phase space distribution functions to model the multiple populations in Sculptor and demonstrated that cored halo models are very strongly preferred over cusped ones. Subsequently, Ref.~\cite{Agnello:2012uc} found a way of recasting the arguments in terms of the virial theorem, which has the benefit of robustness and generality. Ref.~\cite{Amorisco:2012rd} applied the virial method to Fornax and found a core size of $1^{+0.8}_{-0.4}$ kpc. At least for two of the largest dSphs around the Milky Way, therefore, the data point \emph{unambiguously} towards cores. Although~\cite{Strigari:2014yea} did construct fits to Sculptor using NFW haloes, this work remains open to the objection that the data are not well matched in the center, where the velocity dispersions of the two populations are too high (a typical failing of all cusped models).

Given the unambiguous results from observations, a number of investigators have used simulations to explore whether baryonic feedback could erase a cusp and generate a core. Early calculations~\cite{Gnedin:2001ec}, assuming a single burst of star formation which generates a baryonic wind, showed that the energetic requirements are formidable. However, simulations of multiple bursts of star-formation demonstrated that cusps could be smoothed away~\cite{Read:2004xc}. For example, Ref.~\cite{Governato:2012fa} presented SPH and $N$-body simulations including star formation and supernova driven outflows. They showed that repeated gas outflows and bursty star formation can cause the central density of the Dark Matter to become cored. The rapidly changing potential in simulations, caused by outflows of expanding gas from bursts of star formation, can transfer energy into the collisionless Dark Matter particles and generate a Dark Matter core~\cite{Pontzen:2011ty}.  It now seems clear that this process can operate in simulations, but it is not yet evident that the simulations do accurately reproduce the difficult physics of star formation and supernova driven winds. Feedback problems are most acute for LSB galaxies, as star formation is minimal and inefficient when gas density is low. LSBs are typically metal-poor and dust-free, both implying that they form comparatively few stars over a Hubble time. Additionally, supernova winds cannot expel gas from the high mass end of the LSB galaxy sequence. It is therefore a substantial challenge to use feedback to erase cusps in all LSB galaxies, even if the process can be made to work for dwarfs~\cite{deNaray:2011hy}.

WDM (such as sterile neutrinos) does naturally produce cores. The Dark Matter phase space density is finite if the Dark Matter temperature is finite. If the Dark Matter is collisionless, then Liouville's theorem guarantees that the phase space density cannot increase. So, haloes are expected to have cores if the Dark Matter is not cold~\cite{Tremaine:1979we}, though the cores do not appear to be large enough to satisfy the data on LSB galaxies~\cite{VillaescusaNavarro:2010qy,Maccio:2012qf,Shao:2012cg}. WDM also delays the collapse of dwarf-sized haloes and their associated star formation to later epochs. It remains unclear whether WDM is more successful than CDM in reproducing the intrinsic properties of dwarf galaxies.

\subsection{\label{sec:TBTF}Too-big-to-fail (Author: E.~Papastergis)}

\subsubsection{Introduction and background}

The too-big-to-fail (TBTF) problem is a small-scale cosmological challenge to the $\Lambda$CDM paradigm. It posits the issue that it may not be possible to simultaneously reproduce the observed number density and internal kinematics of dwarf galaxies within the $\Lambda$CDM model. The TBTF problem is closely related to the ``missing satellites'' and ``cusp-core'' problems, since it is formulated in terms of both the abundance of galaxies as well as their kinematic properties (see Secs.~\ref{sec:missing-satellites} and~\ref{sec:core-cusp}). At the same time, the TBTF challenge represents a stronger theoretical issue for $\Lambda$CDM than either of the two aforementioned problems on their own. This is why the TBTF problem has become one of the most active topics of research in extragalactic astronomy today.

\begin{figure}
\centering
\includegraphics[scale=0.6]{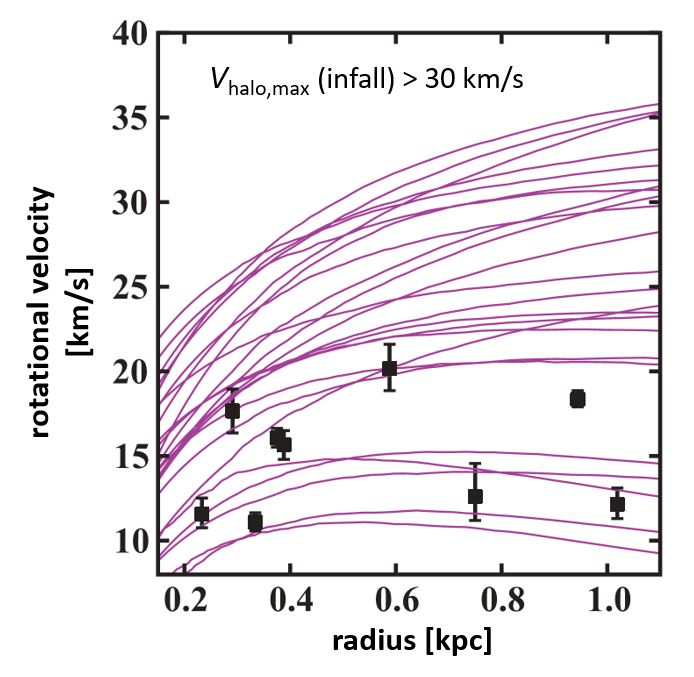}
\caption{\label{BK12}The solid curves represent the rotation curves of massive subhaloes of a MW-sized halo in the Aquarius DM simulation. The datapoints correspond to measurements of the rotational velocity of bright MW satellites at the half-light radius, $V_\mathrm{rot}(r_{1/2})$. Note how most of the expected hosts of the bright MW satellites are ``too massive'' to be accomodated within the measured kinematics of the satellites. (Reproduced with permission from~\cite{BoylanKolchin:2011dk}.)}
\end{figure}

The TBTF problem was first identified in the population of Milky Way (MW) satellites~\cite{BoylanKolchin:2011de,BoylanKolchin:2011dk}. In particular, it is natural to expect that the bright (or ``classical'') MW satellites would be hosted by the most massive subhaloes of the MW. Boylan-Kolchin et al. used the Aquarius~\cite{Springel:2008cc} and Via Lactea~II~\cite{Diemand:2007qr} simulations to show that the largest subhaloes of a MW-sized halo are ``too massive''\footnotemark{} to accommodate the observed kinematics of the bright MW satellites (Fig.~\ref{BK12}). Following the initial formulation of the TBTF problem in the context of the MW, several non-cosmological solutions were quickly identified. For example, the TBTF problem would likely not occur if the virial mass of the MW was relatively low ($M_\mathrm{vir,MW} \lesssim 1 \times 10^{12} \; M_\odot$), or if the MW was a statistical outlier~\cite{Wang:2012sv,VeraCiro:2012na,Purcell:2012kd}.

\footnotetext{To be more precise, the massive subhaloes in the Aquarius and VL~II simulations were found to be ``too dense'' to accommodate the kinematics of the bright MW satellites. That is, the dynamical mass enclosed by the half-light radii of the MW satellites is too low to be compatible with the mass profiles of the largest subhaloes in a MW-sized system.}

However, the MW does not seem to be a unique case. Ref.~\cite{Tollerud:2014zha} showed that the same TBTF problem is encountered in the satellite population of the Andromeda galaxy (M31). In addition,~\cite{Strigari:2011ps} and~\cite{Rodriguez-Puebla:2013aza} have argued based on SDSS data that the MW's satellite population is typical for a galaxy of its size. Overall, it seems that a comprehensive solution of the TBTF problem should be applicable to the satellite population of any MW-like galaxy. Such a solution has recently been put forward by~\cite{Zolotov:2012xd,Brooks:2012ah,Brooks:2012vi}. In particular, they showed that subhaloes in hydrodynamic simulations are more vulnerable to tidal stripping than their DM-only counterparts. Efficient stripping can lower the DM density in the inner parts of massive subhaloes compared to the DM-only case, resulting in kinematics that agree with the measurements for MW satellites, see Fig.~3 in~\cite{Brooks:2012vi}.

This baryonic solution to the satellite TBTF problem has been well received by the community, because it is physically well-motivated and broadly applicable. At the same time, it relies on satellite-specific environmental processes that are not expected to operate on isolated galaxies. As a result, it is of key scientific importance to test whether the TBTF problem is encountered also for dwarf galaxies in the \textit{field}. To this end, Ref.~\cite{Kirby:2014sya} measured the stellar kinematics of a sample of seven non-satellite dwarfs in the Local Group (LG; within $\sim$1.5 Mpc from the MW). They found that their kinematics do not differ from the kinematics of MW satellites of the same luminosity. This constitutes indirect evidence against the Zolotov \& Brooks mechanism, which requires the massive subhaloes of the MW to be heavily stripped. Furthermore,~\cite{Garrison-Kimmel:2014vqa} compared the stellar kinematics of the Kirby et al.~\cite{Kirby:2014sya} dwarfs with the kinematics of the largest haloes present in a simulated LG analog (ELVIS simulation~\cite{Garrison-Kimmel:2013eoa}). They found once again that the expected hosts of the LG dwarfs were to massive to be compatible with the measured dwarf kinematics, see Fig.~7 in~\cite{Garrison-Kimmel:2014vqa}. Their finding demonstrated that the TBTF problem is relevant for both satellites and non-satellite galaxies in the LG.

Even earlier,~\cite{Ferrero:2011au} obtained compelling evidence that the TBTF problem is also present for dwarf galaxies in the field. Ferrero et al.~\cite{Ferrero:2011au} assigned $\Lambda$CDM haloes to galaxies such that the number density of galaxies in SDSS as a function of their stellar mass is reproduced. A quantitative galaxy-halo connection under this constraint can be derived statistically, using the technique of abundance matching \cite[AM;][]{Guo:2009fn}. In a $\Lambda$CDM universe, AM predicts that all detectable field galaxies ($M_\ast \gtrsim 10^6 - 10^7 \; M_\odot$) should be hosted by relatively massive haloes (\Vhalo $\gtrsim 45$ \kmsec). This result is a straightforward consequence of the fact that the mass function of haloes in $\Lambda$CDM rises much faster than the stellar mass function of galaxies. In simpler terms, $\Lambda$CDM predicts that the abundance of haloes rises quickly with decreasing halo mass. Galaxy formation must therefore be restricted to relatively massive haloes with \Vhalo$\gtrsim 45$ \kmsec, because lower mass haloes are too numerous compared to the number of observed dwarf galaxies.

\begin{figure}
\centering
\includegraphics[scale=0.6]{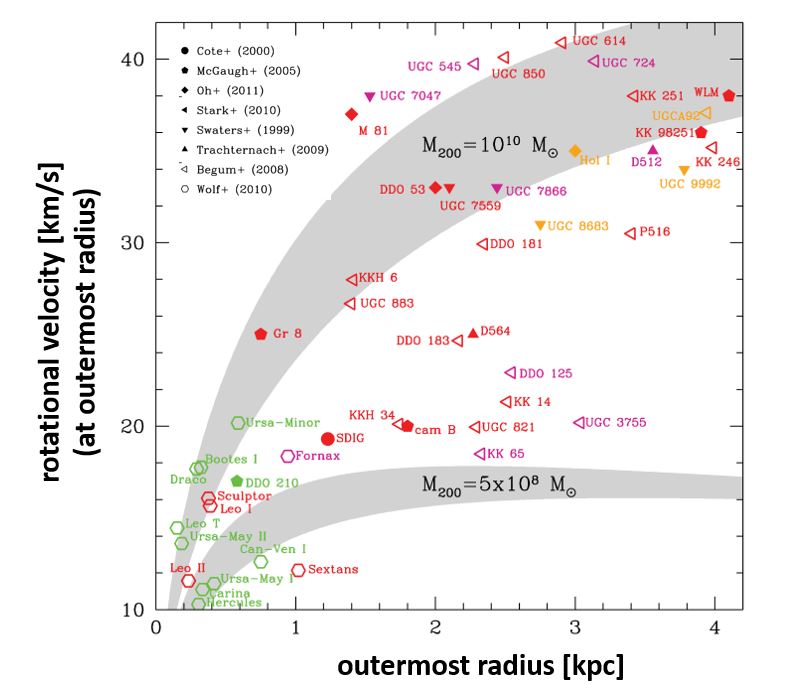}
\caption{\label{F12}The upper gray band represents the typical range for the rotation curves of haloes with $M_\mathrm{vir} = 10^{10} \; M_\odot$. These haloes are the least massive haloes that are expected to host detectable galaxies in a $\Lambda$CDM universe, according to AM. The colored points represent the rotational velocity of a sample of dwarf galaxies, measured at the outermost point of their HI rotation curve ($V_\mathrm{rot}(R_\mathrm{out})$). Similarly to what seen in Fig.~\ref{BK12}, the expected $\Lambda$CDM hosts of dwarf galaxies are ``too massive'' to be compatible with
the galaxies' observed kinematics. Unlike Fig.~\ref{BK12}, most of the dwarfs plotted here are fairly isolated galaxies in the \textit{field}. (Reproduced with permission from~\cite{Ferrero:2011au}.)}
\end{figure}

They then set out to test this theoretical expectation, by compiling a literature sample of dwarf galaxies with measured rotation curves in the 21cm emission line of atomic hydrogen (HI). Surprisingly, they found that the HI rotation curves of a large fraction of field dwarfs suggested that their host haloes have masses well below the expected ``threshold'' for galaxy formation (Fig.~\ref{F12}). The result of Ferrero et al. was subsequently confirmed and strengthened by the work of~\cite{Papastergis:2014aba}. They combined an up-to-date literature sample of galaxies with HI rotation curves and a census of low-mass galaxies from the ALFALFA 21cm survey~\cite{Haynes:2011hi}, to convincingly show that the TBTF problem is also present in the field \cite[Figs. 6 \& B1]{Papastergis:2014aba}. Unlike previous analyses, the work of Papastergis et al.~\cite{Papastergis:2014aba} is based exclusively on HI data, and so it can circumvent many observational uncertainties related to optical measurements present in previous works.

\subsubsection{Possible solutions of the TBTF problem within $\Lambda$CDM}

There is now extensive evidence that the TBTF problem concerns not just satellites, but low-mass dwarf galaxies in general. Accordingly, a generic solution to the problem must be applicable to both the satellite \textit{and} the field context. The most plausible solutions to the problem within $\Lambda$CDM are ``baryonic'', in the sense that they invoke modifications to the abundance and internal structure of DM haloes due to baryonic processes:

\begin{enumerate}

\item[(a)] \textit{Baryonic effects on the abundance of haloes}: There are two baryonic effects that can alter the number density of haloes compared to the DM-only case. First, haloes with \Vhalo$\lesssim 70$ \kmsec \ can lose the majority of their baryonic content to galactic outflows powered by supernova feedback~\cite{Sawala:2012cn}. This effect is not captured by DM-only simulations, because the latter treat the total matter density of the Universe ($\Omega_\mathrm{mat} = \Omega_\mathrm{DM} + \Omega_\mathrm{B} = 0.32$) as a dissipationless fluid. As a result, haloes in this mass range are expected to have about 20\% lower mass than their DM-only counterparts. Second, haloes with \Vhalo$\lesssim 25$ \kmsec \ are subject to cosmic reionization feedback. These lower mass haloes are unable to accrete gas after the Universe became reionized ($z \gtrsim 6-8$), and are therefore expected to be ``dark'' in most cases. The combined consequence of these two effects is to alter the abundance of detectable low-mass haloes compared to the result of a DM-only simulation, see Fig.~2 in~\cite{Sawala:2014baa}. The baryonic modification of the abundance of haloes can alleviate the TBTF problem, because the host haloes assigned through AM to dwarf galaxies are less massive than in the case of a DM-only simulation, see Fig.~3 in~\cite{Sawala:2014baa}.

\item[(b)] \textit{Baryonic effects on the rotation curves of haloes}: Hydrodynamic simulations have shown that the repeated blow-out of gas form the central regions of dwarf galaxies can create shallow ``cores'' in the DM profile of their host haloes~\cite{Governato:2009bg, Pontzen:2011ty}. This effect can alleviate the TBTF problem, because the rotation velocity in the inner parts of a cored halo is lower compared to its ``cuspy'' DM-only counterpart of the same total mass. Core creation through baryonic feedback has two properties that are important in the context of the TBTF problem: (i)~The efficiency of core creation has a characteristic dependence on galaxy mass~\cite{DiCintio:2014xia}. In particular, core formation is most efficient in dwarf galaxies with \vrot$\sim 50$~\kmsec, but the efficiency drops for higher and lower mass galaxies \cite[see][Fig.~6]{DiCintio:2013qxa}. (ii)~Core creation is only effective over the region where the bulk of star formation takes place, typically $R \lesssim 1$ kpc for dwarf galaxies in the field~\cite{Governato:2012fa}. This means that the rotational velocity of a dwarf galaxy can be significantly lower than the DM-only value when measured at a small galactocentric radius, but the difference becomes small at radii $R \gtrsim 2$ kpc.
\end{enumerate}

Overall, it is still not entirely clear whether the baryonic effects described above can fully resolve the TBTF problem. For example, Ref.~\cite{Sawala:2014baa} shows that the effect~(a) can explain the observed kinematics of faint MW and M31 satellites, since most low-mass haloes with \Vhalo$\lesssim 25$ \kmsec \ are expected to be dark (see their Fig.~4). However, this mechanism alone does not seem to be able to resolve the field TBTF problem, which is encountered at a larger mass scale (\Vhalo$\gtrsim 35 - 40$ \kmsec; see~\cite{Papastergis:2014aba}, Figs. 9 \& B1). On the other hand,~\cite{Brook:2014hda} have argued that effect~(b) can be used to explain the kinematics of low-mass dwarfs, both in the case of satellites and isolated objects in the LG (their Fig.~2). At the same time, Ref.~\cite{Papastergis:2014aba} has used the results of a set of hydrodynamic simulations with efficient feedback~\cite{Governato:2012fa,Brooks:2012vi,Christensen:2012zh} to argue that core creation cannot fully resolve the TBTF problem for field dwarfs (see their Fig.~10).

\begin{figure*}[ht]
\centering
\includegraphics[width=\linewidth]{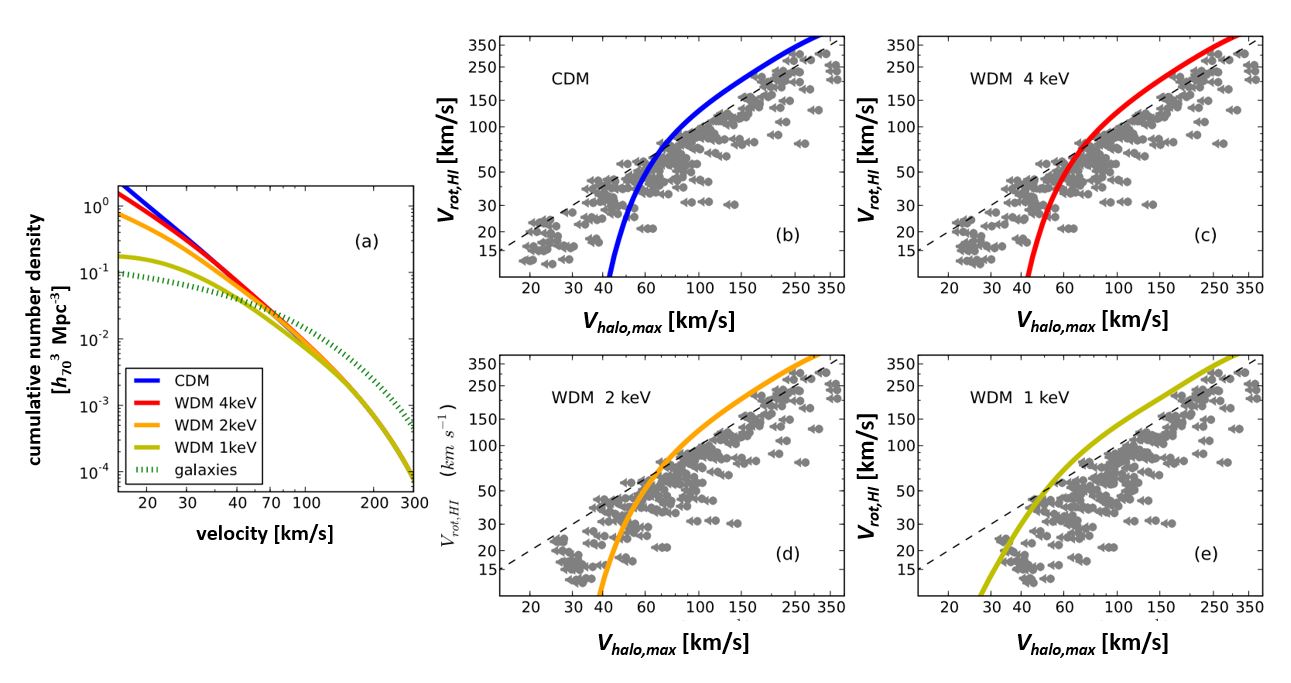}
\caption{\label{Pap15}\emph{Panel}~(a): The cumulative velocity function of galaxies measured by the ALFALFA survey (green dotted line;~\cite{Papastergis:2014aba}), compared to the cumulative velocity function of haloes in four cosmological models: CDM and WDM with \mwdm$= 1,2, \ \& \ 4$ keV (blue, red, orange, and yellow lines respectively; Ref.~\cite{Schneider:2013wwa}). \emph{Panel}~(b)--(e): The solid line in each panel represents the AM relation between the measured HI rotational velocity of a galaxy, \Vrot, and the maximum rotational velocity of the host halo, \Vhalo. The gray datapoints represent a sample of field dwarfs with resolved HI kinematics. Their $y$-axis position is determined by the width of their HI emission line, while their $x$-axis position represents the most massive halo that can be accommodated within their HI rotation curve. Note how the upper limits on \Vhalo \ set by the kinematics of dwarfs shift upwards as \mwdm\ decreases, due to effect (b). Cosmological models where the AM relation (solid line) is inconsistent with the upper limits set by individual galaxies are expected to overestimate the number density of dwarf galaxies detected in the field. (Reproduced with permission from~\cite{Papastergis:2014aba}.)}
\end{figure*}

\subsubsection{Can Warm Dark Matter solve the TBTF problem?}

Warm Dark Matter (WDM) has been regarded as a promising solution to the small-scale challenges of $\Lambda$CDM. The reason is that small-scale power in suppressed in a WDM universe, and as a consequence, structure formation at the mass scales of dwarf galaxies is curbed~\cite{Zavala:2009ms,Polisensky:2010rw,Lovell:2013ola,Schneider:2013wwa,Lovell:2015psz}. In the context of the TBTF problem, WDM has two advantageous features:

\begin{enumerate}

\item[{\bf (1)}] \textbf{There are fewer low-mass haloes in a WDM universe}. The mass (or velocity) function of haloes in a WDM universe is suppressed with respect to a CDM universe. As far as the TBTF problem is concerned, this means that observed galaxies can be assigned to lower mass haloes than in CDM, without overestimating the galaxies' number densities.

\item[{\bf (2)}] \textbf{WDM haloes have low concentrations}. At fixed total mass, a WDM halo has a lower concentration than its CDM counterpart \cite[e.g,][]{Schneider:2011yu,Lovell_eagle}. As a result, the rotation curve of a WDM halo rises more slowly than in CDM, lowering the rotational velocity measured at small galactocentric radii.

\end{enumerate}

\noindent
Fig.~\ref{Pap15} provides a graphic representation of how effects~{\bf (1)} and~{\bf (2)} can alleviate the field TBTF problem. Panels b-e show that both effects become more pronounced as the mass\footnote{For all WDM particle masses quoted in this section, \mwdm, refer to thermal relic particles.} of the WDM particle decreases. In fact, for WDM models with relatively large particle masses, \mwdm$\geq 4$ keV, there is no difference between the CDM and WDM case as far as the TBTF problem is concerned. Even in a 2 keV WDM model the TBTF problem is still present, even though somewhat less severe. According to Fig.~\ref{Pap15}, it takes a \emph{thermal relic} WDM model with a particle mass as low as $\approx$1~keV (and certainly below 2~keV~\cite{Anderhalden:2012jc}) to fully resolve the issue.
The resonantly produce sterile neutrino dark matter models, on the other hand,
can achieve good agreement with the observations for large range of masses~\cite{Lovell_tbtf}.

Fig.~\ref{Pap15} therefore illustrates the main challenge that WDM faces as a
successful solution to the TBTF problem: \textit{the required particle mass is
  quite low}, \mwdm$\lesssim 2$ keV. It is unclear whether WDM models with
such low particle masses are viable in view of the constraints posed by other
astrophysical probes. For example, current constraints from measurements of
the small-scale power spectrum of the Ly-$\alpha$ forest~\cite{Garzilli:2015iwa} put a lower bound at
$2.1$~keV if one does not impose any bias on the temperature of the
intergalactic medium at redshifts $z=5-6$. However, this bound becomes much stronger, $m_{\rm WDM} > 3.3$ keV at 95\% confidence, if one assumes powerlaw scaling of temperature with redshift,~\cite{Viel:2013apy}.\footnote{Exclusion limits are quoted in this article as published in the original reference, and may be subject to limitations (observational or theoretical) of the specific method used to derive them.} The suppression of small-scale power in a WDM model with \mwdm$\lesssim 2$~keV may be ``too much of a good thing''. More specifically, for sufficiently low values of \mwdm\ a MW-sized halo may not possess enough subhaloes to host the number of known ultrafaint satellites in the MW. Based on the consideration above, the authors of~\cite{Polisensky:2010rw} derive a lower limit of \mwdm$> 2.3$ keV, while~\cite{Lovell:2013ola} obtains a value of \mwdm$> 1.5$ keV; see also Ref.~\cite{Horiuchi:2013noa} for an alternative limit from satellite counts. The difference in results is partially due to different realizations of the MW-sized halo in simulations (as discussed e.g.\ in~\cite{Lovell_eagle}). Furthermore,~\cite{Schneider:2013wwa} and~\cite{Klypin:2014ira} have argued that the same is true in the field; WDM models with \mwdm$< 2$ keV severely underestimate the number density of the smallest field dwarfs with \vrot$\approx 10-20$ \kmsec \ (their Figs.~2 and~6, respectively). On the other hand, non-thermal DM settings could provide a game-change, see e.g.~\cite{Lovell_tbtf,Lovell_eagle}.

\subsection{\label{sec:kinematicWDM}The kinematics and formation of subhaloes in Warm Dark Matter simulations (Authors: M.~Wang, L.~Strigari)}

There is debate as to whether the central densities of DM-dominated galaxies are less dense than the haloes that are predicted in DM-only simulations~\cite{Walker:2011zu,Strigari:2014yea}. This issue along with the lack of observed satellites around the Milky Way can be ameliorated by considering alternative Dark Matter models. One particularly intriguing scenario is Warm Dark Matter (WDM), which differs from CDM in that the Dark Matter has a non-negligible velocity dispersion.  Generally, these WDM models come with two distinct observational signatures. First, there is a reduction in clustering and a delay in collapse times for structures on linear scales approximately smaller than the  free-streaming scale  of the WDM particle. Second,
the velocity dispersion of WDM imposes an upper limit on the Dark Matter phase-space density.

Numerical simulations are now rapidly improving, allowing for more precise predictions for Dark Matter halo formation in a WDM universe. Understanding the predictions of these simulations, along with other alternative to $\Lambda$CDM models, might provide clues to discriminating between different particle Dark Matter models. For example, it is found that WDM models with thermal relic mass $\sim$ keV can generate a truncation on matter power spectrum at scale $\sim$ a few Mpc. This results in a suppression of the formation of small structure below the free-streaming scale of WDM and thus result in delayed formation of haloes. As a result these simulations generally indicate that WDM haloes are less concentrated on account of their typically later formation epochs~\cite{Lovell:2011rd,Lovell:2013ola}. These models also imply a different reionization history~\cite{Schultz:2014eia} or star formation history at early times~\cite{Maio:2014qwa}. At present time at subgalactic scales, the stellar abundance and metallicity can thus be quite different from the CDM predictions~\cite{Governato:2014gja}.

WDM simulations also predict different properties for Galactic substructure than in $\Lambda$CDM simulations. It is thus pressing to understand how these predictions can be translated into tests of observations of Milky Way substructure in the form of dwarf satellite galaxies. Here we briefly address this question from the perspective of the kinematics and formation of subhaloes in WDM scenarios using $N$-body numerical simulations.

To address this question, in~\cite{Wang:2015fia} we map subhaloes in WDM simulations to dwarf spheriodal (dSph) galaxies, using kinematic observations of the dSphs. A similar approach was taken in~\cite{Strigari:2010un} for $\Lambda$CDM simulations. With the simple assumptions that these are spherical dynamically equilibrated systems, there is a strong degeneracy between the statistics of stellar orbits (i.e.\ whether velocity dispersions are isotropic, or are radially or tangentially biased) and the shape of the stellar and Dark Matter density profiles. Additional observational constraints like the measurement of stellar proper motion will be necessary to break these degeneracies. Nevertheless, we find that in WDM cosmology the subhaloes that host dSphs satellite galaxies exhibit different properties due to their shallower potential and late formation history.

\begin{figure}[th]
\begin{center}
\includegraphics[width=15cm]{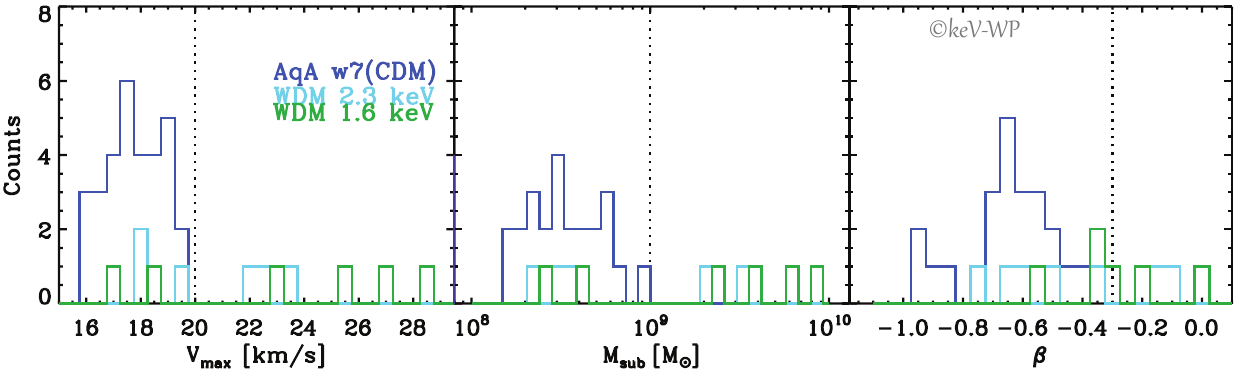}
\caption{\label{fig:hist}Histograms of subhalo properties for subhaloes that fit the kinematics of the Fornax dSph~\cite{Wang:2015fia}. Here we show three properties of subhaloes at the present time: the maximum circular velocity ($V_{\rm max}$, left panels), and the subhalo mass ($M_{sub}$, right panels). The navy histograms show the results from $\Lambda$CDM, the light blue histograms are for the WDM 2.3~keV simulation, the green histograms are for the WDM 1.6~keV simulation.}
\end{center}
\end{figure}

In Fig.~\ref{fig:hist} we show the subhalo properties that provide good-fits to the Fornax galaxy kinematic data~\cite{Wang:2015fia}. Here we show the maximum circular velocity $V_{\rm max}$ at present time, and mass of the subhalo $M_{\rm sub}$. We find that the distribution of these parameters from WDM simulations can be quite different than those from $\Lambda$CDM simulations.

We also find that the mass distributions of the WDM haloes that host the dSphs are different from those that host the dSphs in $\Lambda$CDM simulations. The results are that WDM models tend to reside in more massive subhaloes than in $\Lambda$CDM because its less concentrated profiles. In the case of Fornax, which is the brightest Milky Way dSph with high-quality stellar kinematic data, in $\Lambda$CDM the $V_{max}$ is very well-constrained between $\sim$16--20~km/s with average about 17~km/s. However, for WDM models, the $V_{\rm max}$ can be as large as $\sim$ 28~km/s.

\begin{figure}[th]
\begin{center}
\includegraphics[height=7.5cm]{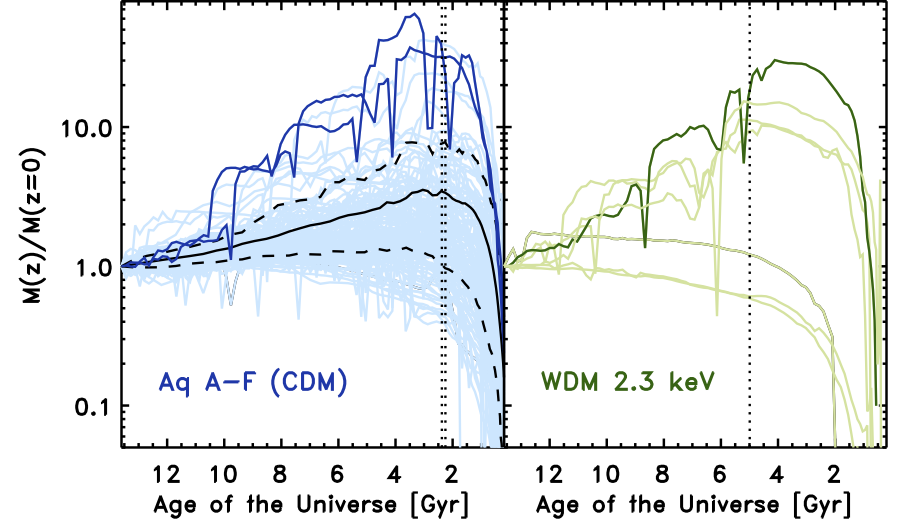}
\caption{\label{fig:mz}Mass assembly history for the subhaloes that fit the kinematics of the Fornax dSph~\cite{Wang:2015fia}. The gray lines show the subhaloes that are less luminosity than Fornax using abundance matching technique from~\cite{Behroozi:2011js}, and the blue lines show those with luminosity that match Fornax. The vertical dotted lines show the infall times of the blue curves. In the left panel the dashed black lines include 68$\%$ of the simulation fits, and the black solid lines show the mean value.}
\end{center}
\end{figure}

It is also interesting to determine the formation histories of subhaloes in WDM and $\Lambda$CDM. In Fig.~\ref{fig:mz} we show the mass assembly history for the Dark Matter subhaloes that match the kinematics of the Fornax dSph in both $\Lambda$CDM and WDM with 2.3~keV mass~\cite{Wang:2015fia}. The blue lines show those subhaloes that match the luminosity of Fornax using abundance matching technique~\cite{Behroozi:2012iw}. Interestingly we see that the WDM subhaloes are usually formed or gain significant amount of mass much later than the $\Lambda$CDM subhaloes.

These same results also can provide an indication of how rare a Fornax-like dSph is in both $\Lambda$CDM and WDM. In the $\Lambda$CDM simulations we found that only two out of 124 subhaloes that match the kinematics of Fornax also predict the luminosity that is observed. If the prediction of abundance matching holds, the bright luminosity of Fornax requires it to be hosted by a progenitor subhalo that is more massive than its present mass. Since Fig.~\ref{fig:hist} shows that the present-time mass of $\Lambda$CDM Fornax candidates are typically less massive than the WDM candidates, this suggests that Fornax subhaloes in the $\Lambda$CDM might suffer from severe tidal stripping and significant loss on Dark Matter halo mass. It is not clear whether this implies visible stellar tidal streams or not, because the stellar distribution is deeply embedded at the center of Dark Matter potential. However, for WDM the amount of Dark Matter stripping may be less significant because the present-time mass is larger than it is in $\Lambda$CDM. Currently, stellar tidal streams have not been observed in Fornax, and from the orbital properties indicate Fornax has not yet gone through many pericenter passages.

The kinematical properties of dSphs may thus provide unique clues to the nature of particle Dark Matter. In particular it may help distinguish between WDM and $\Lambda$CDM models. Future higher resolution hydrodynamical simulations of WDM and $\Lambda$CDM will be important to understand the impact of baryons on the formation of dSphs. Further, more detailed measurements of the kinematics of dSphs, including proper motions of stars within dSphs and proper motions of the dSphs within the Milky Way potential, hold a great amount of promise.

%% file: kevnuwp_section4.tex
\subsection{\label{sec:4-1-Phasespace}Phase space Analysis (Author: D.~Gorbunov)}

\input{Section4-1.tex}

\subsection{\label{sec:4-2-LymanA}Lyman-$\alpha$ forest constraints (Author: M.~Viel)}
\input{Section4-2.tex}

\subsection{\label{sec:xray}X-ray observations (Authors: O.~Ruchayskiy, T.~Jeltema, A.~Neronov, D.~Iakubovskyi)}
\input{Section4-3.tex}

\subsection{Laboratory constraints (Author: O.~Dragoun)}
\input{Section4-4.tex}

%% file: Section4-1.tex


Sterile neutrinos are well-motivated Dark Matter candidates. Since we need a substantial DM component in galaxies, any particles---DM candidates---must be capable of residing there at present. This implies two very general constraints on properties of the DM particles. First, the DM lifetime must exceed the age of the Universe. For the case of sterile neutrinos, this requirement imposes a constraint on the lifetime that is weaker than that from the searches for emission lines from the DM decay (see Sec.~\ref{sec:xray}). Second, the DM ``particle'' $X$ of mass $M_X$ must be slow and ``compact'' enough to be confined inside a galaxy. 

The velocity distribution of DM particles is supposed to be the same as that of galactic stars and cold gas clouds, which may be approximated by the Maxwell distribution 
\[
F_X({\bf v})=\frac{1}{\left(\sqrt{2\,\pi}\, M_X\, \sigma_X\right)^3}\, 
{\rm exp}\left({-\frac{{\bf v}^2}{2\,\sigma_X^2}}\right),  
\] 
with $\sigma_X=\sigma_X({\bf x})$ being the position-dependent velocity dispersion of stars in the galaxy. The dispersion takes the smallest value in the galaxy center and grows outward approaching $10^{-4}$\,--\,$10^{-3}$ for small (dwarf) and large (normal) galaxies, respectively. Dark matter particles of higher velocities escape from the galaxy, whose gravity force fails to keep them inside. As regards to the sterile neutrino, the dispersion determines the width of the monochromatic line expected in the galactic $X$-ray spectrum due to the radiative decay $N\to \nu\,\gamma$ as described in Sec.~\ref{sec:xray}. 

The ``compactness'' refers to the quantum nature of the DM particle. Namely, to locate {\it a boson} inside the galaxy its de Broglie wave length $\lambda_B=2\,\pi/(M_X\,v_X)$ must be smaller than the size of the smallest observed galaxies, the dwarf galaxies, that is about 1\,kpc. This yields a rather eccentric lower limit on the mass of the boson DM particle at the level of $M_X\gtrsim 10^{-22}$\,eV. 

In case of the {\it fermion} DM, a similar limit applicable to sterile neutrinos, is much more constraining. Indeed, fermions are subject to {\it the Pauli principle.} Therefore, for the galactic DM, the maximum of the phase space density with respect to velocity (which happens at ${\bf v}=0$),  
\[
f_X^{\text{max}}({\bf v},{\bf x})=\frac{\rho_X({\bf x})}{M_X}\, F_X(0)
\]
may not exceed the critical value 
\begin{equation}
\label{critical-Pauli}
f^{\text{crit}}_F\equiv\frac{g_X}{\left( 2\,\pi\right)^3}, 
\end{equation}
where $\rho_X({\bf x})$ is the DM mass density and $g_X$ denotes the number of intrinsic degrees of freedom of the DM particle, that is $g_X=2$ for the sterile neutrino (Majorana fermion). Thus the Pauli principle places {\em the lower limit on the fermion DM mass,}  
\begin{equation}
\label{lower-on-fermion-mass}
\frac{\left( 2\,\pi\right)^{3/8}}{g_X^{1/4}}\, \left( \frac{\rho_X}{\sigma_X^3} 
\right)^{1/4} \leq M_X\,.
\end{equation}
This bound is independent of the production mechanism. It gives $M_X\gtrsim 25$\,eV for the Milky Way galaxy, but becomes one order of magnitude stronger when applied to the most compact and dim structures, dwarf spheroidal galaxies. 

Let us note that bound \eqref{lower-on-fermion-mass} refers to the present population of DM particles in galaxies, and hence is {\em applicable to any fermion DM} irrespectively of properties of the DM particles and the DM production mechanism operating in the early Universe. Actually, bound~\eqref{lower-on-fermion-mass} can be strengthened by extending the consideration to the evolution of the phase space density of the DM particles during the structure formation as follows. 

After decoupling from primordial plasma (or production in a non-thermal way) the DM spectrum is fixed, and later on its shape is frozen while the Universe expands: only the physical momenta get redshifted, $p\propto 1/a(t)$ with $a(t)$ being the scale factor, see e.g.~\cite{Gorbunov:2011zz}. Indeed, in the early Universe the plasma is homogeneous to a large extent (the spatial fluctuations are below about $10^{-4}$). Hence the distribution of the DM particles in the phase space is homogeneous too, and initially (e.g. at decoupling) is described in terms of the conformal momentum ${\bf k}={\bf p}/a$ as $f_i({\bf k})$. Inasmuch as the conformal $d^3{\bf x}\,d^3{\bf k}$ and physical $d^3{\bf X}\,d^3{\bf p}$  phase space volumes coincide, 
\[
d^3{\bf x}\,d^3{\bf k}=d^3\left({\bf
  x}\,{a}\right)\,d^3\left(\frac{\bf k}{a}\right) = d^3{\bf
  X}\,d^3{\bf p} \,,
\]
the shape of the momentum distribution remains intact, 
\[
f({\bf k},t)=f_i({\bf k})=f_i\left(\frac{\bf k}{a(t)}\,a(t)\right) 
=f_i({\bf p}\,a(t))\,
 \]
while the inhomogeneities are small. 

\begin{figure}[!htb]
\hskip 0.05\textwidth
\includegraphics[width=0.8\textwidth]{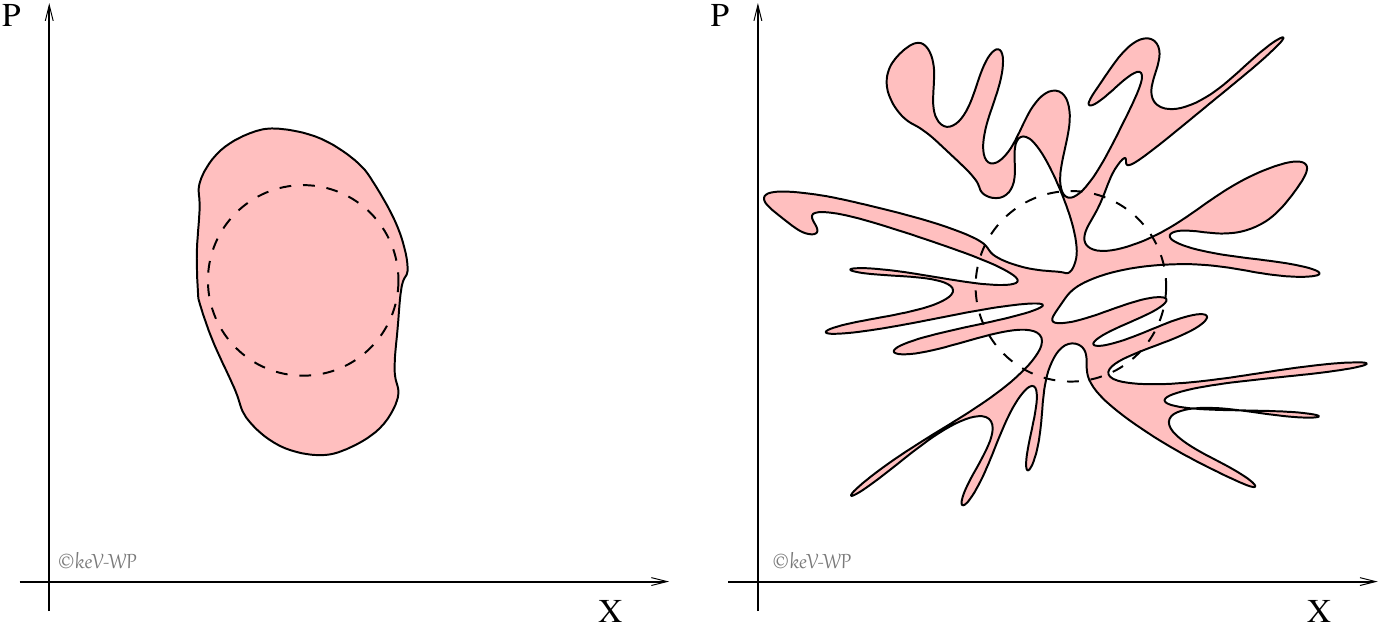}
\caption{Evolution of the phase space density in time: initially occupying the compact region (left panel) particles spread over the phase space (right panel). The volume remains intact, but the coarse grained phase space density decreases in the dense regions. 
\label{Fig:phase-space}
}
\end{figure}

However, later the distribution function becomes inhomogeneous, evolves in time, and looks different for DM particles in voids, filaments, galaxy clusters, dwarf galaxies, etc. Nevertheless, for the collisionless particles (as we believe the DM is and the sterile neutrino is for sure) the Liouville theorem guarantees that the {\em phase space density remains constant in time along any given trajectory,} though flows from one region to another. In practice one is interested not in this, the {\it microscopic density}, but rather in the density averaged over macroscopic regions of the phase space, the so-called {\it coarse grained phase space density}, $\tilde f({\bf k},{\bf x})$. Naturally, it grows in the underdense regions and gets diluted in the overdense regions, as illustrated in Fig.\,\ref{Fig:phase-space}. 
Therefore, it always obeys the following inequality
\begin{equation}
\label{master-inequality}
\tilde f ({\bf k},{\bf x},t)\leq \text{max}_k\, f_i({\bf k})\,,
\end{equation}
including the galactic DM component at present. Then l.h.s.\ of \eqref{master-inequality} can be extracted  from observations, while r.h.s.\ is the prediction of a particular mechanism of the DM production operating in the early Universe. 

The analysis of astronomical data allows to estimate for a particular galaxy the value of the following quantity 
\[
Q\equiv \frac{\rho_0}{\langle {\bf v}^2_\parallel \rangle ^{3/2}} \,,
\]
where $\rho$ is the mass density in the galaxy center region and $\langle{\bf v}^2_\parallel\rangle$ is the averaged (over the objects in that region) squared velocity along the line of sight. For obtaining the strongest constraint on the DM particle mass from Eq.~\eqref{master-inequality} the relevant are the highest values of $Q$, which are provided by observations of the stellar dynamics in the smallest structures, {\it dwarf spheroidal galaxies,} \cite{Simon:2007dq}
\begin{equation}
\label{Q-estimates}
Q= \left( 0.005-0.02\right) 
 \frac{M_\odot /\text{pc}^3}{\left( \text{km}/\text{s}\right)}\,.
\end{equation}

The galaxies are compact and dim with mass strongly dominated even in the central part by the DM component. Hence, one has $\rho_0=M_X\,n_X$ with $n_X$ standing for the number density of the DM particles in the galaxy center. For the spheroidal dwarfs one can substitute $\langle {\bf v}^2_\parallel \rangle =\langle {\bf v}^2 \rangle/3$, where $\langle {\bf v}^2 \rangle$  is the average squared velocity. For the DM particles with average squared momentum $\langle {\bf p}^2\rangle=M_X^2 \langle {\bf  v}^2 \rangle$.\footnote{
In Ref.~\cite{Horiuchi:2013noa} it has been suggested that a Maxwell Boltzmann phase density estimator is more accurate to be applied to dwarf galaxies.
}   

Collecting all terms together we arrive at 
\[
Q=3^{3/2}\,M_X^4\,\frac{n}{\langle {\bf p}^2\rangle^{3/2}} \simeq 
3^{3/2}\,M_X^4\, \tilde f ({\bf p},{\bf X},t_0)\,,
\]
where in the last equality we used the phase space distribution of DM particles in the galaxy center, ${\bf x}\simeq 0$ at present, $t=t_0$. Then from eq.\,\eqref{master-inequality} we obtain the lower limit on the DM particle mass known as the Tremaine--Gunn-type bound \cite{Tremaine:1979we},  
\[
M_X\gtrsim \left( \frac{Q}{3^{3/2}\, \text{max}\tilde f_i} \right)^{1/4}
\]
This bound is valid for {\em both bosons and fermions}, but $\text{max}\tilde f_i$ of course depends on the production mechanism. For the former it supersedes what we have above from the de Broglie waves, while for the latter it coincides with that from the Pauli principle if $\text{max}\tilde f_i$ reaches the critical value~\eqref{critical-Pauli}, but generally is somewhat more restrictive given $\text{max}\tilde f_i$ in the denominator. Applying this limit to the sterile neutrino nonresonantly produced in the early Universe with the corresponding spectrum (see Sec.~\ref{sec:5.thermalproduction}), one obtains:
\begin{equation}
\label{nonresonsnt-lower-limit}
M_X\gtrsim 6\,\text{keV} \, \left( \frac{0.2}{\Omega_{\text{DM}}} 
\right)^{1/3} \left(\frac{Q}{0.005\,\frac{M_\odot /\text{pc}^3}{\left(
    \text{km}/\text{s}\right)} }\right)^{1/3} 
\left( \frac{g_*(T_d)}{43/4}\right)
\end{equation}
with all factors of order one.
Thus, literally using the estimates~\eqref{Q-estimates}, one concludes that the light sterile neutrino, $M_X\lesssim 6$\,keV {\em nonresonantly produced in the early Universe, is disfavored}~\cite{Gorbunov:2008ka,Horiuchi:2013noa} from analysis of the phase space density as the main component of galactic DM. The limit suffers from the systematic uncertainties in the parameter estimates~\eqref{Q-estimates}~\cite{Boyarsky:2008ju}, but is still much stronger as compared to that from the Pauli blocking. Similar limits exist for other fermion warm DM candidates, and are usually relevant for the non-thermal production mechanisms, see ref.~\cite{Gorbunov:2008ui}  for example of light gravitino.

%% file: Section4-2.tex
The Lyman-$\alpha$ forest, the main manifestation of the intergalactic medium (IGM), is produced by intervening filaments of neutral hydrogen along the line-of-sight to a distant source, typically a quasar (QSO). Since its discovery in the early 70s, it has been used to investigate the distribution of matter and the physical conditions of such Lyman-$\alpha$ "clouds'' of gas. However, in the late 90s there has been a paradigm shift: while this observable was in the beginning used as a local probe of baryonic matter and of the physics the absorbers, it became very soon apparent that the Lyma-$\alpha$ forest retains cosmological information and could be used as a tracer of the large scale structure. This new cosmological role for the IGM has been mainly allowed by two factors: on one side the new high-resolution high signal-to-noise quasar spectra and on the other side the progress made by the first set of hydrodynamic simulations~\cite{Meiksin:2007rz}. A unique features of the IGM is that it is very complementary to other cosmological observables: it mainly probes small-medium scales from sub-Mpc up to few hundreds of Mpc and redshift in the range $z=2-6$.

The observable quantity is the transmitted flux along the line-of-sight to a distant  bright source, i.e. $F=\exp(-\tau)$, that needs to be related to underlying matter. In order to get some insight, we  start from the definition of optical depth in redshift-space at $u$ (in km s$^{-1}$):
\begin{equation} 
\tau(u)={\sigma_{0,\alpha} ~c\over H(z)} \int_{-\infty}^{\infty} dy\, n_{\rm HI}(y) ~{\cal V}\left[u-y-v_{\parallel}^{\rm IGM}(y),\,b(y)\right]dy \;, 
\end{equation} 
where $\sigma_{0,\alpha} = 4.45 \times 10^{-18}$ cm$^2$ is the hydrogen Ly$\alpha$ cross-section, $H(z)$ the Hubble constant at $z$, $y$  the real-space coordinate (in km s$^{-1}$), ${\cal V}$ is the standard Voigt profile normalized in real-space, $b=(2k_BT/mc^2)^{1/2}$ is the velocity dispersion in units of $c$. In the highly ionized case (which is of interest here) the local density of neutral hydrogen can be easily related to the local gas/IGM density as:
 \begin{equation}
 n_{\rm HI}({\bf x}, z) \approx 10^{-5} ~{\overline n}_{\rm IGM}(z) \left({\Omega_{0b} h^2 \over 0.019}\right) \left({\Gamma_{-12} \over 0.5}\right)^{-1} \left(T({\bf x},z) \over 10^4 {\rm K} \right)^{-0.7} \left({1+z \over 4}\right)^3 \left(1 + \delta_{\rm IGM}({\bf x},z) \right)^2 \;,
 \end{equation}
with $\Gamma_{-12}$, the hydrogen photoionization rate in units of $10^{-12}$ s$^{-1}$ and $T({\bf x},z) = T_0(z) (1+\delta^{\rm IGM}({\bf x},z))^{\gamma(z)-1}$, where both the temperature at mean density $T_0$ and the adiabatic index $\gamma$ depend on the whole IGM ionization history. From the eqs. above, we have understood how the observed flux is derived from the underlying gas distribution. The latter can be modeled either analytically (with a log-normal model for example) or by means of high resolution hydrodynamic simulations that incorporate baryonic pressure (i.e. Jeans smoothing). In semi-analytical models~\cite{Bi:1996fh,Viel:2001hd} this is usually done in Fourier space where  the IGM component reads:
\begin{equation}
\delta_0^{\rm IGM} ({\bf k}, z) = {\delta_0^{\rm DM}
({\bf k}, z) \over 1 + k^2/k_J^2(z)} \equiv 
W_{\rm IGM}(k,z) D_+(z) \delta_0^{\rm DM}({\bf k})
\end{equation}
with $D_+(z)$ the linear growing mode of DM density fluctuations and $\delta_0^{\rm DM}({\bf k})$ is the Fourier transformed DM linear overdensity at $z=0$. The low-pass filter $W_{\rm IGM}(k,z) = (1 + k^2/k_J^2)^{-1}$ depends on the comoving Jeans length in the following way:
\begin{equation} 
k_J^{-1}(z) \equiv  H_0^{-1} \left[ {2 \gamma k_B T_0(z) \over 3 \mu m_p
\Omega_{0m} (1 + z)}\right]^{1/2} \;,
\end{equation}
with $k_B$ the Boltzmann constant, $T_0$ the temperature at mean density, $\mu$ the molecular weight of the IGM, $\Omega_{0m}$ the present-day matter density parameter and $\gamma$ the ratio of specific heats. 

It is more practical to consider flux fluctuations $\delta F=F/<F>-1$ to those of matter $\delta$ and eqs. above can be simplified to obtain the so-called Fluctuating Gunn-Peterson Approximation:
\begin{equation}
F({\bf x},z)=\exp\left[-A \left(1 + \delta^{\rm IGM} \left({\bf x}, z\right)\right)^\beta\right] \;, 
\end{equation}
where $\beta \approx 2 - 0.7(\gamma-1)$, while $A$ is a normalization constant of order unity which determines the mean flux in the considered redshift interval. It is easy to derive, by Taylor expanding the equation above, that  ${\delta F} ({\bf x}, z) \approx - A \beta \delta_0^{\rm IGM}\left({\bf x},z\right)$. The flux power is usually measured in 1D by using high-resolution or low-resolution spectra, while it has been recently exploited by the SDSS-III/BOSS collaboration in full 3D to allow the discovery of Baryonic Acoustic Oscillations in Lyman-$\alpha$ flux~\cite{Busca:2012bu,Slosar:2013fi}. The relation between 1D and 3D power is:
\begin{equation}
P_{3D}(k) = - {2 \pi \over k} {d \over dk} p_{1D}(k) \;,
\label{eq:recdiff} 
\end{equation} 
which is valid for both flux and matter power spectra and shows that the 1D power is an integral of the 3D and effects like a Warm Dark Matter (WDM) cutoff could impact at relatively larger scale and thereby could be constrained by IGM data.

From the considerations made above, it is pretty clear that the Lyman-$\alpha$ forest can be used to place tight constraints on cold DM coldness~\cite{Markovic:2013iza}, however it must also be noticed that the temperature could affect flux power in a similar way: a hot IGM has less power than a cold one since lines will be broadened more in the first case. This signature could be degenerate with the one induced by WDM. Also, it should be kept in mind that structure formation is not directly sensitive to the properties of DM particles. It is mainly sensitive to their free streaming length. A bound on the free streaming horizon $\lambda_{\rm FS}$ can be converted into a bound on the DM particles' mass if their momentum distribution is known. Heavy CDM particles are in good approximation at rest at all relevant times. Most of the constraints on lighter particles presented so far are obtained by assuming a thermal relic WDM model for which the power is suppressed below a given scale. That is, it has been assumed that the momentum distribution of the DM particles is prortional to a thermal spectrum. In this case, one can estimate~\cite{Bond:1980ha}:
\begin{equation}\label{FreeStreamingVsMass}
\lambda_{\rm FS}\sim 1 {\rm Mpc} \frac{\rm keV}{m_{\rm DM}}\frac{\langle p _{\rm DM}\rangle}{\langle p_\nu \rangle},
\end{equation}
where $\langle p _{\rm DM}\rangle$ is the average momentum of the DM particles and $\langle p_\nu \rangle\sim 1$ keV is the comoving momentum of active neutinos around the time when thermally  produced sterile neutirno DM particles become non-relatvistic. For highly non-thermal distribution functions, which can appear if the sterile neutrinos are produced resonantly (see Sec.~\ref{sec:MSWeffect}) or in a particle decay (see Sec.~\ref{sec:5.decays}), there is no simple relation like (\ref{FreeStreamingVsMass}).

\begin{table}[htb]
\caption{Summary of limits obtained on the mass of a thermal relic (WDM candidate) by using a comprehensive set of Lyman-$\alpha$ forest data both at low and high resolution. All the numbers apart from~\cite{Narayanan:2000tp} quote 2$\sigma$ C.L. lower limits. In~\cite{Boyarsky:2008xj} also a frequentist analysis of the SDSS data is presented. In~\cite{Viel:2005qj} the method used to derive the constraints is based on the effective bias method proposed by~\cite{Croft:2000hs}, in~\cite{Seljak:2006qw} there is an approximate hydrodynamical scheme but the likelihood analysis is performed in a wide parameter space, while in the other works the simulations are full hydro and the method is based on a Taylor expansion of the flux power.}
\begin{center}
\begin{tabular*}{\textwidth}{@{\extracolsep{\fill}} l l l l}
\hline
 reference & limit (keV)  &  method & data\\
 \hline
 \hline
\cite{Narayanan:2000tp} & 0.75 & N-body only, no marginalization & 8 high res.\\ 
\cite{Viel:2005qj} & 0.55 & full hydro & 30 high res.\\
\cite{Viel:2006kd}& 2.5 &  approximate hydro &  3000 low res.\\  
\cite{Seljak:2006qw} & 2 & full hydro, approximate method & 3000 low res.\\    
\cite{Viel:2007mv} & 4 &  full hydro, approximate method & 3000 low res., 55 high res.\\
\cite{Boyarsky:2008xj} & 2.2  & full hydro, frequentist analysis & 3000 low. res \\
\cite{Viel:2013apy} & 3.3 &  full hydro, improved analysis & 3000 low res., 25 high res., high $z$\\
\hline
\end{tabular*}
\end{center}
\end{table}

The first of such limits (notice that strong absorption systems were used before to place constraints on DM nature in~\cite{Klypin:1994nf}),  is presented in~\cite{Narayanan:2000tp} where a thermal relic of masses $> 0.75$ keV is shown to be consistent with the data, however this number was obtained with a limited set of N-body only simulations and does not rely on a full marginalization over the many nuisance and astrophysical parameters. In~\cite{Viel:2005qj}, by using the data points (i.e. linear matter power) derived in~\cite{Viel:2004bf}, it was demonstrated that with full hydro and proper marginalization in the parameter space (although obtained with the effective bias model of~\cite{Croft:2000hs}, on a larger sample of QSO spectra) this number was relaxed to 0.55 keV (2$\sigma$ C.L.), in this paper also a lower limit on a non resonantly produced sterile neutrino was quantified to be  2 keV.

The main new results soon after these analyses were performed, was the advent of SDSS (Sloan Digital Sky Survey), which allowed an unprecedented measurement of linear matter power spectrum at $z=3$ with very good accuracy. This was derived from a set of about $3000$ low resolution low signal-to-noise spectra~\cite{mcdonald05} and by using a new method. In particular, the likelihood space was explored fully by accurately modeling also astrophysical and instrumental effects, while on the other side the calibration of the flux power was determined by running only a few hydro simulations, while the parameter space was explored by using approximate hydro schemes (that were computationally less expensive). The main point to stress is however that for the first time, the relatively wide range explored by the data $z=2.2-4.2$ allowed to break the degeneracies present between cosmology and thermal history, resulting in significantly tighter constraints. In~\cite{Seljak:2006qw}, SDSS data returned the following constraints: 2.5 keV and 14 keV lower limits on a thermal relic and non-resonantly produced sterile neutrinos at a 2$\sigma$ C.L., respectively. The same data set was then analyzed by~\cite{Viel:2006kd} using a completely different method (albeit still in a Bayesian framework): a second-order Taylor expansion of the flux power in a relatively better explored (compared to~\cite{Seljak:2006qw}) hydrodynamical parameter space gave a $\sim 20\%$ weaker constraints of a mass of a thermal relic $> 2$ keV and of a sterile neutrino $>10$ keV at $2\sigma$~C.L., respectively. In~\cite{Abazajian:2006yn} some scenarios in which there is some entropy released is also explored along with non-zero lepton number models and confronted with IGM constraints. The investigation of how the flux power relates to the underlying matter power spectrum emphasized the need for QSO spectra at relatively high redshift, where the non-linear evolution is smaller and possibly closer to the linear cut-off. In~\cite{Viel:2007mv} the SDSS data were complemented by a smaller set of 55 high resolution high redshift QSO spectra that resulted in a lower limit of 4 keV, however the higher redshift bin was explored with hydro simulations that were not fully converged (although there was a resolution correction implemented in the pipeline). This is the tightest limit on WDM coldness ever obtained from IGM data.

After these results, in~\cite{Boyarsky:2008xj} the SDSS data set (without high resolution samples) was used to get constraints in a frequentist framework for a non resonantly produced sterile neutrinos obtaining $m>8$ keV at 99.7\% confidence limit in a pure $\Lambda$CDM model. Most notably, this work also investigated mixed (cold and warm) models and found that any mass for a sterile neutrinos could be allowed if the WDM fraction is below 60\%. An important breakthrough was made in~\cite{Boyarsky:2008mt} where it was realized that mixed models linear power spectra were remarkably similar to those of a resonantly produced sterile neutrinos in the context of minimal extension of the standard model: in such cases the limits above showed that any sterile neutrino mass $> 2$ keV would have been in agreement with the  Lyman-$\alpha$ forest SDSS data.

The most recent and comprehensive analysis of Lyman-$\alpha$ forest data  for WDM models has been made in~\cite{Viel:2013apy}. Compared to the previous works there are several improvements: the hydrodynamic simulations have been run at very high resolution and do not need any resolution correction nor they suffer of numerical fragmentation; the non-linearities both in the matter power and flux power are modeled to unprecedented accuracy~\cite{Viel:2011bk}; the data sets of high-resolution spectra extends the previous ones (Keck HIRES and Magellan spectrograph); the modeling carefully consider instrumental resolution, signal-to-noise and astrophysical nuisance parameters (mainly thermal histories and ultraviolet background inhomogeneities). The conclusion is that the lower limit is 3.3 keV at $2\sigma$ C.L. for a WDM thermal relic; this has to be considered as a robust limit, obtained by marginalizing over a physically motivated set of thermal histories and conservatively considering systematic errors both on the data and modeling side. It is worth noticing that if one allows for a sudden jump in the temperature evolution the constraint above becomes weaker by about 1 keV. These (and almost all other known) bounds rely on the assumption that the DM momentum distribution is not too different from a thermal spectrum (or a superposition of two thermal spectra). Then a relation of the type (\ref{FreeStreamingVsMass}) can be used. Translating this number in a constraint on resonantly produced  sterile neutrinos with a spectrum as shown in Fig.~\ref{DMspectra} is not easy and could be done by comparing the linear transfer function proper of a thermal relic of such mass with the one of the wanted model. Consequences in terms of high redshift structure formation for a thermal relic mass particle compatible with all the constraints have been recently explored in~\cite{Maio:2014qwa}, while forecasts in terms of 21 centimeter intensity mapping are presented in~\cite{Carucci:2015bra} also for lower values of the mass.

\begin{figure*}[!ht]
\begin{center}
\includegraphics[trim=0 3cm 0 14cm,angle=0,width=15cm]{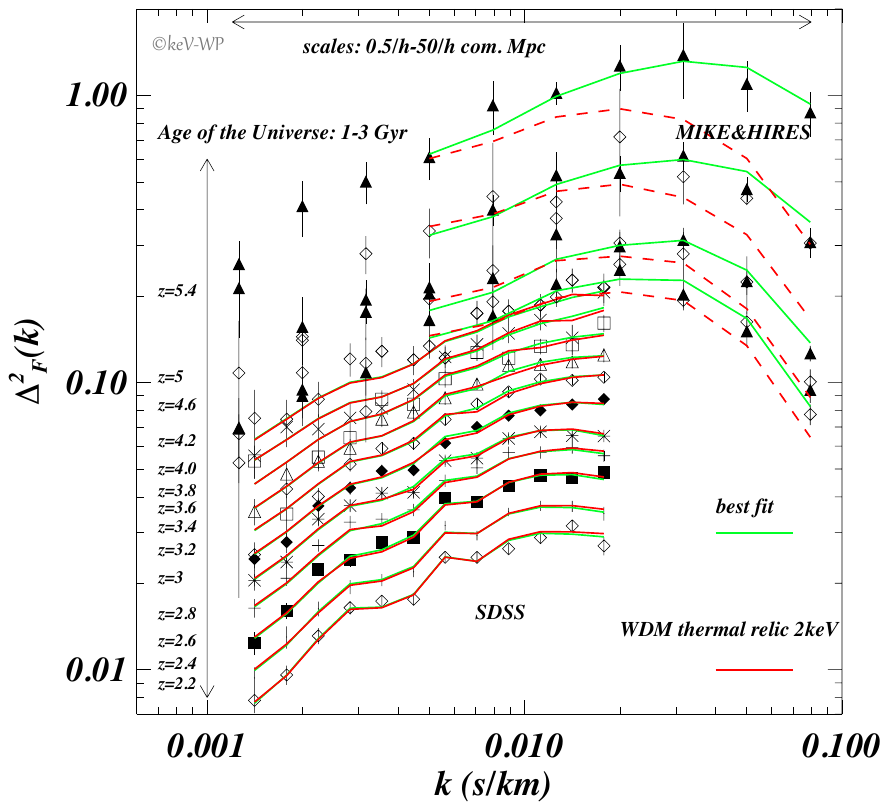}
\end{center}
\vspace{-0.5cm}
\caption{\label{all} Best fit model for the data sets used in the
  analysis (SDSS+HIRES+MIKE) shown as green curves. We also
  show a WDM model that has the  best fit values of the green
  model except for the WDM (thermal relic) mass of 2 keV (red dashed curves). These data span
  about two orders of magnitude in scale and the period 1.1-3.1 Gyrs
  after the Big Bang. From this plot is is apparent how the WDM model does not fit the data at small scales and high redshift.}
\end {figure*}

In Figure \ref{all} we show the difference between a $\Lambda$CDM best fit model and a WDM model of 2~keV in terms of agreement with flux power spectra at low and high resolution. It is clear the power of the high redshift data that allows to severely constrain WDM coldness.

%% file: Section4-3.tex

\begin{figure}[!t]
  \centering
  \includegraphics[width=.5\textwidth]{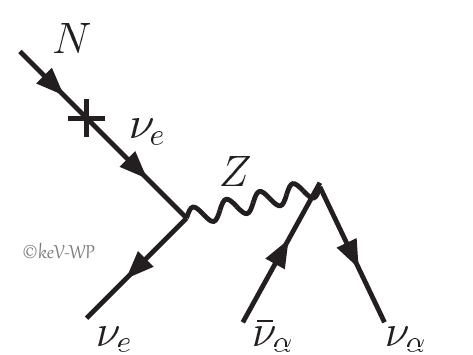}~\includegraphics[width=.5\textwidth]{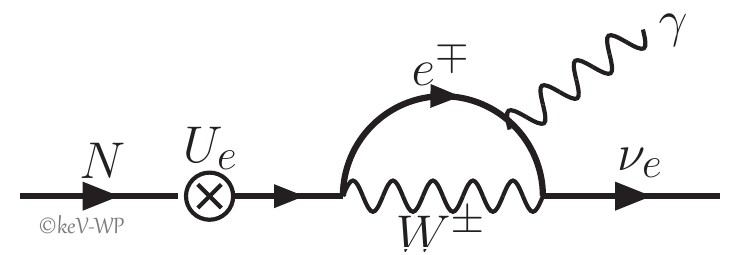}
  \caption{Decay channels of the sterile neutrino $N$ with the mass below twice the electron mass. Left panel: dominant decay channel to three (anti)neutrinos. Right panel shows radiative decay channel that allows to look for the signal of sterile neutrino DM in the spectra of DM dominated objects.}
  \label{fig:decay_DM}
\end{figure}

\subsubsection{X-ray signals - overview}

The main decay channel of sterile neutrinos with the mass below $2 m_e$ is $N\to \nu \nu \bar\nu$ (see Fig.~\ref{fig:decay_DM}, left) with different combinations of flavours. It determines the lifetime of the particle. The sterile neutrino DM has a sub-dominant radiative decay channel $N\to \nu\gamma$ (Fig.~\ref{fig:decay_DM}, right panel). The decay width of this process is about ${128}$ times smaller that the main into active neutrinos $\nu_a$ and photon with energy $E=m_s/2$, with the width~\citep{Shrock:1974nd,Marciano:1977wx,Petcov:1976ff,PhysRevD.16.1444,Pal:82,SHROCK1982359,Barger:1995ty}
\begin{equation}
\label{gamma}
  \Gamma_{N\to\gamma\nu_a} = \frac{9\, \alpha\, G_F^2}
  {256\cdot 4\pi^4}\sin^22\theta\,
  m_s^5 = 
  5.5\times10^{-22}\theta^2
  \left[\frac{m_s}{1\,\mathrm{keV}}\right]^5\;\mathrm{s}^{-1}\:.
\end{equation}
If the sterile neutrino is a main ingredient of the DM, it is potentially detectable in various X-ray observations~\cite{Dolgov:2000ew,Abazajian:2001vt,Herder:2009im}.

There are several types of the X-ray signal produced by the sterile neutrino DM. First, decays of DM particles throughout the history of the Universe should produce a contribution to the diffuse X-ray background (XRB)~\cite{Abazajian:2001vt, Dolgov:2000ew,Mapelli:05,Boyarsky:2005us}.  The DM decay contribution to the XRB is (see~\cite{Boyarsky:2005us} and references therein)
\begin{equation}
  F_{\rm XRB}\simeq \frac{\Gamma_{N\to\gamma\nu_a}  \rho_{DM}^0}{2\pi H_0}\simeq
  8\times 10^{-11}\,\left[\frac{\theta^2}{10^{-11}}\right]\left[\frac{m_s}{7\:
      \mathrm{keV}}\right]^5\frac{\mbox{erg}}{\mathrm{cm^2\cdot s\cdot
      sr}}~, 
\end{equation}
where $\rho^0_{DM}, H_0$ are the DM density in the universe and the Hubble constant today. Neutrinos decaying at different red shifts produce a broad X-ray line with extended ``red'' tail.  Such a feature in the XRB spectrum is, in principle, readily detectable (and distinguishable) from the broad-band continuum of observed XRB~\citep{Gruber:99,Churazov:07}. The non-detection of the DM decay feature in the XRB signal enables to put a bound on $\theta$ and $m_s$ roughly at the level of~\citep{Boyarsky:2005us} (see also~\cite{Abazajian:2006jc}
\begin{equation}
\label{xrbbound}
\text{XRB bound:}\qquad
\Omega_s \sin^2(2\theta)\lesssim 3\times 10^{-5}
\left[\frac{1\mbox{~keV}}{m_s}\right]^{5}~,
\end{equation}
where $\Omega_s\le \Omega_{DM}$ is the present day density of sterile neutrino DM.

As first pointed out Refs.~\cite{Abazajian:2001vt,Abazajian:2001nj}, clustering of the DM at small red shifts results in the enhancement of the DM decay signal in the direction of large mass concentrations, such as galaxy clusters, see also~\cite{Boyarsky:2009rb,Boyarsky:2009af,Cuesta:10,Ibarra:2013cra}. Typical overdensity, ${\cal R} $, (${\cal R} \equiv\rho/\rho_{\rm DM}^0$) of a galaxy cluster is at the level of ${\cal R}\sim 10^3$, while typical cluster size is about $R_{cl}\sim$~Mpc. An estimate for the DM decay flux from a galaxy cluster is
\begin{equation}
  F_{cl}=\frac{\Gamma_{N\to\gamma\nu_a}  M_{DM,FoV}}{4\pi D_L^2m_s} \simeq 1.4\times 10^{-7}\left[\frac{\theta^2}{10^{-11}}\right]\left[\frac{m_s}{7\mbox{~keV}}\right]^4\nonumber\\\left[\frac{D_L}{100\mbox{ Mpc}}\right]^{-2}
  \left[\frac{M_{DM,FoV}}{10^{13}M_\odot}\right]\frac{\mbox{ph}}{\mbox{cm}^2\mbox{s}}
\label{eq:flux}
\end{equation}
($M_{DM,FoV}\simeq {\cal R}\rho^0_{DM}R_{cl}D_\theta^2\Omega_{FoV}$ is the mass of DM within telescope's field of view (FoV), $D_L,D_\theta$ are the luminosity and angular diameter distances to the cluster). The DM decay flux from the direction of individual clusters is typically 
of the same order as 
the DM flux from the XRB contribution:
\begin{equation}
\frac{m_sF_{cl}/(2\Omega_{FoV})}{F_{\rm XRB}}\sim {\cal R}R_{cl}H_0\sim 1\;.
\end{equation}
because numerically ${\cal R}\sim \left(R_{cl}H_0\right)^{-1}$. 
The results in Refs.~\cite{Abazajian:2001vt,Abazajian:2006jc} imply that foreground objects dominate at the level of prefactors.

Although the total DM decay flux from the XRB contribution and from the individual clusters are of the same order, the spectral shape of an expected signal is quite different. This makes stacked observations of galaxy clusters and observations of field galaxies more sensitive~\cite{Abazajian:2001vt}. The flux from the cluster is in the form of a narrow line with the width determined by the spectral resolution of an X-ray detector or by the velocity dispersion of the DM halo of the cluster (if the detector energy resolution is better than $E/\Delta E\sim 10^2$).  At the same time, the DM decay contribution into XRB is produced by the decays at red shifts $z\sim 0\div 1$ and, as a result the DM decay line is broadened to $\Delta E\sim m_s/2$. 
Thus, even if the compact DM sources at $z\simeq 0$ give just moderate enhancement of the DM decay flux, the enhancement of the signal in the narrow energy band centered on the line energy $E=m_s/2$ could be large~\cite{Boyarsky:2005us,Boyarsky:2006fg}.
The results presented in Refs.~\cite{Abazajian:2001vt,Abazajian:2006jc} suggest that there is an even larger enhancement due to the dominant contributino from foreground objects.

This enhancement is however, tempered by the fact that most of the galaxy clusters show strong continuum and line emission from the hot intracluster gas. The temperatures of the intracluster medium are in the range $T_{\rm gas}\sim G\ {\cal R}\rho^0_{DM}R_{cl}^2m_p\sim 1- 10\mbox{~keV}$. The diffuse continuum emission flux from the nearby cluster cores is orders of magnitude stronger than the XRB emission from behind the core~\cite{Abazajian:2001vt,Cavaliere:76,Mushotzky:84,Sarazin:86,Briel:93,Bonamente:02,Boyarsky:2006zi,Pratt:08}. Thus, an increase of the flux in the DM decay line in the narrow energy band of the width about the energy resolution of the X-ray telescope is accompanied by an equally strong increase of the X-ray continuum contribution to the flux in the same energy band.  An additional problem is the presence of emission lines from the hot intracluster medium in the spectra of clusters~\cite{Sarazin:77,Bahcall:78,Sarazin:86,Smith:01a}.  These lines could be confused with the DM decay line, especially while the spectral resolution of the X-ray instruments is much larger than the line broadening.

Nearby DM mass concentrations, like the dwarf spheroidal (dSph) satellites of the Milky Way, have much smaller DM masses and their luminosity in the DM decay line is orders of magnitude lower than that of the galaxy clusters~\cite{Mateo:98}. However, their proximity to the Milky Way makes increases the flux of the line, so that at the end the flux from the direction of individual dSph galaxies is expected to be comparable to the flux from individual galaxy clusters~\cite{Boyarsky:2006fg,Boyarsky:2006ag,Loewenstein:2008yi,RiemerSorensen:2009jp,Loewenstein:2009cm,Boyarsky:2009rb,Boyarsky:2009af,Boyarsky:2010ci,Mirabal:2010jj,Mirabal:2010an,Loewenstein:2012px,Kusenko:2012ch,Malyshev:2014xqa,Sonbas:15}.  Re-using the equation~\eqref{eq:flux} for the flux estimate for the case of the dSph galaxies one finds
\begin{eqnarray}
\label{eq:flux_num}
F_{dSph}\simeq 1.4\times 10^{-7}\left[\frac{\theta^2}{10^{-11}}\right]\left[\frac{m_s}{7\mbox{~keV}}\right]^4\left[\frac{D_L}{100\mbox{ kpc}}\right]^{-2}
\left[\frac{M_{DM,FoV}}{10^{7}M_\odot}\right]\frac{\mbox{ph}}{\mbox{cm}^2\mbox{s}}
\end{eqnarray}

In addition the flux from the DM decays in the largest isolated mass concentrations, galaxy clusters, should produce a potentially detectable signal from individual clusters~\cite{Abazajian:2001vt,Boyarsky:2006zi,RiemerSorensen:2006pi,Boyarsky:2006kc} and from the entire galaxy cluster population~\cite{Bulbul:2014sua}. Finally, the nearest DM halos, which are the Milky Way and other nearby galaxies, are also expected to produce DM decay signal with the strength comparable to the signal from the galaxy clusters, in spite of the much lower mass of the DM in these structures~\cite{Abazajian:2001vt,Abazajian:2006jc,Watson:2006qb,Boyarsky:2006zi,Boyarsky:2006ag,Boyarsky:2007ge,Yuksel:2007xh,Boyarsky:2006hr,Boyarsky:2010ci,RiemerSorensen:2009jp,Loewenstein:2008yi,Loewenstein:2012px,Kusenko:2012ch,Malyshev:2014xqa,Sonbas:15}.

The DM decay flux from the Milky Way halo is also comparable to the flux from isolated distant sources, like dSph galaxies or galaxy clusters~\cite{Boyarsky:2006fg}. The flux from DM decay in the Milky Way within the telescope field-of-view $\Omega_{FoV}$,
\begin{equation}
F=\frac{\Gamma\Omega_{FoV}{\cal S}}{4\pi m_{DM}}\simeq 1.1\times 10^{-2}\left[\frac{\theta^2}{10^{-11}}\right]\left[\frac{S}{10^{22}\mbox{ GeV/cm}^2}\right] \left[\frac{m_s}{7\mbox{~keV}}\right]^4\frac{\mbox{ph}}{\mbox{cm}^2\mbox{s sr}},
\end{equation}
is determined by the column density of the DM
\begin{equation}
{\cal S}=\int\limits_{0}^{\infty}\rho_{DM}\left(\sqrt{r^2_{\odot}-2zr_\odot\cos\phi+z^2}\right)dz
\label{eq:S_MW}
\end{equation}
where $r_\odot=8.5$~kpc is the distance from the Sun to the center of our Galaxy and the "off-Galactic Center" angle $\phi$ is expressed through  the galactic coordinates $(l,b)$ as $\cos\phi=\cos b \cos l$.


Search of the DM decay signal in the~keV--MeV mass range was conducted using all the four types of the signal  (the XRB contribution, the signal from galaxy clusters, from the dSph galaxies and from the Milky Way halo) and a wide range of X-ray telescopes: \xmm~\cite{Boyarsky:2005us,Boyarsky:2006zi,Boyarsky:2006fg,Watson:2006qb,Boyarsky:2006ag,Boyarsky:2007ay,Loewenstein:2012px}, \textit{Chandra}~\cite{RiemerSorensen:2006pi,RiemerSorensen:2006pi,Boyarsky:2006kc,RiemerSorensen:2009jp,Loewenstein:2009cm,Boyarsky:2010ci}, \textit{Suzaku}~\cite{Loewenstein:2008yi,Kusenko:2012ch}, \textit{Swift}~\cite{Mirabal:2010an}, \textit{INTEGRAL}~\cite{Yuksel:2007xh,Boyarsky:2007ge}, HEAO-1~\cite{Boyarsky:2005us} and Fermi/GBM~\cite{Horiuchi:2015pda}, as well as a rocket-borne X-ray microcalorimeter~\cite{Abazajian:2006jc,Boyarsky:2006hr}. A summary of the reported observations is given in Tables~\ref{tab:bounds-summary}--\ref{tab:bounds-summary-clusters}.

\begin{table*}[th]
  \rowcolors{1}{}{lightgray}
\begin{tabularx}{\textwidth}{l|X|X|c}
\hline
Ref. & Object & Instrument & Cleaned exp, ks  \\
\hline 
\cite{Boyarsky:2005us} & Diffuse X-ray background & HEAO-1, \xmm/EPIC & 224, 1450  \\
\cite{Boyarsky:2006zi} & Coma, Virgo & \xmm/EPIC & 20, 40  \\
\cite{Boyarsky:2006fg} & Large Magellanic Cloud & \xmm/EPIC & 20  \\
\cite{RiemerSorensen:2006pi} & Milky Way & \chan/ACIS-S3 & Not specified  \\
\cite{Watson:2006qb} & M31 (central $5'$) & \xmm/EPIC & 35  \\
\cite{RiemerSorensen:2006pi} & Abell~520 & \chan/ACIS-S3 & 67  \\
\cite{Boyarsky:2006ag} & Milky Way, Ursa Minor & \xmm/EPIC & 547, 7  \\
\cite{Abazajian:2006jc} & Milky Way & \chan/ACIS, X-ray microcalorimeter & 1500, 0.1  \\
\cite{Boyarsky:2006kc} & 1E~0657-56 (``Bullet cluster'') & \chan/ACIS-I & 450  \\
\cite{Boyarsky:2006hr} & Milky Way & X-ray microcalorimeter & 0.1  \\
\cite{Yuksel:2007xh} & Milky Way & INTEGRAL/SPI & 5500  \\
\cite{Boyarsky:2007ay} & M31 (central $5-13'$) & \xmm/EPIC & 130  \\
\cite{Boyarsky:2007ge} & Milky Way & INTEGRAL/SPI & 12200 \\
\cite{Loewenstein:2008yi} & Ursa Minor & \suza/XIS & 70  \\
\cite{RiemerSorensen:2009jp} & Draco & \chan/ACIS-S & 32  \\
\cite{Loewenstein:2009cm} & Willman~1 & \chan/ACIS-I & 100  \\
\cite{Prokhorov:2010us,Koyama:06} & Galactic center & \suza/XIS & 190 \\
\cite{Boyarsky:2010ci} & M31, Fornax, Sculptor & \xmm/EPIC , \chan/ACIS & 400, 50, 162   \\
\cite{Mirabal:2010jj} & Willman~1 & \chan/ACIS-I & 100\\
\cite{Mirabal:2010an} & Segue~1 & Swift/XRT & 5  \\

\hline
\end{tabularx} 
\caption{Summary of existing X-ray observations of DM decays in different objects performed
  by different groups.
}
  \label{tab:bounds-summary} 
\end{table*}

\begin{table*}[th]
  \rowcolors{1}{}{lightgray}
\begin{tabularx}{\textwidth}{l|X|X|c}
\hline
Ref. & Object & Instrument & Cleaned exp, ks  \\
\hline 
\cite{Borriello:2011un} & M33 & \xmm/EPIC & 20-30  \\
\cite{Watson:2011dw} & M31 ($12-28'$ off-center)& \chan/ACIS-I & 53 \\
\cite{Loewenstein:2012px} & Willman~1 & \xmm/EPIC & 60 \\
\cite{Kusenko:2012ch} & Ursa Minor, Draco & \suza/XIS & 200, 200 \\
\cite{Iakubovskyi:13} & Stacked galaxies & \xmm/EPIC & 8500 \\
\cite{Horiuchi:2013noa} & M31 & \chan/ACIS-I & 404 \\
\cite{Bulbul:2014sua} & Stacked clusters & \xmm/EPIC, \chan/ACIS & 6000, 1370 \\
\cite{Boyarsky:2014ska} & M31, Perseus, Milky Way & \xmm/EPIC & 1226, 264, 7850\\
\cite{Riemer-Sorensen:2014yda} & Galactic center & \chan/ACIS & 751 \\
\cite{Jeltema:2014qfa} & Galactic center, M31, Tycho & \xmm/EPIC & 690, 490, 175\\
\cite{Malyshev:2014xqa} & Stacked dSphs & \xmm/EPIC & 410 \\
\cite{Anderson:2014tza} & Stacked galaxies & \xmm/EPIC, \chan/ACIS-I & 14600, 15000 \\
\cite{Urban:2014yda} & Perseus, Coma, Ophiuchus, Virgo & \suza/XIS & 740, 164, 83, 90\\
\cite{Carlson:2014lla} & Galactic center, Perseus & \xmm/EPIC & Not specified\\
\cite{Tamura:2014mta} & Perseus & \suza/XIS & 520\\
\cite{Koyama:2014zca} & Galactic Bulge & \suza/XIS & Not specified \\
\cite{Horiuchi:2015pda,Ng:2015gfa} & Milky Way & Fermi/GBM & 4600 \\
\cite{Sekiya:2015jsa} & Milky Way & \suza/XIS & 31500 \\
\cite{Sonbas:15} & Draco & \xmm/EPIC & 87\\
\hline
\end{tabularx} 
\caption{Summary of existing X-ray observations of DM decays in different objects performed
  by different groups (extended).
}
  \label{tab:bounds-summary_B} 
\end{table*}
  
\begin{table*}[th]
  \rowcolors{1}{}{lightgray}
\begin{tabularx}{\textwidth}{l|X|X|c}
\hline
Ref. & Object & Instrument & Cleaned exp, ks  \\
\hline 
\cite{Boyarsky:2006fg} & Large Magellanic Cloud & \xmm/EPIC & 20  \\
\cite{Boyarsky:2006ag} & Ursa Minor & \xmm/EPIC & 7  \\
\cite{Boyarsky:2010ci} & Fornax & \xmm/EPIC & 50   \\
\cite{Malyshev:2014xqa} & Stacked dSphs & \xmm/EPIC & 410 \\
\cite{Loewenstein:2012px} & Willman~1 & \xmm/EPIC & 60 \\
\cite{Sonbas:15} & Draco & \xmm/EPIC & 87\\
\hline
\cite{Loewenstein:2008yi} & Ursa Minor & \suza/XIS & 70  \\
\cite{Kusenko:2012ch} & Ursa Minor, Draco & \suza/XIS & 200, 200 \\
\hline
\cite{RiemerSorensen:2009jp} & Draco & \chan/ACIS-S & 32  \\
\cite{Loewenstein:2009cm} & Willman~1 & \chan/ACIS-I & 100  \\
\cite{Boyarsky:2010ci} & Sculptor & \chan/ACIS & 162   \\
\cite{Mirabal:2010jj} & Willman~1 & \chan/ACIS-I & 100\\
\hline
\cite{Mirabal:2010an} & Segue~1 & Swift/XRT & 5  \\
\cite{Jeltema:2015mee} & Draco & \xmm/EPIC & 2100 (MOS), 580 (PN)\\
\cite{Ruchayskiy:2015onc} & Draco & \xmm/EPIC & 1990 (MOS), 650 (PN)\\
\hline
\end{tabularx} 
\caption{Summary of existing X-ray observations of DM decays in dwarf galaxies performed
  by different groups.
}
  \label{tab:bounds-summary-dwarves} 
  \end{table*}

\begin{table*}[th]
\rowcolors{1}{}{lightgray}
\begin{tabularx}{\textwidth}{l|X|X|c}
\hline
Ref. & Object & Instrument & Cleaned exp, ks  \\
\hline 
\cite{Watson:2006qb} & M31 (central $5'$) & \xmm/EPIC & 35  \\
\cite{Boyarsky:2006ag} & Milky Way & \xmm/EPIC & 547  \\
\cite{Boyarsky:2007ay} & M31 (central $5-13'$) & \xmm/EPIC & 130  \\
\cite{Boyarsky:2010ci} & M31 & \xmm/EPIC & 400   \\
\cite{Borriello:2011un} & M33 & \xmm/EPIC & 20-30  \\
\cite{Iakubovskyi:13} & Stacked galaxies & \xmm/EPIC & 8500 \\
\cite{Horiuchi:2013noa} & M31 & \chan/ACIS-I & 404 \\
\cite{Boyarsky:2014ska} & M31, Milky Way & \xmm/EPIC & 1226, 7850\\
\cite{Jeltema:2014qfa} & Galactic center, M31 & \xmm/EPIC & 690, 490\\
\cite{Anderson:2014tza} & Stacked galaxies & \xmm/EPIC & 14600 \\
\cite{Carlson:2014lla} & Galactic center & \xmm/EPIC & Not specified\\
\hline
\cite{Abazajian:2006jc} & Milky Way & \chan/ACIS & 1500  \\
\cite{RiemerSorensen:2006pi} & Milky Way & \chan/ACIS-S3 & Not specified  \\
\cite{Watson:2011dw} & M31 ($12-28'$ off-center)& \chan/ACIS-I & 53 \\
\cite{Riemer-Sorensen:2014yda} & Galactic center & \chan/ACIS & 751 \\
\cite{Anderson:2014tza} & Stacked galaxies & \chan/ACIS-I & 15000 \\
\hline
\cite{Prokhorov:2010us,Koyama:06} & Galactic center & \suza/XIS & 190 \\
\cite{Koyama:2014zca} & Galactic Buldge & \suza/XIS & Not specified \\
\cite{Sekiya:2015jsa} & Milky Way & \suza/XIS & 31500 \\
\hline
\cite{Boyarsky:2006hr} & Milky Way & X-ray microcalorimeter & 0.1  \\
\hline
\cite{Yuksel:2007xh} & Milky Way & INTEGRAL/SPI & 5500  \\
\cite{Boyarsky:2007ge} & Milky Way & INTEGRAL/SPI & 12200 \\
\hline
\cite{Horiuchi:2015pda,Ng:2015gfa} & Milky Way & Fermi/GBM & 4600 \\
\hline
\cite{Neronov:2016wdd} & Milky Way & NuSTAR & 7500 \\
\hline
\cite{Perez:2016tcq} & Galactic center & NuSTAR & 400 \\
\hline

\hline
\end{tabularx} 
\caption{Summary of existing X-ray observations of DM decays in spiral galaxies and Milky Way halo 
performed by different groups.
}
  \label{tab:bounds-summary-spirals} 
  \end{table*}
  
\begin{table*}[th]
  \rowcolors{1}{}{lightgray}
 \begin{tabularx}{\textwidth}{l|X|X|c}
 \hline
 Ref. & Object & Instrument & Cleaned exp, ks  \\
 \hline 
~\cite{Boyarsky:2005us} & Diffuse X-ray background & HEAO-1 & 224  \\
 \hline
~\cite{Boyarsky:2005us} & Diffuse X-ray background & \xmm/EPIC & 1450  \\
~\cite{Boyarsky:2006zi} & Coma, Virgo & \xmm/EPIC & 20, 40  \\
~\cite{Bulbul:2014sua} & Stacked clusters & \xmm/EPIC & 6000 \\
~\cite{Boyarsky:2014ska} & Perseus & \xmm/EPIC & 264\\
~\cite{Carlson:2014lla} & Perseus & \xmm/EPIC & Not specified\\
 \hline
~\cite{RiemerSorensen:2006pi} & Abell~520 & \chan/ACIS-S3 & 67  \\
~\cite{Boyarsky:2006kc} & 1E~0657-56 (``Bullet cluster'') & \chan/ACIS-I & 450  \\
~\cite{Bulbul:2014sua} & Stacked clusters & \chan/ACIS & 1370 \\
 \hline
~\cite{Urban:2014yda} & Perseus, Coma, Ophiuchus, Virgo & \suza/XIS & 740, 164, 83, 90\\
~\cite{Tamura:2014mta} & Perseus & \suza/XIS & 520\\
 \hline
 \end{tabularx} 
\caption{Summary of existing X-ray observations of DM decays in galaxy clusters and 
diffuse X-ray background performed by different groups.
}
  \label{tab:bounds-summary-clusters} 
  \end{table*}

\begin{figure}
  \includegraphics[width=\linewidth]{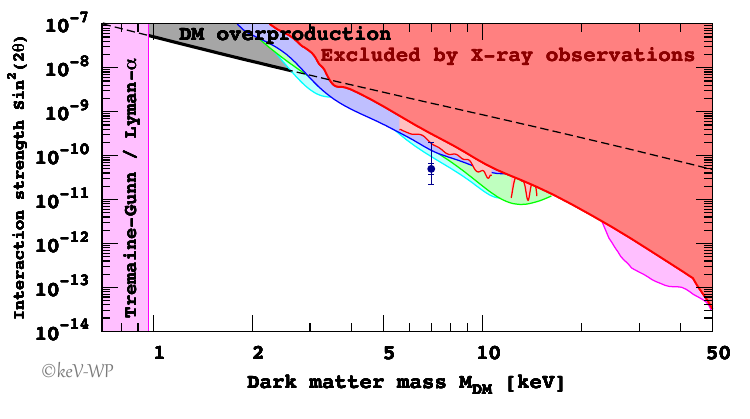} \caption{Limits on the mixing angle $\sin^2(2\theta)$ as a function of sterile neutrino DM mass. The bounds are based on the works~\cite{Boyarsky:2005us,Boyarsky:2006zi,Boyarsky:2006fg,RiemerSorensen:2006pi,Watson:2006qb,RiemerSorensen:2006pi,Boyarsky:2006ag,Abazajian:2006jc,Boyarsky:2006kc,Boyarsky:2006hr,Yuksel:2007xh,Boyarsky:2007ay,Boyarsky:2007ge,Loewenstein:2008yi,RiemerSorensen:2009jp,Loewenstein:2009cm,Mirabal:2010jj,Mirabal:2010an,Borriello:2011un,Watson:2011dw,Loewenstein:2012px,Iakubovskyi:13,Horiuchi:2013noa,Malyshev:2014xqa,Ng:2015gfa}. All bounds are smoothed and additionally divided by a factor of $2$ to take into account possible uncertainties in the DM content of a given object. A lower bound on $\sin^2\theta$ for a given DM mass can be imposed if the DM is produced via active-sterile mixing, see section \ref{sec:prod}, but is model dependent.
}
\label{fig:exclusion_plot}
\end{figure}

Most of the searches of the DM decay line in X-rays did not result in positive detections of the line. The non-detection results are conventionally expressed in the form of the upper limit on the DM sterile neutrino mixing angle as a function of the mass. These limits are shown in Fig. \ref{fig:exclusion_plot}.

\subsubsection{3.5~keV line}
\label{sec:kev-line}

The detection of an unidentified line was reported recently in the stacked spectrum of 
galaxy clusters~\cite{Bulbul:2014sua}, in the individual spectra of nearby galaxy 
clusters~\cite{Bulbul:2014sua,Boyarsky:2014jta,Iakubovskyi:2015dna} (see also~\cite{Urban:2014yda}), 
in the Andromeda galaxy~\cite{Boyarsky:2014jta}, and in the Galactic Center 
region~\cite{Jeltema:2014qfa,Boyarsky:2014ska,Riemer-Sorensen:2014yda}, 
see also~\cite{Iakubovskyi:2015wma} for a recent review. 
The position of the line is $E= 3.55$~keV 
with an uncertainty in position $\sim 0.05$~keV. If the line is interpreted as originating from a 
two-body decay of a DM particle, then the latter has its mass at about $m_s\simeq 7.1$~keV and the 
lifetime $\tau_{DM}\sim 10^{27.8 \pm 0.3}$~sec~\cite{Boyarsky:2014jta} (and more narrow range 
based on the results of~\cite{Bulbul:2014sua}). Using Eq.~(\ref{gamma}) and 
converting the lifetime to the sterile neutrino mixing angle, one finds 
$\sin^2(2\theta)\simeq (2-20)\times 10^{-11}$~\cite{Boyarsky:2014jta} or 
$\sin^2(2\theta)\simeq (4-11)\times 10^{-11}$ (based on results of~\cite{Bulbul:2014sua}).  
The estimates of the errors on the mixing angle $\theta^2$ derived from different types 
of data show a visible scatter, which is, however, consistent with the quoted errors. 
For example, Ref.~\cite[Fig.~14]{Bulbul:2014sua} shows the scatter in the possible mixing 
angles derived from the statistics of the data and not including the systematic error due 
to the uncertainties of the cluster mass estimates.  The 3.5~keV line was also detected in 
the Galactic Center~\cite{Jeltema:2014qfa,Riemer-Sorensen:2014yda,Boyarsky:2014ska} with a 
flux that is be consistent with the DM interpretation of the line in clusters and in the 
Andromeda galaxy~\cite{Boyarsky:2014ska}. Namely, it is sufficiently strong (to be consistent 
with the cluster detection) and sufficiently weak to be consistent with the non-detection from 
the outskirts of the Milky way~\cite{Boyarsky:2014jta}. However, in the Galactic Center the flux of the line is also consistent with the expectations for plasma emission lines with no need for a dark matter component~\cite{Boyarsky:2014ska,Jeltema:2014qfa}.
We note that the line is also recently detected in the \textit{NuStar} blank-sky~\cite{Neronov:2016wdd} 
and Galactic center~\cite{Perez:2016tcq} datasets. 
While it may have an instrumental origin, related to Solar activity~\cite{Neronov:2016wdd},   
it was shown in~\cite{Neronov:2016wdd} that in the `no sun' dataset the 
line is present with the same intensity, and it appears several times stronger near the Galactic center, 
as Fig.~4 of~\cite{Perez:2016tcq} demonstrates.

A difficulty in interpreting the origin of a weak emission line is inherent uncertainty in the astrophysical backgrounds, in particular in the flux of plasma emission lines. The strongest uncertainty comes from two potassium lines, K XVIII at 3.47 and 3.51~keV.  Given the spectral resolution of the XMM-Newton,  within the systematic uncertainty the flux could be attributed to emission from these K XVIII plasma lines. Ref.~\cite{Jeltema:2014qfa} argued that considering a larger range of plasma temperatures reduces the tension between the observed 3.5~keV line flux in clusters and the expectations from known plasma lines (see however the subsequent discussion in~\cite{Boyarsky:2014paa,Bulbul:2014ala,Jeltema:2014mla}). 
The interpretation of the 3.5~keV line as a plasma line would imply that its surface brightness profile must trace the density of the plasma (more precisely: the distribution of potassium). If it is a DM line it should, on the other hand, trace the overall distribution of DM, which dominates the gravitating matter in Perseus and other clusters.
This point is disputed: While the analysis in~\cite{Boyarsky:2014ska} suggests that the line traces the overall matter distribution (pointing towards DM), the authors of~\cite{Carlson:2014lla} conclude that the morphology is incompatible with the DM interpretation.  Explicitly, the authors of Ref.~\cite{Carlson:2014lla} argue that the morphology of the 3.5~keV signal is incompatible with DM in the Galactic Center with a spatial distribution that places strong upper limits on the possible dark matter flux.  The authors of~\cite{Carlson:2014lla} also show that the 3.5~keV emission in Perseus is concentrated on the cool-core and less extended than expected for dark matter decay. Possible flaws in the analysis in~\cite{Carlson:2014lla} are discussed in~\cite{Franse:2016dln}. Specifically,  Ref.~\cite{Franse:2016dln} discusses some challenges with the template analysis of~\cite{Carlson:2014lla}, which the authors of~\cite{Carlson:2014lla} however do not agree upon. The Potassium interpretation also cannot explain the origin of the line in the Andromeda galaxy reported in \cite{Boyarsky:2014jta}, the significance of which is, however, disputed~\cite{Jeltema:2014qfa,Boyarsky:2014paa}. 

Systematic errors in instrumental calibration and/or systematics induced by the analysis procedure may impact the significance of weak lines.  The calibration systematics was explored in~\cite{Boyarsky:2014jta} who demonstrated that no line is detected in an extremely long exposure combination of the off-center observations of the Milky way (``blank sky'' dataset).  A 3.5~keV line is not detected in the stacked dwarf galaxies~\cite{Malyshev:2014xqa} arguing against a global calibration problem. It has also not been seen in a long exposure towards M31 reported in Ref.~\cite{Horiuchi:2013noa}. In order to assess possible systematic effects, the  Ref.~\cite{Jeltema:2014qfa} considered the XMM-Newton spectra of the Tycho supernova remnant, assuming that this system displays similar plasma emission lines to those detected in clusters and the GC but would have no expected DM signal. A line at $\sim 3.55$~keV is detected in Tycho (with a significance of $\Delta \chi^2  = 19$)\footnote{The significance of the signal found in \cite{Carlson:2014lla} was communicated in private by T.~Jeltema.}. Ref. ~\cite{Jeltema:2014qfa} argued for either systematic errors in the estimated significance and flux of the 3.5~keV line or a systematic underprediction of the flux expected from weak plasma lines like K XVIII, claiming that the K XVIII explanation is at odds with theoretical predictions of elemental abundances from Type Ia supernovae. While Tycho does show emission from similar elemental lines as seen in clusters and the GC, supernova remnants have a combination of thermal and non-thermal X-ray emission and the ionization conditions are not necessarily the same as in clusters, making the comparison less than clear cut. However, a 3.5~keV line is not necessarily expected in supernova remnants and the detection of one may point to instrumental or modeling systematics.

Ref.~\cite{Tamura:2014mta} explored the systematic for the Suzaku XIS instrument (somewhat similar to EPIC instruments of the XMM-Newton).  They find that systematic errors in the instrumental response of XIS are at the level of half the detected line flux in clusters; they also suggest that the modeling of a large number of plasma emission lines which overlap given the instrumental energy resolution as in the cluster and GC analyses may artificially suppress the continuum in line free regions creating spurious weak signals (this does not explain, however, the origin of the line in the Andromeda galaxy spectrum, where the 3.5~keV line is the strongest one in the range 3-4~keV, see discussion in~\cite{Jeltema:2014qfa,Boyarsky:2014paa}).

Ref.~\cite{Anderson:2014tza} has developed a new method of data analysis, and applied it to the stacked spectra of galaxies.  It uses a non-physical (spline) model of the background. As a narrow line cannot be modeled by a small number of splines, the line gets simulated using the response matrices of the instrument, subtracted from the spectrum and the remaining signal is fitted by the splines. The sensitivity of this method has only been tested on simulated spectra. The simulations show in particular that the sensitivity is \emph{the lowest} in the range around 3.5~keV (with the exact value of systematic uncertainty not reported by the authors). It can be seen that the simulated line with $\sin^2(2\theta) = 7\times 10^{-12}$ is \emph{not recovered at all}, while the higher values of the mixing angle are recovered with about $\Delta \sin^2(2\theta) \approx 1.6\times 10^{-11}$ lower value (the negative residual in the spectrum propagated linearly). Thus, \emph{the lowest limit from~\cite{Boyarsky:2014jta} cannot be probed with this method.}  In addition, the method of~\cite{Anderson:2014tza} constructs the ``signal estimator'' by dividing an x-ray signal in each pixel by the expected DM signal. The amount of DM in each pixel is roughly estimated by combination of two scaling relations (one is between stellar and halo masses,~\cite{Moster:10} and another concentration--mass relation~\cite{Prada:2011jf}) and it is not clear how strong is the systematic uncertainty from wrong DM estimates. Finally, the recovered continuum residuals shown in~\cite{Anderson:2014tza} indicate a large systematic uncertainty in the continuum that fluctuates up to levels of the signal, which should affect the claimed sensitivity.

Ref.~\cite{Urban:2014yda} has searched for a line in four individual clusters, Perseus, Coma, Virgo, and Ophiuchus is deep Suzaku spectra.  While they detect a line near 3.5~keV in Perseus, they do not detect a line in the other three clusters. Ref.~\cite{Urban:2014yda} claims that this result is inconsistent with a DM interpretation, however, they do not provide sufficient analysis of uncertainties in DM densities of these clusters and therefore it is not clear, how robust this conclusion is. Moreover, the results of Ref.~\cite{Urban:2014yda} appear to be inconsistent with those reported in Refs.~\cite{Tamura:2014mta,Franse:2016dln}.

The authors of Ref.~\cite{Iakubovskyi:2015dna} studied central parts of 19 galaxy clusters observed by XMM-Newton having the largest known DM column density in XMM-Newton Field-of-View. In 8 of them (confirming previous detections in Perseus~\cite{Bulbul:2014sua} and Coma~\cite{Urban:2014yda} clusters),  the new line has been detected at $> 2\sigma$ level. The average position of the new line in these objects is 3.52$\pm$0.08~keV. The decaying DM lifetime imposed from the new line observations and non-observations in these objects is $(3.5-6) \times 10^{27}$~s consistent with previous  detections of~\cite{Bulbul:2014sua,Boyarsky:2014ska,Boyarsky:2014ska}. Recently, Ref.~\cite{Hofmann:2016urz} did not find the line in a stack of 33~clusters observed with Chandra and Hitomi did not have a significant detection in Perseus~\cite{Aharonian:2016gzq}.

More recently, two papers have argued for potential astrophysical origins in the 3.5~keV line~\cite{Phillips:2015wla,Gu:2015gqm}. Ref.~\cite{Phillips:2015wla} analyze high-resolution solar flare spectra and find that the coronal  potassium abundance is a factor of 9-11 higher than the photospheric abundances assumed in the above analyzes, offering a explanation as to why the flux of these lines may be brighter than expected. It is, however, not clear why the solar corona should exhibit similar isotopic abundances to the IGM, as the metallicity is generally expected to be higher near stars. Ref.~\cite{Gu:2015gqm} argue that charge exchange, which has thus far not been included in the spectral modeling, can lead to enhanced S XVI lines near 3.46~keV. Though this is in principle possible, the precise energy of the  S XVI lines does not agree with the unidentified line, and charge exchange has so far not been detected in x-ray cluster plasma.

As discussed above, an observation of the line in the direction of a dSph satellite of the Milky Way would be a `smoking gun' for the DM interpretation of the signal, due to the absence of X-ray emitting gas in these systems. Ref.~\cite{Malyshev:2014xqa} has reported a search of the DM decay line in the stacked sample of dSph galaxies. Non-detection of the signal at 3.5~keV in the staked spectrum of dSph imposes a $2\sigma$ upper limit on the mixing angle at the level $\theta^2<1.1\times 10^{-11}$ and rules out the central value of $\theta^2$ reported in the Ref.~\cite{Bulbul:2014sua}.
The significance of this conclision is claimed to be $3-4\sigma$, depending on the assumption about the contribution of the Milky Way to the line signal. 
The anaylsis in Ref.~\cite{Horiuchi:2013noa}, which is claimed to be more constraining, only leads to a $1-2\sigma$ discrepancy.

Most recently, a data from a very long ($\sim$ 1.4 Msec) XMM-Newton observation of Draco dSph has become available. A statistically significant emission line from Draco dSph has not been detected ~\cite{Jeltema:2015mee,Ruchayskiy:2015onc}. This limits the lifetime of a decaying DM to $\tau>(7-9)\times 10^{27}$~sec at 95\% CL the interval corresponds to the uncertainty of the DM column density in the direction of Draco). However, the analysis in \cite{Ruchayskiy:2015onc} found a positive spectral residual (above the carefully modeled continuum) at $E=3.54\pm0.06$~keV with a $2.3\sigma$ significance in the spectrum of the EPIC PN camera. The two MOS cameras show less-significant or no positive deviations, consistently within $1\sigma$ with PN~\cite{Ruchayskiy:2015onc}. An analysis of the same data by other authors in~\cite{Jeltema:2015mee}, found no excess near 3.5~keV in any of the three EPIC instruments. As a result, the significance of the exclusion is currently under debate.

\begin{figure}[!t]
  \centering
  \includegraphics[width=0.8\textwidth]{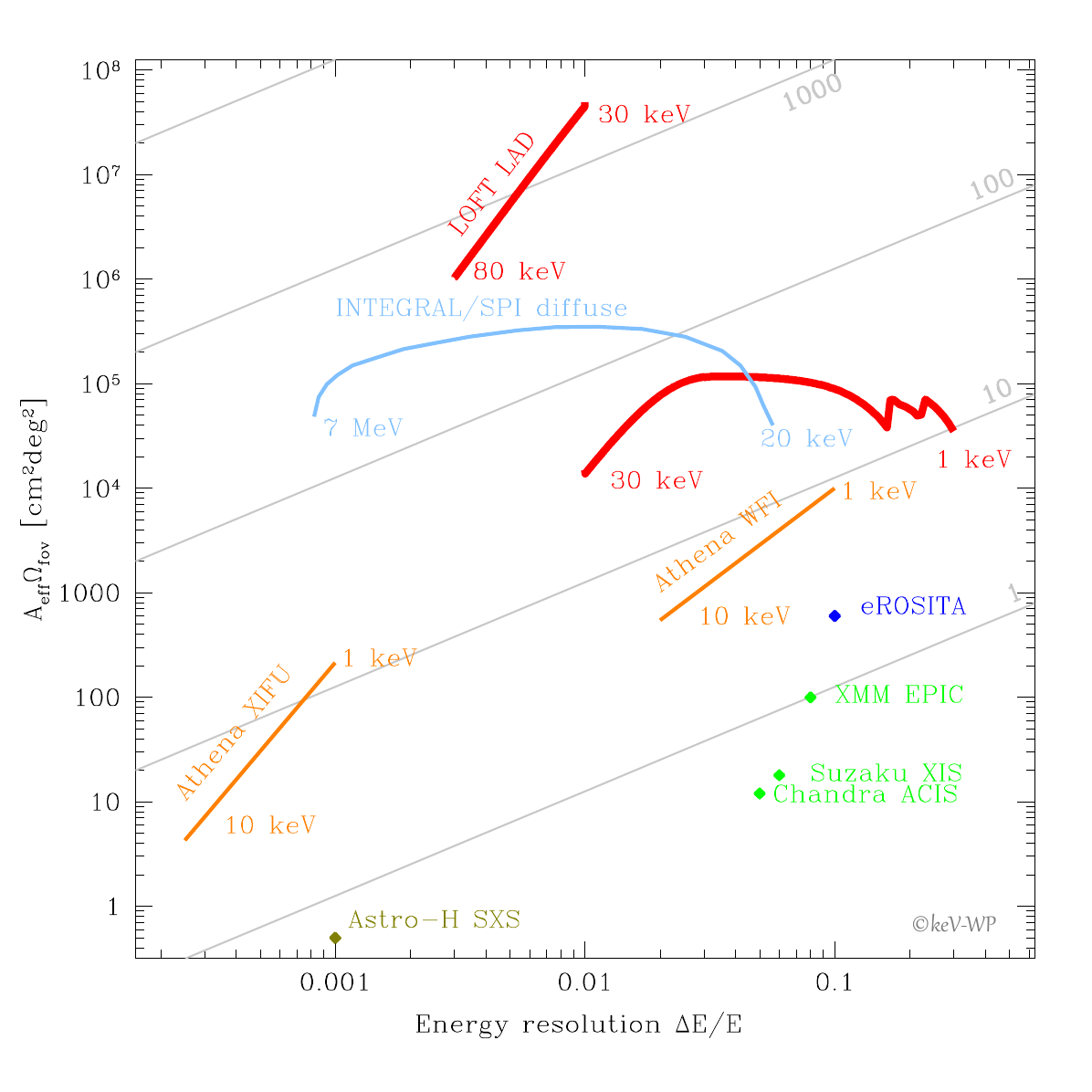}
  \caption{Comparison of sensitivities of current and planned high energy missions towards the searches for weak lines. }
  \label{fig:comparison}
\end{figure}

\subsubsection{Other line candidates in~keV range}
\label{sec:other-kev-lines}

In addition to the new line at 3.5~keV described in Sec.~\ref{sec:kev-line} above,
other potential signs of the radiatively decaying DM have been discussed:
\begin{enumerate}
 
\item According to~\cite{Prokhorov:2010us}, the intensity of \emph{Fe XXVI Ly-$\gamma$ line} at 8.7~keV observed in \suza/XIS spectrum of the Milky Way center~\cite{Koyama:06} cannot be explained by standard ionization and recombination processes, and the DM decay may be a possible explanation of this apparent excess.

\item According to Sec.~1.4 of~\cite{Koyama:2014zca}, two faint extra line-like excesses at 9.4 and 10.1~keV have been detected in the combined \suza/XIS spectrum of Galactic Bulge region. The positions of these excesses do not coincide with any bright astrophysical or instrumental line and their intensities can be explained in frames of decaying DM origin (see right Fig.~8 of~\cite{Koyama:2014zca}).
\end{enumerate}

%% file: Section4-4.tex
The $\beta$-ray spectroscopy is known to provide model independent upper limits on the effective mass of the electron neutrino, $\mathrm{m}_{(\nu_\mathrm{e})}^2 = \sum_i \textbar\mathrm{U}_{\mathrm{e}i}\textbar^2\cdot \mathrm{m}_i^2$, where $\mathrm{U}_{\mathrm{e}i}$ are the elements of the neutrino mixing matrix and $\mathrm{m}_i$ are the neutrino mass states (see e.g.~\cite{Drag15}). However, the measurements turned out to be an uneasy task. The massive neutrino changes the $\beta$-spectrum only it its uppermost part having an extremely low relative intensity. Therefore, the electron spectrometers intended for the neutrino mass determination should exhibit simultaneously the large luminosity L (the product of the relative input solid angle $\Omega$ and the radioactive source area S), high energy resolution $\Delta E_{\mathrm{instr}}$ and sufficiently low background. The large source area is needed to keep the sources as thin as possible. Otherwise the electron energy losses within the source deteriorate the measured $\beta$-spectrum shape. 

The first limit $\mathrm{m}_{\nu_\mathrm{e}}$ < 5~keV was derived from $\beta$-spectrum of $^{35}\mathrm{S}$ (with a $\beta$-decay energy of $\mathrm{Q}_{\beta}$ = 167~keV) taken with a magnetic spectrometer in 1948~\cite{Cook48}. The measurement of $^{3}\mathrm{H}$ $\beta$-spectrum ($\mathrm{Q}_{\beta}$ = 18.6~keV) with a setup of two orders of magnitude higher luminosity and the magnetic spectrometer set to $\Delta E_{\mathrm{instr}}$ = 40~eV yielded in 1972 the limit $\mathrm{m}_{\nu_\mathrm{e}}$ < 55~eV at 90\% CL~\cite{Berg72}. The present limit is $\mathrm{m}_{\nu_\mathrm{e}}$ < 2~eV~\cite{Agashe:2014kda}, while the KATRIN experiment aims at 0.2~eV sensitivity~\cite{Drex13}. 

In 1980, a long time before the undoubted evidence for neutrino oscillations, Shrock examined a possibility of searching for neutrino mass states $\mathrm{m}_i$ also in $\beta$-spectra~\cite{Shro80}. An admixture of each of such states should produce a specific discontinuity (a kink) in the $\beta$-spectrum at energy $\mathrm{E}_0$ - $\mathrm{m}_i$, where $\mathrm{E}_0$ is the endpoint energy of a particular $\beta$-emitter (i.e. the maximum energy of $\beta$-particles in the case when all $\mathrm{m}_i$ = 0). The relative intensity of the kink observed at $\mathrm{E}_0$ - $\mathrm{m}_i$ would determine the value of $\textbar\mathrm{U}_{\mathrm{e}i}\textbar^2$. The idea stimulated several investigators to search for an admixture of heavier neutrinos in their $\beta$-spectra. The decay of $^{64}\mathrm{Cu}$ that proceeds via both $\beta^+$- and $\beta^-$-branches with the endpoints energies of 653 and 579~keV, respectively, was explored with a magnetic spectrometer. The overall resolution including contribution from the energy losses within the radioactive source was determined by means of conversion electron spectroscopy (see e.g.~\cite{Drag83}) and amounted 1~keV at 300~keV. The result was $\textbar\mathrm{U}_{\mathrm{e}4}\textbar^2$ < $8\cdot10^{-3}$ at 90\% CL for 110~keV $\leqq$ $\mathrm{m}_4$ $\leqq$ 450~keV~\cite{Schreckenbach:1983cg}. Also the investigation of the $^{20}\mathrm{F}$ $\beta$-decay ($\mathrm{Q}_{\beta}$ = 7.0~MeV) with a magnetic spectrometer adjusted for the resolution of 3~keV at 4~MeV did not reveal any admixture of sterile neutrinos. The upper limit of $\textbar\mathrm{U}_{\mathrm{e}4}\textbar^2$ ranged from 0.59 to 0.18 percent for 400~keV $\leqq$ $\mathrm{m}_4$ $\leqq$ 2.8~MeV ~\cite{Deutsch:1990ut}.

\begin{figure}
\begin{center}
 	\includegraphics[width=0.9\textwidth]{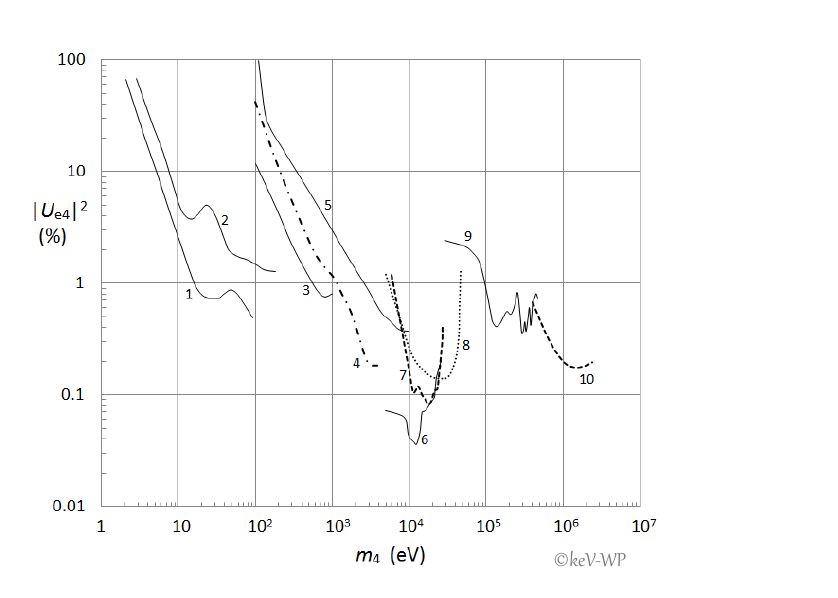} 
 	\caption{The best upper limits on the admixture $\textbar\mathrm{U}_{\mathrm{e}4}\textbar^2$ of sterile neutrinos derived from measured $\beta$-ray spectra. The limits are at the 90\% CL except those for ${}^{64}$Cu~\cite{Schreckenbach:1983cg} and ${}^3$H~\cite{Kraus:2012he} that are at the 90\% CL. The numbers refer to the following $\beta$-emitters and their investigators: 1 $\rightarrow$ ${}^3$H~\cite{Belesev:2013cba}, 2 $\rightarrow$ ${}^3$H~\cite{Kraus:2012he}, 3 $\rightarrow$ ${}^{187}$Re~\cite{Galeazzi:2001py}, 4 $\rightarrow$ ${}^3$H~\cite{Hiddemann:1995ce}, 5 $\rightarrow$ ${}^3$H~\cite{Simp81}, 6 $\rightarrow$ ${}^{63}$Ni~\cite{Holzschuh:1999vy}, 7 $\rightarrow$ ${}^{63}$Ni~\cite{Ohsh93}, 8 $\rightarrow$ ${}^{35}$S~\cite{Mort93}, 9 $\rightarrow$ ${}^{64}$Cu~\cite{Schreckenbach:1983cg}, 10 $\rightarrow$ ${}^{20}$F~\cite{Deutsch:1990ut}. The figure is reproduced from~\cite{Drag15}.}
 	\label{fig:currentlablimits}
\end{center}
\end{figure}

Simpson investigated the $\beta$-spectrum of tritium implanted into a Si(Li) detector and observed a distortion in the spectrum part below 1.5~keV.  He interpreted this distortion as the evidence of a heavy neutrino emission with the mass of about 17.1~keV and mixing probability of 3\%~\cite{Simp85}. This surprising finding prompted numerous studies of $^{3}\mathrm{H}$ and other $\beta$-emitters like $^{14}\mathrm{C}$, $^{35}\mathrm{S}$, and $^{241}\mathrm{Pu}$. For some time, the effect was seen in several $\beta$-spectra taken with semiconductor spectrometers but it has never been recorded by magnetic and electrostatic instruments. Finally, the study of the $\beta$-spectrum of $^{63}\mathrm{Ni}$ ($\mathrm{Q}_{\beta}$ = 66.9~keV) with a magnetic spectrometer put the admixture of the 17~keV neutrino below $5\cdot10^{-4}$ at 95\% CL and found $\textbar\mathrm{U}_{\mathrm{e}4}\textbar^2$ < $1\cdot10^{-3}$ for all $\mathrm{m}_4$ between 4 and 30~keV~\cite{Holzschuh:1999vy}. At present, there are no $\beta$-spectroscopic indications for sterile neutrinos~\cite{OHI1985322, PhysRevC.32.2215, PhysRevLett.55.799, PhysRevC.36.1504, 1992pnap.conf..217R}. The relevant data are summarized in~\cite{Agashe:2014kda}. The best upper limits of $\textbar\mathrm{U}_{\mathrm{e}4}\textbar^2$ are shown in figure~\ref{fig:currentlablimits}.  Current investigators of sterile neutrinos may benefit from a detailed discussion of experimental issues concerning the 17~keV neutrino~\cite{Wiet96}. As an example, the scattering of $\beta$-particles from $^{35}\mathrm{S}$ decay on a thin diaphragm in front of the Si(Li) detector can produce a false 0.3\% admixture of the 17~keV neutrino~\cite{Muel94}.
 
The reliable analysis of measured $\beta$-spectra requires precision knowledge of the spectrometer resolution function (the response to a mono-energetic signal). Although this function can be calculated it is advisable to check it for the actual instrument by means of an electron gun. Even better is to apply a suitable conversion electron line, where the natural width and shake up/off satellites were properly taken into account. Most of the systematic errors in $\beta$-spectroscopic searches for active as well as sterile neutrinos originated from incorrect treatment of the electron energy losses within radioactive sources and spectrometers. Corrections of measured $\beta$-spectra having no clear physical meaning not only decrease sensitivity of the search but can lead to erroneous conclusions. 

\begin{figure}
\center
 	\includegraphics[width=0.7\textwidth]{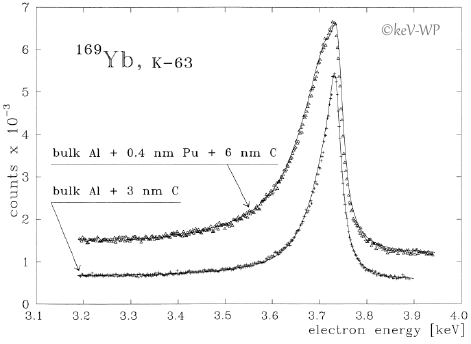} 
 	\caption{The K-shell internal conversion electron line of the 63.1~keV transition in $^{169}_{69}\mathrm{Tm}$ measured with an original and modified $^{169}\mathrm{Yb}$ source~\cite{Drag99}. The original source was prepared by vacuum evaporation on Al backing and was contaminated by a usual hydrocarbon overlayer. Subsequently, the source was covered by a thin plutonium layer in which the 3.7~keV electrons suffered additional elastic and inelastic scattering. The agreement of measured (points) and calculated shapes (smooth curves) demonstrates correctness of the electron loss function in plutonium applied in calculations of the $^{241}\mathrm{Pu}$ $\beta$-spectrum shape. The natural width of this conversion line amounting 32~eV was taken into account.}
 	\label{fig:conversionline}
\end{figure}

In the study of $^{241}\mathrm{Pu}$ $\beta$-spectrum ($\mathrm{Q}_{\beta}$ = 20.8~keV) with two electrostatic spectrometers the electron energy losses within a vacuum evaporated source were incorporated by means of the Monte Carlo simulation of individual elastic and inelastic scattering events. The quality of the approximation was tested by measuring the shape of the 3.7~keV conversion electron line of the $^{169}\mathrm{Yb}$ source covered subsequently with a thin plutonium layer (see figure~\ref{fig:conversionline}). This approach allowed describing the measured part of the $^{241}\mathrm{Pu}$ $\beta$-spectrum (down to 2~keV) without any artificial fitting parameter~\cite{Drag99}.

\begin{figure}
\center
 	\includegraphics[width=0.7\textwidth]{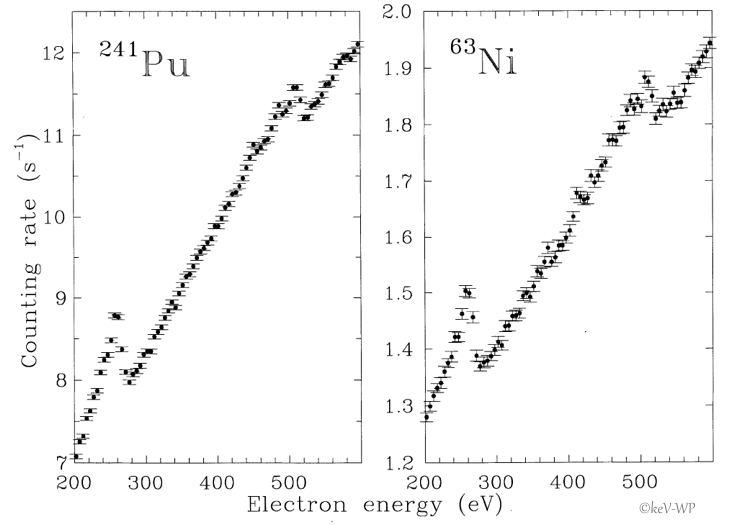} 
 	\caption{The KLL-Auger lines of carbon and oxygen superimposed on continuous $\beta$-spectra of $^{241}\mathrm{Pu}$ and $^{63}\mathrm{Ni}$~\cite{Drag00}. The radioactive sources were prepared by evaporation in vacuum and measured with an electrostatic spectrometer.}
 	\label{fig:augerline}
\end{figure}

Although the kink caused by the keV sterile neutrinos would change the $\beta$-spectrum part of several keV width, it is advisable to apply sufficient spectrometer resolution in order not to miss some unexpected spectrum features. For example the lines at energies of about 260 and 510~eV were observed on $\beta$-spectra of (ground state-to-ground state) decays of $^{241}\mathrm{Pu}$ and $^{63}\mathrm{Ni}$ that corresponded to the KLL Auger lines of carbon and oxygen~\cite{Drag00}. The K-shell vacancies were created by impact of $\beta$-particles emerging from the sources on atoms in the contamination overlayer (see also figure~\ref{fig:augerline}). In $\beta$-spectra taken with a worse resolution the Auger lines would smear and could be misinterpreted as an anomaly.

%% file: kevnuwp_section5.tex
Neutrino states that are considered ``sterile'' can (by definition) not feel any of the known forces of nature except gravity. That is, they carry no charges and are singlets under all gauge groups of the Standard Model. In order to compose the observed DM, they must, however, have some interactions with other particles in order to be produced in the early Universe. One can roughly distinguish three different ways how this can be realized.
\begin{itemize}

\item The sterile states can be produced via the weak interaction if they mix with ordinary neutrinos. This mixing occurs very generically because there is no a priori reason why the neutrino mass and interaction eigenstates should be aligned in flavor space. If the observed neutrino masses are generated via the (type-I) seesaw mechanism (see section~\ref{sub:seesaw} for details), then there is definitely mixing between active and sterile neutrinos. Production mechanisms that are based on this mixing are described in sections~\ref{sec:5.thermalproduction} and~\ref{Sec:ThermalProduction}. 

\item  Many theories beyond the SM predict the existence of additional scalar fields in Nature. If sterile neutrinos have Yukawa couplings to a scalar field, then they can be produced during its decay in the early Universe. This possibility is explored in section~\ref{sec:5.decays}.

\item If the SM is embedded into a more general theory of Nature (such as a Grand Unified Theory), then there are usually additional gauge symmetries that are ``broken'' at some energy scale above the reach of the LHC. This gives masses of the order of the energy scale where the symmetry is broken to the gauge bosons, effectively switching off the corresponding gauge interaction at lower energies. Thus, the neutrino states that appear to be sterile at the energies which can be probed in experiments may interact via new gauge interactions at higher energies. This allows to produce them in the high temperature phase of the early Universe. This idea is discussed in section~\ref{sec:5.dilution}.
\end{itemize}

\input{Section5_12_ml.tex}
\input{3_decay}
\input{4_dilution}

%% file: Section5_12_ml.tex
\subsection{Thermal production: overview (Authors: M.~Drewes, G.~Fuller, A. V.~Patwardhan) \label{sec:5.thermalproduction}} 

\subsubsection{Motivation}

All that we know about DM is consistent with the hypothesis that it is composed of long-lived, massive particles without long-range interactions to each other or to SM particles. Neutrinos are the only particles in the SM that fulfil these requirements. They are, however, too light: thermally produced neutrinos $\nu_\alpha$ are relativistic when structures in the Universe begin to form, and contribute to the total DM density as Hot Dark Matter. This is clearly disfavoured by structure formation, e.g., see Sec.~\ref{sec:indentityDM}. Sterile neutrinos, on the other hand, can be heavier and \lq\lq colder\rq\rq. This makes them natural DM candidates. Their interaction strength with all known particles is \lq\lq weaker than weak\rq\rq, i.e., suppressed with respect to the weak interaction by the active-sterile mixing angles $\theta_{I \alpha}$. Though $N_I$ are unstable particles and decay via the weak interaction of their $\nu_L$-component, a sufficiently small mixing angle $\theta_{I\alpha}$ can guarantee the required longevity. The emission line that is expected from the decay $N_I\rightarrow \nu_\alpha \gamma$ is one of the main testable predictions of sterile neutrino scenarios, and is discussed in Sec.~\ref{sec:xray}.

Whether sterile neutrinos are hot, warm, or cold Dark Matter depends on the way they were produced in the early Universe. 
The sterile neutrino production scenarios all break down into two broad categories: (1) those where sterile neutrinos are produced copiously early on, perhaps even attaining thermal and chemical equilibrium like the active neutrinos, but with early decoupling and subsequent dilution of their relic densities down to acceptable DM values; and (2) those where there are {\it no} sterile neutrinos present to begin with. In the minimal scenario described by the seesaw Lagrangian, eq.~\eqref{eqnuMajorana}, $\nu_R$ are gauge singlets, and their only interactions with other particles at low energies are mediated by their mixing $\theta$ with ordinary neutrinos. Then thermal production via this mixing is the only way to produce the mass eigenstates $N_I$. For values of $M_I$ and $\theta_{I\alpha}$ that are consistent with phase space considerations (see Sec.~\ref{sec:4-1-Phasespace}) and the longevity constraint, the $N_I$ never reach thermal equilibrium in the early Universe. Hence, the minimal scenario falls into category (2). The production by mixing is absolutely unavoidable for any non-zero value of $\theta_{I \alpha}$, therefore the $N_I$ population produced in this way is the minimal Dark Matter content of the Universe for a given parameter choice. If the $N_I$ in addition have other interactions at high energy, then these can produce additional $N_I$ populations. The most studied examples of these are described in detail in Sec.~\ref{sec:5.decays} and Sec.~\ref{sec:5.dilution}.

In this section we concentrate on the minimalist (and unavoidable) model for production of a relic sterile neutrino number density and energy spectrum that could make these particles suitable DM candidates, namely, the thermal production via mixing with ordinary neutrinos~\cite{Dodelson:1993je}.\footnote{Note that a minimum amount of thermally produced right handed neutrinos even exists if neutrinos are Dirac particles \cite{Chen:2015dka}.} This model is \lq\lq minimalist\rq\rq\ in that it requires only a minimal extension of the Standard Model, to massive active and sterile neutrinos which mix in vacuum, but which otherwise have {\it only} Standard Model interactions (i.e., the weak interaction in the case of active flavors). This guarantees that this model falls into the second category, i.e., no sterile neutrinos in the early Universe to begin with.

%
\subsubsection{Active-sterile neutrino oscillations} 

Before going into more details of the thermal production mechanism, we explain the underlying physics in a simple quantum mechanical picture. In the following, we assume that only one species of heavy neutrinos contributes to the Dark Matter density and refer to this mass eigenstate as $\nu_s\equiv N_1$.
   
The earliest theoretical exploration of neutrino flavor mixing was done by Pontecorvo in the 1960s~\cite{Pontecorvo:1967fh}, and it already involved speculation about mixing between active and sterile neutrinos. Let us, for simplicity, assume that there is only one active neutrino flavor $\nu_\alpha$ and a sterile state $\nu_s$, 
with their mixing in vacuum described by
\begin{eqnarray}
\vert \nu_\alpha\rangle & = & \cos\theta\, \vert \nu_1\rangle + \sin\theta\,\vert \nu_2\rangle,\\
\vert \nu_s\rangle & = & -\sin\theta\, \vert \nu_1\rangle + \cos\theta\,\vert \nu_2\rangle,
\end{eqnarray}
where $\vert \nu_1\rangle$ and $\vert \nu_2\rangle$ are the neutrino energy/mass eigenstates (to be identified with $\upnu_i$ and $N_I$ defined in \ref{sub:seesaw}) with corresponding vacuum mass eigenvalues $m_1$ and $m_2$ (to be identified with the light $m_i$ and heavy $M_I$ defined in \ref{sub:seesaw}). Here $\theta$ is the vacuum mixing angle. Of course, the unitary transformation between the three active neutrinos and any sterile states could be quite complicated, with many parameters, but this simple $2\times 2$ example will serve to illustrate the essence of how sterile neutrinos can be produced in the early Universe.

Quantum mechanical systems can be thought of as evolving with time in two ways. A state can evolve coherently in a smooth, continuous fashion according to a Schr\"odinger-like equation. Or, a state could change abruptly when a \lq\lq measurement\rq\rq\ is made, and the state is \lq\lq reduced\rq\rq\ into one of the eigenstates of the observable in question. The latter process is sometimes associated with de-coherence, because quantum mechanical information in the initial state is lost in the reduction/measurement process. Now we understand, of course, that de-coherence is really a manifestation of mixing and transferring quantum information over many states in the environment, a fundamentally many body process. 

Correspondingly, there are two ways that a neutrino can change its flavor. The first way is through coherent neutrino flavor oscillations. As the above transformation equation shows, the neutrino propagation states, i.e., the energy/mass states, are not coincident with the weak interaction eigenstates, i.e., the flavor states. As a neutrino propagates through the plasma of the early Universe it can coherently forward scatter on particles that carry weak charge, for example, electrons and positrons, other charged leptons and their anti-particles, quarks, nucleons, and other neutrinos. In so doing, the neutrino will acquire an effective mass, akin to the way a photon acquires an index of refraction and effective mass as it propagates through glass. Also because of this, the neutrino will typically have an in-medium mixing angle $\theta_m$ that will generally differ from the vacuum mixing angle. At any point along its path a coherently propagating neutrino will have an amplitude to be either active or sterile, which will depend on this mixing angle $\theta_m$, and the oscillation phase. 

The second way neutrinos can change their flavor is through direction- or energy-changing scattering processes. These will act like \lq\lq measurements.\rq\rq\ Active neutrinos with energy $E_\nu$ will have scattering cross sections $\sigma \sim G_{\rm F}^2\,E_\nu^2$, and overall scattering rates $\Gamma_{\nu_\alpha} \sim \sigma \cdot {\rm flux} \sim G_{\rm F}^2\, T^5$, where $G_{\rm F}$ is the Fermi constant that sets the strength of the weak interaction, $E_\nu$ is the typical energy of active neutrinos which is of order the temperature $T$, and the number densities (or fluxes) of thermally equilibrated light particles are $\sim T^3$. A neutrino will scatter only through its active component.

In a single particle sense it works like this. A neutrino is set to a weak interaction (flavor) eigenstate when it suffers a scattering  event. Subsequently, it propagates coherently toward its next scattering target, building up quantum mechanical flavor phase in a way governed by the in-medium unitary transformation between the flavor states and the instantaneous energy eigenstates:
\begin{eqnarray}
\vert \nu_\alpha\rangle & = & \cos\theta_m\left( t\right)\, \vert \nu_1\left( t\right)\rangle + \sin\theta_m\left( t\right)\,\vert \nu_2\left( t\right)\rangle,\\
\vert \nu_s\rangle & = & -\sin\theta_m\left( t\right)\, \vert \nu_1\left( t\right)\rangle + \cos\theta_m\left( t\right)\,\vert \nu_2\left( t\right)\rangle,
\end{eqnarray}
where $t$ is the age of the Universe. When this neutrino comes upon the next scattering target, after one mean free path, the target \lq\lq asks:\rq\rq\ What flavor are you? That collapses the neutrino state into one of the flavor eigenstates, either sterile $\vert \nu_s\rangle$ or active $\vert \nu_\alpha\rangle$. In the latter case, i.e., wave-function collapse into an active flavor state, the neutrino actually suffers a scattering event. If we start out with all active neutrinos, then as the Universe expands, active neutrino scatterings will necessarily build up a reservoir of sterile neutrinos.

Though there are many models for producing a DM relic density of sterile neutrinos~\cite{Dodelson:1993je,Shi:1998km,Abazajian:2001nj,Dolgov:2000ew,Abazajian:2002bh,Asaka:2005pn,Asaka:2005an,Dolgov:2003sg,Abazajian:2005gj,Asaka:2006ek,Kusenko:2006rh,Shaposhnikov:2006xi,Boyanovsky:2006it,Boyanovsky:2007as,Asaka:2006nq,Shaposhnikov:2006nn,Gorbunov:2007ak, Kishimoto:2008ic,Laine:2008pg,Petraki:2008ef,Petraki:2007gq,Kusenko:2009up,Bezrukov:2009th,Kusenko:2010ik,Nemevsek:2012cd,Canetti:2012vf,Canetti:2012kh,Bezrukov:2012as,Merle:2013gea,Abazajian:2014gza,Tsuyuki:2014aia,Merle:2013wta,Horiuchi:2013noa,Bezrukov:2014nza,Roland:2014vba,Abada:2014zra,Lello:2014yha,Humbert:2015epa,Humbert:2015yva,Adulpravitchai:2014xna,Adulpravitchai:2015mna, Drewes:2015eoa}, if active-sterile neutrino flavor mixing exists, then scattering-induced decoherence will necessarily contribute to the relic density~\cite{Merle:2015vzu}. Some of these models run afoul of X-ray observations or large-scale structure considerations or both, as described in Refs.~\cite{Abazajian:2001nj,Abazajian:2001vt,Hansen:2001zv,Boyarsky:2009ix,Abazajian:2009hx,Merle:2014xpa} and also in section~\ref{sec:xray}. However, many of them still remain viable.

%
\subsubsection{De-cohering scatterings} \label{sec:prod}

At each real or potential scattering event, the probability that the active neutrino collapses into a sterile state is $\propto \sin^22\theta_m\left( t\right)$. Of course, as Pontecorvo pointed out, a sterile neutrino really is not \lq\lq sterile,\rq\rq\ since this neutrino will also build up flavor phase as it propagates and it can have a non-zero amplitude to be active. Its interaction rate will be $\Gamma_{\nu_s} \sim G_F^2\,T^5\,\sin^2 [2\theta_m]$, and sterile neutrinos can be converted back to active states via scattering-induced decoherence. 

As mentioned before, active neutrinos in the early Universe can acquire an effective mass through coherent forward scattering off of particles in the background plasma. This leads to a modification in the effective active-sterile mixing angle. For a neutrino state with momentum $p$, the effective in-medium mixing angle at a plasma temperature $T$ is given by
\begin{equation}\label{eq:mixing}
\sin^2(2\theta_m) = \frac{\Delta^2(p) \sin^2(2\theta)}{\Delta^2(p)\sin^2(2\theta) + \left[\Delta(p)\cos(2\theta) - V_D - V_T\right]^2},
\end{equation}
where $\Delta(p) = \Delta m^2/(2p)$, with $\Delta m^2$ being the appropriate mass-squared splitting in vacuum. $V_D$ and $V_T$ are the finite density and finite temperature potentials, respectively, felt by the active neutrino. The finite density potential, $V_D$, arises as a result of asymmetries (particle vs.\ antiparticle) between weakly interacting particles, i.e., baryonic and leptonic asymmetries. The finite temperature potential, $V_T$, on the other hand, results from higher-order corrections and neutrino forward scattering off of thermally created particle-antiparticle pairs. It carries a negative sign, and for $T \lesssim M_W$ (the $W$-boson mass), is given by $V_T = -G_\text{eff}^2\,p\,T^4$, where $G_\text{eff} \sim \mathcal{O}(G_F)$ can be taken to represent some overall neutrino coupling, summed over the various species of particles in the background plasma.

The negative sign and the strong temperature dependence mean that the thermal potential can cause heavy suppression of the effective in-medium mixing angle at high temperatures. For a keV--MeV mass sterile neutrino, $\vert V_T\vert \sim \Delta(p)$ at $T \sim 0.1\text{--}1$ GeV. In the limit of negligible lepton number, i.e., $V_D \sim 0$, the ratio of sterile neutrino interaction rate to the Hubble expansion rate at higher temperatures is $\Gamma_{\nu_s}/H \propto T^{-9}$, whereas at lower temperatures, where the vacuum oscillation term dominates, $\Gamma_{\nu_s}/H \propto T^3$. Owing to this high-temperature suppression, sterile neutrinos with vacuum mixing angles smaller than $\sin^2 (2\theta) \sim 10^{-6} \, (10\text{ keV}/M_1)$ can never attain thermal equilibrium with the plasma, in the absence of additional non-standard-model interactions. Therefore, in such scenarios, their relic density cannot be set via a freeze-out process, unlike ordinary active neutrinos.

Such mixing with active neutrinos can, however, lead to sterile neutrinos being produced in the early Universe via scattering-induced decoherence, as was first pointed out in Refs.~\cite{Barbieri:1989ti,Kainulainen:1990ds}, and later on applied by Dodelson and Widrow~\cite{Dodelson:1993je} to the DM problem. Active neutrino scattering-induced decoherence into sterile states is driven by the active neutrino scattering rate $\Gamma_{\nu_\alpha} \sim G_F^2\,T^5$. However, this scattering also gives rise to the active neutrino matter potentials described above, as well as quantum damping, both of which serve to inhibit active-sterile neutrino conversion. The sterile neutrino production rate via this mechanism can also be shown to peak at temperatures of $\sim 0.1\text{--}1$ GeV. Details of this mechanism are given in subsequent sections.

%
\subsubsection{\label{sec:MSWeffect}MSW-effect and resonant conversion}

If there are considerable lepton numbers ${\cal{L}}_\alpha$ in the plasma, then $V_D \neq 0$, and ordinary neutrinos can have mass level crossings with sterile neutrinos. For a sterile neutrino mass in the range of interest, i.e., keV--MeV, this occurs in the same temperature range $\sim 0.1\text{--}1$ GeV where the scattering-induced production peaks. These mass level crossings are essentially Mikheyev-Smirnov-Wolfenstein (MSW) resonances~\cite{Mikheev:1986gs,Wolfenstein:1977ue}. These resonances, where $\sin^2 (2\theta_m) \rightarrow 1$, occur where the effective in-medium mass-squared of an active neutrino matches the vacuum mass-squared of the sterile species, $M_1^2$. From Eq.~\eqref{eq:mixing}, the MSW resonance condition can be obtained as $\Delta m^2\, \cos (2\theta) = 2\,x\, T\, \left( V_D + V_T \right)$, or,
\begin{equation}
M_1^2 \approx  {\frac{4 \sqrt{2} \zeta\left(  3\right)}{\pi^2}}\,G_F\, {\cal{L}}_\alpha\, x_{\rm res}\ T^4
-2\, r_\alpha\ G_F^2\ x^2_{\rm res}\, T^6,
\label{res}
\end{equation}
where we have used $\Delta m^2\, \cos (2\theta) \approx M_1^2$, and where the first and second terms on the right hand side in Eq.~\eqref{res} are $2 x\, T\, V_D$, and $2 x\, T\, V_T$, respectively. Here the scaled neutrino energy/momentum is $x \equiv p/T$ (the symbol $\epsilon$ for this quantity is also prevalent in the literature). This quantity becomes a co-moving invariant at late times when the neutrinos decouple and freely stream along geodesics \---- in that regime, the active neutrino temperature parameter will scale like the inverse of the scale factor, as will the momentum magnitude of the neutrino. In Eq.~\eqref{res}, omitting the contribution from charged leptons and hadrons, cf.\ Eq.~\eqref{VD}, ${\cal{L}}_\alpha = 2\,L_\alpha+\sum_{\beta\neq \alpha}{L_\beta}$, with $\alpha,\beta = e,\mu,\tau$, corresponding to $\nu_e, \nu_\mu, \nu_\tau$, and where we produce the sterile neutrinos through the $\nu_\alpha \rightleftharpoons \nu_s$ mixing channel, and where the net lepton numbers in the individual neutrino flavors are, $L_{\alpha} \equiv {\left( n_{\nu_\alpha} - n_{\bar\nu_\alpha} \right)}/n_\gamma$. Here, the photon number density is $n_\gamma = 2\,\zeta\left( 3 \right) T^3 /\pi^2$, and the thermal neutrino number densities $n_{\nu_\alpha} = T^3\, F_2\left(\eta_{\alpha}\right)/(2\,\pi^2)$, with $F_2\left( \eta_\alpha\right)$ the relativistic Fermi integral of order 2 and argument $\eta_\alpha$, the $\nu_\alpha$ degeneracy parameter which, in the dilute limit, is related to the corresponding lepton number by $1.46 L_{\alpha} \approx \eta_\alpha$. 
The thermal potential arises from neutrino scattering on seas of thermally-created particle-antiparticle pairs in the early Universe, and its magnitude therefore depends on the extent to which the corresponding charged lepton flavor is populated, which in turn depends on the temperature. For example, at $T \approx 10$~MeV, the second term in Eq.~\eqref{res}, corresponding to the thermal potential, has a parameter $r_\alpha$, which is roughly $79$ for $\alpha=e$, and roughly $22$ for $\alpha = \mu,\tau$ ($r_\mu$ increases to $r_e$ between $T\approx10$ MeV and $T\approx200$ MeV, as the themal $\mu^{\pm}$ seas become increasingly populated). 

\begin{figure}[htb]
\begin{center}
	\includegraphics[width=0.7\linewidth]{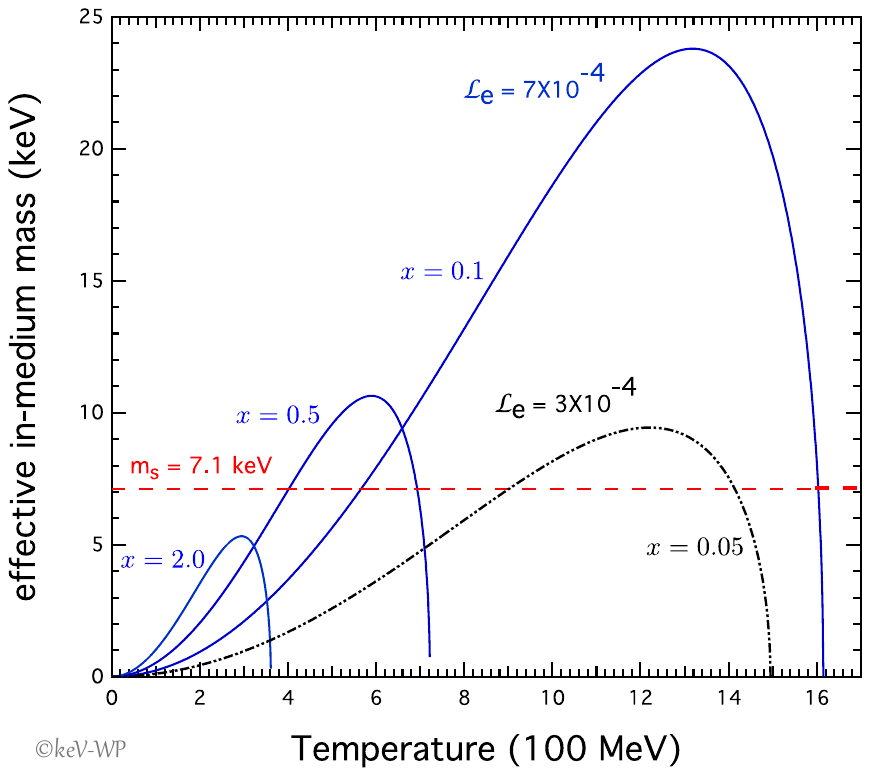}
	\caption{The effective in-medium mass, $\sqrt{2 x T (V_D +V_T)}$, for an electron flavor neutrino, $\nu_e$, is shown as a function of temperature in the early Universe. Two values of initial potential lepton number for the $\nu_e\rightleftharpoons \nu_s$ channel are considered, ${\cal{L}}_e = 7\times {10}^{-4}$ (solid, blue curves) and ${\cal{L}}_e=3\times{10}^{-4}$ (dot-dashed, black curve). Effective mass curves are shown for the labeled values of scaled neutrino energy $x = E_\nu/T$ (also refered to as $\epsilon$ in literature). As an example, the temperature track (dashed, red) of a sterile neutrino with mass $7.1\,{\rm keV}$ is shown \--- because this is sterile the track is independent of temperature. MSW resonances occur at mass level crossings, where the effective in-medium mass of the $\nu_e$ matches the sterile neutrino mass.}
	\label{levelcross}
\end{center}
\end{figure}

Figure~\ref{levelcross} gives examples of possible level crossing scenarios. In this figure we show the temperature dependence of the effective in-medium mass for $\nu_e$ for {\it constant} values of potential lepton number ${\cal{L}}_e$. If $\nu_e$'s are converted to $\nu_s$ at these level crossings, then the potential lepton number ${\cal{L}}_e$ will {\it decrease}, and the shape of the in-medium effective mass curve will change.

As this figure shows, an active neutrino can experience two MSW resonances as the Universe expands. This is because the first, positive term in Eq.~\eqref{res} dominates at lower temperature and scales like $T^4$, whereas the second, negative thermal term scales like $T^6$. Of course, at high enough temperature $T$, this negative term dominates and the potential will decrease with increasing temperature. We can solve Eq.~\eqref{res} for the resonant scaled energies of the active neutrinos,
\begin{equation}
x_{res}\left(t\right) = {\frac{ \sqrt{2}\, \zeta\left( 3\right)}{ \pi^2 r_\alpha G_F}}\,{\frac{ {\cal{L}}\left( t\right)}{ T^2\left( t\right)}}\,{\left( 1\pm \sqrt{1-{\frac{ \pi^4 r_\alpha M_1^2}{ 4 \zeta^2\left( 3\right) {\cal{L}}^2\left( t\right) T^2\left( t\right)}}}  \right)},
\label{epsilonres}
\end{equation}
where $t$ is the age of the Universe. Clearly, resonances can occur only when the argument of the square root in this equation is greater or equal to zero, implying that the following condition be met:
\begin{equation}
\vert {\cal{L}}{\left( t\right)}\vert \cdot T\left( t\right) \ge {\frac{ \pi^2}{2\, \zeta\left( 3\right) }}\, \sqrt{r_\alpha\, M_1^2 }.
\label{cond}
\end{equation}


Whether active neutrinos $\nu_\alpha$ are converted efficiently to sterile neutrinos $\nu_s$ at these resonances depends on several factors. First, scattering-induced de-coherent active-sterile transformation will be proportional to the product of the active neutrino scattering rate and the effective in-medium $\sin^2 (2\theta_m)$, the latter being maximal (unity) at resonance. The number of active neutrinos converted to sterile species at resonance will then depend on a comparison of the scattering rate and the \lq\lq resonance width, \rq\rq\ roughly the length of time the effective in-medium mixing angle is maximal, i.e., $\delta t \sim H^{-1}\, \tan (2\theta)$, where the causal horizon size is $H^{-1} \sim {m_{\rm pl}}/T^2$, with $m_{\rm pl}$ the Planck mass. The point is that gravitation is weak ($m_{\rm pl}$ large), the expansion rate of the Universe is slow, and the causal horizon is therefore large. In turn, this means that active-sterile conversion can {\it begin} efficiently. In coherent MSW language, we would say that the conversion process at resonance is {\it adiabatic} initially. However, efficient, adiabatic conversion $\nu_\alpha \rightarrow \nu_s$ means that the net lepton number ${\cal{L}}$ and, hence, the overall potential is reduced. Rapid reduction of ${\cal{L}}$ renders neutrino evolution through resonances non-adiabatic and, hence, inefficient. As the temperature $T$ drops and lepton number ${\cal{L}}$ is reduced the condition for the existence of resonances in Eq.~\eqref{cond} will eventually be violated. At low enough temperature, a combination of non-adiabaticity or cessation of resonances, plus a reduced active neutrino scattering rate, imply that production of sterile neutrino relic density ceases. In the end, the relic density of sterile neutrinos, and the relic energy spectrum of these, will then be the result of a sometimes complicated nonlinear interplay of the histories of active neutrino scattering rates and in-medium active-sterile mixing angle $\theta_m\left( t\right)$~\cite{Kishimoto:2008ic}.

\begin{figure}[htb]
\begin{center}
	\includegraphics[width=	0.7\textwidth]{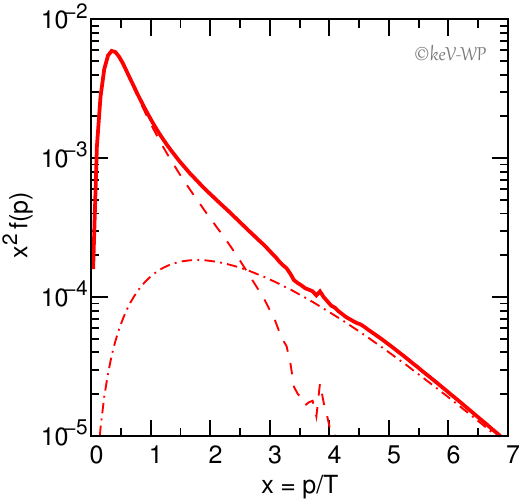}
	\caption{Example for a resonantly produced DM spectrum. The dashed-dotted line represents a momentum distribution that one would expect for $L_\alpha=0$ (Dodelson-Widrow case), the dashed line is the colder contribution from the resonance. The solid line is the combined spectrum.	This illustrative Figure is similar to~\cite{Boyarsky:2009ix}; state of the art spectra can be found in 
Refs.~\cite{Ghiglieri:2015jua,Venumadhav:2015pla}.
	}
	\label{DMspectra}
\end{center}
\end{figure}

Due to the dependence on $x$, different momentum modes go through the resonance at different temperatures. Therefore, resonant production of a sterile neutrino relic density in the right range to be DM leaves the sterile neutrinos with a distorted energy spectrum, with a significant fraction of the sterile neutrino population skewed toward lower $x$ than would be the case for a Fermi-Dirac black body-shaped distribution function. (In the examples in Fig.~\ref{levelcross} note that for a given ${\cal{L}}_e$ only {\it lower} values of $x$ can be resonant!) Therefore, resonantly produced sterile neutrino DM is generally \lq\lq colder\rq\rq\ than sterile neutrino DM particles with the same mass produced non-resonantly via a straight Dodelson-Widrow scenario, e.g., see Fig.~\ref{DMspectra}. 

To give some scale for the parameters involved, accounting for {\it all} of the DM with a 7.1~keV mass sterile neutrino, with a vacuum mixing angle $\sin^2\theta \sim {10}^{-10}$ or ${10}^{-11}$ as suggested by X-ray observations~\cite{Bulbul:2014sua,Boyarsky:2014jta}, via resonant, Shi-Fuller mechanism production, would require an initial net lepton number ${\cal{L}} \sim {10}^{-5}\text{--} {10}^{-4}$ \cite{Abazajian:2001nj,Laine:2008pg,Abazajian:2014gza}. A non-resonant, zero-lepton number Dodelson-Widrow production scenario would need a much larger vacuum mixing angle at a given mass to produce all of the DM, and this would mean a radiative decay rate in conflict with X-ray bounds. Such a scenario could work, however, if the 7.1~keV sterile neutrino were only $\sim 15\%$ of the total DM, thereby allowing for larger active-sterile mixing angles to not be excluded.

Finally, note that a lepton number ${\cal{L}} \sim {10}^{-4}$, though well below what can be probed with light element abundances, i.e., ${\cal{L}} \sim 0.1$, is nevertheless very large compared with the measured baryon number $\eta_B \approx 6.11\times{10}^{-10}$. This brings up a very thorny issue. Where would such a \lq\lq large\rq\rq\ lepton number come from? We have problems in explaining the origin of the baryon number as well. However, the existence of several sterile neutrino generations, the non-equilibrium nature of active-sterile neutrino mixing, and CP-violation in the neutrino sector, can produce the Sakharov conditions required to produce lepton numbers and baryon numbers. Moreover, our ignorance of the sterile neutrino mass scale can be turned on its head, and (large) masses of two of these particles chosen so that the baryon and lepton numbers can be produced. The lepton number so produced could then drive resonant production of a lighter, DM candidate sterile neutrino. One such scenario is the $\nu$MSM, where the required ${\cal{L}}$ can be produced in the decay of heavier sterile neutrinos, which previously generated the baryon asymmetry of the Universe~\cite{Asaka:2005pn,Canetti:2012vf,Canetti:2012kh}.

\subsection{Thermal production: state of the art (Authors: M.~Drewes, M.~Laine)
}\label{Sec:ThermalProduction}

%
\subsubsection{Examples of complete frameworks}\label{sec:DensityMatrix}

The physical picture outlined in Sec.~\ref{sec:5.thermalproduction} is sufficient for understanding most of the effects playing a role, but does as such not yet offer for a quantitative method for determining the DM abundance and spectrum in the early Universe. Indeed the quantities introduced, the potentials $V_T$ and $V_D$ and the active neutrino interaction rate $\Gamma^{ }_{\nu_\alpha}$, need to be put together in a way which accounts for contributions both from active neutrinos and antineutrinos; correctly incorporates Pauli blocking factors; tracks spin dependent effects if present; permits for the inclusion of hadronic contributions that are very important at the temperatures $T \gsim 100$~MeV where the production peaks; and systematically accounts for the dilution of the various lepton asymmetries during the resonant production of DM particles. 

Theoretically speaking, the computation needed in order to complete this task is a problem in non-equilibrium statistical physics. 
In a typical approach that can be applied if most degrees of freedom are in local thermal equilibrium, one attempts to factorize such problems into two parts. The first part is the postulation of equations of motion that should capture the time evolution of the non-equilibrium degrees of freedom. The second part is the determination of the coefficients appearing in these equations (such as $V_T, V_T, \Gamma_{\nu_\alpha}$), which is normally carried out in thermal equilibrium. With both ingredients as well as initial conditions in place, the non-equilibrium equations of motion can be integrated in order to obtain the desired abundance. 

One example of a non-equilibrium framework, often used for freeze-out computations in cosmology, is that of Boltzmann equations. In this case the dynamical degrees of freedom are classical phase space distribution functions, and the coefficients characterizing the equations are vacuum scattering amplitudes. However, since quantum coherence between states of different flavors is crucial to neutrino oscillations, and since the presence of a medium can significantly change the dispersion relations of various particles even in the absence of  ``real'' scatterings, the classical distribution functions in the Boltzmann equations are not necessarily suitable for the present problem. 

Flavor oscillations, thermal modifications of dispersion relations, and real scatterings responsible for de-coherence and the sterile neutrino production, can be taken into account through matrix-valued generalizations of the Boltzmann equations called \emph{density matrix equations}~\cite{Sigl:1992fn}. In this approach, active and sterile neutrinos are thought of as on-shell degrees of freedom, whose non-equilibrium time evolution is governed by a coupled set of linear equations of motion. 
The coefficients parametrizing these equations characterize three  basic properties of neutrinos: the mixing angles; the potentials $V_T$ and $V_D$; and a damping coefficient, reflecting thermal scatterings of left-handed neutrinos with the Standard Model particles constituting the heat bath. 


The basic quantities in the density matrix framework are expectation values of the ladder operators $a_I ,a^\dagger_I$ that appear in the plane wave expansion 
\begin{eqnarray}
N_I=\sum_h\int\frac{d^3\textbf{p}}{(2\pi)^3}\frac{1}{2\sqrt{\textbf{p}^2+M_I^2}}\left(u_{I,\textbf{p}}^h e^{-i\textbf{px}}a_{I,h}(\textbf{p,t})
+ v_{I,\textbf{p}}^h e^{i\textbf{px}}a_{I,h}^\dagger(\textbf{p,t})\right),\nonumber\\
\upnu_i=\sum_h\int\frac{d^3\textbf{p}}{(2\pi)^3}\frac{1}{2\sqrt{\textbf{p}^2+m_i^2}}\left(u_{i,\textbf{p}}^h e^{-i\textbf{px}}b_{i,h}(\textbf{p,t})
+ v_{i,\textbf{p}}^h e^{i\textbf{px}}b_{i,h}^\dagger(\textbf{p,t})\right)\nonumber.\end{eqnarray}
Here $h,h'=\pm$ label the helicities of the particles. The elements of the 
density matrix can be defined as  
\begin{equation}\label{rhoNdef}
(\rho_{NN})_{IJ}^{h h'}\propto
\frac{\langle a_I^h(p)^\dagger a_J^{h'}(p)\rangle}{Vs} \ , \ (\rho_{\nu\nu})_{ij}^{h h'}\propto
\frac{\langle b_i^h(p)^\dagger b_j^{h'}(p)\rangle}{Vs} \ , \ 
(\rho_{\nu N})_{iJ}^{h h'}\propto\frac{\langle b_i^h(p)^\dagger a_J^{h'}(p)\rangle}{Vs}.
\end{equation} 
Here, $V$ is some arbitrarily chosen unit volume and $s$ the entropy density of the Universe. For $I=J$ or $i=j$ these objects simply give the occupation number of particles of flavor $I$ in the quantum state in which the expectation value is calculated, the off-diagonal elements characterize correlations between different flavors. Together, they form a 
``density matrix'' in flavor space for each momentum mode and helicity,
\begin{eqnarray}
\rho=\left(
\begin{tabular}{c c}
$\rho_{\nu\nu}$ & $\rho_{\nu N}$\\
$\rho_{N \nu}$ & $\rho_{N N}$
\end{tabular}
\right)
.\end{eqnarray}
Equations of this type have been widely used to describe the propagation of active neutrinos in a medium, particularly at low temperatures $T \lsim 10$~MeV. 

Let us for a moment assume that the neutrinos reside in a thermal bath with adiabatically changing temperature. It is usually also assumed that one can neglect all correlations with $h\neq h'$ because the Hamiltonian commutes with the helicity operator. Then the density matrix follows an equation of motion of the form
\begin{equation}\label{DensityMatrixEquation}
\frac{1}{\mathcal{H}X}\frac{d}{dX}
\rho = -i[ H_{\rm eff},\rho] - \frac{1}{2}\{\Gamma_{\rm eff}, \rho-\rho^{\rm eq}\}.
\end{equation}
Here, $\rho^{\rm eq}$ is the equilibrium density matrix and $X=M/T$ a dimensionless time variable, where $M$ is an arbitrarily chosen mass scale. It is convenient to identify $M$ with the DM particle's mass. The function
\begin{equation}
\mathcal{H}\equiv -\frac{\partial}{\partial X}\sqrt{\frac{45}{4\pi^3 g}}\frac{m_P}{2M^2} X 
\end{equation}
can be identified with the Hubble parameter if the number of degrees of freedom $g$ is constant during the evolution. 
The commutator contains an effective Hamiltonian $H_{\rm eff}$ that describes coherent flavor oscillations.
In vacuum, this term is simply given by the neutrino mass matrix. Above we have defined $\rho$ in the ``mass basis'' in flavor space; in this basis $H_{\rm eff}$ is diagonal in vacuum. The equation of motion \eqref{DensityMatrixEquation} is invariant under rotations in flavor space and can be transformed into any other basis for convenience. 

The anticommutator comes from the interactions and gives rise to ``dissipation'', i.e.\, processes that change the number of particles of a given species in a given mode. It involves decoherence and leads to scattering-induced sterile neutrino production. The flavor basis in which $\Gamma_{\rm eff}$ is diagonal is called ``interaction basis'', in general it is not identical to the one where $H_{\rm eff}$ is diagonal. It is precisely this misalignment that leads to flavor oscillations: particles are produced in a state that corresponds to a $\rho$ which is diagonal in the interaction basis, then they start oscillating due to the presence of $H_{\rm eff}$. At finite temperature and density, there are corrections to $H_{\rm eff}$ due to $V_D$ and $V_T$. Let us split $H_{\rm eff}=H_{\rm vac} + H_T$ into a vacuum term and a finite temperature/density correction. $H_{\rm vac}$ is determined by the vacuum mass matrix and diagonal in the mass basis, $H_T$ comes from the interactions and is diagonal in the interaction basis. The total $H_{\rm eff}$ defines the effective (thermal) mass basis at a given temperature and density, it rotates when the temperature changes because the relative size of $H_{\rm vac}$ and $H_T$ changes. 
In particular, in the limit of very high temperatures $T\gg M_I$, $H_{\rm vac}\ll H_T$ can be neglected, and the effective mass basis becomes identical to the flavor basis. Then there are no flavor oscillations in spite of the fact that neutrinos have large ``thermal masses'', simply because these thermal mass terms are diagonal in the flavor basis. 

In practice, one usually makes a number of simplifications. Since active neutrino masses are very small, oscillations among active neutrinos can usually be neglected at the temperatures that are relevant for DM production. This allows to set $m_\nu=0$, and $H_{\rm eff}$ takes the form 
\begin{eqnarray}
H_{\rm eff}=
\left(
\begin{tabular}{c c}
$0$ & $0$\\
$0$ & $\frac{M_I^2}{2|\textbf{p}|}$
\end{tabular}
\right) + H_T
\;. 
\end{eqnarray}
Since a rotation among the active flavors does not change $H_{\rm vac}$, it is common to use the flavor basis $(\rho_{\nu \nu})_{\alpha \beta}^{h h}$, where $H_T$ is diagonal, in the active sector. This is done by a block-diagonal rotation matrix ${\rm diag }(U_\nu, 1_{ })$. We can neglect all elements with $\alpha\neq\beta$ in this approximation.
The active neutrinos are highly relativistic. This allows to interpret the helicity states of the Majorana field $\nu_i$ as ``particles'' and ``antiparticles''. Their  sum is the total particle number and their difference is a measure for the lepton asymmetry in the plasma (neglecting the asymmetry carried by charged leptons). 
If the active neutrinos are in thermal equilibrium at $T>1.1$ MeV, the deviation of their phase space distribution function from equilibrium can be neglected, and the occupation numbers for all modes are simply given by a single number $T$. 
In kinetic equilibrium, one in addition needs to specify the asymmetries $L_\alpha$. They are crucial for sterile neutrino DM production because they lead to a non-zero $V_D$ and can trigger the MSW resonance. Since the active neutrino distribution functions remain close to equilibrium, one need not track the occupation numbers of individual momentum modes and can define the integrated quantities: 
\begin{eqnarray}
 n_{\nu_\alpha}&=&\int\frac{d^3\textbf{p}}{(2\pi)^3}\left[(\rho_{\nu\nu})_{\alpha\alpha}^{++} + (\rho_{\nu\nu})_{\alpha\alpha}^{--}\right], \\ 
 L_\alpha&=&\int\frac{d^3\textbf{p}}{(2\pi)^3}\left[(\rho_{\nu\nu})_{\alpha\alpha}^{++} - (\rho_{\nu\nu})_{\alpha\alpha}^{--}\right].
\end{eqnarray}

We are interested in the production of sterile neutrinos with masses $M_I$ in the keV range and mixing angles $U_{I\alpha}^2<10^{-8}$ that are consistent with searches for emission from decaying DM. Comparing a keV mass to the observed $\Omega_{\rm DM}$ shows that the occupation numbers $\rho_{N N}$ at all times must remain significantly below their equilibrium values. The same can be said about $\rho_{\nu N}$ and $\rho_{N \nu}=\rho_{\nu N}^\dagger$. On the other hand, the equilibrium density matrix $\rho^{eq}$ can in good approximation be replaced by a unit matrix as long as the temperature is much larger than all particle masses, which is the case in the regime $T\sim 100$ MeV where thermal sterile neutrino DM production peaks. 
If $L_\alpha=0$, then the off-diagonal elements $\rho_{\nu N}$ oscillate very rapidly due to the  very different effective masses of active and sterile neutrinos. Then their effect on the evolution of the abundances $\rho_{NN}$ averages out. For $L_\alpha\neq0$ the nonzero $V_D$ can lead to a level crossing between the active and sterile neutrino quasiparticle dispersion relations in the plasma, so this argument does not hold. However, in this case the element of $H_{\rm eff}$ that multiplies an element of $\rho_{\nu N}$ on the RHS of \eqref{DensityMatrixEquation} becomes very small (because it is given by the effective mass splitting). That is, the elements of $\rho_{\nu N}$ (which are small numbers) enter the RHS of \eqref{DensityMatrixEquation} only as in a product with another small number, and this product can be neglected.\footnote{Here ``smallness'' can be quantified in powers of the mixing angle $\theta$: the production rate $\Gamma_N$ for sterile neutrinos, which is the lower right block of the
matrix $\Gamma_{\rm eff}$ in \eqref{DensityMatrixEquation}, is of order $\propto\theta^2$. 
This sets the macroscopic time scale $1/\Gamma_N$ in the system. Oscillations can be considered ``fast'' if they are faster than this. If the oscillation time scale is comparable to $\Gamma_N$, then the oscillations in principle have to be tracked. In this case, however, the element of $H_{\rm eff}$ that multiplies an element of $\rho_{\nu N}$ is also $\sim \theta^2$. \label{fnres}
}

Even though the density matrix formalism can describe much of
the physics relevant to the problem, it still omits uncertainties of
order unity. In particular, it assumes that active lepton asymmetries are 
only carried by neutrinos, omitting the fact that weak interactions are in 
equilibrium and lepton asymmetries can also be carried by charged
leptons (with charge neutrality taken care of by leptons of other
generations and/or by light hadrons such as pions). 

A framework which includes all effects 
in Standard Model interactions, with 
the price of restricting to order $\theta^2_{I\alpha}$ in the 
active-sterile mixing angles,  has been presented 
in ref.~\cite{Asaka:2006rw} and generalized to systems
close to equilibrium and with three generations of lepton asymmetries 
in ref.~\cite{Ghiglieri:2015jua}.\footnote{A different approach based on similar physical assumptions was proposed and used to compute DM spectra in ref.~\cite{Venumadhav:2015pla}.} 
In this approximation, we can track the {\em full} density matrix of the system, denoted by $\hat{\rho}$.
Its time evolution is determined by the quantum kinetic equation 
\begin{equation} 
 i \frac{{\rm d} \hat \rho_{\rm I}(t)}{{\rm d} t} =
 [\hat H_{\rm I}(t),\hat\rho_{\rm I}(t)]
 \;, \label{liuv1}
\end{equation}
which is equivalent to the Schr\"odinger equation.
Here   
$ 
 \hat \rho_{\rm I} \equiv \exp(i \hat H_0 t)\hat \rho \exp(-i \hat H_0 t)
$
is the density matrix in the interaction picture and
$
 \hat H_{\rm I}\equiv \exp(i \hat H_0 t) \hat H_{int} \exp(-i \hat H_0 t)
$, where $\hat{H}_{int}$ is the interaction Hamiltonian proportional
to the neutrino Yukawa couplings (i.e.\ $\theta_{Ia}$). The ``free''
part $\hat{H}_0$ includes all other interactions.  
Equation~\eqref{liuv1} is exact and contains all interactions.\footnote{Note that the derivative on the LHS should be a covariant derivative in curved space.} All other kinetic equations that are being used in the literature correspond to different approximations to \eqref{liuv1}. For instance, equation~\eqref{DensityMatrixEquation} can be interpreted as the limit where the sterile neutrinos are on-shell and the RHS has been perturbatively expanded in $\theta$. This approximation allows to include coherent quantum oscillations and corrections to the (quasi)particle dispersion relations from interactions with the plasma. Equation~\eqref{eq:kin} can then be viewed as a limit of \eqref{DensityMatrixEquation} in which coherent oscillations between different flavors have been neglected.

Equation~\eqref{liuv1} can be solved perturbatively, 
\begin{eqnarray}
 \hat \rho_{\rm I}(t) = \hat\rho_0
 - i \int_0^t \! {\rm d} t' \, 
 [\hat H_{\rm I}(t'), \hat\rho_0]
 - 
 \int_0^t \! {\rm d} t' \,
 \int_0^{t'} \! {\rm d} t'' \,
 [\hat H_{\rm I}(t'),[\hat H_{\rm I}(t''), \hat\rho_0]]
 + ... \;,
 \label{pert}
\end{eqnarray}
where $\hat\rho_0 \equiv \hat\rho(0) = \hat\rho_{\rm I}(0)$.
This is equivalent to standard time-dependent perturbation theory.
The strictly perturbative approach breaks down as soon as $\rho_{NN}$ approaches equilibrium, but by an appropriate choice of 
$\hat{\rho}_0$ these so-called secular terms can be avoided, so that
the system can be described both near and far from equilibrium, with 
the former case properly including Pauli blocking.
Subsequently the sterile neutrino production rate can be traced out
from the solution. 
Simultaneously, the lepton number operator evolves as~\cite{Bodeker:2014hqa}:
\be
 \dot{\hat{L}}^{ }_\alpha(t) = 
 \int_\vec{x} i  
  \hat{\bar{N}}^{ }_I F^{\dagger}_{I\alpha}  
  \hat{\tilde{\phi}}^\dagger \hat{\ell}^{ }_\alpha 
  \; + \; \mbox{h.c.}
 \;, 
\ee
where $\tilde\phi \equiv i \sigma^{ }_2 \phi^*$ is the conjugate Higgs doublet,  $\hat{\ell}^{ }_\alpha$ is the left-handed lepton doublet, and $F_{I \alpha}\propto \theta_{I \alpha}$ are Yukawa couplings. Tracing this equation with the density matrix yields an equation for the depletion of lepton asymmetries. Taken together the equations results in a coupled system, parametrized by $V_T$, $V_D$, and $\Gamma_{\nu_\alpha}$, which is valid both near and far from equilibrium to $\mathcal{O}(\theta_{I \alpha}^2)$ but in principle to any order in SM couplings such as $\alpha_s$~\cite{Ghiglieri:2015jua}.

In principle there are also frameworks which can be extended to higher orders
in $\theta_{Ia}^2$. This could be relevant, for instance, for crosschecking 
that resonant production is treated fully consistently 
(cf.\ footnote~\ref{fnres}). 
One possible starting point for this is 
the Schwinger-Keldysh formalism of non-equilibrium quantum field theory~\cite{Schwinger:1960qe,Bakshi:1962dv,Bakshi:1963bn,Keldysh:1964ud}, in which all properties of the system can be expressed in terms of correlation functions of the quantum fields, without reference to particles or asymptotic states. This framework allows for a first principles description of sterile neutrinos in a dense plasma, and significant progress has been made towards this goal~\cite{Buchmuller:2000nd,Anisimov:2008dz,Anisimov:2010aq,Anisimov:2010dk,Fidler:2011yq,Beneke:2010dz,Beneke:2010wd,Drewes:2010pf,Drewes:2012ma,Drewes:2012qw,Cirigliano:2009yt,Garny:2011hg,Millington:2012pf,Dev:2014wsa,Frossard:2012pc,Hohenegger:2014cpa,Drewes:2015eoa,Dev:2015dka,Kartavtsev:2015vto}, in particular in the context of leptogenesis. 
However, even though exact equations can be written down, it remains a challenge to develop a systematic approximation scheme for their practical solution. 






%
\subsubsection{Matter potentials and active neutrino interaction rate \label{ss:coeffs}}

As has been discussed in the previous sections, a key ingredient in the DM production computation is the determination of the potentials $V_T$ and $V_D$ and the active neutrino damping coefficient, or interaction rate, denoted by $\Gamma_{\nu_\alpha}$. We start with a discussion of how these are normally defined in a way which permits for their practical determination, and then discuss the best available values that can be extracted from literature. 

Let us denote by $P \equiv (E,\vec{p})$ the four-momentum of a left-handed (active) neutrino and by $\Sigma$ its self-energy. The matter potentials $V_T$ and $V_D$ parametrize the dispersion relation of the active neutrino. This can be obtained by searching for non-trivial solutions of~\cite{Notzold:1987ik,Enqvist:1990ad,D'Olivo:1992vm,Quimbay:1995jn}: 
\begin{equation}
 \Bigl\{ \bsl{P} - \re \bsl{\Sigma} P_L \Bigr\}
 \, \psi^{ }_{ }(\vec{p}) = 0
 \;, \la{onshell}
\end{equation}
where $\psi$ is a Dirac spinor, $P_L \equiv (1-\gamma_5)/2$ is a chiral projector, and $\re$ denotes the continuous (non-cut) part of the self-energy. Following ref.~\cite{Weldon:1982bn} and for future reference, we have factorized the chiral projector $P_L$ explicitly in Eq.~\nr{onshell}.

Because only left-handed neutrinos experience weak interactions within the Standard Model (the vertices that they attach to are proportional to $\gamma^\mu P_L$), the self-energy must respect chirality. Therefore it is of the form 
$ 
 \bsl{\Sigma} = a \, \bsl{P} + b \, \msl{u}
$~\cite{Weldon:1982bn}, 
guaranteeing that the left and right chiral components decouple from each other. The part $a$ can in general be omitted in comparison with the tree-level term in Eq.~\nr{onshell}, whereas the function $b$ defines the quantities that we are interested in. 

More precisely, we can write~\cite{Notzold:1987ik}  
\begin{equation}
  \bsl{\Sigma} P_L \equiv  
  \Bigl[ a \, \bsl{P} + \Bigl( b_r - i \,\frac{\Gamma_{\nu_{\alpha}}}{2}\, \Bigr)
  \msl{u}  \Bigr]P_L 
  \;, \la{width}
\end{equation}
where $u\equiv (1,\vec{0})$ is the plasma four-velocity; 
$a = a_r + i a_i$ is a complex function; and 
$b_r$ is a real function. The real part can 
be interpreted as 
\be
 b_r = - V_T - V_D
 \;, 
\ee
where $V_D$ is defined to be the part which is odd in chemical potentials. 
Moreover,  
assuming the structure in Eq.~\nr{width}, 
the scattering rate can be projected out as 
\begin{equation}
 \Gamma_{\nu_{\alpha}} \; = \; 
 - 
 \tr \biggl\{ 
   \frac{E \bsl{P} - P^2 \msl{u}}{{p}^2}
     \im \bsl{\Sigma} P_L 
 \biggr\} 
 \;. \la{width2}
\end{equation}
According to ref.~\cite{Tututi:2002gz}, this formula corresponds to the classic results given in refs.~\cite{Enqvist:1991qj,Langacker:1992xk,Tututi:2002gz}. 

The scattering rate has many interpretations. On one hand it is often interpreted as the relaxation rate that drives the active neutrino population to thermal equilibrium~\cite{Weldon:1983jn}; more importantly for us, as discussed above, it directly determines the DM production rate, 
\be
 \Gamma_{\nu_s} \simeq 
 \frac{\sin^2(2\theta)}{4} 
 \frac{M_I^4 \Gamma_{\nu_{\alpha}}}{(M_I^2 + 2 E b_r + b_r^2)^2 
 + E^2 \Gamma^2_{\nu_{\alpha}}} + \mathcal{O}(\theta^4)
 \;.
\ee 
This result can be compared to the expression~\eqref{eq:mixing} for the production in the quantum mechanical limit. General discussions concerning other interpretations of $\Gamma_{\nu_\alpha}$ can be found in refs.~\cite{Drewes:2012qw,Bodeker:2015exa}.

Turning to numerical evaluations, the finite temperature part $V_T$ originates at 1-loop level and has a relatively simple expression, available both for $T \ll M_W$~\cite{Notzold:1987ik} and more generally~\cite{Enqvist:1990ad,D'Olivo:1992vm,Quimbay:1995jn}. In contrast the determination of the finite density potential $V_D$, also addressed in ref.~\cite{Notzold:1987ik}, turns out to be subtle. Because of $Z^0$-boson exchange, which is in thermal equilibrium in the temperature range of interest, it includes a contribution from hadrons; in a notation similar to Eq.~\nr{epsilonres}, 
\ba
 \mathcal{L}_\alpha \; & = & \;  
  2 L^{ }_{\nu_\alpha} + \sum_{\beta\neq \alpha} L^{ }_{\nu_\beta}
 + \Bigl(\fr12 + 2 \sin^2\!\theta^{ }_w\Bigr) L^{ }_{e_\alpha}
 - \Bigl(\fr12 - 2 \sin^2\!\theta^{ }_w\Bigr) 
  \sum_{\beta\neq \alpha} L^{ }_{e_\beta}\nonumber \\ 
 & + & \Bigl(\frac12-\frac43\sin^2\!\theta^{ }_w\Bigr)
 \sum_{i=u,c}L_i
 -\Bigl(\frac12-\frac23\sin^2\!\theta^{ }_w\Bigr)
 \sum_{i=d,s,b}L_i
 \;, \la{VD}
\ea
where $e_\alpha$ denotes charged leptons; $u,c,d,s,b$ stand for quark flavors;
and $\theta^{ }_w$ is the weak mixing angle. Because the hadronic part couples
differently to up and down-type quarks, the hadronic contribution contains a
part which is {\em not} proportional to the baryon number density. Electric
charge neutrality of the Standard Model plasma requires that the hadronic part
is present even if the baryon asymmetry of the plasma were zero, because
charged hadrons are needed to neutralize the contribution from charged
leptons. The way to systematically include these effects, relating in
particular the hadronic contributions to so-called quark number
susceptibilities which can be measured with methods of lattice QCD, has been
worked out in ref.~\cite{Ghiglieri:2015jua}, and  lattice results have been
included in the estimate of $V_D$ in ref.~\cite{Venumadhav:2015pla}. 

Finally, a precise determination of $\Gamma_{\nu_\alpha}$ is considerably more complicated than that of $V_T$ and $V_D$, including a large number of $2\leftrightarrow 2$ scatterings at $T \ll M^{ }_W$. In order to present numerical results, let us define the projections 
\begin{equation}
 I^{ }_P \; \equiv \; 
 - \tr\Bigl\{ \bsl{P}  \im \bsl{\Sigma} P_L \Bigr\}^{ }
   _{E = \sqrt{p^2 + M^2_I}} 
 \;,  \quad
 I^{ }_u \; \equiv \; 
 - \tr\Bigl\{ \bsl{u}  \im \bsl{\Sigma} P_L \Bigr\}^{ }
   _{E = \sqrt{p^2 + M^2_I}} 
 \;. 
\end{equation}
For dimensionless combinations, it is natural to factor out the combination $G_\rmii{F}^2 T^4$ (here $G^{ }_\rmii{F}$ denotes the Fermi constant) as well as appropriate powers of $E$:
\begin{equation}
 \hat{I}^{ }_P  \; \equiv \; \frac{I^{ }_P}{  
  G_\rmii{F}^2\, T^4\, E^2 }   \;, \qquad
 \hat{I}^{ }_u \; \equiv \;  \frac{I^{ }_u}{
  G_\rmii{F}^2\, T^4\, E } \;. \la{hatIu}  
\end{equation}
Then 
\begin{equation}
 \Gamma_{\nu_\alpha} \; = \; 
 G_\rmii{F}^2\, T^4\, E \,
 \biggl\{
   \hat{I}^{ }_P + \frac{M^2_I}{p^2}
   \Bigl( \hat{I}^{ }_P - \hat{I}_u \Bigr) 
 \biggr\}^{ }_{E = \sqrt{{p}^2 + M^2_I} }
 \;. \la{Gammaq}
\end{equation}
Values for $\hat{I}^{ }_P$ and  $\hat{I}_u$, as a function of $M_I$, $p$ and the active neutrino flavor, can be found on the web site {\tt http://www.laine.itp.unibe.ch/neutrino-rate/}, together with an illustration of the magnitude of hadronic effects according to the prescription in ref.~\cite{Asaka:2006nq}. It should be noted that for purely leptonic effects at low temperatures $T \lsim 20$~MeV, these results are {$\sim 5\%$ smaller than previous expressions in the literature~\cite{Notzold:1987ik}, possibly due to the fact that Pauli blocking factors were omitted in ref.~\cite{Notzold:1987ik}}.

%
\subsubsection{Open questions}

Even though the available expressions for $V_T$, $V_D$, and $\Gamma_{\nu_\alpha}$ as explained in section~\ref{ss:coeffs} and one of the frameworks discussed in section~\ref{sec:DensityMatrix} permit to compute DM abundances as a function of $M_I$ and $\sin^2(2\theta)$, it may be questioned what the theoretical accuracy of such computations is. Let us list a number of issues to be kept in mind on this point: 
\begin{itemize}

\item[(i)]
First and foremost, there is the question of how precisely hadronic contributions have been included. Thermal production peaks at temperatures $T \gsim 100$~MeV, where the Standard Model plasma undergoes a smooth crossover from a confined to a deconfined medium. The effective number of degrees of freedom changes by a factor $\sim 3$, which illustrates that hadronic effects dominate the plasma behavior. Roughly speaking, hadronic effects enter the DM production computation at three points: in determining the thermal history of the Universe through the temperature dependence of the equation of state (i.e.\ the temperature dependence of the Hubble rate); through their effect on the finite density potential $V_D$; and through the thermal scatterings that active neutrinos undergo with hadronic plasma constituents. In the most advanced current computations~\cite{Ghiglieri:2015jua,Venumadhav:2015pla}, hadronic effects on the equation of state have been modelled according to ref.~\cite{Laine:2006cp}. Lattice results concerning quark number susceptibilities have been included in the estimate of $V_D$ in ref.~\cite{Venumadhav:2015pla}. The most difficult issue is including hadronic thermal scatterings; even though they can be related to well-defined mesonic spectral functions~\cite{Asaka:2006rw}, it is very hard to extract such spectral functions from first principles lattice measurements. Chiral effective theory inspired model computations have been attempted in refs.~\cite{Venumadhav:2015pla,Lello:2015uma}, whereas in ref.~\cite{Ghiglieri:2015jua} a smooth interpolation between confined and deconfined phases, tracking the hadronic contribution to entropy density, was adopted. In total, the hadronic uncertainties from all these ingredients probably remain on the $\sim 20$\% level or so. 

\item[(ii)]
At temperatures above 10~MeV, where sterile neutrino DM production takes place, the lepton asymmetries carried by the different active lepton generations do not equilibrate with each other. That is, three different lepton asymmetries need to be tracked. Depending on the flavor structure of the neutrino Yukawa couplings, all or only some of these may contribute to resonant DM production. A general framework for describing this rather complicated system, including a proper account of the non-linear backreaction whereby the lepton asymmetries get depleted during resonant DM production, has been developed in ref.~\cite{Ghiglieri:2015jua}. However the results depend significantly ($\sim 100$\%) on the flavor structures of the lepton asymmetries and Yukawa couplings (a collection of spectra for different choices can be found in Fig.~6 of ref.~\cite{Ghiglieri:2015jua}). In order to get a unique answer, these parameters would need to be fixed from other considerations, such as leptogenesis and phenomenological constraints.  

\item[(iii)]
Although possible in principle~\cite{Akhmedov:1998qx,Asaka:2005pn,Canetti:2012kh}, a theoretically reliable computation of the lepton asymmetries generated through the non-DM sterile neutrino oscillations is far from being established in practice. We recall that, since lepton asymmetries much larger than the baryon asymmetry are required, the lepton asymmetries need to be generated below a temperature of 160~GeV, so that the Higgs mechanism is active and sphaleron processes have switched off. Carrying out the computation in this setting is complicated, because the system has multiple mass scales and because two almost degenerate sterile neutrino generations participate in the dynamics.

\item[(iv)]
There are also a number more refined ``theoretical issues'' that may be worth considering. For instance, one may worry about the gauge dependence of the quantities introduced ($V_T, V_D, \Gamma_{\nu_\alpha}$), which is not automatically guaranteed, given that the self-energy is not needed at the active neutrino on-shell point but rather the sterile neutrino one. Or, one may worry about an additional particle branch in the active neutrino dispersion relation, called the plasmino or abnormal or hole branch~\cite{Weldon:1989ys}. That it appears for active neutrinos in the Higgs phase has been demonstrated in ref.~\cite{Quimbay:1995jn}, so one may wonder whether the corresponding quasiparticles should be included as states in the density matrix formalism. Finally, in the context of leptogenesis, where sterile neutrino production is considered at high temperatures, it has been realized that to obtain correct leading-order results requires an all-orders resummation of the loop expansion, in order to account for the so-called Landau-Pomeranchuk-Migdal (LPM) effect~\cite{Anisimov:2010gy,Garbrecht:2013urw}, but so far this has not yet been included in any Higgs phase computations.

\end{itemize}


We remark that solving the theoretical issues of point (iv) 
essentially requires giving up the simple picture of sterile neutrino
production only through active-sterile oscillations, and with it
also frameworks such as the density matrix one. Indeed the way
that gauge independence is restored and the need for LPM resummation 
makes its appearance is that sterile neutrinos can also be produced
via direct $1\leftrightarrow 2$ scatterings involving Higgs bosons or 
longitudinal polarizations of gauge bosons (Goldstone modes). Such 
processes become important once $T \gsim $ a few GeV, and  in general 
their inclusion requires a fully  quantum field theoretic framework.


%% file: 3_decay.tex
\subsection{\label{sec:5.decays}Production by particle decays (Authors: F.~Bezrukov, A.~Merle, M.~Totzauer)}

Sterile neutrino appearance by neutrino flavor transitions, as considered in the previous subsection, is not the only mechanism by which sterile neutrino DM could have been produced in the early Universe.  A cosmological population of these particles can be generated in decays of some heavy particles as well.  A simple example is given by a singlet (with respect to the Standard Model) scalar $S$ that can decay into sterile neutrinos. A generic interaction Lagrangian is given by
\begin{equation}
  \mathcal{L}_\mathrm{int} = \frac{y}{2}\,\overline{(\nu_R)^c}\,\nu_R\,S + \mathrm{h.c.}, 
  \label{eq:decays1}
\end{equation}
where sterile flavor indices are not shown explicitly, but can be added, if necessary.  The general idea is fairly simple: $S$-particles that already exist or are created in some process could decay into sterile neutrinos at some point during the cosmological evolution. Now, the amount of neutrino produced does in general not depend on the active-sterile mixing angle $\theta$, but is instead controlled by the coupling $y$. Furthermore, for a given mass, the sterile DM can be colder or warmer, depending upon details of the particular model used.  Therefore, the parameter range  which is unrestricted by various observational constraints widens up in the mass--mixing angle parameter space.

There are several variations of the described mechanism, which will be described in more detail below. In particular, the ``parent'' $S$-particle can itself be in or out of thermal equilibrium at the time the sterile neutrinos are produced and, depending on the model, the decay into the sterile neutrinos may be the only possible decay channel~\cite{Merle:2015oja} or one of many possible decay channels into light particles~\cite{Shaposhnikov:2006xi,Petraki:2007gq,Bezrukov:2009yw}. Also, the parent particle can be different from a scalar, e.g.\ a vector~\cite{Boyanovsky:2008nc,Shuve:2014doa} or fermion~\cite{Abada:2014zra}. In addition, if the sterile-active neutrino mixing is present, then one has to take into account the mechanisms described in section~\ref{sec:5.thermalproduction} as a competing production mode~\cite{Merle:2015vzu}.

\subsubsection{\label{sec:5.decays_inflaton}Decay in thermal equilibrium}

Let us first analyze what happens if the parent particle $S$ is in thermal equilibrium for all relevant temperatures.  That can happen if the $S$-particle is strongly mixed with the neutral Higgs boson, so that all decay channels in SM particles are present and maintain it in thermal equilibrium until temperatures below its mass.

The scalar $S$ can be either real or complex. If $S$ acquires non-zero vacuum expectation value (VEV), then the interaction eq.~\eqref{eq:decays1} leads not only to the neutrino production, but generates also the Majorana masses for them, $M_1 = y\,\langle S \rangle$. In this case, and if $S$ is complex, the model contains (pseudo) Nambu-Goldstone boson, called the Majoron~\cite{Chikashige:1980ui,Schechter:1981cv}.  As a result, the sterile neutrinos may be not stable enough to form the DM because they decay into a Majoron, if it is lighter,\footnote{The Majoron is not necessarily massless, and may in turn form DM, for the recent discussion see e.g.\ Refs.~\cite{Frigerio:2011in,Queiroz:2014yna}.} and an active neutrino. While this case is not interesting from the point of view of sterile neutrinos being the DM, the case of decaying sterile neutrinos cannot a priori be excluded and by this motivates their laboratory searches in the region of comparatively large mixing angles where they would be considered overabundant otherwise.
For charged scalars, there can be significant thermal corrections to the DM production rate~\cite{Drewes:2015eoa}. These affect both, the equation of motion for the scalar condensate~\cite{Cheung:2015iqa} as well as the properties and kinematics of quasiparticles in the plasma~\cite{Drewes:2013iaa,Drewes:2014pfa}. Moreover, the gauge interactions open up additional channels for DM production in $2\rightarrow 2$ scatterings in addition to the decay~\cite{Drewes:2015eoa}.

If the $S$-field is real, in turn, the Majoron does not exist.  A natural example of such a field $S$ in eq.~\eqref{eq:decays1} would be the inflaton $\phi$. While nothing forbids the inflaton to have a VEV to generate Majorana masses for neutrinos (or even to be complex), in popular ``mainstream'' models the inflaton is a real scalar field with zero VEV. The inflaton field $\phi$ is the driving force behind the Big Bang, and in its decay all matter and entropy in the Universe is produced. Therefore, it is natural to assume that sterile neutrinos can be produced as well~\cite{Shaposhnikov:2006xi}. While in a model considered in Ref.~\cite{Shaposhnikov:2006xi} the inflaton field did have a VEV and by that provided the single source for the scale of electroweak symmetry breaking and for Majorana masses of sterile neutrinos, in general this is not necessary. Sterile neutrino production in the inflaton decay is always possible, subject to the value of the coupling constant $y$.

The remainder of this subsection is devoted to sterile neutrino production from a thermalized population of $\phi$-particles.  Even thermalized quanta of the inflaton field are possible, as in the model of Ref.~\cite{Shaposhnikov:2006xi}.  The population of sterile neutrinos which originate from decays of $\phi$-particles is governed by the standard Boltzmann kinetic description. The result is definite as long as $\phi$-particles are in thermal equilibrium, and in a rough approximation the distribution function of the scalars translates into that of sterile neutrinos.

The distribution function of sterile neutrinos $f(p,t)$, where $p$ is the momentum and $t$ is the time, can be found from the solution of the kinetic equation~\cite{Shaposhnikov:2006xi,Kusenko:2006rh,Petraki:2007gq}:
\begin{equation}
  \frac{\partial f}{\partial t} - H p \frac{\partial f}{\partial p}=
  \frac{2 m_\phi\Gamma}{p^2}\int_{p+m_\phi^2/(4p)}^\infty f_\phi(E)dE,
  \label{eq:kin}
\end{equation}
where it was assumed that the inverse decays $ N_1 N_1 \rightarrow \phi $ can be  neglected (this is true for small Yukawa couplings $y< 10^{-7}$). In eq.~\eqref{eq:kin}, $H$ is the Hubble constant, $f_\phi(E)$ is the distribution of $\phi$-particles, and $\Gamma = m_\phi f^2 /(16\pi)$ is their partial width for the $\phi \rightarrow N_1 N_1$ decay channel. For the case when the effective number of degrees of freedom is time-independent and $f_\phi(E)$ is \emph{thermal}, a semi-analytic solution to eq.~\eqref{eq:kin} can be found easily. At $t \to \infty$ it takes the form
\begin{equation}
f(x) = \frac{16 \Gamma M_0}{3 m_\phi^2} x^2
\int_1^\infty\frac{(y-1)^{3/2} dy}{e^{xy}-1}~,
\label{eq:DistributionSNIE}
\end{equation}
where $x=p/T$ and $M_0 \approx M_{\rm Pl}/(1.66\sqrt{g_*})$, leading to a number density of
\begin{equation}
 n =\int \frac{d^3p}{(2\pi)^3} f(p) = \frac{3\Gamma M_0 \zeta(5)}{2\pi m_\phi^2} T^3
 \label{eq:rho}
\end{equation}
and to an average momentum of created sterile neutrinos immediately after their production equal to $\langle p \rangle = \pi^6/(378\zeta(5))T = 2.45 T$, which is about $20$\% smaller than that for an equilibrium thermal distribution, $p_T=3.15 T$.

The resulting abundance of sterile neutrinos is given by~\cite{Shaposhnikov:2006xi}:
\begin{equation}
\Omega_{N_1} \sim \frac{y^2}{\mathcal{S}_*}\; \frac{M_{\text{Pl}}}{m_\phi} \;   \frac{M_1}  { {\text{keV}}}\;.
\label{eq:decays2}
\end{equation}
It is important to note that a major contribution to $\Omega_{N_1}$ is coming from the epoch when $T\sim m_\phi$, since later on the number density of $\phi$-particles is exponentially suppressed if they remain in thermal equilibrium. The dilution factor
\begin{equation}
  \label{eq:Sstar}
  \mathcal{S}_*\equiv \mathcal{S} \frac{g_*(T_\mathrm{prod})}{\tilde{g}_{*0}} = \mathcal{S}\frac{g_*(T_\mathrm{prod})}{3.9}
\end{equation}
in eq.~\eqref{eq:decays2} is the ratio of the number of effective degrees of freedom at production time, $g_*(T_\mathrm{prod})$, till now, $\tilde{g}_{*0}=3.9$ (this number takes into account the change of the temperature of neutrinos at freezeout). $\mathcal{S}_*$ also contains the entropy production factor $\mathcal{S}$ if this process happens during the QCD transition, see section~\ref{sec:5.dilution} and Ref.~\cite{Asaka:2006ek}.  A more refined analysis is required if the number of d.o.f.\ changes significantly at $T\sim T_\mathrm{prod}\sim m_\phi/3$, see~\cite{Shaposhnikov:2006xi}.

Today, the averaged momentum of sterile neutrinos produced in decays of particles at $T_\mathrm{prod}$ is given by~\cite{Shaposhnikov:2006xi}:
\begin{equation}
\frac{\langle p \rangle}{T_\gamma}  =  \frac{\pi^6\, \mathcal{S}_*^{-1/3}}{378\,\zeta (5)} \approx 2.45\,\mathcal{S}_*^{-1/3}.
\label{eq:decays3}
\end{equation}
Comparing this to non-resonant production with $\langle p_\mathrm{NRP}\rangle/T_\gamma=3.15 (4/11)^{1/3}$,\footnote{It has been pointed out in ref.~\cite{Abazajian:2005gj} that the approximation $\langle p_\mathrm{NRP}\rangle/T_\gamma=3.15 (4/11)^{1/3}$ does not hold exaclty and $\langle p_\mathrm{NRP}\rangle$ is in fact slightly smaller.} 
we can see, that the mass bounds from structure formation\footnote{There are two reference numbers that are usually present in the analysis of the structure formation bounds described in Sections \ref{sec:4-1-Phasespace}, \ref{sec:4-2-LymanA}, bound for the NRP produced sterile neutrino and bound for the thermal relic, which are related as
\[
  m_\mathrm{NRP} = m_\mathrm{TR}\left(\frac{1}{\Omega_{\rm DM}h^2}\frac{m_\mathrm{TR}}{94\eV}\right)^{1/3}.
\]} are weaker than for the NRP case~\cite{Bezrukov:2014nza}
\begin{equation}
  m_{\rm Decay} = \frac{\langle p\rangle}{\langle p_\mathrm{NRP}\rangle}m_\mathrm{NRP} = \frac{2.45}{3.15}\left(\frac{1}{\mathcal{S}}\frac{10.75}{g_*(T_\mathrm{prod})}\right)^{1/3}m_\mathrm{NRP}.
\end{equation}
This is consistent with resonant production being generally threatened by Lyman-$\alpha$ bounds~\cite{Schneider:2016uqi}. However, it should be noted that, taking into account the dilution during the QCD transition, $\langle p_\mathrm{NRP}\rangle/T_\gamma$ can in fact be smaller than $3.15 (4/11)^{1/3}$ by about 20\%~\cite{Abazajian:2005gj}.

\paragraph{Light inflaton model~\cite{Shaposhnikov:2006xi}.}

Let us now turn out attention to one specific example model that makes use of the results described above.  Let us assume, that the field $\phi$ in \eqref{eq:decays1} is the field that leads to the inflationary expansion of the Universe.

First, let us briefly discuss the processes right after inflation and before reheating. They correspond to a particle creation by a coherently oscillating classical scalar field. Using the results of Refs.~\cite{Giudice:1999fb,Chung:1999ve}, we find that the ratio of the number density of light fermions produced to the entropy scales as $n_\nu/n_{\text{rad}} \sim y^{3/2}$. Fermions will be cold at the production, $\langle p \rangle \sim y^{1/2} T$. However, very light fermions in a keV mass range will contribute to the DM abundance appreciably for rather large values of the Yukawa coupling, $y \sim 10^{-2}$. We will not consider such types of models  further in what follows.

The mass of the field and its VEV in this model are related as $m_\phi = \sqrt{2\lambda_\phi}\,\langle \phi \rangle$, where $\lambda_\phi$ is the quartic scalar self-coupling. For a sufficiently light inflaton, with a mass even smaller than the Higgs mass, the inflaton does not decay completely but instead thermalizes and is then in thermal equilibrium down to rather low temperatures, $T \ll m_\phi$. In that case, eq.~\eqref{eq:decays2} yields $\Omega_{N_1} \sim 10^{25} \lambda_\phi \left({M_1}/{m_\phi}\right)^3$. Sterile neutrinos can constitute all the DM e.g.\ for $m_\phi \approx 0.3$~GeV and  $M_1 \approx 20$~keV, if $\lambda_\phi \sim 10^{-13}$. Light inflatons could be specifically searched for at LHC, see Refs.~\cite{Bezrukov:2009yw,Bezrukov:2013fca}. Small values for the inflaton self-coupling, quoted above, are required by the CMB constraints in models with minimal coupling to gravity, and such constraints date back to the epoch when $\phi^4$ inflationary models were not yet ruled out by observations. Otherwise, with a non-minimal coupling to gravity, $\lambda_\phi$ could be larger for the inflaton field~\cite{Bezrukov:2013fca} and may even approach unity~\cite{Bezrukov:2007ep}.

For an inflaton mass of $m_\phi < \mathcal{O}(500)$\,MeV, taking $g_*=\rm{const.}$ is not a good approximation, since exactly in this region $g_*$ changes from $g_* \sim 60$ at $T\sim 1$\,GeV to $g_*\sim 10$ at $T\sim 1$ MeV, because of the disappearance of all the quark and gluon degrees of freedom. In this case a numerical solution of eq.~\eqref{eq:kin} is required, with the input of the hadronic equation of state, which is however not known exactly. In~\cite{Shaposhnikov:2006xi}, numerical calculations have been presented with the use of $g_*(T)$ constructed in~\cite{Asaka:2006nq} on the basis of the hadron gas model at low temperatures, and from the available information on lattice simulations and perturbative computations. It was found that the abundance, eq.~\eqref{eq:rho}, would change by a factor of $\tilde{\mathcal{S}}_*^{-1}(m_\phi) \approx 0.9$ at $m_\phi=70$\,MeV to $0.4$ at $500$~MeV. The average momentum stays almost unchanged in this interval of $m_\phi$. For higher inflaton masses, a good approximation to $\tilde{\mathcal{S}}_*^{-1}(m_\phi)$ is $\tilde{\mathcal{S}}_*^{-1}(m_\phi) \simeq [10.75/g_*(m_\phi/3)]^{3/2}$, leading to an average momentum of $\langle p \rangle \simeq 2.45 T_\nu(10.75/g_*(m_\phi/3))^{1/3}$. The function $g_*(T)$ can be read off from Fig.~2 of ref.~\cite{Asaka:2006nq}, and $\langle p \rangle$ could be as low as $1.5 T$ taking into account entropy dilution by decays of heavier sterile neutrinos.

\subsubsection{\label{sec:5.decays_singlet}Production from generic scalar singlet decays}

Another possibility is to produce the sterile neutrinos from the decay of a general singlet scalar $S$. This idea was first proposed in Refs.~\cite{Kusenko:2006rh,Petraki:2007gq}, where the scalar was assumed to enter thermal equilibrium and either decay immediately or freezing-out before decaying to sterile neutrinos. In this case the distribution of the sterile neutrino can be found approximately as a sum of two components~\cite{Petraki:2007gq}:
\begin{equation}
  f(x,r) = f_1(x,r_f)+f_2(x,r),
\end{equation}
where $x=p/T$ is the neutrino momentum and $r=m_S/T$ is a time-parameter.  At earlier times (lower momenta) the sterile neutrino is produced from in-equilibrium decays of the parent particle.  At larger momenta (later times), sterile neutrinos are produced from the decay of the parent particle which is in thermal equilibrium.  In the most generic situation the process is governed by three dimensionless parameters, $r_f=m_\phi/T_f$ giving the moment of the freeze-out of the parent particle, $\Lambda=\Gamma_\mathrm{tot}M_0/m_\phi^2$ characterizing the total decay width of the parent particle, and by branching of the decay into the sterile neutrino $B=\Gamma(\phi\to N_1 N_1)/\Gamma_\mathrm{tot}$. Then the in-equilibrium decay contributes
\begin{equation}
  \label{eq:Petrakif1}
  f_1(x,r_f) = 2B\Lambda \left[\frac{r_f^3}{3x^2}\ln\frac{1}{1-e^{-x-(r_f^2/4x)}}
  +\frac{8x^2}{3} \int_1^{1+(r_f^2/4x^2)}\frac{(z-1)^{3/2}dz}{e^{xz}-1}\right].
\end{equation}
The out-of-equilibrium component is\footnote{One extra term present in~\cite{Petraki:2007gq} vanishes at late times $r\to\infty$.}
\begin{equation}
  \label{eq:Petrakif2}
  f_2(x,r) = \frac{B}{x^2}\bigg[
  \int_{|(r_f^2/4x)-x|}^\infty x_\phi f_\phi(x_\phi,r_f)dx_\phi -\int_{r_f}^r \frac{r'}{2x}\left(\frac{r'^2}{4x}-x\right)f_\phi\left(\left|\frac{r'^2}{4x}-x\right|,r'\right)dr'\bigg].
\end{equation}
with the concentration of the $\phi$ particle decaying as
\begin{equation}
f_\phi(x_\phi,r) = \frac{1}{e^{\sqrt{x_\phi^2+r_f^2}}-1}
\left(\frac{r+\sqrt{x_\phi^2+r^2}}{r_f+\sqrt{x_\phi^2+r_f^2}}\right)^{\Lambda x_\phi^2}
\times e^{-\Lambda(r\sqrt{x_\phi^2+r^2}-r_f\sqrt{x_\phi^2+r_f^2})}.
\end{equation}
Note, that in analysis of the present day momentum distributions, the overall momenta are divided by the factor $\mathcal{S}_*^{1/3}$ from \eqref{eq:Sstar}. An important observation~\cite{Merle:2015oja} is that these two components can have contributions to the sterile neutrino population with significantly different momenta (see Figure \ref{fig:SnapshotIntermeidateWIMP}).  In general, if the amount of thermal degrees of freedom $g_*$ changed during the production process, an explicit solution of the Boltzman equations is required~\cite{Merle:2015oja}. Another subtlety appears at early times, when the temperature of the plasma exceeds mass of the decaying particle~\cite{Drewes:2015eoa}. This case is mostly relevant if the decay width of the scalar is very large, so that the scalar should be considered as an intermediate state.

An alternative proposal was put forward in Ref.~\cite{Merle:2013wta}, which pointed out that the scalar could also freeze-in and produce a suitable abundance. In Ref.~\cite{Adulpravitchai:2014xna} this mechanism has been extended to the case of frozen-in scalars lighter than the Higgs mass, motivated by the recent hints for an X-ray line at 3.55 keV. The most general numerical study up to now, covering all the cases mentioned, was presented in Ref.~\cite{Merle:2015oja}, which in particular confirmed earlier estimates indicating the sterile neutrinos produced from scalar decays can have a comparatively cold spectrum~\cite{Petraki:2008ef,Boyanovsky:2008nc,Merle:2014xpa}, see also Ref.~\cite{Shakya:2015xnx} for a recent review. While we here focus on the case of a total singlet $S$, note that electrically charged scalars could resemble the case of an equilibrated singlet scalar~\cite{Frigerio:2014ifa}, since for the DM production it does not matter by which interaction the scalar is kept in equilibrium.

\paragraph{The model} From the aforementioned variety of models for scalar decay we pick one which is quite illustrative and allows for a semi-analytic discussion of some relevant aspects. This model is a minimal extension of the SM, featuring one real scalar (which we denote $S$), singlet under the SM and one right handed neutrino (denoted $N_1$). Again, the sterile neutrino is coupled to the scalar via a Yukawa-like interaction, while the scalar is coupled to the Higgs sector via the most general potential $2\lambda\left(H^\dagger H\right) S^2$ allowed after imposing a $\mathbb{Z}_4$-symmetry, thus forbidding the decay to the SM particles and setting the parameter $B=1$ in Eqs.~\eqref{eq:Petrakif1}, \eqref{eq:Petrakif2}.\footnote{For detailed comments on this assumption, see~\cite{Merle:2013wta,Merle:2015oja} and references therein.}

\begin{table}
\caption{Analytically accessible limiting cases of the production of sterile neutrinos from singlet scalar decay. The quantity $M_0$ is the rescaled Planck mass, given by $M_0 = \left(\frac{45 M_{\rm Pl}^2}{4 \pi^3 g_*\left(T_{\rm prod}\right)} \right)^{1/2} = 7.35 g_*^{-1/2}\left(T_{\rm prod}\right) \times 10^{18}$~GeV, where $T_{\rm prod}$ is the temperature at which the production of the scalar and its subsequent decay takes place. Furthermore, $T_{\rm FO}$ is the freeze-out temperature that depends on $m_S$ and $\lambda$, while $K_2$ denotes the second Bessel function of second kind.}
\begin{tabular}{| p{3.5cm}| p{4cm}| p{2.5cm}| p{4cm}|}
 \hline
Regime &  Necessary condition for $\lambda$ & Necessary approximation & Relic yield $Y_{\nu_S}=n/s\left(t\rightarrow \infty\right)$ \\
 \hline\hline
Scalar freeze-in & small enough not to equilibrate the scalar & \multirow{3}{*}{\parbox{2cm}{small enough to ensure $g_* = \text{const.}$ is a good approximation}} & $\frac{135}{1024\pi^5} \frac{M_0\left(T_{\rm prod}\right)}{m_S} \frac{\lambda^2}{g_*\left(T_{\rm prod}\right)}$ \\
\cline{1-2}
\cline{4-4}
Scalar freeze-out + decay in equilibrium & large enough to equilibrate the scalar & & $\frac{135}{64\pi^4} \frac{M_0\left(T_{\rm prod}\right)}{m_S} \frac{y^2}{g_*\left(T_{\rm prod}\right)}$ \\
\cline{1-2}
\cline{4-4}
Scalar freeze-out + decay out of equilibrium & large enough to equilibrate  the scalar & & $\frac{45 m_S^2 K_2\left(m_S/T_{\rm FO}\right)}{4\pi^4 g_*\left(T_{\text{prod}}\right) T{\rm FO}^2}$ \\
 \hline
 \end{tabular}
 \label{tab:OverviewLimitingCases}
\end{table}

We stick to the case where the mass of the scalar is above the EW scale. As in the preceeding part of Sec.~\ref{sec:5.decays}, the production of sterile neutrinos is exclusively governed by the reactions
\begin{align}
 SS & \leftrightarrow hh \quad \text{and} \label{eq:SStohh}\\
 S & \rightarrow N_1 N_1 \label{eq:StoNN} \, ,
\end{align}
to lowest order in perturbation theory.\footnote{For the case of $m_S < v_{\rm EW}$ studies on the level of particle number densitites can be found in~\cite{Adulpravitchai:2014xna}.}
Again we neglect the inverse reaction of two sterile neutrinos producing one scalar. This is justified by kinematical arguments and by the low number density of the sterile neutrinos, which implies that we can use~\eqref{eq:kin} to convert the distribution function of the scalar, once known, into the distribution of the sterile neutrino.

The exact interplay of the reactions~\eqref{eq:SStohh} and~\eqref{eq:StoNN} is governed by two couplings, $\lambda_S$ and $y$. While $\lambda_S$ determines whether or not the scalar ever enters equilibrium and -- if so -- when it decouples from the plasma, the Yukawa coupling $y$ controls the decay width of the scalar and by that virtue has major impact on the production time of the sterile neutrinos.

For different regimes in the parameter space spanned by these couplings, we will calculate the distribution function $f_S$ of the scalar to subsequently obtain the distribution function of the sterile neutrino. Note, that the distribution function contains all relevant information: apart from the number density, the spectral form will also allow to derive consequences for structure formation, BBN, CMB, and so on.

\paragraph{Cosmological dynamics} Let us now turn to the description of the dynamics of the scalar in the different regimes. The abstract Boltzmann equation  is: 
\begin{align}
 \frac{\partial f_S\left(p,t\right)}{\partial t} = &\mathcal{Q}\left(p,t,m_H,m_S, \lambda \right) - \mathcal{P}\left(p,t,m_S,y\right) f_S\left(p,t\right) \nonumber \\ 
 -&\mathcal{R}\left(p,t,m_H,m_S,\lambda\right) f_S\left(p,t\right) \int {\rm d}^3 p' \mathcal{S}\left(p',t,m_H,m_S\right) f_S\left(p',t\right) \, ,
 \label{eq:BoltzmannScalar}
\end{align}
in the limiting case where $g_*$ can be approximated as being constant during production.\footnote{This can directly be mapped to~\cite[eq.~(10)]{Merle:2015oja}, where the authors performed a change of variables on \eqref{eq:BoltzmannScalar}.} The functions $\mathcal{P},\mathcal{Q},\mathcal{R},\mathcal{S}$ encode all the kinematics of the processes. The term $\mathcal{Q}$ describes the production of scalars from Higgs bosons and hence does not depend on $f_S$, while the integral term (containing $\mathcal{R}$ and $\mathcal{S}$) describes the backreaction and therefore scales as $f_S^2$. The depletion of scalars by their decay into two sterile neutrinos is given by the term containing $\mathcal{P}$ and scaling linearly with $f_S$. For three limiting cases, the combination of~\eqref{eq:kin} and~\eqref{eq:BoltzmannScalar} can be solved analytically for the relic abundance of sterile neutrinos. One of these cases is basically equivalent to the one discussed in eqns.~\eqref{eq:DistributionSNIE} to~\eqref{eq:decays2}. For the sake of completeness, we will summarize all three of them in tab.~\ref{tab:OverviewLimitingCases}.

\begin{figure}
 \centering
 \includegraphics[width=0.7 \textwidth]{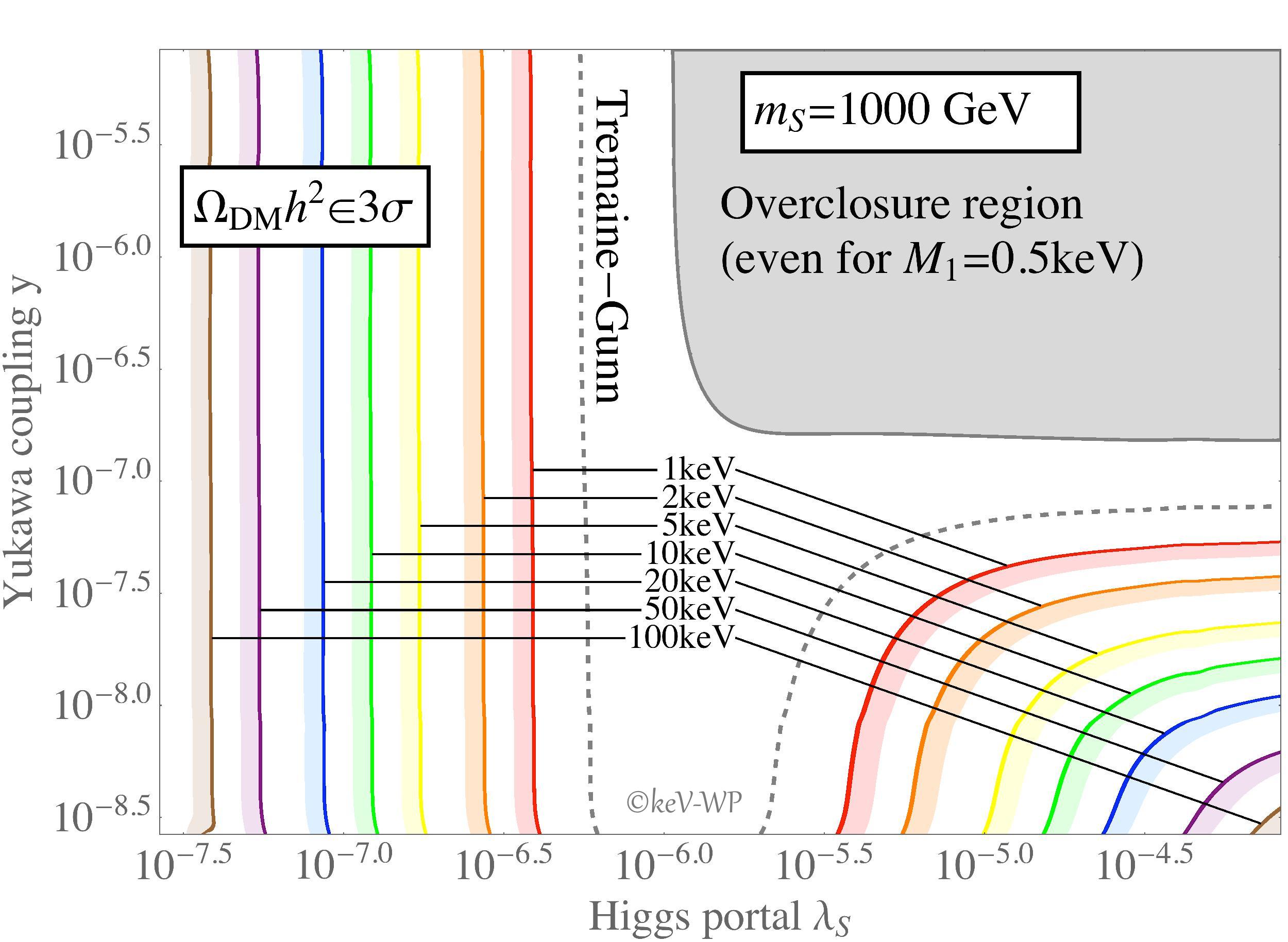}
 \caption{Isoabundance lines in the plane spanned by the Higgs portal $\lambda$ and the Yukawa coupling $y$. The abundance as observed by Planck can be reproduced with masses in the keV-range.}
 \label{fig:AbundanceHiggsPortalYukawa}
\end{figure}

\begin{figure}[th]
 \centering
 \begin{tabular}{lr}
 \includegraphics[width=7cm]{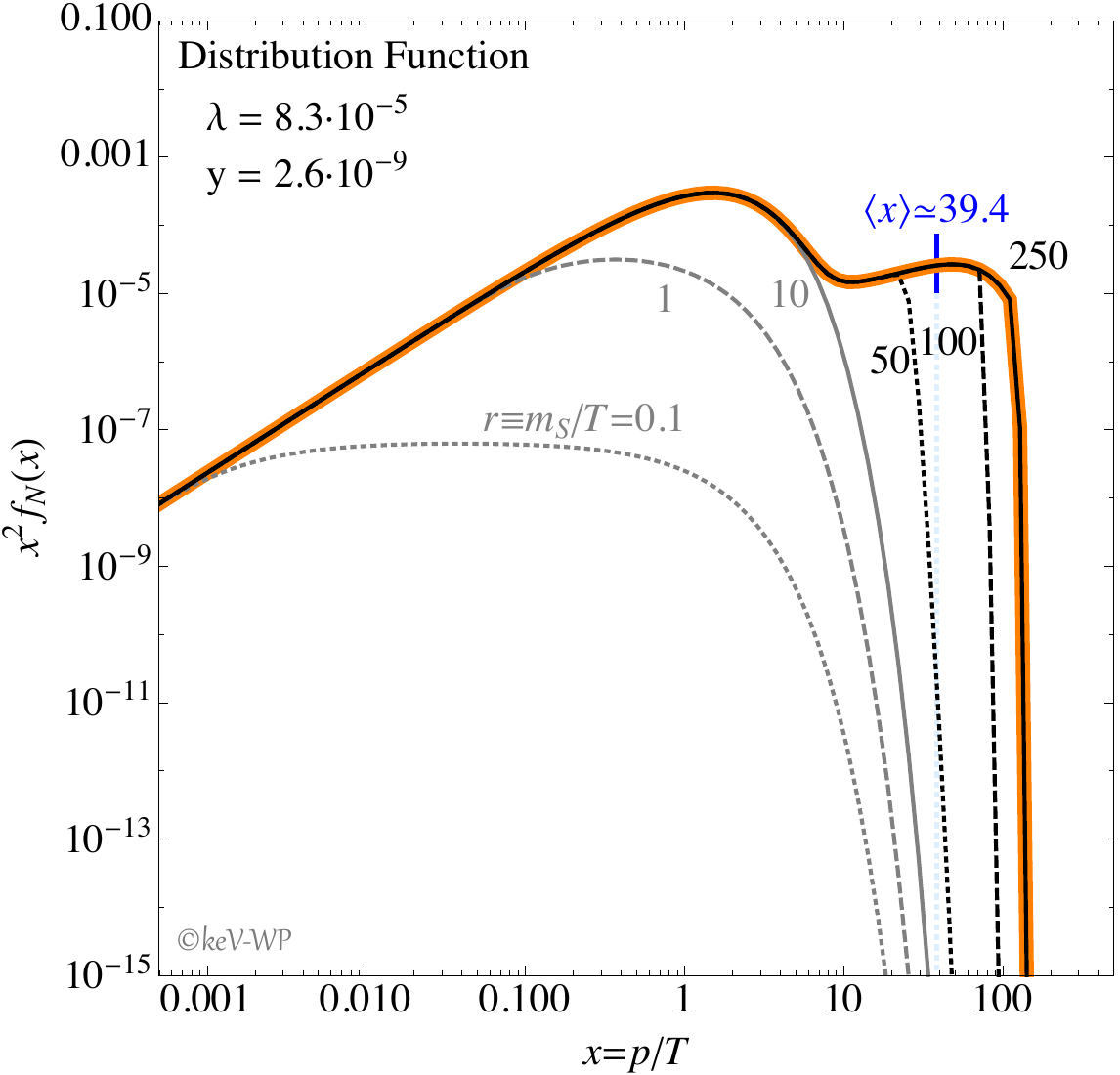} & \includegraphics[width=7cm]{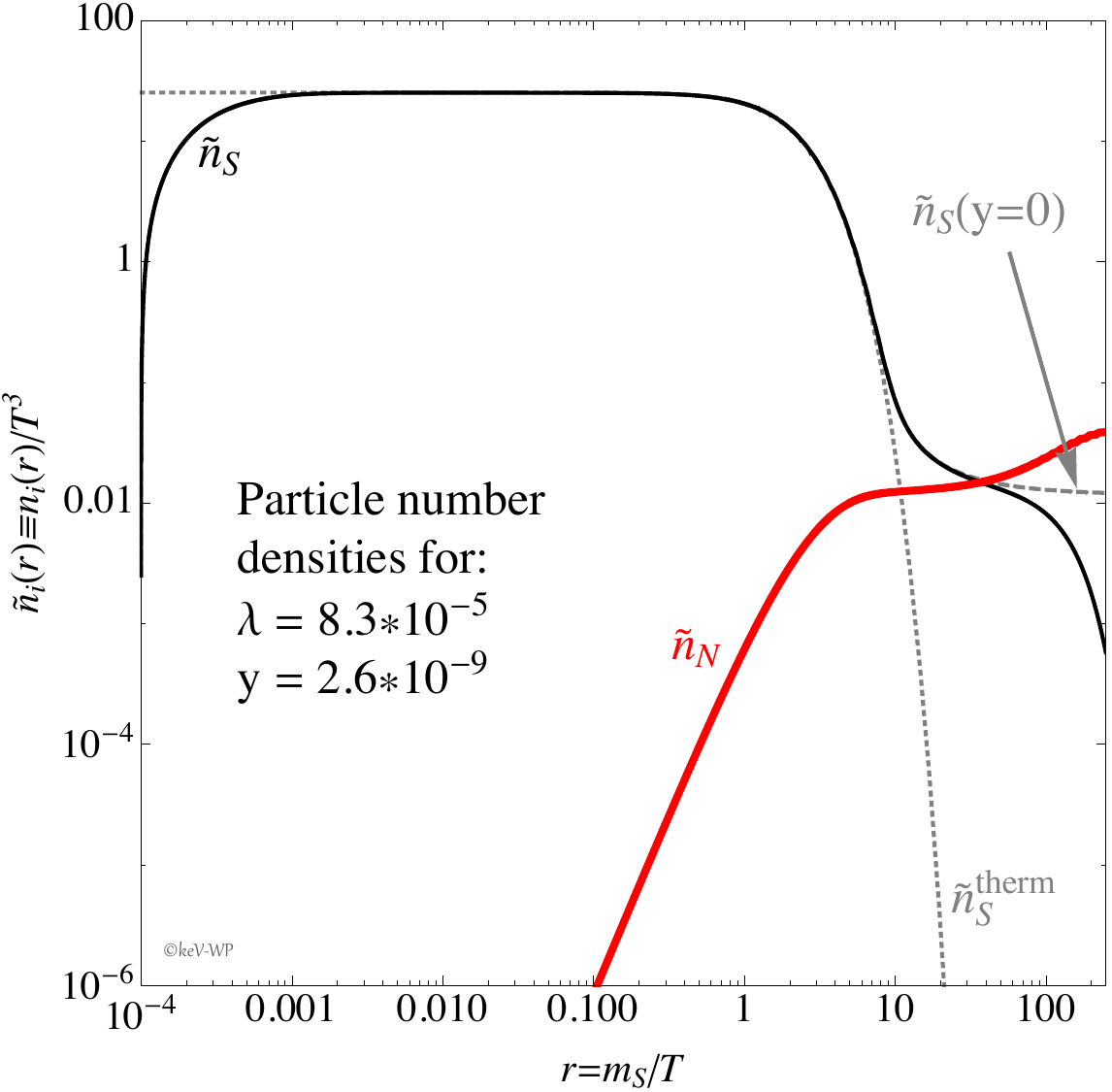}
 \end{tabular}
 \caption{Evolution of the distribution function over the timelike parameter $r=m_S/T$ (left panel) and evolution of the particle number densities (right panel). The plot of the distribution function in the left panel shows $x^2 f_N\left(x\right)$ to give a intuitive picture of the number of particles with the corresponding momentum. In the right panel, $\tilde{n}_S\left(y=0\right)$ shows how the scalar would evolve if it was stable while $\tilde{n}_{S}^{\rm therm}$ depicts how a thermalized particle would evolve. Comparing both panels, it is clearly visible how the decay in equilibrium shapes the peak around lower momenta early on while the second peaks comes from converting the frozen-out relic density of scalars at late times. The two scales result in a rather high average momentum of $\langle x\rangle=\langle p/T \rangle\approx39.4$ which, however, is not a very good measure to describe the sum of two peaks lying at higher / lower momenta.}
 \label{fig:SnapshotIntermeidateWIMP}
\end{figure}

\paragraph{Numerical results}
We have solved eq.~\eqref{eq:BoltzmannScalar} numerically for different regimes of the couplings $\lambda$ and $y$. Fig.~\ref{fig:AbundanceHiggsPortalYukawa} shows lines in the space spanned by $\lambda$ and $y$ reproducing the relic abundance as observed by Planck~\cite{Ade:2015xua} for different masses $m_s$ of the sterile neutrino. It is visible that there is a gap in $\lambda$, around which it is not possible to obtain the required value of the relic density without violating the Tremaine-Gunn (TG) bound~\cite{Tremaine:1979we}. Accordingly, the possible parameter space is clearly divided into the part to the left ($\lambda \lesssim 10^{-6.5}$), where the scalar freezes in and the part with $\lambda \gtrsim 10^{-5.5}$, where the scalar freezes out. In the latter regime, there exists a rather global upper limit on the Yukawa couling $y$. If the scalar equilibrates and the decay width of the scalar, $\propto y^2$, is too large, the particle number density of sterile neutrinos will be too high to be consistent with TG bound.
As one of the major results of the numerical study, we present an intermediary freeze-out case in fig.~\ref{fig:SnapshotIntermeidateWIMP}. In this case, both the decay \emph{in and out} of equilibrium contribute in roughly equal parts to the final abundance. However, since the production times of these two components are significantly different, the spectrum features \emph{two} peaks. This shows that the simple free-streaming estimates might \emph{not} be sufficient and more elaborate anayses have to be done when assessing the observations from structure formation. The most advanced discussion of this mechanism, featuring a detailed account of structure formation bounds, has recently been provided in~\cite{Konig:2016dzg}.

%% file: 4_dilution.tex
\subsection{Dilution of thermally produced DM (Author: F.~Bezrukov)}
\label{sec:5.dilution}

In this section we discuss how to obtain the proper amount of the light keV scale DM if the DM candidate reached thermal equilibrium at some high temperature in the early Universe. This does not happen in $\nu$MSM, where the only source of interactions for $N_1$ is its minuscule Yukawa coupling constant, but is readily achieved in most high energy extensions of SM. Specifically, this happens if $N_1$ participates in gauge interactions of an extension of the SM gauge group. A recent detailed discussion of this possibility without reference to specific particle phyiscs models is given in~\cite{Patwardhan:2015kga}.
Gauge interactions 
If the ``sterile'' neutrinos are charged under some gauge group, their equilibration in the early universe is almost unavoidable unless their freeze-out temperature exceeds the maximal temperature of the primordial plasma.
The freeze-out temperature of the sterile neutrino $T_{f,c}\sim g_{*}^{1/6}(M_{W_R}/M_{W})^{4/3}\times1\MeV$ goes up with the mass $M_{W_R}$ of the new gauge interactions, so for significantly heavy $M_{W_R}$ it may exceed the preheating temperature after inflation.\footnote{This temperature is unknown; it may be constrained by CMB observations in models where reheating is mainly driven by perturbative processes \cite{Martin:2014nya,Drewes:2015coa}.}

According to the standard lore if a light species reaches thermal equilibrium and freezes out while still being relativistic, then its abundance (or number density in comoving volume) is fixed in a unique way at the moment of freeze-out, and its contribution to the energy balance at present, when it becomes non-relativistic, is simply proportional to the mass of the particle.  This simplified analysis gives a bound of $\lesssim10\eV$ for the DM mass.  However, a more detailed analysis~\cite{Bezrukov:2009th} should take into account the subtleties of the expansion history of the Universe after the freeze-out moment, giving the contribution of the sterile neutrino $N_1$ to the present energy density of the Universe
\begin{equation}
  \label{eq5.4:OmegaN}
  \Omega_{N_1} \simeq 0.265
  \frac{1}{\mathcal{S}} \left(\frac{10.75}{g_\text{*}(T_{f,c})}\right)
  \left(\frac{M_1}{1\keV}\right)\times 90.
\end{equation}
Here, $\mathcal{S}$ is the amount of entropy that was released in non-equilibrium processes after the freeze-out\footnote{In fact, a factor of $\mathcal{S}=1-2$ is present in the plain $\nu$MSM as well, and it helps to widen the allowed area of parameters~\cite{Asaka:2006ek}.} (i.e.\ $s_0/a_0^3=\mathcal{S} s_f/a_f^3$), and $g_*(T_{f,c})$ is the effective number of degrees of freedom in the plasma at the moment of the DM neutrino freeze-out. We will not discuss adding 1000 d.o.f.\ at freeze-out, but will consider the possibility of entropy release after the DM freeze-out.

The entropy can be produced in out-of-equilibrium decays of some heavy and relatively long-lived particle~\cite{Kolb:1990vq,Scherrer:1984fd}. In most models there are several sterile neutrinos in addition to the DM $N_1$, and the heavier $N_{2,3}$ can conveniently used for entropy generation (though, one can imagine models with other source of entropy). Assuming that they decay shortly before nucleosynthesis at MeV temperature, their decay leads to the DM abundance
\cite{Nemevsek:2012cd}
\begin{equation}
  \label{eq54:omegaNfull}
  \Omega_{N_1} \simeq
  0.265 \left(\frac{M_1}{1\keV}\right)
  \left(\frac{1.6\GeV}{M_{N_{2,3}}}\right)
  \left(\frac{1\,\text{sec}}{\tau_{N_{2,3}}}\right)^{1/2}
  \left(\frac{g_*(T_{f,2/3})}{g_*(T_{f,1})}\right).
\end{equation}
Thus, for given masses of DM sterile neutrino $N_1$ and entropy generating one $N_{2,3}$ it is possible to define the lifetime $\tau_{N_{2,3}}$, leading to the proper DM abundance today. There are two main constraints on the allowed values of the parameters:

\begin{itemize}

\item Successful nucleosynthesys requires at least $\tau_{N_{2,3}}\lesssim 1\,\text{sec}$, with \eqref{eq54:omegaNfull} leading to the lower bound on the heavy sterile neutrino mass.

\item The diluter should be relativistic at its freeze-out, $T_{f,h}>M_{2,3}$ (otherwise the entropy release is diminished). This can be converted into the lower bound on the scale of the gauge interactions, that lead to the thermalization of the diluter.

\end{itemize}

Simple estimates for these bounds give the heavy sterile neutrinos heavier than $M_{2,3}\gtrsim (M_1/1\keV)\times1.6\GeV$ and the mass of the additional gauge boson, leading to thermalization of $N_{2,3}$, larger than $M_{W_R}\gtrsim(M_1/1\GeV)^{3/4}\times10\TeV$. In~\cite{Bezrukov:2012as} it was shown that the decay of the heavy left neutrinos leading to the entropy dilution can also lead to leptogenesis, which proceeds via usual thermal high temperature scenario.

More careful analysis for a minimal left-right symmetric theory was made in~\cite{Nemevsek:2012cd}.  Taking into account two diluting particles $N_2$ and $N_3$, and tuning the flavour structure to separate the freeze-out temperatures $T_{f,h}$ and $T_{f,c}$ of the dilutors and DM sterile neutrinos, the constraints can be relaxed in a very specific corner of the parameter space.

The momentum distribution of the produced sterile neutrinos is also cooled by the entropy dilution, leading to the weaker bound from the Lyman-$\alpha$.  The spectrum of the DM corresponds to thermal spectrum with the temperature, reduced by a factor of $\mathcal{S}^{-1/3}$. This corresponds to the \emph{thermal relic} case in the terminology of the majority of the Lyman-$\alpha$ paper, and allows to use the bounds from the recent papers without modifications.  The latest bound from~\cite{Viel:2013apy} gives
\[
  M_1 > 3.3\keV(2\sigma)\text{ and }2.5\keV(3\sigma),
\]
while the older very conservative bound from~\cite{Boyarsky:2008xj} gives for thermal relic, $M_1>1.5$~keV at $99.7\%$~C.L.

Last but not least, in the models with entropy dilution one should check that the lifetime of the DM sterile neutrino is still compatible with the X-ray observations, as far as additional interactions often lead to increase in the $N_1\to\nu\gamma$ decay rate. For example in the LR symmetric models a bound on the $W_R-W_L$ mixing appears in addition to the bound on the sterile-active neutrino mixing.

To conclude the section we should note variations of the scenario. In~\cite{Bezrukov:2009th,Nemevsek:2012cd} entropy dilution was achieved in the context of Left-Right symmetric extensions of the SM. A model with radiative generation of active neutrino masses was analyzed in~\cite{Ma:2012if,Hu:2012az}, model with composite neutrinos in~\cite{Robinson:2012wu}. Note, however, that quite generally a large amount of entropy dilution is not easy to get in agreement with big bang nucleosynthesis~\cite{King:2012wg}.

Of course, similar setup with the DM abundance given by \eqref{eq5.4:OmegaN} can be realized with DM candidates that are not sterile neutrinos, but some other light stable particle, see~\cite{Babu:2014uoa} as a recent example.

%% file: kevnuwp_section6.tex

In this section, we will address some of the most generic mechanisms used in the literature to motivate light sterile neutrinos in general and sterile neutrinos with keV masses in particular. While there are some similarities to eV-sterile neutrinos, we will in particular consider the small active-sterile mixing, which is characteristic for sterile neutrinos with keV-scale masses acting as Dark Matter (DM).

\subsection{General principles of keV neutrino model building (Authors: A.~Merle, V.~Niro)}

In the SM of particle physics, sterile neutrinos are viewed as mass eigenstates which have predominantly right-handed chirality, i.e.\ they do \emph{not} couple to SM gauge bosons.\footnote{Note that this might be different in settings with an extended gauge groups, such as left-right models.} These fields are typically called in the literature ``right-handed'' neutrinos if they are heavy (GeV or higher) and ``sterile neutrinos'' if they are light, with mass around eV--keV or MeV. However, this distinction is to some extend artificial and we will use the two terms in this section in an interchangable way. 

These types of fermions can obtain mass from a Majorana mass term, cf.\ sec.~1, which does a priori not have any relation to the electroweak symmetry breaking scale. Thus, it will be natural to expect the mass of sterile neutrinos to be very heavy. For this reason, it is necessary to develop mechanisms which can yield light sterile neutrinos and which can also protect a potentially small sterile neutrino mass from higher order corrections. 

\begin{figure}
\centering
\begin{tabular}{lr}
\includegraphics[scale=0.6]{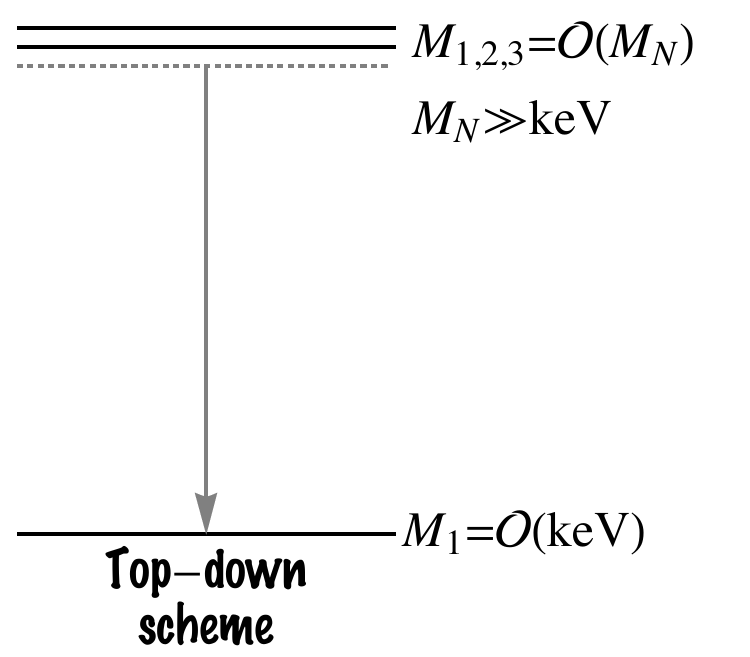} &
\includegraphics[scale=0.6]{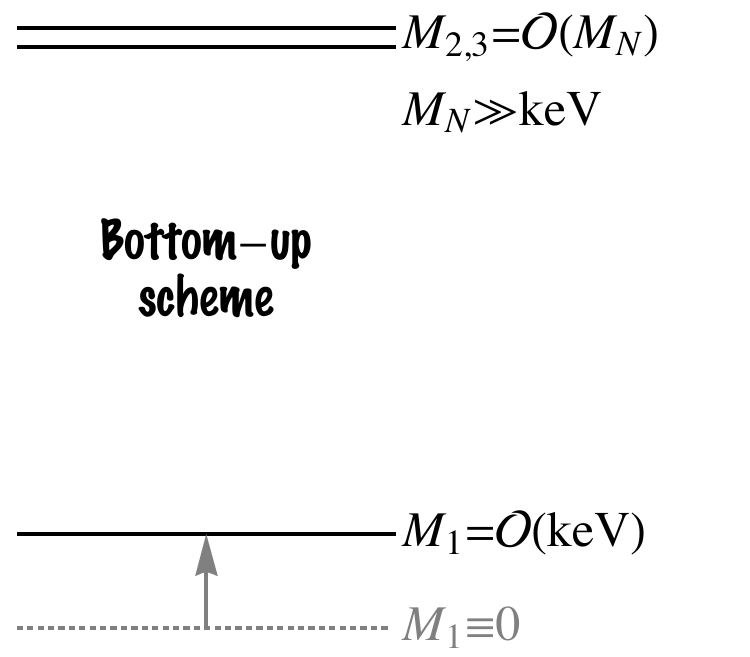}
\end{tabular}
\caption{\label{fig:mass_schemes}
The two generic mass shifting schemes used in models for light sterile neutrinos. Most of the mechanisms proposed ultimately apply some variant of one of those schemes. In both cases, for $M_1$, the ``natural'' mass scales are indicated as well as how the shift to the keV scale. (Figure similar to fig.~4 in ref.~\cite{Merle:2013gea}.)}
\end{figure}

In the literature one can find a number of example mechanisms that achieve this goal. To illustrate the generic principles, we can divide the known mechanisms into two categories: mechanisms which suppress a naturally large sterile neutrino mass and mechanisms which give a small but non-zero mass correction to a sterile neutrino mass that is exactly zero at leading order. In ref.~\cite{Merle:2013gea}, these two categories have been named \emph{top-down} and \emph{bottom-up} scheme, respectively, see fig.~\ref{fig:mass_schemes}. This division may not be entirely exhaustive, as we will discuss in sec.~\ref{sec:VI-other-Heeck} and~sec.~\ref{sec:VI-other-Tsai}, but it nevertheless illustrates a generic principle: in the framework of a quantum field theory it is impossible to obtain absolute mass scales, but we can nevertheless explain strong \emph{mass patterns}, i.e., hierarchies or (quasi-)degeneracies between masses.

The mechanisms presented in sec.~\ref{sec:VI-suppression} all rely on the principle of suppressing one sterile neutrino mass eigenvalue to a value much smaller than its ``natural'' scale (and thus smaller than the masses of other sterile neutrinos present in the model). Such suppressions typically come from new physics affecting the different generations of sterile neutrinos in different ways, which may break mass degeneracies or strongly enhance a mass splitting that is already present. On the other hand, mechanisms based on discrete or continuous flavor symmetries, see sec.~\ref{sec:IV-breaking}, typically predict one strictly massless sterile neutrino in the limit of exact symmetry, along with one or several massive ones. Once (part of) the symmetry is broken, e.g.\ by higher order corrections, this zero mass eigenvalue receives corrections proportional to the symmetry breaking parameter which must be smaller than the mass scale of the heavier sterile neutrinos, as otherwise we would not have the symmetry in the first place. Other models where the suppression is obtained with a more complicated mechanisms are presented in sec.~\ref{sec:VI-other}. In general, all these models generate a hierarchy between the light (nearly massless) sterile neutrino and the heavier sterile neutrinos involved. 

One final point that is important for all mechanisms is the stability of the mass generation mechanism. Whichever mechanism we consider, it will generically receive corrections, e.g.\ from quantum loop contributions or from higher order terms, neglected at the order under consideration. While such corrections are often small in practice, there may be cases in which they can spoil the conditions required for the mechanism to work. This should be cross-checked in order to verify the validity of a certain mechanism. In most of the examples presented, this check has been performed in the respective references, see for example the models presented in secs.~\ref{sec:Q6} and~\ref{sec:A4}, but we will not always explicitly comment on it in this manuscript. However, we invite the reader to consult the original references which provide many more details than we could possible illustrate in the framework of this section.

\subsection{\label{sec:VI-suppression}Models based on suppression mechanisms}

In this section, we will introduce the first class of mechanisms which can motivate a light sterile neutrino mass scale, namely those based on the principle of suppressing a naturally large sterile neutrino mass to a value at the keV scale.

\subsubsection{The split seesaw mechanism and its extensions (Author: R.~Takahashi)}

We start our discussion with two models based on a suppression mechanism for the neutrino Yukawa couplings, both of which rely on the existence of extra spatial dimensions. The models are the split~\cite{Kusenko:2010ik} and the separate~\cite{Takahashi:2013eva} seesaw mechanisms, which are considered in a flat five-dimensional space-time. The five-dimensional space is assumed to be compactified on $S^1/\mathbb{Z}_2$ and the five-dimensional coordinate is denoted as $y\equiv x^5$. The fundamental region is given by $y\in[0,\ell]$. There are two fixed points at $y=0,\ell$. We consider branes (=four-dimensional subspaces) at the fixed points, i.e., one brane at $y=0$ is a standard model (SM) brane and the other at $y=\ell$ is a hidden brane. The SM particles reside at the SM brane.

In both the split and the separate seesaw models, three generations of a Dirac spinor, $\Psi_I^5(x,y)=(\chi_I^5(x,y),\overline{\psi_I^5}(x,y))^T$, are introduced on the bulk with bulk masses $m_{5,I}$ in this flat five-dimensional space-time, where $I$ denotes the generation index as $I=1,2,3$. A fundamental action for the Dirac spinors is
\begin{eqnarray}
  S=\int d^4x\int_0^\ell dy M(i\overline{\Psi_I^5}\Gamma^A\partial_A\Psi_I^5
                             +m_{5,I}\overline{\Psi_I^5}\Psi_I^5),~~~A=0,1,2,3,5,
\end{eqnarray}
where $M$ is a five-dimensional fundamental scale, the Dirac mass matrix $m_5$ is assumed to be diagonal ($m_{5,I}\equiv (m_5)_{II}$) for simplicity, and $\Gamma^A$ are the five-dimensional gamma matrices. The five-dimensional Dirac equations determine wave function profiles of zero modes on the bulk as $e^{\mp m_{5,I}y}$ for $\chi_I^5(x,y)$ and $\overline{\psi_I^5}(x,y)$. Note that only $\overline{\psi_I^5}(x,y)$ can have a zero mode under an orbifold parity as $\Psi_I^5(x,y)\rightarrow P\Psi_I^5(x,y)=+\Psi_I^5(x,y)$ with $P\equiv-i\Gamma_5$. After the canonical normalization of the Dirac spinor in four dimensions, the zero modes of $\Psi_{R_I}^5(x,y)\equiv(0,\overline{\psi_I^5}(x,y))^T$ are given by
\begin{eqnarray}
  \Psi_{R_I}^{(0)}(x,y)=\sqrt{\frac{2m_{5,I}}{(e^{2m_{5,I}\ell}-1)M}}e^{m_{5,I}y}\psi_{R_I}^{(0)}(x),
  \label{0}
\end{eqnarray}
where $\psi_{R_I}^{(0)}(x)$ are identified with the normalized right-handed neutrino fields in four dimensions, $\psi_{R_I}^{(0)}=\nu_{R,I}$. One can see that the five-dimensional wave function profiles of $\Psi_{R_I}^{(0)}(x,y)$ with positive (negative) value of $m_{5,I}$ localize at the hidden (SM) brane since the profile is proportional to $e^{m_{5,I}y}$. Thus, the right-handed neutrinos have exponentially suppressed Yukawa couplings at the SM brane when $m_{5,I}\ell\gg1$ with positive $m_{5,I}$.

The relevant action for the split seesaw model, that uses the above property of the wave function profiles of the right-handed neutrinos, is given by 
\begin{eqnarray}
  S &=& \int d^4x\int_0^\ell dy\Bigg[M
        \left(i\overline{\Psi^{(0)}_{R_I}}\Gamma^A\partial_A\Psi_{R_I}^{(0)} 
               +m_{5,I}\overline{\Psi_{R_I}^{(0)}}\Psi_{R_I}^{(0)}\right) \nonumber \\
    & & \phantom{\int d^4x\int_0^\ell dy\Bigg[}
        -\delta(y)\left((\tilde{F}^\dagger)_{I\alpha}\overline{\Psi^{(0)}_{R_I}}
                        l_{L\alpha}\tilde{\Phi}^\dagger
                        +\frac{(\tilde{M}_M)_{IJ}}{2}\overline{\Psi^{(0)c}_{R_I}}
                         \Psi^{(0)}_{R_J}\right)+h.c.\Bigg].
  \label{split}
\end{eqnarray}
Inserting eq.~\eqref{0} into eq.~\eqref{split}, one obtains the Majorana mass matrix of the right-handed neutrinos and the neutrino Yukawa matrix in four dimensions as $(M_M)_{IJ}=f_If_J(\tilde{M}_M)_{IJ}$ and $F_{I\alpha}=f_I\tilde{F}_{I\alpha}$, respectively, where $f_I\equiv\sqrt{2m_{5,I}/((e^{2m_{5,I}\ell}-1)M)}$. Then, the seesaw formula can be written as $m_\nu=-FM_M^{-1}F^Tv^2=-\tilde{F}\tilde{M}_M^{-1}\tilde{F}^Tv^2$. Note that the factors $f_I$ are canceled out in the seesaw formula. This is one of the important properties of the split seesaw mechanism, which results into the following feature: one can realize strongly hierarchical Majorana mass spectra of the right-handed neutrinos and the neutrino Yukawa couplings due to the exponential suppression factor $f_I$ by taking suitable values of $m_{5,I}$ without introducing strongly hierarchical many-mass scales. For example, one can realize a splitting Majorana mass spectrum of the right-handed neutrinos containing both the keV and intermediate mass scales as $(M_{M,1},M_{M,2},M_{M,3})=(5~{\rm keV},10^{11}~{\rm GeV},10^{12}~{\rm GeV})$ by taking $(m_{5,1}\ell,m_{5,2}\ell,m_{5,3}\ell)\simeq(23.3,3.64,2.26)$, $M=5\times10^{17}~{\rm GeV}$, $\ell^{-1}=10^{16}~{\rm GeV}$, and $(\tilde{M}_M)_{II}=10^{15}~{\rm GeV}$. In this example, the mass scales in the model are only the super heavy ($\mathcal{O}(10^{15-17})~{\rm GeV}$) and the electroweak ones, and the small active neutrino mass can be realized by the (split) seesaw mechanism. Furthermore, the lightest sterile neutrino can be a candidate for DM and the two heavier ones could generate the baryon asymmetry of the Universe via the leptogenesis mechanism.

Next, we comment on the active-sterile mixing angle in the (split) seesaw model. It is well known that the sterile neutrino DM with the keV mass should not contribute to the active neutrino masses through the (split) seesaw mechanism in order to satisfy astrophysical X-ray bounds on the active-sterile mixing angle (equivalent to the corresponding neutrino Yukawa couplings) as $\theta^2\equiv\sum|F_{1\alpha}|^2v^2/M_{M,1}^2\simeq\mathcal{O}(10^{-10})$ in the few keV region of $M_{M,1}$. This bound implies that when the keV sterile neutrino DM gives the small active neutrino mass through the (split) seesaw, the model is in conflict with the astrophysical bound. The simple reason is that the active-sterile mixing angle becomes large for small masses, $\theta^2=\sum |F_{1\alpha}|^2v^2/M_{M,1}^2\simeq m_{\rm sol(atm)}/M_{M,1}\simeq\mathcal{O}(10^{-6}-10^{-5})$, where $m_{\rm sol(atm)}$ denote the solar and the atmospheric neutrino mass scales. In other words, while both the Dirac Yukawa coupling and the sterile neutrino mass are suppressed by the size of the extra dimension, the suppression of the coupling enforced by the split seesaw mechanism is weaker; accordingly, active sterile mixing is somewhat \emph{enhanced}. Thus, an additional suppression for $\theta$ (or, equivalently, $|F_{1\alpha}|$) is required to satisfy the X-ray bound in a realistic model with keV neutrino DM. Such a situation can be easily realized by assuming the Yukawa couplings of the keV sterile neutrino to be additionally small compared with those of the other sterile neutrinos. In this case, the solar and the  atmospheric neutrino mass scales are induced from the (split) seesaw with two heavier right-handed neutrinos. On the other hand, the separate seesaw can give an additional suppression factor for the active-sterile mixing.\\

The separate seesaw model~\cite{Takahashi:2013eva} is a variant of the split seesaw. A fundamental assumption of the separate seesaw is that different generations of right-handed neutrinos can have different localization of the five-dimensional wave function and a Majorana mass of the right-handed neutrino unlike the split seesaw. For a realistic model building of keV neutrino DM, we consider the following action of the separate seesaw model,
\begin{eqnarray}
  S &=& \int d^4x\int_0^\ell dy\Bigg[M
        \left\{i\overline{\Psi^{(0)}_{R_I}}\Gamma^A\partial_A\Psi_{R_I}^{(0)} 
               +\left(m_{5,1}\overline{\Psi_{R_1}^{(0)}}\Psi_{R_1}^{(0)}
                      -m_{5,J}\overline{\Psi_{R_J}^{(0)}}\Psi_{R_J}^{(0)}
                \right)\right\}  \label{separate}\\ 
  && \left.
        -\left\{\delta(y)\left((\tilde{F}^\dagger)_{I\alpha}\overline{\Psi^{(0)}_{R_I}}
                               l_{L\alpha}\tilde{\Phi}^\dagger
                               +\frac{\tilde{M}_{M,1}}{2}
                                \overline{\Psi^{(0)c}_{R_1}}\Psi^{(0)}_{R_1}
                         \right)
                +\delta(y-\ell)\frac{\tilde{M}_{M,J}}{2}
                 \overline{\Psi^{(0)c}_{R_J}}\Psi^{(0)}_{R_J}+h.c.\right\}
        \right], \nonumber
\end{eqnarray}
where $J=2,3$ and we take $\tilde{M}_M={\rm diag}\{\tilde{M}_{M,2},\tilde{M}_{M,3}\}$. Thus, the extra-dimensional wave function profiles of the right-handed neutrinos are given by $\Psi_{R_1}^{(0)}=f_1e^{m_{5,1}y}\psi_{R_1}^{(0)}(x)$ and $\Psi_{R_J}^{(0)}=g_Je^{-m_{5,J}y}\psi_{R_J}^{(0)}(x)$, where $g_J\equiv f_Je^{m_{5,J}\ell}$. Note that the wave functions of the second and third generations of the right-handed neutrinos localize at the SM brane and their Majorana masses localize at the hidden brane, unlike for split seesaw, while the localization for the first generation is the same as in the split seesaw. As a result, the Majorana masses of the right-handed neutrinos and the neutrino Yukawa coupling matrix in four dimensions are $M_{M,I}=f_I^2\tilde{M}_{M,I}$, $F_{1\alpha}=f_1\tilde{F}_{I\alpha}$, and $F_{J\alpha}=g_J\tilde{F}_{J\alpha}$, respectively. The separate seesaw gives the active neutrino mass matrix and a typical mass spectrum as
\begin{eqnarray}
  &&(m_\nu)_{\alpha\beta}=\left(\tilde{F}_{1\alpha}\tilde{F}_{1\beta}\tilde{M}_{M,1}^{-1}+\sum_{J=2,3}e^{2m_{5,J}\ell}\tilde{F}_{J\alpha}\tilde{F}_{J\beta}\tilde{M}_{M,J}^{-1}\right)v^2, \\
  && m_1\sim\mathcal{O}(|\tilde{F}_{1\alpha}|^2v^2\tilde{M}_{M,1}^{-1}),~~~
       m_J\sim\mathcal{O}(e^{2m_{5,J}\ell}|\tilde{F}_{J\alpha}|^2v^2\tilde{M}_{M,J}^{-1}),
\end{eqnarray}
respectively. This separate seesaw can also realize strongly hierarchical mass spectra of the sterile neutrinos (e.g., $(M_{M,1},M_{M,2},M_{M,3})=(5~{\rm keV},10^{11}~{\rm GeV},10^{12}~{\rm GeV})$). Turning to the mass spectrum of the active neutrinos, $m_1$ is suppressed compared to $m_J$, and $m_J$ can realize the solar and the atmospheric neutrino mass scales. This suppression affects on the size of the left-right mixing angle as $\theta^2=\sum|F_{1\alpha}|^2v^2/M_{M,1}^2\simeq m_{\rm sol(atm)}/(e^{2m_{5,J}\ell}M_{M,1})$. Finally, one can realize $\theta^2\sim\mathcal{O}(10^{-10})$ without assuming additional suppressions for the Yukawa couplings of only the first generation of sterile neutrinos.

\subsubsection{\label{sec:FN}Suppressions based on the Froggatt-Nielsen mechanism (Authors: A.~Merle, V.~Niro)}

The Froggatt-Nielsen (FN) mechanism~\cite{Froggatt:1978nt} has been used in the literature to explain very strong mass hierarchies, for example, in the quark sector. In ref.~\cite{Merle:2011yv}, it was shown that the FN mechanism could explain the existence of one keV scale sterile neutrino, whereas the other two sterile neutrinos have masses at least at GeV scale: $M_1 \simeq \mathcal{O}({\rm keV})$ and $M_2, M_3 \gtrsim \mathcal{O}({\rm GeV})$. The most important point of the work was to find out the {\it minimal} assignments for the FN charges to successfully explain the pattern in the sterile neutrino sector, while being in full agreement with the rest of the lepton data. Note that also in ref.~\cite{Barry:2011wb} a FN mechanism, together with a $Z_3$ symmetry, was used in the context of an $A_4$ flavor symmetry. This model will be discussed in sec.~\ref{sec:A4}.

The FN flavon field can acquire a VEV $\langle \Theta \rangle$. Calling $\Lambda$ the cut-off scale at which the heavy sector of the theory is integrated out, the 
parameter $\lambda =\langle \Theta \rangle/\Lambda$ is a small number of the order of the Cabibbo angle: $\lambda \simeq 0.22$~\cite{Datta:2005ci}. 
After the flavon field acquires a VEV, the corresponding mass matrix of right-handed neutrinos, $M_M$, will have the following structure:
\begin{equation}
M_M=
\begin{pmatrix}
\tilde{M}^{11}_M\,\lambda^{|2 g_1|}\hfill \hfill & \tilde{M}^{12}_M\,\lambda^{|g_1+g_2|} & \tilde{M}^{13}_M\, \lambda^{|g_1+g_3|}\\
\bullet & \tilde{M}^{22}_M\,\lambda^{|2 g_2|}\hfill\hfill & \tilde{M}^{23}_M\,\lambda^{|g_2+g_3|}\\
\bullet & \bullet & \tilde{M}^{33}_M\,\lambda^{|2 g_3|}\,\hfill \hfill 
\end{pmatrix}\,,
 \label{eq:MR_struc}
\end{equation}
where $g_i$ are the FN charges of the right-handed neutrinos. 
Calculating the light neutrino mass matrix, it can be seen that the seesaw formula is not spoiled by the presence of a keV neutrino. 
Indeed, the $U(1)_{\rm FN}$ charges of the right-handed fermions cancel out. As a consequence, the basic structure of the neutrino mass matrices is the same for type~I and type~II seesaw scenarios. 

\begin{figure}[t]
\centering
\includegraphics[width=6cm]{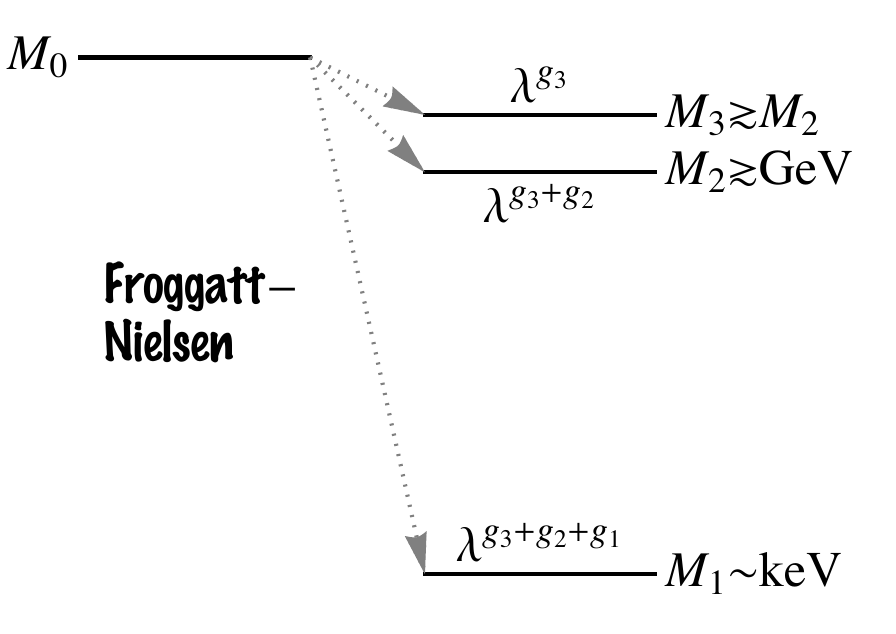}
\caption{\label{fig:FN-scheme} The mass shifting scheme of Froggatt-Nielsen models. (Figure similar to fig.~2 in ref.~\cite{Merle:2011yv} and to fig.~5 in ref.~\cite{Merle:2013gea}.)}
\end{figure}

If we suppose that all the coefficients in the matrix $M_M$ are of the same order, then the explanation of a hierarchical spectrum in the sterile neutrino sector should solely come from the FN charges. Considering $g_1 \geq g_2 \geq g_3$, the minimal conditions to be fulfilled are:
\begin{equation}
\left\{
\begin{array}{rclcrcl}
 g_1 &\geq& g_1|_{\rm min} &\quad {\rm{ with  }} \qquad& g_1|_{\rm min}&=&g_2+3\,,\nonumber\\
 g_2 &\geq& g_2|_{\rm min} &\quad {\rm{ with  }} \qquad& g_2|_{\rm min}&=&g_3\,.
 \label{eq:FN-condition}
\end{array}
\right.
\end{equation}
In fig.~\ref{fig:FN-scheme}, we have schematically depicted the general effect of the FN mechanism, where a certain mass scale $M_0$ is multiplied by powers of $\lambda$ that depend on the fermion generation. A reasonable condition would be to set $g_3=0$ and $g_1=g_2+3$. Two minimal scenarios are given by $(g_1,g_2,g_3)=(3,0,0)$ and $(g_1,g_2,g_3)=(4,1,0)$. In the following, these will be called Scenario~A and Scenario~B, respectively. 

The way the FN charged are combined considerably restricts the allowed paramenter space. Indeed, despite the fact that the FN charge assignment seems to involve great freedom, several scenarios are not able to successfully produce a light neutrino sector compatible with experimental data. In particular, the FN scenarios disagree with the Left-Right symmetric framework for keV sterile neutrino DM. It turns out that an ideal framework is represented by an $SU(5)$ inspired model with two FN fields, see the detailed discussion in ref.~\cite{Merle:2011yv}. 

A problem of models with only one FN field is that they normally lead to small atmospheric neutrino mixing and are thus incompatible with the data, see discussion in refs.~\cite{Kamikado:2008jx,Kanemura:2007yy,Choi:2001rm}. The only way out would be to consider a pseudo-Dirac scenario and, thus, set some elements of the mass matrices equal to zero~\cite{Wolfenstein:1981kw}. However, this would be a specific choice, while the most general approach is to consider two FN fields. The model presented in refs.~\cite{Kamikado:2008jx,Kanemura:2007yy} contains two FN flavon fields $\Theta_{1,2}$ which can obtain complex VEVs:
\begin{equation}
 \lambda=\frac{\langle \Theta_1 \rangle}{\Lambda},\ \ R=\frac{\langle \Theta_1 \rangle}{\langle \Theta_2 \rangle} = 
R_0 e^{i\alpha_0},
 \label{eq:lambda_R}
\end{equation}
with $R_0$ and $\alpha_0$ being real numbers. Besides having two FN fields, it is also necessary to introduce an auxiliary $\mathbb{Z}_2$ symmetry, such that the phase $\alpha_0$ can be responsible for $CP$ violation~\cite{Kanemura:2007yy}. The most general Lagrangian in a seesaw type~II scenario is then given by 
\begin{eqnarray}
 \mathcal{L} &=& 
-\sum_{a,b,\alpha,\beta}^{a+b=k_\alpha+f_\beta} Y_e^{\alpha \beta}\,\overline{e_{\alpha R}}\,H\,L_{\beta L}\,\lambda_1^a \lambda_2^b +h.c.\
-\sum_{a,b,I,\alpha}^{a+b=g_I+f_\alpha} Y_D^{I \alpha}\,\overline{\nu_{I R}}\,\tilde{H}\,L_{\alpha L}\,\lambda_1^a \lambda_2^b +h.c. \\
 && - \sum_{a,b,\alpha,\beta}^{a+b=f_\alpha+f_\beta} \frac{1}{2} \overline{(L_{\alpha L})^C}\,\tilde m_L^{\alpha \beta}\,L_{\beta L}\,\lambda_1^a \lambda_2^b +h.c.\
- \sum_{a,b,I,J}^{a+b=g_I+g_J} \frac{1}{2}\overline{\nu_{I R}}\,\tilde M_M^{IJ}\,(\nu_{J R})^C\,\lambda_1^a \lambda_2^b+h.c.\,,
\nonumber
 \label{eq:FN-Lagrangian}
\end{eqnarray}
where, by using eq.~\eqref{eq:lambda_R}, we have defined
\begin{equation}
 \lambda_1^a \lambda_2^b \equiv \left( \frac{\Theta_1}{\Lambda} \right)^a \left( \frac{\Theta_2}{\Lambda} \right)^b = 
\lambda^{a+b} R^b.
 \label{eq:L-ratios}
\end{equation}
The matrices $Y_e$, $Y_D$, $\tilde M_M$ and $\tilde m_L$ are the charged lepton Yukawa matrix, the Dirac neutrino Yukawa matrix, the uncorrected right-handed Majorana neutrino mass matrix, and the uncorrected left-handed Majorana neutrino mass matrix. Note that the matrix elements of $\tilde M_M$ and $\tilde m_L$ are all of the same order, since they are not yet corrected by FN contributions. Moreover, certain terms in the Lagrangian of Eq.~\eqref{eq:FN-Lagrangian} might violate the $\mathbb{Z}_2$ parity and must be set to zero. We adopt the following FN charge and $\mathbb{Z}_2$ assignments inspired by refs.~\cite{Asaka:2003fp,Kanemura:2007yy}, for the FN fields, for the lepton doublets, and for the right-handed charged leptons and neutrinos:
\begin{eqnarray}
 \Theta_{1,2}: (-1,-1;+,-), &&  L_{1,2,3}: (a+1,a,a;+,+,-),\nonumber\\
 \overline{e_{1,2,3}}: (3,2,0;+,+,-), &&  \overline{\nu_{1,2,3 R}}: (g_1,g_2, g_3; +,+,-),
 \label{eq:FN-assignments}
\end{eqnarray}
where $a=0,1$. Considering this assignment, the mass eigenvalues for the right-handed neutrinos as functions of the right-handed mass scale $M_0$ are:
\begin{eqnarray}
\text{Scenario~A}: && M_1 = M_0 \lambda^6\ 2 R_0^2 \sqrt{1+ R_0^4 + 2 R_0^2 \cos (2\alpha_0)}\,,\quad M_2 = M_0\,,\quad M_3 \simeq M_0 \,, \nonumber\\
\text{Scenario~B}: && M_1 = M_0 \lambda^8\ 2 R_0^4 \sqrt{1 + R_0^8 - 2 R_0^4 \cos (4 \alpha_0)}\,,\quad M_2 = M_0 \lambda^2\,,\quad M_3 \simeq M_0 \,. \nonumber
\end{eqnarray}
If we consider $M_1$ of $\mathcal{O}({\rm keV})$, then $M_0$ is about $10^6~{\rm keV}\sim 1~{\rm GeV}$ for Scenario~A, or about $10^8~{\rm keV}\sim 100~{\rm GeV}$ for Scenario~B. 

Fully democratic matrices in $SU(5)$ inspired FN models are, in general, not in agreement with the experimental constraints on the leptonic mixing angles, as has already been discussed in ref.~\cite{Sato:2000kj}. However, the agreement with the data strongly improves considering slightly non-democratic matrices. 
Considering, for example, the Dirac Yukawa couplings $Y_D^{12}=Y_D^{22}=\delta_0 y_D$, instead of $Y_D^{12}=Y_D^{22}=y_D$, is possible to increase the consistency with light neutrino data. In order to find fully working models, however, a numerical analysis is necessary, see ref.~\cite{Merle:2011yv} for details.

\subsubsection{The minimal radiative inverse seesaw mechanism (Authors: A.~Pilaftsis, B.~Dev)}

A technically natural realization for the seesaw scale to be in the TeV range is  the  so-called   Inverse  Seesaw  Model (ISM)~\cite{Mohapatra:1986aw, Nandi:1985uh, Mohapatra:1986bd}, where  in addition  to RH  neutrinos  $\{\nu_{R,I}\}$,  another  set of  SM singlet fermions  $\{S_{L,a}\}$ are introduced. After electroweak symmetry breaking, the neutrino Yukawa  sector of  a general ISM is described by the Lagrangian
\begin{eqnarray}
{\cal L}_Y &=& - \overline{\nu_L} m_D \nu_R - \overline{S_L} m_N \nu_R - \frac{1}{2} \overline{\nu^c_R} M_M \nu_R - \frac{1}{2} \overline{S_L} \mu_S S^c_L + h.c. \nonumber \\
\label{lagISM}
\end{eqnarray}
For a symmetric extension  of the SM  with three pairs of  singlet neutrinos, eq.~\eqref{lagISM} gives rise to the following $9\times 9$ neutrino mass matrix in the basis $\{(\nu_{L,\alpha})^c,\nu_{R, I},(S_{L,a})^c \}$: 
\begin{eqnarray}
{\cal M}_\nu\ =\ \left(\begin{array}{ccc}
0 & m_D & 0\\
m_D^T & M_M &  m_N^T\\
0 & m_N & \mu_S 
\end{array}\right)\; .
\label{eq:inverse1}
\end{eqnarray}
Here we have  not  included  in  ${\cal L}_Y$  the  dimension-four lepton-number  breaking term $\bar  l_L \Phi  S_L^c$ which  appears, for instance, in  linear seesaw models~\cite{Wyler:1982dd, Akhmedov:1995vm, Malinsky:2005bi},  since the resulting neutrino mass matrix in presence of this term can always be rotated to the form given  in~\eqref{eq:inverse1}~\cite{Ma:2009du}. In the limit  of $M_{M}, \mu_S \to 0$, lepton   number  symmetry   is  restored   and  the   light  neutrinos $\nu_{L,\alpha}$  are massless  to all  orders in  perturbation, whereas $\{\nu_{R,I},S_{L,a}\}$  form  three SM-singlet massive  Dirac   neutrinos.

In the original ISM presented in ref.~\cite{Mohapatra:1986bd},  the  RH neutrino Majorana mass is zero, $M_M = 0$  in~\eqref{eq:inverse1}.  In this  case, for  $|\mu_S|  \ll |m_N|, |m_D|$, it  is possible to  have very light sterile  neutrinos, e.g.\ in  the  presence  of  a  $\mu-\tau$  symmetry~\cite{Mohapatra:2005wk}  or  in theories with warped extra dimensions~\cite{Fong:2011xh}.

Another interesting realization of  ISM arises when $M_M\neq 0$ but  $\mu_S=0$~\cite{Dev:2012sg}. This may occur, for instance, in  models where the  Majorana masses  for the $S$-fields are forbidden due to specific flavor symmetries~\cite{Chun:1995bb, Barry:2011wb, Zhang:2011vh, Heeck:2012bz}. In this case,  the rank of the mass matrix given  by~\eqref{eq:inverse1}  reduces to six and  the  light neutrinos are {\it exactly} massless  at tree-level. However, they acquire a  non-zero mass  at the one-loop  level which is  {\em directly} proportional  to  the Majorana  mass  matrix $M_M$  and results from standard electroweak  radiative corrections involving  the neutral gauge and  Higgs  bosons~\cite{Pilaftsis:1991ug}.   This  scenario is known as  the  Minimal Radiative  Inverse  Seesaw  Model  (MRISM)~\cite{Dev:2012sg} which is very economical, since it does not require the existence of other non-standard scalar or gauge fields or other fermionic matter beyond  the singlet neutrinos~$\{\nu_{R, I},S_{L,a}\}$ necessary for a general ISM.

The possibility  of  having light  sterile neutrinos  in  the  MRISM can be realized in  the  limit $|M_M|\gg |m_D|,|m_N|$~\cite{Dev:2012bd}.  In this limit, the  $\nu_{R,I}$-fields decouple  below the  mass  scale $M_M$,  resulting  in an  effective theory   with   six   neutrino   states:  $\nu_{L, \alpha}$   and $S_{L,a}$.  At tree-level, the effective $6\times 6$ neutrino mass matrix in the weak basis becomes
\begin{eqnarray}
{\cal M}_{\rm eff}^{\rm tree} \ = \ \left(\begin{array}{c} 
m_D \\ m_N \end{array}\right) M_M^{-1} \left(\begin{array}{cc} 
m_D^T & m_N^T \end{array}\right) \ = \ \left(\begin{array}{cc} 
m_D M_M^{-1} m_D^T & m_D M_M^{-1}m_N^T \\
m_N M_M^{-1}m_D^T & m_N M_M^{-1}m_N^T
\end{array}\right)\; .
\label{eq:Meff}
\end{eqnarray}
Note  that  one  of  the  block eigenvalues  must  vanish,  since  the effective mass matrix ${\cal M}_{\rm eff}^{\rm tree}$ is of rank 3. 

At one-loop level, the mass  matrix  given  by eq.~\eqref{eq:Meff} receives   an  electroweak   radiative   correction  proportional   to $M_M$~\cite{Dev:2012sg}: 
\begin{eqnarray}
	M_{\nu_L}^{\rm 1-loop} &=&\frac{\alpha_w}{16\pi m_W^2}{\cal
          M}_D{\cal M}_S\left[m_H^2\left({\cal M}_S^2-m_H^2 \mathbb{1}\right)^{-1}\ln\left(\frac{ {\cal M}_S^2}{m_H^2}\right)\right.  \nonumber \\ 
&& \qquad \qquad \qquad \qquad \left. +\:   
	3m_Z^2\left({\cal M}_S^2-m_Z^2\mathbb{1}\right)^{-1}\ln\left(\frac{
          {\cal M}_S^2}{m_Z^2}\right)\right]{\cal M}_D^T\; , 
\label{eq:nuself}
\end{eqnarray} 
where   $\alpha_w\equiv   g^2/\left(4\pi\right)$  is   the   weak coupling strength, $m_{H,W,Z}$ are the masses of the SM Higgs, $W$ and $Z$ bosons respectively, and we have defined a  $3\times 6$ mass matrix ${\cal M}_D=(m_D,0)$  and a $6\times  6$ mass matrix 
	${\cal M}_S = \left(\begin{array}{cc}
M_M & m_N^T\\
m_N & \mu_S
\end{array}\right)$.
In    the   limit    $|M_M|   \gg
|m_D|,|m_N|$, eq.~\eqref{eq:nuself} simplifies to~\cite{Dev:2012bd} 
\begin{eqnarray}
{\cal M}_{\rm eff}^{1-{\rm loop}} &\simeq & 
\left(\begin{array}{cc} 
m_D M_M^{-1} x_Rf(x_R) m_D^T & 0 \\
0 & 0
\end{array}\right) \equiv \left(\begin{array}{cc} 
\Delta M & 0 \\
0 & 0
\end{array}\right) 
\; ,
\end{eqnarray}
where the one-loop function $f(x_R)$ is defined as 
\begin{eqnarray}
f(x_R) =
\frac{\alpha_W}{16\pi}\left[\frac{x_H}{x_R-x_H}\ln\left(\frac{x_R}{x_H}\right)
  + \frac{3x_Z}{x_R-x_Z}\ln\left(\frac{x_R}{x_Z}\right) \right] ,
\end{eqnarray} 
with   $x_R   \equiv\hat{M}_M^2/m_W^2$,   $x_H\equiv   m_H^2/m_W^2$, $x_Z\equiv m_Z^2/m_W^2$,  assuming $M_M =  \hat{M}_M \mathbb{1}$ for simplicity. 

Thus, the  full effective neutrino mass  matrix in the  basis $\{\nu_{L,\alpha},S_{L,a}\}$ is given by
\begin{eqnarray}  
  \label{eq:Meffl}
{\cal M}_{\rm eff} = {\cal M}_{\rm eff}^{\rm tree}+{\cal M}_{\rm
  eff}^{\rm 1-loop} = \left(\begin{array}{cc}  
m_D\, M_M^{-1}\Big(\mathbb{1}+x_Rf(x_R)\Big)\, m_D^T & 
m_D M_M^{-1}m_N^T \\[2mm] 
m_N M_M^{-1}m_D^T & m_NM_M^{-1}m_N^T 
\end{array}\right)\; .
\end{eqnarray}
This can be diagonalized by a general $6\times 6$ unitary matrix parametrized in terms of 15~Euler angles and 10~Dirac phases. Of  the resulting six light  mass eigenstates, three should describe  the mainly active  neutrinos and the  remaining three will  be  predominantly sterile states. In ref.~\cite{Dev:2012bd}, it was shown that one of the sterile states could be at the keV-scale, having very small mixing $\lesssim 10^{-5}$ with the active neutrinos,

thus serving as a potentially WDM candidate. Depending on the masses of the other two sterile states, one could consider several possibilities, e.g.\ (i) if both of them are in the eV range, possessing a non-zero mixing with the active states, one could explain the LSND+MiniBooNE+reactor neutrino data, or (ii) if one of the light sterile states is in the eV range and the second one is superlight and almost mass-degenerate with the solar neutrinos, it could also give rise to potentially observable effects in solar neutrino experiments~\cite{deHolanda:2003tx, deHolanda:2010am, Bakhti:2013ora}. Moreover, this scenario is more compatible with the cosmological constraints on the sum of the neutrino masses.

Thus, the MRISM provides a minimal extension of the SM accommodating neutrino masses and keV-scale sterile neutrino Dark Matter, while being consistent with {\em all} the existing experiments, including the anomalies in the neutrino oscillation data. In addition, the heavy singlet neutrinos in this model with a degenerate mass spectrum could be used to explain the observed matter-antimatter asymmetry in the Universe through the mechanism of resonant leptogenesis~\cite{Pilaftsis:2003gt, Pilaftsis:2005rv, Dev:2014laa}.

\subsubsection{Models based on loop-suppressions (Authors: D.~Borah, R.~Adhikari)}

Tiny neutrino masses can also arise at one-loop level within the class of the so-called ``scotogenic'' models first proposed by Ma~\cite{Ma:2006km}. These models can not only motivate tiny neutrino masses at one-loop level~\cite{Sierra:2008wj,Suematsu:2009ww,Ahn:2012cga}, but they can also give rise to at least one stable DM candidate~\cite{LopezHonorez:2006gr,Sierra:2008wj,Suematsu:2009ww,Dolle:2009fn,Gelmini:2009xd} -- and in general, the fact that the neutrino mass is generated at loop-level strongly affects their phenomenology~\cite{Kubo:2006yx,Sierra:2008wj,Adulpravitchai:2009gi,Adulpravitchai:2009re,Bouchand:2012dx,Toma:2013zsa,Vicente:2014wga,Merle:2015gea,Merle:2015ica}. An Abelian gauge extended version of the original scotogenic model was proposed in~\cite{Adhikari:2008uc} to explain tiny active netrino masses as well as DM. Suitable modifications of this model accommodating tiny sterile neutrino masses have appeared in~\cite{Borah:2013waa,Adhikari:2014nea,Adhikari:2015woo}. Models with dynamical symmetry breaking, loop suppression of neutrino masses, and a resultant low-scale seesaw have been studied in~\cite{Appelquist2002204,PhysRevLett.90.201801,PhysRevD.69.015002}.

\begin{figure}[htb]
\centering
\includegraphics[scale=0.5]{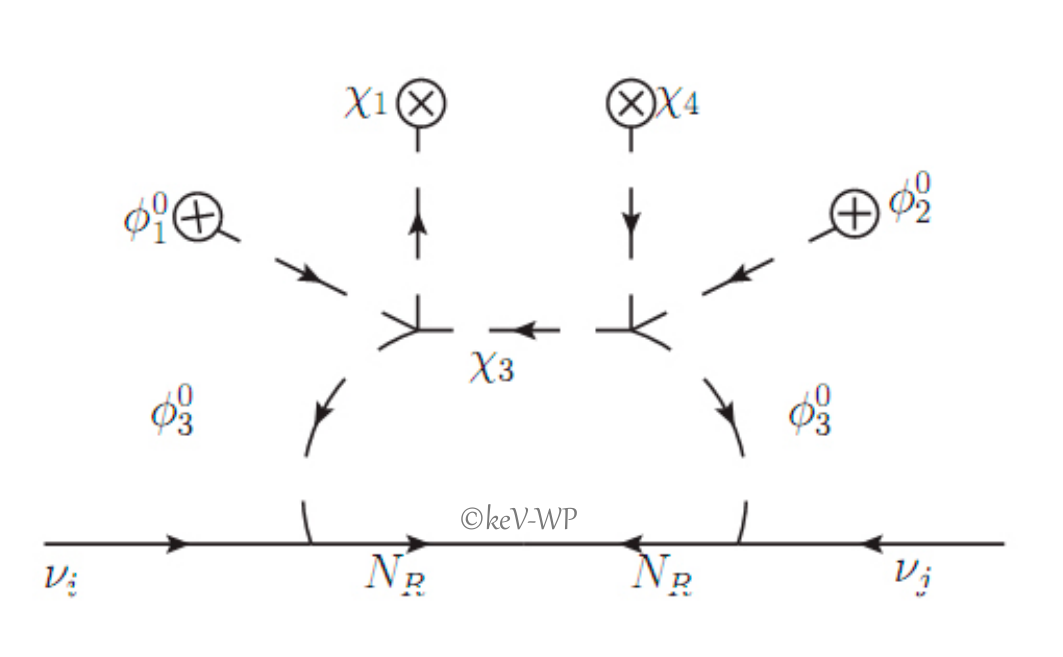}
\caption{One-loop contribution to active neutrino mass. The figure is inspired from ref.~\cite{Borah:2012qr}.}
\label{numass}
\end{figure}

The field content of the Abelian gauge model~\cite{Adhikari:2008uc} is shown in tab.~\ref{table1}. The third column shows the quantum numbers of the fields under a new $U(1)_X$ gauge symmetry. The fourth column indicates the transformations under the remnant $\mathbb{Z}_2$ symmetry, such that the lightest $\mathbb{Z}_2$-odd particle is stable. The combination of one singlet fermion $\nu_R$ and two triplet fermions $\Sigma_{1R}, \Sigma_{2R}$ with suitable Higgs scalars is chosen such that tiny neutrino masses arise at one-loop level. Two more singlets $S_{1R}, S_{2R}$ are required to be present in order to satisfy the anomaly freedom conditions. The Higgs doublets $\Phi_1, \Phi_2$ give masses to quarks and charged leptons, respectively. The Yukawa Lagrangian relevant for the neutrino mass is 
$$ \mathcal{L}_Y \supset y \bar{L} \Phi^{\dagger}_1 S_{1R} + h_N \bar{L} \Phi^{\dagger}_3 \nu_R + h_{\Sigma}  \bar{L}\Phi^{\dagger}_3 \Sigma_R + f_N \nu_R \nu_R \chi_4+ f_S S_{1R} S_{1R} \chi_1 $$
\begin{equation}
+ f_{\Sigma} \Sigma_R \Sigma_R \chi_4 + f_{NS} \nu_R S_{2R} \chi^{\dagger}_2 + f_{12} S_{1R} S_{2R} \chi^{\dagger}_3,
\label{Yukawa} 
\end{equation}
with VEVs $\langle \phi^0_{1,2} \rangle = v_{1,2}, \; \langle \chi^0_{1,4} \rangle  =u_{1,4}$. The tree-level neutrino mass from eq.~\eqref{Yukawa} is
\begin{equation}
m_{\nu 3} \approx \frac{ 2y^2 v_1^2}{f_S u_1},
\label{neutmass}
\end{equation}
while the other two active neutrinos are massless and only become massive at one-loop, see fig.~\ref{numass} and ref.~\cite{Borah:2012qr} for details. The sub-eV scale for active neutrino masses is generated naturally by adjusting the Yukawa couplings involved in fig.~\ref{numass}.

\begin{table}
\caption{\label{table1}Particle content of the model.}
\centering
\begin{tabular*}{\textwidth}{@{\extracolsep{\fill}} l c c c}
\hline
Particle & $SU(3)_c \times SU(2)_L \times U(1)_Y$ & $U(1)_X$ & $\mathbb{Z}_2$ \\
\hline\hline
$ (u,d)_L $ & $(3,2,\frac{1}{6})$ & $n_1$ & + \\
$ u_R $ & $(\bar{3},1,\frac{2}{3})$ & $\frac{1}{4}(7 n_1 -3 n_4)$ & + \\
$ d_R $ & $(\bar{3},1,-\frac{1}{3})$ & $\frac{1}{4} (n_1 +3 n_4)$ & +\\
$ (\nu, e)_L $ & $(1,2,-\frac{1}{2})$ & $n_4$ & + \\
$e_R$ & $(1,1,-1)$ & $\frac{1}{4} (-9 n_1 +5 n_4)$ & + \\
\hline
$\nu_R$ & $(1,1,0)$ & $\frac{3}{8}(3n_1+n_4)$ & - \\
$\Sigma_{1R,2R} $ & $(1,3,0)$ & $\frac{3}{8}(3n_1+n_4)$ & - \\
$ S_{1R}$ & $(1,1,0)$ & $\frac{1}{4}(3n_1+n_4)$ & + \\
$ S_{2R}$ & $(1,1,0)$ & $-\frac{5}{8}(3n_1+n_4)$ & - \\
\hline
$ (\phi^+,\phi^0)_1 $ & $(1,2,-\frac{1}{2})$ & $\frac{3}{4}(n_1-n_4)$ & + \\
$ (\phi^+,\phi^0)_2 $ & $(1,2,-\frac{1}{2})$& $\frac{1}{4}(9n_1-n_4)$ & + \\
$(\phi^+,\phi^0)_3 $ & $(1,2,-\frac{1}{2})$& $\frac{1}{8}(9n_1-5n_4)$ & - \\
\hline
$ \chi_1 $ & $(1,1,0)$ & $-\frac{1}{2}(3n_1+n_4)$ & + \\
$ \chi_2 $ & $(1,1,0)$ & $-\frac{1}{4}(3n_1+n_4)$ & + \\
$ \chi_3 $ & $(1,1,0)$ & $-\frac{3}{8}(3n_1+n_4)$ & - \\
$ \chi_4 $ & $(1,1,0)$ & $-\frac{3}{4}(3n_1+n_4)$ & + \\
\hline
\end{tabular*}
\end{table}

\begin{figure}[htb]
\hspace{-1cm}
$
\begin{array}{ccc}
\includegraphics[width=6cm]{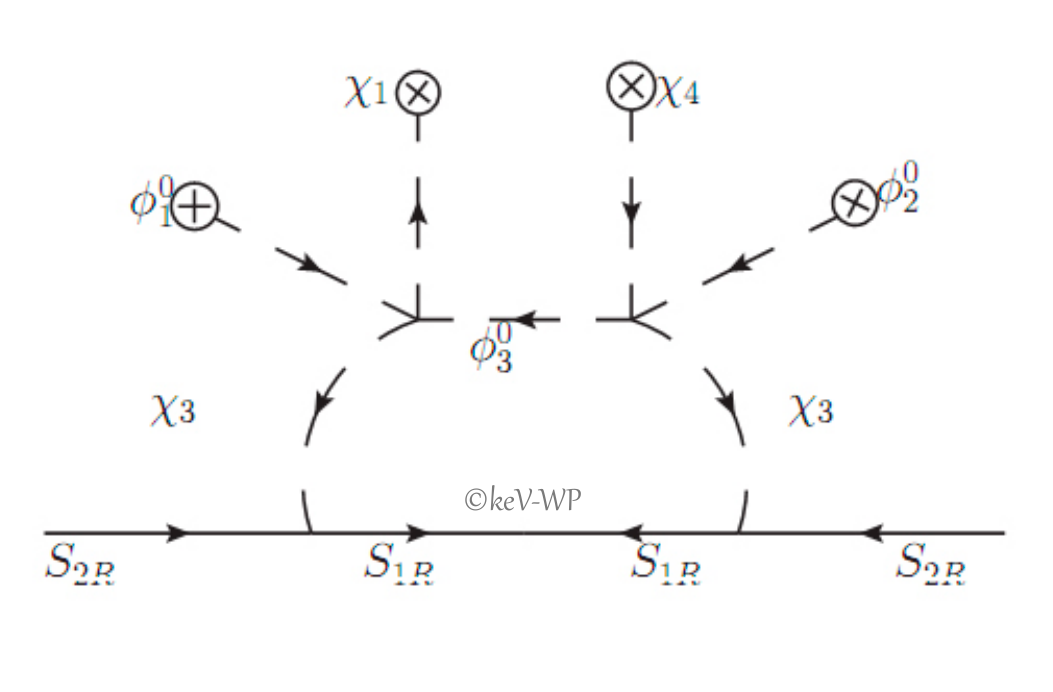} &
\includegraphics[width=6cm]{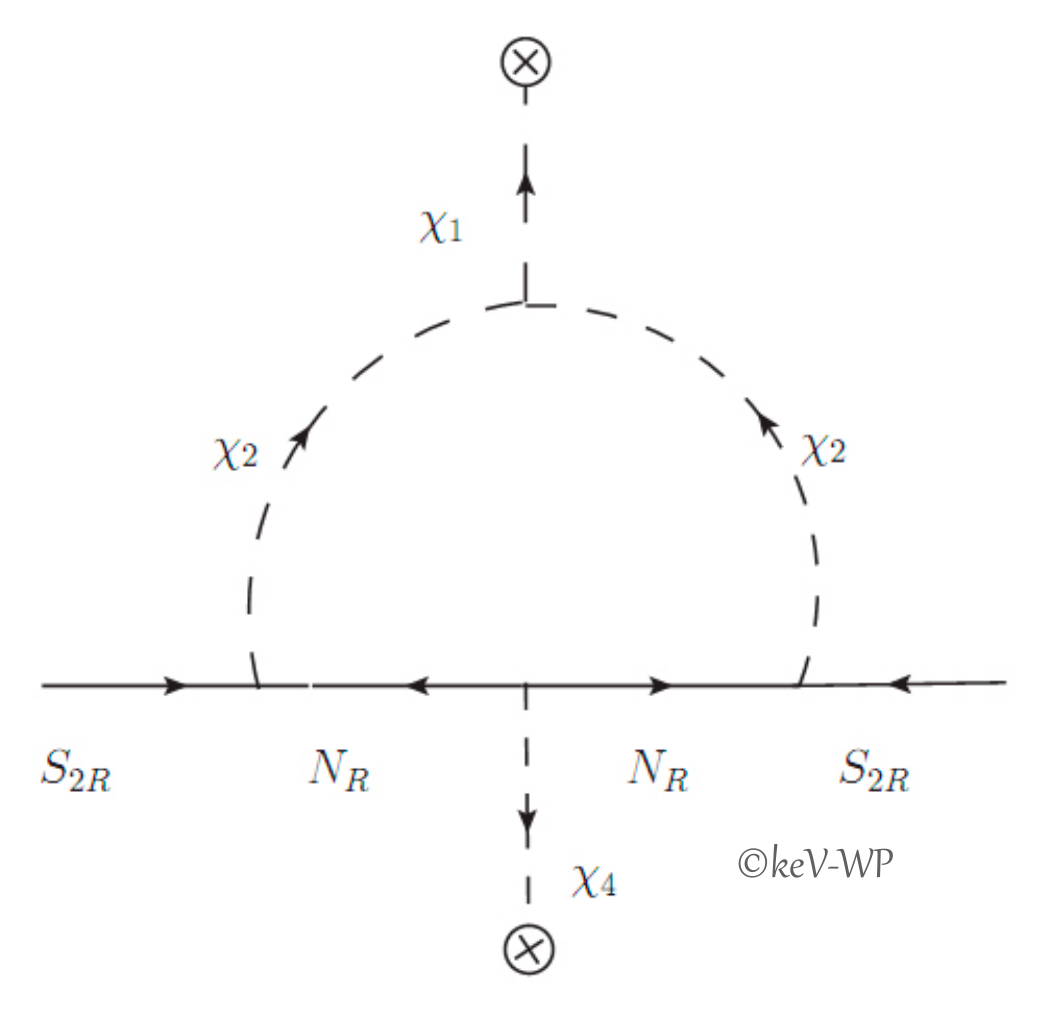} &
\includegraphics[width=5.5cm]{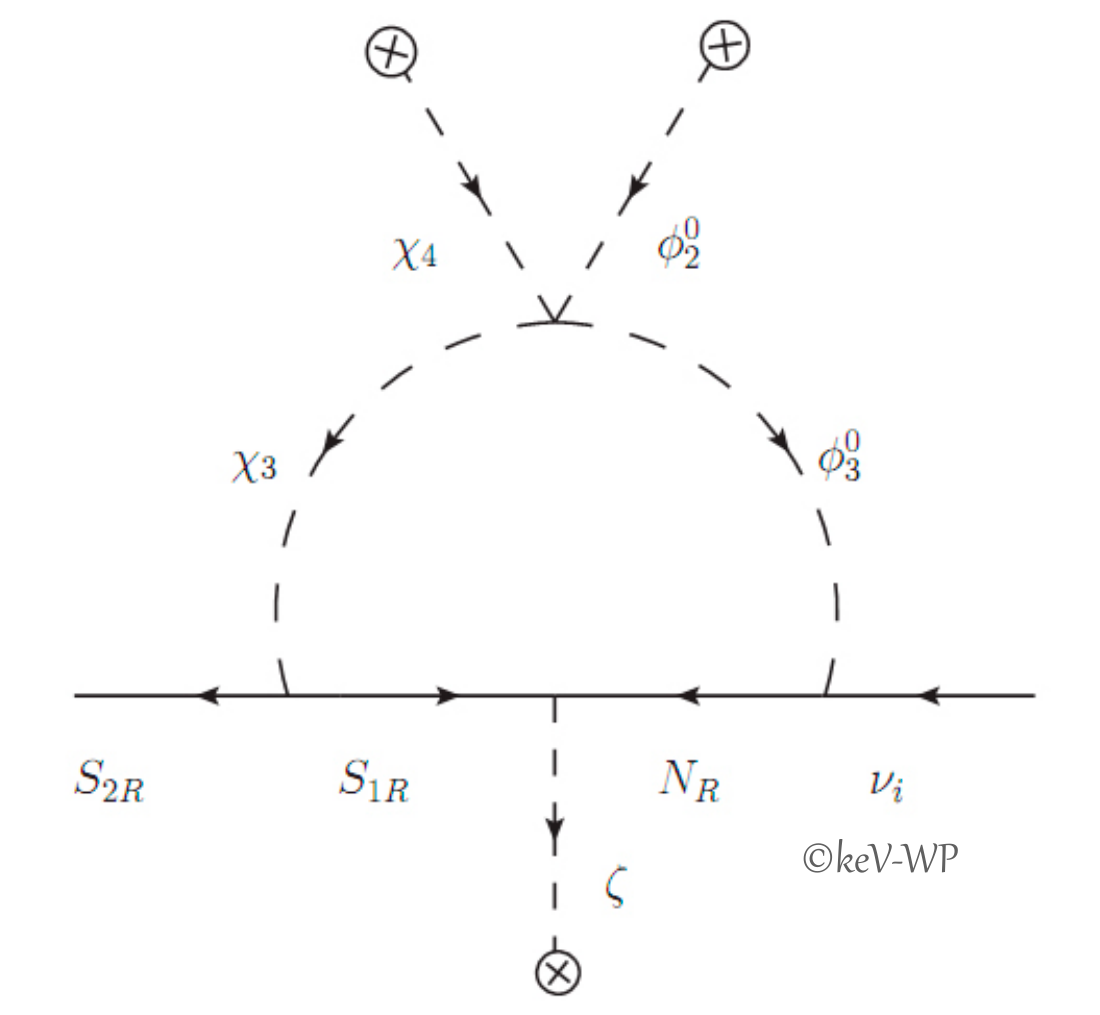}
\end{array}
$
\caption{One-loop contributions to the sterile neutrino mass (first and second diagram) and to active-sterile mixing (third diagram). 
Figures are inspired from~\cite{Borah:2013waa,Adhikari:2014nea}.}
\label{sterile1}
\end{figure}

The model shown in tab.~\ref{table1} has three sterile neutrinos $S_{1R}, S_{2R}, \nu_R$ out of which $S_{1R}$ couples to the active neutrinos and contributes to the tree-level mass term of $\nu_3$. Thus, for generic Dirac Yukawa couplings of the neutrinos, the sterile neutrino $S_{1R}$ is expected to be much heavier than the eV scale. As seen from the Lagrangian in eq.~\eqref{Yukawa}, $S_{1R}$ and $\nu_R$ acquire non-zero masses from the VEVs of $\chi_1$, $\chi_4$, respectively. Since the Abelian gauge symmetry is spontaneously broken down to the standard model by the VEVs of $\chi_1$, $\chi_4$ at a scale above the electroweak scale, one can take the VEVs to be at least at the TeV scale. Thus, for generic $\mathcal{O}(1)$ Yukawa couplings, the masses of $S_{1R}, \nu_R$ also lie at the TeV scale. Unlike for $S_{1R}$ and $\nu_R$, there is \emph{no} tree-level mass term for the singlet fermion $S_{2R}$. However, there are mixing terms of $S_{2R}$ with $\nu_R, S_{1R}$ through the Higgs fields $\chi_2$, $\chi_3$ respectively. If these Higgs fields do not acquire non-zero VEVs, then $S_{2R}$ remains massless at tree-level. Thus, the tree-level neutrino mass matrix in the basis $(\nu_e, \nu_{\mu}, \nu_{\tau}, S_{1R}, S_{2R}, \nu_R)$ is
\begin{equation}
M_f =
\left(\begin{array}{cccccc}
\ 0 & 0 & 0 & y_1v_1 & 0 & 0 \\
\ 0 & 0 & 0 & y_2v_2 & 0 & 0 \\
\ 0 & 0 & 0 & y_3 v_1 & 0 & 0 \\
\ y_1v_1 & y_2v_1 & y_3v_1 & f_S u_1 & 0 & 0 \\
\ 0 & 0 & 0 & 0 & 0 & 0 \\
\ 0 & 0 & 0 & 0 & 0 & f_N u_4
\end{array}\right).
\label{numassmatrix}
\end{equation}
The sterile neutrino $S_{2R}$, which remains massless at tree-level, can aquire a small eV scale mass at one-loop level from the diagrams shown in fig.~\ref{sterile1}, which corresponds to a non-zero $(M_f)_{55}$ element in the above basis. With suitable choices of the couplings and the VEVs, it is possible to get the desired loop-suppression for the mass of the sterile neutrino, no matter if eV or keV.

Apart from the mass for sterile neutrinos, one should also address the issue of their mixing with active neutrinos. In particular astrophysical observations can constrain the mixing of keV neutrinos with active neutrinos. In the original version of this model~\cite{Adhikari:2008uc} with the minimal field content, there is no mixing between active and sterile neutrinos. In the later work~\cite{Borah:2013waa}, the field content was extended by a new scalar singlet field $\zeta$ with $U(1)_X$ charge $\frac{5}{8}(3n_1+n_4)$, which allows for a tree-level mixing term of $N_R$ and $S_{1R}$ and also for one-loop mixing between active and sterile neutrinos, see the third diagram shown in fig.~\ref{sterile1}. 

In the original model~\cite{Adhikari:2008uc},  the singlet fermion $S_{2R}$, which was taken to be the light sterile neutrino in~\cite{Borah:2013waa}, is $\mathbb{Z}_2$-odd. The one-loop mass of this sterile neutrino can be adjusted at eV or keV scale, with a mixing that vanishes at leading order. However, allowing for mixing between the $\mathbb{Z}_2$-odd singlet fermion $S_{2R}$ with the $\mathbb{Z}_2$-even active neutrinos breaks the $\mathbb{Z}_2$ symmetry, making the sterile neutrino unstable. This motivates $S_{2R}$ as decaying keV sterile neutrino DM.

Finally, let us remark that further modifications of the model can lead to even richer phenomenology, such as multi-component DM with a mixed WDM/CDM setting~\cite{Adhikari:2014nea}.

\subsection{\label{sec:IV-breaking}Models based on symmetry breaking}

The next class of mechanisms are those based on some type of flavor symmetry. These symmetries can either be used to force one sterile neutrino to be massless at leading order --~which 
is then corrected at next-to-leading order to a finite value~-- or they can complement a suppression mechanism already present in the model by enforcing the mixing angle between active and sterile neutrinos to be small enough in order not to contradict observations.

\subsubsection{$L_e - L_\mu - L_\tau$ symmetry (Authors: A.~Merle, V.~Niro)}

In refs.~\cite{Shaposhnikov:2006nn,Lindner:2010wr,Lindner:2010wr_Erratum,Merle:2012ya}, a $L_e - L_\mu - L_\tau$ flavor symmetry was applied at the same time to active and sterile neutrinos. If the symmetry is unbroken, the model predicts a massless neutrino and two neutrinos with exactly the same mass, both in the active and in the sterile sectors. For this reason, the $L_e-L_\mu-L_\tau$ flavor symmetry should be softly broken. This leads to a non-zero mass for one of the active and one of the sterile neutrinos. The other four neutrinos (two active and two sterile), after the introduction of soft-breaking terms in the Lagrangian, are not anymore degenerate in mass. This model could naturally accomodate for a keV sterile neutrino in the  spectrum, since the soft symmetry breaking terms will be much smaller than the masses of the almost degenerate sterile neutrinos. Note that an important feature of this model is the prediction of a strongly hierarchical neutrino spectra. 

Let us denote the symmetry as $\mathcal{F}\equiv L_e-L_\mu-L_\tau$. The particle content and the charge assignments of our model are given in tab.~\ref{tab:table_LeLmuLtau}. We extended the model previously proposed in ref.~\cite{Lavoura:2000ci} by a Higgs triplet~$\Delta$, in order to accommodate a type~II seesaw, which is present, for example, in the context of a left-right symmetric framework, like the one presented in ref.~\cite{Bezrukov:2009th}, required also for a sufficient entropy production. Considering the limit of vanishing left-handed mass matrix and/or triplet Yukawa coupling, the model can also be applied to scenarios where the 
Higgs triplet is not present.

\begin{table}[t]
\caption{\label{tab:table_LeLmuLtau} $\mathcal{F}\equiv L_e-L_\mu-L_\tau$ charges, where $L_{\alpha L}=(\nu_{\alpha L}, \alpha_L)^T$, and $\alpha=e, \mu, \tau$. The scalars are defined as: $\phi=(\phi^+, \phi^0)^T$, $\Delta= ((\Delta^+/\sqrt{2}, \Delta^{++}), (\Delta^0, -\Delta^+/\sqrt{2}))$.}
 \centering
\begin{tabular*}{\textwidth}{@{\extracolsep{\fill}} l c c c c c c c c c c c }
\hline
\textbf{Fields}  & $L_{e L}$ & $L_{\mu L}$ & $L_{\tau L}$ & 
$e_R$ & $\mu_R$ & $\tau_R$ &
$\nu_{1 R}$ & $\nu_{2 R}$ & $\nu_{3 R}$ & 
$\phi$ & $\Delta$ 
\\\hline\hline
 $\mathcal{F}$ & $1$ & $-1$ & $-1$ &
$1$ & $-1$ & $-1$ &
$1$ & $-1$ & $-1$ &
$0$ & $0$   \\
\hline
 \end{tabular*}
\end{table}

The full neutrino mass term is given by 
\begin{equation}
\mathcal{L}_{\rm mass}=-\frac{1}{2} 
\overline{\Psi^C} \mathcal{M}_\nu \Psi +h.c.\,,\quad 
\text{with} \quad
\mathcal{M}_\nu=
\begin{pmatrix}
\begin{array}{c|c}
\begin{matrix}
0 & m^{e \mu}_L & m^{e \tau}_L \\
m^{e \mu}_L & 0 & 0 \\
m^{e \tau}_L & 0 & 0 
\end{matrix}
&
\begin{matrix}
m^{e 1}_D & 0 & 0 \\
0 & m^{\mu 2}_D & m^{\mu 3}_D \\
0 & m^{\tau 2}_D & m^{\tau 3}_D
\end{matrix}\\\hline
\begin{matrix}
m^{e 1}_D & 0 & 0 \\
0 & m^{\mu 2}_D & m^{\tau 2}_D \\
0 & m^{\mu 3}_D & m^{\tau 3}_D
\end{matrix}
&
\begin{matrix}
0 & M^{12}_M & M^{13}_M \\
M^{12}_M & 0 & 0 \\
M^{13}_M & 0 & 0
\end{matrix}
\end{array}
\end{pmatrix},
\label{eq:nu_mass_tot}
\end{equation}
and $\Psi \equiv \left((\nu_{e L})^C,(\nu_{\mu L})^C,(\nu_{\tau L})^C,\nu_{1 R},\nu_{2 R},\nu_{3 R}\right)^T$. We have defined $m^{\alpha i}_D \equiv v_\phi Y^{\alpha i}_D$ and $m^{\alpha \beta}_L \equiv v_\Delta Y^{\alpha \beta}_L$, with $v_\phi$ and $v_\Delta$ being the vacuum expectation values of the scalar doublet $\phi$ and the scalar triplet $\Delta$.

\begin{figure}[t]
\centering
\includegraphics[width=5.5cm]{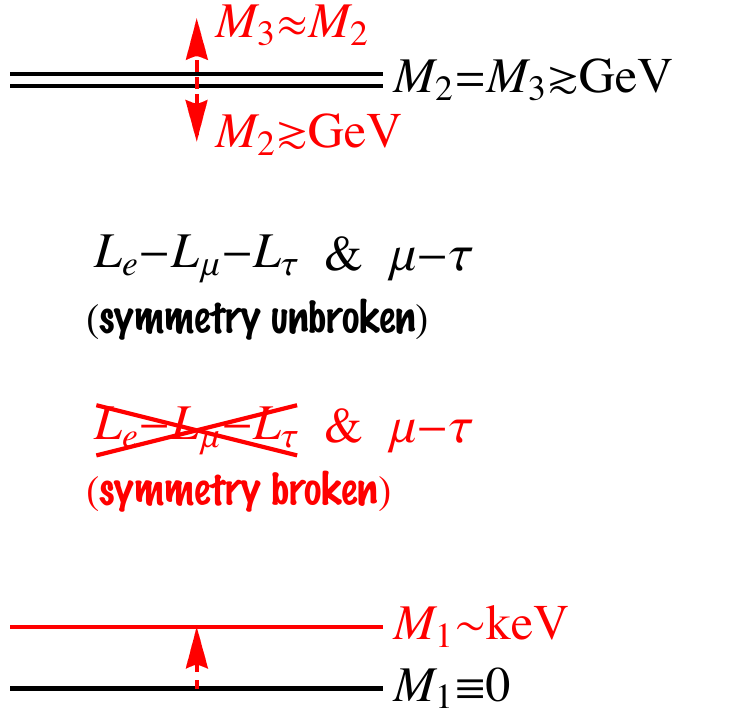}
\caption{\label{fig:schemesLeLmuLtau} 
The mass shifting schemes due to soft breaking of a global $L_e-L_\mu-L_\tau$ flavor symmetry. (Figure similar to fig.~6 of ref.~\cite{Merle:2013gea} and fig.~1 of ref~\cite{Merle:2011yv}.)}
\end{figure}

The exact symmetry predicts a bimaximal mixing in the neutrino sector:
\begin{equation}
 \mathcal{U}_\nu=
\begin{pmatrix}
\frac{1}{\sqrt{2}} & \frac{1}{\sqrt{2}} & 0\\
-\frac{1}{2} & \frac{1}{2} & \frac{1}{\sqrt{2}} \\
\frac{1}{2} & -\frac{1}{2} & \frac{1}{\sqrt{2}}
\end{pmatrix}\,.
 \label{eq:bimax}
\end{equation} 
This matrix predicts a zero $\theta_{13}$ and a maximal mixing angle $\theta_{12}$, which are both ruled out experimentally. However, it is still possible to obtain a PMNS matrix $\mathcal{U}_{\rm PMNS}= \mathcal{U}_L^\dag \mathcal{U}_\nu$ compatible with experimental data, depending on the actual form of the charged lepton mixing matrix $\mathcal{U}_L$. Considering CP conservation and a hierarchical relation between $\lambda_{ij}\equiv \sin \theta^\prime_{ij}$, of the type $\lambda_{12}=\lambda$, $\lambda_{13}\simeq\lambda^3$, $\lambda_{23}\simeq \lambda^2$, with $\lambda\simeq 0.20$ being the parameter that describes the deviation of $\theta_{12}$ from $\pi/4$~\cite{Frampton:2004ud}, we find
\begin{equation}
 \mathcal{U}_L=
\begin{pmatrix}
1-\lambda^2/2 & \lambda & \lambda^3\\
-\lambda & 1-\lambda^2/2 & \lambda^2\\
\lambda^3 & -\lambda^2 & 1
\end{pmatrix}+\mathcal{O}(\lambda^4)\,.
 \label{eq:UL}
\end{equation}
This form of $\mathcal{U}_L$, combined with the bimaximal matrix from eq.~\eqref{eq:bimax}, could lead to a $\mathcal{U}_{\rm PMNS}$ matrix compatible with the experimental values~\cite{Lindner:2010wr,Lindner:2010wr_Erratum}, as it was also shown in ref.~\cite{Frampton:2004ud}. The matrix $\mathcal{U}_L$ is associated to the diagonalization of the matrix
\begin{equation}
 \mathcal{M}_l \mathcal{M}^\dagger_l \simeq
\begin{pmatrix}
m^2_e+m^2_\mu\lambda^2 & m^2_\mu\lambda & 0\\
m^2_\mu \lambda & m^2_\mu & 0\\
0 & 0 & m^2_\tau
\end{pmatrix}\,+ \mathcal{O}(\lambda^3)\,,
\label{eq:Mlsymmetry}
\end{equation}
with $\mathcal{M}_l$ being the charged leptons mass matrix. 

As stated before, in this scenario we need soft breaking terms to avoid the presence of two zero eigenvalues. For this reason, we add to the neutrino mass in eq.~\eqref{eq:bimax}, a diagonal part containing symmetry-breaking terms: 
\begin{equation}
M_\nu\rightarrow
M_\nu + {\rm diag}(s^{e e}_L, s^{\mu \mu}_L, s^{\tau \tau}_L, S^{11}_R, S^{22}_R, S^{33}_R)\,.
\end{equation}
In the case in which $m^{\alpha \beta}_L \ll m^{\alpha I}_D \ll M^{I J}_M$ we can block-diagonalize the matrix $\mathcal{M}_\nu$, like in the usual type II scenario. Considering an additional unbroken  extra $\mu-\tau$ symmetry~\cite{Mohapatra:2001ns}, we have $M^{12}_M\simeq M^{13}_M \sim M_M$  and $m_L^{e \mu} \simeq m_L^{e \tau} \sim m_L$, and setting, for simplicity, $m_D^{\alpha I} \simeq m_D$, $s^{\alpha \alpha}_L \simeq s$, $S^{I I}_R \simeq S$, we obtain the eigenvalues $\mathcal{E}'=\{ \lambda^\prime_+, \lambda^\prime_-, \lambda_s, \Lambda^\prime_+, \Lambda^\prime_-, \Lambda_s \}$, with $\lambda^\prime_\pm \simeq \pm \sqrt{2} \left[ m_L - \frac{m_D^2}{M_M} \right] $, $\lambda_s=s$, $\Lambda^\prime_\pm=S\pm\sqrt{2} M_M$, and $\Lambda_s=S$. In this case, we could explain the presence of one keV sterile neutrino ($S \simeq$ keV) and two heavier sterile neutrinos ($M_M \gg S$) and the neutrino mixing matrix, even in the presence of soft-breaking terms, results compatible with the bimaximal form. 

The form of the mass matrix $\mathcal{M}_l$ for the charged leptons will also experience restrictions by the $\mathcal{F}$-symmetry. We can consider soft breaking terms, $s^{\alpha I}_D=v_\phi Y^{\alpha I}_D$ with $\alpha=e,\mu,\tau$ and $I=1,2,3$, and assume, for simplicity, a symmetric form of the charged lepton mass matrix ($s^{\tau e}_D=s^{e \tau}_D$, $s^{\mu e}_D=s^{e \mu}_D$, and $m^{\tau \mu}_D=m^{\mu \tau}_D$). Under the assumptions of small $s^{e \tau}_D$ and $m^{\mu \tau}_D$, we obtain 
\begin{equation}
\mathcal{M}_l \mathcal{M}^\dagger_l=
\begin{pmatrix}
\left(m^{e e}_D\right)^2 + \left(s^{e \mu}_D\right)^2 & s^{e \mu}_D\left(m^{e e}_D+m^{\mu \mu}_D\right) & 0 \\
s^{e \mu}_D\left(m^{e e}_D+m^{\mu \mu}_D\right) & \left(m^{\mu \mu}_D\right)^2 + \left(s^{e \mu}_D\right)^2 & 0 \\
0 & 0 & \left(m^{\tau \tau}_D\right)^2
\end{pmatrix}.
\end{equation}
If we identify $s^{e \mu}_{D}=m^{\mu \mu}_D \lambda$, $m^{e e}_{D}=m_e$, $m^{\mu \mu}_{D}=m_\mu$, and $m^{\tau \tau}_{D}=m_\tau$, we get a matrix similar to eq.~\eqref{eq:Mlsymmetry} and a charged lepton mixing matrix like the one in eq.~\eqref{eq:UL}. Thus, in the case of softly broken $\mathcal{F}$-symmetry, it is possible to obtain a PMNS matrix that is compatible with experimental data~\cite{Lindner:2010wr,Lindner:2010wr_Erratum}. The model based on a $\mathcal{F}-$symmetry is then a successful model that can explain a hierarchical spectrum in the sterile neutrino sector and, at the same time, reproduce the light neutrino experimental data and even give testable predictions in the form of neutrino mass sum rules~\cite{King:2013psa,Agostini:2015dna,Gehrlein:2015ena}. 

As a final note, the mixing angle in this type of models will scale with the parameter $m^{e 1}_D$ in the mass matrix, devided by the keV-scale mass. Thus, the X-ray bound for sterile neutrino DM does pose a constraint on $m^{e 1}_D$.

\subsubsection{\label{sec:Q6}$Q_6$ symmetry (Author: T.~Araki)}

We address the origin of the small active-sterile mixing and the mass hierarchy of the sterile neutrinos  ($M_1 \ll M_{2,3}$) by exploiting a non-Abelian discrete flavor symmetry. We also demand that masses of the heavier sterile neutrinos are quasi-degenerate ($M_2 \simeq M_3$), which may be desirable for successful baryogenesis. The mass degeneracy of the heaver sterile neutrinos could easily be realized by embedding them into a doublet representation of a given flavor symmetry, while it may be reasonable to assign a singlet representation to the lightest one. Furthermore, if the singlet representation is complex, one can prohibit a bare mass term for the lightest sterile neutrino because of its Majorana nature. Then, a small $M_1$ and a slight mass difference between $M_2$ and $M_3$ would arise after the flavor symmetry is spontaneously broken. In view of these facts, we employ $Q_6$ as our flavor symmetry since it is the smallest discrete group which contains both complex singlet and a real doublet representations.

$Q_6$ consists of four singlet and two doublet irreducible representations: ${\bf 1}, {\bf 1}^{'}, {\bf 1}^{''},{\bf 1}^{'''},{\bf 2}, {\bf 2}^{'}$, 
where ${\bf 1}$, ${\bf 1}^{'}$ and ${\bf 2}^{'}$ are real, while ${\bf 2}$ and ${\bf 1}^{'''} = ({\bf 1}^{''})^*$ are complex. The multiplication rules of the irreducible representations are summarized in tab.~\ref{q6}. Especially, those of the doublets are defined as follows:
\begin{eqnarray}
\begin{array}{ccccccccc}
\left(\begin{array}{c} x_1 \\ x_2 \end{array}\right) & \otimes & \left(\begin{array}{c} y_1 \\ y_2 \end{array}\right) & = &
(x_1y_2-x_2y_1) & \oplus & (x_1y_2+x_2y_1) & \oplus &
\left(\begin{array}{c} x_1y_1 \\ -x_2y_2 \end{array}\right), \\
{\bf 2} & \otimes & {\bf 2} & = & {\bf 1} & \oplus &
{\bf 1}^{'} & \oplus & {\bf 2}^{'} \\
&&&&&&&& \\
\left(\begin{array}{c} x_1 \\ x_2 \end{array}\right) & \otimes & \left(\begin{array}{c} y_1 \\ y_2 \end{array}\right) & = &
(x_1y_2+x_2y_1) & \oplus & (x_1y_2-x_2y_1) & \oplus &
\left(\begin{array}{c} x_2y_2 \\ x_1y_1 \end{array}\right), \\
{\bf 2}^{'} & \otimes & {\bf 2}^{'} & = & {\bf 1} & \oplus &
{\bf 1}^{'} & \oplus & {\bf 2}^{'}\\
&&&&&&&& \\
\left(\begin{array}{c} x_1 \\ x_2 \end{array}\right) & \otimes & \left(\begin{array}{c} y_1 \\ y_2 \end{array}\right) & = &
(x_1y_1-x_2y_2) & \oplus & (x_1y_1+x_2y_2) & \oplus &
\left(\begin{array}{c} x_2y_1 \\ x_1y_2 \end{array}\right). \\
{\bf 2} & \otimes & {\bf 2}^{'} & = & {\bf 1}^{''} & \oplus & {\bf 1}^{'''} & \oplus & {\bf 2}
\end{array}
\end{eqnarray}
The complex conjugation of ${\bf 2}$, i.e.\ ${\bf 2}^*$, is obtained by $i\tau_2 {\bf 2}$, where $\tau_2$ is the second Pauli matrix.

\begin{table}
\caption{The multiplication rules of $Q_6$.}
\begin{center}
\begin{tabular}{|l||c|c|c|c|c|}\hline
$Q_6$        & ${\bf 1}^{'}$   & ${\bf 1}^{''}$  &
${\bf 1}^{'''}$ & ${\bf 2}$  & ${\bf 2}^{'}$ \\ \hline\hline
${\bf 1}^{'}$   & ${\bf 1}$ & ${\bf 1}^{'''}$ & ${\bf 1}^{''}$  & ${\bf 2}$  & ${\bf 2}^{'}$ \\ \hline
${\bf 1}^{''}$  & ${\bf 1}^{'''}$ & ${\bf 1}^{'}$   & ${\bf 1}$ & ${\bf 2}^{'}$   & ${\bf 2}$ \\ \hline
${\bf 1}^{'''}$ & ${\bf 1}^{''}$ & ${\bf 1}$ & ${\bf 1}^{'}$   & ${\bf 2}^{'}$   & ${\bf 2}$ \\ \hline
${\bf 2}$  & ${\bf 2}$ & ${\bf 2}^{'}$ & ${\bf 2}^{'}$ & ${\bf 1} \oplus {\bf 1}^{'} \oplus {\bf 2}^{'}$ & ${\bf 1}^{''} \oplus {\bf 1}^{'''} \oplus {\bf 2}$ \\ \hline 
${\bf 2}^{'}$   & ${\bf 2}^{'}$ &${\bf 2}$ & ${\bf 2}$ & ${\bf 1}^{''} \oplus {\bf 1}^{'''} \oplus {\bf 2}$ & ${\bf 1} \oplus {\bf 1}^{'} \oplus {\bf 2}^{'}$ \\
\hline
\end{tabular}

\label{q6}
\end{center}
\end{table}

We demonstrate how the mass hierarchy and the mass degeneracy can be realized in the presence of $Q_6$ with a simple example. The $Q_6$ representations are assigned as 
\begin{eqnarray}
l_L ~\sim~ {\bf 1},~~~
\nu_{R,1} ~\sim~ {\bf 1}^{'''},~~~
\nu_{R,D}=(\nu_{R,2}~\nu_{R,3})  ~\sim~ {\bf 2}^{'},~~~
\Phi ~\sim~ {\bf 1},~~~
D ~\sim~ {\bf 2},
\nonumber
\end{eqnarray}
so that a mass term  of $\overline{\nu^c_{R,1}} \nu_{R,1}$ is not allowed, and $\overline{\nu^c_{R,D}} \nu_{R,D}$ results in degenerate masses. The left-handed lepton doublet, $l_L$, and the SM Higgs doublet, $\Phi$, are assumed to be ${\bf 1}$; therefore, the charged lepton sector is the same as that of the SM. In order to spontaneously break the $Q_6$ symmetry, we introduce a $Q_6$ doublet scalar $D$ which is SM gauge singlet, and we also take into account non-renormalizable effective operators with $D$. Then, the Dirac and Majorana mass terms are given by 
\begin{eqnarray}
&&-{\cal L}_D = 
\frac{f_\alpha}{\Lambda^2} \overline{l_{L,\alpha}} \nu_{R,D} \tilde{\Phi}DD
+\frac{\tilde{f_\alpha}}{\Lambda^2} \overline{l_{L,\alpha}} \nu_{R,D} \tilde{\Phi}DD^*
\nonumber \\
&&\hspace{3cm}
+\frac{F_{\alpha 1}}{\Lambda^3} \overline{l_{L,\alpha}} \nu_{R,1} \tilde{\Phi}DDD
+\cdots
+h.c.~, \\
&&-{\cal L}_M = 
\frac{1}{2} \left(
 M_d \overline{\nu^c_{R,D}} \nu_{R,D}
+ \frac{M_a}{\Lambda}\overline{\nu^c_{R,1}} \nu_{R,D} D
\right. 
\nonumber \\
&&\hspace{3cm}
\left.
+2\frac{M_b}{\Lambda^2}\overline{\nu^c_{R,D}} \nu_{R,D} DD
+ \frac{M_c}{\Lambda^2}\overline{\nu^c_{R,1}} \nu_{R,1}  DD
+\cdots
+h.c.
\right),
\end{eqnarray}
respectively, where $\alpha=e,\mu,\tau$, $\Lambda$ stands for the $Q_6$ symmetry breaking scale, which is supposed to be very high, and $\cdots$ represents higher order terms. Note that we have omitted several terms whose contributions can be absorbed into other terms and implicitly assumed a mechanism which provides $M_{a,b,c,d} = {\cal O}(10)$ GeV, e.g., spontaneous breaking of a lepton number symmetry at the GeV scale. After $D$ (and $\Phi$) acquire vacuum expectation values, $\langle D \rangle = (d_1~d_2)^T$, the Dirac and Majorana mass matrices are obtained as
\begin{eqnarray}
&&m_D=
v\lambda^2
\left(\begin{array}{ccc}
0 & F_{e2} & F_{e3} \\
0 & F_{\mu 2} & F_{\mu 3} \\
0 & F_{\tau 2} & F_{\tau 3}
\end{array}\right)
+v\lambda^3
\left(\begin{array}{ccc}
F_{e1} & 0 & 0 \\
F_{\mu 1} &  0 & 0\\
F_{\tau 1} &  0 & 0
\end{array}\right),
\\
&&M_M=
\left(\begin{array}{ccc}
0 & 0 & 0 \\
0 & 0 & M_d \\
0 & M_d & 0
\end{array}\right)
+\lambda
\left(\begin{array}{ccc}
0 & M_a & -M_a \\
M_a & 0 & 0 \\
-M_a & 0 & 0
\end{array}\right)
+\lambda^2
\left(\begin{array}{ccc}
2M_c & 0 & 0 \\
0 & M_b & 0 \\
0 & 0 & -M_b
\end{array}\right).
\end{eqnarray}
In the above equations, we assume $d_1 = d_2=d$ for simplicity and define $\lambda=d/\Lambda$, $F_{\alpha 2}=-f_\alpha + \tilde{f}_\alpha$, and $F_{\alpha 3}=f_\alpha + \tilde{f}_\alpha$. Given $\lambda \ll 1$, the masses of the sterile neutrinos are found to be
\begin{eqnarray}
M_1 \simeq 2\lambda^2 \left| M_c + \frac{M_a^2}{M_d} \right|,~~~
M_{2,3} \simeq \left| M_d \pm \lambda^2 M_b \right|.
\end{eqnarray}
As can be seen, $M_1$ is suppressed by $\lambda^2$ in comparison with $M_{2,3}$, and nearly degenerate $M_2$ and $M_3$ are realized. Moreover, very small $(m_D)_{\alpha 1}$ are obtained, which may result in a small active-sterile mixing. 

Suppose $M_1 = {\cal O}(1)$ keV and $M_{2,3}={\cal O}(10)$ GeV, then $\lambda \simeq10^{-3.5}$ is required, leading to $|M_2- M_3|={\cal O}(10^{-7})$ GeV. The mass scale of $M_{2,3}$ is chosen to be suitable for baryogenesis via flavor oscillation~\cite{Asaka:2005pn,Asaka:2010kk,Canetti:2012kh}. Interestingly, $|M_2- M_3|={\cal O}(10^{-7})$ GeV coincides with the mass difference obtained in ref.~\cite{Canetti:2012kh}. Thanks to the suppression by $\lambda$, the active neutrino masses, $m_\nu = {\cal O}(10^{-11})$ GeV, are reproduced with $F_{\alpha 2},F_{\alpha 3}={\cal O}(10^{-1}-1)$ even for $M_{2,3}={\cal O}(10~{\rm GeV})$. Lastly, a small active-sterile mixing
\begin{eqnarray}
\theta_{\alpha 1} 
= \frac{(m_D)_{\alpha 1}}{M_1}
= \frac{v \lambda^3 F_{\alpha 1}}{M_1} 
\simeq 10^{-5}-10^{-4}, 
\end{eqnarray}
is successfully realized by demanding $F_{\alpha 1}= {\cal O}(10^{-2.5}-10^{-1.5})$.

Last of all, let us comment on flavor mixing of the active neutrinos. $Q_6$ includes $\mu - \tau$ symmetry as subgroup, and one can exploit it to derive the observed large atmospheric and small reactor mixing angles~\cite{Araki:2011zg}.

\subsubsection{\label{sec:A4}$A_4$ symmetry (Author: A.~Merle)}

An alternative route to pursue with flavor symmetries is to use them to explain only the correct active and active-sterile mixing patterns, while using a different mechanism for the mass suppression. This means that, contrary to the approach taken in sec.~\ref{sec:Q6}, one could take an existing mass suppression mechanism and supplement it by a flavor symmetry. 
Here we focus on an example present in the literature~\cite{Barry:2011fp,Barry:2011wb} which uses the FN mechanism~\cite{Froggatt:1978nt}, along with an auxiliary $A_4$ discrete symmetry. 
The FN mechanism has been shown in sec.~\ref{sec:FN} to be able to generate the desired mass patterns. The whole course from the group theory of $A_4$ over developing concrete models 
down to experimental tests is outlined for example in ref.~\cite{King:2014nza}. Here, we will simply focus on how to use this symmetry as an auxiliary tool to tweak a setting with a ``pre-existing'' mass suppression mechanism in such a way that a phenomenologically acceptable model is obtained.

The symmetry group $A_4$ is a subgroup of $SU(3)$~\cite{Merle:2011vy}, and it consists of twelve different elements and it admits four irreducible representations: ${\bf 1}$, ${\bf 1}^{'}$, ${\bf 1}^{''}$, and ${\bf 3}$. The group has two generators $S$ and $T$ which obey the relations $S^2 = T^3 = (S T)^3 = \mathbb{1}$~\cite{King:2014nza}. The model presented here already contains a suitable mass pattern generated purely by the FN mechanism, with one light sterile neutrino $\nu_{R,1}$ at the keV scale (with FN charge $F_1$), while two sterile (right-handed) neutrinos $\nu_{R,2/3}$ with FN charges $F_{2,3}$ are considerably heavier (GeV or higher). This model needs two Higgs doublets $H_{u,d}$, too, as well as seven flavons $\varphi$, $\varphi'$, $\varphi''$, $\xi$, $\xi'$, $\xi''$, and $\Theta$. 
Among these fields, $\xi$ obtains a VEV $u=\langle \xi \rangle$ which generates the lightest sterile neutrino mass according to $M_1 \propto u \lambda^{2 F_1}$, where $u \simeq 10^{12}$~GeV and $\lambda \sim 0.22$, such that the choice $F_1 = 9$ seems appropriate to get a keV scale mass for $\nu_{R,1}$.

The most general Lagrangian at leading order in the family symmetry breaking scale, using the charge assignment of tab.~\ref{tab:A4BRZ}, is given by
\begin{eqnarray}
 \mathcal{L}_Y &=&  -\frac{y_e}{\Lambda}\lambda^3 \overline{e_R} \left(\varphi H_d L \right)_{\mathbf{1}} - \frac{y_\mu}{\Lambda} \lambda \overline{\mu_R} \left(\varphi H_d L \right)_{\mathbf{1'}} - \frac{y_\tau}{\Lambda} \overline{\tau_R} \left( \varphi H_d L \right)_{\mathbf{1''}} \label{eq:BRZ_lag} \\
 && -  \frac{y_1}{\Lambda}\lambda^{F_1} \overline{\nu_{R,1}} (\varphi H_u L)_{\mathbf{1}} - \frac{y_2}{\Lambda}\lambda^{F_2} \overline{\nu_{R,2}} (\varphi' H_u L)_{\mathbf{1''}} - \frac{y_3}{\Lambda}\lambda^{F_3} \overline{\nu_{R,3}} (\varphi'' H_u L)_{\mathbf{1}} \nonumber \\
 &&- \frac{1}{2} \left[w_1^* \lambda^{2 F_1}\xi^* \overline{(\nu_{R,1})^c} \nu_{R,1} - w_2^* \lambda^{2 F_2} {\xi'}^* \overline{(\nu_{R,2})^c} \nu_{R,2} - w_3^* \lambda^{2 F_3} {\xi''}^* \overline{(\nu_{R,3})^c} \nu_{R,3}
 \right] + h.c., \nonumber
\end{eqnarray}
where the notation $(...)_{\bf x}$ indicates the irreducible $A_4$ representation ${\bf x}$ to which the respective terms belong to and $\Lambda$ denotes a high scale. With a VEV $u=\langle \xi \rangle$, the first term in the last line indeed leads to a keV neutrino mass of
\begin{equation}
 M_1 = w_1 u \lambda^{2 F_1}.
 \label{eq:keV-mass_A4}
\end{equation}
Looking again at tab.~\ref{tab:A4BRZ}, one can see that the field $\nu_{R,1}$ is, in fact, a total singlet under all symmetries but the $U(1)_{\rm FN}$. Thus, as anticipated, its mass is exclusively determined by the FN symmetry, except for the rather indirect influence of the VEV $u$ of $\xi$. 

What is then the actual use of the $A_4$ symmetry? This can be seen by considering the active neutrino mixing. Illustrating the case of normal mass ordering, one can use the alignments $\langle \varphi' \rangle = (v'_\varphi,v'_\varphi,v'_\varphi)$ and $\langle \varphi'' \rangle = (0,v''_\varphi,-v''_\varphi)$ to obtain a phenomenologically viable $5\times 5$ neutrino mass matrix.\footnote{One can show that the keV neutrino practically decouples, in the sense that its presence does hardly affect the masses or mixings of the other neutrino mass eigenstates, so that it is sufficient to take into account $\nu_{2,3 R}$.} Calculating the corresponding leptonic mixing to next-to-leading order, one obtains a mixing matrix given by
\begin{equation}
 U_\nu^{({\rm NO})} \simeq
 \begin{pmatrix}
 \frac{2}{\sqrt{6}} & \frac{1}{\sqrt{3}} & 0 & 0 & 0 \\
 -\frac{1}{\sqrt{6}} & \frac{1}{\sqrt{3}} & -\frac{1}{\sqrt{2}} & 0 & 0 \\
 -\frac{1}{\sqrt{6}} & \frac{1}{\sqrt{3}} & \frac{1}{\sqrt{2}} & 0 & 0 \\ 
 0 & 0 & 0 & 1 & 0 \\ 
 0 & 0 & 0 & 0 & 1 
 \end{pmatrix} + 
 \begin{pmatrix} 
 0 & 0 & 0 & \epsilon_1 & 0\\ 
 0 & 0 & 0 & \epsilon_1 & -\epsilon_2 \\ 
 0 & 0 & 0 & \epsilon_1 & \epsilon_2 \\ 
 0 & -\sqrt{3} \epsilon_1 & 0 & 0 & 0\\ 
 0 & 0 & -\sqrt{2}\epsilon_2 & 0 & 0
 \end{pmatrix} + \mathcal{O}(\epsilon_i^2),
 \label{eq:mix_BRZ}
\end{equation}
where $\epsilon_{1,2}$ are small parameters arising only at higher order which features the other two FN charges $F_{2,3}$. It is thus possible to see the actual effect of the $A_4$ symmetry: it leads to phenomenologically acceptable leptonic mixing. While the leading order terms generate a tri-bimaximal mixing matrix, which in particular features $\theta_{13} \equiv 0$ that is excluded as well as maximal $\theta_{23}$, the terms arising at next to leading order correct the mixing pattern such that a non-zero $\theta_{13}$, which can indeed arise from the sterile neutrino sector~\cite{Merle:2014eja}, and a deviation of $\theta_{23}$ from $\pi/4$ are obtained at the same time. Thus, while the smallness of the keV neutrino mass is achieved by a completely separated sector, the $A_4$ symmetry adds another important ingredient.

\begin{table}[t]
\caption{\label{tab:A4BRZ}Particle content and representation/charge assignments model from ref.~\cite{Barry:2011fp}. In addition to $A_4$ and the FN symmetry, another auxiliary symmetry $\mathbb{Z}_3$ is used in order for CP violation to survive. 
For $SU(2)_L$ and $A_4$ the representations are reported, while for $\mathbb{Z}_3$ and $U(1)_{\rm FN}$ the charge is given, where $\omega = e^{2\pi i /3}$.}
\begin{tabular}{|l|ccccccccccccccc|} \hline
Field & $L_{1,2,3}$ & $\overline{e_R}$ & $\overline{\mu_R}$ & $\overline{\tau_R}$ & $\overline{\nu_{R,1}}$ & $\overline{\nu_{R,2}}$ & $\overline{\nu_{R,3}}$ & $H_{u,d}$ & $\varphi$ & $\varphi'$ & $\varphi''$ & $\xi$ & $\xi'$ & $\xi''$ & $\Theta$ \\ \hline \hline
$SU(2)_L$ & $\mathbf{2}$ & $\mathbf{1}$ & $\mathbf{1}$ & $\mathbf{1}$ & $\mathbf{1}$ & $\mathbf{1}$ & $\mathbf{1}$ & $\mathbf{2}$ & $\mathbf{1}$ & $\mathbf{1}$ & $\mathbf{1}$ & $\mathbf{1}$ & $\mathbf{1}$ & $\mathbf{1}$ & $\mathbf{1}$ \\
$A_4$ & $\mathbf{3}$ & $\mathbf{1}$ & $\mathbf{1''}$ & $\mathbf{1'}$ & $\mathbf{1}$ & $\mathbf{1'}$ & $\mathbf{1}$ & $\mathbf{1}$ & $\mathbf{3}$ & $\mathbf{3}$ & $\mathbf{3}$ & $\mathbf{1}$ & $\mathbf{1'}$ & $\mathbf{1}$ & $\mathbf{1}$ \\
$\mathbb{Z}_3$ & $\omega$ & $\omega^2$ & $\omega^2$ & $\omega^2$ & $1$ & $1$ & $\omega$ & $\omega^2$ & $\omega^2$ & $\omega$ & $1$ & $1$ &  $\omega^2$ & $\omega$ & $1$  \\
$U(1)_{\rm FN}$ & $0$ & $3$ & $1$ & $0$ & $F_1$ & $F_2$ & $F_3$ & $0$ & $0$ & $0$ & $0$ & $0$ & $0$ & $0$ & $-1$ \\ \hline
\end{tabular}
\end{table}

Let us finally comment on an aspect which relies on both ingredients. While a keV neutrino does not affect the parameters of the light neutrino sector, it mixes with the active neutrinos via the 
mixing angles $\theta_{\alpha 1}$ -- whose values are tightly bound by the X-ray limit. This mixing is \emph{enhanced} by the relatively high FN charge according to
\begin{equation}
 \theta_{e1} \simeq \frac{y_1 v_\varphi v_u}{w_1 u \Lambda} \lambda^{-F_1},
 \label{eq:A_4_AS-enhancement}
\end{equation}
with $\theta_{\mu 1, \tau 1}$ being of similar order. Thus, a too large value of $F_1$ might destroy the validity of the model. However, in practice, there is typically enough freedom to vary the couplings and VEVs involved, such that this constraint is avoided. 
On the other hand, if the sterile neutrino sector was known better, the link between the different sectors could be exploited to probe the model.

In this section, we could only give a glance of the aspects of the $A_4$ model presented and the benefits of combining different ingredients. To learn more about the detailed aspects, we refer the reader to the original references~\cite{Barry:2011fp,Barry:2011wb} as well as to the pedagogical review~\cite{Merle:2013gea}.

\subsection{\label{sec:VI-other}Models based on other principles}

Finally, there also exist models based on other principles which do not belong to one of the categories discussed up to now. 

\subsubsection{\label{sec:VI-other-Heeck}Extended seesaw (Author: J.~Heeck)}

The extended seesaw (ES)~\cite{Ma:1995gf, Chun:1995js} aims to explain the smallness of one or more sterile neutrinos by the very same seesaw mechanism that is behind small active neutrino masses.\footnote{While most often employed to generate sterile neutrinos at the eV scale~\cite{Abazajian:2012ys}, the ES is also applicable to keV sterile neutrinos, since their mass is still far below the generic seesaw scale.} In order for the light sterile neutrino(s) to mimic the seesaw couplings of active neutrinos, a structure in the mass matrix of neutral fermions is necessary, to be invoked by a symmetry.

On top of the standard seesaw scenario with $n_R$ right-handed singlets $\nu_R$ with Majorana mass matrix $M_M$ and Dirac mass terms $m_D\propto \langle \Phi\rangle$ from electroweak symmetry breaking
\begin{align}
\mathcal{L} \supset - \overline{\nu_L} m_D \nu_R - \tfrac12 \overline{\nu_R^c} M_M \nu_R + \mathrm{h.c.},
\end{align}
the ES introduces $n_S$ left-handed SM-singlet fermions $S_L$ with \emph{ad hoc} $\nu_L$-like couplings
\begin{align}
 \mathcal{L}_S =- \overline{S_L} m_N \nu_R + \mathrm{h.c.}
\label{eq:allowed_extended_seesaw_couplings}
\end{align} 
Here, $m_N$ is a mass scale in principle unrelated to electroweak symmetry breaking but typically also generated by a vacuum expectation value (see discussion of models below). 
The full $(3+n_R+n_S )\times (3+n_R+n_S)$ ES mass matrix for $(\nu_L^c, \nu_R, S^c_L)$ then takes the form~\cite{Barry:2011wb, Zhang:2011vh}
\begin{align}
M^{3+n_R+n_S}_\text{ES} = \begin{pmatrix} 0 & m_D & 0 \\ m_D^T & M_M & m_N^T \\ 0 & m_N & 0 \end{pmatrix} ,
\label{eq:extended_seesaw_structure}
\end{align}
leading, in the seesaw limit $M_M \gg m_D, m_N$, to a low-energy mass matrix for the $3+n_S$ light neutrinos suppressed by the high seesaw scale $M_M$
\begin{align}
M_\nu^{3+n_S} \simeq - \begin{pmatrix}  m_D M_M^{-1} m_D^T & m_D M_M^{-1} m_N^T  \\ m_N M_M^{-1} m_D^T & m_N M_M^{-1} m_N^T \end{pmatrix} ,
\end{align}
as intended. Since the $S_L$ mimic left-handed neutrinos by construction, standard arguments~\cite{Schechter:1980gr} tell us that at most $n_R$ of the $3+ n_S$ light neutrinos gain mass by this mechanism. 

We will in the following consider the \emph{minimal} extended seesaw~\cite{Barry:2011wb, Zhang:2011vh} ($n_S = 1$ and $n_R = 3$) for simplicity, which leads to one massless (active) neutrino. In the case of interest for a keV sterile neutrino, we consider $m_N \gg m_D$, allowing us to use the seesaw formula once more on $M_\nu^{3+1}$ and find the keV neutrino mass, $m_s \simeq |m_N M_M^{-1} m_N^T|$, and active-sterile mixing angle,
\begin{align}
\theta_{\rm as} \equiv m_D M_M^{-1} m_N^T \, (m_N M_M^{-1} m_N^T)^{-1} = \mathcal{O}(m_D/m_N) = \mathcal{O}(\sqrt{m_\nu}/\sqrt{m_s})\,.
\end{align}
From the naive scaling in the last two equations we see that small cancellations in the matrix structure $ m_D M_M^{-1} m_N^T$ are necessary to obtain $m_s = \mathcal{O}(\text{keV})$ with a small enough mixing~$\theta_{\rm as}$ to satisfy X-ray bounds~\cite{Zhang:2011vh}. All in all, the minimal ES can however quite naturally accommodate a weakly coupled keV sterile neutrino without the need of unnaturally small couplings. 
The price to pay is the introduction of a third mass scale, $m_N \sim m_D \sqrt{m_s/m_\nu}$, somewhat above the electroweak scale, plus a mechanism/symmetry to actually generate the desired ES structure of eq.~\eqref{eq:extended_seesaw_structure}. 
$S_L$ must carry some (broken) quantum numbers to distinguish it from the $\nu_R$, because a \emph{full} singlet would also allow for the unwanted couplings
\begin{align}
- \overline{l_L} \left(\frac{m_D^S}{v}\right) \tilde{\Phi} S^c_L - \tfrac12 \overline{S_L} \mu_S S_L^c + \mathrm{h.c.} \supset - \overline{\nu_L} m_D^S S_L^c - \tfrac12 \overline{S_L} \mu_S S_L^c + \mathrm{h.c.}
\label{eq:forbidden_extended_seesaw_couplings}
\end{align}
The model-building exercise for ES is hence to forbid the terms in eq.~\eqref{eq:forbidden_extended_seesaw_couplings} and only allow for those in eq.~\eqref{eq:allowed_extended_seesaw_couplings}, which can be accomplished by imposing symmetries.

The most straightforward approach is to introduce a SM-singlet scalar $\phi$ that carries the same non-trivial charge as $S_L$ under some \emph{global symmetry} group $\mathcal{G}$, allowing for the coupling $\overline{S_L} \phi \nu_R$ and hence generating $m_N \propto \langle \phi \rangle$ upon symmetry breaking of $\mathcal{G}$ via the VEV of $\phi$. $\mathcal{G}$ has to be chosen such that the couplings in eq.~\eqref{eq:forbidden_extended_seesaw_couplings} remain forbidden at the renormalizable level, which excludes the use of $\mathbb{Z}_2$ and $\mathbb{Z}_3$, as they 
would allow for the naturally large Majorana mass terms $\overline{S_L} \mu_S S^c_L$ and $\overline{S_L} \langle \overline{\phi}\rangle S^c_L$, respectively. $\mathcal{G} = \mathbb{Z}_4$ emerges as the simplest choice to realize the ES structure with global symmetries~\cite{Sayre:2005yh}; the $\mathbb{Z}_4$ invariant term $(\phi^4 + \overline{\phi}^4)$ in the scalar potential breaks the accidental $U(1)$ symmetry one encounters for $\mathcal{G} = \mathbb{Z}_{n\geq 5}$ and thus removes a potentially problematic Goldstone boson from the spectrum.\footnote{The Goldstone boson $J\propto \text{Im}(\phi)$ would enable the fast 
decay $\Gamma (\nu_s\to\nu_a J) \propto \theta_{\rm as}^2 m_s^3/\langle \phi\rangle^2$, unless the $U(1)$ breaking scale is much higher than expected from $m_N \sim \sqrt{m_s/m_\nu} m_D$, say $\langle \phi \rangle > 10^9$\,GeV.} In ref.~\cite{Zhang:2011vh}, it was further shown how this symmetry can be incorporated in the flavor symmetry group $A_4\times \mathbb{Z}_4$ in order to additionally shed light on the lepton mixing pattern. Other choices of $\mathcal{G}$ without Goldstone bosons can be realized with non-minimal scalar sectors~\cite{Ma:1995gf}.

In a different approach, one can choose $\mathcal{G}$ to be a \emph{gauge symmetry} in order to avoid potential drawbacks of discrete global symmetries such as incompatibility with quantum gravity~\cite{Banks:2010zn} or domain walls~\cite{Zeldovich:1974uw}. Since non-Abelian symmetries $\mathcal{G}$ generate (typically) more than one light sterile neutrino~\cite{Babu:2003is, Babu:2004mj, Sayre:2005yh}, we focus on $\mathcal{G} = U(1)$ here. Renormalizability requires the introduction of additional chiral fermions $T$ which cancel the gauge anomaly from the chiral $S_L$~\cite{Babu:2003is, Babu:2004mj, Sayre:2005yh, Heeck:2012bz}. In order for the $T$ not to spoil the ES matrix structure of eq.~\eqref{eq:extended_seesaw_structure}, $S$ and $T$ can be chosen to form an anomaly-free \emph{chiral set} of $U(1)$ charges~\cite{Batra:2005rh}, somewhat reminiscent of the hypercharge assignment in the SM, 
for example as~\cite{Heeck:2012bz}
\begin{align}
(S,\, T_1,\, T_2,\, T_3,\, T_4,\, T_5,\, T_6) \sim (11,\,  -5,\,  -6,\,  1,\,  -12,\,  2,\,  9) \,.
\end{align}
The $U(1)$ symmetry breaking by a scalar $\phi \sim 11$ couples $S$ to the right-handed neutrinos in the ES manner and also generates a mass for the new gauge boson $Z'$, as well as for the $T_j$ via
\begin{align}
 \mathcal{L}_T = y_1 \overline{T}_1^c T_2 \phi + y_2 \overline{T}_3^c T_4 \phi + y_3 \overline{T}_5^c T_6 \overline{\phi} + \mathrm{h.c.} \,.
\end{align}
The six anomaly-canceling $T_j$ thus form three massive Dirac fermions and have no impact on the ES mechanism. As emphasized in ref.~\cite{Heeck:2012bz}, these Dirac fermions are stabilized 
by the remaining $\mathbb{Z}_{11}$ symmetry, after $U(1)$ symmetry breaking (similar to ref.~\cite{Batell:2010bp}), and are coupled to the SM by (kinetic) $Z$--$Z'$ mixing --the Higgs portal $|H|^2 |\phi|^2$-- 
and via active-sterile neutrino mixing. A consistent realization of the ES structure by means of a gauge symmetry thus automatically leads to (multi-component and self-interacting) (C)DM candidates. 
The keV neutrino $\nu_s \sim S$ is charged under the $U(1)$ gauge symmetry and hence strictly speaking not sterile, but can be consistent with observational constraints. More than one light sterile neutrinos 
can be obtained using different chiral sets~\cite{Babu:2003is, Babu:2004mj, Sayre:2005yh}, which can also modify the phenomenology of the CDM component~\cite{Heeck:2012bz}.

\subsubsection{\label{sec:VI-other-Tsai}Dynamical mass generation and composite neutrinos (Authors: D.~Robinson, Y.~Tsai)}

Most keV neutrino model-building frameworks require Majorana neutrinos. Here, in contrast, we outline a framework that dynamically produces both sub-eV active and keV sterile \emph{Dirac} neutrinos. The central idea~\cite{Robinson:2012wu,Robinson:2014bma} is that the right-handed neutrinos are composite states of a hidden confining sector~\cite{ArkaniHamed:1998pf,Okui:2004xn,Grossman:2008xb,Grossman:2010iq,McDonald:2010jm,Duerr:2011ks}, while the keV steriles are elementary spectators that gain their masses in a similar fashion to the quarks in extended Technicolor theories. The Dirac-type mass structure of the active and sterile states typically fixes their mixing angle $\theta_{\alpha I} \sim m_{\alpha}/m_{I}$. This feature is particularly appealing in light of the recent, putative $3.55$~keV X-ray line result~\cite{Boyarsky:2014jta,Bulbul:2014sua}. Should the source turn out to be sterile neutrino DM with $m_I \simeq 7.1$~keV, then the observed flux implies a mixing angle $\theta_{\alpha I} \simeq 4\times 10^{-6}$: this is comparable to the active-sterile mass ratio $m_{\alpha}/m_{I} \sim 7\times 10^{-6}$ for a typical active neutrino mass of $m_{\alpha} \sim 0.05$~eV.

In addition to the SM field content, $\psi_{\rm SM}$, suppose that there exists a hidden sector of chiral fermions, $\chi$ and $\xi$ -- there are generically multiple species of each --  charged under a hidden confining group (hidden flavor symmetry) $G_{\rm c}$ ($G_{\rm F}$), such that $\psi_{\rm SM} \sim G_{\rm SM}\times G_{\rm F}$, $\chi \sim G_{\rm c}\times G_{\rm F}$, and $\xi \sim G_{\rm F}$.  We consider this to be a low-energy effective field theory below some UV completion scale, $M$, so that the SM and hidden sectors interact via $M$-scale irrelevant operators. At a scale $\Lambda \ll M$, $G_{\rm c}$ becomes strongly coupled, inducing $\chi$ confinement into hidden bound states.  The $\xi$ are spectators, required to cancel the $G_{\rm F}$ anomalies. 

In order to avoid flavor constraints, we necessarily need $M \gg v$. It is therefore convenient to define two parameters $\varepsilon \equiv \Lambda/M \ll1$ and $\vartheta \equiv v/M \ll 1$.  With appropriate choices for the hidden sector, one may ensure: (i) Confinement of the $\chi$ induces the breaking $G_{\rm F} \to U(1)_{\rm F}$;  (ii) Anomaly matching between the free and confined phases requires that there are precisely three massless chiral bound states, $\nu_R$, all with the same $F$ charge; (iii) $F$ is a linear combination of $B-L$ and hypercharge such that $U(1)_{\rm EM}\times U(1)_{B-L}$ survives electroweak symmetry breaking (EWSB); (vi) The $\nu_R$ have appropriate $F$ charge to be right-handed neutrinos, i.e.\ they form the $\overline{l_L}\Phi^\dagger \nu_R$ Yukawas; (v) Along with the three $\nu_R$, there are fermionic bound states (spectators) $N_L^c$ and $N_R$ ($\xi_L^c$ and $\xi_R$) which furnish Dirac multiplets, and have the correct $F$ charge to be sterile neutrinos; (vi) The composite $\nu_R$ and $N_{R,L}$ contain three $\chi$'s, i.e.\ $\nu_R \sim \chi^3$ etc., and the symmetry breaking condensate has two $\chi$'s, i.e.\ $\langle \chi^2 \rangle \not=0$. Examples of toy models that incorporate these features are included in ref.~\cite{Robinson:2014bma}.

After $\chi$-confinement, suppressed Yukawa couplings between the active lepton doublet and composite neutrinos are generated via dimensional transmutation, viz.
\begin{equation}
	\label{eqn:RobinsonTsaiEFTC}
	\frac{\overline{l_L}\Phi^\dagger \chi^3}{M^3} \to \frac{\varepsilon^3}{(4\pi)^2} \overline{l_L}\Phi^\dagger \nu_R~~\mbox{or} ~~\frac{\varepsilon^3}{(4\pi)^2} \overline{l_L}\Phi^\dagger N_R\,, \quad \mbox{and} \quad  \frac{\overline{l_L} \Phi^\dagger\chi^2\xi}{M^3} \to  \frac{\varepsilon^3}{(4\pi)^2} \overline{l_L}\Phi^\dagger \xi_R\,.
\end{equation}
The spectators, however, obtain Dirac masses through irrelevant couplings to the condensate VEV, 
\begin{equation}
	\frac{\xi \langle \chi^2 \rangle \xi}{M^2}  \to  \frac{\Lambda\varepsilon^2}{(4\pi)^2} \overline{\xi_L}\xi_R\,.
\end{equation}
One may require that the $G_{\rm F}$ structure does not admit Dirac mass cross-terms of composites with $\xi$'s. By analogy with large-$N_c$ QCD combinatorics the three-$\chi$ massive bound states $N_{L,R}$ typically have a Dirac mass $M_N \sim 3\Lambda$.  After EWSB, one then obtains a mass term of block-matrix form~\cite{Robinson:2012wu, Robinson:2014bma},
\begin{equation}
	\label{eqn:RobinsonTsaiDMM}
	\frac{\Lambda}{(4\pi)^2} \begin{pmatrix}	\overline{\nu_L} & \overline{\xi_L} & \overline{N_L} \end{pmatrix} \begin{pmatrix} \vartheta \varepsilon^2 & \vartheta \varepsilon^2 & \vartheta \varepsilon^2\\ 0 &\varepsilon^2  & 0 \\ 0 & 0 & 3(4\pi)^2\end{pmatrix}\begin{pmatrix}\nu_R \\ \xi_R \\ N_R\end{pmatrix}\,.
\end{equation}

Up to $\mathcal{O}(1)$ factors, the neutrino masses and mixing angles are now fixed by the two dynamical scales $M$ and $\Lambda$ via eq.~\eqref{eqn:RobinsonTsaiDMM}. The spectrum corresponding to eq.~\eqref{eqn:RobinsonTsaiDMM} is
\begin{equation}
	\label{eqn:RobinsonTsaiNMS}
	m_{\alpha} \sim v\varepsilon^3/ (16\pi^2), \qquad {m_{\xi}}_I \sim \Lambda \varepsilon^2/ (16\pi^2), \qquad {m_N}_I \sim 3\Lambda~.
\end{equation}
Hence there are Dirac neutrinos $\nu^\alpha_{L,R}$ ($\xi_{L,R}^I$) whose masses are dynamically $\varepsilon^3$-($\varepsilon^2$-)suppressed compared to the confinement scale.  The corresponding left-handed neutrino mass basis
\begin{equation}
\begin{pmatrix} \nu^\alpha_L \\ \xi^I_L \\ N^I_L \end{pmatrix} 
	\sim \begin{pmatrix} 1 & \vartheta & \frac{\vartheta\epsilon^2}{48\pi^2} \\ \vartheta  & 1 & 0 \\ \frac{\vartheta\epsilon^2}{48\pi^2} & 0 & 1 \end{pmatrix}\begin{pmatrix} \nu_L \\ \xi_L \\ N_L \end{pmatrix}~,
\end{equation}
from which one sees that the active-keV sterile mixing angle $\theta_{\alpha I}$ (i.e.\ the angle spanned by  $\nu^\alpha_L$ and $\xi^I_L$) is given by $\theta_{\alpha I} \sim \vartheta \equiv v/M$. Constraining the active neutrino mass $m_{\alpha} < 0.05$~eV immediately requires $\varepsilon \lesssim 5 \times 10^{-4}$. For the benchmark choices $\varepsilon= 3 \times 10^{-4}$ and $\Lambda = 15$~TeV -- i.e.\ $M \sim 5\times 10^4$~TeV, certainly flavor safe -- then $m_{\alpha} \sim 0.05$~eV,  ${m_{\xi}}_I \sim 5$~keV and $\sin^2(2\theta_{\alpha I}) \sim 5 \times 10^{-11}$. 

With a keV scale mass and a mixing angle of this order, the elementary spectators $\xi$ serve as potential WDM candidates. We emphasize that in this composite neutrino framework, the two dynamical scales $\Lambda$ and $M$ -- or, equivalently, $\Lambda$ and $\varepsilon$ -- fix the scales of the spectrum and couplings, and hence also the approximate cosmological history of $\xi$. For example, the $\xi$, $\chi$, SM sectors are thermalized via $M$-scale irrelevant interactions, with  decoupling temperature $T_{\rm dec} \sim (\Lambda^4/\varepsilon^4 M_{\rm pl})^{1/3}$~\cite{Robinson:2012wu}. Remarkably, $T_{\rm dec} \sim \Lambda$ for the benchmark $\Lambda$ and $\varepsilon$. The keV $\xi$'s are therefore thermally produced while ultrarelativistic, resulting in an overclosed Universe. Significant entropy production is then required to obtain a DM-like relic abundance. 

Intriguingly, since $\Lambda \sim T_{\rm dec}$, one possible source of such entropy could be a strongly supercooled $\chi$-confining phase transition just after $\xi$ decoupling, but just before hidden bound state-SM decoupling. Such supercooling deposits significant entropy into the hidden bound state and SM sectors, diluting the decoupled $\xi$'s~\cite{Robinson:2012wu}. However, this option generally requires the ratio of the confinement and phase transition temperatures to be $\sim 6$, which seems implausibly large.

An alternative is the late, out-of-equilibrium decay of a heavy state. Conveniently, at least one heavy composite neutrino species, $N_h$($\equiv N_{L,R}$), is generically present in this framework. If the decay of $N_h$ proceeds dominantly through $N_h \to\nu^\alpha_L h$, the $\varepsilon^3$-suppressed coupling~\eqref{eqn:RobinsonTsaiEFTC} produces a sufficiently small decay rate $\Gamma_{N_h} \sim 9 \Lambda \varepsilon^6/4096\pi^5$ -- the benchmark lifetime is $\lesssim 10^{-2}$~s -- such that sizable entropy production occurs right before the BBN epoch. The resulting entropy production factor, $\gamma$, and the $\xi$ relic density are respectively
\begin{equation}
\gamma \sim m_{N_h}Y_{N_h}~/ ~\Gamma^{1/2}_{N_h}M^{1/2}_{\text{pl}}\,, \qquad \Omega_{\xi} / \Omega_{\text{DM}} = m_{\xi}(Y_{\xi}/\gamma) s_0 \rho_{\text{DM}}^{-1} \,.
\end{equation}
Dominance of the $N_h \to\nu^\alpha_L h$ decay channel can happen, e.g., if the composite neutrino species have nearly democratic couplings and suppressed hidden-pion couplings. A more detailed discussion can be found in ref.~\cite{Robinson:2014bma}. The yield $Y_{N_h}$ is then determined by $\Lambda$ and $\varepsilon$ up to $\mathcal{O}(1)$ factors, just as were its mass and decay rate. The $\xi$ relic density is then highly predictable: at the benchmark values, one finds that $\Omega_{\xi} \lesssim \Omega_{\rm DM}$. 

Being a Dirac fermion, note that $\xi$ has four degrees of freedom per species, making $\xi$ WDM slightly cooler compared to typical sterile Majorana neutrino DM. This permits to evade structure formation bounds slightly more easily, including 
the constraints from Lyman-$\alpha$ bounds~\cite{Viel:2005qj,Markovic:2013iza,Viel:2013apy} and N-body simulations of subhalo formation~\cite{Schultz:2014eia}, see ref.~\cite{Robinson:2014bma} for details.

\subsubsection{3-3-1-models (Authors: A.G.~Dias, N.~Anh~Ky, C.A.~de~S.~Pires, P.S.~Rodrigues da Silva, N.~Thi Hong Van)}

The 3-3-1 models~\cite{Singer:1980sw,Valle:1983dk,Montero:1992jk,Foot:1994ym} are distinct SM gauge extensions which can naturally contain RH neutrinos at keV scale~\cite{Dias:2005yh,Dinh:2006ia,Cogollo:2009yi,Ky:2005yq}. These models belong to a class of constructions where the SM gauge symmetry $SU(2)_L\times U(1)_Y$ is extended to $SU(3)_L\times U(1)_X$ in such a way that the cancellation of the anomalies involves all three families, offering an explanation for the long standing family number problem~\cite{Singer:1980sw,Pisano:1991ee,Frampton:1992wt,Pleitez:1992xh}.  As it happens, RH neutrinos are predicted in 3-3-1 models by the reason that they form, together with the SM leptons, representations of  $SU(3)_L$. We will take here one version of the models in which the RH neutrinos do  not carry lepton number from the beginning and, therefore,  they will not be related to active neutrinos by mean of charge conjugation. The ordinary leptons and the RH neutrino are then arranged in left-handed triplets,  $L_{\alpha}=[\nu_{L,\alpha},\, e_{L,\alpha},\,N^{ c}_{R,\alpha}]^T$$\sim$ $(3,\,-1/3)$, and  RH singlets $e_{R,\alpha}$$\sim$$(1\,,\,-1)$, $\alpha=e,\,\mu,\,\tau$, with the numbers in parentheses being the ($SU(3)_L$, $U(1)_X$) quantum numbers.

To generate mass consistently for all fermions in this model, the following three scalar triplets are considered: $\chi=[\chi^0,\, \chi^-,\,\chi^{\prime0}]^T$ $\sim$ $(3,\,-1/3)$, $\eta=[\eta^0,\, \eta^-,\,\eta^{\prime0}]^T$ $\sim$ $(3,\,-1/3)$,  and  $\rho=[\rho^+,\, \rho^0,\,\rho^{\prime+}]^T$ $\sim$ $(3\,,\,2/3)$. The symmetry breaking pattern is assumed as $SU(3)_{L}\times U(1)_X/SU(2)_L\times U(1)_Y$, occurring through the VEV $\langle\chi\rangle=[0,\, 0,\,v_{\chi^{\prime}}]^T$ with  $v_{\chi^{\prime}}\sim{\cal O}(1)$ TeV, and $SU(2)_L\times U(1)_Y/U(1)_{\rm em}$ occurring through the VEVs $\langle\eta\rangle=[v_\eta,\, 0,\,0]^T$, $\langle\rho\rangle=[0,\, v_\rho,\,0]^T$, with the scales related according to  $v_\eta^2+v_\rho^2=v^2$  ($v=174$ GeV).  The  scale $v_{\chi^\prime}$  gives the main contribution to the mass of the new gauge 
bosons which are two electrically charged bosons, $V^{\pm}$, two neutral complex gauge bosons $U^0$ and $U^{0*}$, and a real neutral one, the  $Z^{\prime}$. Some of them, $V^\pm$ and $U^0$, carry lepton number, connecting the charged leptons to the RH neutrinos.  Thus, if produced at the LHC, the  decays of those new gauge bosons might imprint distinct signatures involving RH neutrinos plus SM charged leptons, for example.

Majorana mass terms for six neutrinos are obtained from dimension-five effective ope\-ra\-tors involving the leptons and scalar triplets~\cite{Dias:2005yh,Dinh:2006ia,Cogollo:2009yi}. This can be seen by considering a simpler case than the one analyzed in refs.~\cite{Dias:2005yh,Dinh:2006ia}, just assuming a parity symmetry defined by  $P=(-1)^{3(B-L)+2s}$, where the following fields are odd $(N_R\,,\,\rho^{\prime +}\,,\,\eta^{\prime 0}\,,\,\chi^{0}\,,\,\chi^-\,,\, V\,,\,U) \rightarrow -1$~\cite{Dong:2013wca,Kelso:2013nwa}.\footnote{The new quarks predicted by the model are all odd under this symmetry, too.} Such a symmetry forbids the $\nu_{L}-N^{ c}_{R}$ mixing which could happen  through the effective operator $\left( \overline{L^c} \chi^* \right)\left(\eta^{\dagger}L\right)$. In this scenario  the RH neutrinos are completely sterile with respect to $W^{\pm}$ and $Z^0$. Therefore, the  effective Lagrangian containing the  main operators leading to neutrino masses is
\begin{equation}
{\cal{L}} \supset  \left( \overline{L^c} \eta^* \right)\frac{f}{\Lambda}\left( \eta^{\dagger}L \right)+\left( \overline{L^c} \chi^* \right)\frac{h}{\Lambda}\left( \chi^{\dagger}L \right)+h.c\,,
\label{leff331}
\end{equation}
where $f\equiv f_{\alpha\beta}$ and $h\equiv h_{\alpha\beta}$ are taken as arbitrary symmetric matrices, and $\Lambda$ as a very high energy scale.  Majorana mass terms for the active and RH sterile neutrinos are generated with the VEVs $\langle \eta\rangle$ and  $\langle \chi^{\prime}\rangle$, leading to the mass matrices
\begin{eqnarray}
M_{L}=0.6\,f\left(\frac{v_\eta}{25\,{\rm GeV}}\right)^2\left(\frac{10^{12}\,{\rm GeV}}{\Lambda}\right){\rm eV},
\hspace{0.3 cm}
M_{M}=10\,h\left(\frac{v_{\chi^{\prime}}}{3.2\,{\rm TeV}}\right)^2\left(\frac{10^{12}\,{\rm GeV}}{\Lambda}\right){\rm keV}.\,\,\,\,
\label{mmas331}
\end{eqnarray}
Assuming $f_{\alpha\beta}$, $h_{\alpha\beta}$ of $\mathcal{O}(1)$ and $\Lambda \approx 10^{12}$ GeV, with the symmetry breaking scales $v_{\chi^{\prime}}$,  $v_\eta$ at TeV and tens of GeV, respectively, masses are generated at the keV scale for the RH sterile neutrinos and at the sub-eV scale for the active neutrinos. The eigenstate of the smaller eigenvalue of  $M_M$ in eq.~\eqref{mmas331} corresponds to lightest RH neutrino and is stable, due the parity symmetry, as required for a particle acting as WDM~\cite{Dias:2005yh}. 
Another result in this scenario is that the next-to-lightest RH neutrino, $N^{c }_{R,2}$, obtained from diagonalization of $M_M$ in eq.~\eqref{mmas331}, decays into the lightest one plus a 
photon, $N^{c }_{R,2}\rightarrow N^{c}_{R,1}+\gamma$, through a one-loop process involving the  $V^{\pm}$. This process might be relevant for explaining  astrophysical phenomena involving X-ray lines, as for example the recent observation of the 3.5 keV line~\cite{Bulbul:2014sua,Boyarsky:2014jta}.

A realization of how the effective operators in eq.~\eqref{leff331} can be induced at low energies was done in ref.~\cite{Cogollo:2009yi}, assuming the existence of a heavy scalar sextet $S\sim (6,\,{-2}/{3})$. In this case Majorana mass terms for neutrinos are generated through the Yukawa interaction~\cite{Dinh:2006ia,Ky:2005yq},
\begin{equation}
{\cal{L}}_Y\supset \overline{L^c}\, y\,S^* L+h.c.\,,
\label{lS}
\end{equation}
where $y\equiv y_{\alpha\beta}$, with the VEV of the sextet components $\langle S_{11}\rangle\equiv\langle\Delta^0\rangle=v_{\Delta^0}$ and $\langle S_{33}\rangle\equiv\langle\sigma^0\rangle=v_{\sigma^0}$. This leads to the mass matrices $\left[M_{L}\right]_{\alpha\beta}=y_{\alpha\beta}v_{\Delta^0}$ and $\left[M_{R}\right]_{\alpha\beta}=y_{\alpha\beta}v_{\sigma^0}$ for the active and RH neutrinos, respectively. The small values of $v_{\Delta^0}$ and $v_{\sigma^0}$ are due to a seesaw mechanism 
involving the sextet\footnote{This mechanism resembles the known type II seesaw mechanism~\cite{Magg:1980ut,Mohapatra:1980yp,Ma:1998dx,Schechter:1981bd}.} and an explicit lepton number violation. Such violation occurs mainly through the terms in the potential $V\supset  M_1\eta^{T}S^{\dagger}\eta + M_2\chi^{T}S^{\dagger}\chi+h.c.$, where it is assumed that $M_1\approx M_2\approx\Lambda$ are very high energy scales in which lepton number is violated. Thus, considering this and the fact that the mass scale of the sextet is also of order of $\Lambda\approx 10^{12}$ GeV, minimization of the potential leads to constraint equations furnishing small values for the VEVs, $v_{\Delta^0}\approx v_\eta^2/M_1$ and $v_{\sigma^0}\approx v_\eta^2/M_2$, so that $\left[M_{L}\right]_{\alpha\beta}\approx y_{\alpha\beta}\,v_\eta^2/\Lambda$ and $\left[M_{R}\right]_{\alpha\beta}\approx y_{\alpha\beta}\,v_{\chi^{\prime}}^2/\Lambda$. These mass matrices have the same suppression factors as the ones in eq.~\eqref{mmas331}. Besides, the mechanism predicts a mass relation for these two types of neutrinos, as a result of the fact that their masses arise from the same coupling $g_{\alpha\beta}$ in eq.~\eqref{lS}. 
It can be seen that the RH/active neutrino mass ratio is $m_{N_{iR}}/m_{\nu_{iL}}=v_{\chi^{\prime}}^2/v_\eta^2$ ~\cite{Dinh:2006ia,Ky:2005yq}, so that for $v_{\chi^{\prime}}\approx 1$ TeV, $v_\eta\approx10$ GeV masses at the keV scale for RH neutrinos are predicted, taking into account that the active neutrinos have masses of order  $m_{\nu_{iL}}\approx0.1$ eV~\cite{Cogollo:2009yi}. The use of a scalar sextet was also developed in other ways to implement the  seesaw mechanisms in 3-3-1 models~\cite{Palcu:2006ti,Dong:2008sw,Cogollo:2008zc}. 

Another model extending the SM lepton sector only by neutrino fields trans\-for\-ming nontrivially under a non-Abelian gauge group is the $SU(3)_L\times SU(3)_R\times U(1)_X$ model~\cite{Dias:2010vt}. It has the essence of the left-right models~\cite{Pati:1974yy,Senjanovic:1975rk}, plus the peculiarity that the number of families is restricted by gauge anomaly cancellation and by the embedding of the 3-3-1 models of refs.~\cite{Montero:1992jk,Foot:1994ym}.  One of the main features of the $SU(3)_L\times SU(3)_R\times U(1)_X$ model is that its rich neutrino content allows to address, simultaneously, problems related to neutrino oscillation, WDM, and baryogenesis through leptogenesis~\cite{Dias:2010vt}. In its particle spectrum the model might have, in addition to the light active neutrinos, three neutrinos at keV mass scale, and six heavy ones having masses around $10^{11}$ GeV. Such  neutrinos result from the leptonic field content  $\Psi_{L,\alpha}=\left[\nu_{L,\alpha},\, e_{L,\alpha},\, N_{L,\alpha}\right]^{T}$$\sim$$(3,\,1,\,-1/3)$ and $\Psi_{R,\alpha}=\left[\nu_{R,\alpha},\, e_{R,\alpha},\, N_{R,\alpha}\right]^{T}$$\sim$$(1,\,3,\,-1/3)$, with the numbers in parenthesis being the ($SU(3)_L$, $SU(3)_R$, $U(1)_X$) quantum numbers, and  $N_{\alpha}$ being new neutral fermions.

Two sets of scalar triplets, $\eta_L$, $\chi_L$$\sim$$(3,\,1,\,-1/3)$ and $\eta_R$, $\chi_R$$\sim$$(1,\,3,\,-1/3)$, break  the gauge symmetries and participate in the neutrinos mass generation mechanism. This sets up by means of the following effective operators~\cite{Dias:2010vt}, when the scalar fields get VEV as $\langle\eta_{L,R}\rangle=[v_{\eta_{L,R}},\,0,\, 0]^{T}$ and $\langle\chi_{L,R}\rangle=[0,\, 0,\, v_{\chi_{L,R}}]^{T}$,
\begin{equation}
{\cal{L}} \supset  \left( \overline{\Psi_L} \chi_L \right)\frac{g^D_\chi}{\Lambda_D}\left( \chi_R^{\dagger}\Psi_{R} \right)+ \left( \overline{\Psi^c_L} \chi^*_L \right)\frac{g^M_\chi}{\Lambda_M}\left( \chi_L^{\dagger}\Psi_{L} \right) +
 \left( \overline{\Psi^c_R} \chi^*_R \right)\frac{g^M_\chi}{\Lambda_M}\left( \chi_R^{\dagger}\Psi_{R} \right) +(\chi\rightarrow \eta)\,,
\label{leff331LR}
\end{equation}
where a $\mathbb{Z}_2$ symmetry,  $\chi_{L,R} \rightarrow -\chi_{L,R} $, is assumed for simplicity. The sterile neutrinos within this model can be fully stabilized with the previous parity symmetry. The suppression scales in eq.~\eqref{leff331LR} are such that $\Lambda_M\approx 10^{19}$ GeV, meaning that lepton number is broken by gravitational interactions at  the Planck scale, and $\Lambda_D\approx 10^{15}$ GeV, which could be related to a grand unification scale. The breaking of $SU(3)_L\times SU(3)_R\times U(1)_X/SU(3)_L\times U(1)_N$ is taken to occur with the VEVs around this last scale, $v_{\eta_{R}}\simeq  v_{\chi_{R}}\simeq \Lambda_D$, leaving at low energy an effective 3-3-1 model plus very heavy singlets~\cite{Dias:2010vt}. Given that $\Lambda_M\gg\Lambda_D$, it follows from eq.~\eqref{leff331LR} that choosing  $v_{\eta_{L}}\approx 10$ GeV,  $v_{\chi_{L}}\approx 1$ TeV, the mass matrix of the active neutrinos is $M_{\nu_L}\approx  g^D_\eta(g_\eta^M)^{-1}(g^D_\eta)^T$ eV. There is also a mass matrix $M_{N_L}\approx10\, g^D_\chi(g_\chi^M)^{-1}(g^D_\chi)^T$ keV for three sterile neutrinos, with the lightest one being a  WDM candidate. Additionally, there are six very heavy neutrinos following from the mass matrix $M_M\approx \, 10^{11}{\rm diag}\,(g_\eta^M,\,g_\chi^M)$ GeV, which can act in the baryogenesis through leptogenesis for explaining the Universe matter-antimatter asymmetry~\cite{Fukugita:1986hr}.

As a final note, it is worth to point out that there are also more complete GUT settings where light sterile neutrinos can exist and be well-motivated, see the works on $E_6$~\cite{Rosner:2014cha} as prominent example.

\subsubsection{Anomalous Majorana Neutrino  Masses from Torsionful Quantum Gravity (Authors: N.~Mavromatos, A.~Pilaftsis)}

A potentially interesting radiative mechanism for generating gauge-invariant  fermion  masses  at  three  loops  has  been  studied in~\cite{Pilaftsis:2012hq}. It uses  global anomalies  triggered by the possible existence of scalar  or pseudoscalar fields in $U(1)$ gauge theories and  by heavy fermions $F$  whose masses may  not result from spontaneous symmetry  breaking. In~\cite{Mavromatos:2012cc}, whose contents are reviewed here, this  mass-generating mechanism has been ``geometrized'' by applying it to appropriate field theories of gravity with torsion~\cite{Hehl:1976kj,Shapiro:2001rz,Kibble:1961ba,Sciama:1964wt}, to create low-scale fermion  masses by pure quantum  gravity effects. Here, the role  of the $U(1)$ gauge field strength tensor~$F_{\mu\nu}$ is played by the Riemann curvature tensor~$R_{\mu\nu\rho\sigma}$,  and hence the  role of the gauge fields by the gravitons.  Such a gravitationally-assisted mass mechanism could  straightforwardly be  applied to fermions  without SM quantum  charges, such  as Majorana  right-handed neutrinos. Our goal is to review briefly this mechanism and provide  reliable estimates of the gravitationally induced   right-handed Majorana   mass  $M_M$. As we shall see, masses in the keV region can be naturally provided. Although   quantum gravitational interactions  are non-renormalizable, nevertheless there are aspects of the theory that can be exact, in a path integral sense, and these are related to some aspects of torsionful manifolds. Torsion appears as a non-propagating  3-form in a quantum gravity  path integral and as such it  can be integrated out exactly~\cite{Hehl:1976kj,Shapiro:2001rz,Kibble:1961ba,Sciama:1964wt}.

To start, consider Dirac QED fermions in a torsionful space-time. The Dirac  action reads:  $S_\psi = \frac{i}{2}  \int d^4  x \sqrt{-g} \Big(  \overline{\psi}  \gamma^\mu  \overline{\mathcal{D}}_\mu \psi  - (\overline{\mathcal{D}}_\mu \overline{\psi}  ) \gamma^\mu \psi \Big)$, where $\overline{\mathcal{D}}_\mu =  \overline{\nabla}_\mu - i e A_\mu$, with  $e$ being the electron charge and  $A_\mu$ the photon  field.  The overline above the covariant derivative, i.e.~$\overline{\nabla}_\mu$, denotes  the presence  of  torsion, which  is  introduced through  the torsionful spin connection $\overline{\omega}_{a b \mu} $. The  presence of  torsion in  the covariant derivative  in the Dirac-like  action leads,  apart from  the standard terms in  manifolds without torsion,  to an additional  term involving the axial current  $J^\mu_5 \equiv \overline{\psi} \gamma^\mu \gamma^5 \psi  $:  $S_\psi  \ni  -  \frac{3}{4} \int  d^4  \sqrt{-g}  \,  S_\mu \overline{\psi} \gamma^\mu \gamma^5 \psi = - \frac{3}{4} \int S \wedge {}^\star\! J^5 $, where $\textbf{S} = {}^\star\! \, \textbf{T}$ is   the   dual    of   the torsion tensor $\textbf{T}= \textbf{d   e}^a   +   \overline{\omega}^a   \wedge   \textbf{e}^b   $,   $S_d   =   \frac{1}{3!} \epsilon^{abc}_{\quad   d}  T_{abc}$. It is an axial pseudovector field, one of the three types of structures appearing in the irreducible decomposition of \textbf{T}~\cite{Shapiro:2001rz,Duncan:1992vz}. The gravitational part of the  action can be written as $S_G =\frac{1}{2\kappa^2}  \,  \int   d^4  x  \sqrt{-g}  \Big(R  + \widehat{\Delta} \Big)  + \frac{3}{4\kappa^2} \int \textbf{S} \wedge  {}^\star\!  \, \textbf{S}$,  where $R$ is the scalar curvature without torsion and $\widehat  \Delta  $ contains quadratic forms of irreducible parts of the torsion tensor other than \textbf{S}. In   a  quantum   gravity setting,  where one  integrates over  all  fields, the torsion  terms appear  as non-propagating  fields and hence can be integrated out  exactly. The authors of~\cite{Duncan:1992vz} have  observed though that the classical    equations   of   motion    identify  the axial-pseudovector  torsion  field  $S_\mu$  with  the axial  current. It follows that $\textbf{d}\,{}^\star\! \, \textbf{S} = 0$,  leading to a conserved ``torsion  charge'' $Q =\int   {}^\star\! \,  \textbf{S}$. To   maintain  this  conservation   in  the quantum  theory,   they  postulated $\textbf{d}\,{}^\star\!\,\textbf{S} =  0$ at the quantum level -- which  can  be  achieved by  the  addition  of judicious  counter  terms. This  constraint,  in  a path-integral formulation of  quantum gravity, is then implemented via  a $\delta$-function, $\delta(\textbf{d}\,{}^\star\! \, \mathbf{S})$,  and the latter  via the well-known trick of  introducing a (pseudoscalar) Lagrange multiplier field  $\Phi   (x)  \equiv  (3/\kappa^2)^{1/2}  b(x)$. Hence,    the   relevant    torsion   part    of   the quantum-gravity  path  integral ${\mathcal Z}$ includes a factor                           
\begin{eqnarray}
 \label{brr}
 {\mathcal Z} &=& \int e^{-iS_G + iS_\psi} \propto \, \int D \textbf{S} \, D b \, \exp \Big[ i \int  \Big(
    \frac{3}{4\kappa^2} \textbf{S} \wedge {}^\star\! \,\textbf{S} -
      \frac{3}{4} \textbf{S} \wedge {}^\star\! \, \textbf{J}^5  +
      \Big(\frac{3}{2\kappa^2}\Big)^{1/2} \, b \, \textbf{d}\, {}^\star\! \,\textbf{S}\, \Big)
      \Big]\nonumber \\  
&=& \, \int D b \, \exp\Big[ -i \int  \Big( \frac{1}{2}
      \textbf{d} b\wedge {}^\star\! \, \textbf{d} b + \frac{1}{f_b}\textbf{d}b 
\wedge {}^\star\! \, \textbf{J}^5 + \frac{1}{2f_b^2}
    \textbf{J}^5\wedge{}^\star \,\textbf{J}^5 \,\Big) \Big] \nonumber\\
& = & \, \int D b \,\exp\Big[ - i \int \Big( \frac{1}{2}
    \textbf{d} b\wedge {}^\star\!\, \textbf{d} b  - \frac{1}{f_b} b
    G(\textbf{A}, \omega)  
+ \frac{1}{2f_b^2} \textbf{J}^5 \wedge {}^\star \, \textbf{J}^5 \Big)\Big],
\end{eqnarray} 
where $f_b = (3\kappa^2/8)^{-1/2} = \frac{M_P}{\sqrt{3\pi}}$, and in the last line we have partially integrated the second term in the exponent in the middle line of~\eqref{brr} and took into account the well-known field theoretic result that in QED the axial current is not conserved at the quantum level, due to anomalies, but its divergence is obtained by the one-loop result~\cite{Weinberg:1996kr}: $\nabla_\mu J^{5\mu} \! =\! \frac{e^2}{8\pi^2} {F}^{\mu\nu} \widetilde{F}_{\mu\nu} - \frac{1}{192\pi^2} {R}^{\mu\nu\rho\sigma} \widetilde {R}_{\mu\nu\rho\sigma} \equiv G(\textbf{A}, \omega)$, with the tilde denoting appropriate tensor duals.

An interesting example of a torsionful space-time which also provides a concrete UV-complete theoretical framework for quantum gravity comes from string theory.  In string theories, torsion is introduced as a consequence of the existence of the (spin one) anti-symmetric tensor field $B_{\mu\nu} = - B_{\nu\mu}$ existing in the gravitational multiplet of the string. Indeed, as a result of the stringy ``gauge'' symmetry $B_{\mu\nu} \rightarrow B_{\mu\nu} + \partial_{[\mu }B_{\nu]}$, the low-energy string effective action depends only on the field strength $H_{\mu\nu\rho} = \partial_{[\mu} B_{\nu\rho]}$, where the symbol $[\dots ]$ denotes anti-symmetrization of the appropriate indices. In fact, it can be shown~\cite{Metsaev:1987zx,Gross:1986mw} that the terms involving the field strength perturbatively to each order in the Regge slope parameter $\alpha^\prime$ can be assembled in such a way that only torsionful Christoffel symbols, such as $\overline{\Gamma}^\mu_{\nu\rho} = \Gamma^\mu_{\nu\rho} + \frac{\kappa}{\sqrt{3}}\, H^\mu_{\nu\rho}\ \ne\ \overline{\Gamma}^\mu_{\rho\nu}$, appear in the low-energy effective action.  In four dimensions, we may define the dual of $\textbf{H}$, $Y_\sigma\ =\ - 3\, \sqrt{2} \partial_\sigma b\ =\ \sqrt{-g}\, \epsilon_{\mu\nu\rho\sigma} H^{\mu\nu\rho}$, after adopting the normalizations of~\cite{Duncan:1992vz}. The field $b(x)$ is a form-valued pseudoscalar field, with a canonically normalized kinetic term which is termed the Kalb-Ramond (KR) axion~\cite{Kalb:1974yc}, in order to distinguish it from other axion-like field coming, e.g.\ from the moduli sector of string theory. It plays a role entirely analogous to the gravitational axion field we have encountered above in our generic discussion of torsional field theories.

An important aspect of the coupling  of the field $b(x)$ to the  fermionic   matter  discussed   above  is  its   shift  symmetry, characteristic of axion fields. Indeed, by shifting the field $b(x)$ by  a constant, $b(x)  \to b(x)  +  c$, the  action~\eqref{brr}  only changes by  total derivative terms, such  as $c\, R^{\mu\nu\rho\sigma} \widetilde{R}_{\mu\nu\rho\sigma}$ or $c\, F^{\mu\nu}\widetilde{F}_{\mu\nu}$. These terms are irrelevant for the equations  of motion and  the induced  quantum dynamics,  provided the fields fall off to zero sufficiently fast  at infinity. The anomalous  Majorana mass generation  through torsion consists of augmenting the  effective action~\eqref{brr} by terms that break the shift symmetry. To illustrate this last point, we  first couple the gravitational axion $b(x)$ to another pseudoscalar  axion field $a(x)$.  The  proposed   coupling  occurs through  kinetic mixing. To  be specific, we consider the action
\begin{eqnarray} 
  \label{bacoupl}
    \mathcal{S} \!&=&\! \int d^4 x \sqrt{-g} \, \Big[\frac{1}{2}
      (\partial_\mu b)^2 + \frac{b(x)}{192 \pi^2 f_b}
      {R}^{\mu\nu\rho\sigma} \widetilde{R}_{\mu\nu\rho\sigma} 
       + \frac{1}{2f_b^2} J^5_\mu {J^5}^\mu \nonumber \\ &+ & \gamma
      (\partial_\mu b )\, (\partial^\mu a ) + \frac{1}{2}
      (\partial_\mu a)^2 
- y_a i a\, \Big( \overline{\psi}_R^{\ C} \psi_R - \overline{\psi}_R
\psi_R^{\ C}\Big) \Big]\;, \qquad 
\end{eqnarray}
where $\psi_R^{\ C} = (\psi_R)^C$ is the charge-conjugated right-handed fermion $\psi_R$, $J_\mu^5  = \overline{\psi} \gamma_\mu \gamma_5 \psi$ is the  axial current of  the four-component Majorana fermion  $\psi = \psi_R  +  (\psi_R)^C$,  and  $\gamma$  is  a  real  parameter  to  be constrained later.  We have ignored gauge  fields, which are not of interest to us,  as well as the possibility of a non-perturbative mass $M_a$  for the  pseudoscalar $a(x)$.  Moreover,  we remind  the reader that the {\em repulsive} self-interaction fermion terms are due to the existence of torsion in the Einstein-Cartan theory.  The Yukawa coupling $y_a$ of  the axion moduli field $a$  to right-handed sterile neutrino $\psi_R$  may  be due  to  non-perturbative  effects. These \emph{break} the shift symmetry, $a \to a + c$. 

To evaluate the anomalous Majorana mass, it is convenient to diagonalize  the axion kinetic terms by redefining the KR axion field, $b(x) \rightarrow {b^\prime}(x) \equiv b(x) + \gamma a(x)$, and to appropriately redefine the axion field to appear with a canonically normalized kinetic term. In this way, the $b^\prime  $ field  decouples and can be integrated out  in the  path integral, leaving  behind an  axion field $a(x)$ coupled both to matter fermions  and  to   the  operator $R^{\mu\nu\rho\sigma}   {\widetilde   R}_{\mu\nu\rho\sigma}$, thereby playing the  role of the torsion field.   We observe through~\cite{Mavromatos:2012cc} that the approach is only valid for $ |\gamma|\ <\ 1$, since otherwise the  axion field would  appear as a  ghost, i.e.\ with the  wrong  sign  of  its  kinetic  terms,  which  would  indicate  an instability  of  the  model.  This  is the  only  restriction  of  the parameter $\gamma$. The resulting effective action reads:
\begin{eqnarray} 
  \label{bacoupl3} 
\mathcal{S}_a \!\! &=& \!\! \int d^4 x
    \sqrt{-g} \, \Big[\frac{1}{2} (\partial_\mu a )^2 - \frac{\gamma
        a(x)}{192 \pi^2 f_b \sqrt{1 - \gamma^2}}
      {R}^{\mu\nu\rho\sigma} \widetilde{R}_{\mu\nu\rho\sigma} \nonumber\\ 
&& \hspace{-5mm} - \frac{iy_a}{\sqrt{1 - \gamma^2}} \, 
a\, \Big( \overline{\psi}_R^{\ C} \psi_R - \overline{\psi}_R
\psi_R^{\ C}\Big) + \frac{1}{2f_b^2} J^5_\mu {J^5}^\mu
      \Big]\; .
\end{eqnarray}

\begin{figure}[t]
 \centering
  \includegraphics[clip,width=0.40\textwidth,height=0.15\textheight]{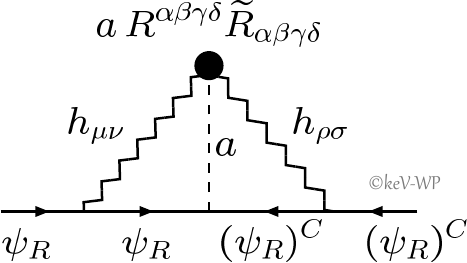} 
\caption{\label{fig:feyn}\it Typical Feynman graph giving rise to anomalous fermion mass generation.  The black circle denotes the operator $a(x)\, R_{\mu\nu\lambda\rho}\widetilde{R}^{\mu\nu\lambda\rho}$ induced by torsion. (Similar to Fig.~1 in ref.~\cite{Mavromatos:2012cc}.)}
\end{figure}

The mechanism for  the anomalous Majorana mass generation is shown in fig.~\ref{fig:feyn}.   Adopting  the   effective   field theory  framework of~\cite{Donoghue:1994dn}, one may estimate  the  two-loop  Majorana neutrino mass $M_M$ in  torsional quantum gravity with an effective  UV energy cut-off $\Lambda$~\cite{Mavromatos:2012cc},  
\begin{equation}
  \label{MR}
M_M \sim \frac{1}{(16\pi^2)^2}\;
\frac{y_a\, \gamma\  \kappa^4 \Lambda^6}{192\pi^2 f_b (1 - \gamma^2 )} =
\frac{\sqrt{3}\, y_a\, \gamma\,  \kappa^5 \Lambda^6}{49152\sqrt{8}\,
\pi^4 (1 - \gamma^2 )}\; . 
\end{equation}
In a UV complete theory  such as  strings, $\Lambda$  and $M_P$  are proportional to each other, since $\Lambda$ is proportional to  the string scale $M_s=1/\sqrt{\alpha^\prime}$ ($\alpha^\prime$ the Regge slope) and the latter is related to $M_P$  through $\frac{1}{g_s^2} \, M_s^2 \, V^{(c)} = \frac{M_P^2}{16\pi}$, with $g_s$ being the string coupling, and $V^{(c)}$ the compactification volume in units of $\alpha^\prime$. It is interesting  to provide a numerical estimate  of the anomalously generated Majorana mass $M_M$. Assuming that $\gamma \ll 1$, the size of $M_M$ may be estimated from~\eqref{MR} to be $M_M \sim (3.1\times 10^{11}~{\rm GeV})\bigg(\frac{y_a}{10^{-3}}\bigg)\; \bigg(\frac{\gamma}{10^{-1}}\bigg) \bigg(\frac{\Lambda}{2.4 \times 10^{18}~{\rm GeV}}\bigg)^6$. Obviously, the generation of $M_M$ is highly model dependent.  Taking, for example, the quantum gravity  scale to be $\Lambda = 10^{17}$~GeV, we  find that  $M_M$ is  at the  TeV scale,  for $y_a  =  10^{-3}$ and $\gamma =  0.1$. However, if we  take the quantum gravity  scale to be close  to the  GUT scale,  i.e.~$\Lambda =  10^{16}$~GeV, we  obtain a right-handed neutrino  mass $M_M \sim  16$~keV, for the choice  $y_a = \gamma = 10^{-3}$.   This is in the preferred  ballpark region for the sterile   neutrino    $\psi_R \equiv \nu_R$ to qualify as a WDM~\cite{Asaka:2006ek}, e.g.\ in the framework of the $\nu$MSM model~\cite{Asaka:2005an}, of interest in this white paper.

We  conclude this  subsection  by noting  that  in a  string-theoretic framework,  many   axions  might  exist  that  could   mix  with  each other~\cite{Arvanitaki:2009fg,Cicoli:2012sz}.  Such  a mixing can give rise  to  reduced  UV  sensitivity  of the  two-loop  graph  shown  in fig.~\ref{fig:feyn}.   For  instance,   in  models  with  three  axion species, it is possible to  obtain two-loop induced masses that are UV finite, and as such independent  of the UV-cutoff $\Lambda$.  For  more details on neutrino mass  estimates in  such string-theoretic settings we refer the reader to~\cite{Mavromatos:2012cc}.


%% file: kevnuwp_section7.tex

\subsection{\label{section7.1}Previous Bounds (Authors: A.~Boyarsky, J.~Franse, A.~Garzilli, D.~Iakubovskyi)}

When looking for fermionic DM particles, one should in general search above the Tremaine-Gunn limit~\cite{Tremaine:1979we} of about $ 300 \unit{eV}$). Essentially, below this mass, the phase space density of Dark Matter particles that would be required by the observed amount of Dark Matter in dwarf galaxies, would violate the Pauli-exclusion principle. If the mass is below $2m_e c^2$, such a fermion can decay to neutrinos and photons with energy $E_\gamma=\frac12 m_\dm$~\cite{Pal:82}. One can search for such particles in X-rays~\cite{Abazajian:2001vt}, either by aiming to detect the diffuse X-ray background~\cite{Dolgov:2000ew} or by looking at nearby DM-dominated astrophysical objects~\cite{Abazajian:2001vt}. As a side note, Ref.~\cite{Abazajian:2001vt} also proposed the stacking method which was later on involved in the detection of the 3.5~keV hint~\cite{Bulbul:2014sua}.

The omni-presence of Dark Matter in galaxies and galaxy clusters opens the way to check the decaying Dark Matter hypothesis~\cite{Boyarsky:2010ci}. The decaying Dark Matter signal is proportional to the \emph{column density} $\S_\dm=\int \rho_\dm d\ell$ -- the integral along the line of sight of the DM density distribution (unlike the case of annihilating Dark Matter, where the signal is proportional to $\int \rho_\dm^2 d\ell$). As long as the angular size of an object is larger than the field-of-view, the distance to the object drops out, which means that distant objects can give fluxes comparable to those of nearby ones~\cite{Boyarsky:2009af,Boyarsky:2009rb}. To be explicit, the flux in a Dark Matter decay signal is given by
\begin{equation}
  F=\frac{ M_\text{fov}}{4\pi D_L^2}\frac{\Gamma_{\dm}}{m_{\dm}} 
\end{equation}
where $m_{\dm}$ is the DM particle mass, $D_L$ the luminosity distance and $M_\text{fov}$ the amount of Dark Matter in the field of view of the telescope. The decay rate $\Gamma_\dm$ is related to the particle lifetime as $1/\tau_\dm$ and is given by
\begin{small}
  \begin{displaymath} \tau_\dm = \frac{1024\pi^{4}}{9 \alpha
      G_{F}^{2}\sin^{2}( 2\theta) m_{\dm}^{5}} = 7.2 \times 10^{29} \sec
    \left[\frac{10^{-8}}{\sin^{2}(2\theta)}\right]\left[\frac{1\keV}
      {m_\dm}\right]^{5}\, ,
  \end{displaymath}
\end{small}
if DM is made of right-handed (sterile) neutrinos~\cite{Dodelson:1993je}, with $\sin^2 (2\theta)$ the interaction strength (\emph{mixing angle}).

In the case of decaying DM, the signal strength also decreases more slowly with increasing distance from the centers of objects as compared to the case of annihilating DM where the expected signal is concentrated towards the centers of DM-dominated objects.  This in principle allows one to check the Dark Matter origin of a signal by comparison between objects and/or by studying the angular dependence of the signal within one object, rather than trying to exclude all possible astrophysical explanations for each target~\cite{Boyarsky:2006fg,Herder:2009im,Abazajian:2009hx}.

The DM decay line is much narrower than the spectral resolution of the existing X-ray telescopes and, as previous searches have shown, should be rather weak. The X-ray spectra of astrophysical objects are crowded with weak atomic and instrumental lines, not all of which may be known. Therefore, even if the exposure of available observations continues to increase, it is hard to exclude an astrophysical or instrumental origin of any weak line found in the spectrum of individual object. However, if the same feature is present in the spectra of many different objects, and its surface brightness and relative normalization between objects are consistent with the expected behavior of the DM signal, this can provide much more convincing evidence about its nature.

Many studies have indeed been performed on the widest range of objects. From objects as small as dwarf galaxies and as nearby as the center and halo of our own Milky Way, to other galaxies and clusters of galaxies, and more specialized searches such as through stacking of many objects or studying the peculiar `dark blobs' like the Bullet cluster and A520 (\cite{Abazajian:2001vt,Dolgov:2000ew,Boyarsky:2006fg,Boyarsky:2005us,Boyarsky:2006zi,RiemerSorensen:2006fh,Watson:2006qb,  RiemerSorensen:2006pi,Boyarsky:2006ag,Abazajian:2006yn,Boyarsky:2006kc,Boyarsky:2006hr,Yuksel:2007xh,Boyarsky:2007ay,  Boyarsky:2007ge,Loewenstein:2008yi,RiemerSorensen:2009jp,Loewenstein:2009cm,Boyarsky:2010ci,Mirabal:2010jj,Mirabal:2010an,  Prokhorov:2010us,Borriello:2011un,Watson:2011dw,Loewenstein:2012px,Kusenko:2012ch,Horiuchi:2013noa}).

In reality, of course, after years of systematic search for this signal, any candidate line can be detected only at the edge of the possible sensitivity of the method. Therefore, one needs comparatively long-exposure data to cross-check the signal. Moreover, even an uncertainty at the level of a factor of a few in the expected signal, which is very hard to avoid, can result in the necessity to have significantly more statistics than in the initial data set in which the candidate signal was found. The parameter space that is available for a DM sterile neutrino within the $\nu$MSM is shown in fig.~\ref{fig:exclusion_plot}.

\subsection{\label{ssc:Xray355keV}X-ray telescopes  and observation of the 3.55 keV Line (Authors: A.~Boyarsky, J.~Franse, A.~Garzilli, D.~Iakubovskyi)}

Early in 2014, two independent groups reported a detection of an unidentified X-ray line at an energy of $\sim$3.55~keV in the long-exposure X-ray observations of a number of Dark Matter-dominated objects: first in a stack of 73 galaxy clusters~\cite{Bulbul:2014sua} and in the Andromeda galaxy and the Perseus galaxy cluster one week later~\cite{Boyarsky:2014jta}.

The possibility that this spectral feature may be the signal from decaying Dark Matter has sparked a lot of interest in the community as the signal has passed a number of `sanity checks' expected for Dark Matter decay: it is consistent with the expected mass scaling between the Andromeda galaxy, the Milky Way center and the upper bound from non-detection in the blank sky data~\cite{Boyarsky:2014jta,Boyarsky:2014ska}, and also between different sub-samples of clusters~\cite{Bulbul:2014sua}. The signal has radial surface brightness profiles in the Perseus cluster, Andromeda galaxy~\cite{Boyarsky:2014jta} and in the Milky Way~\cite{Boyarsky:2014ska} that are consistent with our expectations for decaying Dark Matter and the mass distribution in these objects. Instrumental origins of this signal have been shown to be highly unlikely for a variety of reasons: the signal's presence in all of \xmm's detectors and in \chan's; the correct behavior with redshift (especially in the stacking procedure employed by ref.~\cite{Bulbul:2014sua}); its absence in a very-long exposure blank-sky dataset. It has also not proved possible to attribute the signal plausibly and consistently between datasets to any known atomic emission line or lines~\cite{Boyarsky:2014jta,Boyarsky:2014paa,Bulbul:2014sua,Bulbul:2014ala}. 

The center of our Galaxy is a classical target for DM searches. Its proximity allows the observations to concentrate on the very central part and therefore, one can expect a significant gain in the signal if the DM distribution in the Milky Way happens to be steeper than a cored profile. The Galactic Center (GC) region has been extensively studied with \xmm\ and \chan\ and several mega-seconds of raw exposure have been accumulated to date. Three recent studies~\cite{Riemer-Sorensen:2014yda,Boyarsky:2014ska,Jeltema:2014qfa} argued that this line may be a potassium K~\textsc{XVIII} transition (as it is difficult to put an upper bound on the potassium abundance in the GC). However, as was argued in ref.~\cite{Boyarsky:2014ska}, the correct question is: \emph{Are the observations of the Galactic Center consistent with the Dark Matter interpretation of the 3.55~keV line of} ref.~\cite{Bulbul:2014sua,Boyarsky:2014jta}?  
  
This is a non-trivial question, as the DM interpretation of the 3.55 keV line in M31 and the Perseus cluster puts a \emph{lower limit on the flux} expected from the GC. On the other hand, the non-detection of any signal in the off-center observations of the Milky Way halo (the blank sky dataset of ref.~\cite{Boyarsky:2014jta}) \emph{provides an upper limit} on the possible flux from the GC, given observational constraints on the DM distribution in the Galaxy. ref.~\cite{Boyarsky:2014ska} argued that the flux from the GC falls into this range, while a consistent interpretation of the line in all the datasets as an atomic transition is problematic~\cite{Bulbul:2014ala,Boyarsky:2014paa}.

\begin{figure}
\centering
\includegraphics[width=0.9\linewidth]{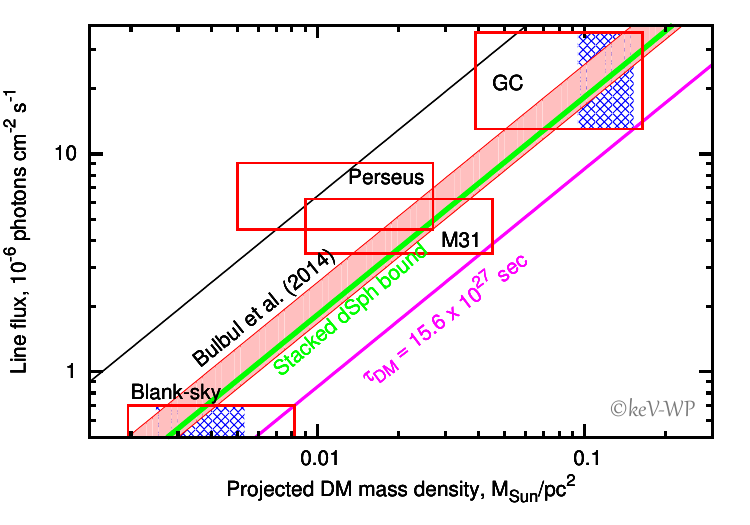}
\caption{\label{fig:flux-mproj}\footnotesize The flux in the 3.55~keV line in the spectra of the Galactic center, the Perseus cluster, and M31, as well as the $2\sigma$ upper bound from the blank sky dataset (from ref.~\cite{Boyarsky:2014jta}) as a function of the expected DM signal; see also Ref.~\cite{Horiuchi:2013noa} which provides one of the strongest constraints on the line from the Chandra of M31. The results from ref.~\cite{Bulbul:2014sua} are shown as the diagonal shaded strip (width $\pm 1\sigma$). The vertical sizes of the boxes are $\pm 1 \sigma$ statistical error on the line's flux. The horizontal sizes of the boxes represent systematic errors in mass modeling: the projected mass density calculated using different literature models of the DM distribution. Diagonal lines correspond to different lifetimes of DM particles, assuming a decaying DM interpretation of the signal. Blue shaded regions give an example of the relation between the projected DM mass densities in the GC and in the blank-sky dataset based on the Milky Way mass model of ref.~\cite{Weber:2009pt}. The  magenta line shows the \emph{estimated} sensitivity of 1.4~Msec of Draco observations~\cite{Lovell:2014lea}. The actually derived limits, however, differ~\cite{Jeltema:2015mee,Ruchayskiy:2015onc}, see the discussion in sec.~\ref{sec:xray}. The values of the lifetime above the \textbf{green line} ($\tau_\dm = 7.3\times 10^{27}$~sec) are excluded at $3\sigma$ level, based on ref.~\cite{Malyshev:2014xqa}.}
\end{figure}

Even though the recent studies of stacked spectra of dwarf spheroidal satellites of the Milky Way by~ref.~\cite{Malyshev:2014xqa}, and galaxy clusters with \suza by~ref.~\cite{Urban:2014yda} and~ref.~\cite{Tamura:2014mta} are in tension with the initial value of the DM lifetime from~ref.~\cite{Bulbul:2014sua} (which is based on their estimates of the DM content of their galaxy clusters), all of this work still leaves parameter-space for a common interpretation of all detections and non-detections in terms of DM. An overview of the detections and how they relate to each other and to the Dark Matter interpretation can be found in figure~\ref{fig:flux-mproj}. Analyses by~ref.~\cite{Anderson:2014tza} and~ref.~\cite{Carlson:2014lla} make strong claims for ruling out the DM interpretation using sophisticated techniques. However, uncertainties in the DM content and distributions are still being debated~\cite{Boyarsky:2014jta,Boyarsky:2014ska}; moreover, Ref.~\cite{Franse:2016dln} revealed problems in the morphological analysis of~\cite{Carlson:2014lla}. We have to refrain from making comment beyond this and would like to refer the reader to these references. 

\subsection*{Future Searches for the 3.55~keV Line}

As mentioned in the previous section, the existing non-detections limit the range of DM-lifetimes associated with the 3.55 keV excess; specifically to $\tau_\dm=(7.3-15.6)\times 10^{27}$~sec (green and magenta lines in Fig.~\ref{fig:flux-mproj}). Given that the comparison of the results from different objects depends on assumptions about their DM content, the convincing ultimate test of the DM interpretation of the 3.55~keV signal requires an observation that allows for a detection of at least $3\sigma$ for the longest admissible lifetime $\tau_\dm = 15.6 \times 10^{27}$~sec.

By the beginning of 2016, the forthcoming \textit{Astro-H}~\cite{Takahashi:2012jn} mission and the planned \textit{Micro-X} sounding rocket experiment~\cite{Figueroa-Feliciano:2015gwa} were expected have sufficient spectral resolution to resolve the line against other nearby features and to detect the candidate line in the ``strong line'' regime~\cite{Boyarsky:2006hr}. In particular, the general hope was \textit{Astro-H} should be able to resolve the Milky Way halo's DM decay signal and therefore all its observations can be used, which unfortunately did not happen -- see paragraph below.

The large field-of-view of the upcoming \textit{eROSITA} X-ray observatory \citep{2012arXiv1209.3114M} will allow for tight constraints on the line by homogeneously covering nearby X-ray bright clusters to large radii. Cross-correlation power spectra of the cosmic X-ray background in the \textit{eROSITA} all-sky survey will deliver additional competitive constraints \citep{Zandanel:2015xca}.

\subsection*{A short wrap-up of the sad Hitomi story}

The Japanese Space Agency (JAXA) successfully launched the Astro-H satellite from Tanegashima Space Center in Japan on the 16th of February 2016.  After it was placed in orbit and its solar panels deployed, the spacecraft was renamed Hitomi. However, JAXA lost contact with the satellite on March 26th, while the it was executing its first test observations in orbit. Several sources, including the U.S.\ Joint Space Operation Center and various amateur observers, have reported that the satellite has broken up into several pieces of debris and changed orbit. In a press briefing on April 8th, JAXA stated that the Hitomi satellite appears to rotate with a period of 5.2~s, and that this rotation may have led to the separation of vulnerable pieces from the spacecraft's main body. It also confirmed that 10 other objects have been detected nearby on April 1st. Already at that time, it seemed possible that Hitomi (or Astro-H) would lost and could be recovered, which would make XMM-Newton and Chandra the only operational X-ray telescopes.  In spite of the difficulties, the Hitomi team had been working hard to stabilize the satellite orbit with the hope of recovering part of the instrumentation. However, despite their best efforts, JAXA ceased efforts to rescue the satellite on April 28th, as it became clear that the spacecraft could not be recovered. From the updated information released by JAXA, it seems that the reason behind the failure was a too fast rotation that was caused unintentionally to compensate for a non-existing rotation in the opposite direction indicated by the software. This rotation was so fast that it took the satellite beyond its design limit, therefore causing main parts of the spacecraft to separate. See \url{http://global.jaxa.jp/projects/sat/astro_h/} for more details. It is worth to not that, despite the satellite being lost, some data was collected from the Perseus cluster~\cite{Aharonian:2016gzq}; unfortunately, while yielding en even stronger constraint on a possible line signal, the statistics achieved had not been large enough to fully clear up the sitution.

\subsection{\label{sec:LymanAlphaBounds}Lyman-$\alpha$ Methods for keV-scale Dark Matter (Authors: A.~Boyarsky, J.~Franse, A.~Garzilli, D.~Iakubovskyi)}

The study of the matter power spectrum in the intergalactic medium (IGM) has
been extensively used in the past to constrain the nature of DM -- see for
example~\cite{Hansen:2001zv,Viel:2005qj,Viel:2006kd,Seljak:2006qw,Viel:2007mv,Boyarsky:2008mt,Boyarsky:2008xj,Markovic:2013iza,Viel:2013apy},
see also Section~\ref{sec:4-2-LymanA}. The IGM is constituted by the under-dense gas that fills the otherwise empty space between galaxies. Because the IGM is gravitationally dominated by the DM distribution, it is sensitive to the spatial distribution of DM; on the other hand its neutral fraction interacts with light. Hence, the IGM can be used as a probe of the DM power spectrum.

The physics of the IGM is relatively straight-forward with respect to other probes considered currently in cosmology, being constituted mostly by gas with almost primordial composition, and being organized in large scale structures that lie just outside the linear regime. Because of the relative simplicity of its physical description, the IGM is the natural choice for constraining cosmology, after the study of the cosmic microwave background. In particular, we are interested to get new limits on the nature of DM, from the observable of the IGM in the ultraviolet and optical bands, the so-called Lyman-$\alpha$ forest. This is the signature of the neutral component of the IGM on the spectra of distant quasars, and it is due to the scattering of the quasar's light by the neutral hydrogen atoms through the Lyman-$\alpha$ transition. The ideal time for investigating the nature of DM through the Lyman-$\alpha$ forest is at high redshift, when the effect of AGN feedback is not as important as it can be at lower redshift.

Because we are interested in the smallest scales observable with the Lyman-$\alpha$ forest, where the difference between the CDM and the sterile neutrinos is the largest, we are susceptible to the unknown IGM temperature. In fact, the thermal history of the IGM is largely unknown, and in~\cite{Garzilli:2015iwa} the significant effect that arbitrary assumptions on the cosmic thermal history can have on SNe constraints have been made explicitly. Moreover, the authors have revised the current constraints on SNe, and we recognized that the latest analysis on high redshift and high resolution Lyman-$\alpha$ spectra, presented in~\cite{Viel:2013apy}, does not constrain the warmness of DM better than previous works (see~\cite{Seljak:2006qw} and~\cite{Boyarsky:2008xj}).

The temperature of the IGM has been measured from the broadening of the lines that constitute the Lyman-$\alpha$ forest, but one of the obstacles has been to discern between the size of the filaments that constitute the IGM, and the IGM temperature, as both the temperature and the physical extent of the clouds affect the broadening. In recent work~\cite{Garzilli:2015bha}, a new method for measuring both the size of the filaments and the temperature of the IGM has been outlined for the first time. Previous works on constraining the nature of DM through the Lyman~$\alpha$ forest considered the flux power spectrum, in which the information about the initial matter power spectrum is mixed with astrophysical uncertainties.  We believe that the line decomposition method considered in~\cite{Garzilli:2015bha} can be adapted to the case of WDM cosmologies, and that in this way we can obtain new constraints on the nature of the DM, with an approach that is radically different from the one used in the previous works.

\subsection{Pulsar kicks (Author: S.~Hansen)}

When massive stars end their life in a supernova explosion a neutron star is born, and radio pulsars are a small fraction of the billions of neutron stars in the Milky Way.  Observations show that these radio pulsars have large velocities, with averages of several hundred km/sec, and with a non-Gaussian tail extending up to 1600~km/sec~\cite{Arzoumanian:2001dv,Hobbs:2005yx}. It is non-trivial to achieve such large velocities in standard core collapse supernova explosions~\cite{Janka:2012wk}. In addition, any model predicting these large velocities should preferably also explain the observed correlation between the spin axis and the velocity of the pulsars~\cite{Wang:2005jg,Ng:2007aw}.

Standard neutrinos carry away up to 99$\%$ of the $\sim 10^{53} {\rm erg}$ of gravitational energy released in the collapse of the supernova, however, due to the large temperatures and densities, the neutrino scattering rate is in fact extremely high, and standard neutrinos diffuse spherically from the supernova core. Instead, if sterile neutrinos could be produced asymmetrically, then they would leave the supernova non-spherically. Indeed, with only a few percent asymmetry in the neutrino distribution, one could explain both the observed velocities and the correlation with the spin axis. The magnetic field inside the proto-neutron star provides exactly such an asymmetry by polarizing the electrons~\cite{Kusenko:1997sp}, and as we will describe below the combination of a large magnetic field and the mixing between a sterile neutrino with a $\mu$- or $\tau$-neutrino allows for a plausible explanation of both the pulsar velocities as well as the correlation with the spin axis~\cite{Kusenko:2009up}.

The sterile neutrinos are created in the standard reactions
\begin{eqnarray}
p + e^- &\leftrightarrow& n + \nu_e, \nonumber \\
n+e^+ &\leftrightarrow& p +\bar \nu_e, \, 
\end{eqnarray}
reduced by the mixing angle, $1/2 \times {\rm sin}^2 \left(2\theta\right)$. In vacuum the production rate is negligible when the mixing angle is small, however, the matter-dependent potential may lead to the an MSW resonance, which implies that the active and sterile neutrinos are maximally mixed. This means that, for a given large magnetic field, there may exist a specific radius in the proto-neutron star where the temperature and density produce sterile neutrinos with a very specific momentum. And these neutrinos may then escape the supernova unhindered.

The surface of this resonance point is non-spherical due to the magnetic field, and it can be written as
\begin{equation}
r_{\rm res} (\phi) = r_0 + \delta \, {\rm cos}\phi \, ,
\end{equation}
where $\phi$ is the angle between the neutrino momentum and the direction of the magnetic field and $\delta$ depends on the magnitude of the magnetic field and the chemical potential of the degenerate electron gas. The different momenta of sterile neutrinos are therefore created at different radii (with different temperatures) and this leads to an effective momentum transfer to the entire proto-neutron star and the asymmetry in the momentum distribution may be written~\cite{Kusenko:2009up}
\begin{equation}
\frac{\Delta k_s}{k} = 0.01 \, 
\left( \frac{\mu_e}{100 {\rm MeV}} \right)
\left( \frac{\mu_n}{80 {\rm MeV}} \right)^{1/2}
\left( \frac{20 {\rm MeV}}{T} \right)^2
\left( \frac{B}{3 \times 10^{16} {\rm G}} \right) \, ,
\end{equation}
where $\mu_e$ and ${\mu_n}$ are the chemical potentials of electrons and neutrons, respectively, while $T$ and $B$ are averaged core temperatures and magnetic field strengths.

Detailed calculations~\cite{Kusenko:2008gh,Kishimoto:2011mw} find that sterile neutrinos with masses about $5-20$ keV and mixing angles in the range of $10^{-11} < {\rm sin^2 \theta} < 5 \times 10^{-10}$ may provide the pulsar a sufficiently large kick, for proto-neutron star parameters in the range of $B> {\rm few} \, 10^{16} G$, $T>15 {\rm MeV}$, and central density $\rho > 10^{12} {\rm g/cm}^3$.

An alternative non-resonant production mechanism was described in~\cite{Fuller:2003gy} (see also~\cite{Kusenko:2008gh}) which allows one to kick the pulsars sufficiently for slightly smaller neutrino masses, between 2~and 10~keV, and larger mixing angles up to ${\rm sin^2 \theta} \sim \times 10^{-7}$.


\subsection{Supernovae (Authors: S. Hansen and S. Zhou}
When a massive star collapses to produce a supernova, the vast amount of
gravitational energy is released mainly in neutrinos. During and
immediately after the collapse the active neutrinos are trapped in the
core of the supernova and they diffuse out slowly on a timescale of
the order 10 seconds. The observed duration of neutrino signals from SN1987A matched well with the theoretical predictions. However, if one of the active neutrinos is mixed with a sterile state, then the sterile neutrino may carry
away too much of the energy, and thereby unduly shorten the neutrino
burst. The discussion in this section uses exactly this energy-loss argument to constrain the allowed mass and mixing angle of a new sterile state.

\subsubsection{The vacuum limit}
When the supernova core emits energy in an invisible channel, the
duration of neutrino signals shortens, potentially leading to
disagreement with the observation of SN1987A. The result of
\cite{Raffelt:1999tx} is that such exotic energy loss should obey
\begin{equation}
\varepsilon < 10^{19} \, {\rm erg} \, {\rm g}^{-1}\, {\rm s}^{-1} \, ,
\label{eq:snconstraint}
\end{equation}
where $\varepsilon$ must be calculated at typical values for the
density and temperature of the supernova core, here taken as $\rho=3
\times 10^{14} \, {\rm g} \, {\rm cm }^{-3}$ and $T=30 \, {\rm MeV}$.

If one of the active neutrinos is mixed with a sterile state, then
the sterile neutrinos may be produced through both oscillations and frequent scattering with background particles, and carry away energies. The neutrino oscillation length in matter $\lambda_{\rm osc}$ is much shorter than the mean free path $\lambda_{\rm mfp}$ at typical supernova core
temperatures~\cite{Raffelt:2011nc}
\begin{equation}
\frac{\lambda _{\rm mfp}}{\lambda _{\rm osc}} \approx 10^3 \, \left(
\frac{{\rm sin} 2\theta}{10^{-4} }\right) \, \left( \frac{m_s}{ 10~
  {\rm keV}} \right)^2 \, ,
\end{equation}
which implies that once an active neutrino is produced, it will
oscillate many times before collisions interrupt the coherent
development of the flavor amplitude.  For high mass sterile neutrinos,
$m_s > 100 ~{\rm keV}$, the matter effects can be neglected. The
sterile neutrino production rate is given by
\begin{equation}
\Gamma_s = \frac{1}{2} \, {\rm sin} ^2 2 \theta \, \Gamma_a \, ,
\end{equation}
where $\Gamma_a$ is the collision rate of active neutrinos on free nucleons. For $\nu _\mu$ or $\nu_\tau$, the neutral-current collisions are dominant and we obtain $\Gamma_a = G_F^2 E_\nu^2 n_B /\pi$ with $n_B$ being the number density of baryons. For $\nu _e$, the production and absorption on nucleons via the charged-current interaction are important, and the collision rate is $\Gamma_a = (C_V^2+3C_A^2) G_F^2 E_\nu^2 n_B /\pi$. In the case where one can
ignore significant chemical potentials, the energy-loss rate for
mixing with $\nu _\mu$ or $\nu_\tau$
becomes~\cite{Dolgov:2000jw,Dolgov:2000pj,Raffelt:2011nc}
\begin{equation}
\varepsilon_{\nu_s} = \left( \frac{{\rm sin}^2 2\theta}{3 \times 10^{-8}} \right) \, 10^{19} \, {\rm erg} \, {\rm g}^{-1}\, {\rm s}^{-1} \, .
\end{equation}
Together with eq.~(\ref{eq:snconstraint}) this leads to
\begin{equation}
{\rm sin}^2 2 \theta_{s\tau} < 3 \times 10^{-8} \, .
\end{equation}

For mixing with electron-neutrinos the calculation is slightly more
complicated due to the chemical potential of the electron neutrinos,
and the result is~\cite{Kainulainen:1990bn,Raffelt:1992bs}
\begin{equation}
{\rm sin}^2 2 \theta_{s e} < 10^{-10} \, .
\end{equation}

\begin{figure}[]
\includegraphics[width=0.9\linewidth]{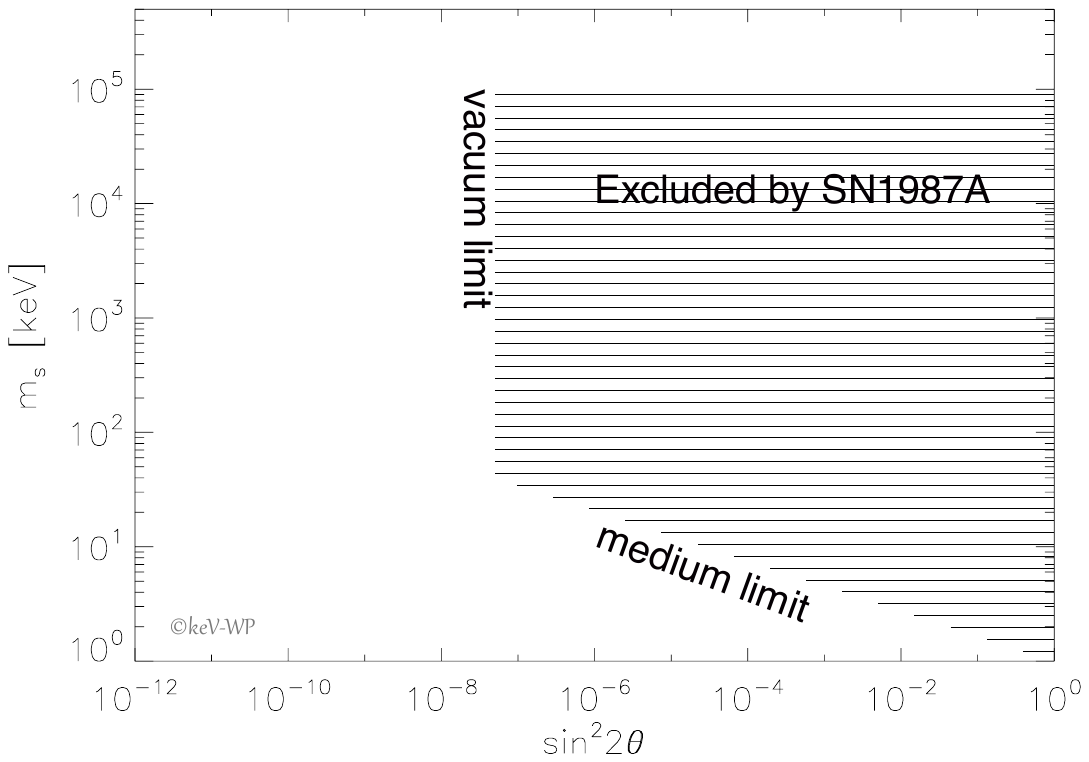}
\caption{Constraints on the mixing parameters of a sterile neutrino
  mixed with a $\tau$ neutrino. The limits are rather similar for
  mixing with a $\mu$ neutrino. For mixing with electron neutrinos the
  vacuum limit (high mass) is stronger, ${\rm sin}^2 2 \theta <
  10^{-10}$, however, the calculations are more involved and less
  accurately known for smaller masses ($m_s < 100 {\rm keV}$). }
\label{supernovafig}
\end{figure}

The constraints above are valid for sterile neutrinos heavier than
$m_s > 100 \, {\rm keV}$, where the matter effects can be neglected. However, they are only valid for masses smaller than $m_s < 100 \, {\rm MeV}$: for typical core-collapse temperatures of $30 \, {\rm MeV}$, the average thermal neutrino energy is about $100 \, {\rm MeV}$, and thus heavier sterile states cannot be produced abundantly. For such massive neutrinos and maximal mixing the lifetime is about $10 \, \mu
{\rm s}$, corresponding to distances of few km, i.e. of the order of
the supernova core, and much longer than the mean free path of the active
neutrinos of the order 10 meters.  Thus, all the parameter space of
massive neutrinos is excluded for mixing angles larger than the
constraints above~\cite{Dolgov:2000jw,Raffelt:2011nc}.  This is shown
as ``vacuum limit'' in fig.~\ref{supernovafig}.

\subsubsection{Matter effects}

The scattering of neutrinos off background particles will modify the
dispersion relation of neutrinos in matter. This matter effect can be
described by an effective potential, $V_{\nu_\alpha}$ for each kind of
active neutrinos. One therefore has an effective mixing angle which
depends on the properties of the background
\begin{equation}
{\rm sin}^2 2 \theta_M = \frac{{\rm sin}^2 2 \theta}{{\rm sin}^2 2 \theta
+ \left( {\rm cos}2 \theta + E/E_r\right)^2} \,
\label{eq:resonantangle}
\end{equation}
where the resonant energy is defined as $E_r = \Delta m^2 /2\left| V_{\nu_\alpha}\right|$. The sign in the second term of the denominator is taken to be positive for neutrinos and a negative $V_{\nu_\alpha}$, and it becomes negative either for antineutrinos or for a positive potential. For $\nu_\tau$, the resonant energy is given by~\cite{Raffelt:2011nc}
\begin{equation}
E_r = 3.25 \, {\rm MeV} \, \left( \frac{m_s}{10 \, {\rm keV}} \right) ^2
\rho_{14} ^{-1} \left| Y_0-Y_{\nu_\tau} \right|^{-1} \, ,
\end{equation}
where $\rho_{14}$ is the matter density in units of $10^{14}
{\rm g} \, {\rm cm}^{-3}$, and $Y_0 = \left( 1-Y_e-2Y_{\nu_e}
\right)/4$ using $Y_x = (n_x - n_{\bar x})/n_B$, where $n_x$ and $n_{\bar x}$ stand for the number density of particles $x$ and that of their antiparticles, respectively. Given $Y_e = 0.3$ and $Y_{\nu_e} = 0.07$, the effective potential $V_{\nu_\tau}$ is negative, so the sign in the denominator in eq.~(\ref{eq:resonantangle}) is positive and the resonant oscillations happen in the antineutrino sector.

For very large masses, $m_s > 100 \, {\rm keV}$, the vacuum limit is
obtained whereas for small masses, $m_s < 1 \, {\rm keV}$ the medium
limit reached \cite{Kainulainen:1990bn,Raffelt:1992bs}, and the smooth
transition between these two limits lies around $40 \, {\rm keV}$
\cite{Dolgov:2000ew}.  One effectively has a shift from the vacuum
limit calculation of
\begin{equation}
{\rm sin} ^2 2 \theta \rightarrow {\rm sin} ^2 2
\theta \times \left( \frac{m_s}{40 {\rm keV}} \right) ^4 \, ,
\end{equation}
for small masses.  The constraints in
the matter dominated region is indicated by ``medium limit'' in
fig.~\ref{supernovafig}.

With the help of eq.~(\ref{eq:resonantangle}) one sees that the mixing angle in matter is always suppressed for neutrinos, whereas antineutrinos may experience matter enhancement when $E \sim E_r \, {\rm cos}2 \theta$. This
happens for masses between 20 and 80 keV. The mixing angle of
antineutrinos is always larger than that of the neutrinos, and hence
the emission rate of the antineutrinos always exceeds that of the
neutrinos. Hence, an asymmetry between the active neutrinos will therefore
build up in the supernova core. Some of the effects hereof have been
discussed \cite{Abazajian:2001nj, Raffelt:2011nc} and the conclusion
is, that for masses near the resonance ($m_s \sim 50 {\rm keV}$) the
exclusion region increases by almost an order of magnitude for mixing
with a tau neutrino. For mixing with electron neutrinos the
calculation is significantly more complicated due to the high $\nu_e$
degeneracy and the large electron number. More explicitly, the resonant energy in this case is \cite{Abazajian:2001nj, Zhou:2015jha}
\begin{equation}
E_r = 13 \, {\rm MeV} \, \left( \frac{m_s}{10 \, {\rm keV}} \right) ^2
\rho_{14} ^{-1} \left| 1 - 3 Y_e - 4 Y_{\nu_e}\right|^{-1} \, .
\end{equation}
The effective potential $V^{}_{\nu_e}$ is initially positive for $Y_e = 0.3$ and $Y_{\nu_e} = 0.07$, indicating the resonance in the neutrino sector. However, as the lepton number decreases due to the emission of sterile neutrinos, the resonant energy is rapidly pushed to infinity corresponding to the condition $1 - 3 Y_e - 4 Y_{\nu_e} = 0$. As a consequence, the constraints in the vacuum limit become applicable to the cases of $m_s \ll 100 \, {\rm keV}$ as well \cite{Abazajian:2001nj, Zhou:2015jha}.

The mixing with electron neutrinos may also affect the explosion of
the supernova itself. Intuitively one would expect that mixing would
reduce the energy of the supernova, and hence make it even more
difficult to make supernovae explode.  Several papers have suggested
alternatives, for instance through a reconversion which could
transport energy from the deep core to the region just below the
neutrino sphere~\cite{Hidaka:2007se}. This mechanism may be
operational for masses of few keV.  Alternatively, an anisotropic
neutrino emission is also found to be able to enhance the energy of
the exploding supernova~\cite{Fryer:2005sz}.

An interesting possibility for boosting the supernova explosion
includes a neutrino with a mass above the pion
mass~\cite{Fuller:2009zz}.  Such neutrinos will preferentially decay
to a pion, which immediately decays to two photons
\cite{Dolgov:2000jw} $\nu_s \rightarrow \nu_\alpha \pi^0\rightarrow
\nu_\alpha \gamma \gamma$.  If the mixing angle is extremely small,
such a decay may help re-ionize the Universe~\cite{Hansen:2003yj},
however, if the mixing angle is just right, ${\rm sin}^2 2 \theta \sim
10^{-7}$, then the sterile neutrinos may decay before leaving the
supernova.  In this way one may produce the sterile neutrino in the
hot core of the supernova, and its decay may deposit energy near the
shock front. Thereby one alters the supernova energetics which may
increase the possibilities of a supernova
explosion~\cite{Fuller:2009zz}.


%% file: kevnuwp_section8.tex
\subsection{Introduction}
Direct dark matter experiments are an essential probe to complement indirect searches in a largely model-independent way. In analogy to WIMP searches, two distinct paths can be considered. 

First, one may produce the dark matter particle in a laboratory setup and detect its presence via kinematic considerations. Section~\ref{ssc:tritium} focuses on this approach and discusses in detail a possible realization through tritium $\beta$ decay. Section~\ref{ssc:holmium} explores the sensitivity of this detection techniques with electron capture on holmium and other isotopes. 

The second option is a direct detection of the dark matter particle present in our galaxy by using large-scale detectors. The density, energy distribution, and cross-sections of sterile neutrino dark matter is significantly different from WIMPs in typical cold dark matter scenarios. Nevertheless, the ability of existing direct DM search experiments (such as Xenon~\cite{Aprile:2012nq} and LUX~\cite{Akerib:2013tjd}) to detect sterile neutrinos via elastic scattering is currently being investigated~\cite{Werner}. Alternatively, one may consider dedicated experiments directly detecting sterile neutrino dark matter for example via inverse $\beta$ decay. A detailed discussion of this approach is given in section~\ref{ssc:Direct}.

Current laboratory limits for the mixing of sterile neutrinos with the electron neutrino flavour are of the order of $|\mathrm{U}_{\mathrm{e}4}|^2\approx 10^{-3}$ in a large mass range of $m_4 \approx 1 - 100$~keV~\footnotemark, see figure~\ref{fig:currentlablimits}. The experiments discussed here will provide much stronger sensitivity, possibly allowing to start probing the cosmological allowed region at $|\mathrm{U}_{\mathrm{e}4}|<10^{-6}$. No current technology, however, allows to reach a sensitivity of the order of $ |\mathrm{U}_{\mathrm{e}4}| \approx 10^{-10}$, which would allow to test indirect hints for sterile neutrinos of 7~keV mass, see section~\ref{sec:xray}.

Finally, we consider the specific case of the $\nu$MSM~\cite{Asaka:2005an,Asaka:2005pn,Boyarsky:2009ix}, where a keV-scale sterile neutrino and two GeV-scale sterile neutrinos are proposed as a minimal extension of the Standard Model. We discuss the scientific reach of the proposed SHiP experiment at CERN to detect the latter heavy sterile neutrinos and thereby to probe the $\nu$MSM model.

\footnotetext{Throughout this section we assume the $4\times4$ or $2\times2$ mixing scenarios interchangeably. The first, considers the three active neutrino mass eigenstates $m_{1,2,3}$ plus one additional, mostly sterile, mass eigenstate $m_4$. The latter groups the light neutrino mass eigenstates together to a single effective neutrino mass $m(\nu_{\mathrm{e}})$ and assumes a single effective heavy neutrino mass $m_\mathrm{s}$. In the case of $2\times2$ mixing the mixing amplitude can be written as $|\mathrm{U}_{\mathrm{e}4}|^2 = \sin^2\theta$.}

\subsection{Tritium Beta Decay Experiments (Author: S.\ Mertens)}
\label{ssc:tritium}
\input{Tritium.tex}

\subsubsection{The Troitsk Experiment (Authors:  V.\ S.\ Pantuev, I.\ I.\ Tkachev, A.\ A.\ Nozik)}
\input{troitsk.tex}

\subsubsection{The KATRIN Experiment (Authors: S. Mertens, J. Behrens, K. Dolde, V. Hannen, A. Huber, M. Korcekzek, T. Lasserre, D. Radford, P. C.-O. Ranitzsch, O. Rest, N. Steinbrink, C. Weinheimer) }

\input{KATRIN.tex}

\subsubsection{The Project 8 Experiment (Author: B.\ Monreal)}
\input{project8.tex}

\subsubsection{PTOLEMY Experiment (Authors: B.\ Suerfu, C. G.\ Tully)}
\input{ptolemy.tex}

\subsubsection{Full kinematic reconstruction of the beta decay (Authors: F.\ Bezrukov, E.\ Otten)}
\input{coltrims.tex}

\subsection{Electron Capture Experiments (Author: L.\ Gastaldo)}
\label{ssc:holmium}
\input{EC_nuclides.tex}

\subsubsection{The Electron Capture in $^{163}$Ho experiment ECHo (Authors: L.\ Gastaldo, T.\ Lasserre, A.\ Faessler)}
\label{sssc:echo}
\input{Echo.tex}

\subsubsection{Other nuclides from the electron capture sector (Authors: L.\ Gastaldo, Y.\ Novikov)}
\input{OtherNuclides.tex}
\subsection{Direct Detection }
\label{ssc:Direct}
A measurement of {\it relic sterile neutrinos in the laboratory} will ultimately provide the strongest evidence for this interpretation of the dark matter.  
Whereas other experiments, {\it e.g.} the search for kinks in beta decay spectra, may establish sterile neutrinos as a part of the fundamental theory of Nature, such measurements do not directly assess the role of sterile neutrinos as the dark matter.  
Similarly astrophysical observations such as X-ray spectroscopy may reveal the dark matter to be an unstable particle, but they leave open the possibility that the dark matter is not a neutrino.  
Thus it is necessary to explore strategies for the direct detection of relic sterile neutrinos in the laboratory.  

\subsubsection{Direct Detection via inverse $\beta$ decay (Authors: Y.\ Li, W.\ Liao, and Z.\ Xing) }
\input{DirectDetection.tex}

\subsubsection{Prospects for Sterile Neutrino Dark Matter Direct Detection (Author: A. J.\ Long)}
\input{ALong_Subsection.tex}

\subsection{Search for heavy sterile neutrinos with SHiP (Author: R.\ Jacobsson on behalf of SHiP) }
\label{ssc:Ship}

\input{Ship.tex}

%% file: Tritium.tex
Tritium $\beta$-decay is a platform to probe the existence of keV-scale massive sterile neutrinos with electron-flavor mixing.  A small admixture of a keV-scale sterile neutrino mass eigenstate to the electron neutrino $\nu_\mathrm{e}$ would manifest itself in high resolution $\beta$ spectroscopy: At a specific energy below the endpoint corresponding to the keV-scale sterile neutrino mass $m_\mathrm{s}$ the reaction kinematics provides enough energy for emission of a heavy sterile neutrino mass eigenstate along with the electron~\cite{Shro80, de2013role, Rod14, Rod14b}.

The $\beta$-decay spectrum is given as a superposition of the spectra corresponding to each mass eigenstate $m(\nu_i)$, weighted by its mixing amplitude $|U_{ei}|$ to the electron flavor. Since the mass splittings between the three light mass eigenstates are so small, no current $\beta$-decay experiment can resolve them. Instead, a single effective light neutrino mass $m(\nu_{\mathrm{e}})^2 = \sum_{i=1}^{3} | U_{ei} |^2 m(\nu_i)^2$ is assumed. 

\begin{figure}
  \centering
  \subfigure[]{\includegraphics[width = 0.49\textwidth]{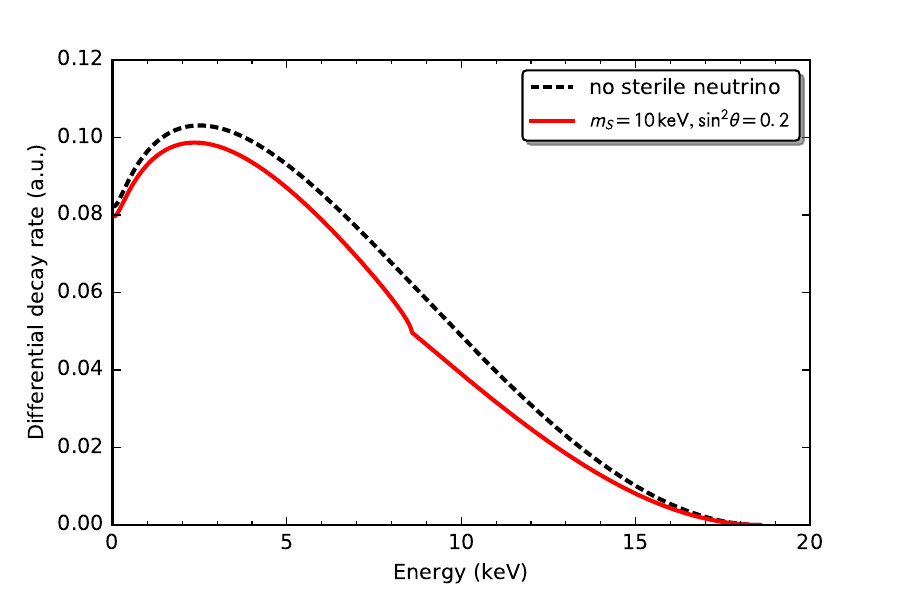}}
  \subfigure[]{\includegraphics[width = 0.49\textwidth]{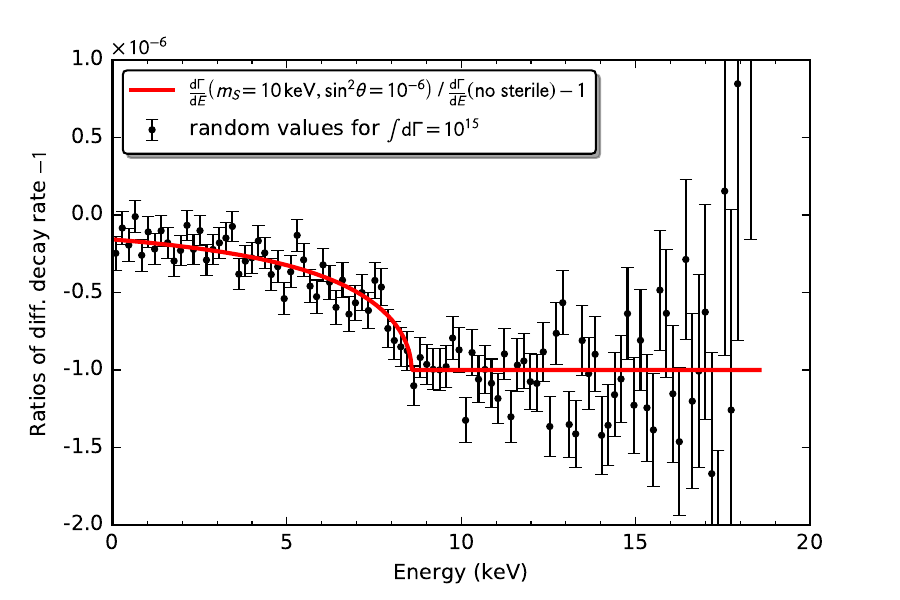}}
  \caption{a: A tritium $\beta$-decay spectrum with no mixing (dashed black line) compared to a spectrum with a keV sterile neutrino mass of 10~keV and a mixing angle of $\sin^{2}\theta=0.2$ (solid red line). One can clearly see a kink-like signature of the keV-scale sterile neutrino at the electron energy $E = E_0 - m_{\mathrm{s}}$ and its influence on the spectral shape below the kink energy. b: Ratio of a tritium $\beta$-decay spectrum without mixing and a spectrum with a 10~keV neutrino mass and a mixing amplitude of $\sin^{2}\theta=10^{-7}$. The error bars correspond to a total statistics of $\sim 10^{18}$ electrons.}
 \label{fig:Kink}
\end{figure}

If the electron neutrino contains an admixture of a neutrino mass eigenstate with a mass $m_{\mathrm{s}}$ in the keV range, the different mass eigenstates will no longer form one effective neutrino mass term. In this case, due to the large mass splitting, the superposition of the $\beta$-decay spectra corresponding to the light effective mass term $m(\nu_{\mathrm{e}})$ and the heavy mass eigenstate $m_{\mathrm{s}}$, can be detectable. The differential spectrum can be written as
\begin{equation}
\label{eq:tritiumspec}
 \frac{d\Gamma}{dE} = \cos^2\theta \frac{d\Gamma}{dE}(m(\nu_{\mathrm{e}})) + \sin^2\theta \frac{d\Gamma}{dE}(m_{\mathrm{s}}),
\end{equation} 
where $\theta$ describes the active-sterile neutrino mixing, and predominantly determines the size of the effect on the spectral shape~\cite{Shro80}. Figure~\ref{fig:Kink} shows a qualitative example with perfect energy resolution and no energy smearing from atomic, thermal or scattering effects.

Tritium $\beta$ decay provides distinct advantages when search for the signature of a keV-scale sterile neutrino. First, tritium $\beta$-decay is of super-allowed type, and therefore a precise theoretical description of the spectral shape is possible. Second, the 12.3-year half-life of tritium is relatively short, allowing for high signal rates with low source densities, which in turn minimizes source-related systematic effects such as inelastic scattering. Finally, with an endpoint energy of $E_0=18.575$~keV, the phase space of tritium provides access to a search for heavy sterile neutrinos in a mass range of astrophysical interest. 

In this section the possibility of future experiments to measure the full phase space of tritium $\beta$-decay spectra will be discussed. The Troitsk neutrino mass experiment, which provided together with the Mainz Experiment the current best limits on the effective electron anti-neutrino mass~\cite{Kraus:2004zw, Aseev:2011dq}, is planning to utilize their apparatus to search for sterile neutrinos. The upcoming KATRIN Experiment, which is designed to probe the neutrinos mass with a sensitivity of 200~meV (90\%CL)~\cite{Angrik:2005ep, Drex13} provides a ultra-luminous tritium source, allowing for a  high statistics search for keV-scale sterile neutrinos. Promising novel detection techniques based on a measurement of the cyclotron radiation of the $\beta$ electron or cryogenic techniques and the usage of an atomic tritium source (as opposed to molecular tritium) will further push the sensitivity of $\beta$-decay experiments and are successfully investigated by the Project~8~\cite{Monreal:PhysRevD80051301:2009, PhysRevLett.114.162501} and Ptolemny~\cite{betts2013development} collaborations. Finally, we discuss the advantages and limitations of a full kinematic reconstruction~\cite{Bezrukov:2006cy} of $\beta$ decay, which in principle would allow to detect a heavy sterile neutrino as missing energy in the decay.

%% file: troitsk.tex
%
%

\begin{figure}
  \begin{minipage}[t]{0.55\textwidth}
    \vspace{0pt}
    \includegraphics[width = \textwidth]{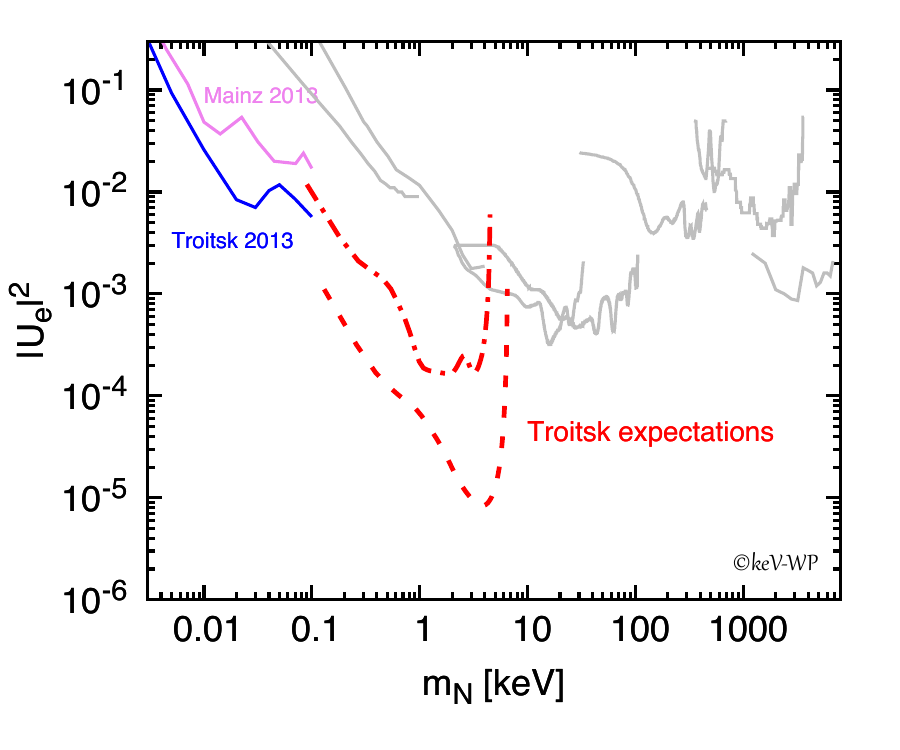}
  \end{minipage}
  \hfill
  \begin{minipage}[t]{0.45\textwidth}
	\vspace{0.3pt}
	\caption{An estimate of sensitivity to the presence of additional sterile neutrinos depending on the neutrino mass at ``Troitsk nu-mass''. The dot-dashed curve corresponds to measurements at the existing equipment, while the lower dotted curve is a projection for the result to be obtained after an anticipated upgrad. Gray lines correspond to the existing limits from Fig.~\ref{fig:currentlablimits}.}
	\label{expect}
  \end{minipage}
\end{figure}

The ``Troitsk nu-mass'' laboratory is located in the Institute for Nuclear Research (INR) in Moscow, Troitsk, Russia.  The initial intention of the experiment was to set up limits  on the effective electron anti-neutrino  mass or to measure it in tritium beta decay. This program has been conducted from 1985 to 2009 and resulted in the currently best upper limit~\cite{Aseev:2011dq,Agashe:2014kda} on the effective electron anti-neutrino neutrino mass, $m(\nu_{\mathrm{e}}) < 2.05$ eV. Later on the same data were used to search for sterile neutrinos with masses from 3 to 100 eV, see Refs.~\cite{Belesev:2012hx,Belesev:2013cba}. Limits on the mixing are shown in Fig.~\ref{expect} as ``Troitsk 2013''. 
Currently, the measurements of the $\beta$-spectrum are continued in INR in a much wider energy range with the intention to set up limits on sterile neutrinos in the keV mass range~\cite{Abdurashitov:2015jha}.

\begin{figure}
\center
	\includegraphics[width=0.8\linewidth]{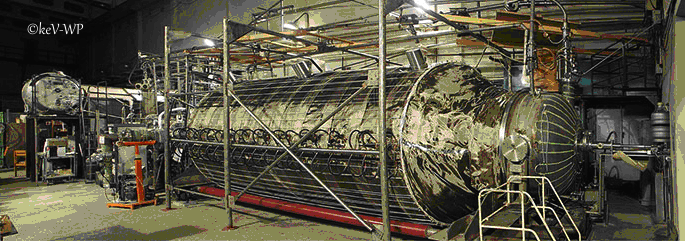} 
	\caption{``Troitsk nu-mass'' spectrometer.}
	\label{spectrom}
\end{figure}

\paragraph{Current experimental setup}
At the moment the ``Troitsk nu-mass" installation is unique, as it is the only one in the world (before commissioning of KATRIN) combining a windowless gaseous tritium source with an electrostatic spectrometer with adiabatic magnetic collimation. After a recent upgrade and the manufacturing of the new spectrometer, which has been commissioned in 2012, the best resolution of about 1-2 eV (depending on the settings) in the range up to 30 keV was obtained. The length of the spectrometer is about 10 meters, it is shown in Fig.~\ref{spectrom}.

The major systematical limitations at ``Troitsk nu-mass'' are related to:
\begin{itemize}
\item insufficiently accurate determination of the electron energy losses in the scattering of electrons inside the molecular tritium source;
\item distortion of the spectrum, caused by the deviation of the transmission function of the spectrometer from the estimated or inaccurate account of the dependence on the energy efficiency of the registration;
\item electronics dead time and pile up uncertainty;
\item gas column density fluctuations;
\item insufficient high voltage stability.
\end{itemize}
In order to achieve a statistical error comparable to the known systematics quoted for the existing equipment, Troitsk needs 15 days of live data acquisition with an integral source intensity of $10^6$ decays per second (which corresponds to the maximum count rate of $2\cdot 10^4$ at 14 kV).
The upper 95\% C.L. on sterile neutrino mixing parameter, which can be achieved at the existing equipment with minimal modifications is shown in fig.~\ref{expect} by the dot-dashed red curve. 

%

\paragraph{Planned upgrades}
After these measurements are finalized, necessary upgrades to reach better sensitivity are planned at Troitsk. In particular this concerns: 
\begin{itemize}
\item the absolute stability of the high voltage system will be better than 0.1 V;
\item the relative stability of the intensity of the electron gun will be better than 0.1\%; 
\item therelative accuracy of the calibration of the spectrometer transmission function will be better than 0.1\%; 
\item accuracy of determination of the effective dead time of the data acquisition system will be of the order of 1 ns (at rates up to 200 kHz); 
\item the absolute accuracy of the determination of the effective thickness of the source will be of the order of or better than 0.001 in units of inelastic scattering probability; 
\item the intensity of the tritium source should be increased to get better statistics without loss of sensitivity at high rates. 
\end{itemize}
The expected limits are approximately two orders of magnitude stronger than the existing ones, see fig.~\ref{expect}, lower dotted curve.

Finally, in order to extend the probed range of  neutrino masses, another new experiment is in preparation at Troitsk~\cite{Abdurashitov:2014vqa}.  This new experiment will be employing the conventional technique of proportional counters filled with tritium gas, but combine it  with a modern fast signal digitization at high count rates -- up to $10^6$ Hz. 


%% file: KATRIN.tex
The KATRIN experiment is a next-generation, large-scale, single $\beta$-decay experiment~\cite{Osipowicz:2001sq, Drex13}. It is currently under construction at the Tritium Laboratory at the Karlsruhe Institute of Technology (KIT) and will prospectively start taking data in 2016. The key goal of the experiment is to probe the effective light neutrino mass $m(\nu_{\mathrm{e}})$ with a sensitivity of 200~meV at 90\% confidence level (CL) by analyzing the shape of the tritium $\beta$-spectrum in a narrow region below the $\beta$-decay endpoint energy, where the impact of the light neutrino mass is maximal. 

\begin{figure}[]
\begin{center}
\includegraphics[width = \textwidth]{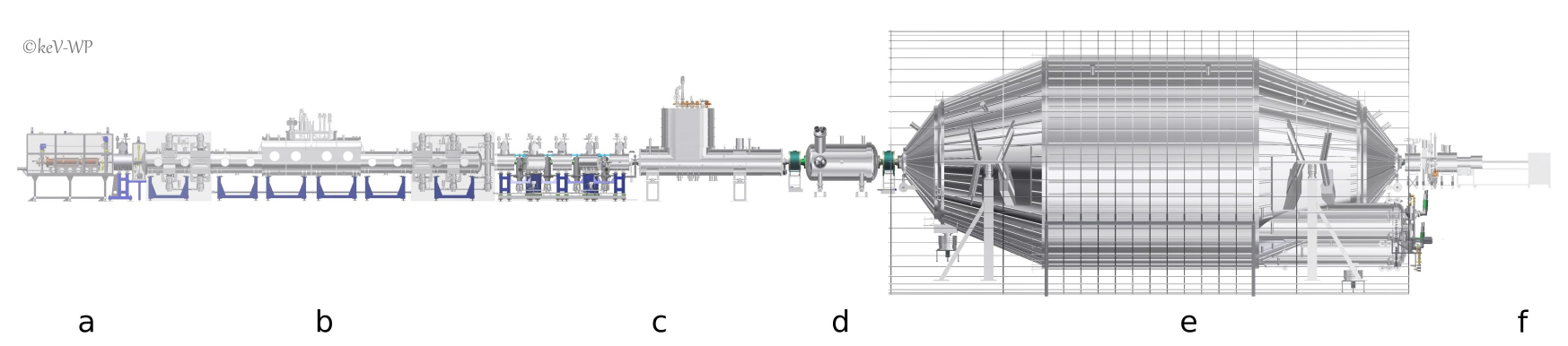}
\caption{Main components of the KATRIN experimental setup. a: rear section, b: windowless gaseous tritium source, c: differential and cryogenic pumping section, d: prespectrometer, e: main spectrometer, f: focal plane detector.}
\label{fig:KATRIN}
\end{center}
\end{figure}

Considering that only $10^{-13}$ of the $\beta$-decay electrons are created within an energy in the last 1~eV of the tritium $\beta$-decay spectrum, an extremely high decay rate is needed to reach the desired light neutrino mass sensitivity. KATRIN makes use of a gaseous molecular tritium source of very high activity ($\lambda_d \approx 1\cdot10^{11}$ decays per second) and stability (at the level of < $10^{-3}$ per hour). These unique source properties may allow KATRIN to extend its physics reach to look for contributions of possible heavy neutrinos in the eV to multi-keV range. In the following we discuss the properties of the KATRIN main components (see figure~\ref{fig:KATRIN}) in the light of searching for keV-scale sterile neutrinos in the entire tritium $\beta$-decay spectrum.

\paragraph{Advantages and Limitations of KATRIN with Respect to a keV-Scale Neutrino Search}
The key component of KATRIN in view of the keV-scale sterile neutrino search is the high luminosity of its molecular tritium source. Its activity of $\lambda_d \approx 1\cdot10^{11}$ decays per second corresponds to a count rate of $\lambda_r \approx 1.5\cdot10^{10}$~counts per second (cps) (taking into account acceptance angle and the transmitted cross section of the source) and a total statistics of $N_{\text{decays}} \approx 1.4 \cdot 10^{18}$ electrons after 3 years of measurement time. Beyond that, the KATRIN source features high isotopic purity (95\% T$_2$), which is constantly monitored by Laser Raman spectroscopy~\cite{Stu10, Sch11, Fis11}. With temperature variations much smaller than 30~mK and a precise monitoring of the tritium column density~\cite{Babutzka:2012xd, Grohmann:2011zz}, the decay rate is expected to be extremely stable. 

On the other hand, the gaseous tritium source entails a number of systematic effects. Energy losses of the $\beta$ electrons due to inelastic scattering in the source are unavoidable~\cite{Aseev:2000}. Furthermore, the source section poses magnetic traps for electrons emitted under a large angle with respect to the beam axis. These initially trapped electrons can escape eventually by scattering into the beam. At the cost of reduced statistics, systematic effects due to scattering may partly be reduced by lowering the source strength.

Another systematic effect related to the KATRIN setup is due to backscattering of $\beta$ electrons and the emissiono of Auger electrons from the gold surface of the rear wall of the KATRIN setup~\cite{Babutzka:2012xd}. Placing the rear wall in a low magnetic field would be a possibility to reduce the backscattering and the returning of backscattered electrons into the source. In depth study of these energy-dependent source and rear wall effects are ongoing. 

KATRIN provides a sophisticated tritium retention system, reducing the tritium flow by 14 orders of magnitude from the windowless gaseous tritium source (WGTS) to the spectrometer. This system is based on differential~\cite{Lukic:2011fw} and cryogenic~\cite{Eichelhardt:2011zz, Luo08} pumping obviating any tritium blocking membrane. This fact is extremely advantageous for the search for keV-scale sterile neutrinos, as it avoids systematic effects due to energy losses of the electrons in such a membrane. A scenario in which a detector would be installed at the rear section of KATRIN is disfavored since the tritium flow is not reduced as efficiently at this location.

The KATRIN main spectrometer works as a MAC-E filter, where a magnetic adiabatic collimation is combined with electrostatic filtering~\cite{Angrik:2005ep}, providing an unsurpassed energy resolution of less than 1~eV at the endpoint energy and linearly decreasing with the electron kinetic energy. The basic working principle is to apply a certain retarding potential to the spectrometer, which allows only electrons with a kinetic energy higher than this potential to pass and reach a counting detector. Hence, by recording the count rate for different retarding potentials the \textit{integral} shape of the energy spectrum is measured.

In the case of a keV-scale sterile neutrino search, the entire tritium $\beta$-decay spectrum is of interest and therefore the main spectrometer would operate at very small retarding energies to allow the electrons of the interesting part of the spectrum to reach the detector. To guarantee an adiabatic transport of electrons with high surplus energy through the spectrometer~\cite{Pra12}, the magnetic field at the center of the spectrometer has to be increased by a factor of $\sim$3 while at the same time the magnetic field in the source section has to be decreased by a factor of $\sim$10.

One of the biggest challenges in using the present KATRIN experimental setup as an apparatus to search for keV-scale sterile neutrinos arises from the high count rates of $\lambda_r = 1.5\cdot10^{10}$~cps when aiming to measure the entire $\beta$-decay spectrum. The 148 pixel silicon detector~\cite{Amsbaugh:2014uca}, as well as the electronics and DAQ systems, presently in use for normal KATRIN operation, are not designed to deal with these electron rates. 

\paragraph{Pre-KATRIN measurement}
Despite the challenges listed in the previous section, a measurement with the first light of KATRIN might be feasible. The idea here is, to significantly reduce the tritium source strength, to achieve a counting rate of the order of 100~kHz which can be handles by the current KATRIN detector system. This reduction can be achieved in a 3-fold way: 1) less tritium is inserted into the WGTS, 2) the acceptance angle is reduced by reducing the source magnetic field relative to the maximal field, 3) the visible fluxtube is reduced by reducing the detector magnetic field. 

Two measurement schemes are foreseen: 1) a differential measurement, making use of the current detector energy resolution of about 2~keV (FWHM) at 18.6~keV. In this case, the spectrometer would be set to a fixed low retarding potential at all times, allowing the entire part of the spectrum of interest to reach the detector. 2) An integral measurement, as in the standard KATRIN operation, could be performed, by measuring the count rate for different retarding potentials of the main spectrometer. The spectrometer would provide an sharp (100~eV) energy cut-off. The latter requires high source stability whereas the first method requires a good understanding of the detector energy response. A combination of both will help reducing systematic uncertainties. 

\begin{figure}
  \centering
  \begin{minipage}{0.65\textwidth}
    \includegraphics[width = \textwidth]{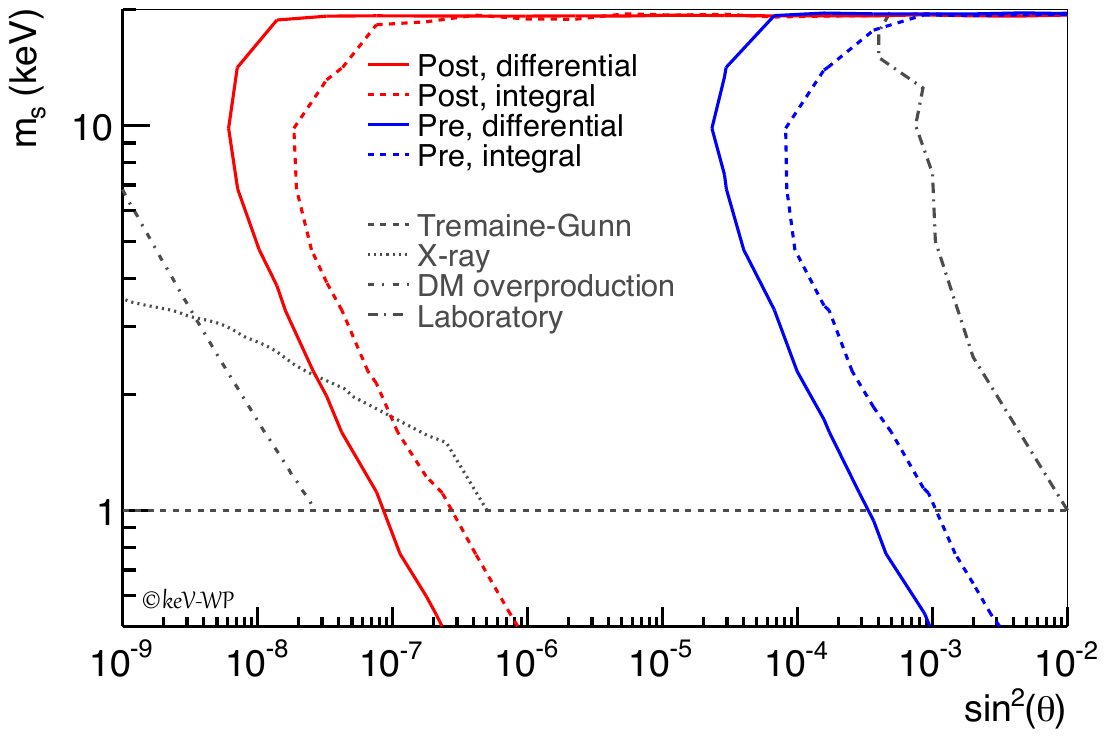}
  \end{minipage}
  \hfill
  \begin{minipage}{0.34\textwidth}
	\caption{90\% statistical exclusion limit of a Pre- and Post-KATRIN-like experiment. The Post-KATRIN measurement is based on a 3-years measurement with the full KATRIN source strength. The Pre-KATRIN measurement assumes a factor $10^5$ reduction of count rate and a measurement time of 7 days. The gray lines represent the current laboratory limits~\cite{Holzschuh:1999vy} and the parameter space excluded by astrophysical observations~\cite{Canetti:2012vf, merle}.}
	\label{fig:sensi}
  \end{minipage}
\end{figure}

In general, for a low statistics Pre-KATRIN measurement the requirements on systematic uncertainties will less stringent, furthermore a major systematic effect related to inelastic scattering will be mitigated by reducing the source gas density. Figure~\ref{fig:sensi} shows that from a statistical point of view, a 7-day measurement at reduced source strength will allow to significantly improve current laboratory limits. Detailed studies to asses the achievable sensitivity and to lay out a detailed measurement strategy are ongoing.

\paragraph{Generic sensitivity studies for Post-KATRIN measurement}
Assuming the full KATRIN source strength being available for a keV-scale sterile neutrino search, a high statistical sensitivity can be reached. Figure~\ref{fig:sensi} shows the statistical sensitivity assuming the number of tritium decays KATRIN would provide after three years of measurement time. However, before addressing the technical realization of such experiment generic sensitivity studies investigating the effect of theoretical and generic experimental uncertainties have been performed.

In ~\cite{Mertens:2014nha} most importantly, the question whether theoretical uncertainties may impact on the sensitivity of the search was raised. By including the state-of-the-art theoretical corrections available in the literature and allowing for an normalization uncertainty in these corrections it was shown that the sensitivity is reduced by about a factor of five. This can be understood by the fact that the considered corrections are smooth functions of energy and therefore are not prone to mimic a sterile neutrino discontinuous signature. Furthermore, it could be shown that, by using the full information of the spectral shape, an energy resolution of ~300~eV (FWHM) is sufficient to detect the signature of a sterile neutrino. This can be explained by the fact that the imprint of a sterile neutrino on the tritium $\beta$-decay spectrum is not only a local kink-like effect but it leads to a characteristic distortion of the spectrum in a rather large energy range below the ``kink''. 

In a second study ~\cite{Mertens:2014osa} an alternate approach with the goal of detecting the characteristic kink-like signature of a sterile neutrino without perfect knowledge of the spectral shape was investigated. This approach is based on wavelet transformation, which detects which frequencies are present in a tritium $\beta$-decay spectrum at a given energy. Hence, in contrast to Fourier transformation, this method allows for both frequency and energy resolution. The studies revealed a high potential of this method, allowing to probe mixing angles of sin$^2 \theta > 10^{-6}$. However, in contrast to the spectral fit approach, a very good energy resolution of the order of 100 eV (FWHM) is necessary in order to not smear out the kink-like signal, which is what the wavelet technique is sensitive to.

\paragraph{Novel Detector System for Post-KATRIN measurement}
To make use of the full KATRIN source strength, a novel detector system is needed. The combination of high rates and the smallness of the sterile neutrino signature make the design of a suitable detector system a very challenging effort. The main physics requirements are the following:

In order to minimize and control systematic effects (such as non-linearities, pile-up, energy-scale etc.) a sophisticated read-out system will be needed, possibly with a full digitization of each signal waveform. As a consequence, the number of pixels might be limited to < 100 000. On the other hand, in order to minimize the rate per pixel and pile-up related systematic effects, the minimal number of pixels is > 10 000. In order to minimize charge sharing between pixels the minimal pixel size is 0.3~mm, which together with the number of pixels determines the minimal size of the detector.

One of the most dangerous systematic effects will be related to backscattering of electrons from the detector surface. To mitigate this effect the detector location needs to be optimized. By placing the detector at a distance from the maximal magnetic field at the exit of the main spectrometer, most of the backscattered electrons are reflected back to the detector in short time intervals of less than a few hundreds of nanosecond. Since the magnetic flux area is increased at lower magnetic field, the detector area needs to be increased accordingly. A minimal detector radius is $\mathrm{r}_\mathrm{det}$ > 100~mm. Furthermore, the energy threshold needs to be low enough (a few hundreds of~eV) in order to detect backscattered electrons, which deposit only a small fraction of their energy in the detector. This in turn, requires a very thin deadlayer of the order of 10~nm.

To allow for a differential measurement of the tritium $\beta$-decay spectrum, a good energy resolution of ~300~eV at 20~keV is necessary. This requires a thin deadlayer, low capacity and low leakage current. To combine the requirement of large pixel size with low capacity, a design with small (point-like) read-out contact and steering electrodes is being considered.

Currently, detector prototyping in collaboration with the Max-Planck Semiconductor Laboratory in Munich, Commissariat a l'\'energie atomique et aux \'energies alternatives (CEA), Lawrence Berkeley National Laboratory (LBNL), Oak Ridge National Laboratory, and Karlsruhe Institute of Technology is ongoing.

\paragraph{Time-of-Flight (TOF) mode}
An alternative to an upgrade of the detector system might be the introduction of a TOF mode. In a similar fashion to a novel detector, a TOF mode also changes the measurement strategy towards obtaining a differential spectrum. However, this approach is based on using the existing detector/DAQ infrastructure while applying minor modifications in the spectrometer section.

\begin{figure}[]
\begin{center}
\subfigure[]{\includegraphics[width=0.52\textwidth]{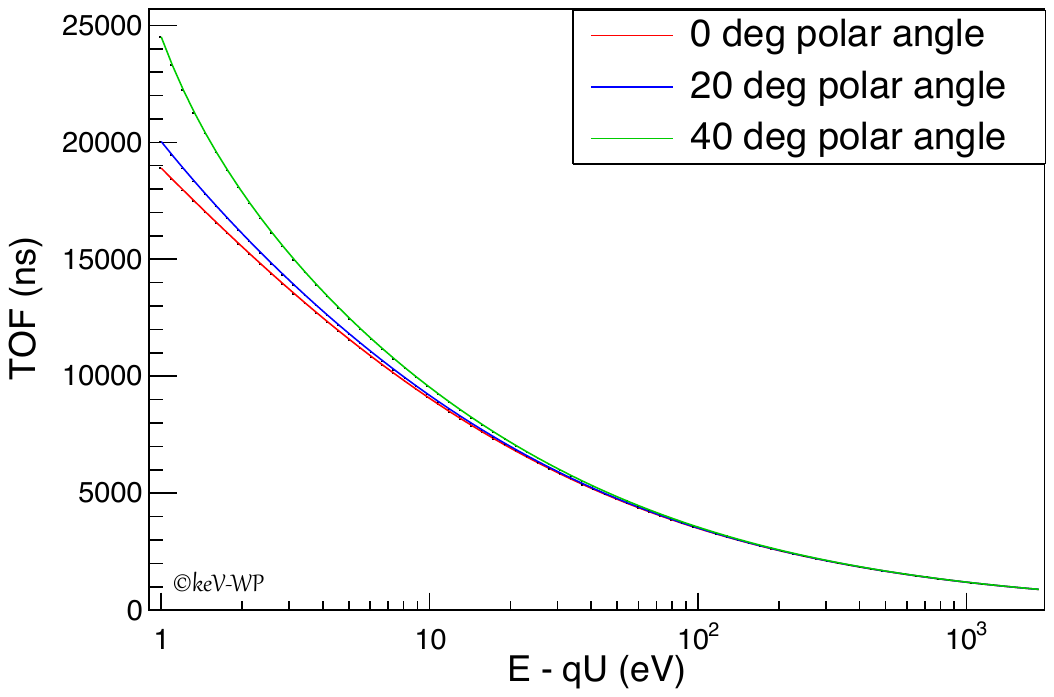}}
\subfigure[]{\includegraphics[width=0.47\textwidth]{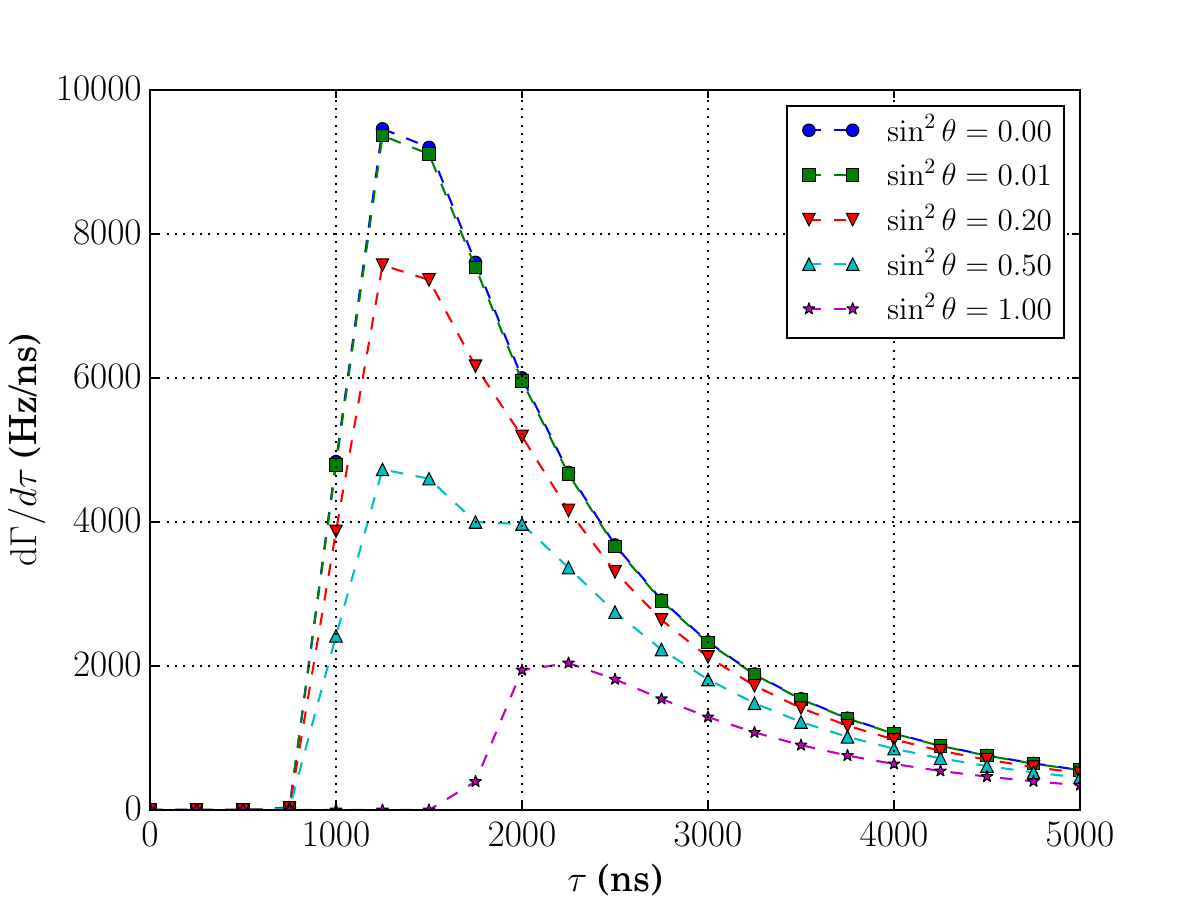}}
\caption{a: Electron TOF as function of the surplus energy relative to the retarding potential for different starting polar angles and starting radius $r=0$~m. The slope is especially higher for low surplus energies, meaning that a small change in energy will produce a significant change in TOF. Thus, the TOF spectrum gives information about the energy distribution close to the retarding potential. b: TOF spectra for sterile neutrino with $m = 1$~keV, a fixed retarding potential of 17~keV and different mixing angles. The values of the mixing angles are strongly exaggerated to demonstrate the signature of a sterile neutrino in the TOF spectrum \cite{Mertens:2014nha}.}
\label{fig:katrin-tof-e}
\end{center}
\end{figure}

The principal sensitivity enhancement of such a strategy has already been shown in the context of light neutrino mass measurements in~\cite{Steinbrink:2013ska}. The basic idea is to measure and fit the TOF distribution of the electrons rather than the integrated energy spectrum. Since the TOF is dominantly a function of the electron surplus energy relative to the retarding potential as well as the polar emission angle of the electron, the measurement contains important information about the energy distribution closely to the retarding potential besides the raw count-rate (see figure~\ref{fig:katrin-tof-e}). Having this supplementary information is statistically equivalent to a differential energy spectrum, thus enabling all the benefits of a differential measurement explained above as for instance better signal reconstruction of the 'kink' and increased tolerance against systematics. 

As for the TOF measurement method, there are two methods being investigated right now. The first method is based on periodic blocking of the electron flux e.g. by switching the pre-spectrometer potential between two settings with full and and zero transmission. If the pulses are sharp enough (in the order of a few micro seconds) and the time interval between the pulses is sufficiently large, the distribution of electron arrival times at the detector follows approximately the TOF spectrum. This method has already been tested in the past in the context of active neutrino mass measurements and proven technically realizable~\cite{Bonn1999256}. The shortcoming of it, however, is the tradeoff between statistics and TOF resolution. In order to have a good sensitivity on the actual TOF spectrum, the duty cycle needs to be low, resulting in loss of events. A too high duty cycle leads to a smearing of the TOF distribution. For small repetition intervals the danger of attributing electrons to the wrong cycle increases. On the other hand, if the existing detector/DAQ infrastructure is to be used, the count-rate needs to be lowered, which could be accomplished by using pulses with low duty cycle, giving sharp timing. Furthermore, positive effects concerning systematics are expected which could possibly compensate for the loss of statistics \cite{Steinbrink:prep}.

To address these issues, a second option is being investigated. In this strategy the time dependency is introduced by periodic post-acceleration and -deceleration. This process must be performed in a delay line behind the analysis plane at low magnetic field to avoid an effect on the energy filtering. The arrival time of electrons at the detector then depends on the pulse phase and the electron velocity at the time of entry into the delay line. By this 'time-focussing' \cite{Behrens:prep} it can be arranged, that for a specific electron energy, all electrons within one period are arriving approximately at the same time at the detector. To ensure the latter condition it is most likely beneficial to use the pre-spectrometer instead of the main spectrometer for the electrostatic energy analysis (as the energy resolution of the pre-spectrometer is sufficient  for keV scale sterile neutrino search) and use the main spectrometer as delay line.

\paragraph{Conclusion}
It seems feasible to operate KATRIN with changed source and electro-magnetic design parameters in order to conduct a Pre-KATRIN keV-scale sterile neutrino search. This measurement has the potential to significantly improve current laboratory limits and will be provide valuable information for a Post-KATRIN measurement with higher statistics. Detector R\&D, and simulations are currently being conducted to assess the feasibility and final sensitivity of the latter. A major goal is to ascertain whether a sensitivity of the order of $\sin^2 \theta = 10^{-6}$ is possible with a KATRIN-like approach.

%% file: project8.tex

A relativistic electron in a uniform magnetic field will emit cyclotron radiation.   The frequency $f_\gamma$ of this radiation depends on the magnetic field strength $B$ and, due to a relativistic effect, on the electron kinetic energy $K$: 

\[
f_\gamma = \frac{f_c}{\gamma} = \frac{eB}{2\pi \gamma m_e} \;,
\]

where $\gamma = \left(1 + K/\left(m_ec^2\right)\right)$ is the Lorentz factor and $f_c$ is the nonrelativistic electron cyclotron frequency, approximately 28 GHz per Tesla.   This radiation is powerful enough that a modern cryogenic amplifier can detect single electrons.   The relativistic shift allows us to use cyclotron-radiation measurements to measure $K$, which we refer to as cyclotron radiation emission spectroscopy (CRES).  This was first proposed~\cite{Monreal:PhysRevD80051301:2009} for a very high resolution (sub-eV resolution) measurement of the tritium endpoint shape.   The first CRES electron detections were obtained~\cite{PhysRevLett.114.162501} by the Project 8 collaboration, which showed $\sim$10~eV resolution on $^{83\mathrm{m}}$Kr conversion electrons held in a magnetic bottle trap. 

To search for keV-scale sterile neutrinos in the tritium spectrum, we expect to need  the highest possible statistics but an energy resolution of only $\Delta E_e \sim 100$~eV ($\sim$5 MHz/T when measured as frequency).  The electron energy resolution of CRES is determined by several factors.   One of these is the gas pressure in the source.  Electron scattering on residual gas can be thought of as pressure-broadening of the cyclotron emission line, so we choose a density where the scattering rate is below the desired frequency resolution.    The magnetic field needs to be uniform at a level of $\Delta B/B =  \Delta E_e/m_e$; specifically, this precision applies to the time-averaged field experienced by trapped  electrons (which are trapped along different B field lines and explore them with a range of different pitch angles).

Design parameters for such an experiment are shown in tab.~\ref{tab_cres_params}.   In contrast to an endpoint experiment, we will see numerous electrons radiating into each frequency bin at any time, unless the experiment is very highly segmented.  This does not cause a ``pileup''-like problem if we simply measure the average cyclotron radiation power as a function of frequency.   Over a broad range of parameters, electron counting statistics (rather than, e.g., thermal noise statistics) are the main source of fluctuations, just like in a counting experiment.   We can, of course, obtain a counting-experiment behavior by subdividing the tritium source into independently read-out subvolumes.

A mean-power measurement has notably different instrumental effects than a counting measurement.   Frequency-dependent variations are expected in the noise temperature, cavity resonances, gain, and digitizer dynamic range.  In a counting measurement, these effects may result in counting-efficiency corrections.  In a power measurement, gain variations simply rescale the observed power.  In either case, to measure the tritium spectrum in the presence of these variations, we need to exploit the variable $B$ field.   In a CRES experiment, true electron-related spectral features will move in frequency space when $B$ changes.   Gain shape variations and other equipment-dependent effects are fixed in frequency space and independent of $B$.   Therefore, on top of whatever other gain calibrations are possible, a careful $B$ scan ought to allow reconstruction of an instrument-independent measurement of the electron emission spectrum.   

An electron may undergo inelastic scattering (losing 10--20 eV) but remain trapped and continue radiating at its lower energy, or it may undergo an angle change that allows it to escape from the magnetic mirror.   The magnetic mirror configuration determines the range of pitch angles which are trapped, and hence the average number of energy-degraded electrons which contribute to the power spectrum.   A strong mirror will retain an electron through a large number of scatters, resulting in a resolution function with a very long low-energy tail.  In this case the instrument behaves somewhat like an integrating spectrometer.   A weak mirror will allow for a narrower resolution function but at a cost in acceptance.    

The scattering resolution function would itself need to be known very precisely in order to measure the overall tritium spectrum shape.   In a sterile neutrino search, looking for a fairly sharp spectral discontinuity, the only requirement is that the resolution function varies smoothly with energy.   We have not identified any likely sources of discontinuities in this function, but the question can be answered experimentally using higher-resolution, single-scatter-resolving measurements with CRES.  More generally, any source of sharp energy-dependent behavior (in radiated power, scattering, trapping, non-tritium backgrounds, etc.) might be a source of systematic error; since the CRES technique is very new we do not have a complete understanding of such systematics.  

\begin{table}
\caption{Parameters of the search for a 7 keV sterile neutrino with a 100 eV resolution  CRES instrument.   ``Source density'' is the tritium density compatible with this resolution. ``Bandwidth'' is the required readout bandwidth for a single 100 bin.   ``Rate per bin'' is the rate of detectable electrons in 100~eV bins at the energy of interest.   ``Occupancy'' is the number of (new) electrons apppearing to each sample, assuming a non-subdivided experiment. ``Stat error, 1y''  is the fractional statistical precision on the spectrum obtained in 1~y.   We give explicit numbers for (a) small experiment which could be realized within a few years and (b) an ambitious experiment aimed at approaching astrophysically-motivated mixing angles.}
\begin{center}
\begin{tabular*}{\textwidth}{@{\extracolsep{\fill}} l l l l } 
\hline 
 & General case & Small  & Large \\
 & & experiment & experiment\\
\hline
\hline
field $B$ & &  1 T & 4 T \\
volume $v$ & &  0.01 m$^3$ & 10 m$^3$ \\
pitch angle $\theta_{m}$ & & 5$^\circ$ &  5$^\circ$ \\
\hline 
Source density, Bq/cm$^3$ & $2{\times}10^7$ T$^{-1}$ &  $2{\times}10^7$ &  $8{\times}10^7$\\
Bandwidth, MHz & 5 $\cdot \frac{B}{\mathrm{T}}$ & 5.0  & 20 \\
rate, s$^{-1}$  & $8{\times}10^4 \frac{B t v}{\mathrm{cm}^3\cdot\mathrm{s}\cdot\mathrm{T}} \mathrm{sin}(\theta_{m})$& $7{\times}10^{7}$  & $3{\times}10^{11}$ \\
Pileup occupancy & $0.008\cdot \frac{v}{\mathrm{cm}^3}\cdot\mathrm{sin}(\theta_m) $& 7 & 7000 \\
Stat error, 1 y & $3.5{\times}10^{-3} / \sqrt{\frac{B t v \mathrm{sin}(\theta_{m})}{\mathrm{cm}^3\cdot\mathrm{T}\cdot\mathrm{y}}}$ &  $2.1{\times}10^{-8}$ &  $3{\times}10^{-10}$ \\
\hline
\end{tabular*}
\end{center}
\label{tab_cres_params}
\end{table}

%% file: ptolemy.tex

The Princeton Tritium Observatory for Light,
Early-universe, Massive-neutrino Yield, \linebreak PTOLEMY, proposes a novel search strategy that 
combines ultra-high resolution calorimetric-based methods with simultaneous high-precision 
side-band integration made possible through a multi-band spectroscopic 
MAC-E selection.  The systematic uncertainties from knowledge of the energy resolution of the 
calorimetry limit the sensitivity in tritium $\beta$-decay spectrum analyses and make the need
for high resolution information unavoidable for sterile neutrino mixings below $|\mathcal{U}_{e1}|^2$ $\sim$ $10^{-6}$.
Studies of the PTOLEMY multi-band spectroscopic selection
with simulation achieve sterile neutrino search sensitivities down to $|\mathcal{U}_{e1}|^2$ $\sim$ $10^{-8}$
in the mass range of 2--14~keV for a tritium exposure of 300~ $\mu$ g $\cdot$ years.

\paragraph{Introduction}


PTOLEMY
is a prototype cosmic neutrino background experiment located at the Princeton Plasma Physics
Laboratory~(PPPL)~\cite{betts2013development}.  The energy of tritium $\beta$-decay electrons is measured with a high-resolution
calorimeter with a target resolution of 0.15~eV at 100~eV.  In a tritium end-point analysis, the 18.6~keV
electrons from tritium are decelerated down to 100~eV following a 10$^{-2}$--10$^{-3}$ precision 
MAC-E filter.  The electrons are guided into a thin absorber layer thermally coupled to a transition-edge sensor~(TES) for calorimetric measurements.  Assuming that an anomalous X-ray observation preceeds
the sterile neutrino search in the tritium spectrum, the energy window of interest, centered on the
kink in the tritium $\beta$-decay spectrum, is known in advance.
The candidate sterile neutrino mass is twice the X-ray energy, for X-rays emitted in the rest
frame of the sterile neutrino, and the energy resolutions achieved with X-ray observatories based on the 
TES-technology are typically a few eV~\cite{takahashi2012astro}. 

%

This study investigates an approach to fit the tritium spectrum that can yield high sensitivity to low
active-sterile mixing angles across a broad range of keV-scale sterile neutrino masses.  The most sensitive signal shape information for the sterile neutrino search is within a narrow energy window in the vicinity of the predicted kink.  The side-bands of the kink region provide predominantly normalization constraints which can be integrated at lower resolution.  The separation of the electrons with energies in the vicinity of the kink region from the side-band region requires spectroscopic selection.  Spectroscopic selection has a vital
impact on the rates seen by calorimetry in the central energy window.
In the central energy window, the $10^{-2}$ reduction in rate will allow calorimeter pixels to 
maintain a relatively slow readout rate, 10~kHz, compatible with ultra-high resolution TES.  For a tritium mass of 100~$\mu$g, roughly 10$^4$ high-resolution calorimeter cells are needed.  The reduction of the number readout channels could be achieved with microwave multiplexing techniques, allocating MHz analog bandwidth 
per channel with resonance frequencies of a few GHz.  The side-band calorimeter readout would benefit from windowless avalanche photo-diode~(APD) detectors with 40~MHz readout, as currently used in the PTOLEMY prototype~\cite{betts2013development}.  The resolution in the side-bands is expected to be roughly 300~eV.  To meet rate requirements with low cell occupancy, approximately 10$^{3}$ low-resolution calorimetry 
channels would be needed.

\paragraph{Method}

\begin{figure}[h]
	\centering
	\subfigure[] {\includegraphics[width=0.49\textwidth]{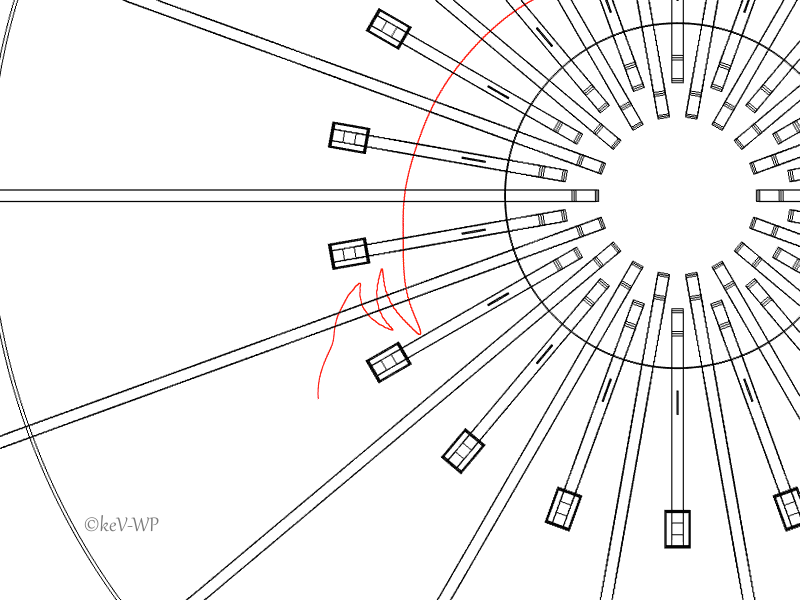}}
	\subfigure[] {\includegraphics[width=0.49\textwidth]{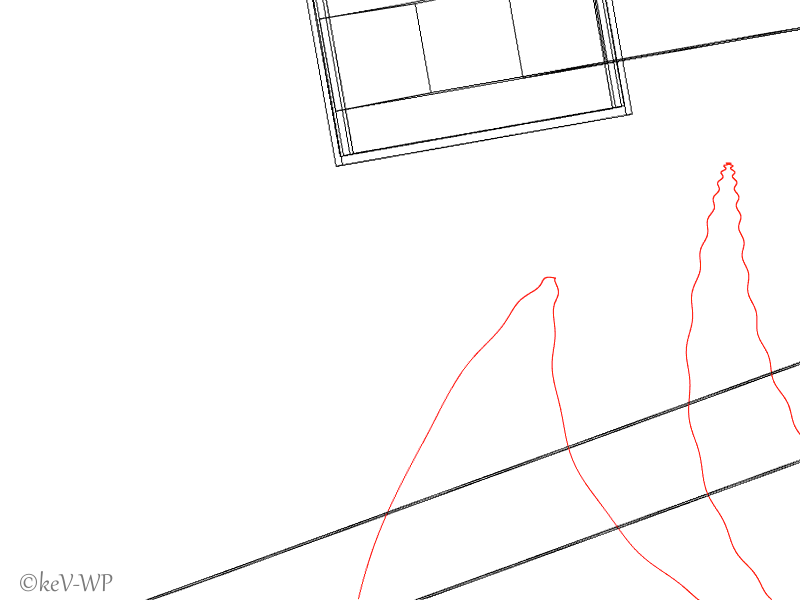}}a
	\caption{Spectroscopic multi-band MAC-E filter simulation of a tritium endpoint electron
trajectory showing radial drift (on the left).  On the right, a zoomed view of the same trajectory 
shows the cyclotron motion and magnetic bouncing near the coils.\label{dualtorus}}
\end{figure}

Initial simulations of tritium endpoint electrons in a multi-band MAC-E filter show
that spectroscopic selection is indeed possible.  An example electron trajectory with toroidal coils and
$E \times B$ drifting is plotted in fig.~\ref{dualtorus}.
The magnetic adiabatic collimation of the spiral trajectory of the $\beta$-decay electron
drifts the electron from a high $B$~field region in the inner torus to a low field outer torus.
The phase space of the system grows radially, allowing for narrow selection of the energy
band of interest within an azimuthal slice of the coil and electrode geometry. 
Details of the multi-band MAC-E spectroscopic selection are not described here.  Instead, we
assume that it will be possible to select a 100~eV energy window centered on a candidate
sterile neutrino mass, based on mass information from an observation of an anomalous
X-ray line.  A mass uncertainty of 20~eV or smaller is sufficient for setting the energy window.
We also assume that neighboring bands will be instrumented with lower resolution
calorimetry with a stable and well-known efficiency to provide a high-precision side-band shape
and integration.
The purpose of this study is to examine whether the multi-band approach 
with high resolution calorimetry in the kink region has the sensitivity
required to probe relevant mixing angles for WDM models.

\begin{figure}[h]
	\centering
	\subfigure[] {\includegraphics[width=0.49\textwidth]{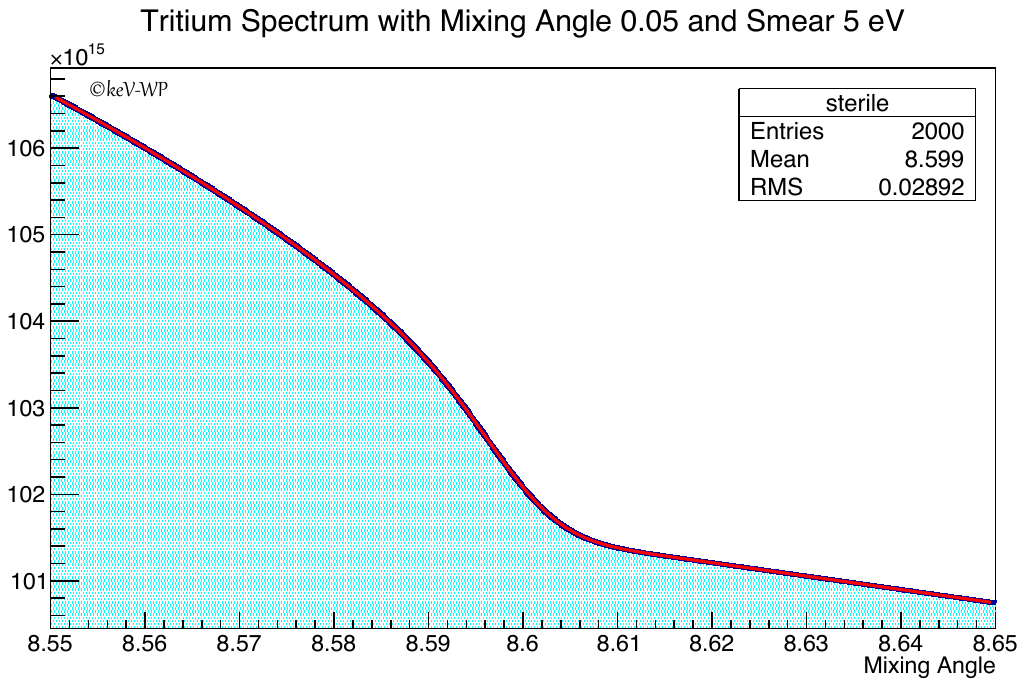}}
	\subfigure[] {\includegraphics[width=0.49\textwidth]{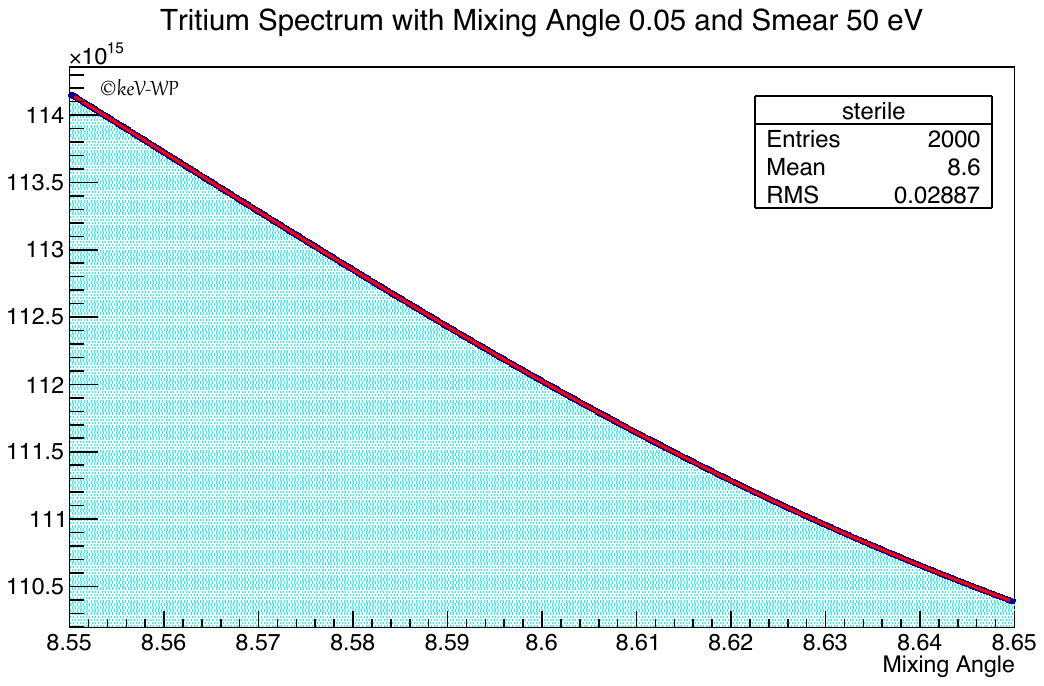}}
	\caption{A comparison of the tritium $\beta$-decay spectra with 5~eV (left) and 50~eV (right)
	energy smearing in the presence of a 10~keV sterile neutrino with a large mixing ($\sin^2 \theta = 0.05$) with electron neutrinos.  The detailed shape information is washed out in the rapidly falling region near the $Q-M_1$ endpoint.\label{smear5and50eV}}
\end{figure}

The relative importance of calorimeter energy resolution against sources of energy smearing in the tritium decay and electron propagation to the calorimeter are considered.  We take typical scales for atomic
and Doppler shifting and assign a 5~eV energy smearing to the $\beta$-spectrum.  We then compare
the 5~eV smeared spectrum to an energy smearing of 50~eV, as shown side-by-side in 
fig.~\ref{smear5and50eV}.  A smearing of 50~eV models the approximate performance of a low resolution calorimeter based on semi-conductor technology and limited by counting statistics from a large conduction-valence band-gap.  This scale is also typical
for the energy loss from inelastic electron scattering on tritium atoms or diatomic hydrogen.  One can clearly 
see that fine features of the shape information are washed out in the rapidly falling region near 
the $Q-M_1$ endpoint.  Large resolution smearing lowers the average signal-to-noise by spreading the sharp part of the signal over a broad range of energies.

The statistical uncertainties in the bins of the tritium $\beta$-decay spectrum are modeled directly,
as opposed to throwing individual trials for on order 10$^{18}$ events.  This is achieved by constructing
the underlying distribution, either with perfect resolution or convoluted with Gaussian energy resolutions,
and then by finely binning the distribution into 0.1~eV bins in a 100~eV energy window centered around the kink.  The height of the bin contents are then smeared with a Gaussian counting statistics uncertainty corresponding to 300~$\mu$g$\cdot$years of tritium exposure.  We have not considered the exact atomic energy smearing and excitations in this study.  In the PTOLEMY calorimetry-based approach, we plan to explicitly measure and unfold the atomic smearing distributions with sub-eV precision.
In particular, we expect to be able to differentiate between the $\sim$2~eV scale smearing present in diatomic tritium from the sub-eV smearing expected from weakly bound tritium on the surface of graphene
through explicit measurement at the endpoint.

The direct measurement of the endpoint smearing with sub-eV resolution will reduce systematics on the knowledge of the signal shape for the heavy sterile neutrino search.  Similarly, a narrow window shape
analysis is less sensitive to $\beta$-decay spectrum shape variations that span many keV in electron energy, or correspondingly a few eV in outgoing $^3$He kinetic energy.  
The side-band integrals using lower resolution
calorimetry will provide complementary sensitivity to the narrow window fit and with different systematics.
Slowly varying shape information is the largest contributor to the sensitivity in the side-bands, especially when one considers that a single bin of 0.1~eV is roughly 10$^{-5}$ of the full integration range.  The step
in the side-band normalization on either side of the narrow window
is an important closure test to validate the self-consistency of the presence of a kink in the spectrum using
independent data over the full phase
space of heavy neutrino production from tritium $\beta$-decays.

\paragraph{Result}

\begin{figure}[h]
	\centering
	\includegraphics[width=0.70\textwidth]{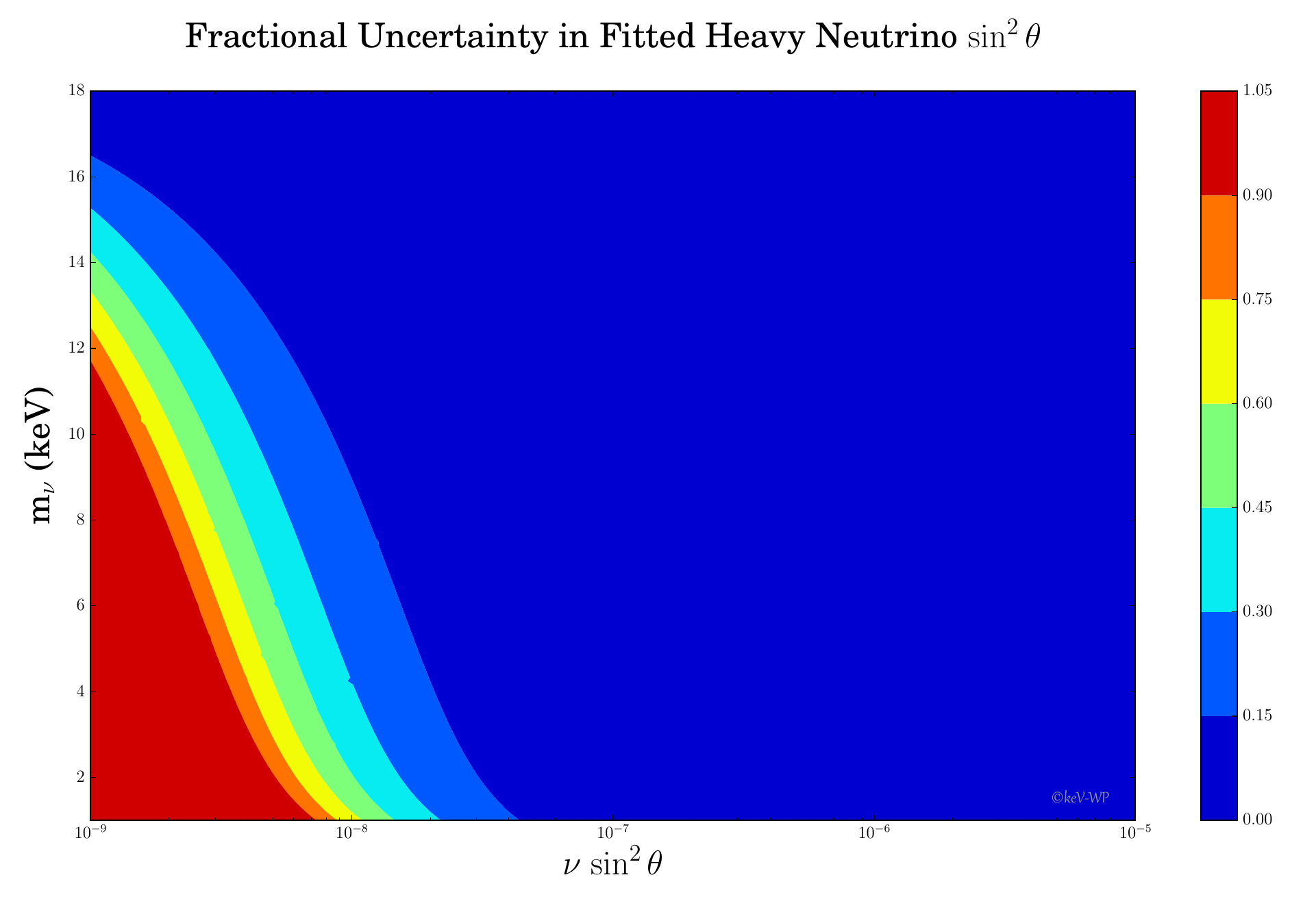}
	\caption{The fractional uncertainty on the fitted
	heavy neutrino mixing is plotted as a function of the mixing and heavy neutrino mass in conditions of perfect resolution.  Over 3$\sigma$ sensitivity is achieved 
	at $\sin^2 \theta \sim 10^{-8}$ for sterile neutrino masses above 4~keV for 300~$\mu$g$\cdot$years
	of tritium exposure.\label{mixingerr}}
\end{figure}

\begin{figure}[h]
	\centering
	\subfigure[] {\includegraphics[width=0.49\textwidth]{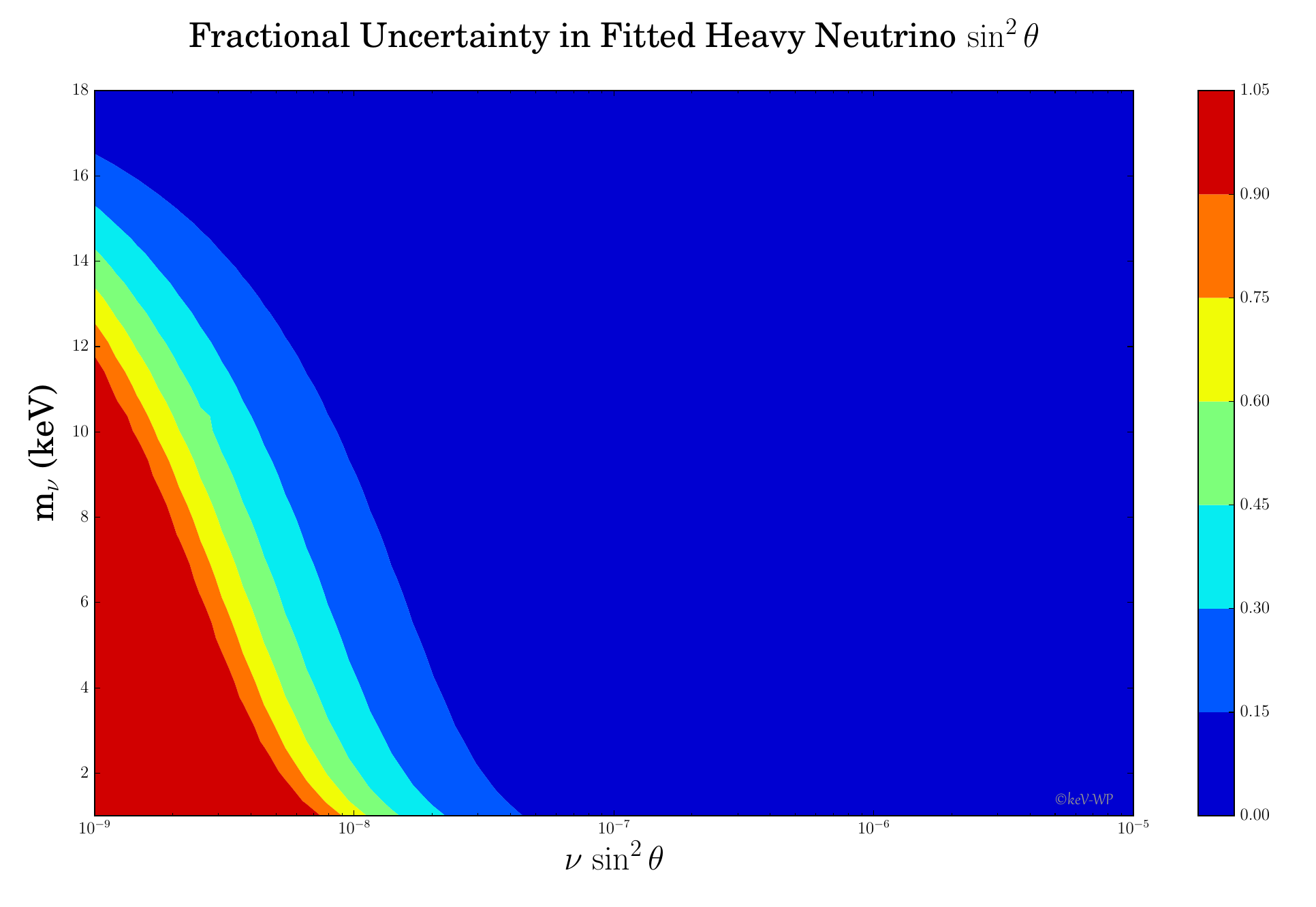}}
	\subfigure[] {\includegraphics[width=0.49\textwidth]{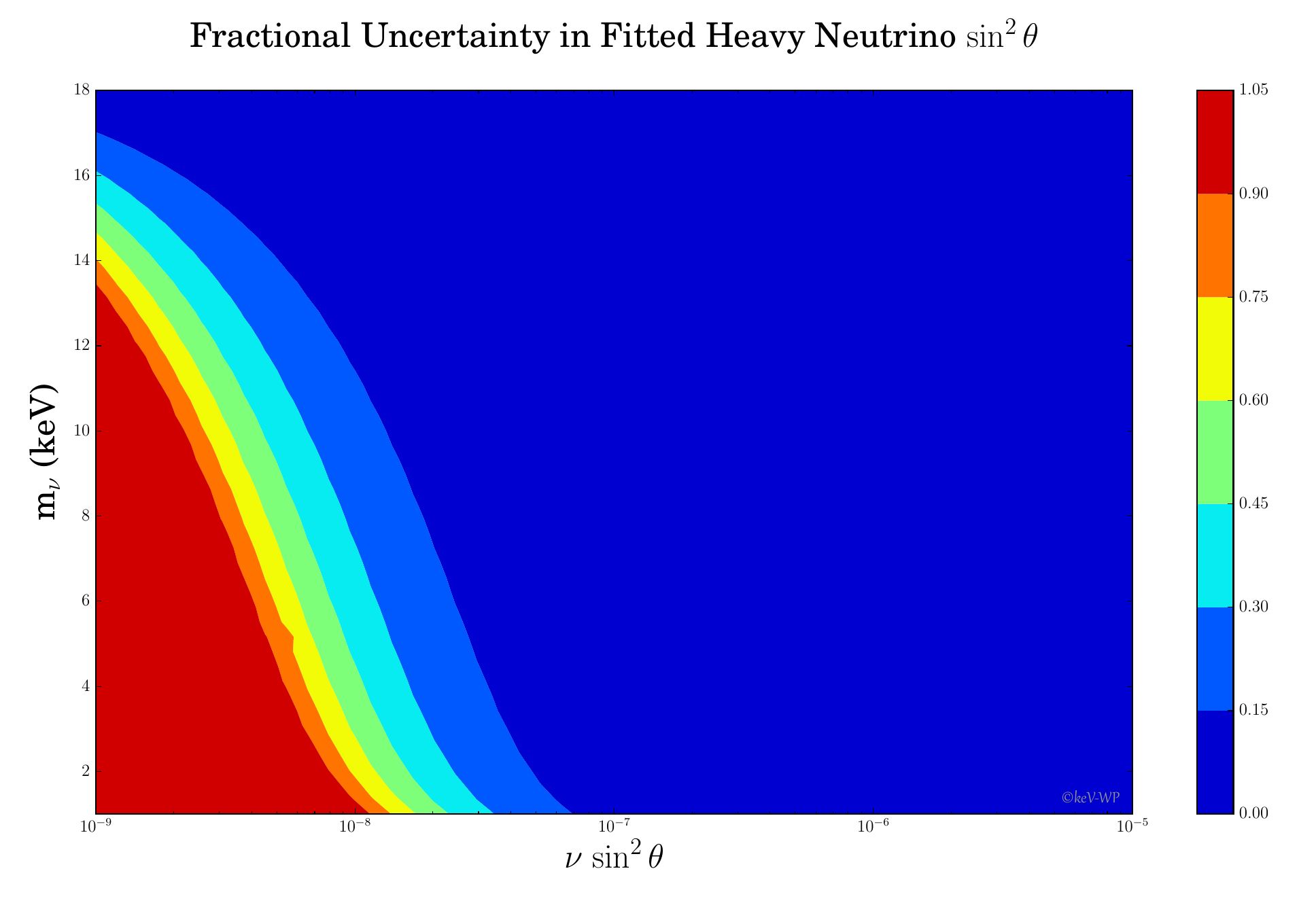}}
	\caption{A comparison of the fractional uncertainty on the fitted
	heavy neutrino mixing is plotted as a function of the mixing and heavy neutrino mass with 5~eV (a) and 50~eV (b)
	energy smearing.\label{mixingerr5and50eV}}
\end{figure}

The narrow window fits for the heavy neutrino mixing as a function of sterile neutrino mass 
reproduce the input mixing parameters down to $\sin^2 \theta \sim 10^{-8}$ in conditions where the
exact energy smearing is known.  The fractional error 
on the fitted heavy neutrino mixing for these fits is plotted in fig.~\ref{mixingerr} in conditions of
perfect resolution.  Over 3$\sigma$ sensitivity is achieved at $\sin^2 \theta \sim 10^{-8}$ for 
sterile neutrino masses above 4~keV for 300~$\mu$g$\cdot$years of tritium exposure.
For finite resolution the results are degraded, as shown in fig.~\ref{mixingerr5and50eV}, 
for energy resolutions of 5~eV and 50~eV.
Over 3$\sigma$ sensitivity is achieved at $\sin^2 \theta \sim 10^{-8}$ for 
sterile neutrino masses above 5~keV(8~keV) for an energy smearing of 5~eV(50~eV) in the case of 300~$\mu$g$\cdot$years of tritium exposure.

\begin{figure}[h]
	\centering
	\includegraphics[width=0.70\textwidth]{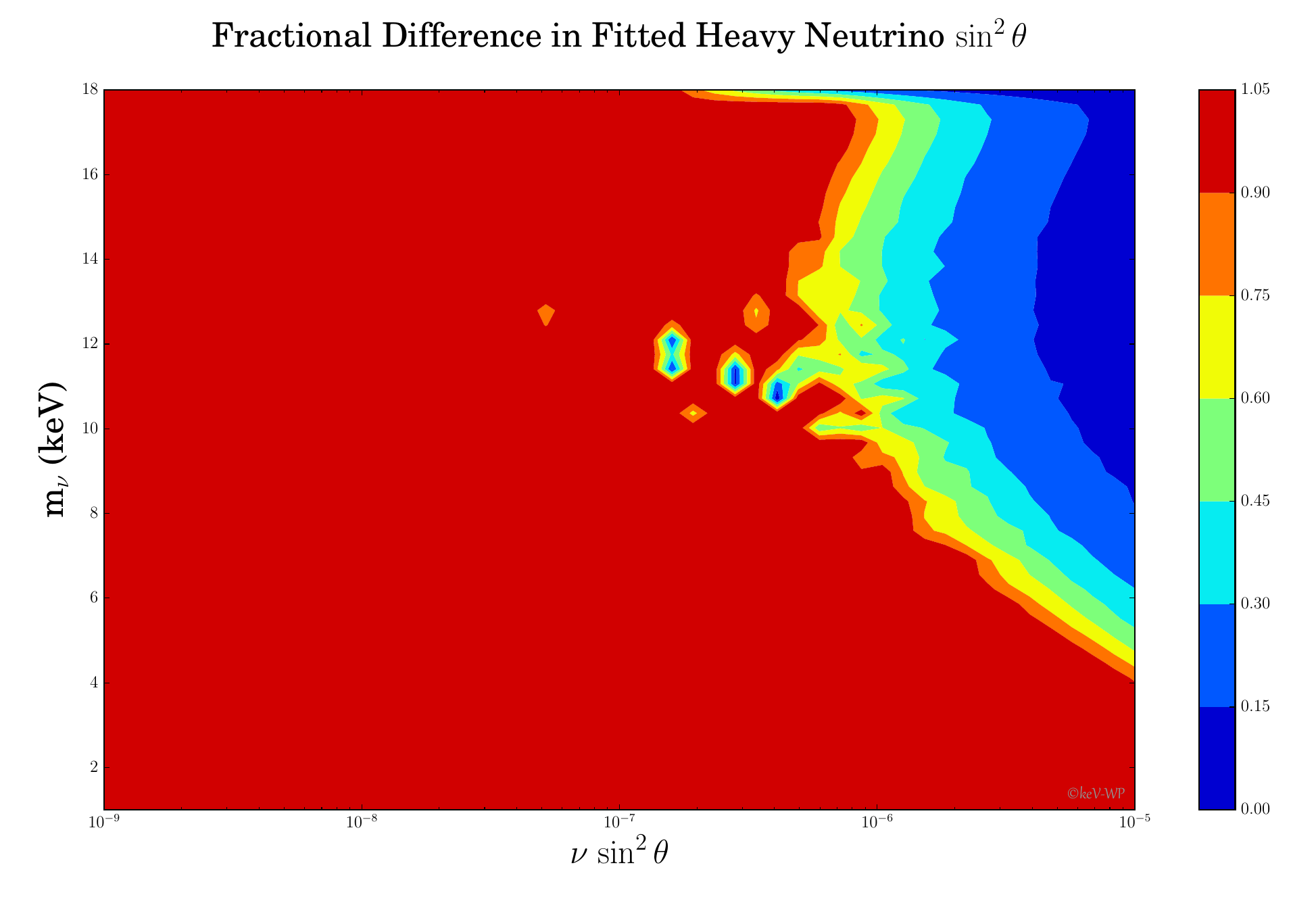}
	\caption{The fractional difference on the fitted
	heavy neutrino mixing is plotted as a function of the mixing and heavy neutrino mass in conditions of 50~eV energy resolution and 10\% uncertainty on knowledge of the energy smearing.  Below $\sin^2 \theta$ $\sim$ $10^{-6}$, the extraction of the mixing parameter from the fit fails.\label{smearuncertainty}}
\end{figure}

One would argue that the degradation from 5~eV to 50~eV is not that significant relative to the order
of magnitude sensitivity achieved in the mixing.  However, the energy smearing
is known to be a leading systematic for such searches.  If we vary the knowledge of the energy
smearing by 10\%, then the mixing sensitivity for 50~eV energy smearing deteriorates
rapidly in the sterile neutrino search.  Below $\sin^2 \theta$ $\sim$ $10^{-6}$, the extraction of the mixing parameter
from the fit fails for 10\% uncertainty on a 50~eV energy smearing, as shown in 
fig.~\ref{smearuncertainty}.

\paragraph{Conclusion}

The general approach of using a narrow energy search window with high-resolution calorimetry measurements have been applied to the heavy sterile neutrino search in the vicinity of the expected
kinematic edge.  The results show promising sensitivity for resolutions of 5~eV in a 100~eV window
when the energy smearing uncertainty is constrained at this precision by direct data measurements.
Over 3$\sigma$ sensitivity is achieved at $\sin^2 \theta \sim 10^{-8}$ for 
sterile neutrino masses above 5~keV for 300~$\mu$g$\cdot$years of tritium exposure.
The information from side-band integration from lower resolution calorimetry and the slow
variation in the shape predicted by heavy sterile neutrinos would provide important independent
confirmation of the heavy sterile neutrino hypothesis over the full production phase space of the
$\beta$-decay spectrum.


%% file: coltrims.tex
%
%
%
%


The mass ($m_{\mathrm{s}}$) of the unobserved sterile neutrino can be determined in principle from a single $\beta$-decay event, if one can reconstruct its energy and full momentum from the measured observables, namely the full momenta $\mathbf{k}$ and $\mathbf{p}$ of the emitted electron and the recoiling daughter, \emph{assuming that the parent nucleus was at rest}~\cite{Pontecorvo:47, PhysRevD.46.R888, PhysRevD.46.R6, PhysRevLett.90.012501}.

\begin{equation} 
  m_{\mathrm{s}}^2 = (Q- E_{\mathrm{e}}- E_{\mathrm{d}})^2 - (\mathbf{p}+\mathbf{k})^2 ;
\end{equation} 
where $E_{\mathrm{e}}=\sqrt{m_e^2+\mathbf{k}^2}-m_e$ and $E_{\mathrm{d}}=\sqrt{M^2+\mathbf{p}^2}-M$ are electron and recoil ion kinetic energies.  The search should be limited to moderate neutrino energies not surpassing the expected mass
by too large factors. On the other hand, the energy limit cuts back the phase space of the emitted neutrino and hence the rate of accepted events; it shrinks in proportion to the third power of the
energy. This dilemma is particular painful in any attempt to search for the mass of the light neutrinos in $\beta$-decay, which so far have lead to an upper limit $m_\nu<2$\,eV~\cite{Lobashev:1999tp,Kraus:2004zw}. This result -- like all preceding ones -- has  been obtained from investigating the $\beta$-spectrum of tritium near its endpoint at 18.6~keV. The various ideas for determining the mass of the light neutrinos alternatively by measuring the full $\beta$-decay kinematics have not made it to the floor, yet. The difficulties of measuring energy and
momentum of daughter and electron with uncertainty $<$ eV, together with the rate problem near the endpoint, seem yet insurmountable.

In case of searching for sterile neutrino emission in tritium decay in the mass range of keV, however, the problems of reaching sufficiently small uncertainties of the observables, as well as that of avoiding
high $\gamma$-factors of the neutrino, are relaxed by 3 orders of magnitude. Hence it makes sense to check the chances of a missing mass experiment under this aspect anew. The ColTRIMS technique (\textbf{Col}d \textbf{T}arget \textbf{R}ecoil \textbf{I}on \textbf{M}omentum \textbf{S}pectroscopy), pioneered and reviewed by the atomic physics group at Frankfurt~\cite{doi:10.1088/0953-4075/30/13/006,Doerner200095} seems to offer a viable experimental ansatz, since it is capable of measuring the tiny recoil ion momentum occurring in tritium decay with sufficient precision. While for measuring the mass of an active, light neutrino precision of momenta measurements of order 1~eV or better is required, for the case of a keV neutrino precision of about 0.5~keV in momenta measurement would be sufficient.  For the example of $^3$H decay this corresponds to a relative precision in momentum measurement of a few per mill.

There are a set of theoretical and experimental obstacles that limit the sensitivity of the full kinematic reconstruction technique.  They concern the problem of luminosity and finite temperature of the source and the precision of the reconstruction of the event kinematics necessary to avoid contamination of the signal by active neutrino events. In the following we will present the principle components of a particular ColTRIMS scheme which might suit our purpose and analyze a few typical decay events with respect to the uncertainties of the results which are imposed by the instrument. We will then discuss the three major sources of systematic uncertainties of the final state kinematics which cause background from light neutrino events in the kinematic domain of heavy sterile ones. They stem from (i) the finite temperature of the source, (ii) the emission of an additional, unobserved $\gamma$-particle, and (iii) scattering of the recoiling daughter-ion within the source. Whereas the temperature problem seems manageable, the two others will cut back the sensitivity towards small mixing probabilities $\theta^2$ seriously.

\paragraph{ Cold ${}^3$H-source}
The luminosity problem arises in particular at small  $\theta^2$ and requires either a rather large source or a rather dense source. The former choice will deteriorate the angular resolution of the momenta $\mathbf{k}$ and $\mathbf{p}$, as well as the time of flight measurement (TOF) of the recoil ion by which the amount of its momentum is determined. The latter choice leads to enhanced scattering of the electron and in particular of the slow recoil ion from source atoms, affecting the angular distribution again.

\begin{figure}
  \centering
  \includegraphics[trim=0 1cm 0 3cm,width=\textwidth]{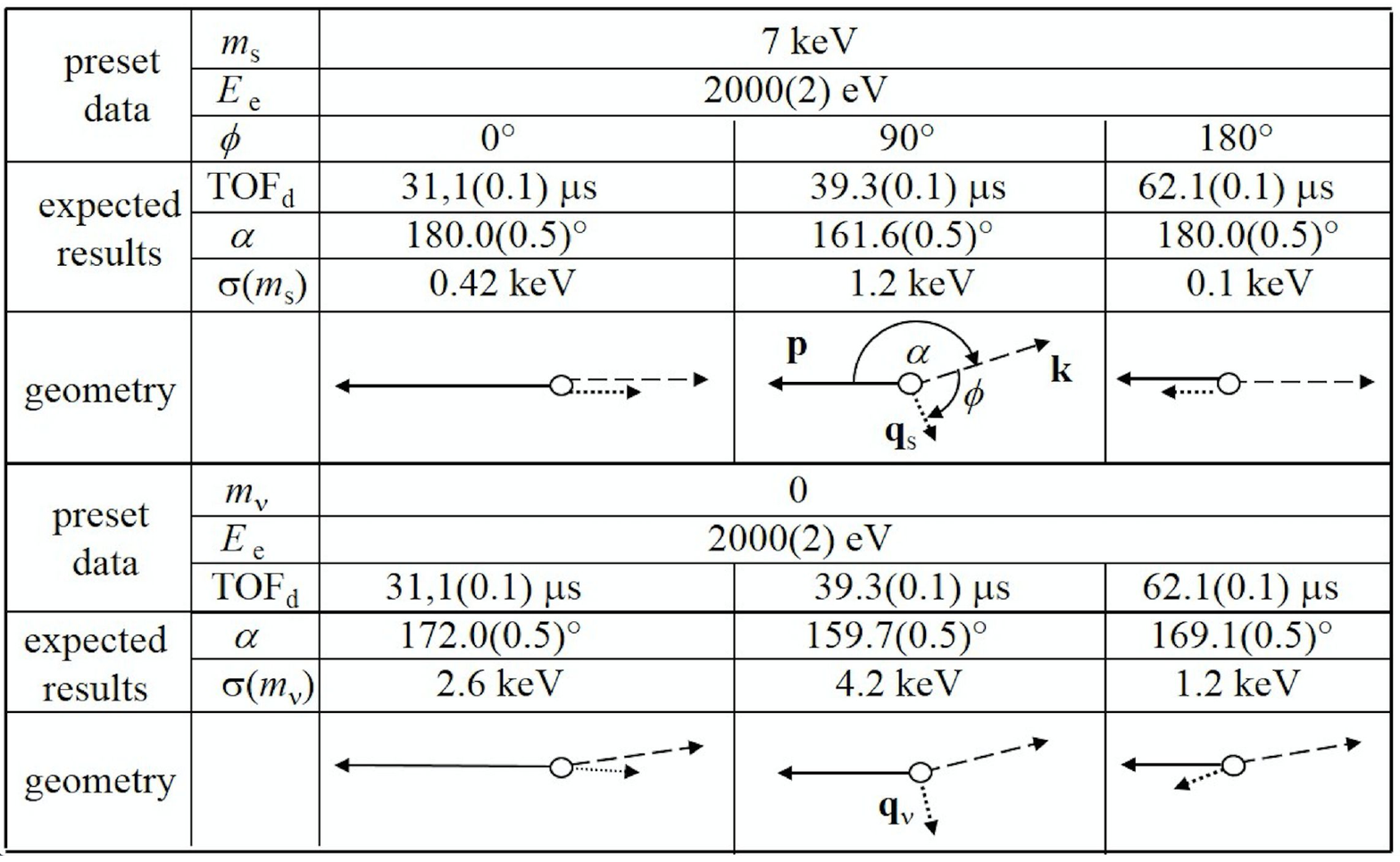}
  \caption{Analysis of selected heavy and light neutrino events as explained in the text}
  \label{fig:OttenTable}
\end{figure}

Currently \cite{doi:10.1088/0953-4075/30/13/006}, one can build supersonic gas jets with particle densities of about $10^{11}-10^{12}\mathrm{\,cm}^{-3}$ and internal temperature of the moving jet of the order of 0.1~K. Alternatively one may choose  the well known optical traps either of the magneto-optical  or of the electric dipole type reaching densities of $10^{10}\mathrm{\,cm}^{-3}$ and temperatures below 0.1~mK.  These methods provide $10^6-10^8$ beta decays per year for the source size of about 1~mm$^3$. Still higher rates would be problematic in view of the rather wide TOF-window of the recoil ions (see Figure~\ref{fig:OttenTable}), during which accidental coincidences of different decay events will occur. Part of them may fall into the kinematic domain of heavy neutrino decay and hence produce indistinguishable background events.

The ideal source material would be a gas of spin polarized atoms since they do not recombine. They could be trapped for instance by an electric dipole trap, i.\,e. within the focus of a laser cavity by the electric polarization potential $V=-\frac{1}{2}\alpha\mathbf{E}^2$ where $\alpha$ is here the electric polarizability and $\mathbf{E}$ the electric field strength of the laser field. Taking for $\alpha$ the estimate of the static
polarizability of the hydrogen atom $\alpha_H=4\pi\epsilon_0(9/2)a_0^2$ (in SI units, $a_0$ is the Bohr radius) one finds that a laser power of 100\,kW, focused onto a focal spot of $10^{-4}$~cm$^2$ within the cavity of a CO$_2$ laser, would be sufficient for trapping a gas at a temperature of 0.01~K.

\paragraph{Detectors for recoil-ions and electrons}
In view of the extremely small mixing angle as suggested by X-ray astronomy, one has to aim on optimized detection efficiency of the rare sterile neutrino events. The ColTRIMS technique allows projecting the total flux of the  slow recoil ions onto a multichannel plate detector by help of a weak electric drift field. $\mathbf{p}$ is reconstructed from the coordinates of the hit channel and the time of flight with high precision thanks to the excellent angular and temporal resolution of this device. This option is not given, however, for the energetic and very fast decay electrons. Instead one has to look out for a detector array which offers sufficient angular and energy resolution as well as a reasonably large solid angle. Covering a fraction of of $\Delta\Omega/4\pi=10\%$ of the full solid angle with angular resolution of $\simeq0.5^\circ$ would require segmentation into $\simeq6000$~pixels. In view of the marginal resolution of commercial items one may resort here to cutting edge R\&D results. An appropriate detector might be built from the sophisticated calorimetric 64-pixel chips described in ~\cite{Gastaldo:2013wha, Kem13b}. At 6~keV particle energy they feature signals of 2~eV full half-width within 90~ns rise time. The electron signal would be fast enough as to determine the moment of decay with sufficient precision.

\paragraph{ Analysis of selected heavy and light neutrino events }
In order to get an impression about the precision of results expected from  such devices we have analysed a few characteristic events (see Fig.~\ref{fig:OttenTable}). $E_\mathrm{e}$ has been set to 2000(2)~eV, close to the maximum of the $\beta$-spectrum, $m_\mathrm{s}$ to 7~keV as suggested from X-ray astronomy, and the angle $\phi$ enclosed by $\mathbf{p}$ and $\mathbf{k}$ has been chosen to $0^{\circ}$, $90^{\circ}$, and $180^{\circ}$ respectively. The time of flight of the daughter, $\mathrm{TOF}_{\mathrm{d}}$, has been calculated for a flight path of 0.2~m. Choosing a TOF-uncertainty $\sigma(\mathrm{TOF}_{\mathrm{d}})$ = 0.1~$\mu$s and an uncertainty $\sigma$($\alpha$) = $0.5^{\circ}$ for the angle $\phi$ enclosed by $\mathbf{p}$ and $\mathbf{k}$, we calculate for the 7~keV sterile neutrino uncertainties $\sigma(m_\mathrm{s})$ as given in line 6 of Fig.~\ref{fig:OttenTable}. For the stretched events, given in row 3 and 5 of Fig.~\ref{fig:OttenTable}, $\sigma(m_{\mathrm{s}})$ is mainly due to $\sigma$($\mathrm{TOF}_{\mathrm{d}}$). $\sigma(\mathbf{p}_e)$ is negligible in view of the high energy resolution of the envisaged electron detector. The mass uncertainties of 0.42~keV and 0.1~keV are satisfactory and could be improved further by choosing a longer flight path. However, stretched events are quite rare since the differential solid angle vanishes there in first order, whereas it maximizes at $\phi$ = $90^{\circ}$. But towards $\phi$ = $90^{\circ}$,  $\sigma(m_{\mathrm{s}})$ rises up to 1.2~keV (row 4), since it is dominated there by $\sigma$($\alpha$).

In the lower half of Fig.~\ref{fig:OttenTable} we have analyzed competing events with emission of light neutrinos, choosing $E_{\mathrm{e}} = 2000(2) eV$ again and $\mathrm{TOF}_{\mathrm{d}}$-values as calculated in line 4. The before stretched events are now kinked by $8^{\circ}$ and $11^{\circ}$ degrees, respectively. But in the region of favorable rates around $\phi$ = $90^{\circ}$, the $\theta$-angle differs by only $1.9^{\circ}$ corresponding to a distance of 3 pixels. Hence separation of these events can only work if ion optics is perfect and any scattering excluded. 

\paragraph{Background induced by  source temperature and unobserved $\gamma$ - emission.}
We turn now to the discussion of the before mentioned three major effects which cause irreducible broadening of the momentum distribution such that the kinematic domains of the dominant light and the rare heavy neutrino decays might overlap. In this case one would face an indistinguishable background from light neutrino emission in the kinematic domain of heavy sterile neutrino emission.  The first stems from the thermal
velocity of the decaying nucleus; it has been treated already in an earlier publication (c.\,f.~\cite{Bezrukov:2006cy}). The second concerns the possible emission of a soft photon during the decay process, which escapes detection. If the decaying nucleus was moving with thermal velocity $\mathbf{v}$, and a photon of momentum $\mathbf{k}_\gamma$ escaped detection, then the uncertainty  of the determination of the neutrino mass induced by these two unobserved momenta is

\begin{equation}
\label{eq:deltamnu}
  \sigma^2({m_{\nu,(\mathrm{s})}})\sim
  \mathbf{q}(M\mathbf{v}+\mathbf{k}_\gamma)
  - (M\mathbf{v}+\mathbf{k}_\gamma)^2
  + 2(Q-\frac{\mathbf{k}^2}{2m_e})|\mathbf{k}_\gamma|,
\end{equation}

where $\mathbf{q}\equiv \mathbf{p}+\mathbf{k}$ is the reconstructed neutrino momentum, $M$ is the mass of the decaying nucleus, and we neglected all terms suppressed by $M$. The danger of this error is the possibility of misidentifying  a massless neutrino event (which are abundant) as a  massive neutrino event. In so far as the latter are rare, only a fraction of $\theta^2$  of the total events, this background should be strongly suppressed. Hence the temperature should be low enough in order  that only the far wings of the Maxwell distribution can interfere.  The rough, conservative bound on the temperature of the source as function of $\Theta^2$, $m_{\mathrm{s}}$, $M$, and $Q$ is then \cite{Bezrukov:2006cy}

\begin{equation}
  T\lesssim
  \left(\frac{\ln 10^{-4}}{\ln(\theta^2)}\right)
  \left(\frac{m_{\mathrm{s}}}{7\keV}\right)^4
  \left(\frac{3\GeV}{M}\right)
  \left(\frac{18.6\keV}{Q}\right)^2
  (0.3\,\mathrm{K}).
\end{equation}
It marks the temperature above which the background rate from light neutrino decays would exceed the one from heavy neutrino decays searched for. The bound relaxes rather fast with increasing sterile neutrino mass, and becomes slightly (logarithmically) stricter for smaller mixing angles. Note, that one can put a constraint on the reconstructed neutrino momentum $|\mathbf{q}|\equiv|\mathbf{p}+\mathbf{k}|<C$, making the temperature requirements weaker but suppressing the overall event count.

\begin{figure}
  \centering
  \includegraphics[width=0.7\textwidth]{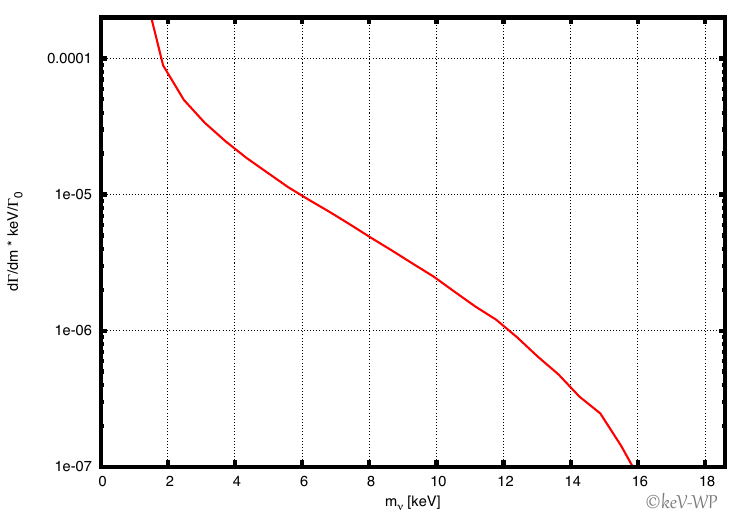}
  \caption{Estimate of the relative branching per keV for undetected photon
    emission imitating the neutrino mass $m_{\mathrm{s}}$ in an event
    ${}^3\mathrm{H}\to{}^3\mathrm{He}+e+\bar\nu_e+\gamma$.}
  \label{fig:FCRdecaybg}
\end{figure}

The second contribution to the mass uncertainty in Eq.~\ref{eq:deltamnu} is due to the missing photon momentum $\mathbf{k}$. The probability that such an unobserved photon emission occurs in a light neutrino event and hence simulates via Eq.~\ref{eq:deltamnu} a heavy neutrino event with mass $m_{\mathrm{s}}$ has been estimated and is plotted in Fig~\ref{fig:FCRdecaybg}. Within the interval $2~\mathrm{keV} < m_{\mathrm{s}} < 16~\mathrm{keV}$  it drops from  $10^{-4}$~keV to $10^{-7}$~keV. For the most interesting case of a 7~keV sterile neutrino it amounts to about $10^{-5}$~keV. No attempt has been made up to date to study the possibility of reducing this background by efficient detection of this radiation tail. It would be very hard. 


\paragraph{Background by scattering of recoil ions from source atoms}
Another obstacle to the experiment is the large scattering probability of the slow recoil ions within the source. The total elastic cross section for ${}^3$H on ${}^3$H collisions is about $10^{-14}$~cm$^2$ \cite{http://dx.doi.org/10.1103/PhysRevA.73.012710}. A similar value will hold for ${}^3$He$^+$ on ${}^3$H. Assuming a source of e.\,g. $2\cdot10^{10}$\,${}^3$H-atoms and a 10\% efficiency of the detector, one would collect about $10^8$ events per year which seems necessary for a competitive sterile neutrino search. Trapping this source by laser in a volume of order $10^{-6}$~cm$^3$ would present a column density of $2\cdot10^{14}$~cm$^2$, causing a fatal scattering probability of order 1. If one would observe ${^3}$H decay from a jet source instead, the same number of atoms may occupy a 1000 times larger volume. This would ease the scattering problem by a factor 100 which is still insufficient for a sensitive search.

\paragraph{Summary}
Todays experimental technology allows in principle to search for a missing mass of order keV in tritium beta decay by full kinematic reconstruction of the event, although such an undertaking would require a very extensive design and R\&D phase in order to arrive at a functioning setup. Still the sensitivity for detecting a very small admixture of heavy neutrino events in the decay will be handicapped by  residual shortcomings in the full kinematic reconstruction. They concern predominantly two effects: (i) the missing momentum of  an unobserved photon accompanying the decay, (ii) the scattering of the recoil ions from source atoms.

%% file: EC_nuclides.tex
The study of the calorimetrically measured electron capture (EC) spectrum of $^{163}$Ho is presently a promising method for the investigation of the electron neutrino mass in the sub-eV range. With experiments based on $^{163}$Ho it appears possible to reach, for the electron neutrino mass, the same sensitivity which tritium based experiments, as KATRIN \cite{Osipowicz:2001sq,Angrik:2005ep} can achieve for the electron anti-neutrino mass. 

\noindent $^{163}$Ho is considered the best candidate among all nuclides undergoing electron capture processes to be used in an experiment for the investigation of the neutrino mass because of its extremely low energy available to the decay, $Q_{\mathrm{EC}}\,=\,2.833\, \pm\, 0.030_{\mathrm{stat}}\, \pm\, 0.015_{\mathrm{syst}}$ keV \cite{PhysRevLett.115.062501,Kopp:2009yp}. Such a low $Q_{\mathrm{EC}}$ allows for a reasonable fraction of counts in the endpoint region of the spectrum to be analyzed for identifying effects due to a finite effective neutrino mass.

The idea to use the analysis of the calorimetric spectrum of $^{163}$Ho to determine the electron neutrino mass was first proposed by De Rujula and Lusignoli in 1982~\cite{De_Rujula_1982}. After about 30 years the feasibility of such an experiment was demonstrated by the ECHo collaboration~\cite{Gastaldo:2013wha, ECHo_web} by showing the possibility to perform high resolution measurements of the $^{163}$Ho electron capture spectrum using low temperature metallic magnetic calorimeters~\cite{Ranitzsch:2014kma}. Meanwhile there are three large international collaborations which aim to reach the sub-eV sensitivity on the electron neutrino mass by the analysis of high precision and high statistics EC spectrum of $^{163}$Ho: the already mentioned "Electron Capture in $^{163}$Ho" (ECHo) collaboration~\cite{Gastaldo:2013wha, ECHo_web}, the "Electron Capture Decay of $^{163}$Ho to Measure the Electron Neutrino Mass with sub-eV sensitivity" (HOLMES) collaboration \cite{HOLMES} and the "Neutrino Mass via $^{163}$Holmium Electron Capture Spectroscopy" (NuMECS) collaboration \cite{NuMECS}. Intrinsic to the calorimetric measurement approach is the fact that all events occurring in the detectors give a measurable signal. Because of that, all these three experiments would have the possibility to analyze the data not only in the endpoint region to look for a deviation of the shape in respect to the case of massless neutrinos, but also over the full spectrum to look for a small kink which would be the signature of the existence of a fourth mass eigenstate, $m_4$, and therefore proving the existence of sterile neutrinos. In the following one of these experiments, ECHo, will be described in more details as well as the approach to investigate the existence of sterile neutrinos.

\paragraph{$^{163}$Ho electron capture spectrum}

In an EC process a nucleus $^A _Z$X decays by capturing an electron from the inner atomic shells and emitting an electron neutrino to $^A _{Z-1}$X. Considering only the leading first order excitation, the daughter atom after the capture of the electron is left with one hole in the internal shell and one electron more in an external shell. Higher order excitations would include the presence of more than one holes in the atomic shells which happens with a much lower probability. The structures due to the higher order excitations could make the analysis of the spectrum more complicated both for the determination of the electron neutrino mass as well as for the identification of a kink due to a heavy neutrino mass state. Preliminary work to investigate the effects of higher order excitations in the analysis of the $^{163}$Ho spectrum has already been done by the ECHo collaboration and will be discussed in the following paragraph.
The atomic de-excitation is a complex process, which includes cascades of both x-rays and electron emissions (Auger electrons and Coster-Kronig transitions). The possibility to measure all the energy released in the decay minus the energy taken away by the neutrino simplifies the description of the spectrum. The expected shape of the calorimetrically measured EC spectrum, considering only first order excitations with only one hole in the internal shell, is:
\begin{equation}
\frac{dN}{dE_{\mathrm{C}}}\,=\,A(Q_{\mathrm{EC}}-E_{\mathrm{C}})^2\sqrt{1-\frac{m_{\nu}^2}{(Q_{\mathrm{EC}}-E_{\mathrm{C}})^2}}\sum{C_{\mathrm{H}}n_{\mathrm{H}}B_{\mathrm{H}}\,\phi^2_{\mathrm{H}}(0)\frac{\frac{\Gamma_{\mathrm{H}}}{2\pi}}{(E_{\mathrm{C}}-E_{\mathrm{H}})^2+\frac{\Gamma_{\mathrm{H}}^2}{4}}}	 
\label{ECspectrum}
\end{equation}
and shows Breit-Wigner resonances centered at an energy, $E_{\mathrm{H}}$, where $H$ indicates the level from which the electron has been captured, given in first approximation by the difference between the energy of the electron that has been captured and the energy of the extra-electron in the outer shell, in respect to the daughter atom. The resonances have an intrinsic width $\Gamma_{\mathrm{H}}$ related to the half-life of the excited states. The intensities of these lines are given mainly by the squared wave-function of the captured electron calculated at the nucleus $\phi^2_{\mathrm{H}}(0)$, with a small correction, $B_{\mathrm{H}}$, due to exchange and overlap. These factors are then multiplied by the nuclear shape factors $C_{\mathrm{H}}$ and the fraction of occupancy of the $H$-atomic shell $n_{\mathrm{H}}$. The Breit-Wigner resonances are then modulated by the phase space factor, which depends on the square of the electron neutrino mass $m(\nu_{\mathrm{e}})^2$ and the energy available to the decay $Q_{\mathrm{EC}}$. $A$ is a constant. In case of $^{163}$Ho, due to the extremely low energy available to the decay, only capture from the 3s and higher shells are allowed and due to the overlap of the electron wavefunctions with the nucleus, only electrons from the s and p$_{1/2}$ shells can be captured. Therefore the $^{163}$Ho spectrum will consist of the following lines: MI (3s electrons), MII (3p$_{1/2}$ electrons),NI (4s electrons), NII (4p$_{1/2}$), OI (5s electrons), OII (5p$_{1/2}$ electrons) and PI (6s electrons). The binding energy of these electrons, $E_{\mathrm{H}}^{\mathrm{bin}}$, and the intrinsic linewidths $\Gamma_{\mathrm{H}}$ of the transition are given in Table~\ref{tab:lines}.
\begin{table}[t]
\begin{center}
\caption{The binding energies $E_{\mathrm{H}}^{\mathrm{bin}}$ of the electron in Dy and the linewidths $\Gamma_{\mathrm{H}}^{\mathrm{lit}}$ as reported in \cite{lusignoli_2012}.}
\begin{tabular*}{\textwidth}{@{\extracolsep{\fill}}lrrrrr}
\hline
H & 	MI		& MII		& NI		& NII		& OI	
\\ \hline \hline
$E_{\mathrm{H}}^{\mathrm{bin}}\ [\mathrm{eV}]$ \cite{Deslattes2003} & 2046.9 & 1844.6 & 420.3 & 340.6 & 49.9~\cite{XRay2009} \\
$\Gamma_{\mathrm{H}}^{\mathrm{lit}}$ $[\mathrm{eV}]$ \cite{Campbell2001} & 13.2 & 6.0 & 5.4 & 5.3 & 3.7~\cite{Cohen1972} \\
\hline
\end{tabular*}
\label{tab:lines}
\end{center}
\end{table}

The electron neutrino mass that can be investigated with $^{163}$Ho-based experiments can be written in terms of the single mass eigenstates, in the scenario of only three active neutrinos, as:
\begin{equation}
m(\nu_{\mathrm{e}})^2=\sum_{i=1}^3|U_{\mathrm{ei}}^2|m(\nu_{\mathrm{i}})^2
\label{e_neutrino}
\end{equation}

In case keV sterile neutrinos would exist, at least a fourth neutrino mass eigenstate, $m_4$, would mix with the three "active" mass eigenstates to give the electron neutrino emitted in EC processes. In such a case, and considering the approximation that the three light neutrino mass eigenstates are much lighter than $m_4$, the electron neutrino state can be written as:

\begin{equation}
m(\nu_{\mathrm{e}})^2=\sum_{i=1}^3|U_{\mathrm{ei}}^2|m(\nu_{\mathrm{i}})^2+|U_{\mathrm{e4}}^2| m(\nu_{\mathrm{4}})^2 = |U_{\mathrm{ea}}^2| m_{\mathrm{a}}^2+ |U_{\mathrm{e4}}^2| m_{\mathrm{4}}^2
\label{e_neutrino2}
\end{equation}

By substituting this equation for the electron neutrino mass in the formula of the spectrum and by using the approximation that $m_{\mathrm{a}}=0$ eV, the following expression is obtained:
\begin{equation}
\label{Amplitude_sterile2}
\lambda_{\mathrm{H}}\,=\,A \left( (Q_{\mathrm{EC}}-E_{\mathrm{H}})^2 (1-U_{\mathrm{e4}}^2) + (Q_{\mathrm{EC}}-E_{\mathrm{H}})^2 U_{\mathrm{e4}}^2 \sqrt{1-\frac{m_{\mathrm{4}}^2}{(Q_{\mathrm{EC}}-E_{\mathrm{H}})^2}}\right ) \,C_{\mathrm{H}}n_{\mathrm{H}}B_{\mathrm{H}}\,\phi^2_{\mathrm{H}}(0)
\end{equation}

\begin{figure}
  \centering
  \subfigure[]{\includegraphics[width = 0.45\textwidth]{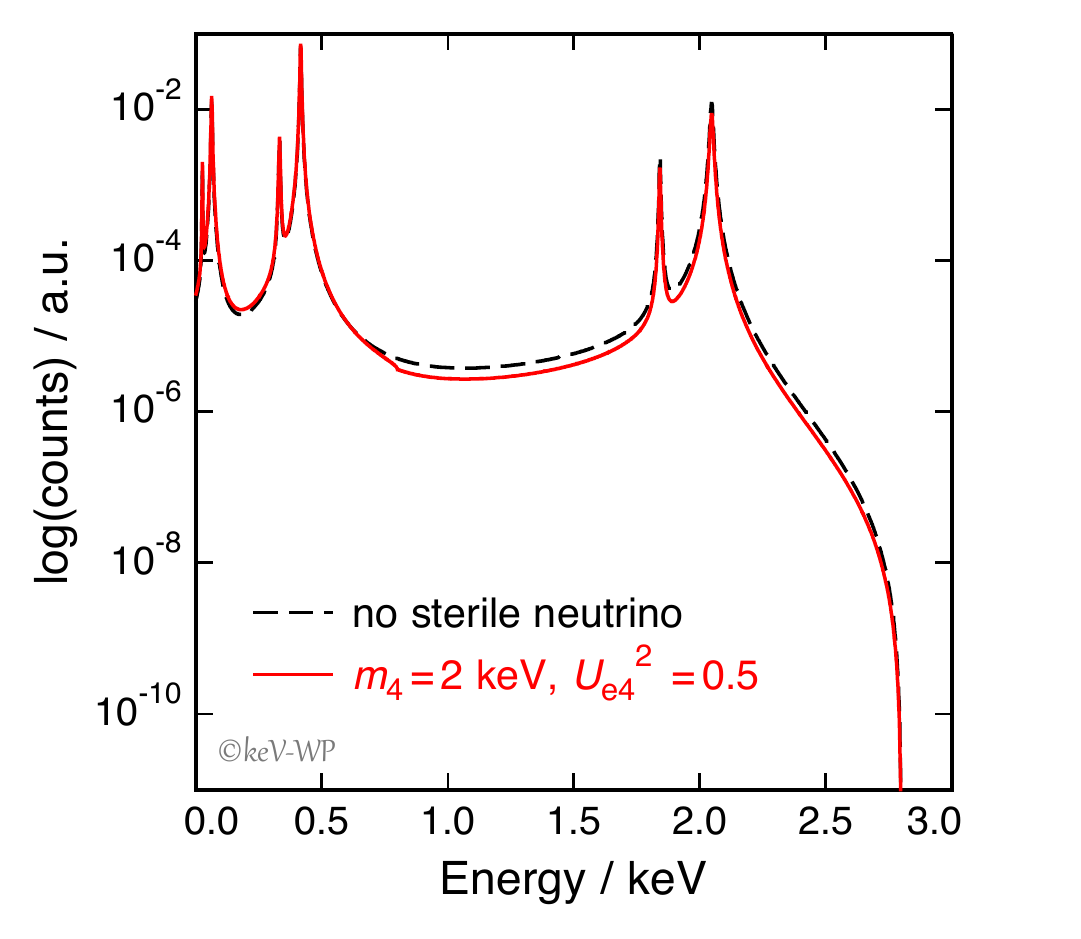}}
  \subfigure[]{\includegraphics[width = 0.44\textwidth]{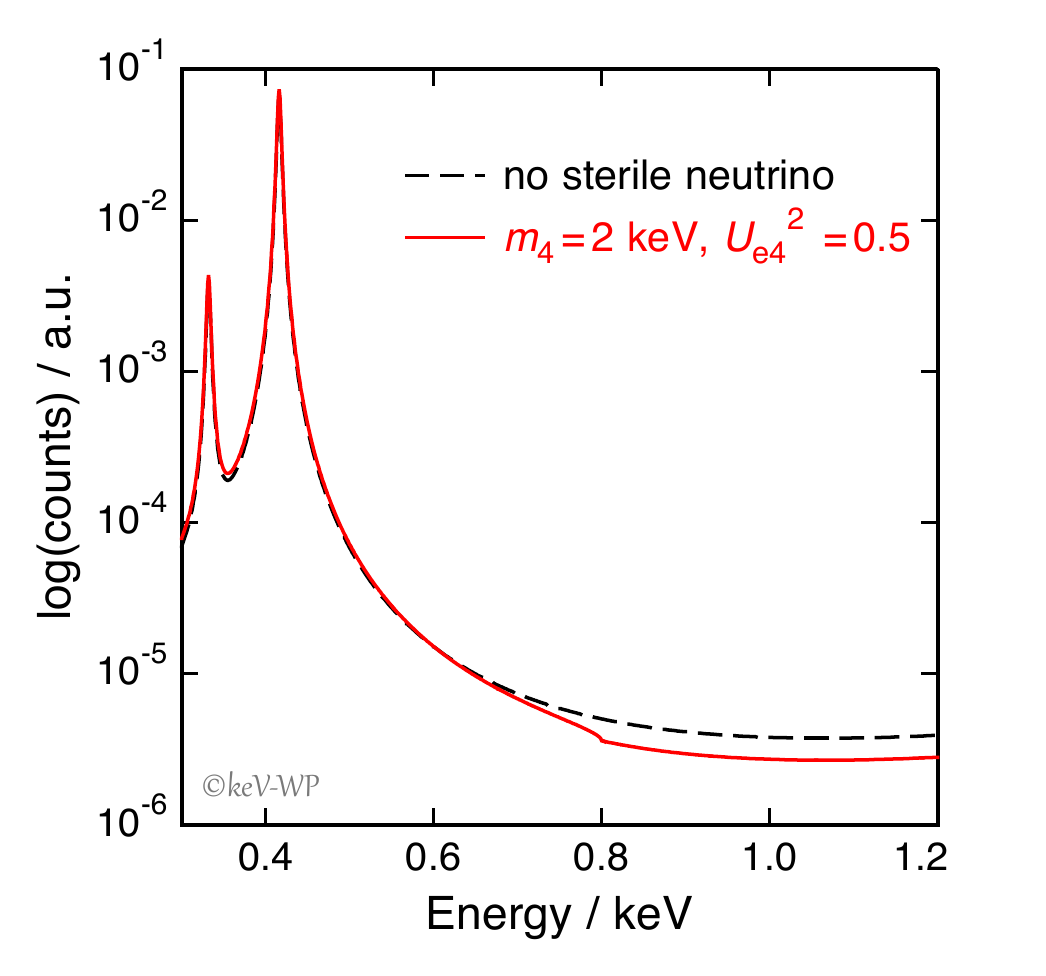}}
\caption{a) Comparison between the expected calorimetrically measured $^{163}$Ho spectrum in the case of no sterile neutrino (black dashed line) and in the case of a heavy neutrino mass $m_4=2$~keV with a mixing of $U_{\mathrm{e4}}^2 = 0.5$. b) A magnification of (a) in the region of the kink.}
 \label{performance}
\end{figure}

It follows that the evidence in the spectrum of the existence of the fourth neutrino mass eigenstate is a kink positioned at the energy $Q_{\mathrm{EC}}$ - $m_4$ and with an amplitude proportional to $|U_{\mathrm{e4}}^2|$.
With the analysis of the electron capture spectrum of $^{163}$Ho it is possible to investigate the existence of sterile neutrinos only for $m_4$ smaller than the $Q_{\mathrm{EC}}$ value of the decay of about 2.5~keV. Figure~\ref{performance} shows the comparison between the expected calorimetrically measured $^{163}$Ho spectrum calculated using Equation~\ref{Amplitude_sterile2} in the case of no sterile neutrino (black dashed line) and in the case of a heavy neutrino mass $m_4\,=\, 2$ keV with an unrealistic mixing of $U_{\mathrm{e4}}^2 \, = \, 0.5$. The two spectra have been normalized to both have an integral equal to 1. The magnification in Figure~\ref{performance} shows, in more details, the region of the kink.

%

%% file: Echo.tex
The Electron Capture in $^{163}$Ho Experiment, ECHo, has been designed to investigate the electron neutrino mass in the energy range below 1 eV by a high precision and high statistics calorimetric measurement of the $^{163}$Ho electron capture spectrum~\cite{Gastaldo:2013wha}.

\begin{figure}
  \centering
  \begin{minipage}{0.5\textwidth}
    \includegraphics[width = \textwidth]{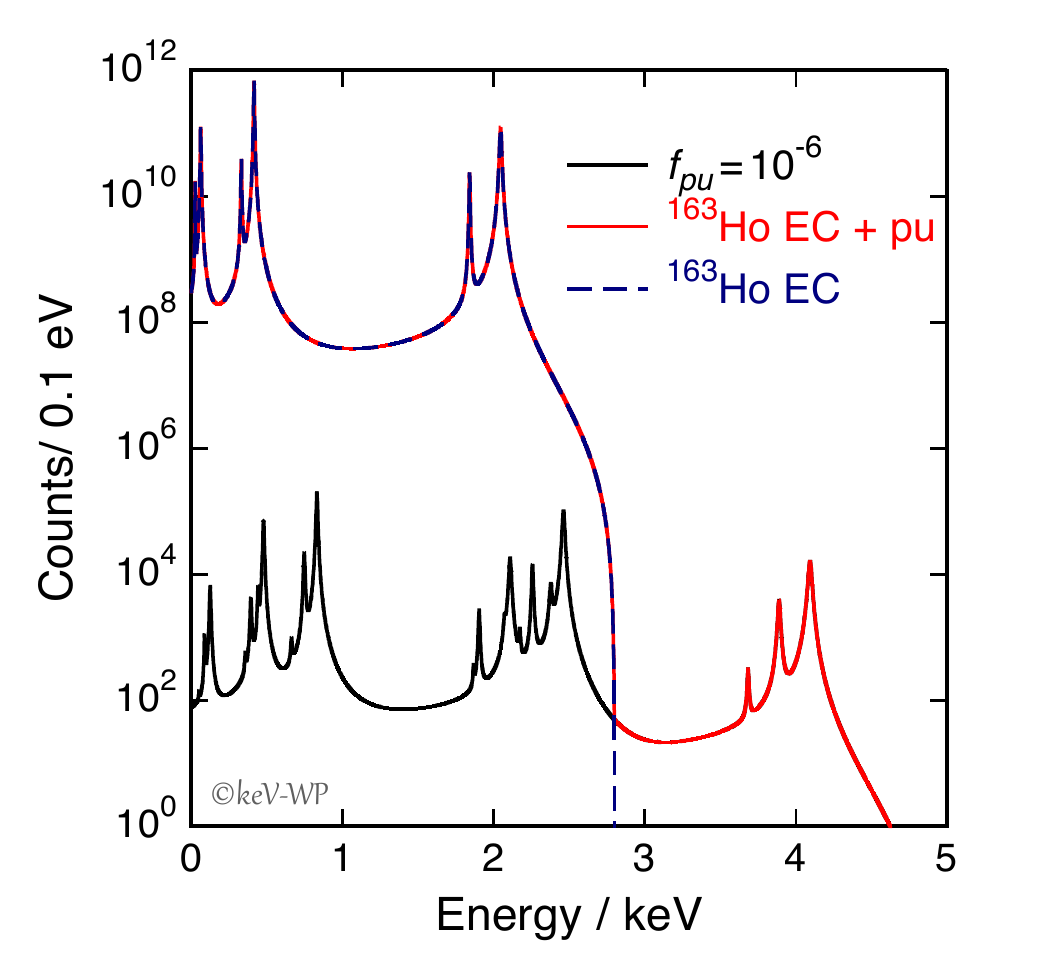}
  \end{minipage}
  \hfill
  \begin{minipage}{0.49\textwidth}
	\caption{The $^{163}$Ho spectrum with no unresolved pile up (blue dashed line) is compared with the $^{163}$Ho spectrum in the case of an unresolved pileup fraction of $10^{-6}$ (red solid line) for a total statistics of $10^{14}$ events.}
	\label{pu}
  \end{minipage}
\end{figure}

\noindent Presently the detectors that can detect with the highest precision energy input below 3 keV, as is needed for the calorimetric measurement of the $^{163}$Ho, are low temperature micro-calorimeters~\cite{Enss:2005md}. They are energy dispersive detectors typically operated at temperatures below $50\,$mK. These detectors can be classified on the basis of the temperature sensor. Within the ECHo experiment, low temperature metallic magnetic calorimeters (MMCs) will be used~\cite{Fleischmann_LTD13}.
The spectral resolving power of a state of the art MMCs for soft X-rays is above 3000. For completely micro-structured detectors, an energy resolution of $\Delta E_{\mathrm{FWHM}}\,=\,1.6\,$eV at $6\,$keV has been achieved~\cite{Fleischmann_tbs}. Sub-eV energy resolution is expected to be reached in future design with SQUID readout integrated on the detector chip. Such an energy resolution will allow to have a precise characterization of the endpoint region of the spectrum with a minimal spread of the events.
The typical signal rise-time is $\tau_{\mathrm{r}}\,=\,90\,$ns~\cite{Pies_LTD14}. Among the different temperature sensors presently used to read-out the temperature of micro-calorimeters, MMCs show the fastest risetime.
This feature is extremely important to reduce a source of background which is intrinsic to the calorimetric measurements: un-resolved pile-up events.
The spectral shape of this background is given by the autoconvolution of the $^{163}$Ho spectrum with the integral given by the so-called unresolved pileup fraction which can be written in first approximation as the activity in the detector $A$ times the risetime of the signal $\tau_{\mathrm{r}}$. For the ECHo experiment an unresolved pile-up fraction below $10^{-5}$ is required. This unresolved pile up fraction, combined with the typical risetime of MMCs, implies a limit in the maximal activity per pixel to a few tens of Bequerel. Figure~\ref{pu} compares the $^{163}$Ho spectrum without unresolved pile up (blue dashed line) with the $^{163}$Ho spectrum in the case of an unresolved pileup fraction of $10^{-6}$ (red solid line).

\noindent In the ECHo experiment, the required activity of $^{163}$Ho to reach a sub-eV sensitivity is a few MBq. The reduced activity per pixel, in order to limit the unresolved pile-up events, implies that in the order of $10^5$ single detectors have to be simultaneously measured. A convenient approach to read out so many detectors is the use of a multiplexing scheme. The choice within the ECHo collaboration is to use the microwave SQUID multiplexing scheme~\cite{Mates} which allows to have a large bandwidth for each pixel and a reduced degradation of the detectors performance respect to the single pixel readout~\cite{Kempf_LTD15}.

\noindent The ECHo collaboration has already demonstrated the possibility to perform a calorimetric measurement of the $^{163}$Ho with high energy resolution using MMC detectors with ion implanted $^{163}$Ho~\cite{Ranitzsch:2014kma}.
The $^{163}$Ho activity per pixel was about $10^{-2}$ Bq. The implantation process did not degrade the performance of the MMC~\cite{Gastaldo:2012nv}. An energy resolution of $\Delta E_{\mathrm{FWHM}}\,\simeq\,7.6\,$eV, at a working temperature of about 30 mK, and the rise-time $\tau_{\mathrm{r}}\,\simeq\,130\,$ns have been measured.
One of the next important goals within ECHo is to demonstrate that very good energy resolutions, $\Delta E_{\mathrm{FWHM}}\,< \, 2\, $eV  can be achieved for MMCs having $^{163}$Ho ions embedded in the absorber and read out using the microwave multiplexing technique.

\noindent The information on the neutrino mass is obtained by the analysis of the events in a very small energy range around the endpoint of the spectrum. Therefore to have an independent measurement of the $Q_{\mathrm{EC}}$ for the $^{163}$Ho decay to be compared to the value that can be extracted by fitting the spectrum is of outmost importance for the reduction of systematic errors. By the analysis of calorimentrically measured spectra the best values for $Q_{\mathrm{EC}}$ are approximately 2.8 keV~\cite{Meunier:1996ge,Ranitzsch:2014kma}. These results do not agree with the recommended value $Q_{\mathrm{EC}}$ = (2.55 $\pm$ 0.016) keV published by the Atomic Mass Evaluation edited in 2012~\cite{Q_EC}.

\noindent One of the priority of the ECHo collaboration was therefore to perform and independent measurement of the $Q_{\mathrm{EC}}$ by precisely measuring the mass of the parent atom, $^{163}$Ho, and daughter atom $^{163}$Dy. This has been performed by high precision Penning traps mass spectrometry~\cite{PT-MS}. The measurement has been performed at the Penning-trap mass spectrometer SHIPTRAP~\cite{SHIPTRAP} applying the novel phase-imaging ion-cyclotron-resonance technique~\cite{ICRT} and the laser ionization of the source developed at the TRIGA-TRAP~\cite{Ketelaer:2008by,Fabian_2015}. This measurement has been able to strongly reduce the uncertainties on $Q_{\mathrm{EC}}$. The value that has been obtained, $Q_{\mathrm{EC}}= (2.833 \pm 0.030^{\mathrm{stat}} \pm 0.015^{\mathrm{syst}})~\mathrm{keV}$~\cite{PhysRevLett.115.062501} is in very good agreement with the $Q_{\mathrm{EC}}$-value obtained by the analysis of calorimetrically measured $^{163}$Ho spectra.

\noindent In order to remove systematic uncertainties, at a few eV level, on the $Q_{\mathrm{EC}}$-value as determined by the analysis of the calorimetrically measured spectra, which could derive from solid state effects, an independent measurement at 1 eV precision level is of outmost importance. This precision will be achieved by measuring the masses of $^{163}$Ho and $^{163}$Dy with the newly developed PENTATRAP~\cite{Repp:2011hm,Roux:2011hn} at the Max Planck Institute for Nuclear Physics in Heidelberg.

\noindent The spectral shape as described in equation~\ref{ECspectrum} considers only first order transitions where the excited states of the daughter atom $^{163}$Dy can be characterized by a hole at the level of the captured electron and an additional electron in the $4f-$shell. In order to describe small structures in the spectrum above the NI-line, second and third order excitations have been theoretically investigated~\cite{Robertson:2014fka,Faessler:2015pka,Faessler:2015txa}. Presently there is no good agreement between the experimentally observed structures and the ones expected by the theoretical models. Though, some of the theoretically calculated structures coincide in energy with observed ones. New models for the description of atomic de-excitation are under investigation \cite{ADR_ML_2015} as well as dedicated experiments will be performed to better characterize the electron capture decay in $^{163}$Ho.

\noindent The aim of the ECHo collaboration is to gain a precise understanding of background sources which could affect the $^{163}$Ho spectrum and to develop methods to reduce them to a level where their contribution will be smaller compared to the unresolved pile-up events in the endpoint region of the spectrum. To reach this limit the background level should be smaller than $10^{-5}$ events/eV/day/det.
The next step of the ECHo collaboration is to perform a medium scale experiment, ECHo-1k, which will run during the years 2015-2018. In ECHo-1k first prototypes of MMC array with implanted $^{163}$Ho will be fabricated. The $^{163}$Ho source will be produced through neutron irradiation of a $^{162}$Er-enriched target and dedicated chemical purifications step along with mass separation will be used to remove radioactive contamination to a not detectable level. The sum of the activity of all the pixels will be approximately 1 kBq. With this first medium scale experiment the ECHo collaboration aims to reduce the present limit on the electron neutrino mass by more than one order of magnitude, from a present upper limit of $m(\nu_{\mathrm{e}})$ = 225~eV~\cite{Springer_1987} to below $m(\nu_{\mathrm{e}})$ = 20 eV and show the scalability of the developed techniques for the construction of the next generation experiment, ECHo-1M. The goal of ECHo-1M is to measure a $^{163}$Ho EC spectrum with a statistics of more than $10^{14}$ events using a $^{163}$Ho source of the order of 1 MBq distributed over a very large number of single pixels. With such an experiment it will be possible to reach the sub-eV sensitivity on the electron neutrino mass.

\paragraph{keV sterile neutrino program in ECHo}
The calorimetric technique to measure the $^{163}$Ho spectrum has the intrinsic property that all the events generated in the detectors produce a signal. Therefore, in respect to an experiment as KATRIN, where a MAC-E filter selects only electrons close to the endpoint of the Tritium spectrum, in ECHo there is the possibility to acquire all the events occurring in the detector, even if for the determination of the electron neutrino mass mainly the events at the end point are important.
The ECHo program to investigate the effect of keV sterile neutrino on the $^{163}$Ho EC spectrum can be divided into two parts:
\begin{itemize}
\item assess systematic effects and analysis of high statistics $^{163}$Ho EC spectra searching for a kink at $Q_{\mathrm{EC}}-m_4$
\item develop efficient methods to acquire and store large number of events
\end{itemize}
The first part of this investigation will be carried on during the first stage of the ECHo experiment, ECHo-1k. The assess of systematic effects due to the not precise knowledge of the calorimetrically measured spectrum of $^{163}$Ho is a fundamental milestone to reach the sub-eV sensitivity. The precise knowledge of the $Q_{\mathrm{EC}}$ is of outmost importance not only for the investigation of the "standard" effective electron neutrino mass, but also to define with high precision the value of the heavy neutrino mass $m_4$ through the difference between the position of the possible kink and the endpoint of the spectrum. The uncertainties on the $Q_{\mathrm{EC}}$ have already been strongly reduced by the measurements performed using Penning traps \cite{PhysRevLett.115.062501} and this achievement paved the way for future experiments aiming to the 1 eV precision.

\noindent The effects due to higher order excitation states in the $^{163}$Dy atom after the electron capture in $^{163}$Ho is presently a topic of high interest. In the high resolution spectrum measured within the ECHo experiment, additional structures above the NI-line, which can hardly be explained as being due to background sources, have been detected~\cite{Ranitzsch_2015}. This experimental evidence has already triggered the work of different groups~\cite{Robertson:2014fka,Faessler:2015pka,Faessler:2015txa,ADR_ML_2015}. In these works, some of them performed within the ECHo experiment, the modifications to the calorimentrically measured spectrum due to excitations in the $^{163}$Dy atoms including two and three holes have been calculated. Presently the agreement between theoretical calculations and the ECHo data is still not satisfactory, but there is already evidence that this approach can be used to refine the description of the spectrum. Within ECHo-1k several experiments will be performed to better understand the shape of the $^{163}$Ho spectrum. In particular spectra with higher statistics and with a reduced background level, in a way that the shape of the additional structures will be better characterized, will be available to the theorists.

%

\noindent As already mentioned, one of the milestones of ECHo-1k is the identification and reduction of the background in the $^{163}$Ho spectrum deriving from contaminations in the $^{163}$Ho source, from natural radioactivity and from cosmic rays. The goal is to reduce all these contributions to a negligible level with respect to the intrinsic background due to unresolved pile-up events. This requirement, in the case of an unresolved pileup fraction of $10^{-6}$, leads to an allowed background activity per pixel of less than $10^{-5}$ events eV/det/day in the energy region below the endpoint of the $^{163}$Ho spectrum.

\noindent A good theoretical description of the $^{163}$Ho spectrum as well as of the background spectrum will allow also for a precise description of the unresolved pile up spectrum. The presence of additional small structures in the spectrum due to unresolved pileup could complicate the identification of a kink due to the existence of sterile neutrinos. Even if the choice made by the ECHo collaboration to keep the level of the unresolved pileup fraction below $10^{-5}$ implies that over a large part of the spectrum, the number of the expected unresolved pile up events is smaller than the statistical error for the counts in the $^{163}$Ho spectrum, a precise knowledge of this background spectrum is very important to push the sensitivity to detect a kink to heavy neutrinos having a very small mixing element.
One of the aims of ECHo-1k is therefore to define a precise paremeterization for the calorimetrically measured $^{163}$Ho spectrum.

\begin{figure}
  \centering
  \subfigure[]{\includegraphics[width = 0.48\textwidth]{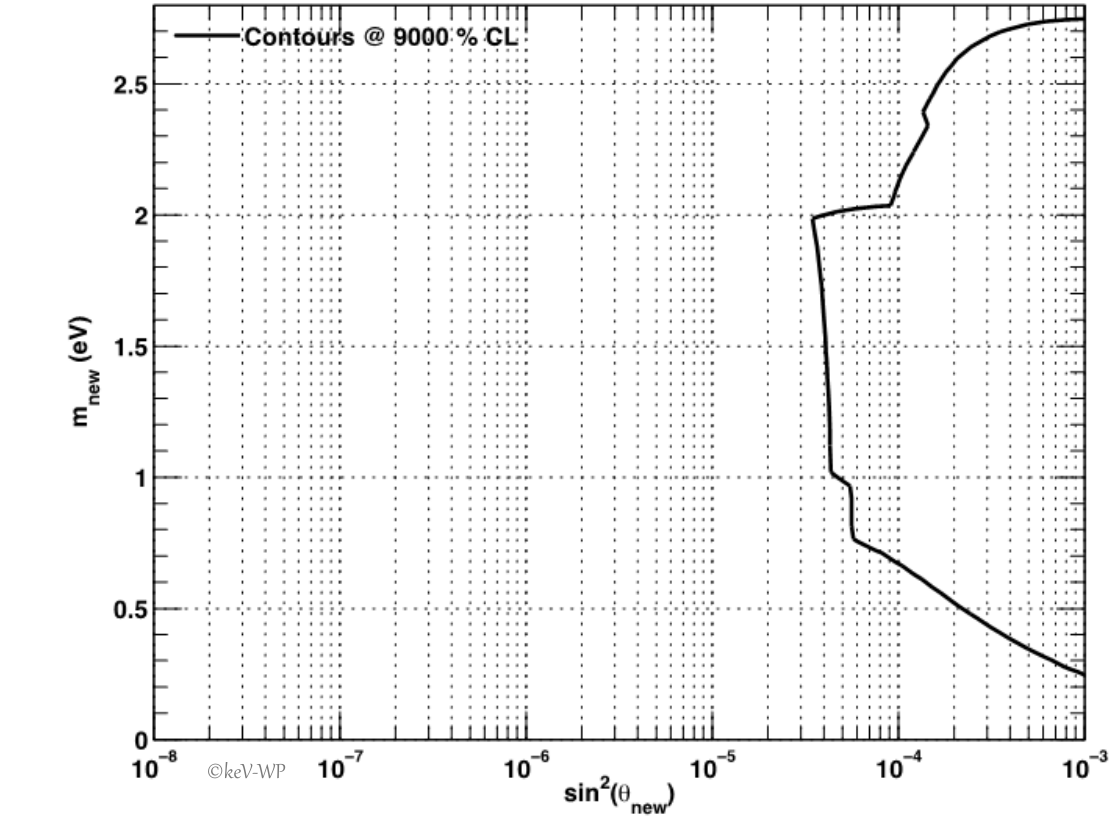}}
  \subfigure[]{\includegraphics[width = 0.49\textwidth]{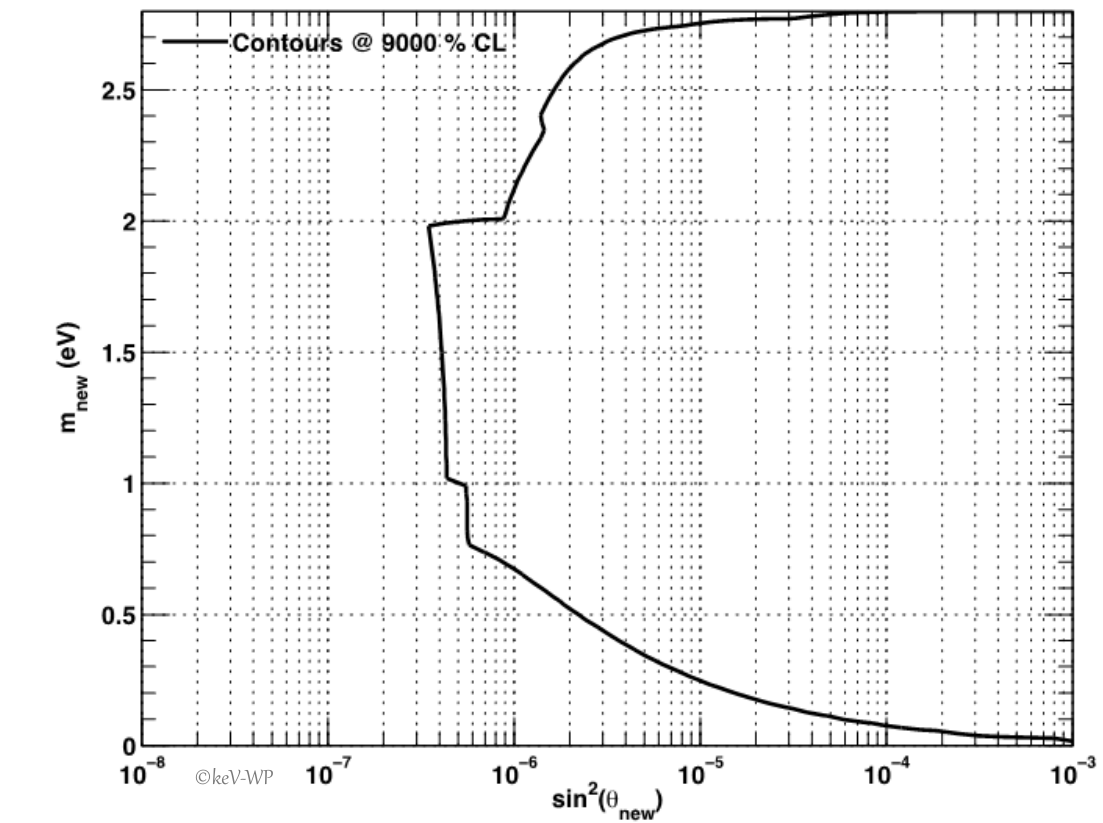}}
\caption{Achievable sensitivity 90$\%$ C.L. for the detection of a kink in the calorimetrically measured spectrum in the case of a perfect knowledge of the expected shape and for a total statistics of a) $10^{10}$ and b) $10^{14}$ events.}
 \label{fig:sensiEcho}
\end{figure}

\noindent With the planned activity of 1 kBq and a measuring time of about one year a total statistics in the spectrum of about $10^{10}$ events will be acquired at the end of the first phase of the ECHo experiment. Figure~\ref{fig:sensiEcho}a shows the achievable sensitivity for the detection of a kink in the calorimetrically measured spectrum in the case of a perfect knowledge of the expected shape of the spectrum and for the total statistics foreseen for the ECHo-1k experiment. For these calculations, $Q_{\mathrm{EC}}$ = 2.8~keV has been used. The best sensitivity to detect a kink in the spectrum is achieved for a heavy neutrino mass $1\,\mathrm{keV}\, <\,m_4 \,<\, 2\,\mathrm{keV}$, which corresponds to the part of the spectrum between the M- and N-lines. In this energy range it will be possible to test the mixing element down to about $|U_{\mathrm{e4}}|^2 \,=\,4\cdot10^{-5}$. For the second phase of the ECHo experiment, ECHo-1M, an activity of 1~MBq to a few MBq is foreseen. With such an activity it will be possible to acquire, within one year of measuring time, about $10^{14}$ events in the $^{163}$Ho spectrum. With this statistics it will be possible to improve the sensitivity to find evidence of sterile neutrinos. Figure~\ref{fig:sensiEcho}b shows the achievable sensitivity with a total statistics of $10^{14}$ events and a perfect knowledge of the spectral shape. In this case, for $1\, \mathrm{keV}\, <\,m_4 \,<\, 2\,\mathrm{keV}$, the mixing element could be tested down to about $|U_{\mathrm{e4}}|^2 \,=\,4\cdot10^{-7}$.

The sensitivity to detect a kink due to sterile neutrinos achieved considering only statistical errors on the counts per bin is very promising. More investigations are going on to define the influence on the reachable sensitivity of the finite energy resolution of the detectors and of the presence of background, in particular of the unresolved pile-up events.

From the experimental point of view it is important to point out that the acquisition and the storage of the $^{163}$Ho events over the full spectrum is not a negligible aspect of the investigation. During ECHo-1k about 1000 $^{163}$Ho events per seconds will occur. For this moderate rate, the acquisition of the full signal shape for all the events will still be possible. With the corresponding data it will be possible to develop routines to be applied for the on-line analysis of the data in the next stage of ECHo, ECHo-1M. In this experiment only a few characteristic parameters of each signal will be stored for the building of the complete spectrum. Nevertheless the aim is to store information for the pulses all over the energy range. This is not only important to reduce systematics in the analysis of the end point region of the spectrum, but it will allow for the analysis, over the complete spectrum for the search of a kink related to the existence of sterile neutrinos.

%% file: OtherNuclides.tex
Several models and observations allow the sterile neutrino mass $m_{\mathrm{s}}$ to cover a large range. In the electron capture sector, the analysis of the $^{163}$Ho electron capture spectrum allows only to investigate heavy neutrino masses up to the $Q_{\mathrm{EC}}$ value which, even allowing for uncertainties due to solid state effects , will hardly exceed 3 keV. On the other hand, the same methods applied within the ECHo experiment \cite{Gastaldo:2013wha} could be adapted to other nuclides showing higher $Q_{\mathrm{EC}}$ therefore allowing for investigating massive neutrinos in a larger mass region. The idea is to perform a calorimetric measurement of the electron capture spectrum of the interesting isotopes and search for the signature of a heavy neutrino mass. The method consists in embedding the source in the absorber of low temperature micro-calorimeters in a way that all the energy released in the electron capture process is measured by the detector. In low temperature micro-calorimeters the relatively small energy release in electron capture processes leads to an increase of temperature, which is measured by very sensitive thermometers. For the ECHo experiment, low temperature metallic magnetic calorimeters \cite{Fleischmann_LTD13} will be used. These detectors are also suitable to be used for the calorimetric measurement of the electron capture spectrum of a large variety of other nuclides. The possibility to use metallic magnetic calorimeters for the search of the signature of sterile neutrinos in calorimetrically measured electron capture spectra was first investigated by Filianin et al. \cite{Filianin_2014}. In that work, the achievable sensitivity to detect the effect of massive neutrinos of different masses and mixing was discussed for several nuclides in the electron capture sector. The proposed isotopes were selected on the base of their $Q_{\mathrm{EC}}$ value and/or half-life. Table \ref{nuclides} shows a list of nuclides which have been selected from the list proposed in \cite{Filianin_2014}. With respect to the other nuclides presented in \cite{Filianin_2014} the ones listed in Table \ref{nuclides} are more suitable, due to their decay schemes, to be investigated with micro-calorimetric techniques in future experiments, as discussed in the same paper.
\begin{center}
\begin{table}
\caption{Nuclides with relevant energy balance and suitable half-life to be used in potential future experiments to investigate keV sterile neutrinos by the analysis of the calorimetrically measured electron capture spectrum}
\begin{tabular*}{\textwidth}{@{\extracolsep{\fill}} l c c c c c}
\hline
Nuclide & $T_{1/2}$ & $Q_{\mathrm{EC}}$ [keV] & $E_{\mathrm{i}}$ [keV] & $E_{\mathrm{j}}$ [keV] & $\phi^2_{\mathrm{i}}(0)/\phi^2_{\mathrm{j}}(0)$ \\
&  \cite{halflife} & \cite{Q_EC-values}& \cite{binding_E}& \cite{binding_E} & \cite{amplitudes}\\
\hline
\hline
$^{157}$Tb & 71 a & 60.04 & K: 50.2391 & LI: 8.3756 & 7.124 \\
$^{163}$Ho & 4570 a & 2.833 \cite{PhysRevLett.115.062501} & MI: 2.0468 & NI: 0.4163  & 4.151 \\
$^{179}$Ta & 1.82 a & 105.6 & K: 65.3508 & LI: 11.2707 & 6.711 \\
$^{193}$Pt & 50 a & 56.63 & LI: 13.4185 & MI: 3.1737 & 4.077 \\
$^{205}$Pb &  15 Ma & 50.6 & K: 15.3467  & LI: 3.7041  & 4.036  \\
\hline
\end{tabular*}
\label{nuclides}
\end{table}
\end{center}

The sensitivity to investigate the mass and mixing of sterile neutrinos can be determined by considering the fact that the presence of a massive neutrino will reduce the phase space for energy above $(Q_{\mathrm{EC}}-m_{\mathrm{s}})$. If the value $(Q_{\mathrm{EC}}-m_{\mathrm{s}})$ were between the binding energies of two electronic states from which electrons can be captured, then the ratio between the amplitude of the two corresponding lines in the calorimetrically measured spectrum would be affected.
From Eq. \ref{ECspectrum}, where the expected shape of the calorimetrically measured EC spectrum was given considering only first order transitions, the amplitude for each line can be calculated:
\begin{equation}
\lambda_{\mathrm{H}}\,=\,A(Q_{\mathrm{EC}}-E_{\mathrm{H}})^2\sqrt{1-\frac{m_{\nu}^2}{(Q_{\mathrm{EC}}-E_{\mathrm{H}})^2}}\,C_{\mathrm{H}}n_{\mathrm{H}}B_{\mathrm{H}}\,\phi^2_{\mathrm{H}}(0)	 \label{Amplitude}
\end{equation}
with a scaling constant $A$, the peak energy $E_{\mathrm{H}}$, the nuclear shape factors $C_{\mathrm{H}}$, the fraction of occupancy of the $H$-atomic shell $n_{\mathrm{H}}$, the squared wave-function of the captured electron calculated at the nucleus $\phi^2_{\mathrm{H}}(0)$ and the exchange and overlap correction, $B_{\mathrm{H}}$.  The Breit-Wigner resonances are then modulated by the phase space factor which depends on the square of the electron neutrino mass $m(\nu_{\mathrm{e}})^2$ and the energy available to the decay $Q_{\mathrm{EC}}$. Second order transitions which require the formation of two or more hole states in the final atom are here neglected since at the present level of investigation their effect would be marginal.

\noindent By explicitly introducing the dependency on the neutrino masses and reducing this to a 1+1 model, the electron neutrino state is given by the sum of a massless neutrino $|\nu_{\mathrm{a}}\rangle$, to represent the three small mass active neutrino masses, and one massive neutrino $|\nu_{\mathrm{4}}\rangle$:
\begin{equation}
|\nu_{\mathrm{e}}\rangle=\sum_{i=1}^{3} U_{\mathrm{ei}} |\nu_{\mathrm{i}}\rangle + U_{\mathrm{e4}} |\nu_{\mathrm{4}}\rangle = U_{\mathrm{ea}} |\nu_{\mathrm{a}}\rangle + U_{\mathrm{e4}} |\nu_{\mathrm{4}}\rangle
\label{e_neutrino3}
\end{equation}
With this approximation the amplitude of the line H can be written as:
\begin{equation}
\lambda_{\mathrm{H}}\,=\,A \left( (Q_{\mathrm{EC}}-E_{\mathrm{H}})^2 (1-U_{\mathrm{e4}}^2) + (Q_{\mathrm{EC}}-E_{\mathrm{H}})^2 U_{\mathrm{e4}}^2 \sqrt{1-\frac{m_{\mathrm{4}}^2}{(Q_{\mathrm{EC}}-E_{\mathrm{H}})^2}}\right ) \,C_{\mathrm{H}}n_{\mathrm{H}}B_{\mathrm{H}}\,\phi^2_{\mathrm{H}}(0)
\label{Amplitude_sterile}
\end{equation}
If no keV sterile exists, then the amplitude of the line H in the approximation of vanishing active masses can be written as:
\begin{equation}
\lambda_{\mathrm{H}}\,=\,A(Q_{\mathrm{EC}}-E_{\mathrm{H}})^2 \, C_{\mathrm{H}}n_{\mathrm{H}}B_{\mathrm{H}}\,\phi^2_{\mathrm{H}}(0)
\label{Amplitude_active}
\end{equation}
Using Equation~\ref{Amplitude_sterile} and Equation~\ref{Amplitude_active} the ratio between the amplitude of two lines in the spectrum can be written as:
\begin{equation}
\left ( \frac{\lambda_{\mathrm{i}}}{\lambda_{\mathrm{j}}} \right ) = \left ( \frac{\lambda_{\mathrm{i}}}{\lambda_{\mathrm{j}}} \right )_{\mathrm{act}} \frac{U_{\mathrm{e4}}^2\left ( H\left [ (Q_{\mathrm{EC}}- E_{\mathrm{i}}) - m_{\mathrm{4}} \right ] \sqrt{1-m_{\mathrm{4}}^2/(Q_{\mathrm{EC}}- E_{\mathrm{i}})} - 1 \right ) +1}{U_{\mathrm{e4}}^2\left ( H\left [ (Q_{\mathrm{EC}}- E_{\mathrm{j}}) - m_{\mathrm{4}} \right ] \sqrt{1-m_{\mathrm{4}}^2/(Q_{\mathrm{EC}}- E_{\mathrm{j}})} - 1 \right ) +1}
\label{Amplitude_ste_ratio}
\end{equation}
with $H\left [ (Q_{\mathrm{EC}}- E_{\mathrm{i}}) - m_{\mathrm{4}} \right ]$ the Heaviside step function and
\begin{equation}
\left ( \frac{\lambda_{\mathrm{i}}}{\lambda_{\mathrm{j}}}\right )_{\mathrm{act}}\,=\,\frac{(Q_{\mathrm{EC}}-E_{\mathrm{i}})^2 \, C_{\mathrm{i}}n_{\mathrm{i}}B_{\mathrm{i}}\,\phi^2_{\mathrm{i}}(0)}{(Q_{\mathrm{EC}}-E_{\mathrm{j}})^2 \, C_{\mathrm{j}}n_{\mathrm{j}}B_{\mathrm{j}}\,\phi^2_{\mathrm{j}}(0)}
\label{Amplitude_act_ratio}
\end{equation}
From Eq. \ref{Amplitude_ste_ratio} it appears evident that the signature of the existence of a sterile neutrino is a deviation of the ratio between the amplitudes of two lines of the electron capture spectrum with respect to the value expected in the case of only three active neutrinos. The magnitude of this deviation depends on the position of $(Q_{\mathrm{EC}}-m_{\mathrm{s}})$ with respect to the energy of the two lines and of course on the mixing of the sterile neutrino $U_{\mathrm{e4}}^2$.

In order to be able to identify the effect of the existence of a sterile neutrino by the analysis of the ratio between two lines of the electron capture spectrum, it is important to minimize all systematic effects which could mimic a deviation from the expected amplitude ratio for only active neutrinos. The most relevant aspects that need to be considered are:
\begin{itemize}
 \item a precise knowledge of the $Q_{\mathrm{EC}}$-value
 \item a precise knowledge of the de-excitation processes after the electron capture
 \item a reduced unresolved pile-up fraction
 \item a reduced unknown background
\end{itemize}

\begin{figure}
\begin{center}
 \includegraphics[height=.35\textheight]{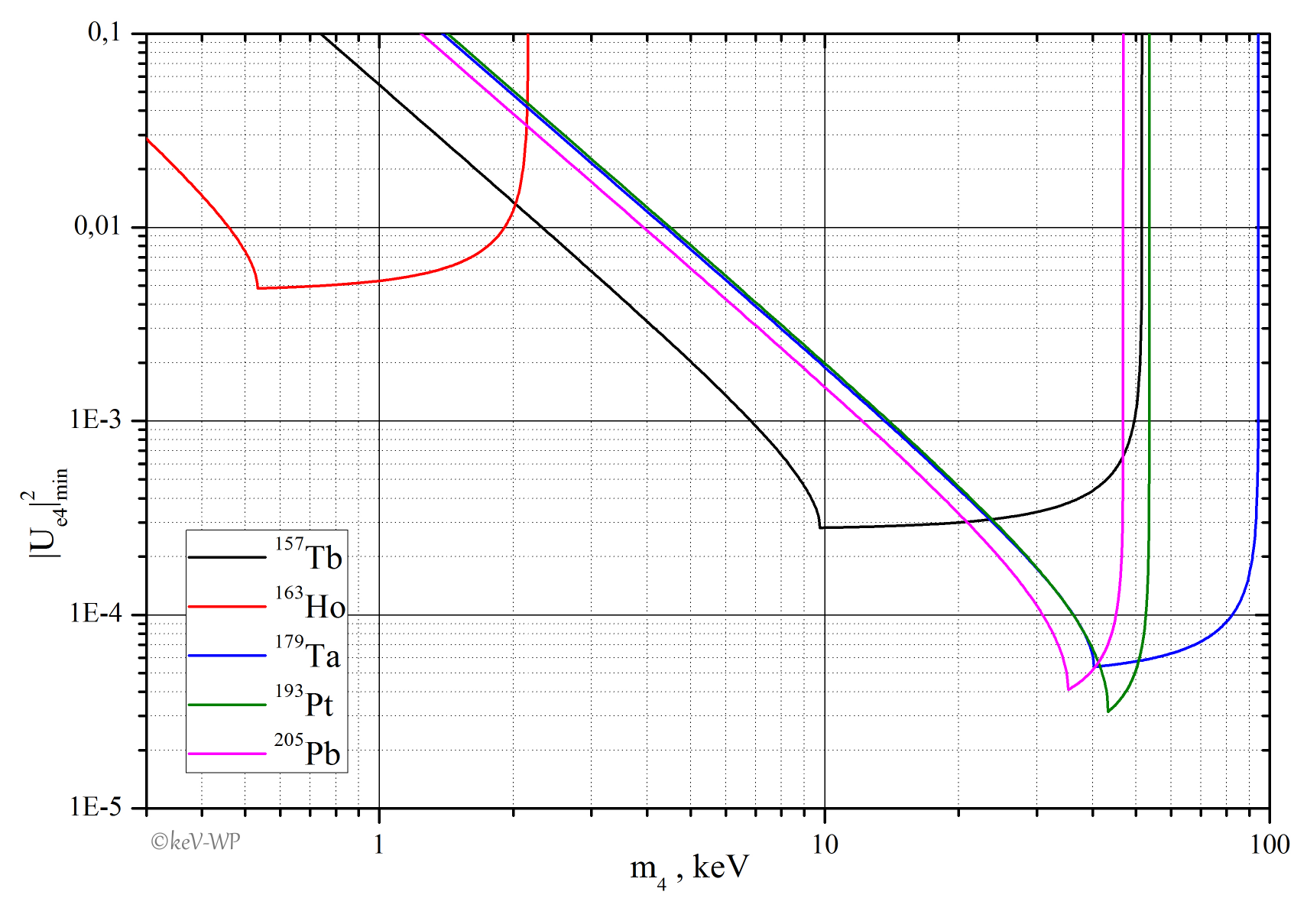}
  \caption{Minimal sterile neutrino mixing matrix element to the square that can be deduced at 90 $\%$ confidence level versus the mass $m_4$. The curves are calculated for a total statistics in the spectrum of $10^{13}$ events under the assumptions that the uncertainties on the absolute atomic mass difference are 1 eV and that the capture probability for each line is perfectly known while the position of the lines is given with an uncertainty of 0.1 eV.}
  \label{nuclide_sensitivity}
\end{center}
\end{figure}

A precise knowledge of the $Q_{\mathrm{EC}}$-value is very important since the amplitude of the lines in the electron capture spectrum is modulated by the phase space factor. The more a line in the spectrum is close to the end-point, the more the amplitude of this line is reduced with respect to the factor $C_{\mathrm{H}}n_{\mathrm{i}}B_{\mathrm{H}}\,\phi^2_{\mathrm{H}}(0)$. A large uncertainty in this value leads to a large uncertainty in the expected ratio between the amplitude of two lines of the spectrum. The most promising method to determine the $Q_{\mathrm{EC}}$-value of radioactive nuclide is by the precise measurement of the mass of parent and daughter atoms whose difference provides the energy available to the decay. Penning trap mass spectrometry \cite{PT-MS} is presently the best technique to precisely measure the mass of atoms. In these measurements, the cyclotron frequencies $\nu_{\mathrm{C}}$ of the initial and final ionic states of the transition are measured in the crossing electrostatic and strong homogeneous magnetic field $B$. The mass $M$ of an ion having charge $q$ can be determine using the following relation:
\begin{equation}
\nu_{\mathrm{C}}= \frac{qB}{2\pi M}
\label{cyc_freq1}
\end{equation}
In order to reach eV precision for the $Q_{\mathrm{EC}}$-value of heavy nuclides, a novel cryogenic Penning-trap mass spectrometer, PENTATRAP, is currently under construction at the Max-Planck Institute for Nuclear Physics in Heidelberg \cite{Repp:2011hm,Roux:2011hn}. With this mass spectrometer a relative mass accuracy of $10^{-11}$ will be reached allowing for a few eV precision on the $Q_{\mathrm{EC}}$-value of the discussed nuclides. Figure~\ref{nuclide_sensitivity} shows, at the 90$\%$ confidence level, the sensitivity for the sterile neutrino mass and mixing that can be reached by the analysis of the calorimetrically measured spectrum of the nuclides listed in Table \ref{nuclides}, under the assumptions that the uncertainties of the absolute atomic mass difference are 1 eV and that the capture probability for each line is perfectly known while the position of the lines is known with an uncertainty of 0.1 eV .
In this approach, the limiting factor for the achievable minimal sterile neutrino mixing is the relative error for the $Q_{\mathrm{EC}}$-value.

\noindent The curves in Figure~\ref{nuclide_sensitivity} have been calculated by considering that the de-excitation processes after the electron capture, which define the intrinsic amplitude of the lines, are perfectly known. Recently the case of the electron capture in $^{163}$Ho received attention due to the identification of additional small lines above the line corresponding to the capture of 4s electrons, the NI-line, in calorimetrically measured spectra \cite{Ranitzsch:2014kma}. A possible explanation of these lines seems to be related to higher order transitions which include the creation of two holes in the atom after the capture process \cite{Robertson:2014fka,Faessler:2015pka}. Presently there is still not satisfactory match between models and data. More theoretical investigations are presently on going to determine the mechanism which leads to the satellite lines observed in the $^{163}$Ho spectra.

\noindent A possible way to remove the uncertainty given by the limited knowledge of the intrinsic amplitude of the lines is to compare the same ratios for the intensities in different isotopes  of the same chemical element. As the authors of \cite{Filianin_2014} discussed, the influence of the uncertainties on the intrinsic amplitudes cancel to a large extent.
The capture intensity ratios for the isotopes 1 and 2 can be written as:
\begin{equation}
\Theta = \frac {(\lambda_{\mathrm{i}}/\lambda_{\mathrm{j}})_1}{(\lambda_{\mathrm{i}}/\lambda_{\mathrm{j}})_2} 
\end{equation}
\begin{equation}
\Theta = \Theta_{\mathrm{act}} \, \frac{U_{\mathrm{e4}}^2\left ( H_{\mathrm{1i}} \sqrt{1- \frac{m_{\mathrm{4}}^2}{(Q_{\mathrm{EC1}}- E_{\mathrm{i}})^2}} - 1 \right ) +1}{U_{\mathrm{e4}}^2\left ( H_{\mathrm{1j}} \sqrt{1-\frac{m_{\mathrm{4}}^2}{(Q_{\mathrm{EC1}}- E_{\mathrm{j}})^2}} - 1 \right ) +1}\;\;
\frac {U_{\mathrm{e4}}^2\left ( H_{\mathrm{2j}} \sqrt{1-\frac{m_{\mathrm{4}}^2}{(Q_{\mathrm{EC2}}- E_{\mathrm{j}})^2}} - 1 \right ) +1}{U_{\mathrm{e4}}^2\left ( H_{\mathrm{2i}} \sqrt{1-\frac{m_{\mathrm{4}}^2}{(Q_{\mathrm{EC2}}- E_{\mathrm{j}})^2}} - 1 \right ) +1}
\label{cyc_freq2}
\end{equation}
with
\begin{equation}
\Theta_{\mathrm{act}} = \left [ \frac{(Q_{\mathrm{EC1}}- E_{\mathrm{i}})(Q_{\mathrm{EC2}}- E_{\mathrm{j}})}{(Q_{\mathrm{EC1}}- E_{\mathrm{j}})(Q_{\mathrm{EC2}}- E_{\mathrm{i}})} \right ]
\label{cyc_freq3}
\end{equation}
and
\begin{equation}
H_{\mathrm{ki}} = H\left [ (Q_{\mathrm{ECk}}- E_{\mathrm{i}}) - m_{\mathrm{4}} \right ]
\end{equation}
the Heaviside functions to ensure real solutions for the square roots.
In \cite{Filianin_2014} two interesting cases have been discussed: the pair $^{157}$Tb-$^{158}$Tb and the pair $^{202}$Pb-$^{205}$Pb. While the calorimetric measurements of $^{157}$Tb and $^{205}$Pb do not present particular difficulties, the analysis of the $^{158}$Tb and $^{202}$Pb spectra is complicated by their decay modes. 

\noindent In the case of $^{158}$Tb there are two aspects which could complicate the analysis. First of all, a completely calorimetric measurement on the full spectrum can not be performed due to the high $Q_{\mathrm{EC}}$-value of about 1.2 MeV. On the other hand, for the important energy range, up to the KI-line, the total energy released in the decay minus the one of the neutrino can be measured, and therefore the study of the effects due to the existence of heavy neutrinos will still be possible after estimating the effects on the spectrum of EC processes of higher energy. A more important aspect to consider is the fact that $^{158}$Tb undergoes in 16.6 $\%$ of the cases a beta decay to $^{158}$Dy. The energy released by the electrons in the microcalorimeters should then be precisely modeled in order to extract the features of the spectrum due to the EC processes.   

\noindent In the case of the $^{202}$Pb it is important to understand the background in the spectrum due to the partial $\alpha$-decay to $^{198}$Hg and due to fact that the daughter atom following the EC in $^{202}$Pb, $^{202}$Tl, is not stable and decays via EC/$\beta^+$ to stable $^{202}$Hg.

\noindent The cases of $^{158}$Tb and $^{202}$Pb show that decay schemes which are different from the one branch mode, where the daughter atom is a stable nuclide, lead to the presence of additional events in the measured spectra, which do not belong to the interesting EC branch. Therefore, before planning experiments with such nuclides, it is important to characterize the calorimetrically measured spectra related to the different decay modes. In this way the contribution to the background related to these events can be reduced.

\noindent On the other hand the calorimetric measurements of the electron capture spectra have an intrinsic background which consists in the un-resolved pile-up events. Such events are generated when two or more EC decays occur within a time interval which is shorter than the rise-time of the detector signal. The fraction of these events is given by the product of the activity in the detector and the rise-time. As was discussed in the case of $^{163}$Ho the reduction of this background leads to a reduced allowed activity in the detectors which in turn leads to a large number of single detectors to host the required activity. The shape of the unresolved pile-up spectrum can be calculated with confidence if all the parameters describing the EC spectrum of the decay under study are also know with confidence.
Other sources of background due to natural radioactivity and cosmic rays should also be characterized with high precision and reduced to minimum to avoid that possible structures of the background spectrum could mimic additional counts in one of the interesting lines of the EC spectrum.

\noindent The nuclides discussed here, besides the case of $^{163}$Ho, are presently not considered to be employed in large experiments for the search of sterile neutrino signatures. Figure~\ref{nuclide_sensitivity} shows, for the different nuclides, the mass region for the heavy sterile neutrino state where the highest sensitivity to the mixing element can be reached. The minimum mixing factors showed in Figure~\ref{nuclide_sensitivity} are mainly limited by the uncertainties on the $Q_{\mathrm{EC}}$-values. If future observations would indicate that an interesting mass region for the heavy neutrino state would be one of those where the discussed nuclides have higher sensitivity, a different method of analysis should be considered which reduces the effects of the uncertainties on the spectrum parameters. As already demonstrated by the KATRIN collaboration \cite{Angrik:2005ep} as well as by the ECHo collaboration (see Section~\ref{sssc:echo}) the fitting of the kind at $Q_{\mathrm{EC}}-m_4$ leads to higher sensitivity. In this case the uncertainties on the $Q_{\mathrm{EC}}$-value will lead to an uncertainty on the value of $m_4$ while the uncertainty on the mixing element will be related to the precise knowledge of the spectral shape and of the background.

%% file: DirectDetection.tex
\renewcommand{\baselinestretch}{1.15}
\def\thefootnote{\fnsymbol{footnote}}
\addtolength{\arraycolsep}{-3pt}

One of the most promising methods for direct detection of keV sterile neutrino DM in the laboratory is the captures of cosmic electron neutrinos on radioactive $\beta$-decaying nuclei \cite{Liao:2010yx,Li:2010vy}.
Through the mixing with the active electron (anti)neutrino $\nu_{\rm e}$($\bar{\nu}_{e}$),
the DM candidate $N^{}_{1}$ at the keV mass scale can undergo the capture reactions
\begin{eqnarray}
{N}^{}_{1} + {\cal N}(A,Z) \to {\cal  N}^\prime (A, Z\mp1) + e^{\pm}\,,
\end{eqnarray}
where $A$ and $Z$ are the mass and atomic numbers of the parent nucleus respectively.
The signatures of the capture reactions are measured by the outgoing mono-energetic electrons (positrons), which are located beyond the corresponding
$\beta$-decay endpoint $Q_{\beta}$. A measurement of the distance between the decay and capture processes will directly probe keV sterile neutrino DM 
and determine or constrain the corresponding mass and mixing parameters.
It should be noted that similar capture methods have been considered for detection of the cosmic neutrino background \cite{Cocco:2007za,Lazauskas:2007da,Li:2010sn,Li:2011px,Li:2015koa}.

The differential capture rate of keV sterile neutrino DM is the product of the capture cross-section times neutrino velocity
(i.e., $\sigma_{\nu}\times v_{\nu}$), the neutrino number density (i.e., $n_{\nu}$),
and the squared norm of the active-sterile mixing (i.e., $|U^{}_{e4}|^2$). Among the above mentioned factors, the cross-section times neutrino velocity can be derived as
\cite{Cocco:2007za}
\begin{eqnarray}
\sigma_{\nu}\times v_{\nu}=\frac{2\pi^{2}}{\cal{A}}\cdot\frac{\ln2
}{T_{1/2}}\,,
\end{eqnarray}
where $\cal{A}$ is the nuclear factor and $T_{1/2}$ the half-life of the decaying nuclei, respectively.
$\cal{A}$ is the function of the neutrino energy and characterized by $Q_{\beta}$ and $Z$.
Here the half-life $T_{1/2}$ is introduced to express the nuclear matrix element
of this process using the experimental data of the associated beta decay, which is of the same
energy scale as the process under discussion.

Assuming the mass density of DM around the Earth is $0.3 ~{\rm GeV} \ {\rm
cm}^{-3}$ \cite{Kusenko:2009up}, we can calculate the DM number density as $n^{}_{\nu^{}}\simeq 10^{5} \ (3 ~{\rm
keV}/m^{}_4) ~{\rm cm}^{-3}$,
where ${m^{}_4}$ is the mass of the sterile neutrino DM candidate.
The estimation of the mixing parameter $|U^{}_{e4}|^2$ depends on the production mechanisms at the early Universe. A search for the X-ray flux from the
radiative decay of $N^{}_{1}$ can set a model-independent bound as follows:
\begin{equation}
|U^{}_{e4}|^2 \lesssim \;
1.8 \times 10^{-5} \left(\frac{1 \ {\rm keV}}{m^{}_4}\right)^5 \,.
\end{equation}
Finally another important factor for the nuclei candidate selection is
the average target number during the detecting time interval $t$,
\begin{eqnarray}
{\bar N}_{\rm T}=N(0)\cdot\frac{1}{\lambda\,t^{}_{}
}\cdot\left(1-e^{-\lambda\,t}\right)\,,
\end{eqnarray}
where $N(0)$ is the initial target number at $t=0$ and $\lambda={\ln2}/{T_{1/2}}\,$.

To maximize the observability of the neutrino capture process, one should find the nucleus candidate with a larger cross-section times neutrino velocity,
and require the half-life to be at least compatible with the exposure time. Therefore,
we consider the ruthenium ($^{106}{\rm Ru}$) and tritium ($^3{\rm H}$) nuclei as the candidate targets to capture keV sterile
neutrino DM.
The typical values of $Q^{}_{\beta}$, $T^{}_{1/2}$ and $\sigma^{}_{\nu^{}}\times v^{}_{\nu^{}}$ for these two kinds of
nuclei are quoted as follows \cite{Cocco:2007za}: $Q^{}_{\beta} = 39.4 ~{\rm keV}$,
$T^{}_{1/2} = 3.2278\times 10^7 ~{\rm s}$ and $\sigma^{}_{\nu^{}}\times
v^{}_{\nu^{}} = 5.88\times 10^{-45} ~{\rm cm}^2$ for $^{106}{\rm
Ru}$ or $Q^{}_{\beta} = 18.59~{\rm keV}$, $T^{}_{1/2} = 3.8878\times 10^8 ~{\rm s}$ and
$\sigma^{}_{\nu^{}}\times v^{}_{\nu^{}} = 7.84\times 10^{-45} ~ {\rm
cm}^2$ for $^3{\rm H}$. Taking 10 kg $^3{\rm H}$ and 1 ton $^{106}{\rm Ru}$
as the target masses, and assuming $m^{}_4 = 2 ~{\rm keV}$ and $|U^{}_{e4}|^2 \simeq 5 \times 10^{-7}$,
we calculate the event rates of $^{106}{\rm Ru}$ and $^3{\rm H}$ respectively as
\begin{eqnarray}
R^{\rm keV}_{}(1\; {\rm ton}\; ^{106}{\rm Ru}) \simeq 1.8 \;{\rm year}^{-1}\,,\quad\quad
R^{\rm keV}_{}(10\; {\rm kg}\; ^3{\rm H}) \simeq 1.3 \;{\rm year}^{-1}\,.
\end{eqnarray}

\begin{figure}[t]
\begin{center}
\begin{tabular}{cc}
\includegraphics*[bb=18 18 275 216, width=0.46\textwidth]{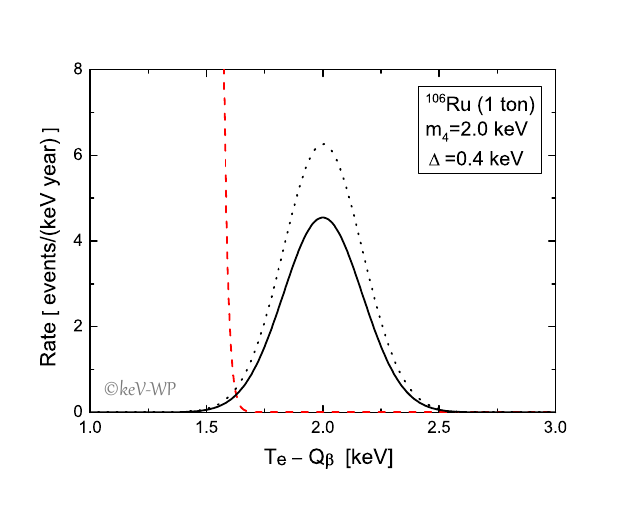}
&
\includegraphics*[bb=18 18 275 216, width=0.46\textwidth]{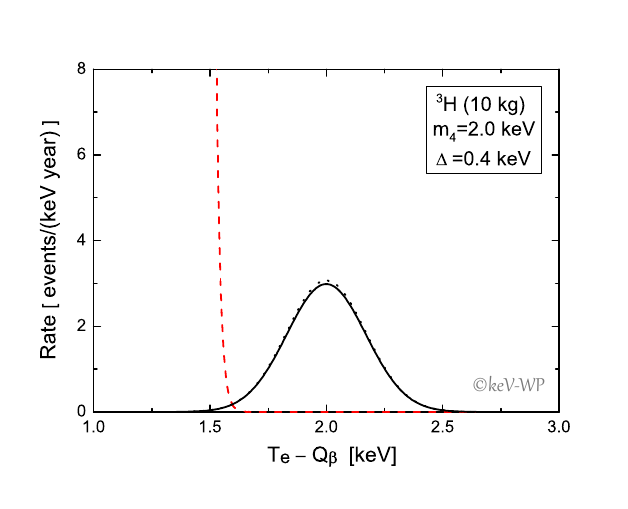}
\end{tabular}
\end{center}
\vspace{-0.5cm}
\caption{The keV sterile neutrino capture rate as a function of the
kinetic energy of electrons with $^{106}{\rm Ru}$ (left panel) and $^3{\rm H}$
(right panel) as the capture targets \cite{Li:2010vy}. The solid (or
dotted) curves denote the signals with (or without) the half-life
effect.}
\label{fig:signature}
\end{figure}

Next we shall discuss the relevant backgrounds for neutrino capture signals. The corresponding $\beta$-decay events can extend
to higher energy regions due to the finite energy resolution, and therefore mimic the desired neutrino capture signals.
To suppress the $\beta$-decay background, the energy resolution $\Delta$ (i.e., full width at half maximum)
should be smaller than $m^{}_4/4$, which is $\Delta \lesssim 0.5\rm ~{\rm keV}$ for the above assumption of $m^{}_4 = 2 ~{\rm keV}$.
In the numerical illustration of Figure~\ref{fig:signature}, we take $\Delta = 0.4 \rm ~{\rm keV}$ as an example, and present the keV sterile neutrino DM capture rate as a function of the kinetic energy of electrons with $^{106}{\rm Ru}$ (left panel) and $^3{\rm H}$
(right panel) as the capture targets \cite{Li:2010vy}. The solid (or
dotted) curves denote the signals with (or without) the half-life
effect. The lifetime effect is negligible for $^3{\rm H}$, but important for $^{106}{\rm Ru}$. It can reduce around $30\%$ of
the capture rate on $^{106}{\rm Ru}$.

%
%

Another potential background is the electron events produced by the scattering of low energy solar $pp$ neutrinos with
electrons in the target matter \cite{Liao:2013jwa}. Taking account of the bound state features of orbital electrons in the $^{106}{\rm Ru}$
atom, one can estimate the background rate of $^{106}{\rm Ru}$ as \cite{Liao:2013jwa}
\begin{eqnarray}
R^{pp}_{}(1\; {\rm ton}\; ^{106}{\rm Ru}) \simeq 0.02 \times \frac{\Delta}{10\;\rm eV}\;{\rm year}^{-1}\,.
\end{eqnarray}
In comparison, the background rate of $^{3}{\rm H}$ is two orders of magnitude smaller because of the much smaller target mass of $^{3}{\rm H}$.
From Eq.~(6), one find a better energy resolution is desirable to suppress the scattering background of solar $pp$ neutrinos.
The background generated by captures of low energy solar $pp$ neutrinos on $^{106}{\rm Ru}$ or $^{3}{\rm H}$ is $10^{-3}$ per year \cite{Liao:2010yx}
for the target masses of Eq.~(5), and is negligible in our studies.

As can be seen in the nuclear factor ${\cal A}$ of Eq.~(2), a great virtue of the neutrino capture reaction
is that the energy scale of this process is determined by the energy release of the corresponding $\beta$-decay,
not by the mass of keV sterile neutrino DM, i.e., $\sigma_\nu \propto p_e E_e$, where $p_e$ and $E_e$ are the momentum and energy of the
outgoing electron respectively, with $E_e=m_e+Q_\beta +m_4 \approx m_e+Q_\beta\,$.
As a consequence, the cross section of this process can be several orders of magnitude larger than
the neutrino scattering process $N^{}_1 + e^- \to \nu^{}_i + e^-$ ($i=1,2,3$), in which the
energy scale is characterized by the mass of keV sterile neutrino DM \cite{Ando:2010ye,Campos:2016gjh}. 

$\beta$-decaying nuclei can only be used to capture the neutrino component of DM. One need consider the electron-capture
decaying nuclei as possible capture targets of keV sterile antineutrino DM. In this regard the isotope $^{163}{\rm Ho}$ is an interesting
candidate \cite{Li:2011mw}. A signal rate of one per year requires a target mass of 600 ton $^{163}{\rm Ho}$ in
assumption of $m^{}_4 = 2 ~{\rm keV}$ and $|U^{}_{e4}|^2 \simeq 5 \times 10^{-7}$.
One should stress that captures of both the neutrino and antineutrino components
are important to test the symmetric or asymmetric nature of the warm DM candidate, which is otherwise impossible
with only the $\beta$-decaying nuclei.

In summary, we have discussed direct detection of keV sterile neutrino DM using $\beta$-decaying or electron-capture
decaying nuclei. We calculated the capture signal rates for the $^{106}{\rm Ru}$ and $^3{\rm H}$ nuclei and identified the main
sources of backgrounds. We stress that although the current experimental approach
meets several technical challenges, but it should not be hopeless in the long run.


%% file: ALong_Subsection.tex

\begin{figure}[t]
\begin{center}
\includegraphics[width=0.48\textwidth]{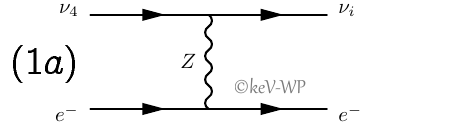} \hfill
\includegraphics[width=0.48\textwidth]{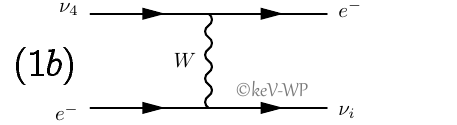} \\ \vspace{0.4cm}
\includegraphics[width=0.48\textwidth]{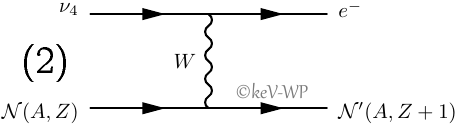}
\caption{
\label{fig:graphs}
The leading order Feynman graphs contributing to the neutrino conversion (1a and 1b) and neutrino capture (2) processes.  
}
\end{center}
\end{figure}

Two strategies for direct detection will be discussed: neutrino conversion on atomic electrons and neutrino capture on beta decaying nuclei.  
Conversion refers to the scattering $\nu_4 + e^- \to \nu_i + e^-$ in which a sterile neutrino (heavy mass eigenstate) $\nu_4$ becomes an active neutrino (light mass eigenstate) $\nu_i$, and the signal is an electron that recoils with momentum of order the sterile's mass $p_{e} \approx m_4$.  
Capture may occur when a neutrino is incident on a beta decaying nucleus, $\nu_4 + \mathcal{N}(A,Z) \to e^- + \mathcal{N}^{\prime}(A,Z+ 1)$, and the signal is a peak in the beta decay spectrum at $\Delta E_e \approx m_{4}$ above the endpoint.  
The leading-order Feynman graphs are shown in Fig.~\ref{fig:graphs}.  

The prospects for relic sterile neutrino detection through the neutrino conversion process was studied by Ando \& Kusenko (2010) \cite{Ando:2010ye}.  
In the laboratory frame, the spin-averaged scattering cross section is given by \cite{Ando:2010ye}
\begin{align}\label{ALong:sigma_convert}
	\sigma_{\rm conv} = Z^2 \frac{G_{F}^2 |U_{\rm e4}|^2}{2\pi v} m_{4}^2 \bigl( c_V^2 + 3 c_A^2 \bigr) \ , 
\end{align}
where $v \sim 10^{-3} c$ is the sterile neutrino velocity, and the active-sterile mixing is assumed to be dominated by the matrix element $U_{\rm e4}$.  
Averaging the spin of the non-relativistic neutrino $\nu_4$ brings a factor of $1/2$ that had not been included in Ref.~\cite{Ando:2010ye}.  
The momentum transfer is $p_{e} \approx m_4$, and if the corresponding de Broglie wavelength exceeds the atomic radius, $h/(p_e) \gtrsim 10^{-8} \, {\rm cm}$ or $m_4 \lesssim O(10 \, {\rm keV})$, then the neutrino interacts coherently with all the electrons in the atom, which enhances the cross section by a factor of the squared atomic number, $Z^2$.  

Consider an experiment consisting of $N$ target atoms that each have atomic number $A$, and thus the total mass of the target is $M \approx N A u$ with $u$ the atomic mass unit.  
The rate of sterile neutrino conversions is given by $\Gamma_{\rm conv} \approx \sigma_{\rm conv} v n_{4} N$ where $n_{4}$ is the density of $\nu_4$ at the Earth.  
If sterile neutrinos make up all of the dark matter, then $n_4 \approx \rho_{\text{\sc dm}} / m_4$ where $\rho_{\text{\sc dm}} \simeq 0.3 \, {\rm GeV} \, {\rm cm}^{-3}$ is the local dark matter energy density inferred from stellar motions \cite{Bovy:2012tw}.  
The expected signal rate at a Xenon-based detector ($Z_{\rm Xe} = 54$ and $A_{\rm Xe}=131$) using $M\approx 1000 \, {\rm kg}$ of target material is approximately 
\begin{align}\label{ALong:Gamma_convert}
	\Gamma_{\rm conv} \simeq (0.5 \, {\rm yr}^{-1}) \left( \frac{|U_{\rm e4}|^2}{10^{-6}} \right) \left( \frac{m_4}{10 \, {\rm keV}} \right) \left( \frac{Z}{54} \right)^2 \left( \frac{A}{131} \right)^{-1} \left( \frac{M}{10^3 \, {\rm kg}} \right) \ .
\end{align}
The recoiling electron carries a kinetic energy $T_e \approx m_4^2 / (2m_e) \simeq (98 \, {\rm eV}) (m_4 / 10 \, {\rm keV})^2$.  
With sufficient energy, the electron can ionize nearby atoms producing a detectable signal.  
It has recently been suggested that a WIMP direct detection experiment could have sensitivity to sterile neutrino dark matter in this channel \cite{Werner}.  

The capture of neutrinos on beta decaying nuclei was originally proposed as a strategy to measure the cosmic neutrino background \cite{Weinberg:1962zza,Irvine:1983nr}.  
The same channel may be used to detect the heavier sterile neutrinos, and this possibility has been explored recently by Refs.~\cite{Li:2010sn,Li:2010vy,Liao:2010yx}.  
In the laboratory frame, the spin-averaged cross section for neutrino capture is given by \cite{Long:2014zva}
\begin{align}\label{ALong:sigma_capture}
	\sigma_{\rm cap} = |V_{ud}|^2 F(Z,E_e) \frac{G_F^2 |U_{\rm e4}|^2}{2\pi v} E_e p_e \Bigl( \langle f_F \rangle^2 + (g_A / g_V)^2 \langle g_{GT} \rangle^2 \Bigr) \ .
\end{align}
Here $V_{ud} \simeq 0.974$ is the first element of the CKM matrix, $F(Z,E_e) \approx 1$ is the Fermi function, and the parenthetical factor contains combinations of nuclear matrix elements that evaluate to $\simeq 5.5$.  
The cross section in Eq.~(\ref{ALong:sigma_capture}) differs by a factor of two from earlier calculations \cite{Cocco:2007za, Lazauskas:2007da}, and it can be understood to arise from the non-relativistic nature of the neutrinos \cite{Long:2014zva}.  
Reference~\cite{Cocco:2007za} surveyed various beta-decaying nuclei and found that tritium is favorable for a neutrino capture experiment because of its relatively large cross section ($\sigma_{\rm cap} \simeq (4 \times 10^{-45} \, {\rm cm}^2) |U_{\rm e4}|^2 (c/v)$ for $m_4 \lesssim Q_{\beta} \simeq 18.6 \, {\rm keV}$) and long lifetime ($\tau \simeq 12 \, {\rm yr}$).  
We will focus on this case for the following discussion.  
In an experiment using $M_{\rm T} \approx 10 \, {\rm kg}$ of tritium for target material, the expected capture rate is 
\begin{align}\label{ALong:Gamma_capture}
	\Gamma_{\rm cap} \simeq \bigl( 0.3 \, {\rm yr}^{-1} \bigr) \left( \frac{|U_{\rm e4}|^2}{10^{-6}} \right) \left( \frac{m_4}{10 \, {\rm keV}} \right)^{-1} \left( \frac{M_{\rm T}}{10 \, {\rm kg}} \right) \ .
\end{align}
In comparing neutrino conversion and neutrino capture, Eq.~(\ref{ALong:Gamma_convert})~and~(\ref{ALong:Gamma_capture}), the former is better suited to the detection of heavy neutrinos and the latter to lighter ones.  
It should be noted that the fiducial detector size, $M_{\rm T} \approx 10 \, {\rm kg}$, represents a significant fraction of the global tritium supply, but since most tritium is manmade the amount can increase in the future.  

In a neutrino capture experiment, one would seek to measure the beta decay spectrum, and the signal of the relic sterile neutrinos would be a peak in the spectrum at an energy $\Delta E_e \approx m_4$ above the beta decay endpoint.  
Such a measurement poses a number of experimental challenges.  
First, the electron energy must be measured sufficiently precisely so as to distinguish the signal electrons from the background of beta decay electrons.  
The energy resolution of the detector should satisfy $\delta E_e \lesssim m_4$ \cite{Long:2014zva,Cocco:2007za, Lazauskas:2007da}.  
Second, efforts should be made to detect every capture event since the signal rate is expected to be low.  
This concern favors an experimental design with large angular coverage.  
Additionally it disfavors a cut-and-count strategy, such as measuring the spectrum with a MAC-E filter, which might eliminate potential signal.  
Third, solar $pp$ neutrinos will scatter from atomic electrons producing a smooth component to the beta decay spectrum that extends above the endpoint.  
Generalizing the result of Ref.~\cite{Liao:2013jwa} for tritium, the rate of $pp$ background events in an energy bin of width $\Delta E_e$ is estimated as $\Gamma_{pp} \sim (10^{-4} \, {\rm yr}^{-1}) \bigl( \Delta E_e / 10 \, {\rm eV} \bigr) (M_{\rm T} / 10 \, {\rm kg})$.  
This background is smaller than the signal rate in Eq.~(\ref{ALong:Gamma_capture}) provided that $|U_{\rm e4}|^2 \gtrsim 10^{-10} ( m_{4} / 10 \, {\rm keV})$.  

Perhaps the most confounding experimental challenge is the issue of backscatter.  
While backscatter is already a problem for beta decay experiments aiming to measure the endpoint with high precision, the problem is only exacerbated for a neutrino capture experiment, which requires many more nuclei to act as target material.  
After the capture event has occurred, the recoiling electron might scatter on one (or many) of the tritium atoms before reaching the detector where its energy is measured.  
The cross section for inelastic electron-tritium scattering is $\sigma_{\rm eT} \sim 10^{-18} \, {\rm cm}^2$ \cite{Aseev:2000} and assuming a homogenous gas of tritium in cubical volume $L^3$, the mean free path is given by $\lambda = 1 / (\sigma_{\rm eT} n_{\rm T}) \sim (10^{-4} {\rm cm}) (L/100 \, {\rm cm})^3 (M_{\rm T} / 10 \, {\rm kg})^{-1}$.  
If $\lambda \gg L$ a large number of scatterings can occur, and their collective effect is to broaden the electron spectrum.  
Determining an acceptable degree of broadening depends on a number of factors, such as the anticipated signal rate, background rate, and energy resolution.  
Even if the electron-tritium scattering is avoided, one may worry about the scattering of signal electrons on the beta decay electrons, which are produced at a rate $\sim 10^{18} \, {\rm s}^{-1} (M_{\rm T} / 10 \, {\rm kg})$.  

For both the conversion and capture processes discussed here, the scattering cross section is inversely proportional to the relative neutrino-detector velocity, {\it i.e.} $\sigma \propto 1/v$ in Eqs~(\ref{ALong:sigma_convert})~and~(\ref{ALong:sigma_capture}).  
This should be contrasted with the elastic scattering of a weakly interacting massive particle (WIMP) on a nucleus for which the cross section is independent of $v$.  
Unlike WIMP scattering, the predicted neutrino signal rates in Eqs~(\ref{ALong:Gamma_convert})~and~(\ref{ALong:Gamma_capture}) are insensitive to the velocity distribution of sterile neutrinos in the galaxy (halo model).  
For the same reason, the capture rate is not expected to display an annular modulation due to the relative motions of the Earth and the solar system \cite{Lazauskas:2007da}.  

In closing, it should be noted that the fiducial neutrino mass and mixing used here, $m_4 \simeq 10 \, {\rm keV}$ and $|U_{\rm e4}|^2 \simeq 10^{-6}$, are already excluded by X-ray observations, which impose $|U_{\rm e4}|^2 \lesssim 10^{-10}$ at $m_4 \simeq 10 \, {\rm keV}$.  
The X-ray bound weakens for lighter neutrinos, $|U_{\rm e4}|^2 \lesssim 10^{-5}$ at $m_4 \simeq 1 \, {\rm keV}$, where the allowed region of parameter space may be probed using the neutrino capture strategy.  
Nevertheless, a direct detection even in the excluded parameter regime would be an interesting possibility:  not only would it provide evidence for the sterile neutrino interpretation of dark matter, but it would also hint at new physics, which is required to bring the measurement in agreement with the X-ray bounds.

%% file: Ship.tex
The keV range sterile neutrino is a cornerstone in the $\nu$MSM \cite{Asaka:2005an,Asaka:2005pn,Boyarsky:2009ix} 
which is one of the models that will be explored by the Search for Hidden Particles (SHiP) experiment. 
The $\nu$MSM is a low-scale seesaw model which is based on introducing three right-handed sterile neutrinos, 
also referred to as Heavy Neutral Leptons. It assumes masses for the sterile neutrinos which are similar to 
those in the quark and charged lepton sector. The keV range sterile neutrino provides a decaying form of dark matter. 
The other two sterile neutrino are in the GeV range allowing them to produce the expected pattern of neutrino flavor 
oscillations and masses, and allowing them to generate baryon asymmetry of the Universe via leptogenesis. 
In order to guarantee sufficient production of keV sterile neutrinos at around temperatures of 100~MeV in the Universe, 
the heavier sterile neutrinos are required to produce a large lepton asymmetry which survives down to these temperatures. 
This allows making a natural connection 
in the $\nu$MSM between the GeV masses for the two heavier sterile neutrinos and the constraints on the keV 
DM sterile neutrino. In these respects the $\nu$MSM is one of the most economical extensions of the Standard 
Model which simultaneously allows accounting for neutrino masses and oscillations, baryogenesis, and dark matter. 

In the $\nu$MSM the sterile-active neutrino mixing leads to production of the heavier sterile neutrinos in weak 
decays of hadrons, making them accessible in an accelerator based experiment. The same mixing gives rise 
to decays to SM particles. As a consequence of the small mixing angles allowed in the $\nu$MSM and the interesting 
mass range, the lifetimes are in the order microseconds to milliseconds. For the keV sterile neutrinos,
the upper limit on the Yukawa coupling means that its production is completely negligible in an accelerator 
based experiment.

More generally, the $\nu$MSM is part of a large class of models~\cite{Alekhin:2015byh} with portals to a Hidden 
Sector which addresses the shortcomings of the Standard Model including inflation, the hierarchy problem etc, 
without involving a new scale. Instead they are based on introducing very weakly interacting particles such as 
Majorana leptons, dark photons, dark scalars or axion-like particles (ALPs) with masses below the electroweak 
scale. Even in the scenarios in which BSM physics is related to high mass scales such as SUSY, many models 
contain degrees of freedom with suppressed couplings that stay relevant at much lower energies. For example, 
in the extensively studied MSSM, the existence of light neutralinos and some other light SUSY particles has not 
been excluded. Motivated by dark matter phenomenology, hidden sectors have also been introduced in weak-scale 
supersymmetric models with gauge mediated or gravity mediated SUSY breaking. If the SUSY breaking feeds 
into the hidden sector only via some suppressed mechanism such as gauge kinetic mixing, a GeV scale mass spectrum 
for the hidden sector may be dynamically generated~\cite{Morrissey:2009ur, Chun:2008by}. 

Given the small couplings and mixings, and hence typically long lifetimes, the hidden particles have not been 
significantly constrained by previous experiments, and the reach at current experiments is limited by both 
luminosity and acceptance. The strongest bounds on the interaction strength of new light particles exist up to 
the mass of the kaon. Above this scale the bounds weaken significantly. SHiP is a new type of intensity frontier 
experiment motivated by the possibility of searching for any type of hidden particles with masses from sub-GeV 
up to $\cal{O}$(10)~GeV with super weak couplings down to $10^{-10}$. Consequently, 
SHiP has also a complementary sensitive to a part of the SUSY low-energy parameter space.

As with the GeV range sterile neutrino, many of the hidden particles in the same mass range are produced in the 
decays of heavy hadrons. In addition, the coupling of the dark photons and the ALPs to gauge bosons means that 
they may also be produced through photons: dark photons from proton bremsstrahlung, electromagnetic decays of 
mesons, as well as through direct QCD production, and ALPs through Primakoff production.

\begin{figure}
\begin{center}
\includegraphics[width=0.8\linewidth]{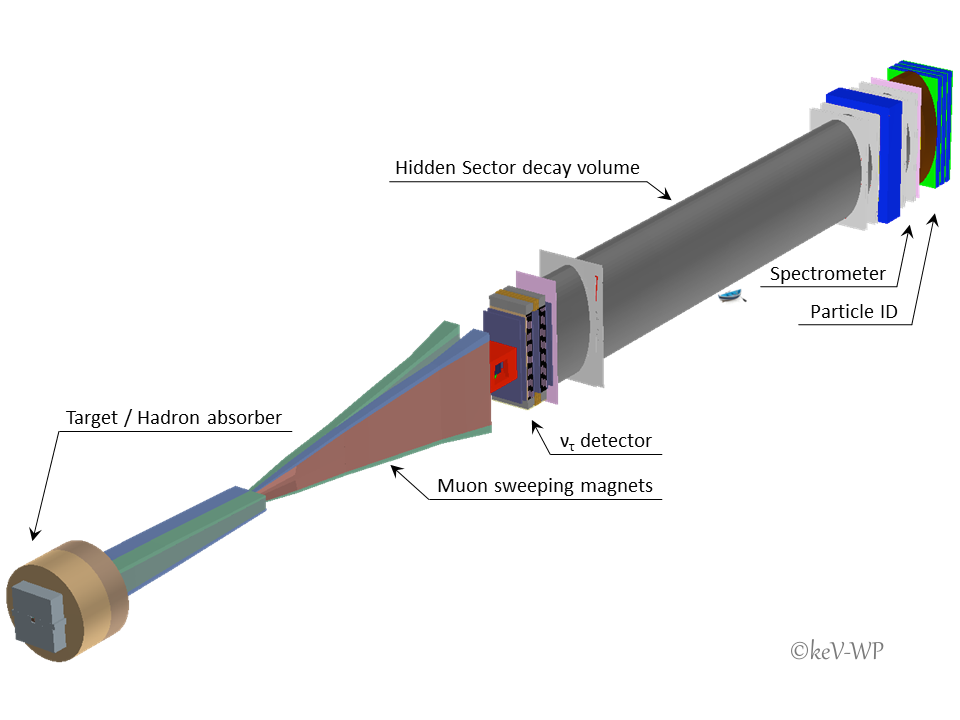}
\caption{Overview of the SHiP experimental facility.}
\label{fig:SHiP_overview}
\end{center}
\end{figure}

The key experimental parameters in the phenomenology of the Hidden Sector models are relatively similar. This 
allows a common optimization of the design of the experimental facility and of the SHiP detector 
(Figure \ref{fig:SHiP_overview})~\cite{Anelli:2015pba}. Since the hidden particles are expected to be predominantly accessible 
through the decays of heavy hadrons and photon interactions, the facility is designed to maximize their production 
and the detector acceptance while providing the cleanest possible environment. The proposal locates the SHiP experiment on a 
new beam extraction line which branches off from the CERN SPS transfer line to the North Area. The high intensity 
of the 400~GeV beam and the unique operational mode of the SPS provide ideal conditions. The current 
design of the experimental facility and the estimates of the physics sensitivities assume the SPS accelerator in its 
present state. Sharing the SPS beam time with the other SPS fixed target experiments and the LHC allows producing 
$2\times10^{20}$ protons on target in five years of nominal operation. In order to maximize charm and beauty 
production, and the production and interactions of photons, while minimizing the neutrino background from pions 
and kaons, the choice of the target material is driven by the requirement of high atomic mass number, high atomic 
number, and short interaction length. Currently the target is a hybrid design of a molybdenum alloy and pure tungsten. 
As a result, with $2\times10^{20}$ protons on target, the expected yields of different hidden particles greatly 
exceed those of any other existing or planned facility in decays of both charm and beauty hadrons.

The target is followed by a hadron stopper and an active muon shield
which deflects the high flux of muon background away from the detector. The detector for the hidden 
particles is designed to fully reconstruct their exclusive decays and to reject the background down to below 0.1 events in 
$2\times10^{20}$ protons on target. The detector consists of a large magnetic spectrometer located downstream of a 
50m long and 5m~$\times$~10m wide decay volume. In order to suppress the background from neutrinos interacting in the 
fiducial volume, it is maintained under vacuum. The spectrometer is designed to accurately reconstruct the decay 
vertex, the mass, and the impact parameter of the decaying particle at the target. A set of calorimeters and 
muon chambers provide identification of electrons, photons and muons, and pions.  A dedicated high resolution 
timing detector measures the coincidence of the decay products which allows rejecting combinatorial backgrounds. 
The decay volume is surrounded by background taggers to veto neutrino and muon inelastic scattering in the surrounding 
structures which may produce long-lived SM $V^0$ particles, such as $K_L$ etc. 

\begin{figure}[htb]
\begin{center}
\includegraphics[width=0.6\linewidth]{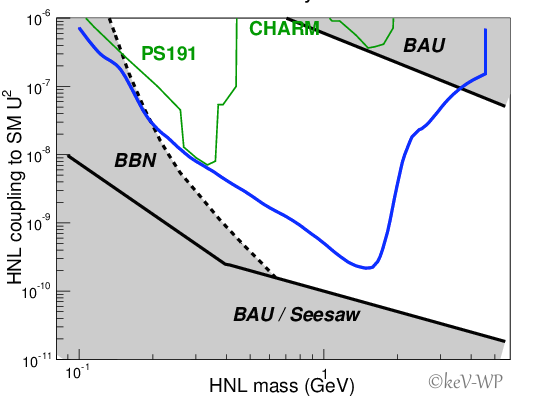}
\caption{Sensitivity to the heavier sterile neutrino (HNL) as function of the mass in the scenario with inverted mass 
hierarchy for the active neutrinos and Yukawa couplings with electron flavour dominance \cite{Anelli:2015pba}.
The region left of the ``BBN'' line is excluded by constraints from primordial nucleosynthesis, 
the region below the ``seesaw'' line is inconsistent with neutrino oscillation data. 
The baryon asymmetry of the universe can be explained in the region between the ``BAU'' lines,
The green lines indicate upper bounds from past experiments.
An updated global analysis of these constraints can be found in Ref.~\cite{Drewes:2016jae}.
}
\label{fig:HNL_sensitivity}
\end{center}
\end{figure}

With $2\times 10^{20}$ protons on target, the experiment is able to achieve sensitivities which are up to four orders 
of magnitude better than previous searches. As shown in Figure \ref{fig:HNL_sensitivity}, for sterile neutrinos 
below the mass of the D-meson, SHiP can probe the cosmologically interesting region of parameters and approach 
the lower limit in couplings, determined by the neutrino oscillations. Such an experiment would clearly be an 
essential complement to the searches for Dark Matter in astroparticle physics experiments and for new physics in
accelerator based experiments.

%% file: kevnuwp_section9.tex
Although well established by many astrophysical observations the nature of Dark Matter (DM) still remains a mystery. Only a modest fraction of (hot) DM is made of neutrinos and the Standard Model (SM) of elementary particle physics does not provide any relevant non-baryonic candidate. Hence theories beyond the SM have to be explored.

The most popular candidates are Weakly Interacting Massive Particles (WIMPs), often associated with 100~MeV -- 10~TeV neutralinos in supersymmetric extensions of the SM. So far, no direct DM search experiment has conclusively reported any WIMP detection but the experimental methods are progressing fast. Indeed, direct detection experiments may exclude WIMPs as a relevant DM candidate in the coming years. This hypothetical but plausible scenario is adding a further motivation for in-depth investigation of alternative DM candidates.

In this White Paper we have attempted to provide an up-to-date comprehensive document reviewing the case of keV-scale sterile neutrinos as DM candidates. In particular we cross-linked the expertise from astrophysicists, cosmologists, nuclear and particle physicists, reviewing particle physics theories as well as cosmological models, and confronting them with astrophysical observations.\\

\emph{What is the current situation?}\\

Sterile neutrinos could indeed account for the DM. Section~\ref{Section1} shows that their existence is well motivated in several extensions of the SM. These new elementary particles act as right-handed neutral fermions only interacting through their mixing with active neutrinos (expected to be tiny, however) or if further new particles exist to which they could couple. Although right-handed neutrinos are well-motivated, no theory can predict their exact mass scale. It is an important point to underline that the DM relevant keV-scale only arises when taking into account astrophysical and cosmological considerations.

Over the past 15~years a robust cosmological model has been established, combining the theory of the General Relativity and the SM of elementary particles. This model includes a DM component which is still unknown but characterized by its lack of electromagnetic interaction. Its velocity dispersion is not too large, which is why it is typically assumed to be cold DM (CDM) -- although in fact we have not sufficient observational evidence to conclude anything beyond DM not being hot. Finally, to reproduce all large-scales observations, a non-vanishing vacuum energy is mandatory, leading to the so-called $\Lambda$CDM cosmological model. Section~\ref{sec:neutrino-cosmology} first reviews the role of the three standard model neutrino flavors in this cosmological framework and then discusses the consequences of incorporating sterile neutrinos. Indeed cosmology could accommodate for their existence, especially at the keV-scale and beyond.

In relation to keV-scale sterile neutrino motivations, the entanglement between the type of DM (particularly its velocity) and the formation of structures in the Universe, at all-scales, is of upmost importance. While the current cosmological model perfectly fits the data for galaxy clusters and scales beyond, Section~\ref{sec:DMGalactic} discusses observational at galactic scales and their comparison with the $\Lambda$CDM predictions. In CDM models every DM halo contains ever increasing number of smaller and smaller mass substructures. However, when compared to observations some discrepancies arise at scales smaller than 10~kpc. Too few dwarf satellite galaxies are yet observed compared to what CDM settings predict, leading to the so-called missing satellite problem. Furthermore, the number of the largest dwarf satellite galaxies is less than expected from CDM-based simulations, making the astrophysical feedback explanation problematic and leading to the too-big-to-fail issue. Finally, some observations seem to favor a galactic center with a cored profile, while CDM simulations predict a cuspy matter distribution with a large rise in matter density at the innermost regions. Although astrophysical feedback effects or refinement of the simulations to fully include baryonic feedback could perhaps solve these CDM issues, as-of-yet no consensus has been reached in the community. Interestingly, the apparent discrepancy could also arise from alternative DM physics if the candidate particle velocity distribution is not cold but rather warm or even non-thermal. This fact motivates the keV mass scale for sterile neutrinos, as it marks a departure from the standard CDM scenario.

Astrophysical observations strongly constrain sterile neutrino DM, as shown in Section~\ref{sec:kev-neutrino-observables}. The solid Tremaine-Gunn bound indicates that sterile neutrinos need to have masses greater than $\sim 0.5$~keV. Putting aside the tentative signal for now, another strong astrophysical bound comes from the non-observation of the monoenergetic X-rays induced by the decay of sterile neutrinos. This limit generically forces the relevant mass range to be within a few tens of keV. Nevertheless, as just hinted, two groups have independently reported a hint for a 3.5~keV X-ray emission line that could be related to a 7~keV relic sterile neutrino with an active-sterile mixing angle of $\sin^2(2\theta)\sim10^{-10}$, a tiny value that has important consequences when it comes to experimental considerations. This observation, at the cutting edge of the best X-ray telescope sensitivities, is being discussed and debated meticulously in Section~\ref{sec:kev-line}. Further interesting constraints arise from the analysis of Lyman-$\alpha$ forest data. These bounds however rely heavily on the actual DM velocity spectrum, as they ultimately track the structures grown in space. The reader should be aware that, in some of models, the DM velocity spectrum significantly deviates from a thermal spectrum. Consequently, keV-sterile neutrinos could act as warm, cool, or even cold DM, thus possibly circumventing current Lyman-$\alpha$ constraints and/or X-ray bounds, as published.

Section~\ref{sec:production-mechanisms} reviews how sterile neutrinos could have been produced in the early Universe, within the framework of extensions of the Standard Models of particle physics and cosmology. The so-called Dodelson-Widrow (DW) mechanism is the simplest way to produce sterile neutrinos, but in contradiction to astrophysical observations. However, resonantly enhanced oscillations in the early Universe driven by a net lepton number asymmetry in the early Universe (Shi-Fuller mechanism) are easier to bring into accordance with observations, and they could also explain the 3.5~keV X-ray signal. Another alternative relies in the decay of a new particle at early times. This could be the inflaton or also a more general electrically neutral or charged particle that may be produced from the primordial plasma. Again, current X-ray and Lyman-$\alpha$ bounds could be evaded for some of these alternative production mechanisms. Finally, beyond the SM of elementary particle physics, thermal production of sterile neutrinos could have happened if accompanied with a strong subsequent entropy dilution. The important message of this section is that several possibilities beyond ordinary thermal freeze-out do exist to produce sterile neutrino DM. They may partially lead to rich non-thermal spectra beyond the simple cold, warm, or hot scenarios. While this complicates the conversion of the astrophysical observables into sterile neutrino mass and mixing parameters, it may also help to distinguish between the various models.

Section~\ref{sec:kev-neutrino-theory} turns to particle physics, discussing mechanisms generating keV-scale sterile neutrinos. Typically this scale it not only regarded as ``unnatural'', but it is also not desired for all the sterile neutrinos in order to match the phenomenology of active neutrino masses and mixings. This is achieved by forcing the mass of only one sterile neutrino to be smaller than those of the others, either by perturbing a framework in which this mass is naturally zero or by suppressing one mass compared to the others. Different models achieving a suitable mass spectrum exist in the literature. The most interesting ones link the properties of the sterile neutrinos to those of the active neutrinos, e.g.\ by predicting correlations between light neutrino observables, such that a more precise determination of the neutrino oscillation parameters could be used to indirectly test sterile neutrino scenarios.

The final sections look at the current and future searches for keV-scale sterile neutrinos, both with astrophysical (Section~\ref{sec:current-future-astro-exp}) and with laboratory (Section~\ref{sec:current-future-lab-exp}) experiments. On the astrophysical side, two direct observations are extremely promising: X-ray searches and the Lyman-$\alpha$ forest data. Both are going to be improved significantly in the near future, with the delivery of new data awaited from more sensitive instruments. But the interpretation of these results could still remain somewhat ambiguous, since it ultimately relies on the DM model being considered. Other indirect observations may further constrain the current panorama, like pulsar kicks or supernova explosions, but they unfortunately lack a smoking gun signature to corroborate any positive detection. While many experiments are currently searching for WIMPs, there is no ground-based laboratory to test the existence of keV sterile neutrinos of cosmological interest. However, several studies or R\&D have been initiated in this direction. In particular experiments based on single beta decays, electron capture, or neutrino capture processes could probably reach reasonable sensitivities, although only at the upper bound of the cosmologically relevant region in terms of mixing. Nevertheless future experimental advances may lead to serious new insights into the field in the future, but the question remains open whether a laboratory searches could eventually explore sterile neutrino mixings of less than 1~part per million.

\begin{figure}[th]
\centering
\includegraphics[width=14cm]{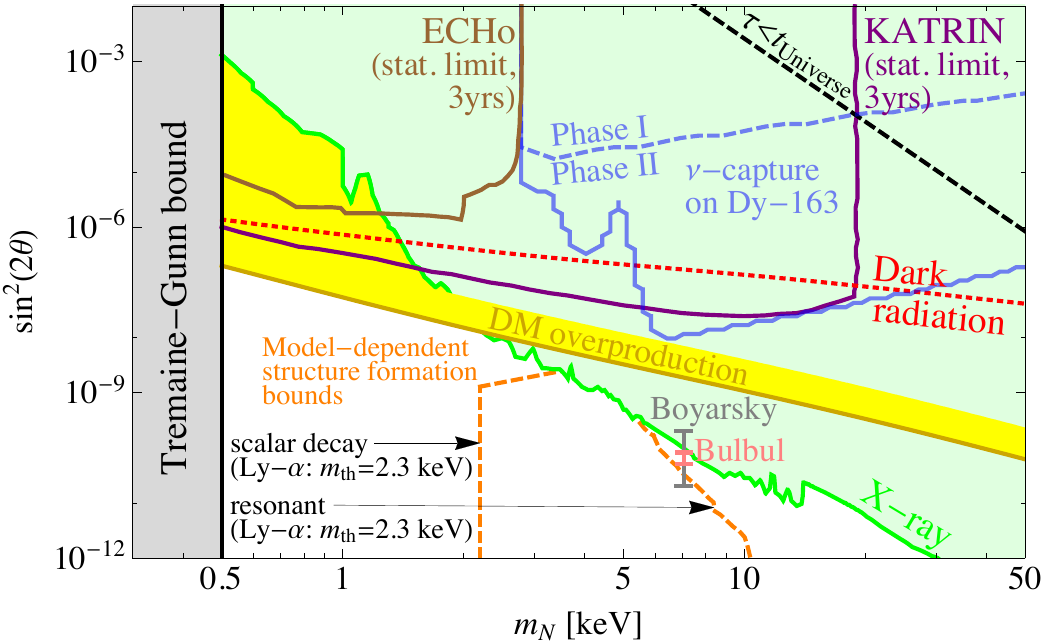}
\caption{\label{fig:ConstraintSummaryEnd}Summary of experimental and observational constraints on keV sterile neutrino Dark Matter.
}
\end{figure}

In this White Paper we have presented state-of-the-art insights and we have summarized different viewpoints, sometimes conflicting within the community. A summary of all generic constraints on keV sterile neutrino DM is presented in Fig.~\ref{fig:ConstraintSummaryEnd}. The figure shows the $\sin^2(2\theta)$--$M_1$ plane, where the sterile neutrino mass $M_1$ is displayed in units of keV and the total mixing angle is defined by $\theta^2 \equiv \sum_{\alpha=e,\mu,\tau} |\theta_{\alpha 1}|^2$. The strongest bound on $\sin^2(2\theta)$ is derived from the non-observation of X-ray photons from sterile neutrino decay~(e.g.\ \cite{Boyarsky:2006fg,Boyarsky:2006hr,Boyarsky:2007ay,Abazajian:2006jc,Boyarsky:2007ge,Watson:2011dw,Sekiya:2015jsa,Ng:2015gfa}, see Secs.~\ref{sec:xray},  \ref{section7.1}, and~\ref{ssc:Xray355keV} for a complete list of references). This excludes the region above the green line. For very small sterile neutrino masses, the main upper bound on the mixing angle arises from not producing too much DM by non-resonant production~\cite{Dodelson:1993je,Asaka:2006nq}, cf.\ Secs.~\ref{sec:5.thermalproduction} and~\ref{Sec:ThermalProduction}. The correct DM density is produced for any combination of $M_1$ and ${\rm sin}^2(2\theta)$ along the yellow-gold line, above this line DM is overproduced. Weaker bounds arise  from constraints on the effective number of relativistic species $N_{\rm eff}$ at the time of CMB decoupling (``dark radiation'', red dashed line)~\cite{Ade:2015xua}. A deviation of $N_{\rm eff}$ from its SM prediction $N_{\rm eff}=3.046$ \cite{Mangano:2005cc} appears in case the keV sterile neutrino is not actually DM, but instead decays into some type of invisible radiation (e.g., lighter sterile neutrinos)~\cite{DiBari:2013dna}. This region is in fact contained in the parameter space, as for large enough active-sterile mixing, already the decay modes $N_1 \to 3\nu$ and $N_1 \to \nu\gamma$ induced by active-sterile mixing enforce a sterile neutrino lifetime smaller than the age of the Universe (black dashed line). The most optimistic regions for the KATRIN/TRISTAN~\cite{Mertens:2014osa,Mertens:2014nha} (purple line, cf.\ Sec.~\ref{ssc:tritium}) and ECHo~\cite{Filianin_2014} (brown line, statistical sensitivity for 3 years, cf.\ Sec.~\ref{ssc:holmium}) are displayed, this time assuming that the full mixing is contained in the electron sector ($\theta\equiv \theta_{e1}$). The newest idea for a possible lab detection is based on relic keV sterile neutrinos enabling sterile neutrino capture transitions of the otherwise \emph{stable} isotope Dy-163~\cite{Lasserre:2016eot}, which may be the most promising given that it is scalable without the use of radioactive material. Example sensitivities are indicated by the blue lines in the plot.

The most model-independent constraint on $M_1$ comes from the phase space constraint on fermionic DM~\cite{Boyarsky:2008ju,Gorbunov:2008ka}, i.e., from the Tremaine-Gunn bound~\cite{Tremaine:1979we}, see the gray-shaded area on the left of the plot (this bound becomes stronger when a particular production mechanism is assumed, see \cite{Boyarsky:2008ju}). Taking into account the Lyman-$\alpha$ forest data (see Sec.~\ref{sec:LymanAlphaBounds}), the lower bound can be considerably strengthened. These tighter bounds, however, have  theoretical and observational uncertainties~\cite{Boyarsky:2008xj,Viel:2013apy,Garzilli:2015iwa}. In particular, at small scales the neutral hydrogen stops following the underlying DM distribution due to its pressure and temperature. The limits on non-cold DM settings thus become strongly degenerate with the temperature of the intergalactic medium that has not been independently constrained at high redshifts ~\cite{Garzilli:2015iwa}. By the orange dashed lines, we indicated bounds based on the Lyman-$\alpha$ constraints based on~\cite{Garzilli:2015iwa}, with primordial distribution functions from scalar decay production, see Sec.~\ref{sec:5.decays}~\cite{Shaposhnikov:2006xi,Kusenko:2006rh,Petraki:2007gq,Merle:2015oja,Merle:2015vzu,Konig:2016dzg} and resonant production~\cite{Shi:1998km,Laine:2008pg,Ghiglieri:2015jua,Venumadhav:2015pla}, see Secs.~\ref{sec:5.thermalproduction} and~\ref{Sec:ThermalProduction}. If in the future the temperature of the intergalactic medium will be measured (e.g.\ via the broadening of individual lines, \cite{Garzilli:2015bha}) and found sufficiently high, the considerably stronger bounds from~\cite{Viel:2013apy} would be applicable (found e.g.\ in~\cite{Schneider:2016uqi}). The bounds from halo counting (comparison between the observed number of dwarf satellite galaxies around the Milky Way and theoretical predictions) provide restrictions on sterile neutrino parameters, similar to those from the Lyman-$\alpha$ bounds shown in Fig.~\ref{fig:ConstraintSummaryEnd}. However, such bounds have intrinsic systematic uncertainty, due to the strong scatter in galactic halo realizations in numerical simulations (see e.g.\ the discussion in~\cite{Lovell:2015psz,Lovell_eagle}) and the fact that we do not know observationally how typical the Milky Way (or the Local Group) and its satellite distribution is.

We note in passing that the statement in the recent paper from \emph{NuSTAR}~\cite{Perez:2016tcq} of these bounds practically closing the parameter space for sterile neutrino Dark Matter is clearly too strong, as such a bound depends on the production mechanism. Since resonant production is driven by the presence of a lepton asymmetry in the primordial plasma that is much larger than the baryon asymmetry, a lower bound on ${\sin}^2(2\theta)$ can in principle be derived by estimating the asymmetry that can realistically be generated \cite{Canetti:2012kh} and is consistent with known constraints from BBN \cite{Mangano:2011ip,Castorina:2012md}. We do not display such a bound because the lepton asymmetry is strongly model-dependent. Based on the results displayed here, scalar production leads to slightly cooler spectra -- hence the less restrictive lower bounds on $M_1$ for that production mechanism. This means that, if the 3.5~keV line signal survives after all, or if a new signal shows up, one could possibly distinguish the different production mechanisms by constraints from cosmic structure formation~\cite{Merle:2014xpa}.

We would like to end this journey by emphasizing that the field of keV-scale sterile neutrino DM is still in development. Many observations, simulations, or theoretical results have to be clarified and it remains an open question whether keV-scale sterile neutrinos exist and if they account for a significant part of the DM in the Universe. Our common goal for the future shall thus be to continue improving the current picture by corroborating theoretical and observational information from all sides: nuclear and particle physics, astrophysics, and cosmology.


%% file: acknowledgments.tex


The work of C.~Giunti was partially supported by the research grant Theoretical Astroparticle Physics number 2012CPPYP7 under the program PRIN 2012 funded by the Ministero dell'Istruzione, Universit\`a e della Ricerca (MIUR). The authors of Section~2 would like to thank A.~Vincent for discussion and comments. The work of A.~Mirizzi is supported by the Italian Ministero dell'Istruzione, Universit\`a e Ricerca (MIUR) and Istituto Nazionale di Fisica Nucleare (INFN) through the ``Theoretical Astroparticle Physics'' projects. The work of A.~Ibarra was supported in part by the DFG cluster of excellence ``Origin and Structure of the Universe'' and by the ERC Advanced Grant project ``FLAVOUR'' (267104). M.~Viel is supported by the ERC StG ``cosmoIGM''. The work of O.~Dragoun was supported by the GACR (contract P203/12/1896) and the ASCR (contract IRP AV02 10480585). This work was supported in part by the Gottfried Wilhelm Leibniz Programme of the Deutsche Forschungsgemeinschaft (DFG) and the NSF grant PHY-1307372 at UCSD. M.~Laine was supported in part by the Swiss National Science Foundation (SNF) under grant 200020-155935. A.~Merle acknowledges partial support by the Micron Technology Foundation, Inc. V.~Niro acknowledges support by Spanish MINECO through project FPA2012-31880, by Spanish MINECO (Centro de excelencia Severo Ochoa Program) under grant SEV-2012-0249 as well as the Spanish grants FPA2014-58183-P, Multidark CSD2009-00064 and SEV-2014-0398 (MINECO) and by the PROMETEOII/2014/084 grant from Generalitat Valenciana; the work of V.~Niro was also supported by the Deutsche Forschungsgemeinschaft (DFG) through the Collaborative Research Centre SFB 676 ``Particles, Strings and the Early Universe'' (through an SFB fellowship). M.~ Drewes acknowledges support by the Gottfried Wilhelm Leibniz program of the Deutsche Forschungsgemeinschaft (DFG) and the DFG cluster of excellence Origin and Structure of the Universe. D.~Robinson is supported by the NSF under grant No.~PHY-1002399. Y.~Tsai acknowledges support by the NSF under grant No. PHY-1315155 and the Maryland Center for Fundamental Physics. This research was  partially supported by  Conselho Nacional de Desenvolvimento Cient\'{\i}fico e Tecnol\'ogico (CNPq), (A.~Gomes Dias, C.A.~de S.~Pires, and P.S.~Rodrigues da Silva,), by Funda\c{c}\~{a}o de Amparo \`{a} Pesquisa do Estado de S\~ao Paulo (FAPESP), (A.~Gomes Dias), and by the National Foundation for Science and Technology Development (NAFOSTED) of Vietnam under grant 103.03-2012.49 (N.~Anh Ky and N.~T.~Hong Van). The work of J.~Heeck is funded in part by IISN and by Belgian Science Policy (IAP VII/37). The  work of  N.E.~Mavromatos is  supported in  part by the London Centre for Terauniverse Studies (LCTS), using  funding from the European Research Council  via  the  Advanced  Investigator  Grant 267352,  and  by  the U.K.~Science  and  Technology  Facilities  Council  (STFC)  under  the research grants ST/J002798/1 and ST/L000326/1. The  work of A.~Pilaftsis is supported in part by the Lancaster--Manchester--Sheffield  Consortium for  Fundamental Physics, under STFC (UK) research grant: ST/J000418/1. L.~Gastaldo  would like to acknowledge the support by the DFG Research Unit FOR 2202 ``Neutrino Mass Determination by Electron Capture in $^{163}$Ho, ECH'' (funding under GA 2219/2-1). T.~Lasserre would like to acknowledge the support of the Technische Universit\"at M\"unchen Institute for Advanced Study, funded by the German Excellence Initiative and the European Union Seventh Framework Programme under grant agreement No.~291763, as well as the European Union Marie Curie COFUND program. S.~Mertens gratefully acknowledges support of a Feodor Lynen fellowship by the Alexander von Humboldt Foundation and support by the Helmholtz Association. R.~Shrock acknowledges support from the U.S.~NSF grants NSF-PHY-13-16617 and NSF-PHY-16-1620628. The work of Y.~F.~Li, W.~Liao, and Z.~Xing is supported in part by the National Natural Science Foundation of China under Grant No.~11135009. Several authors acknowledge (partial) support by the European Union through the FP7 Marie Curie Actions ITN INVISIBLES (PITN-GA-2011-289442). Furthermore, his project has received funding from the European Union's Horizon 2020 research and innovation programme under the Marie Sklodowska-Curie grant agreements No.~690575 (InvisiblesPlus RISE) and No.~674896 (Elusives ITN).\\

\paragraph{History of the White Paper} This White Paper was born at the NIAPP (Neutrinos in Astro- and Particle Physics) Workshop of the MIAPP (Munich Institute for Astro- and Particle Physics), organized by S.~Sch\"onert, G.~G.~Raffelt, A.~Smirnov, and T.~Lasserre and held at MIAPP in Garching, Germany, from 30~June -- 25~July 2014. The lively discussions at this workshop and the interest of so many participants ultimately motivated us to undergo this enterprise of writing a White Paper on keV sterile neutrinos as Dark Matter, as we could see that many scientific communities are deeply interested in the topic and there is a strong need for a document including information on all aspects involved. This we have also experienced at the $\nu$-Dark 2015 Workshop held at the Institute for Advanced Study of TUM, Garching, Germany from 7~--9~December 2015, where we for the first time gathered representatives from all the relevant communities, many of which are co-authors of this document, for a critical discussion of the topic. At this stage we would like to acknowledge the efforts of the Chalonge School, Colloquiums, and Workshops that have been held in the same spirit. The discussions, focused on sterile neutrino WDM greatly advanced the collaboration and exchange between theory, observations, numerical simulations, particle physics, and experimental detection plans~\cite{deVega:2013hpa,Biermann:2013nxa,deVega:2012jm,deVega:2011si,
deVega:2010zk,deVega:2010wj}

\paragraph{Note on the figures} All authors of the different texts have been explicitly informed that figures used within the White Paper must not be protected under copyright of any third party. The editors have made their best effort to ensure that no figure has been used in the same way anywhere else. However, should any dispute about copyright arise, the responsibility lies within the author(s) of the respective section where the decisive figure has been used. The editors take no responsibility for any of these cases, unless in cases where one of the editors is simultaneously an author of the respective section.

%% file: WhitePaperKevSterileNeutrinos.bbl
\let\jnlstyle=\rm\def\jref#1{{\jnlstyle#1}}\def\aj{\jref{AJ}}
  \def\araa{\jref{ARA\&A}} \def\apj{\jref{ApJ}\ } \def\apjl{\jref{ApJ}\ }
  \def\apjs{\jref{ApJS}} \def\ao{\jref{Appl.~Opt.}} \def\apss{\jref{Ap\&SS}}
  \def\aap{\jref{A\&A}} \def\aapr{\jref{A\&A~Rev.}} \def\aaps{\jref{A\&AS}}
  \def\azh{\jref{AZh}} \def\baas{\jref{BAAS}} \def\jrasc{\jref{JRASC}}
  \def\memras{\jref{MmRAS}} \def\mnras{\jref{MNRAS}\ }
  \def\pra{\jref{Phys.~Rev.~A}\ } \def\prb{\jref{Phys.~Rev.~B}\ }
  \def\prc{\jref{Phys.~Rev.~C}\ } \def\prd{\jref{Phys.~Rev.~D}\ }
  \def\pre{\jref{Phys.~Rev.~E}} \def\prl{\jref{Phys.~Rev.~Lett.}}
  \def\pasp{\jref{PASP}} \def\pasj{\jref{PASJ}} \def\qjras{\jref{QJRAS}}
  \def\skytel{\jref{S\&T}} \def\solphys{\jref{Sol.~Phys.}}
  \def\sovast{\jref{Soviet~Ast.}} \def\ssr{\jref{Space~Sci.~Rev.}}
  \def\zap{\jref{ZAp}} \def\nat{\jref{Nature}\ } \def\iaucirc{\jref{IAU~Circ.}}
  \def\aplett{\jref{Astrophys.~Lett.}}
  \def\apspr{\jref{Astrophys.~Space~Phys.~Res.}}
  \def\bain{\jref{Bull.~Astron.~Inst.~Netherlands}}
  \def\fcp{\jref{Fund.~Cosmic~Phys.}} \def\gca{\jref{Geochim.~Cosmochim.~Acta}}
  \def\grl{\jref{Geophys.~Res.~Lett.}} \def\jcp{\jref{J.~Chem.~Phys.}}
  \def\jgr{\jref{J.~Geophys.~Res.}}
  \def\jqsrt{\jref{J.~Quant.~Spec.~Radiat.~Transf.}}
  \def\memsai{\jref{Mem.~Soc.~Astron.~Italiana}}
  \def\nphysa{\jref{Nucl.~Phys.~A}} \def\physrep{\jref{Phys.~Rep.}}
  \def\physscr{\jref{Phys.~Scr}} \def\planss{\jref{Planet.~Space~Sci.}}
  \def\procspie{\jref{Proc.~SPIE}} \let\astap=\aap \let\apjlett=\apjl
  \let\apjsupp=\apjs \let\applopt=\ao \def\jcap{\jref{JCAP}}
\providecommand{\href}[2]{#2}\begingroup\raggedright\begin{thebibliography}{100}

\bibitem{Ade:2015xua}
{\scshape Planck} collaboration, P.~A.~R. Ade et~al., \emph{{Planck 2015
  results. XIII. Cosmological parameters}},
  \href{http://dx.doi.org/10.1051/0004-6361/201525830}{\emph{Astron.
  Astrophys.} {\bf 594} (2016) A13},
  [\href{http://arxiv.org/abs/1502.01589}{{\tt 1502.01589}}].

\bibitem{Persic:1995ru}
M.~Persic, P.~Salucci and F.~Stel, \emph{{The Universal rotation curve of
  spiral galaxies: 1. The Dark matter connection}},
  \href{http://dx.doi.org/10.1093/mnras/278.1.27}{\emph{Mon. Not. Roy. Astron.
  Soc.} {\bf 281} (1996) 27},
  [\href{http://arxiv.org/abs/astro-ph/9506004}{{\tt astro-ph/9506004}}].

\bibitem{Faber:1976sn}
S.~M. Faber and R.~E. Jackson, \emph{{Velocity dispersions and mass to light
  ratios for elliptical galaxies}},
  \href{http://dx.doi.org/10.1086/154215}{\emph{Astrophys. J.} {\bf 204} (1976)
  668}.

\bibitem{Kaiser:1992ps}
N.~Kaiser and G.~Squires, \emph{{Mapping the dark matter with weak
  gravitational lensing}},
  \href{http://dx.doi.org/10.1086/172297}{\emph{Astrophys. J.} {\bf 404} (1993)
  441--450}.

\bibitem{Clowe:2003tk}
D.~Clowe, A.~Gonzalez and M.~Markevitch, \emph{{Weak lensing mass
  reconstruction of the interacting cluster 1E0657-558: Direct evidence for the
  existence of dark matter}},
  \href{http://dx.doi.org/10.1086/381970}{\emph{Astrophys. J.} {\bf 604} (2004)
  596--603}, [\href{http://arxiv.org/abs/astro-ph/0312273}{{\tt
  astro-ph/0312273}}].

\bibitem{Percival:2007yw}
W.~J. Percival, S.~Cole, D.~J. Eisenstein, R.~C. Nichol, J.~A. Peacock, A.~C.
  Pope et~al., \emph{{Measuring the Baryon Acoustic Oscillation scale using the
  SDSS and 2dFGRS}},
  \href{http://dx.doi.org/10.1111/j.1365-2966.2007.12268.x}{\emph{Mon. Not.
  Roy. Astron. Soc.} {\bf 381} (2007) 1053--1066},
  [\href{http://arxiv.org/abs/0705.3323}{{\tt 0705.3323}}].

\bibitem{Dave:1998gm}
R.~Dave, L.~Hernquist, N.~Katz and D.~H. Weinberg, \emph{{The Low redshift
  Lyman alpha forest in cold dark matter cosmologies}},
  \href{http://dx.doi.org/10.1086/306722}{\emph{Astrophys. J.} {\bf 511} (1999)
  521--545}, [\href{http://arxiv.org/abs/astro-ph/9807177}{{\tt
  astro-ph/9807177}}].

\bibitem{Paczynski:1985jf}
B.~Paczynski, \emph{{Gravitational microlensing by the galactic halo}},
  \href{http://dx.doi.org/10.1086/164140}{\emph{Astrophys. J.} {\bf 304} (1986)
  1--5}.

\bibitem{Griest:1990vu}
K.~Griest, \emph{{Galactic Microlensing as a Method of Detecting Massive
  Compact Halo Objects}},
  \href{http://dx.doi.org/10.1086/169575}{\emph{Astrophys. J.} {\bf 366} (1991)
  412--421}.

\bibitem{Lasserre:2000xw}
{\scshape EROS} collaboration, T.~Lasserre, \emph{{Not enough stellar mass
  machos in the galactic halo}}, {\emph{Astron. Astrophys.} {\bf 355} (2000)
  L39--L42}, [\href{http://arxiv.org/abs/astro-ph/0002253}{{\tt
  astro-ph/0002253}}].

\bibitem{Bennett:2005at}
D.~P. Bennett, \emph{{Large Magellanic Cloud microlensing optical depth with
  imperfect event selection}},
  \href{http://dx.doi.org/10.1086/432830}{\emph{Astrophys. J.} {\bf 633} (2005)
  906--913}, [\href{http://arxiv.org/abs/astro-ph/0502354}{{\tt
  astro-ph/0502354}}].

\bibitem{Clowe:2006eq}
D.~Clowe, M.~Bradac, A.~H. Gonzalez, M.~Markevitch, S.~W. Randall, C.~Jones
  et~al., \emph{{A direct empirical proof of the existence of dark matter}},
  \href{http://dx.doi.org/10.1086/508162}{\emph{Astrophys. J.} {\bf 648} (2006)
  L109--L113}, [\href{http://arxiv.org/abs/astro-ph/0608407}{{\tt
  astro-ph/0608407}}].

\bibitem{Yoo:2003fr}
J.~Yoo, J.~Chaname and A.~Gould, \emph{{The end of the MACHO era: limits on
  halo dark matter from stellar halo wide binaries}},
  \href{http://dx.doi.org/10.1086/380562}{\emph{Astrophys. J.} {\bf 601} (2004)
  311--318}, [\href{http://arxiv.org/abs/astro-ph/0307437}{{\tt
  astro-ph/0307437}}].

\bibitem{Griest:2013esa}
K.~Griest, A.~M. Cieplak and M.~J. Lehner, \emph{{New Limits on Primordial
  Black Hole Dark Matter from an Analysis of Kepler Source Microlensing Data}},
  \href{http://dx.doi.org/10.1103/PhysRevLett.111.181302}{\emph{Phys. Rev.
  Lett.} {\bf 111} (2013) 181302}.

\bibitem{Pani:2014rca}
P.~Pani and A.~Loeb, \emph{{Tidal capture of a primordial black hole by a
  neutron star: implications for constraints on dark matter}},
  \href{http://dx.doi.org/10.1088/1475-7516/2014/06/026}{\emph{JCAP} {\bf 1406}
  (2014) 026}, [\href{http://arxiv.org/abs/1401.3025}{{\tt 1401.3025}}].

\bibitem{Milgrom:1983ca}
M.~Milgrom, \emph{{A Modification of the Newtonian dynamics as a possible
  alternative to the hidden mass hypothesis}},
  \href{http://dx.doi.org/10.1086/161130}{\emph{Astrophys. J.} {\bf 270} (1983)
  365--370}.

\bibitem{Kraus:2004zw}
C.~Kraus et~al., \emph{{Final results from phase II of the Mainz neutrino mass
  search in tritium beta decay}},
  \href{http://dx.doi.org/10.1140/epjc/s2005-02139-7}{\emph{Eur. Phys. J.} {\bf
  C40} (2005) 447--468}, [\href{http://arxiv.org/abs/hep-ex/0412056}{{\tt
  hep-ex/0412056}}].

\bibitem{Lobashev:1999tp}
V.~M. Lobashev et~al., \emph{{Direct search for mass of neutrino and anomaly in
  the tritium beta spectrum}},
  \href{http://dx.doi.org/10.1016/S0370-2693(99)00781-9}{\emph{Phys. Lett.}
  {\bf B460} (1999) 227--235}.

\bibitem{Kolb:1990vq}
E.~W. Kolb and M.~S. Turner, \emph{{The Early Universe}}, {\emph{Front. Phys.}
  {\bf 69} (1990) 1--547}.

\bibitem{White:1984yj}
S.~D.~M. White, C.~S. Frenk and M.~Davis, \emph{{Clustering in a Neutrino
  Dominated Universe}},
  \href{http://dx.doi.org/10.1086/161425}{\emph{Astrophys. J.} {\bf 274} (1983)
  L1--L5}.

\bibitem{Gondolo:1990dk}
P.~Gondolo and G.~Gelmini, \emph{{Cosmic abundances of stable particles:
  Improved analysis}},
  \href{http://dx.doi.org/10.1016/0550-3213(91)90438-4}{\emph{Nucl. Phys.} {\bf
  B360} (1991) 145--179}.

\bibitem{Jungman:1995df}
G.~Jungman, M.~Kamionkowski and K.~Griest, \emph{{Supersymmetric dark matter}},
  \href{http://dx.doi.org/10.1016/0370-1573(95)00058-5}{\emph{Phys. Rept.} {\bf
  267} (1996) 195--373}, [\href{http://arxiv.org/abs/hep-ph/9506380}{{\tt
  hep-ph/9506380}}].

\bibitem{Gelmini:2006pw}
G.~B. Gelmini and P.~Gondolo, \emph{{Neutralino with the right cold dark matter
  abundance in (almost) any supersymmetric model}},
  \href{http://dx.doi.org/10.1103/PhysRevD.74.023510}{\emph{Phys. Rev.} {\bf
  D74} (2006) 023510}, [\href{http://arxiv.org/abs/hep-ph/0602230}{{\tt
  hep-ph/0602230}}].

\bibitem{Belanger:2005kh}
G.~Belanger, F.~Boudjema, C.~Hugonie, A.~Pukhov and A.~Semenov, \emph{{Relic
  density of dark matter in the NMSSM}},
  \href{http://dx.doi.org/10.1088/1475-7516/2005/09/001}{\emph{JCAP} {\bf 0509}
  (2005) 001}, [\href{http://arxiv.org/abs/hep-ph/0505142}{{\tt
  hep-ph/0505142}}].

\bibitem{Gunion:2005rw}
J.~F. Gunion, D.~Hooper and B.~McElrath, \emph{{Light neutralino dark matter in
  the NMSSM}}, \href{http://dx.doi.org/10.1103/PhysRevD.73.015011}{\emph{Phys.
  Rev.} {\bf D73} (2006) 015011},
  [\href{http://arxiv.org/abs/hep-ph/0509024}{{\tt hep-ph/0509024}}].

\bibitem{Servant:2002aq}
G.~Servant and T.~M.~P. Tait, \emph{{Is the lightest Kaluza-Klein particle a
  viable dark matter candidate?}},
  \href{http://dx.doi.org/10.1016/S0550-3213(02)01012-X}{\emph{Nucl. Phys.}
  {\bf B650} (2003) 391--419}, [\href{http://arxiv.org/abs/hep-ph/0206071}{{\tt
  hep-ph/0206071}}].

\bibitem{Kong:2005hn}
K.~Kong and K.~T. Matchev, \emph{{Precise calculation of the relic density of
  Kaluza-Klein dark matter in universal extra dimensions}},
  \href{http://dx.doi.org/10.1088/1126-6708/2006/01/038}{\emph{JHEP} {\bf 01}
  (2006) 038}, [\href{http://arxiv.org/abs/hep-ph/0509119}{{\tt
  hep-ph/0509119}}].

\bibitem{Bonnevier:2011km}
J.~Bonnevier, H.~Melbeus, A.~Merle and T.~Ohlsson, \emph{{Monoenergetic
  Gamma-Rays from Non-Minimal Kaluza-Klein Dark Matter Annihilations}},
  \href{http://dx.doi.org/10.1103/PhysRevD.85.109902,
  10.1103/PhysRevD.85.043524}{\emph{Phys. Rev.} {\bf D85} (2012) 043524},
  [\href{http://arxiv.org/abs/1104.1430}{{\tt 1104.1430}}].

\bibitem{Melbeus:2012wi}
H.~Melbeus, A.~Merle and T.~Ohlsson, \emph{{Higgs Dark Matter in UEDs: A Good
  WIMP with Bad Detection Prospects}},
  \href{http://dx.doi.org/10.1016/j.physletb.2012.07.037}{\emph{Phys. Lett.}
  {\bf B715} (2012) 164--169}, [\href{http://arxiv.org/abs/1204.5186}{{\tt
  1204.5186}}].

\bibitem{LopezHonorez:2006gr}
L.~Lopez~Honorez, E.~Nezri, J.~F. Oliver and M.~H.~G. Tytgat, \emph{{The Inert
  Doublet Model: An Archetype for Dark Matter}},
  \href{http://dx.doi.org/10.1088/1475-7516/2007/02/028}{\emph{JCAP} {\bf 0702}
  (2007) 028}, [\href{http://arxiv.org/abs/hep-ph/0612275}{{\tt
  hep-ph/0612275}}].

\bibitem{Dolle:2009fn}
E.~M. Dolle and S.~Su, \emph{{The Inert Dark Matter}},
  \href{http://dx.doi.org/10.1103/PhysRevD.80.055012}{\emph{Phys. Rev.} {\bf
  D80} (2009) 055012}, [\href{http://arxiv.org/abs/0906.1609}{{\tt
  0906.1609}}].

\bibitem{Cirelli:2012tf}
M.~Cirelli, \emph{{Indirect Searches for Dark Matter: a status review}},
  \href{http://dx.doi.org/10.1007/s12043-012-0419-x}{\emph{Pramana} {\bf 79}
  (2012) 1021--1043}, [\href{http://arxiv.org/abs/1202.1454}{{\tt 1202.1454}}].

\bibitem{Goodman:2010ku}
J.~Goodman, M.~Ibe, A.~Rajaraman, W.~Shepherd, T.~M.~P. Tait and H.-B. Yu,
  \emph{{Constraints on Dark Matter from Colliders}},
  \href{http://dx.doi.org/10.1103/PhysRevD.82.116010}{\emph{Phys. Rev.} {\bf
  D82} (2010) 116010}, [\href{http://arxiv.org/abs/1008.1783}{{\tt
  1008.1783}}].

\bibitem{Baudis:2012ig}
L.~Baudis, \emph{{Direct dark matter detection: the next decade}},
  \href{http://dx.doi.org/10.1016/j.dark.2012.10.006}{\emph{Phys. Dark Univ.}
  {\bf 1} (2012) 94--108}, [\href{http://arxiv.org/abs/1211.7222}{{\tt
  1211.7222}}].

\bibitem{Angloher:2014myn}
{\scshape CRESST-II} collaboration, G.~Angloher et~al., \emph{{Results on low
  mass WIMPs using an upgraded CRESST-II detector}},
  \href{http://dx.doi.org/10.1140/epjc/s10052-014-3184-9}{\emph{Eur. Phys. J.}
  {\bf C74} (2014) 3184}, [\href{http://arxiv.org/abs/1407.3146}{{\tt
  1407.3146}}].

\bibitem{Aprile:2012nq}
{\scshape XENON100} collaboration, E.~Aprile et~al., \emph{{Dark Matter Results
  from 225 Live Days of XENON100 Data}},
  \href{http://dx.doi.org/10.1103/PhysRevLett.109.181301}{\emph{Phys. Rev.
  Lett.} {\bf 109} (2012) 181301}, [\href{http://arxiv.org/abs/1207.5988}{{\tt
  1207.5988}}].

\bibitem{Akerib:2013tjd}
{\scshape LUX} collaboration, D.~S. Akerib et~al., \emph{{First results from
  the LUX dark matter experiment at the Sanford Underground Research
  Facility}},
  \href{http://dx.doi.org/10.1103/PhysRevLett.112.091303}{\emph{Phys. Rev.
  Lett.} {\bf 112} (2014) 091303}, [\href{http://arxiv.org/abs/1310.8214}{{\tt
  1310.8214}}].

\bibitem{Accardo:2014lma}
{\scshape AMS} collaboration, L.~Accardo et~al., \emph{{High Statistics
  Measurement of the Positron Fraction in Primary Cosmic Rays of 0.5�500 GeV
  with the Alpha Magnetic Spectrometer on the International Space Station}},
  \href{http://dx.doi.org/10.1103/PhysRevLett.113.121101}{\emph{Phys. Rev.
  Lett.} {\bf 113} (2014) 121101}.

\bibitem{Ackermann:2013uma}
{\scshape Fermi-LAT} collaboration, M.~Ackermann et~al., \emph{{Search for
  gamma-ray spectral lines with the Fermi large area telescope and dark matter
  implications}},
  \href{http://dx.doi.org/10.1103/PhysRevD.88.082002}{\emph{Phys. Rev.} {\bf
  D88} (2013) 082002}, [\href{http://arxiv.org/abs/1305.5597}{{\tt
  1305.5597}}].

\bibitem{Adriani:2013uda}
{\scshape PAMELA} collaboration, O.~Adriani et~al., \emph{{Cosmic-Ray Positron
  Energy Spectrum Measured by PAMELA}},
  \href{http://dx.doi.org/10.1103/PhysRevLett.111.081102}{\emph{Phys. Rev.
  Lett.} {\bf 111} (2013) 081102}, [\href{http://arxiv.org/abs/1308.0133}{{\tt
  1308.0133}}].

\bibitem{ATLAS:2012ky}
{\scshape ATLAS} collaboration, G.~Aad et~al., \emph{{Search for dark matter
  candidates and large extra dimensions in events with a jet and missing
  transverse momentum with the ATLAS detector}},
  \href{http://dx.doi.org/10.1007/JHEP04(2013)075}{\emph{JHEP} {\bf 04} (2013)
  075}, [\href{http://arxiv.org/abs/1210.4491}{{\tt 1210.4491}}].

\bibitem{Chatrchyan:2012me}
{\scshape CMS} collaboration, S.~Chatrchyan et~al., \emph{{Search for dark
  matter and large extra dimensions in monojet events in $pp$ collisions at
  $\sqrt{s}=7$ TeV}},
  \href{http://dx.doi.org/10.1007/JHEP09(2012)094}{\emph{JHEP} {\bf 09} (2012)
  094}, [\href{http://arxiv.org/abs/1206.5663}{{\tt 1206.5663}}].

\bibitem{Khachatryan:2014qwa}
{\scshape CMS} collaboration, V.~Khachatryan et~al., \emph{{Searches for
  electroweak production of charginos, neutralinos, and sleptons decaying to
  leptons and W, Z, and Higgs bosons in pp collisions at 8 TeV}},
  \href{http://dx.doi.org/10.1140/epjc/s10052-014-3036-7}{\emph{Eur. Phys. J.}
  {\bf C74} (2014) 3036}, [\href{http://arxiv.org/abs/1405.7570}{{\tt
  1405.7570}}].

\bibitem{Aad:2014vma}
{\scshape ATLAS} collaboration, G.~Aad et~al., \emph{{Search for direct
  production of charginos, neutralinos and sleptons in final states with two
  leptons and missing transverse momentum in $pp$ collisions at $\sqrt{s} =$ 8
  TeV with the ATLAS detector}},
  \href{http://dx.doi.org/10.1007/JHEP05(2014)071}{\emph{JHEP} {\bf 05} (2014)
  071}, [\href{http://arxiv.org/abs/1403.5294}{{\tt 1403.5294}}].

\bibitem{Springel:2005nw}
V.~Springel et~al., \emph{{Simulating the joint evolution of quasars, galaxies
  and their large-scale distribution}},
  \href{http://dx.doi.org/10.1038/nature03597}{\emph{Nature} {\bf 435} (2005)
  629--636}, [\href{http://arxiv.org/abs/astro-ph/0504097}{{\tt
  astro-ph/0504097}}].

\bibitem{Klypin:1999uc}
A.~A. Klypin, A.~V. Kravtsov, O.~Valenzuela and F.~Prada, \emph{{Where are the
  missing Galactic satellites?}},
  \href{http://dx.doi.org/10.1086/307643}{\emph{Astrophys. J.} {\bf 522} (1999)
  82--92}, [\href{http://arxiv.org/abs/astro-ph/9901240}{{\tt
  astro-ph/9901240}}].

\bibitem{Moore:1999nt}
B.~Moore, S.~Ghigna, F.~Governato, G.~Lake, T.~R. Quinn, J.~Stadel et~al.,
  \emph{{Dark matter substructure within galactic halos}},
  \href{http://dx.doi.org/10.1086/312287}{\emph{Astrophys. J.} {\bf 524} (1999)
  L19--L22}, [\href{http://arxiv.org/abs/astro-ph/9907411}{{\tt
  astro-ph/9907411}}].

\bibitem{Dubinski:1991bm}
J.~Dubinski and R.~G. Carlberg, \emph{{The Structure of cold dark matter
  halos}}, \href{http://dx.doi.org/10.1086/170451}{\emph{Astrophys. J.} {\bf
  378} (1991) 496}.

\bibitem{Navarro:1995iw}
J.~F. Navarro, C.~S. Frenk and S.~D.~M. White, \emph{{The Structure of cold
  dark matter halos}}, \href{http://dx.doi.org/10.1086/177173}{\emph{Astrophys.
  J.} {\bf 462} (1996) 563--575},
  [\href{http://arxiv.org/abs/astro-ph/9508025}{{\tt astro-ph/9508025}}].

\bibitem{Maccio':2009dx}
A.~V. Maccio', X.~Kang, F.~Fontanot, R.~S. Somerville, S.~E. Koposov and
  P.~Monaco, \emph{{On the origin and properties of Ultrafaint Milky Way
  Satellites in a LCDM Universe}},
  \href{http://dx.doi.org/10.1111/j.1365-2966.2009.16031.x}{\emph{Mon. Not.
  Roy. Astron. Soc.} {\bf 402} (2010) 1995},
  [\href{http://arxiv.org/abs/0903.4681}{{\tt 0903.4681}}].

\bibitem{GarrisonKimmel:2013aq}
S.~Garrison-Kimmel, M.~Rocha, M.~Boylan-Kolchin, J.~Bullock and J.~Lally,
  \emph{{Can Feedback Solve the Too Big to Fail Problem?}},
  \href{http://dx.doi.org/10.1093/mnras/stt984}{\emph{Mon. Not. Roy. Astron.
  Soc.} {\bf 433} (2013) 3539}, [\href{http://arxiv.org/abs/1301.3137}{{\tt
  1301.3137}}].

\bibitem{Geen:2011fj}
S.~Geen, A.~Slyz and J.~Devriendt, \emph{{How Does Feedback Affect Milky Way
  Satellite Formation?}},
  \href{http://dx.doi.org/10.1051/eas/1148096}{\emph{EAS Publ. Ser.} {\bf 48}
  (2011) 441}, [\href{http://arxiv.org/abs/1101.2232}{{\tt 1101.2232}}].

\bibitem{BoylanKolchin:2011de}
M.~Boylan-Kolchin, J.~S. Bullock and M.~Kaplinghat, \emph{{Too big to fail? The
  puzzling darkness of massive Milky Way subhaloes}}, {\emph{Mon. Not. Roy.
  Astron. Soc.} {\bf 415} (2011) L40},
  [\href{http://arxiv.org/abs/1103.0007}{{\tt 1103.0007}}].

\bibitem{BoylanKolchin:2011dk}
M.~Boylan-Kolchin, J.~S. Bullock and M.~Kaplinghat, \emph{{The Milky Way's
  bright satellites as an apparent failure of LCDM}},
  \href{http://dx.doi.org/10.1111/j.1365-2966.2012.20695.x}{\emph{Mon. Not.
  Roy. Astron. Soc.} {\bf 422} (2012) 1203--1218},
  [\href{http://arxiv.org/abs/1111.2048}{{\tt 1111.2048}}].

\bibitem{Brooks:2012vi}
A.~M. Brooks and A.~Zolotov, \emph{{Why Baryons Matter: The Kinematics of Dwarf
  Spheroidal Satellites}},
  \href{http://dx.doi.org/10.1088/0004-637X/786/2/87}{\emph{Astrophys. J.} {\bf
  786} (2014) 87}, [\href{http://arxiv.org/abs/1207.2468}{{\tt 1207.2468}}].

\bibitem{Zhu:2015}
Q.~Zhu, F.~Marinacci, M.~Maji, Y.~Li, V.~Springel and L.~Hernquist,
  \emph{{Baryonic impact on the dark matter distribution in Milky Way-size
  galaxies and their satellites}},  \href{http://arxiv.org/abs/1506.05537}{{\tt
  1506.05537}}.

\bibitem{Herpich:2013yga}
J.~Herpich, G.~S. Stinson, A.~V. Macci�, C.~Brook, J.~Wadsley, H.~M.~P.
  Couchman et~al., \emph{{MaGICC-WDM: the effects of warm dark matter in
  hydrodynamical simulations of disc galaxy formation}},
  \href{http://dx.doi.org/10.1093/mnras/stt1883}{\emph{Mon. Not. Roy. Astron.
  Soc.} {\bf 437} (2014) 293--304}, [\href{http://arxiv.org/abs/1308.1088}{{\tt
  1308.1088}}].

\bibitem{Lovell:2011rd}
M.~R. Lovell, V.~Eke, C.~S. Frenk, L.~Gao, A.~Jenkins, T.~Theuns et~al.,
  \emph{{The Haloes of Bright Satellite Galaxies in a Warm Dark Matter
  Universe}},
  \href{http://dx.doi.org/10.1111/j.1365-2966.2011.20200.x}{\emph{Mon. Not.
  Roy. Astron. Soc.} {\bf 420} (2012) 2318--2324},
  [\href{http://arxiv.org/abs/1104.2929}{{\tt 1104.2929}}].

\bibitem{Boyarsky:2008xj}
A.~Boyarsky, J.~Lesgourgues, O.~Ruchayskiy and M.~Viel, \emph{{Lyman-alpha
  constraints on warm and on warm-plus-cold dark matter models}},
  \href{http://dx.doi.org/10.1088/1475-7516/2009/05/012}{\emph{JCAP} {\bf 0905}
  (2009) 012}, [\href{http://arxiv.org/abs/0812.0010}{{\tt 0812.0010}}].

\bibitem{Majorana:1937vz}
E.~Majorana, \emph{{Theory of the Symmetry of Electrons and Positrons}},
  \href{http://dx.doi.org/10.1007/BF02961314}{\emph{Nuovo Cim.} {\bf 14} (1937)
  171--184}.

\bibitem{Ibarra:2011xn}
A.~Ibarra, E.~Molinaro and S.~T. Petcov, \emph{{Low Energy Signatures of the
  TeV Scale See-Saw Mechanism}},
  \href{http://dx.doi.org/10.1103/PhysRevD.84.013005}{\emph{Phys. Rev.} {\bf
  D84} (2011) 013005}, [\href{http://arxiv.org/abs/1103.6217}{{\tt
  1103.6217}}].

\bibitem{Ruchayskiy:2012si}
O.~Ruchayskiy and A.~Ivashko, \emph{{Restrictions on the lifetime of sterile
  neutrinos from primordial nucleosynthesis}},
  \href{http://dx.doi.org/10.1088/1475-7516/2012/10/014}{\emph{JCAP} {\bf 1210}
  (2012) 014}, [\href{http://arxiv.org/abs/1202.2841}{{\tt 1202.2841}}].

\bibitem{Abada:2012mc}
A.~Abada, D.~Das, A.~M. Teixeira, A.~Vicente and C.~Weiland, \emph{{Tree-level
  lepton universality violation in the presence of sterile neutrinos: impact
  for $R_K$ and $R_\pi$}},
  \href{http://dx.doi.org/10.1007/JHEP02(2013)048}{\emph{JHEP} {\bf 02} (2013)
  048}, [\href{http://arxiv.org/abs/1211.3052}{{\tt 1211.3052}}].

\bibitem{Ruchayskiy:2011aa}
O.~Ruchayskiy and A.~Ivashko, \emph{{Experimental bounds on sterile neutrino
  mixing angles}}, \href{http://dx.doi.org/10.1007/JHEP06(2012)100}{\emph{JHEP}
  {\bf 1206} (2012) 100}, [\href{http://arxiv.org/abs/1112.3319}{{\tt
  1112.3319}}].

\bibitem{Abada:2013aba}
A.~Abada, A.~M. Teixeira, A.~Vicente and C.~Weiland, \emph{{Sterile neutrinos
  in leptonic and semileptonic decays}},
  \href{http://dx.doi.org/10.1007/JHEP02(2014)091}{\emph{JHEP} {\bf 02} (2014)
  091}, [\href{http://arxiv.org/abs/1311.2830}{{\tt 1311.2830}}].

\bibitem{Merle:2013gea}
A.~Merle, \emph{{keV Neutrino Model Building}},
  \href{http://dx.doi.org/10.1142/S0218271813300206}{\emph{Int. J. Mod. Phys.}
  {\bf D22} (2013) 1330020}, [\href{http://arxiv.org/abs/1302.2625}{{\tt
  1302.2625}}].

\bibitem{Drewes:2013gca}
M.~Drewes, \emph{{The Phenomenology of Right Handed Neutrinos}},
  \href{http://dx.doi.org/10.1142/S0218301313300191}{\emph{Int.J.Mod.Phys.}
  {\bf E22} (2013) 1330019}, [\href{http://arxiv.org/abs/1303.6912}{{\tt
  1303.6912}}].

\bibitem{Hernandez:2014fha}
P.~Hernandez, M.~Kekic and J.~Lopez-Pavon, \emph{{$N_{\rm eff}$ in low-scale
  seesaw models versus the lightest neutrino mass}},
  \href{http://dx.doi.org/10.1103/PhysRevD.90.065033}{\emph{Phys. Rev.} {\bf
  D90} (2014) 065033}, [\href{http://arxiv.org/abs/1406.2961}{{\tt
  1406.2961}}].

\bibitem{Fernandez-Martinez:2015hxa}
E.~Fernandez-Martinez, J.~Hernandez-Garcia, J.~Lopez-Pavon and M.~Lucente,
  \emph{{Loop level constraints on Seesaw neutrino mixing}},
  \href{http://dx.doi.org/10.1007/JHEP10(2015)130}{\emph{JHEP} {\bf 10} (2015)
  130}, [\href{http://arxiv.org/abs/1508.03051}{{\tt 1508.03051}}].

\bibitem{Drewes:2015jna}
M.~Drewes, \emph{{Theoretical Status of Neutrino Physics}}, {\emph{PoS} {\bf
  NUFACT2014} (2015) 001}, [\href{http://arxiv.org/abs/1502.06891}{{\tt
  1502.06891}}].

\bibitem{Drewes:2015iva}
M.~Drewes and B.~Garbrecht, \emph{{Experimental and cosmological constraints on
  heavy neutrinos}},  \href{http://arxiv.org/abs/1502.00477}{{\tt 1502.00477}}.

\bibitem{Antusch:2015mia}
S.~Antusch and O.~Fischer, \emph{{Testing sterile neutrino extensions of the
  Standard Model at future lepton colliders}},
  \href{http://dx.doi.org/10.1007/JHEP05(2015)053}{\emph{JHEP} {\bf 05} (2015)
  053}, [\href{http://arxiv.org/abs/1502.05915}{{\tt 1502.05915}}].

\bibitem{deGouvea:2015euy}
A.~de~Gouvea and A.~Kobach, \emph{{Global Constraints on a Heavy Neutrino}},
  \href{http://arxiv.org/abs/1511.00683}{{\tt 1511.00683}}.

\bibitem{Deppisch:2015qwa}
F.~F. Deppisch, P.~S. Bhupal~Dev and A.~Pilaftsis, \emph{{Neutrinos and
  Collider Physics}},
  \href{http://dx.doi.org/10.1088/1367-2630/17/7/075019}{\emph{New J. Phys.}
  {\bf 17} (2015) 075019}, [\href{http://arxiv.org/abs/1502.06541}{{\tt
  1502.06541}}].

\bibitem{Tremaine:1979we}
S.~Tremaine and J.~E. Gunn, \emph{{Dynamical Role of Light Neutral Leptons in
  Cosmology}}, \href{http://dx.doi.org/10.1103/PhysRevLett.42.407}{\emph{Phys.
  Rev. Lett.} {\bf 42} (1979) 407--410}.

\bibitem{Bellini:2013wra}
G.~Bellini, L.~Ludhova, G.~Ranucci and F.~Villante, \emph{{Neutrino
  oscillations}}, \href{http://dx.doi.org/10.1155/2014/191960}{\emph{Adv.High
  Energy Phys.} {\bf 2014} (2014) 191960},
  [\href{http://arxiv.org/abs/arXiv:1310.7858}{{\tt arXiv:1310.7858}}].

\bibitem{Wang:2015rma}
Y.~Wang and Z.~zhong Xing, \emph{{Neutrino Masses and Flavor Oscillations}},
  \href{http://arxiv.org/abs/arXiv:1504.06155}{{\tt arXiv:1504.06155}}.

\bibitem{Pontecorvo:1967fh}
B.~Pontecorvo, \emph{{Neutrino Experiments and the Problem of Conservation of
  Leptonic Charge}}, {\emph{Sov.Phys.JETP} {\bf 26} (1968) 984--988}.

\bibitem{Pontecorvo:1957cp}
B.~Pontecorvo, \emph{{Mesonium and anti-mesonium}}, {\emph{Sov.Phys.JETP} {\bf
  6} (1957) 429}.

\bibitem{Pontecorvo:1957qd}
B.~Pontecorvo, \emph{{Inverse beta processes and nonconservation of lepton
  charge}}, {\emph{Sov.Phys.JETP} {\bf 7} (1958) 172--173}.

\bibitem{Maki:1962mu}
Z.~Maki, M.~Nakagawa and S.~Sakata, \emph{{Remarks on the unified model of
  elementary particles}},
  \href{http://dx.doi.org/10.1143/PTP.28.870}{\emph{Prog.Theor.Phys.} {\bf 28}
  (1962) 870--880}.

\bibitem{Nakagawa01111963}
M.~Nakagawa, H.~Okonogi, S.~Sakata and A.~Toyoda, \emph{Possible existence of a
  neutrino with mass and partial conservation of muon charge},
  \href{http://dx.doi.org/10.1143/PTP.30.727}{\emph{Progress of Theoretical
  Physics} {\bf 30} (1963) 727--729}.

\bibitem{Ahmed:2003kj}
{\scshape SNO} collaboration, S.~Ahmed et~al., \emph{{Measurement of the total
  active B-8 solar neutrino flux at the Sudbury Neutrino Observatory with
  enhanced neutral current sensitivity}},
  \href{http://dx.doi.org/10.1103/PhysRevLett.92.181301}{\emph{Phys.Rev.Lett.}
  {\bf 92} (2004) 181301}, [\href{http://arxiv.org/abs/nucl-ex/0309004}{{\tt
  nucl-ex/0309004}}].

\bibitem{Gonzalez-Garcia:2014bfa}
M.~C. Gonzalez-Garcia, M.~Maltoni and T.~Schwetz, \emph{{Updated fit to three
  neutrino mixing: status of leptonic CP violation}},
  \href{http://dx.doi.org/10.1007/JHEP11(2014)052}{\emph{JHEP} {\bf 1411}
  (2014) 052}, [\href{http://arxiv.org/abs/1409.5439}{{\tt 1409.5439}}].

\bibitem{Agashe:2014kda}
{\scshape Particle Data Group} collaboration, K.~Olive et~al., \emph{{Review of
  Particle Physics}},
  \href{http://dx.doi.org/10.1088/1674-1137/38/9/090001}{\emph{Chin.Phys.} {\bf
  C38} (2014) 090001}.

\bibitem{Rodejohann:2011vc}
W.~Rodejohann and J.~W.~F. Valle, \emph{{Symmetrical Parametrizations of the
  Lepton Mixing Matrix}},
  \href{http://dx.doi.org/10.1103/PhysRevD.84.073011}{\emph{Phys. Rev.} {\bf
  D84} (2011) 073011}, [\href{http://arxiv.org/abs/1108.3484}{{\tt
  1108.3484}}].

\bibitem{Mena:2003ug}
O.~Mena and S.~J. Parke, \emph{{Unified graphical summary of neutrino mixing
  parameters}},
  \href{http://dx.doi.org/10.1103/PhysRevD.69.117301}{\emph{Phys.Rev.} {\bf
  D69} (2004) 117301}, [\href{http://arxiv.org/abs/hep-ph/0312131}{{\tt
  hep-ph/0312131}}].

\bibitem{Gribov:1968kq}
V.~N. Gribov and B.~Pontecorvo, \emph{{Neutrino astronomy and lepton charge}},
  \href{http://dx.doi.org/10.1016/0370-2693(69)90525-5}{\emph{Phys. Lett.} {\bf
  B28} (1969) 493}.

\bibitem{GonzalezGarcia:2007ib}
M.~C. Gonzalez-Garcia and M.~Maltoni, \emph{{Phenomenology with Massive
  Neutrinos}},
  \href{http://dx.doi.org/10.1016/j.physrep.2007.12.004}{\emph{Phys. Rept.}
  {\bf 460} (2008) 1--129}, [\href{http://arxiv.org/abs/0704.1800}{{\tt
  0704.1800}}].

\bibitem{skatm:nu2014}
{\scshape Super-Kamiokande} collaboration, R.~Wendell, \emph{{Atmospheric
  Results from Super-Kamiokande}}, {\emph{XXVI International Conference on
  Neutrino Physics and Astrophysics, Boston, USA, June 2--7, 2014} (2014) }.

\bibitem{Cleveland:1998nv}
B.~T. Cleveland, T.~Daily, J.~Davis, Raymond, J.~R. Distel, K.~Lande et~al.,
  \emph{{Measurement of the solar electron neutrino flux with the Homestake
  chlorine detector}}, \href{http://dx.doi.org/10.1086/305343}{\emph{Astrophys.
  J.} {\bf 496} (1998) 505--526}.

\bibitem{Kaether:2010ag}
F.~Kaether, W.~Hampel, G.~Heusser, J.~Kiko and T.~Kirsten, \emph{{Reanalysis of
  the GALLEX solar neutrino flux and source experiments}},
  \href{http://dx.doi.org/10.1016/j.physletb.2010.01.030}{\emph{Phys. Lett.}
  {\bf B685} (2010) 47--54}, [\href{http://arxiv.org/abs/1001.2731}{{\tt
  1001.2731}}].

\bibitem{Abdurashitov:2009tn}
{\scshape SAGE} collaboration, J.~N. Abdurashitov et~al., \emph{{Measurement of
  the solar neutrino capture rate with gallium metal. III: Results for the
  2002--2007 data-taking period}},
  \href{http://dx.doi.org/10.1103/PhysRevC.80.015807}{\emph{Phys. Rev.} {\bf
  C80} (2009) 015807}, [\href{http://arxiv.org/abs/0901.2200}{{\tt
  0901.2200}}].

\bibitem{Hosaka:2005um}
{\scshape Super-Kamiokande} collaboration, J.~Hosaka et~al., \emph{{Solar
  neutrino measurements in super-Kamiokande-I}},
  \href{http://dx.doi.org/10.1103/PhysRevD.73.112001}{\emph{Phys. Rev.} {\bf
  D73} (2006) 112001}, [\href{http://arxiv.org/abs/hep-ex/0508053}{{\tt
  hep-ex/0508053}}].

\bibitem{Cravens:2008aa}
{\scshape Super-Kamiokande} collaboration, J.~P. Cravens et~al., \emph{{Solar
  neutrino measurements in Super-Kamiokande-II}},
  \href{http://dx.doi.org/10.1103/PhysRevD.78.032002}{\emph{Phys. Rev.} {\bf
  D78} (2008) 032002}, [\href{http://arxiv.org/abs/0803.4312}{{\tt
  0803.4312}}].

\bibitem{Abe:2010hy}
{\scshape Super-Kamiokande} collaboration, K.~Abe et~al., \emph{{Solar neutrino
  results in Super-Kamiokande-III}},
  \href{http://dx.doi.org/10.1103/PhysRevD.83.052010}{\emph{Phys. Rev.} {\bf
  D83} (2011) 052010}, [\href{http://arxiv.org/abs/1010.0118}{{\tt
  1010.0118}}].

\bibitem{sksol:nu2014}
{\scshape Super-Kamiokande} collaboration, Y.~Koshio, \emph{{Solar Results from
  Super-Kamiokande}}, {\emph{XXVI International Conference on Neutrino Physics
  and Astrophysics, Boston, USA, June 2--7, 2014} (2014) }.

\bibitem{Aharmim:2011vm}
{\scshape SNO} collaboration, B.~Aharmim et~al., \emph{{Combined Analysis of
  all Three Phases of Solar Neutrino Data from the Sudbury Neutrino
  Observatory}},
  \href{http://dx.doi.org/10.1103/PhysRevC.88.025501}{\emph{Phys. Rev.} {\bf
  C88} (2013) 025501}, [\href{http://arxiv.org/abs/1109.0763}{{\tt
  1109.0763}}].

\bibitem{Bellini:2011rx}
G.~Bellini, J.~Benziger, D.~Bick, S.~Bonetti, G.~Bonfini et~al.,
  \emph{{Precision measurement of the 7Be solar neutrino interaction rate in
  Borexino}},
  \href{http://dx.doi.org/10.1103/PhysRevLett.107.141302}{\emph{Phys. Rev.
  Lett.} {\bf 107} (2011) 141302}, [\href{http://arxiv.org/abs/1104.1816}{{\tt
  1104.1816}}].

\bibitem{Bellini:2008mr}
{\scshape Borexino} collaboration, G.~Bellini et~al., \emph{{Measurement of the
  solar 8B neutrino rate with a liquid scintillator target and 3 MeV energy
  threshold in the Borexino detector}},
  \href{http://dx.doi.org/10.1103/PhysRevD.82.033006}{\emph{Phys. Rev.} {\bf
  D82} (2010) 033006}, [\href{http://arxiv.org/abs/0808.2868}{{\tt
  0808.2868}}].

\bibitem{Adamson:2013whj}
{\scshape MINOS} collaboration, P.~Adamson et~al., \emph{{Measurement of
  Neutrino and Antineutrino Oscillations Using Beam and Atmospheric Data in
  MINOS}}, \href{http://dx.doi.org/10.1103/PhysRevLett.110.251801}{\emph{Phys.
  Rev. Lett.} {\bf 110} (2013) 251801},
  [\href{http://arxiv.org/abs/1304.6335}{{\tt 1304.6335}}].

\bibitem{Abe:2014ugx}
{\scshape T2K} collaboration, K.~Abe et~al., \emph{{Precise Measurement of the
  Neutrino Mixing Parameter $\theta_{23}$ from Muon Neutrino Disappearance in
  an Off-Axis Beam}},
  \href{http://dx.doi.org/10.1103/PhysRevLett.112.181801}{\emph{Phys. Rev.
  Lett.} {\bf 112} (2014) 181801}, [\href{http://arxiv.org/abs/1403.1532}{{\tt
  1403.1532}}].

\bibitem{Adamson:2013ue}
{\scshape MINOS} collaboration, P.~Adamson et~al., \emph{{Electron neutrino and
  antineutrino appearance in the full MINOS data sample}},
  \href{http://dx.doi.org/10.1103/PhysRevLett.110.171801}{\emph{Phys. Rev.
  Lett.} {\bf 110} (2013) 171801}, [\href{http://arxiv.org/abs/1301.4581}{{\tt
  1301.4581}}].

\bibitem{Abe:2013hdq}
{\scshape T2K} collaboration, K.~Abe et~al., \emph{{Observation of Electron
  Neutrino Appearance in a Muon Neutrino Beam}},
  \href{http://dx.doi.org/10.1103/PhysRevLett.112.061802}{\emph{Phys. Rev.
  Lett.} {\bf 112} (2014) 061802}, [\href{http://arxiv.org/abs/1311.4750}{{\tt
  1311.4750}}].

\bibitem{An:2012eh}
{\scshape Daya Bay} collaboration, F.~P. An et~al., \emph{Observation of
  electron-antineutrino disappearance at daya bay}, {\emph{Phys. Rev. Lett.}
  {\bf 108} (2012) 171803}, [\href{http://arxiv.org/abs/1203.1669}{{\tt
  1203.1669}}].

\bibitem{Ahn:2012nd}
{\scshape RENO} collaboration, J.~K. Ahn et~al., \emph{{Observation of Reactor
  Electron Antineutrino Disappearance in the RENO Experiment}},
  \href{http://dx.doi.org/10.1103/PhysRevLett.108.191802}{\emph{Phys. Rev.
  Lett.} {\bf 108} (2012) 191802}, [\href{http://arxiv.org/abs/1204.0626}{{\tt
  1204.0626}}].

\bibitem{An:2015rpe}
{\scshape Daya Bay} collaboration, F.~P. An et~al., \emph{{A new measurement of
  antineutrino oscillation with the full detector configuration at Daya Bay}},
  \href{http://arxiv.org/abs/1505.03456}{{\tt 1505.03456}}.

\bibitem{reno:nu2014}
{\scshape RENO} collaboration, S.-H. Seo, \emph{{New Results from RENO}},
  {\emph{XXVI International Conference on Neutrino Physics and Astrophysics,
  Boston, USA, June 2--7, 2014} (2014) }.

\bibitem{Abe:2012tg}
{\scshape Double Chooz} collaboration, Y.~Abe et~al., \emph{{Reactor electron
  antineutrino disappearance in the Double Chooz experiment}},
  \href{http://dx.doi.org/10.1103/PhysRevD.86.052008}{\emph{Phys. Rev.} {\bf
  D86} (2012) 052008}, [\href{http://arxiv.org/abs/1207.6632}{{\tt
  1207.6632}}].

\bibitem{Abe:2014bwa}
{\scshape Double Chooz} collaboration, Y.~Abe et~al., \emph{{Improved
  measurements of the neutrino mixing angle $\theta\_{13}$ with the Double
  Chooz detector}}, \href{http://dx.doi.org/10.1007/JHEP02(2015)074,
  10.1007/JHEP10(2014)086}{\emph{JHEP} {\bf 10} (2014) 086},
  [\href{http://arxiv.org/abs/1406.7763}{{\tt 1406.7763}}].

\bibitem{Gando:2010aa}
{\scshape KamLAND} collaboration, A.~Gando et~al., \emph{{Constraints on
  $\theta_{13}$ from A Three-Flavor Oscillation Analysis of Reactor
  Antineutrinos at KamLAND}},
  \href{http://dx.doi.org/10.1103/PhysRevD.83.052002}{\emph{Phys. Rev.} {\bf
  D83} (2011) 052002}, [\href{http://arxiv.org/abs/1009.4771}{{\tt
  1009.4771}}].

\bibitem{Capozzi:2013csa}
F.~Capozzi, G.~L. Fogli, E.~Lisi, A.~Marrone, D.~Montanino et~al.,
  \emph{{Status of three-neutrino oscillation parameters, circa 2013}},
  \href{http://dx.doi.org/10.1103/PhysRevD.89.093018}{\emph{Phys. Rev.} {\bf
  D89} (2014) 093018}, [\href{http://arxiv.org/abs/1312.2878}{{\tt
  1312.2878}}].

\bibitem{Forero:2014bxa}
D.~V. Forero, M.~Tortola and J.~W.~F. Valle, \emph{{Neutrino oscillations
  refitted}}, \href{http://dx.doi.org/10.1103/PhysRevD.90.093006}{\emph{Phys.
  Rev.} {\bf D90} (2014) 093006}, [\href{http://arxiv.org/abs/1405.7540}{{\tt
  1405.7540}}].

\bibitem{Mueller:2011nm}
T.~A. Mueller, D.~Lhuillier, M.~Fallot, A.~Letourneau, S.~Cormon et~al.,
  \emph{{Improved Predictions of Reactor Antineutrino Spectra}},
  \href{http://dx.doi.org/10.1103/PhysRevC.83.054615}{\emph{Phys. Rev.} {\bf
  C83} (2011) 054615}, [\href{http://arxiv.org/abs/1101.2663}{{\tt
  1101.2663}}].

\bibitem{Huber:2011wv}
P.~Huber, \emph{{On the determination of anti-neutrino spectra from nuclear
  reactors}}, \href{http://dx.doi.org/10.1103/PhysRevC.85.029901,
  10.1103/PhysRevC.84.024617}{\emph{Phys. Rev.} {\bf C84} (2011) 024617},
  [\href{http://arxiv.org/abs/1106.0687}{{\tt 1106.0687}}].

\bibitem{Mention:2011rk}
G.~Mention, M.~Fechner, T.~Lasserre, T.~A. Mueller, D.~Lhuillier et~al.,
  \emph{{The Reactor Antineutrino Anomaly}},
  \href{http://dx.doi.org/10.1103/PhysRevD.83.073006}{\emph{Phys. Rev.} {\bf
  D83} (2011) 073006}, [\href{http://arxiv.org/abs/1101.2755}{{\tt
  1101.2755}}].

\bibitem{Declais:1994ma}
Y.~Declais, H.~de~Kerret, B.~Lefievre, M.~Obolensky, A.~Etenko et~al.,
  \emph{{Study of reactor anti-neutrino interaction with proton at Bugey
  nuclear power plant}},
  \href{http://dx.doi.org/10.1016/0370-2693(94)91394-3}{\emph{Phys. Lett.} {\bf
  B338} (1994) 383--389}.

\bibitem{Kuvshinnikov:1990ry}
A.~A. Kuvshinnikov, L.~A. Mikaelyan, S.~V. Nikolaev, M.~D. Skorokhvatov and
  A.~V. Etenko, \emph{{Measuring the $\bar{\nu}_e + p \to n + e^+$
  cross-section and beta decay axial constant in a new experiment at Rovno NPP
  reactor. (In Russian)}}, {\emph{JETP Lett.} {\bf 54} (1991) 253--257}.

\bibitem{Declais:1994su}
Y.~Declais, J.~Favier, A.~Metref, H.~Pessard, B.~Achkar et~al., \emph{{Search
  for neutrino oscillations at 15-meters, 40-meters, and 95-meters from a
  nuclear power reactor at Bugey}},
  \href{http://dx.doi.org/10.1016/0550-3213(94)00513-E}{\emph{Nucl. Phys.} {\bf
  B434} (1995) 503--534}.

\bibitem{Vidyakin:1987ue}
G.~S. Vidyakin, V.~N. Vyrodov, I.~I. Gurevich, Y.~V. Kozlov, V.~P. Martemyanov
  et~al., \emph{{Detection of Anti-neutrinos in the Flux From Two Reactors}},
  {\emph{Sov. Phys. JETP} {\bf 66} (1987) 243--247}.

\bibitem{Vidyakin:1994ut}
G.~S. Vidyakin, V.~N. Vyrodov, Y.~V. Kozlov, A.~V. Martemyanov, V.~P.
  Martemyanov et~al., \emph{{Limitations on the characteristics of neutrino
  oscillations}}, {\emph{JETP Lett.} {\bf 59} (1994) 390--393}.

\bibitem{Kwon:1981ua}
H.~Kwon, F.~Boehm, A.~A. Hahn, H.~E. Henrikson, J.~Vuilleumier et~al.,
  \emph{{Search for Neutrino Oscillations at a Fission Reactor}},
  \href{http://dx.doi.org/10.1103/PhysRevD.24.1097}{\emph{Phys. Rev.} {\bf D24}
  (1981) 1097--1111}.

\bibitem{Zacek:1986cu}
{\scshape CALTECH-SIN-TUM} collaboration, G.~Zacek et~al., \emph{{Neutrino
  Oscillation Experiments at the Gosgen Nuclear Power Reactor}},
  \href{http://dx.doi.org/10.1103/PhysRevD.34.2621}{\emph{Phys. Rev.} {\bf D34}
  (1986) 2621--2636}.

\bibitem{Greenwood:1996pb}
Z.~D. Greenwood, W.~R. Kropp, M.~A. Mandelkern, S.~Nakamura, E.~L. Pasierb-Love
  et~al., \emph{{Results of a two position reactor neutrino oscillation
  experiment}}, \href{http://dx.doi.org/10.1103/PhysRevD.53.6054}{\emph{Phys.
  Rev.} {\bf D53} (1996) 6054--6064}.

\bibitem{Afonin:1988gx}
A.~I. Afonin, S.~N. Ketov, V.~I. Kopeikin, L.~A. Mikaelyan, M.~D. Skorokhvatov
  et~al., \emph{{A Study of the Reaction $\bar{\nu}_e + P \to e^+ + N$ on a
  Nuclear Reactor}}, {\emph{Sov. Phys. JETP} {\bf 67} (1988) 213--221}.

\bibitem{Jarlskog:1985ht}
C.~Jarlskog, \emph{{Commutator of the Quark Mass Matrices in the Standard
  Electroweak Model and a Measure of Maximal CP Violation}},
  \href{http://dx.doi.org/10.1103/PhysRevLett.55.1039}{\emph{Phys. Rev. Lett.}
  {\bf 55} (1985) 1039}.

\bibitem{Shro80}
R.~E. Shrock, \emph{New tests for and bounds on neutrino masses and lepton
  mixing}, {\emph{Phys. Lett B} {\bf 96} (1980) 159--64}.

\bibitem{Aseev:2011dq}
V.~N. Aseev et~al., \emph{{Upper limit on the electron antineutrino mass from
  the Troitsk experiment}},
  \href{http://dx.doi.org/10.1103/PhysRevD.84.112003}{\emph{Phys. Rev.} {\bf
  D84} (2011) 112003}, [\href{http://arxiv.org/abs/1108.5034}{{\tt
  1108.5034}}].

\bibitem{Angrik:2005ep}
{\scshape KATRIN} collaboration, J.~Angrik et~al., \emph{{KATRIN design report
  2004}}, .

\bibitem{Schechter:1981bd}
J.~Schechter and J.~W.~F. Valle, \emph{{Neutrinoless Double beta Decay in SU(2)
  x U(1) Theories}},
  \href{http://dx.doi.org/10.1103/PhysRevD.25.2951}{\emph{Phys. Rev.} {\bf D25}
  (1982) 2951}.

\bibitem{Duerr:2011zd}
M.~Duerr, M.~Lindner and A.~Merle, \emph{{On the Quantitative Impact of the
  Schechter-Valle Theorem}},
  \href{http://dx.doi.org/10.1007/JHEP06(2011)091}{\emph{JHEP} {\bf 06} (2011)
  091}, [\href{http://arxiv.org/abs/1105.0901}{{\tt 1105.0901}}].

\bibitem{Lindner:2005kr}
M.~Lindner, A.~Merle and W.~Rodejohann, \emph{{Improved limit on theta(13) and
  implications for neutrino masses in neutrino-less double beta decay and
  cosmology}}, \href{http://dx.doi.org/10.1103/PhysRevD.73.053005}{\emph{Phys.
  Rev.} {\bf D73} (2006) 053005},
  [\href{http://arxiv.org/abs/hep-ph/0512143}{{\tt hep-ph/0512143}}].

\bibitem{Merle:2006du}
A.~Merle and W.~Rodejohann, \emph{{The Elements of the neutrino mass matrix:
  Allowed ranges and implications of texture zeros}},
  \href{http://dx.doi.org/10.1103/PhysRevD.73.073012}{\emph{Phys. Rev.} {\bf
  D73} (2006) 073012}, [\href{http://arxiv.org/abs/hep-ph/0603111}{{\tt
  hep-ph/0603111}}].

\bibitem{gerda}
{\scshape GERDA} collaboration, M.~Agostini et~al., \emph{{Results on
  Neutrinoless Double-$\beta$ Decay of $^{76}$Ge from Phase I of the GERDA
  Experiment}},
  \href{http://dx.doi.org/10.1103/PhysRevLett.111.122503}{\emph{Phys. Rev.
  Lett.} {\bf 111} (2013) 122503}, [\href{http://arxiv.org/abs/1307.4720}{{\tt
  1307.4720}}].

\bibitem{kzen}
{\scshape KamLAND-Zen} collaboration, A.~Gando et~al., \emph{{Limit on
  Neutrinoless $\beta\beta$ Decay of Xe-136 from the First Phase of KamLAND-Zen
  and Comparison with the Positive Claim in Ge-76}},
  \href{http://dx.doi.org/10.1103/PhysRevLett.110.062502}{\emph{Phys. Rev.
  Lett.} {\bf 110} (2013) 062502}, [\href{http://arxiv.org/abs/1211.3863}{{\tt
  1211.3863}}].

\bibitem{exo}
{\scshape EXO-200} collaboration, J.~B. Albert et~al., \emph{{Search for
  Majorana neutrinos with the first two years of EXO-200 data}},
  \href{http://dx.doi.org/10.1038/nature13432}{\emph{Nature} {\bf 510} (2014)
  229--234}, [\href{http://arxiv.org/abs/1402.6956}{{\tt 1402.6956}}].

\bibitem{bbothers}
J.~J. Gomez-Cadenas, J.~Martin-Albo, M.~Mezzetto, F.~Monrabal and M.~Sorel,
  \emph{{The Search for neutrinoless double beta decay}},
  \href{http://dx.doi.org/10.1393/ncr/i2012-10074-9}{\emph{Riv. Nuovo Cim.}
  {\bf 35} (2012) 29--98}, [\href{http://arxiv.org/abs/1109.5515}{{\tt
  1109.5515}}].

\bibitem{Maneschg:2008sf}
W.~Maneschg, A.~Merle and W.~Rodejohann, \emph{{Statistical Analysis of future
  Neutrino Mass Experiments including Neutrino-less Double Beta Decay}},
  \href{http://dx.doi.org/10.1209/0295-5075/85/51002}{\emph{Europhys. Lett.}
  {\bf 85} (2009) 51002}, [\href{http://arxiv.org/abs/0812.0479}{{\tt
  0812.0479}}].

\bibitem{Patterson:2012zs}
{\scshape NOvA} collaboration, R.~Patterson, \emph{{The NOvA Experiment: Status
  and Outlook}},
  \href{http://dx.doi.org/10.1016/j.nuclphysbps.2013.04.005}{\emph{Nucl.Phys.Proc.Suppl.}
  {\bf 235-236} (2013) 151--157}, [\href{http://arxiv.org/abs/1209.0716}{{\tt
  1209.0716}}].

\bibitem{Abe:2011ks}
{\scshape T2K} collaboration, K.~Abe et~al., \emph{{The T2K Experiment}},
  \href{http://dx.doi.org/10.1016/j.nima.2011.06.067}{\emph{Nucl.Instrum.Meth.}
  {\bf A659} (2011) 106--135}, [\href{http://arxiv.org/abs/1106.1238}{{\tt
  1106.1238}}].

\bibitem{Aartsen:2013aaa}
{\scshape IceCube, PINGU} collaboration, M.~Aartsen et~al., \emph{{PINGU
  Sensitivity to the Neutrino Mass Hierarchy}},
  \href{http://arxiv.org/abs/1306.5846}{{\tt 1306.5846}}.

\bibitem{He:2014zwa}
{\scshape JUNO} collaboration, M.~He, \emph{{Jiangmen Underground Neutrino
  Observatory}},  \href{http://arxiv.org/abs/1412.4195}{{\tt 1412.4195}}.

\bibitem{deGouvea:2005hk}
A.~de~Gouvea, J.~Jenkins and B.~Kayser, \emph{{Neutrino mass hierarchy, vacuum
  oscillations, and vanishing |U(e3)|}},
  \href{http://dx.doi.org/10.1103/PhysRevD.71.113009}{\emph{Phys.Rev.} {\bf
  D71} (2005) 113009}, [\href{http://arxiv.org/abs/hep-ph/0503079}{{\tt
  hep-ph/0503079}}].

\bibitem{Nunokawa:2005nx}
H.~Nunokawa, S.~J. Parke and R.~Zukanovich~Funchal, \emph{{Another possible way
  to determine the neutrino mass hierarchy}},
  \href{http://dx.doi.org/10.1103/PhysRevD.72.013009}{\emph{Phys.Rev.} {\bf
  D72} (2005) 013009}, [\href{http://arxiv.org/abs/hep-ph/0503283}{{\tt
  hep-ph/0503283}}].

\bibitem{Avignone:2007fu}
I.~Avignone, Frank~T., S.~R. Elliott and J.~Engel, \emph{{Double Beta Decay,
  Majorana Neutrinos, and Neutrino Mass}},
  \href{http://dx.doi.org/10.1103/RevModPhys.80.481}{\emph{Rev.Mod.Phys.} {\bf
  80} (2008) 481--516}, [\href{http://arxiv.org/abs/0708.1033}{{\tt
  0708.1033}}].

\bibitem{deGouvea:2013zba}
A.~de~Gouvea and P.~Vogel, \emph{{Lepton Flavor and Number Conservation, and
  Physics Beyond the Standard Model}},
  \href{http://dx.doi.org/10.1016/j.ppnp.2013.03.006}{\emph{Prog.Part.Nucl.Phys.}
  {\bf 71} (2013) 75--92}, [\href{http://arxiv.org/abs/1303.4097}{{\tt
  1303.4097}}].

\bibitem{Adams:2013qkq}
{\scshape LBNE} collaboration, C.~Adams et~al., \emph{{The Long-Baseline
  Neutrino Experiment: Exploring Fundamental Symmetries of the Universe}},
  \href{http://arxiv.org/abs/1307.7335}{{\tt 1307.7335}}.

\bibitem{Kearns:2013lea}
{\scshape Hyper-Kamiokande Working Group} collaboration, E.~Kearns et~al.,
  \emph{{Hyper-Kamiokande Physics Opportunities}},
  \href{http://arxiv.org/abs/1309.0184}{{\tt 1309.0184}}.

\bibitem{Sakharov:1967dj}
A.~D. Sakharov, \emph{{Violation of CP Invariance, c Asymmetry, and Baryon
  Asymmetry of the Universe}},
  \href{http://dx.doi.org/10.1070/PU1991v034n05ABEH002497}{\emph{Pisma Zh.
  Eksp. Teor. Fiz.} {\bf 5} (1967) 32--35}.

\bibitem{Canetti:2012zc}
L.~Canetti, M.~Drewes and M.~Shaposhnikov, \emph{{Matter and Antimatter in the
  Universe}}, \href{http://dx.doi.org/10.1088/1367-2630/14/9/095012}{\emph{New
  J. Phys.} {\bf 14} (2012) 095012},
  [\href{http://arxiv.org/abs/1204.4186}{{\tt 1204.4186}}].

\bibitem{Langacker:2011bi}
P.~Langacker, \emph{{Neutrino Masses from the Top Down}},
  \href{http://dx.doi.org/10.1146/annurev-nucl-102711-094925}{\emph{Ann.Rev.Nucl.Part.Sci.}
  {\bf 62} (2012) 215--235}, [\href{http://arxiv.org/abs/1112.5992}{{\tt
  1112.5992}}].

\bibitem{deGouvea:2009fp}
A.~de~Gouvea, W.-C. Huang and J.~Jenkins, \emph{{Pseudo-Dirac Neutrinos in the
  New Standard Model}},
  \href{http://dx.doi.org/10.1103/PhysRevD.80.073007}{\emph{Phys.Rev.} {\bf
  D80} (2009) 073007}, [\href{http://arxiv.org/abs/0906.1611}{{\tt
  0906.1611}}].

\bibitem{Minkowski:1977sc}
P.~Minkowski, \emph{{$\mu \to e \gamma$ at a Rate of One Out of 1-Billion Muon
  Decays?}}, \href{http://dx.doi.org/10.1016/0370-2693(77)90435-X}{\emph{Phys.
  Lett.} {\bf B67} (1977) 421}.

\bibitem{Yanagida:1979as}
T.~Yanagida, \emph{{HORIZONTAL SYMMETRY AND MASSES OF NEUTRINOS}}, {\emph{Conf.
  Proc.} {\bf C7902131} (1979) 95}.

\bibitem{GellMann:1980vs}
M.~Gell-Mann, P.~Ramond and R.~Slansky, \emph{{Complex Spinors and Unified
  Theories}}, {\emph{Conf. Proc.} {\bf C790927} (1979) 315--321},
  [\href{http://arxiv.org/abs/1306.4669}{{\tt 1306.4669}}].

\bibitem{Glashow:1979nm}
S.~L. Glashow, \emph{{The Future of Elementary Particle Physics}},
  \href{http://dx.doi.org/10.1007/978-1-4684-7197-7_15}{\emph{NATO Sci. Ser. B}
  {\bf 61} (1980) 687}.

\bibitem{Mohapatra:1979ia}
R.~N. Mohapatra and G.~Senjanovic, \emph{{Neutrino Mass and Spontaneous Parity
  Violation}}, \href{http://dx.doi.org/10.1103/PhysRevLett.44.912}{\emph{Phys.
  Rev. Lett.} {\bf 44} (1980) 912}.

\bibitem{Schechter:1980gr}
J.~Schechter and J.~W.~F. Valle, \emph{{Neutrino Masses in $SU(2) \times U(1)$
  Theories}}, \href{http://dx.doi.org/10.1103/PhysRevD.22.2227}{\emph{Phys.
  Rev.} {\bf D22} (1980) 2227}.

\bibitem{Abazajian:2012ys}
K.~N. Abazajian et~al., \emph{{Light Sterile Neutrinos: A White Paper}},
  \href{http://arxiv.org/abs/1204.5379}{{\tt 1204.5379}}.

\bibitem{Conrad:2013mka}
J.~M. Conrad, W.~C. Louis and M.~H. Shaevitz, \emph{{The LSND and MiniBooNE
  Oscillation Searches at High $\Delta m^2$}},
  \href{http://dx.doi.org/10.1146/annurev-nucl-102711-094957}{\emph{Ann.Rev.Nucl.Part.Sci.}
  {\bf 63} (2013) 45--67}, [\href{http://arxiv.org/abs/1306.6494}{{\tt
  1306.6494}}].

\bibitem{Kopp:2013vaa}
J.~Kopp, P.~A.~N. Machado, M.~Maltoni and T.~Schwetz, \emph{{Sterile Neutrino
  Oscillations: The Global Picture}},
  \href{http://dx.doi.org/10.1007/JHEP05(2013)050}{\emph{JHEP} {\bf 1305}
  (2013) 050}, [\href{http://arxiv.org/abs/1303.3011}{{\tt 1303.3011}}].

\bibitem{Giunti:2013aea}
C.~Giunti, M.~Laveder, Y.~Li and H.~Long, \emph{{Pragmatic View of
  Short-Baseline Neutrino Oscillations}},
  \href{http://dx.doi.org/10.1103/PhysRevD.88.073008}{\emph{Phys.Rev.} {\bf
  D88} (2013) 073008}, [\href{http://arxiv.org/abs/1308.5288}{{\tt
  1308.5288}}].

\bibitem{Conrad:2012qt}
J.~Conrad, C.~Ignarra, G.~Karagiorgi, M.~Shaevitz and J.~Spitz, \emph{{Sterile
  Neutrino Fits to Short Baseline Neutrino Oscillation Measurements}},
  \href{http://dx.doi.org/10.1155/2013/163897}{\emph{Adv.High Energy Phys.}
  {\bf 2013} (2013) 163897}, [\href{http://arxiv.org/abs/1207.4765}{{\tt
  1207.4765}}].

\bibitem{Ade:2013zuv}
{\scshape Planck Collaboration} collaboration, P.~Ade et~al., \emph{{Planck
  2013 results. XVI. Cosmological parameters}},
  \href{http://dx.doi.org/10.1051/0004-6361/201321591}{\emph{Astron.Astrophys.}
  (2014) }, [\href{http://arxiv.org/abs/1303.5076}{{\tt 1303.5076}}].

\bibitem{Mirizzi:2013kva}
A.~Mirizzi, G.~Mangano, N.~Saviano, E.~Borriello, C.~Giunti et~al., \emph{{The
  strongest bounds on active-sterile neutrino mixing after Planck data}},
  \href{http://dx.doi.org/10.1016/j.physletb.2013.08.015}{\emph{Phys. Lett.}
  {\bf B726} (2013) 8--14}, [\href{http://arxiv.org/abs/1303.5368}{{\tt
  1303.5368}}].

\bibitem{deGouvea:2011zz}
A.~de~Gouvea and W.-C. Huang, \emph{{Constraining the (Low-Energy) Type-I
  Seesaw}},
  \href{http://dx.doi.org/10.1103/PhysRevD.85.053006}{\emph{Phys.Rev.} {\bf
  D85} (2012) 053006}, [\href{http://arxiv.org/abs/1110.6122}{{\tt
  1110.6122}}].

\bibitem{Donini:2012tt}
A.~Donini, P.~Hernandez, J.~Lopez-Pavon, M.~Maltoni and T.~Schwetz, \emph{{The
  minimal 3+2 neutrino model versus oscillation anomalies}},
  \href{http://dx.doi.org/10.1007/JHEP07(2012)161}{\emph{JHEP} {\bf 1207}
  (2012) 161}, [\href{http://arxiv.org/abs/1205.5230}{{\tt 1205.5230}}].

\bibitem{Langacker:1998ut}
P.~Langacker, \emph{{A Mechanism for ordinary sterile neutrino mixing}},
  \href{http://dx.doi.org/10.1103/PhysRevD.58.093017}{\emph{Phys.Rev.} {\bf
  D58} (1998) 093017}, [\href{http://arxiv.org/abs/hep-ph/9805281}{{\tt
  hep-ph/9805281}}].

\bibitem{Sayre:2005yh}
J.~Sayre, S.~Wiesenfeldt and S.~Willenbrock, \emph{{Sterile neutrinos and
  global symmetries}},
  \href{http://dx.doi.org/10.1103/PhysRevD.72.015001}{\emph{Phys.Rev.} {\bf
  D72} (2005) 015001}, [\href{http://arxiv.org/abs/hep-ph/0504198}{{\tt
  hep-ph/0504198}}].

\bibitem{Atre:2009rg}
A.~Atre, T.~Han, S.~Pascoli and B.~Zhang, \emph{{The Search for Heavy Majorana
  Neutrinos}},
  \href{http://dx.doi.org/10.1088/1126-6708/2009/05/030}{\emph{JHEP} {\bf 0905}
  (2009) 030}, [\href{http://arxiv.org/abs/0901.3589}{{\tt 0901.3589}}].

\bibitem{Kusenko:2009up}
A.~Kusenko, \emph{{Sterile neutrinos: The Dark side of the light fermions}},
  \href{http://dx.doi.org/10.1016/j.physrep.2009.07.004}{\emph{Phys. Rept.}
  {\bf 481} (2009) 1--28}, [\href{http://arxiv.org/abs/0906.2968}{{\tt
  0906.2968}}].

\bibitem{Shrock:1980ct}
R.~E. Shrock, \emph{{General Theory of Weak Leptonic and Semileptonic Decays.
  1. Leptonic Pseudoscalar Meson Decays, with Associated Tests For, and Bounds
  on, Neutrino Masses and Lepton Mixing}},
  \href{http://dx.doi.org/10.1103/PhysRevD.24.1232}{\emph{Phys. Rev.} {\bf D24}
  (1981) 1232}.

\bibitem{Shrock:1981wq}
R.~E. Shrock, \emph{{General Theory of Weak Processes Involving Neutrinos. 2.
  Pure Leptonic Decays}},
  \href{http://dx.doi.org/10.1103/PhysRevD.24.1275}{\emph{Phys. Rev.} {\bf D24}
  (1981) 1275}.

\bibitem{Barry:2011wb}
J.~Barry, W.~Rodejohann and H.~Zhang, \emph{{Light Sterile Neutrinos: Models
  and Phenomenology}},
  \href{http://dx.doi.org/10.1007/JHEP07(2011)091}{\emph{JHEP} {\bf 1107}
  (2011) 091}, [\href{http://arxiv.org/abs/1105.3911}{{\tt 1105.3911}}].

\bibitem{Faessler:2014kka}
A.~Faessler, M.~Gonzalez, S.~Kovalenko and F.~Simkovic, \emph{{Arbitrary mass
  Majorana neutrinos in neutrinoless double beta decay}},
  \href{http://dx.doi.org/10.1103/PhysRevD.90.096010}{\emph{Phys.Rev.} {\bf
  D90} (2014) 096010}, [\href{http://arxiv.org/abs/1408.6077}{{\tt
  1408.6077}}].

\bibitem{Escrihuela:2015wra}
F.~J. Escrihuela, D.~V. Forero, O.~G. Miranda, M.~Tortola and J.~W.~F. Valle,
  \emph{{On the description of non-unitary neutrino mixing}},
  \href{http://dx.doi.org/10.1103/PhysRevD.92.053009}{\emph{Phys. Rev.} {\bf
  D92} (2015) 053009}, [\href{http://arxiv.org/abs/1503.08879}{{\tt
  1503.08879}}].

\bibitem{Dodelson:1993je}
S.~Dodelson and L.~M. Widrow, \emph{{Sterile-neutrinos as dark matter}},
  \href{http://dx.doi.org/10.1103/PhysRevLett.72.17}{\emph{Phys.Rev.Lett.} {\bf
  72} (1994) 17--20}, [\href{http://arxiv.org/abs/hep-ph/9303287}{{\tt
  hep-ph/9303287}}].

\bibitem{deVega:2009ku}
H.~J. de~Vega and N.~G. Sanchez, \emph{{Model independent analysis of dark
  matter points to a particle mass at the keV scale}},
  \href{http://dx.doi.org/10.1111/j.1365-2966.2010.16319.x}{\emph{Mon. Not.
  Roy. Astron. Soc.} {\bf 404} (2010) 885},
  [\href{http://arxiv.org/abs/0901.0922}{{\tt 0901.0922}}].

\bibitem{deVega:2010yk}
H.~J. de~Vega, P.~Salucci and N.~G. Sanchez, \emph{{The mass of the dark matter
  particle from theory and observations}},
  \href{http://dx.doi.org/10.1016/j.newast.2012.04.001}{\emph{New Astron.} {\bf
  17} (2012) 653--666}, [\href{http://arxiv.org/abs/1004.1908}{{\tt
  1004.1908}}].

\bibitem{deVega:2011gg}
H.~J. de~Vega and N.~G. Sanchez, \emph{{Cosmological evolution of warm dark
  matter fluctuations I: Efficient computational framework with Volterra
  integral equations}},
  \href{http://dx.doi.org/10.1103/PhysRevD.85.043516}{\emph{Phys. Rev.} {\bf
  D85} (2012) 043516}, [\href{http://arxiv.org/abs/1111.0290}{{\tt
  1111.0290}}].

\bibitem{deVega:2011gs}
H.~J. de~Vega and N.~G. Sanchez, \emph{{Cosmological evolution of warm dark
  matter fluctuations II: Solution from small to large scales and keV sterile
  neutrinos}}, \href{http://dx.doi.org/10.1103/PhysRevD.85.043517}{\emph{Phys.
  Rev.} {\bf D85} (2012) 043517}, [\href{http://arxiv.org/abs/1111.0300}{{\tt
  1111.0300}}].

\bibitem{Destri:2012yn}
C.~Destri, H.~J. de~Vega and N.~G. Sanchez, \emph{{Fermionic warm dark matter
  produces galaxy cores in the observed scales because of quantum mechanics}},
  \href{http://dx.doi.org/10.1016/j.newast.2012.12.003}{\emph{New Astron.} {\bf
  22} (2013) 39--50}, [\href{http://arxiv.org/abs/1204.3090}{{\tt 1204.3090}}].

\bibitem{Destri:2013pt}
C.~Destri, H.~J. de~Vega and N.~G. Sanchez, \emph{{Quantum WDM fermions and
  gravitation determine the observed galaxy structures}},
  \href{http://dx.doi.org/10.1016/j.astropartphys.2013.04.004}{\emph{Astropart.
  Phys.} {\bf 46} (2013) 14--22}, [\href{http://arxiv.org/abs/1301.1864}{{\tt
  1301.1864}}].

\bibitem{deVega:2013jfy}
H.~J. de~Vega, P.~Salucci and N.~G. Sanchez, \emph{{Observational rotation
  curves and density profiles versus the Thomas-Fermi galaxy structure
  theory}}, \href{http://dx.doi.org/10.1093/mnras/stu972}{\emph{Mon. Not. Roy.
  Astron. Soc.} {\bf 442} (2014) 2717--2727},
  [\href{http://arxiv.org/abs/1309.2290}{{\tt 1309.2290}}].

\bibitem{Destri:2013hha}
C.~Destri, H.~J. de~Vega and N.~G. Sanchez, \emph{{Warm dark matter primordial
  spectra and the onset of structure formation at redshift $z$}},
  \href{http://dx.doi.org/10.1103/PhysRevD.88.083512}{\emph{Phys. Rev.} {\bf
  D88} (2013) 083512}, [\href{http://arxiv.org/abs/1308.1109}{{\tt
  1308.1109}}].

\bibitem{Boyarsky:2009ix}
A.~Boyarsky, O.~Ruchayskiy and M.~Shaposhnikov, \emph{{The Role of sterile
  neutrinos in cosmology and astrophysics}},
  \href{http://dx.doi.org/10.1146/annurev.nucl.010909.083654}{\emph{Ann.Rev.Nucl.Part.Sci.}
  {\bf 59} (2009) 191--214}, [\href{http://arxiv.org/abs/0901.0011}{{\tt
  0901.0011}}].

\bibitem{Merle:2012xq}
A.~Merle, \emph{{Constraining models for keV sterile neutrinos by
  quasi-degenerate active neutrinos}},
  \href{http://dx.doi.org/10.1103/PhysRevD.86.121701}{\emph{Phys. Rev.} {\bf
  D86} (2012) 121701}, [\href{http://arxiv.org/abs/1210.6036}{{\tt
  1210.6036}}].

\bibitem{Asaka:2005pn}
T.~Asaka and M.~Shaposhnikov, \emph{{The nuMSM, dark matter and baryon
  asymmetry of the universe}},
  \href{http://dx.doi.org/10.1016/j.physletb.2005.06.020}{\emph{Phys.Lett.}
  {\bf B620} (2005) 17--26}, [\href{http://arxiv.org/abs/hep-ph/0505013}{{\tt
  hep-ph/0505013}}].

\bibitem{Asaka:2005an}
T.~Asaka, S.~Blanchet and M.~Shaposhnikov, \emph{{The nuMSM, dark matter and
  neutrino masses}},
  \href{http://dx.doi.org/10.1016/j.physletb.2005.09.070}{\emph{Phys.Lett.}
  {\bf B 631} (2005) 151--156},
  [\href{http://arxiv.org/abs/hep-ph/0503065}{{\tt hep-ph/0503065}}].

\bibitem{Bulbul:2014sua}
E.~Bulbul, M.~Markevitch, A.~Foster, R.~K. Smith, M.~Loewenstein and S.~W.
  Randall, \emph{{Detection of An Unidentified Emission Line in the Stacked
  X-ray spectrum of Galaxy Clusters}},
  \href{http://dx.doi.org/10.1088/0004-637X/789/1/13}{\emph{Astrophys. J.} {\bf
  789} (2014) 13}, [\href{http://arxiv.org/abs/1402.2301}{{\tt 1402.2301}}].

\bibitem{Boyarsky:2014jta}
A.~Boyarsky, O.~Ruchayskiy, D.~Iakubovskyi and J.~Franse, \emph{{Unidentified
  Line in X-Ray Spectra of the Andromeda Galaxy and Perseus Galaxy Cluster}},
  \href{http://dx.doi.org/10.1103/PhysRevLett.113.251301}{\emph{Phys. Rev.
  Lett.} {\bf 113} (2014) 251301}, [\href{http://arxiv.org/abs/1402.4119}{{\tt
  1402.4119}}].

\bibitem{Bezrukov:2005mx}
F.~L. Bezrukov, \emph{{nu MSM-predictions for neutrinoless double beta decay}},
  \href{http://dx.doi.org/10.1103/PhysRevD.72.071303}{\emph{Phys. Rev.} {\bf
  D72} (2005) 071303}, [\href{http://arxiv.org/abs/hep-ph/0505247}{{\tt
  hep-ph/0505247}}].

\bibitem{Asaka:2011pb}
T.~Asaka, S.~Eijima and H.~Ishida, \emph{{Mixing of Active and Sterile
  Neutrinos}}, \href{http://dx.doi.org/10.1007/JHEP04(2011)011}{\emph{JHEP}
  {\bf 04} (2011) 011}, [\href{http://arxiv.org/abs/1101.1382}{{\tt
  1101.1382}}].

\bibitem{Merle:2013ibc}
A.~Merle and V.~Niro, \emph{{Influence of a keV sterile neutrino on
  neutrinoless double beta decay: How things changed in recent years}},
  \href{http://dx.doi.org/10.1103/PhysRevD.88.113004}{\emph{Phys. Rev.} {\bf
  D88} (2013) 113004}, [\href{http://arxiv.org/abs/1302.2032}{{\tt
  1302.2032}}].

\bibitem{Yanagida:1980xy}
T.~Yanagida, \emph{{Horizontal Symmetry and Masses of Neutrinos}},
  \href{http://dx.doi.org/10.1143/PTP.64.1103}{\emph{Prog. Theor. Phys.} {\bf
  64} (1980) 1103}.

\bibitem{Ma:1998dn}
E.~Ma, \emph{{Pathways to naturally small neutrino masses}},
  \href{http://dx.doi.org/10.1103/PhysRevLett.81.1171}{\emph{Phys. Rev. Lett.}
  {\bf 81} (1998) 1171--1174}, [\href{http://arxiv.org/abs/hep-ph/9805219}{{\tt
  hep-ph/9805219}}].

\bibitem{Magg:1980ut}
M.~Magg and C.~Wetterich, \emph{{Neutrino Mass Problem and Gauge Hierarchy}},
  \href{http://dx.doi.org/10.1016/0370-2693(80)90825-4}{\emph{Phys. Lett.} {\bf
  B94} (1980) 61}.

\bibitem{Cheng:1980qt}
T.~P. Cheng and L.-F. Li, \emph{{Neutrino Masses, Mixings and Oscillations in
  SU(2) x U(1) Models of Electroweak Interactions}},
  \href{http://dx.doi.org/10.1103/PhysRevD.22.2860}{\emph{Phys. Rev.} {\bf D22}
  (1980) 2860}.

\bibitem{Lazarides:1980nt}
G.~Lazarides, Q.~Shafi and C.~Wetterich, \emph{{Proton Lifetime and Fermion
  Masses in an SO(10) Model}},
  \href{http://dx.doi.org/10.1016/0550-3213(81)90354-0}{\emph{Nucl. Phys.} {\bf
  B181} (1981) 287--300}.

\bibitem{Mohapatra:1980yp}
R.~N. Mohapatra and G.~Senjanovic, \emph{{Neutrino Masses and Mixings in Gauge
  Models with Spontaneous Parity Violation}},
  \href{http://dx.doi.org/10.1103/PhysRevD.23.165}{\emph{Phys. Rev.} {\bf D23}
  (1981) 165}.

\bibitem{Foot:1988aq}
R.~Foot, H.~Lew, X.~G. He and G.~C. Joshi, \emph{{Seesaw Neutrino Masses
  Induced by a Triplet of Leptons}},
  \href{http://dx.doi.org/10.1007/BF01415558}{\emph{Z. Phys.} {\bf C44} (1989)
  441}.

\bibitem{Froggatt:1978nt}
C.~D. Froggatt and H.~B. Nielsen, \emph{{Hierarchy of Quark Masses, Cabibbo
  Angles and CP Violation}},
  \href{http://dx.doi.org/10.1016/0550-3213(79)90316-X}{\emph{Nucl. Phys.} {\bf
  B147} (1979) 277}.

\bibitem{Chikashige:1980ui}
Y.~Chikashige, R.~N. Mohapatra and R.~D. Peccei, \emph{{Are There Real
  Goldstone Bosons Associated with Broken Lepton Number?}},
  \href{http://dx.doi.org/10.1016/0370-2693(81)90011-3}{\emph{Phys. Lett.} {\bf
  B98} (1981) 265}.

\bibitem{Gelmini:1980re}
G.~B. Gelmini and M.~Roncadelli, \emph{{Left-Handed Neutrino Mass Scale and
  Spontaneously Broken Lepton Number}},
  \href{http://dx.doi.org/10.1016/0370-2693(81)90559-1}{\emph{Phys. Lett.} {\bf
  B99} (1981) 411}.

\bibitem{Wyler:1982dd}
D.~Wyler and L.~Wolfenstein, \emph{{Massless Neutrinos in Left-Right Symmetric
  Models}}, \href{http://dx.doi.org/10.1016/0550-3213(83)90482-0}{\emph{Nucl.
  Phys.} {\bf B218} (1983) 205}.

\bibitem{GonzalezGarcia:1988rw}
M.~C. Gonzalez-Garcia and J.~W.~F. Valle, \emph{{Fast Decaying Neutrinos and
  Observable Flavor Violation in a New Class of Majoron Models}},
  \href{http://dx.doi.org/10.1016/0370-2693(89)91131-3}{\emph{Phys. Lett.} {\bf
  B216} (1989) 360}.

\bibitem{Branco:1996bq}
G.~C. Branco, W.~Grimus and L.~Lavoura, \emph{{Relating the scalar flavor
  changing neutral couplings to the CKM matrix}},
  \href{http://dx.doi.org/10.1016/0370-2693(96)00494-7}{\emph{Phys. Lett.} {\bf
  B380} (1996) 119--126}, [\href{http://arxiv.org/abs/hep-ph/9601383}{{\tt
  hep-ph/9601383}}].

\bibitem{Abada:2007ux}
A.~Abada, C.~Biggio, F.~Bonnet, M.~B. Gavela and T.~Hambye, \emph{{Low energy
  effects of neutrino masses}},
  \href{http://dx.doi.org/10.1088/1126-6708/2007/12/061}{\emph{JHEP} {\bf 12}
  (2007) 061}, [\href{http://arxiv.org/abs/0707.4058}{{\tt 0707.4058}}].

\bibitem{Gavela:2009cd}
M.~B. Gavela, T.~Hambye, D.~Hernandez and P.~Hernandez, \emph{{Minimal Flavour
  Seesaw Models}},
  \href{http://dx.doi.org/10.1088/1126-6708/2009/09/038}{\emph{JHEP} {\bf 09}
  (2009) 038}, [\href{http://arxiv.org/abs/0906.1461}{{\tt 0906.1461}}].

\bibitem{Zee:1980ai}
A.~Zee, \emph{{A Theory of Lepton Number Violation, Neutrino Majorana Mass, and
  Oscillation}},
  \href{http://dx.doi.org/10.1016/0370-2693(80)90349-4}{\emph{Phys. Lett.} {\bf
  B93} (1980) 389}.

\bibitem{Witten:1979nr}
E.~Witten, \emph{{Neutrino Masses in the Minimal O(10) Theory}},
  \href{http://dx.doi.org/10.1016/0370-2693(80)90666-8}{\emph{Phys. Lett.} {\bf
  B91} (1980) 81}.

\bibitem{Zee:1985id}
A.~Zee, \emph{{Quantum Numbers of Majorana Neutrino Masses}},
  \href{http://dx.doi.org/10.1016/0550-3213(86)90475-X}{\emph{Nucl. Phys.} {\bf
  B264} (1986) 99}.

\bibitem{Babu:1988ki}
K.~S. Babu, \emph{{Model of 'Calculable' Majorana Neutrino Masses}},
  \href{http://dx.doi.org/10.1016/0370-2693(88)91584-5}{\emph{Phys. Lett.} {\bf
  B203} (1988) 132}.

\bibitem{Ma:2006km}
E.~Ma, \emph{{Verifiable radiative seesaw mechanism of neutrino mass and dark
  matter}}, \href{http://dx.doi.org/10.1103/PhysRevD.73.077301}{\emph{Phys.
  Rev.} {\bf D73} (2006) 077301},
  [\href{http://arxiv.org/abs/hep-ph/0601225}{{\tt hep-ph/0601225}}].

\bibitem{FileviezPerez:2009ud}
P.~Fileviez~Perez and M.~B. Wise, \emph{{On the Origin of Neutrino Masses}},
  \href{http://dx.doi.org/10.1103/PhysRevD.80.053006}{\emph{Phys. Rev.} {\bf
  D80} (2009) 053006}, [\href{http://arxiv.org/abs/0906.2950}{{\tt
  0906.2950}}].

\bibitem{Khoze:2013oga}
V.~V. Khoze and G.~Ro, \emph{{Leptogenesis and Neutrino Oscillations in the
  Classically Conformal Standard Model with the Higgs Portal}},
  \href{http://dx.doi.org/10.1007/JHEP10(2013)075}{\emph{JHEP} {\bf 10} (2013)
  075}, [\href{http://arxiv.org/abs/1307.3764}{{\tt 1307.3764}}].

\bibitem{King:2014uha}
S.~F. King, A.~Merle and L.~Panizzi, \emph{{Effective theory of a doubly
  charged singlet scalar: complementarity of neutrino physics and the LHC}},
  \href{http://dx.doi.org/10.1007/JHEP11(2014)124}{\emph{JHEP} {\bf 11} (2014)
  124}, [\href{http://arxiv.org/abs/1406.4137}{{\tt 1406.4137}}].

\bibitem{Geib:2015tvt}
T.~Geib, S.~F. King, A.~Merle, J.~M. No and L.~Panizzi, \emph{{Probing the
  Origin of Neutrino Masses and Mixings via Doubly Charged Scalars:
  Complementarity of the Intensity and the Energy Frontiers}},
  \href{http://arxiv.org/abs/1512.04391}{{\tt 1512.04391}}.

\bibitem{King:2014nza}
S.~F. King, A.~Merle, S.~Morisi, Y.~Shimizu and M.~Tanimoto, \emph{{Neutrino
  Mass and Mixing: from Theory to Experiment}},
  \href{http://dx.doi.org/10.1088/1367-2630/16/4/045018}{\emph{New J. Phys.}
  {\bf 16} (2014) 045018}, [\href{http://arxiv.org/abs/1402.4271}{{\tt
  1402.4271}}].

\bibitem{Appelquist2002204}
T.~Appelquist and R.~Shrock, \emph{Neutrino masses in theories with dynamical
  electroweak symmetry breaking},
  \href{http://dx.doi.org/http://dx.doi.org/10.1016/S0370-2693(02)02854-X}{\emph{Physics
  Letters B} {\bf 548} (2002) 204 -- 214}.

\bibitem{PhysRevLett.90.201801}
T.~Appelquist and R.~Shrock, \emph{Dynamical symmetry breaking of extended
  gauge symmetries},
  \href{http://dx.doi.org/10.1103/PhysRevLett.90.201801}{\emph{Phys. Rev.
  Lett.} {\bf 90} (May, 2003) 201801}.

\bibitem{PhysRevD.69.015002}
T.~Appelquist, M.~Piai and R.~Shrock, \emph{Fermion masses and mixing in
  extended technicolor models},
  \href{http://dx.doi.org/10.1103/PhysRevD.69.015002}{\emph{Phys. Rev. D} {\bf
  69} (Jan, 2004) 015002}.

\bibitem{Barry:2010yk}
J.~Barry and W.~Rodejohann, \emph{{Neutrino Mass Sum-rules in Flavor Symmetry
  Models}},
  \href{http://dx.doi.org/10.1016/j.nuclphysb.2010.08.015}{\emph{Nucl. Phys.}
  {\bf B842} (2011) 33--50}, [\href{http://arxiv.org/abs/1007.5217}{{\tt
  1007.5217}}].

\bibitem{Dorame:2011eb}
L.~Dorame, D.~Meloni, S.~Morisi, E.~Peinado and J.~W.~F. Valle,
  \emph{{Constraining Neutrinoless Double Beta Decay}},
  \href{http://dx.doi.org/10.1016/j.nuclphysb.2012.04.003}{\emph{Nucl. Phys.}
  {\bf B861} (2012) 259--270}, [\href{http://arxiv.org/abs/1111.5614}{{\tt
  1111.5614}}].

\bibitem{King:2013psa}
S.~F. King, A.~Merle and A.~J. Stuart, \emph{{The Power of Neutrino Mass Sum
  Rules for Neutrinoless Double Beta Decay Experiments}},
  \href{http://dx.doi.org/10.1007/JHEP12(2013)005}{\emph{JHEP} {\bf 12} (2013)
  005}, [\href{http://arxiv.org/abs/1307.2901}{{\tt 1307.2901}}].

\bibitem{Agostini:2015dna}
M.~Agostini, A.~Merle and K.~Zuber, \emph{{Probing flavor models with
  Ge-76-based experiments on neutrinoless double-beta decay}},
  \href{http://arxiv.org/abs/1506.06133}{{\tt 1506.06133}}.

\bibitem{Petcov:2004rk}
S.~T. Petcov and W.~Rodejohann, \emph{{Flavor symmetry $L_e - L_{\mu} -
  L{\tau}$, atmospheric neutrino mixing and CP violation in the lepton
  sector}}, \href{http://dx.doi.org/10.1103/PhysRevD.71.073002}{\emph{Phys.
  Rev.} {\bf D71} (2005) 073002},
  [\href{http://arxiv.org/abs/hep-ph/0409135}{{\tt hep-ph/0409135}}].

\bibitem{Marzocca:2013cr}
D.~Marzocca, S.~T. Petcov, A.~Romanino and M.~C. Sevilla, \emph{{Nonzero
  $|U_{e3}|$ from Charged Lepton Corrections and the Atmospheric Neutrino
  Mixing Angle}}, \href{http://dx.doi.org/10.1007/JHEP05(2013)073}{\emph{JHEP}
  {\bf 05} (2013) 073}, [\href{http://arxiv.org/abs/1302.0423}{{\tt
  1302.0423}}].

\bibitem{Ballett:2013wya}
P.~Ballett, S.~F. King, C.~Luhn, S.~Pascoli and M.~A. Schmidt, \emph{{Testing
  atmospheric mixing sum rules at precision neutrino facilities}},
  \href{http://dx.doi.org/10.1103/PhysRevD.89.016016}{\emph{Phys. Rev.} {\bf
  D89} (2014) 016016}, [\href{http://arxiv.org/abs/1308.4314}{{\tt
  1308.4314}}].

\bibitem{Harrison:1999cf}
P.~F. Harrison, D.~H. Perkins and W.~G. Scott, \emph{{A Redetermination of the
  neutrino mass squared difference in tri - maximal mixing with terrestrial
  matter effects}},
  \href{http://dx.doi.org/10.1016/S0370-2693(99)00438-4}{\emph{Phys. Lett.}
  {\bf B458} (1999) 79--92}, [\href{http://arxiv.org/abs/hep-ph/9904297}{{\tt
  hep-ph/9904297}}].

\bibitem{Harrison:2002er}
P.~F. Harrison, D.~H. Perkins and W.~G. Scott, \emph{{Tri-bimaximal mixing and
  the neutrino oscillation data}},
  \href{http://dx.doi.org/10.1016/S0370-2693(02)01336-9}{\emph{Phys. Lett.}
  {\bf B530} (2002) 167}, [\href{http://arxiv.org/abs/hep-ph/0202074}{{\tt
  hep-ph/0202074}}].

\bibitem{Ma:2004zv}
E.~Ma, \emph{{A(4) symmetry and neutrinos with very different masses}},
  \href{http://dx.doi.org/10.1103/PhysRevD.70.031901}{\emph{Phys. Rev.} {\bf
  D70} (2004) 031901}, [\href{http://arxiv.org/abs/hep-ph/0404199}{{\tt
  hep-ph/0404199}}].

\bibitem{Altarelli:2005yx}
G.~Altarelli and F.~Feruglio, \emph{{Tri-bimaximal neutrino mixing, A(4) and
  the modular symmetry}},
  \href{http://dx.doi.org/10.1016/j.nuclphysb.2006.02.015}{\emph{Nucl. Phys.}
  {\bf B741} (2006) 215--235}, [\href{http://arxiv.org/abs/hep-ph/0512103}{{\tt
  hep-ph/0512103}}].

\bibitem{Xing:2002sw}
Z.-z. Xing, \emph{{Nearly tri bimaximal neutrino mixing and CP violation}},
  \href{http://dx.doi.org/10.1016/S0370-2693(02)01649-0}{\emph{Phys. Lett.}
  {\bf B533} (2002) 85--93}, [\href{http://arxiv.org/abs/hep-ph/0204049}{{\tt
  hep-ph/0204049}}].

\bibitem{Hall:1999sn}
L.~J. Hall, H.~Murayama and N.~Weiner, \emph{{Neutrino mass anarchy}},
  \href{http://dx.doi.org/10.1103/PhysRevLett.84.2572}{\emph{Phys. Rev. Lett.}
  {\bf 84} (2000) 2572--2575}, [\href{http://arxiv.org/abs/hep-ph/9911341}{{\tt
  hep-ph/9911341}}].

\bibitem{Weinberg:1979sa}
S.~Weinberg, \emph{{Baryon and Lepton Nonconserving Processes}},
  \href{http://dx.doi.org/10.1103/PhysRevLett.43.1566}{\emph{Phys. Rev. Lett.}
  {\bf 43} (1979) 1566--1570}.

\bibitem{Mohapatra:1986bd}
R.~Mohapatra and J.~W.~F. Valle, \emph{{Neutrino Mass and Baryon Number
  Nonconservation in Superstring Models}},
  \href{http://dx.doi.org/10.1103/PhysRevD.34.1642}{\emph{Phys. Rev.} {\bf D34}
  (1986) 1642}.

\bibitem{Malinsky:2005bi}
M.~Malinsky, J.~C. Romao and J.~W.~F. Valle, \emph{{Novel supersymmetric SO(10)
  seesaw mechanism}},
  \href{http://dx.doi.org/10.1103/PhysRevLett.95.161801}{\emph{Phys. Rev.
  Lett.} {\bf 95} (2005) 161801},
  [\href{http://arxiv.org/abs/hep-ph/0506296}{{\tt hep-ph/0506296}}].

\bibitem{Shaposhnikov:2006nn}
M.~Shaposhnikov, \emph{{A Possible symmetry of the $\nu$MSM}},
  \href{http://dx.doi.org/10.1016/j.nuclphysb.2006.11.003}{\emph{Nucl. Phys.}
  {\bf B763} (2007) 49--59}, [\href{http://arxiv.org/abs/hep-ph/0605047}{{\tt
  hep-ph/0605047}}].

\bibitem{Araki:2011zg}
T.~Araki and Y.~F. Li, \emph{{$Q_6$ flavor symmetry model for the extension of
  the minimal standard model by three right-handed sterile neutrinos}},
  \href{http://dx.doi.org/10.1103/PhysRevD.85.065016}{\emph{Phys. Rev.} {\bf
  D85} (2012) 065016}, [\href{http://arxiv.org/abs/1112.5819}{{\tt
  1112.5819}}].

\bibitem{Antusch:2014woa}
S.~Antusch and O.~Fischer, \emph{{Non-unitarity of the leptonic mixing matrix:
  Present bounds and future sensitivities}},
  \href{http://dx.doi.org/10.1007/JHEP10(2014)094}{\emph{JHEP} {\bf 10} (2014)
  094}, [\href{http://arxiv.org/abs/1407.6607}{{\tt 1407.6607}}].

\bibitem{Fixsen:1996nj}
D.~J. Fixsen, E.~S. Cheng, J.~M. Gales, J.~C. Mather, R.~A. Shafer et~al.,
  \emph{{The Cosmic Microwave Background spectrum from the full COBE FIRAS data
  set}}, \href{http://dx.doi.org/10.1086/178173}{\emph{Astrophys. J.} {\bf 473}
  (1996) 576}, [\href{http://arxiv.org/abs/astro-ph/9605054}{{\tt
  astro-ph/9605054}}].

\bibitem{Fixsen:2009ug}
D.~J. Fixsen, \emph{{The Temperature of the Cosmic Microwave Background}},
  \href{http://dx.doi.org/10.1088/0004-637X/707/2/916}{\emph{Astrophys. J.}
  {\bf 707} (2009) 916--920}, [\href{http://arxiv.org/abs/0911.1955}{{\tt
  0911.1955}}].

\bibitem{Adam:2015rua}
{\scshape Planck Collaboration} collaboration, R.~Adam et~al., \emph{{Planck
  2015 results. I. Overview of products and scientific results}},
  \href{http://arxiv.org/abs/1502.01582}{{\tt 1502.01582}}.

\bibitem{Ade:2015ava}
{\scshape Planck Collaboration} collaboration, P.~A.~R. Ade et~al.,
  \emph{{Planck 2015 results. XVII. Constraints on primordial
  non-Gaussianity}},  \href{http://arxiv.org/abs/1502.01592}{{\tt 1502.01592}}.

\bibitem{Bond:1997wr}
J.~R. Bond, G.~Efstathiou and M.~Tegmark, \emph{{Forecasting cosmic parameter
  errors from microwave background anisotropy experiments}}, {\emph{Mon. Not.
  Roy. Astron. Soc.} {\bf 291} (1997) L33--L41},
  [\href{http://arxiv.org/abs/astro-ph/9702100}{{\tt astro-ph/9702100}}].

\bibitem{Delubac:2014aqe}
{\scshape BOSS Collaboration} collaboration, T.~Delubac et~al., \emph{{Baryon
  Acoustic Oscillations in the Ly$\alpha$ forest of BOSS DR11 quasars}},
  \href{http://dx.doi.org/10.1051/0004-6361/201423969}{\emph{Astron.
  Astrophys.} {\bf 574} (2015) A59},
  [\href{http://arxiv.org/abs/1404.1801}{{\tt 1404.1801}}].

\bibitem{Riess:2011yx}
A.~G. Riess, L.~Macri, S.~Casertano, H.~Lampeitl, H.~C. Ferguson et~al.,
  \emph{{A 3
  Space Telescope and Wide Field Camera 3}},
  \href{http://dx.doi.org/10.1088/0004-637X/732/2/129,
  10.1088/0004-637X/730/2/119}{\emph{Astrophys. J.} {\bf 730} (2011) 119},
  [\href{http://arxiv.org/abs/1103.2976}{{\tt 1103.2976}}].

\bibitem{Efstathiou:2013via}
G.~Efstathiou, \emph{{H0 Revisited}},
  \href{http://dx.doi.org/10.1093/mnras/stu278}{\emph{Mon. Not. Roy. Astron.
  Soc.} {\bf 440} (2014) 1138--1152},
  [\href{http://arxiv.org/abs/1311.3461}{{\tt 1311.3461}}].

\bibitem{Riess:2016jrr}
A.~G. Riess et~al., \emph{{A 2.4\% Determination of the Local Value of the
  Hubble Constant}},  \href{http://arxiv.org/abs/1604.01424}{{\tt 1604.01424}}.

\bibitem{Heymans:2013fya}
C.~Heymans, E.~Grocutt, A.~Heavens, M.~Kilbinger, T.~D. Kitching et~al.,
  \emph{{CFHTLenS tomographic weak lensing cosmological parameter constraints:
  Mitigating the impact of intrinsic galaxy alignments}},
  \href{http://dx.doi.org/10.1093/mnras/stt601}{\emph{Mon. Not. Roy. Astron.
  Soc.} {\bf 432} (2013) 2433}, [\href{http://arxiv.org/abs/1303.1808}{{\tt
  1303.1808}}].

\bibitem{Ade:2015fva}
{\scshape Planck} collaboration, P.~A.~R. Ade et~al., \emph{{Planck 2015
  results. XXIV. Cosmology from Sunyaev-Zeldovich cluster counts}},
  \href{http://arxiv.org/abs/1502.01597}{{\tt 1502.01597}}.

\bibitem{Samushia:2013yga}
L.~Samushia, B.~A. Reid, M.~White, W.~J. Percival, A.~J. Cuesta et~al.,
  \emph{{The Clustering of Galaxies in the SDSS-III Baryon Oscillation
  Spectroscopic Survey (BOSS): measuring growth rate and geometry with
  anisotropic clustering}},
  \href{http://dx.doi.org/10.1093/mnras/stu197}{\emph{Mon. Not. Roy. Astron.
  Soc.} {\bf 439} (2014) 3504--3519},
  [\href{http://arxiv.org/abs/1312.4899}{{\tt 1312.4899}}].

\bibitem{Wyman:2013lza}
M.~Wyman, D.~H. Rudd, R.~A. Vanderveld and W.~Hu, \emph{{Neutrinos Help
  Reconcile Planck Measurements with the Local Universe}},
  \href{http://dx.doi.org/10.1103/PhysRevLett.112.051302}{\emph{Phys. Rev.
  Lett.} {\bf 112} (2014) 051302}, [\href{http://arxiv.org/abs/1307.7715}{{\tt
  1307.7715}}].

\bibitem{Hamann:2013iba}
J.~Hamann and J.~Hasenkamp, \emph{{A new life for sterile neutrinos: resolving
  inconsistencies using hot dark matter}},
  \href{http://dx.doi.org/10.1088/1475-7516/2013/10/044}{\emph{JCAP} {\bf 1310}
  (2013) 044}, [\href{http://arxiv.org/abs/1308.3255}{{\tt 1308.3255}}].

\bibitem{Battye:2013xqa}
R.~A. Battye and A.~Moss, \emph{{Evidence for Massive Neutrinos from Cosmic
  Microwave Background and Lensing Observations}},
  \href{http://dx.doi.org/10.1103/PhysRevLett.112.051303}{\emph{Phys. Rev.
  Lett.} {\bf 112} (2014) 051303}, [\href{http://arxiv.org/abs/1308.5870}{{\tt
  1308.5870}}].

\bibitem{Leistedt:2014sia}
B.~Leistedt, H.~V. Peiris and L.~Verde, \emph{{No new cosmological concordance
  with massive sterile neutrinos}},
  \href{http://dx.doi.org/10.1103/PhysRevLett.113.041301}{\emph{Phys. Rev.
  Lett.} {\bf 113} (2014) 041301}, [\href{http://arxiv.org/abs/1404.5950}{{\tt
  1404.5950}}].

\bibitem{Palanque-Delabrouille:2014jca}
N.~Palanque-Delabrouille, C.~Y{\`e}che, J.~Lesgourgues, G.~Rossi, A.~Borde
  et~al., \emph{{Constraint on neutrino masses from SDSS-III/BOSS Ly$\alpha$
  forest and other cosmological probes}},
  \href{http://dx.doi.org/10.1088/1475-7516/2015/02/045}{\emph{JCAP} {\bf 1502}
  (2015) 045}, [\href{http://arxiv.org/abs/1410.7244}{{\tt 1410.7244}}].

\bibitem{Ade:2014xna}
{\scshape BICEP2} collaboration, P.~A.~R. Ade et~al., \emph{{Detection of
  $B$-Mode Polarization at Degree Angular Scales by BICEP2}},
  \href{http://dx.doi.org/10.1103/PhysRevLett.112.241101}{\emph{Phys. Rev.
  Lett.} {\bf 112} (2014) 241101}, [\href{http://arxiv.org/abs/1403.3985}{{\tt
  1403.3985}}].

\bibitem{Ade:2015tva}
{\scshape BICEP2, Planck} collaboration, P.~A.~R. Ade et~al., \emph{{A Joint
  Analysis of BICEP2/Keck Array and Planck Data}},
  \href{http://dx.doi.org/10.1103/PhysRevLett.114.101301}{\emph{Phys. Rev.
  Lett.} {\bf 114} (2015) 101301}, [\href{http://arxiv.org/abs/1502.00612}{{\tt
  1502.00612}}].

\bibitem{Mangano:2011ip}
G.~Mangano, G.~Miele, S.~Pastor, O.~Pisanti and S.~Sarikas, \emph{{Updated BBN
  bounds on the cosmological lepton asymmetry for non-zero $\theta_{13}$}},
  \href{http://dx.doi.org/10.1016/j.physletb.2012.01.015}{\emph{Phys. Lett.}
  {\bf B708} (2012) 1--5}, [\href{http://arxiv.org/abs/1110.4335}{{\tt
  1110.4335}}].

\bibitem{Mangano:2005cc}
G.~Mangano, G.~Miele, S.~Pastor, T.~Pinto, O.~Pisanti et~al., \emph{{Relic
  neutrino decoupling including flavor oscillations}},
  \href{http://dx.doi.org/10.1016/j.nuclphysb.2005.09.041}{\emph{Nucl. Phys.}
  {\bf B729} (2005) 221--234}, [\href{http://arxiv.org/abs/hep-ph/0506164}{{\tt
  hep-ph/0506164}}].

\bibitem{Grohs:2015tfy}
E.~Grohs, G.~M. Fuller, C.~T. Kishimoto, M.~W. Paris and A.~Vlasenko,
  \emph{{Neutrino energy transport in weak decoupling and big bang
  nucleosynthesis}},
  \href{http://dx.doi.org/10.1103/PhysRevD.93.083522}{\emph{Phys. Rev.} {\bf
  D93} (2016) 083522}, [\href{http://arxiv.org/abs/1512.02205}{{\tt
  1512.02205}}].

\bibitem{NuCosmo}
J.~Lesgourgues, G.~Mangano, G.~Miele and S.~Pastor, \emph{{Neutrino
  cosmology}}.
\newblock Cambridge Univ. Press, Cambridge, 2013.

\bibitem{Dolgov:2002wy}
A.~D. Dolgov, \emph{{Neutrinos in cosmology}},
  \href{http://dx.doi.org/10.1016/S0370-1573(02)00139-4}{\emph{Phys. Rept.}
  {\bf 370} (2002) 333--535}, [\href{http://arxiv.org/abs/hep-ph/0202122}{{\tt
  hep-ph/0202122}}].

\bibitem{Sarkar:1995dd}
S.~Sarkar, \emph{{Big bang nucleosynthesis and physics beyond the standard
  model}}, \href{http://dx.doi.org/10.1088/0034-4885/59/12/001}{\emph{Rept.
  Prog. Phys.} {\bf 59} (1996) 1493--1610},
  [\href{http://arxiv.org/abs/hep-ph/9602260}{{\tt hep-ph/9602260}}].

\bibitem{Primack:2001ib}
J.~R. Primack, \emph{{Whatever happened to hot dark matter?}}, {\emph{SLAC Beam
  Line} {\bf 31N3} (2001) 50--57},
  [\href{http://arxiv.org/abs/astro-ph/0112336}{{\tt astro-ph/0112336}}].

\bibitem{Lesgourgues:2006nd}
J.~Lesgourgues and S.~Pastor, \emph{{Massive neutrinos and cosmology}},
  \href{http://dx.doi.org/10.1016/j.physrep.2006.04.001}{\emph{Phys. Rept.}
  {\bf 429} (2006) 307--379},
  [\href{http://arxiv.org/abs/astro-ph/0603494}{{\tt astro-ph/0603494}}].

\bibitem{Bird:2011rb}
S.~Bird, M.~Viel and M.~G. Haehnelt, \emph{{Massive Neutrinos and the
  Non-linear Matter Power Spectrum}},
  \href{http://dx.doi.org/10.1111/j.1365-2966.2011.20222.x}{\emph{Mon. Not.
  Roy. Astron. Soc.} {\bf 420} (2012) 2551--2561},
  [\href{http://arxiv.org/abs/1109.4416}{{\tt 1109.4416}}].

\bibitem{Jimenez:2010ev}
R.~Jimenez, T.~Kitching, C.~Pena-Garay and L.~Verde, \emph{{Can we measure the
  neutrino mass hierarchy in the sky?}},
  \href{http://dx.doi.org/10.1088/1475-7516/2010/05/035}{\emph{JCAP} {\bf 1005}
  (2010) 035}, [\href{http://arxiv.org/abs/1003.5918}{{\tt 1003.5918}}].

\bibitem{Cuesta:2015iho}
A.~J. Cuesta, V.~Niro and L.~Verde, \emph{{Neutrino mass limits: robust
  information from the power spectrum of galaxy surveys}},
  \href{http://arxiv.org/abs/1511.05983}{{\tt 1511.05983}}.

\bibitem{Gil-Marin:2014baa}
H.~Gil-Marin, L.~Verde, J.~Norena, A.~J. Cuesta, L.~Samushia et~al., \emph{{The
  power spectrum and bispectrum of SDSS DR11 BOSS galaxies II: cosmological
  interpretation}},  \href{http://arxiv.org/abs/1408.0027}{{\tt 1408.0027}}.

\bibitem{Battye:2014qga}
R.~A. Battye, T.~Charnock and A.~Moss, \emph{{Tension between the power
  spectrum of density perturbations measured on large and small scales}},
  \href{http://dx.doi.org/10.1103/PhysRevD.91.103508}{\emph{Phys. Rev.} {\bf
  D91} (2015) 103508}, [\href{http://arxiv.org/abs/1409.2769}{{\tt
  1409.2769}}].

\bibitem{Audren:2012vy}
B.~Audren, J.~Lesgourgues, S.~Bird, M.~G. Haehnelt and M.~Viel, \emph{{Neutrino
  masses and cosmological parameters from a Euclid-like survey: Markov Chain
  Monte Carlo forecasts including theoretical errors}},
  \href{http://dx.doi.org/10.1088/1475-7516/2013/01/026}{\emph{JCAP} {\bf 1301}
  (2013) 026}, [\href{http://arxiv.org/abs/1210.2194}{{\tt 1210.2194}}].

\bibitem{Hall:2012kg}
A.~C. Hall and A.~Challinor, \emph{{Probing the neutrino mass hierarchy with
  CMB weak lensing}},
  \href{http://dx.doi.org/10.1111/j.1365-2966.2012.21493.x}{\emph{Mon. Not.
  Roy. Astron. Soc.} {\bf 425} (2012) 1170--1184},
  [\href{http://arxiv.org/abs/1205.6172}{{\tt 1205.6172}}].

\bibitem{Pritchard:2008wy}
J.~R. Pritchard and E.~Pierpaoli, \emph{{Constraining massive neutrinos using
  cosmological 21 cm observations}},
  \href{http://dx.doi.org/10.1103/PhysRevD.78.065009}{\emph{Phys. Rev.} {\bf
  D78} (2008) 065009}, [\href{http://arxiv.org/abs/0805.1920}{{\tt
  0805.1920}}].

\bibitem{Wu:2014hta}
W.~L.~K. Wu, J.~Errard, C.~Dvorkin, C.~L. Kuo, A.~T. Lee, P.~McDonald et~al.,
  \emph{{A Guide to Designing Future Ground-based Cosmic Microwave Background
  Experiments}},
  \href{http://dx.doi.org/10.1088/0004-637X/788/2/138}{\emph{Astrophys. J.}
  {\bf 788} (2014) 138}, [\href{http://arxiv.org/abs/1402.4108}{{\tt
  1402.4108}}].

\bibitem{Font-Ribera:2013rwa}
A.~Font-Ribera, P.~McDonald, N.~Mostek, B.~A. Reid, H.-J. Seo and A.~Slosar,
  \emph{{DESI and other dark energy experiments in the era of neutrino mass
  measurements}},
  \href{http://dx.doi.org/10.1088/1475-7516/2014/05/023}{\emph{JCAP} {\bf 1405}
  (2014) 023}, [\href{http://arxiv.org/abs/1308.4164}{{\tt 1308.4164}}].

\bibitem{Alpher:1948ve}
R.~A. Alpher, H.~Bethe and G.~Gamow, \emph{{The origin of chemical elements}},
  \href{http://dx.doi.org/10.1103/PhysRev.73.803}{\emph{Phys. Rev.} {\bf 73}
  (1948) 803--804}.

\bibitem{Wagoner:1966pv}
R.~V. Wagoner, W.~A. Fowler and F.~Hoyle, \emph{{On the Synthesis of elements
  at very high temperatures}},
  \href{http://dx.doi.org/10.1086/149126}{\emph{Astrophys. J.} {\bf 148} (1967)
  3--49}.

\bibitem{Kawano:1992ua}
L.~Kawano, \emph{{Let's go: Early universe. 2. Primordial nucleosynthesis: The
  Computer way}}, .

\bibitem{Pisanti:2007hk}
O.~Pisanti, A.~Cirillo, S.~Esposito, F.~Iocco, G.~Mangano et~al.,
  \emph{{PArthENoPE: Public Algorithm Evaluating the Nucleosynthesis of
  Primordial Elements}},
  \href{http://dx.doi.org/10.1016/j.cpc.2008.02.015}{\emph{Comput. Phys.
  Commun.} {\bf 178} (2008) 956--971},
  [\href{http://arxiv.org/abs/0705.0290}{{\tt 0705.0290}}].

\bibitem{Serpico:2004gx}
P.~D. Serpico, S.~Esposito, F.~Iocco, G.~Mangano, G.~Miele et~al.,
  \emph{{Nuclear reaction network for primordial nucleosynthesis: A Detailed
  analysis of rates, uncertainties and light nuclei yields}},
  \href{http://dx.doi.org/10.1088/1475-7516/2004/12/010}{\emph{JCAP} {\bf 0412}
  (2004) 010}, [\href{http://arxiv.org/abs/astro-ph/0408076}{{\tt
  astro-ph/0408076}}].

\bibitem{Iocco:2008va}
F.~Iocco, G.~Mangano, G.~Miele, O.~Pisanti and P.~D. Serpico, \emph{{Primordial
  Nucleosynthesis: from precision cosmology to fundamental physics}},
  \href{http://dx.doi.org/10.1016/j.physrep.2009.02.002}{\emph{Phys. Rept.}
  {\bf 472} (2009) 1--76}, [\href{http://arxiv.org/abs/0809.0631}{{\tt
  0809.0631}}].

\bibitem{Aver:2013wba}
E.~Aver, K.~A. Olive, R.~L. Porter and E.~D. Skillman, \emph{{The primordial
  helium abundance from updated emissivities}},
  \href{http://dx.doi.org/10.1088/1475-7516/2013/11/017}{\emph{JCAP} {\bf 1311}
  (2013) 017}, [\href{http://arxiv.org/abs/1309.0047}{{\tt 1309.0047}}].

\bibitem{Cooke:2013cba}
R.~Cooke, M.~Pettini, R.~A. Jorgenson, M.~T. Murphy and C.~C. Steidel,
  \emph{{Precision Measures of the Primordial Abundance of Deuterium}},
  \href{http://dx.doi.org/10.1088/0004-637X/781/1/31}{\emph{\apj} {\bf 781}
  (Jan., 2014) 31}, [\href{http://arxiv.org/abs/1308.3240}{{\tt 1308.3240}}].

\bibitem{Karagiorgi:2012usa}
G.~Karagiorgi, \emph{{Toward Solution of the MiniBooNE-LSND Anomalies}},
  \href{http://dx.doi.org/10.1016/j.nuclphysbps.2012.09.008}{\emph{Nucl. Phys.
  Proc. Suppl.} {\bf 229-232} (2012) 50--54}.

\bibitem{Abazajian:2002bj}
K.~N. Abazajian, \emph{{Telling three from four neutrinos with cosmology}},
  \href{http://dx.doi.org/10.1016/S0927-6505(02)00204-9}{\emph{Astropart.
  Phys.} {\bf 19} (2003) 303--312},
  [\href{http://arxiv.org/abs/astro-ph/0205238}{{\tt astro-ph/0205238}}].

\bibitem{Mangano:2011ar}
G.~Mangano and P.~D. Serpico, \emph{{A robust upper limit on $N_{\rm eff}$ from
  BBN, circa 2011}},
  \href{http://dx.doi.org/10.1016/j.physletb.2011.05.075}{\emph{Phys. Lett.}
  {\bf B701} (2011) 296--299}, [\href{http://arxiv.org/abs/1103.1261}{{\tt
  1103.1261}}].

\bibitem{Hou:2011ec}
Z.~Hou, R.~Keisler, L.~Knox, M.~Millea and C.~Reichardt, \emph{{How Massless
  Neutrinos Affect the Cosmic Microwave Background Damping Tail}},
  \href{http://dx.doi.org/10.1103/PhysRevD.87.083008}{\emph{Phys. Rev.} {\bf
  D87} (2013) 083008}, [\href{http://arxiv.org/abs/1104.2333}{{\tt
  1104.2333}}].

\bibitem{Archidiacono:2013fha}
M.~Archidiacono, E.~Giusarma, S.~Hannestad and O.~Mena, \emph{{Cosmic dark
  radiation and neutrinos}},
  \href{http://dx.doi.org/10.1155/2013/191047}{\emph{Adv. High Energy Phys.}
  {\bf 2013} (2013) 191047}, [\href{http://arxiv.org/abs/1307.0637}{{\tt
  1307.0637}}].

\bibitem{Archidiacono:2014apa}
M.~Archidiacono, N.~Fornengo, S.~Gariazzo, C.~Giunti, S.~Hannestad et~al.,
  \emph{{Light sterile neutrinos after BICEP-2}},
  \href{http://dx.doi.org/10.1088/1475-7516/2014/06/031}{\emph{JCAP} {\bf 1406}
  (2014) 031}, [\href{http://arxiv.org/abs/1404.1794}{{\tt 1404.1794}}].

\bibitem{Bergstrom:2014fqa}
J.~Bergstrom, M.~C. Gonzalez-Garcia, V.~Niro and J.~Salvado, \emph{{Statistical
  tests of sterile neutrinos using cosmology and short-baseline data}},
  \href{http://dx.doi.org/10.1007/JHEP10(2014)104}{\emph{JHEP} {\bf 10} (2014)
  104}, [\href{http://arxiv.org/abs/1407.3806}{{\tt 1407.3806}}].

\bibitem{Maltoni:2001bc}
M.~Maltoni, T.~Schwetz and J.~W.~F. Valle, \emph{{Status of four neutrino mass
  schemes: A Global and unified approach to current neutrino oscillation
  data}}, \href{http://dx.doi.org/10.1103/PhysRevD.65.093004}{\emph{Phys. Rev.}
  {\bf D65} (2002) 093004}, [\href{http://arxiv.org/abs/hep-ph/0112103}{{\tt
  hep-ph/0112103}}].

\bibitem{Sigl:1992fn}
G.~Sigl and G.~Raffelt, \emph{{General kinetic description of relativistic
  mixed neutrinos}},
  \href{http://dx.doi.org/10.1016/0550-3213(93)90175-O}{\emph{Nucl. Phys.} {\bf
  B406} (1993) 423--451}.

\bibitem{Dolgov:1980cq}
A.~D. Dolgov, \emph{{Neutrinos in the Early Universe}}, {\emph{Sov. J. Nucl.
  Phys.} {\bf 33} (1981) 700--706}.

\bibitem{Dolgov:2002ab}
A.~D. Dolgov, S.~H. Hansen, S.~Pastor, S.~T. Petcov, G.~G. Raffelt et~al.,
  \emph{{Cosmological bounds on neutrino degeneracy improved by flavor
  oscillations}},
  \href{http://dx.doi.org/10.1016/S0550-3213(02)00274-2}{\emph{Nucl. Phys.}
  {\bf B632} (2002) 363--382}, [\href{http://arxiv.org/abs/hep-ph/0201287}{{\tt
  hep-ph/0201287}}].

\bibitem{Mirizzi:2012we}
A.~Mirizzi, N.~Saviano, G.~Miele and P.~D. Serpico, \emph{{Light sterile
  neutrino production in the early universe with dynamical neutrino
  asymmetries}},
  \href{http://dx.doi.org/10.1103/PhysRevD.86.053009}{\emph{Phys. Rev.} {\bf
  D86} (2012) 053009}, [\href{http://arxiv.org/abs/1206.1046}{{\tt
  1206.1046}}].

\bibitem{Hannestad:2012ky}
S.~Hannestad, I.~Tamborra and T.~Tram, \emph{{Thermalisation of light sterile
  neutrinos in the early universe}},
  \href{http://dx.doi.org/10.1088/1475-7516/2012/07/025}{\emph{JCAP} {\bf 1207}
  (2012) 025}, [\href{http://arxiv.org/abs/1204.5861}{{\tt 1204.5861}}].

\bibitem{Saviano:2013ktj}
N.~Saviano, A.~Mirizzi, O.~Pisanti, P.~D. Serpico, G.~Mangano et~al.,
  \emph{{Multi-momentum and multi-flavour active-sterile neutrino oscillations
  in the early universe: role of neutrino asymmetries and effects on
  nucleosynthesis}},
  \href{http://dx.doi.org/10.1103/PhysRevD.87.073006}{\emph{Phys. Rev.} {\bf
  D87} (2013) 073006}, [\href{http://arxiv.org/abs/1302.1200}{{\tt
  1302.1200}}].

\bibitem{Hannestad:2013ana}
S.~Hannestad, R.~S. Hansen and T.~Tram, \emph{{How Self-Interactions can
  Reconcile Sterile Neutrinos with Cosmology}},
  \href{http://dx.doi.org/10.1103/PhysRevLett.112.031802}{\emph{Phys. Rev.
  Lett.} {\bf 112} (2014) 031802}, [\href{http://arxiv.org/abs/1310.5926}{{\tt
  1310.5926}}].

\bibitem{Dasgupta:2013zpn}
B.~Dasgupta and J.~Kopp, \emph{{Cosmologically Safe eV-Scale Sterile Neutrinos
  and Improved Dark Matter Structure}},
  \href{http://dx.doi.org/10.1103/PhysRevLett.112.031803}{\emph{Phys. Rev.
  Lett.} {\bf 112} (2014) 031803}, [\href{http://arxiv.org/abs/1310.6337}{{\tt
  1310.6337}}].

\bibitem{Archidiacono:2013dua}
M.~Archidiacono and S.~Hannestad, \emph{{Updated constraints on non-standard
  neutrino interactions from Planck}},
  \href{http://dx.doi.org/10.1088/1475-7516/2014/07/046}{\emph{JCAP} {\bf 1407}
  (2014) 046}, [\href{http://arxiv.org/abs/1311.3873}{{\tt 1311.3873}}].

\bibitem{Mirizzi:2014ama}
A.~Mirizzi, G.~Mangano, O.~Pisanti and N.~Saviano, \emph{{Collisional
  production of sterile neutrinos via secret interactions and cosmological
  implications}},
  \href{http://dx.doi.org/10.1103/PhysRevD.91.025019}{\emph{Phys. Rev.} {\bf
  D91} (2015) 025019}, [\href{http://arxiv.org/abs/1410.1385}{{\tt
  1410.1385}}].

\bibitem{Saviano:2014esa}
N.~Saviano, O.~Pisanti, G.~Mangano and A.~Mirizzi, \emph{{Unveiling secret
  interactions among sterile neutrinos with big-bang nucleosynthesis}},
  \href{http://dx.doi.org/10.1103/PhysRevD.90.113009}{\emph{Phys. Rev.} {\bf
  D90} (2014) 113009}, [\href{http://arxiv.org/abs/1409.1680}{{\tt
  1409.1680}}].

\bibitem{Bringmann:2013vra}
T.~Bringmann, J.~Hasenkamp and J.~Kersten, \emph{{Tight bonds between sterile
  neutrinos and dark matter}},
  \href{http://dx.doi.org/10.1088/1475-7516/2014/07/042}{\emph{JCAP} {\bf 1407}
  (2014) 042}, [\href{http://arxiv.org/abs/1312.4947}{{\tt 1312.4947}}].

\bibitem{Chu:2014lja}
X.~Chu and B.~Dasgupta, \emph{{Dark Radiation Alleviates Problems with Dark
  Matter Halos}},
  \href{http://dx.doi.org/10.1103/PhysRevLett.113.161301}{\emph{Phys. Rev.
  Lett.} {\bf 113} (2014) 161301}, [\href{http://arxiv.org/abs/1404.6127}{{\tt
  1404.6127}}].

\bibitem{Boehm:2014vja}
C.~Boehm, J.~A. Schewtschenko, R.~J. Wilkinson, C.~M. Baugh and S.~Pascoli,
  \emph{{Using the Milky Way satellites to study interactions between cold dark
  matter and radiation}},
  \href{http://dx.doi.org/10.1093/mnrasl/slu115}{\emph{Mon. Not. Roy. Astron.
  Soc.} {\bf 445} (2014) L31--L35}, [\href{http://arxiv.org/abs/1404.7012}{{\tt
  1404.7012}}].

\bibitem{Archidiacono:2014nda}
M.~Archidiacono, S.~Hannestad, R.~S. Hansen and T.~Tram, \emph{{Cosmology with
  self-interacting sterile neutrinos and dark matter - A pseudoscalar model}},
  \href{http://arxiv.org/abs/1404.5915}{{\tt 1404.5915}}.

\bibitem{Walker:13}
M.~G. Walker, \emph{{Dark Matter in the Milky Way's Dwarf Spheroidal
  Satellites}},  \href{http://arxiv.org/abs/1205.0311}{{\tt 1205.0311}}.

\bibitem{Boyarsky:2008ju}
A.~Boyarsky, O.~Ruchayskiy and D.~Iakubovskyi, \emph{{A Lower bound on the mass
  of Dark Matter particles}},
  \href{http://dx.doi.org/10.1088/1475-7516/2009/03/005}{\emph{JCAP} {\bf 0903}
  (2009) 005}, [\href{http://arxiv.org/abs/0808.3902}{{\tt 0808.3902}}].

\bibitem{Gorbunov:2008ka}
D.~Gorbunov, A.~Khmelnitsky and V.~Rubakov, \emph{{Constraining sterile
  neutrino dark matter by phase-space density observations}},
  \href{http://dx.doi.org/10.1088/1475-7516/2008/10/041}{\emph{JCAP} {\bf 0810}
  (2008) 041}, [\href{http://arxiv.org/abs/0808.3910}{{\tt 0808.3910}}].

\bibitem{Horiuchi:2013noa}
S.~Horiuchi, P.~J. Humphrey, J.~Onorbe, K.~N. Abazajian, M.~Kaplinghat and
  S.~Garrison-Kimmel, \emph{{Sterile neutrino dark matter bounds from galaxies
  of the Local Group}},
  \href{http://dx.doi.org/10.1103/PhysRevD.89.025017}{\emph{Phys. Rev.} {\bf
  D89} (2014) 025017}, [\href{http://arxiv.org/abs/1311.0282}{{\tt
  1311.0282}}].

\bibitem{Shi:1998km}
X.-D. Shi and G.~M. Fuller, \emph{{A New dark matter candidate: Nonthermal
  sterile neutrinos}},
  \href{http://dx.doi.org/10.1103/PhysRevLett.82.2832}{\emph{Phys. Rev. Lett.}
  {\bf 82} (1999) 2832--2835},
  [\href{http://arxiv.org/abs/astro-ph/9810076}{{\tt astro-ph/9810076}}].

\bibitem{Abazajian:2001nj}
K.~Abazajian, G.~M. Fuller and M.~Patel, \emph{{Sterile neutrino hot, warm, and
  cold dark matter}},
  \href{http://dx.doi.org/10.1103/PhysRevD.64.023501}{\emph{Phys. Rev.} {\bf
  D64} (2001) 023501}, [\href{http://arxiv.org/abs/astro-ph/0101524}{{\tt
  astro-ph/0101524}}].

\bibitem{Laine:2008pg}
M.~Laine and M.~Shaposhnikov, \emph{{Sterile neutrino dark matter as a
  consequence of nuMSM-induced lepton asymmetry}},
  \href{http://dx.doi.org/10.1088/1475-7516/2008/06/031}{\emph{JCAP} {\bf 0806}
  (2008) 031}, [\href{http://arxiv.org/abs/0804.4543}{{\tt 0804.4543}}].

\bibitem{Venumadhav:2015pla}
T.~Venumadhav, F.-Y. Cyr-Racine, K.~N. Abazajian and C.~M. Hirata,
  \emph{{Sterile neutrino dark matter: A tale of weak interactions in the
  strong coupling epoch}},  \href{http://arxiv.org/abs/1507.06655}{{\tt
  1507.06655}}.

\bibitem{Mikheev:1986gs}
S.~P. Mikheev and A.~Y. Smirnov, \emph{{Resonance Amplification of Oscillations
  in Matter and Spectroscopy of Solar Neutrinos}}, {\emph{Sov. J. Nucl. Phys.}
  {\bf 42} (1985) 913--917}.

\bibitem{Wolfenstein:1977ue}
L.~Wolfenstein, \emph{{Neutrino Oscillations in Matter}},
  \href{http://dx.doi.org/10.1103/PhysRevD.17.2369}{\emph{Phys. Rev.} {\bf D17}
  (1978) 2369--2374}.

\bibitem{Asaka:2006nq}
T.~Asaka, M.~Laine and M.~Shaposhnikov, \emph{{Lightest sterile neutrino
  abundance within the nuMSM}},
  \href{http://dx.doi.org/10.1088/1126-6708/2007/01/091,
  10.1007/JHEP02(2015)028}{\emph{JHEP} {\bf 0701} (2007) 091},
  [\href{http://arxiv.org/abs/hep-ph/0612182}{{\tt hep-ph/0612182}}].

\bibitem{PhysRevD.16.1444}
B.~W. Lee and R.~E. Shrock, \emph{Natural suppression of symmetry violation in
  gauge theories: Muon- and electron-lepton-number nonconservation},
  \href{http://dx.doi.org/10.1103/PhysRevD.16.1444}{\emph{Phys. Rev. D} {\bf
  16} (Sep, 1977) 1444--1473}.

\bibitem{Pal:82}
P.~B. {Pal} and L.~{Wolfenstein}, \emph{{Radiative decays of massive
  neutrinos}}, \href{http://dx.doi.org/10.1103/PhysRevD.25.766}{\emph{\prd}
  {\bf 25} (Feb., 1982) 766--773}.

\bibitem{Barger:1995ty}
V.~D. Barger, R.~J.~N. Phillips and S.~Sarkar, \emph{{Remarks on the KARMEN
  anomaly}}, \href{http://dx.doi.org/10.1016/0370-2693(95)00486-5}{\emph{Phys.
  Lett.} {\bf B352} (Feb., 1995) 365--371},
  [\href{http://arxiv.org/abs/hep-ph/9503295}{{\tt hep-ph/9503295}}].

\bibitem{Dolgov:2000ew}
A.~D. Dolgov and S.~H. Hansen, \emph{{Massive sterile neutrinos as warm dark
  matter}},
  \href{http://dx.doi.org/10.1016/S0927-6505(01)00115-3}{\emph{Astropart.
  Phys.} {\bf 16} (2002) 339--344},
  [\href{http://arxiv.org/abs/hep-ph/0009083}{{\tt hep-ph/0009083}}].

\bibitem{Boyarsky:2006jm}
A.~Boyarsky, A.~Neronov, O.~Ruchayskiy and M.~Shaposhnikov, \emph{{The Masses
  of active neutrinos in the nuMSM from X-ray astronomy}},
  \href{http://dx.doi.org/10.1134/S0021364006040011}{\emph{JETP Lett.} {\bf 83}
  (2006) 133--135}, [\href{http://arxiv.org/abs/hep-ph/0601098}{{\tt
  hep-ph/0601098}}].

\bibitem{Hansen:2001zv}
S.~H. Hansen, J.~Lesgourgues, S.~Pastor and J.~Silk, \emph{{Constraining the
  window on sterile neutrinos as warm dark matter}},
  \href{http://dx.doi.org/10.1046/j.1365-8711.2002.05410.x}{\emph{Mon. Not.
  Roy. Astron. Soc.} {\bf 333} (2002) 544--546},
  [\href{http://arxiv.org/abs/astro-ph/0106108}{{\tt astro-ph/0106108}}].

\bibitem{Viel:2006kd}
M.~Viel, J.~Lesgourgues, M.~G. Haehnelt, S.~Matarrese and A.~Riotto, \emph{{Can
  sterile neutrinos be ruled out as warm dark matter candidates?}},
  \href{http://dx.doi.org/10.1103/PhysRevLett.97.071301}{\emph{Phys. Rev.
  Lett.} {\bf 97} (2006) 071301},
  [\href{http://arxiv.org/abs/astro-ph/0605706}{{\tt astro-ph/0605706}}].

\bibitem{Viel:2005qj}
M.~Viel, J.~Lesgourgues, M.~G. Haehnelt, S.~Matarrese and A.~Riotto,
  \emph{{Constraining warm dark matter candidates including sterile neutrinos
  and light gravitinos with WMAP and the Lyman-alpha forest}},
  \href{http://dx.doi.org/10.1103/PhysRevD.71.063534}{\emph{Phys. Rev.} {\bf
  D71} (2005) 063534}, [\href{http://arxiv.org/abs/astro-ph/0501562}{{\tt
  astro-ph/0501562}}].

\bibitem{Seljak:2006qw}
U.~Seljak, A.~Makarov, P.~McDonald and H.~Trac, \emph{{Can sterile neutrinos be
  the dark matter?}},
  \href{http://dx.doi.org/10.1103/PhysRevLett.97.191303}{\emph{Phys. Rev.
  Lett.} {\bf 97} (2006) 191303},
  [\href{http://arxiv.org/abs/astro-ph/0602430}{{\tt astro-ph/0602430}}].

\bibitem{Boyarsky:2008mt}
A.~Boyarsky, J.~Lesgourgues, O.~Ruchayskiy and M.~Viel, \emph{{Realistic
  sterile neutrino dark matter with keV mass does not contradict cosmological
  bounds}}, \href{http://dx.doi.org/10.1103/PhysRevLett.102.201304}{\emph{Phys.
  Rev. Lett.} {\bf 102} (2009) 201304},
  [\href{http://arxiv.org/abs/0812.3256}{{\tt 0812.3256}}].

\bibitem{Viel:2013apy}
M.~Viel, G.~D. Becker, J.~S. Bolton and M.~G. Haehnelt, \emph{{Warm dark matter
  as a solution to the small scale crisis: New constraints from high redshift
  Lyman-α forest data}},
  \href{http://dx.doi.org/10.1103/PhysRevD.88.043502}{\emph{Phys. Rev.} {\bf
  D88} (2013) 043502}, [\href{http://arxiv.org/abs/1306.2314}{{\tt
  1306.2314}}].

\bibitem{Viel:2007mv}
M.~Viel, G.~D. Becker, J.~S. Bolton, M.~G. Haehnelt, M.~Rauch and W.~L.~W.
  Sargent, \emph{{How cold is cold dark matter? Small scales constraints from
  the flux power spectrum of the high-redshift Lyman-alpha forest}},
  \href{http://dx.doi.org/10.1103/PhysRevLett.100.041304}{\emph{Phys. Rev.
  Lett.} {\bf 100} (2008) 041304}, [\href{http://arxiv.org/abs/0709.0131}{{\tt
  0709.0131}}].

\bibitem{Bode:2000gq}
P.~Bode, J.~P. Ostriker and N.~Turok, \emph{{Halo formation in warm dark matter
  models}}, \href{http://dx.doi.org/10.1086/321541}{\emph{Astrophys. J.} {\bf
  556} (2001) 93--107}, [\href{http://arxiv.org/abs/astro-ph/0010389}{{\tt
  astro-ph/0010389}}].

\bibitem{Schneider:2013wwa}
A.~Schneider, D.~Anderhalden, A.~Maccio and J.~Diemand, \emph{{Warm dark matter
  does not do better than cold dark matter in solving small-scale
  inconsistencies}}, \href{http://dx.doi.org/10.1093/mnrasl/slu034}{\emph{Mon.
  Not. Roy. Astron. Soc.} {\bf 441} (2014) 6},
  [\href{http://arxiv.org/abs/1309.5960}{{\tt 1309.5960}}].

\bibitem{Maccio:2012qf}
A.~V. Maccio, S.~Paduroiu, D.~Anderhalden, A.~Schneider and B.~Moore,
  \emph{{Cores in warm dark matter haloes: a Catch 22 problem}},
  \href{http://dx.doi.org/10.1111/j.1365-2966.2012.21284.x}{\emph{Mon. Not.
  Roy. Astron. Soc.} {\bf 424} (Feb., 2012) 1105--1112},
  [\href{http://arxiv.org/abs/1202.1282}{{\tt 1202.1282}}].

\bibitem{Maccio':2009rx}
A.~V. Maccio' and F.~Fontanot, \emph{{How cold is Dark Matter? Constraints from
  Milky Way Satellites}}, {\emph{Mon. Not. Roy. Astron. Soc.} {\bf 404} (2010)
  16}, [\href{http://arxiv.org/abs/0910.2460}{{\tt 0910.2460}}].

\bibitem{Schneider:2011yu}
A.~Schneider, R.~E. Smith, A.~V. Maccio and B.~Moore, \emph{{Nonlinear
  Evolution of Cosmological Structures in Warm Dark Matter Models}},
  \href{http://dx.doi.org/10.1111/j.1365-2966.2012.21252.x}{\emph{Mon. Not.
  Roy. Astron. Soc.} {\bf 424} (2012) 684},
  [\href{http://arxiv.org/abs/1112.0330}{{\tt 1112.0330}}].

\bibitem{Goerdt:06}
T.~Goerdt, B.~Moore, J.~I. Read, J.~Stadel and M.~Zemp, \emph{{Does the fornax
  dwarf spheroidal have a central cusp or core?}},
  \href{http://dx.doi.org/10.1111/j.1365-2966.2006.10182.x}{\emph{Mon. Not.
  Roy. Astron. Soc.} {\bf 368} (May, 2006) 1073--1077},
  [\href{http://arxiv.org/abs/astro-ph/0601404}{{\tt astro-ph/0601404}}].

\bibitem{Moore:1999gc}
B.~Moore, T.~R. Quinn, F.~Governato, J.~Stadel and G.~Lake, \emph{{Cold
  collapse and the core catastrophe}},
  \href{http://dx.doi.org/10.1046/j.1365-8711.1999.03039.x}{\emph{Mon. Not.
  Roy. Astron. Soc.} {\bf 310} (Dec., 1999) 1147--1152},
  [\href{http://arxiv.org/abs/astro-ph/9903164}{{\tt astro-ph/9903164}}].

\bibitem{Kennedy:2013uta}
R.~Kennedy, C.~Frenk, S.~Cole and A.~Benson, \emph{{Constraining the warm dark
  matter particle mass with Milky Way satellites}},
  \href{http://dx.doi.org/10.1093/mnras/stu719}{\emph{Mon. Not. Roy. Astron.
  Soc.} {\bf 442} (2014) 2487--2495},
  [\href{http://arxiv.org/abs/1310.7739}{{\tt 1310.7739}}].

\bibitem{Lovell:2013ola}
M.~R. Lovell, C.~S. Frenk, V.~R. Eke, A.~Jenkins, L.~Gao et~al., \emph{{The
  properties of warm dark matter haloes}},
  \href{http://dx.doi.org/10.1093/mnras/stt2431}{\emph{Mon. Not. Roy. Astron.
  Soc.} {\bf 439} (2014) 300--317}, [\href{http://arxiv.org/abs/1308.1399}{{\tt
  1308.1399}}].

\bibitem{Shao:2012cg}
S.~Shao, L.~Gao, T.~Theuns and C.~S. Frenk, \emph{{The phase space density of
  fermionic dark matter haloes}},
  \href{http://dx.doi.org/10.1093/mnras/stt053}{\emph{Mon. Not. Roy. Astron.
  Soc.} {\bf 430} (2013) 2346}, [\href{http://arxiv.org/abs/1209.5563}{{\tt
  1209.5563}}].

\bibitem{Boyarsky:2012rt}
A.~Boyarsky, D.~Iakubovskyi and O.~Ruchayskiy, \emph{{Next decade of sterile
  neutrino studies}},
  \href{http://dx.doi.org/10.1016/j.dark.2012.11.001}{\emph{Phys. Dark Univ.}
  {\bf 1} (2012) 136--154}, [\href{http://arxiv.org/abs/1306.4954}{{\tt
  1306.4954}}].

\bibitem{Abazajian:2005gj}
K.~Abazajian, \emph{{Production and evolution of perturbations of sterile
  neutrino dark matter}},
  \href{http://dx.doi.org/10.1103/PhysRevD.73.063506}{\emph{Phys. Rev.} {\bf
  D73} (2006) 063506}, [\href{http://arxiv.org/abs/astro-ph/0511630}{{\tt
  astro-ph/0511630}}].

\bibitem{Boyarsky:2006fg}
A.~Boyarsky, A.~Neronov, O.~Ruchayskiy, M.~Shaposhnikov and I.~Tkachev,
  \emph{{Where to find a dark matter sterile neutrino?}},
  \href{http://dx.doi.org/10.1103/PhysRevLett.97.261302}{\emph{Phys. Rev.
  Lett.} {\bf 97} (dec, 2006) 261302},
  [\href{http://arxiv.org/abs/astro-ph/0603660}{{\tt astro-ph/0603660}}].

\bibitem{Boyarsky:2007ay}
A.~Boyarsky, D.~Iakubovskyi, O.~Ruchayskiy and V.~Savchenko, \emph{{Constraints
  on decaying Dark Matter from XMM-Newton observations of M31}},
  \href{http://dx.doi.org/10.1111/j.1365-2966.2008.13266.x}{\emph{Mon. Not.
  Roy. Astron. Soc.} {\bf 387} (2008) 1361},
  [\href{http://arxiv.org/abs/0709.2301}{{\tt 0709.2301}}].

\bibitem{Boyarsky:2007ge}
A.~Boyarsky, D.~Malyshev, A.~Neronov and O.~Ruchayskiy, \emph{{Constraining DM
  properties with SPI}},
  \href{http://dx.doi.org/10.1111/j.1365-2966.2008.13003.x}{\emph{Mon. Not.
  Roy. Astron. Soc.} {\bf 387} (2008) 1345},
  [\href{http://arxiv.org/abs/0710.4922}{{\tt 0710.4922}}].

\bibitem{SHROCK1982382}
R.~E. Shrock, \emph{Pure leptonic decays with massive neutrinos and arbitrary
  lorentz structure},
  \href{http://dx.doi.org/http://dx.doi.org/10.1016/0370-2693(82)91074-7}{\emph{Physics
  Letters B} {\bf 112} (1982) 382 -- 386}.

\bibitem{Kusenko:2004qc}
A.~Kusenko, S.~Pascoli and D.~Semikoz, \emph{{New bounds on MeV sterile
  neutrinos based on the accelerator and Super-Kamiokande results}},
  \href{http://dx.doi.org/10.1088/1126-6708/2005/11/028}{\emph{JHEP} {\bf 0511}
  (2005) 028}, [\href{http://arxiv.org/abs/hep-ph/0405198}{{\tt
  hep-ph/0405198}}].

\bibitem{PIENU:2011aa}
{\scshape PIENU} collaboration, M.~Aoki et~al., \emph{{Search for Massive
  Neutrinos in the Decay $\pi \to e \nu$}},
  \href{http://dx.doi.org/10.1103/PhysRevD.84.052002}{\emph{Phys. Rev.} {\bf
  D84} (2011) 052002}, [\href{http://arxiv.org/abs/1106.4055}{{\tt
  1106.4055}}].

\bibitem{Artamonov:2014urb}
{\scshape E949} collaboration, A.~V. Artamonov et~al., \emph{{Search for heavy
  neutrinos in $K^+\to\mu^+\nu_H$ decays}},
  \href{http://dx.doi.org/10.1103/PhysRevD.91.059903,
  10.1103/PhysRevD.91.052001}{\emph{Phys. Rev.} {\bf D91} (2015) 052001},
  [\href{http://arxiv.org/abs/1411.3963}{{\tt 1411.3963}}].

\bibitem{Dolgov:2000pj}
A.~D. Dolgov, S.~H. Hansen, G.~Raffelt and D.~V. Semikoz, \emph{{Cosmological
  and astrophysical bounds on a heavy sterile neutrino and the KARMEN
  anomaly}}, \href{http://dx.doi.org/10.1016/S0550-3213(00)00203-0}{\emph{Nucl.
  Phys.} {\bf B580} (2000) 331--351},
  [\href{http://arxiv.org/abs/hep-ph/0002223}{{\tt hep-ph/0002223}}].

\bibitem{Dolgov:2000jw}
A.~D. Dolgov, S.~H. Hansen, G.~Raffelt and D.~V. Semikoz, \emph{{Heavy sterile
  neutrinos: Bounds from big bang nucleosynthesis and SN1987A}},
  \href{http://dx.doi.org/10.1016/S0550-3213(00)00566-6}{\emph{Nucl. Phys.}
  {\bf B590} (2000) 562--574}, [\href{http://arxiv.org/abs/hep-ph/0008138}{{\tt
  hep-ph/0008138}}].

\bibitem{Fuller:2009zz}
G.~M. Fuller, A.~Kusenko and K.~Petraki, \emph{{Heavy sterile neutrinos and
  supernova explosions}},
  \href{http://dx.doi.org/10.1016/j.physletb.2008.11.016}{\emph{Phys. Lett.}
  {\bf B670} (2009) 281--284}, [\href{http://arxiv.org/abs/0806.4273}{{\tt
  0806.4273}}].

\bibitem{Vincent:2014rja}
A.~C. Vincent, E.~F. Martinez, P.~Hern�ndez, M.~Lattanzi and O.~Mena,
  \emph{{Revisiting cosmological bounds on sterile neutrinos}},
  \href{http://dx.doi.org/10.1088/1475-7516/2015/04/006}{\emph{JCAP} {\bf 1504}
  (2015) 006}, [\href{http://arxiv.org/abs/1408.1956}{{\tt 1408.1956}}].

\bibitem{Gelmini:2008fq}
G.~Gelmini, E.~Osoba, S.~Palomares-Ruiz and S.~Pascoli, \emph{{MeV sterile
  neutrinos in low reheating temperature cosmological scenarios}},
  \href{http://dx.doi.org/10.1088/1475-7516/2008/10/029}{\emph{JCAP} {\bf 0810}
  (2008) 029}, [\href{http://arxiv.org/abs/0803.2735}{{\tt 0803.2735}}].

\bibitem{Fukugita:1986hr}
M.~Fukugita and T.~Yanagida, \emph{{Baryogenesis Without Grand Unification}},
  \href{http://dx.doi.org/10.1016/0370-2693(86)91126-3}{\emph{Phys. Lett.} {\bf
  B174} (1986) 45}.

\bibitem{Ibarra:2013cra}
A.~Ibarra, D.~Tran and C.~Weniger, \emph{{Indirect Searches for Decaying Dark
  Matter}}, \href{http://dx.doi.org/10.1142/S0217751X13300408}{\emph{Int. J.
  Mod. Phys.} {\bf A28} (2013) 1330040},
  [\href{http://arxiv.org/abs/1307.6434}{{\tt 1307.6434}}].

\bibitem{SHROCK1982359}
R.~E. Shrock, \emph{Electromagnetic properties and decays of dirac and majorana
  neutrinos in a general class of gauge theories},
  \href{http://dx.doi.org/http://dx.doi.org/10.1016/0550-3213(82)90273-5}{\emph{Nuclear
  Physics B} {\bf 206} (1982) 359 -- 379}.

\bibitem{Abramowski:2013ax}
{\scshape HESS} collaboration, A.~Abramowski et~al., \emph{{Search for
  Photon-Linelike Signatures from Dark Matter Annihilations with H.E.S.S.}},
  \href{http://dx.doi.org/10.1103/PhysRevLett.110.041301}{\emph{Phys. Rev.
  Lett.} {\bf 110} (2013) 041301}, [\href{http://arxiv.org/abs/1301.1173}{{\tt
  1301.1173}}].

\bibitem{Aleksic:2013xea}
J.~Aleksic, S.~Ansoldi, L.~Antonelli, P.~Antoranz, A.~Babic et~al.,
  \emph{{Optimized dark matter searches in deep observations of Segue 1 with
  MAGIC}}, \href{http://dx.doi.org/10.1088/1475-7516/2014/02/008}{\emph{JCAP}
  {\bf 1402} (2014) 008}, [\href{http://arxiv.org/abs/1312.1535}{{\tt
  1312.1535}}].

\bibitem{Buchmuller:1991tu}
W.~Buchmuller and C.~Greub, \emph{{Heavy Majorana neutrinos in electron -
  positron and electron - proton collisions}},
  \href{http://dx.doi.org/10.1016/0550-3213(91)80024-G}{\emph{Nucl. Phys.} {\bf
  B363} (1991) 345--368}.

\bibitem{Ackermann:2012qk}
{\scshape Fermi-LAT} collaboration, M.~Ackermann et~al., \emph{{Fermi LAT
  Search for Dark Matter in Gamma-ray Lines and the Inclusive Photon
  Spectrum}}, \href{http://dx.doi.org/10.1103/PhysRevD.86.022002}{\emph{Phys.
  Rev.} {\bf D86} (2012) 022002}, [\href{http://arxiv.org/abs/1205.2739}{{\tt
  1205.2739}}].

\bibitem{Dugger:2010ys}
L.~Dugger, T.~E. Jeltema and S.~Profumo, \emph{{Constraints on Decaying Dark
  Matter from Fermi Observations of Nearby Galaxies and Clusters}},
  \href{http://dx.doi.org/10.1088/1475-7516/2010/12/015}{\emph{JCAP} {\bf 1012}
  (2010) 015}, [\href{http://arxiv.org/abs/1009.5988}{{\tt 1009.5988}}].

\bibitem{Garny:2012vt}
M.~Garny, A.~Ibarra and D.~Tran, \emph{{Constraints on Hadronically Decaying
  Dark Matter}},
  \href{http://dx.doi.org/10.1088/1475-7516/2012/08/025}{\emph{JCAP} {\bf 1208}
  (2012) 025}, [\href{http://arxiv.org/abs/1205.6783}{{\tt 1205.6783}}].

\bibitem{Cholis:2010xb}
I.~Cholis, \emph{{New Constraints from PAMELA anti-proton data on Annihilating
  and Decaying Dark Matter}},
  \href{http://dx.doi.org/10.1088/1475-7516/2011/09/007}{\emph{JCAP} {\bf 1109}
  (2011) 007}, [\href{http://arxiv.org/abs/1007.1160}{{\tt 1007.1160}}].

\bibitem{Adriani:2010rc}
{\scshape PAMELA} collaboration, O.~Adriani et~al., \emph{{PAMELA results on
  the cosmic-ray antiproton flux from 60 MeV to 180 GeV in kinetic energy}},
  \href{http://dx.doi.org/10.1103/PhysRevLett.105.121101}{\emph{Phys. Rev.
  Lett.} {\bf 105} (2010) 121101}, [\href{http://arxiv.org/abs/1007.0821}{{\tt
  1007.0821}}].

\bibitem{Ibarra:2013zia}
A.~Ibarra, A.~S. Lamperstorfer and J.~Silk, \emph{{Dark matter annihilations
  and decays after the AMS-02 positron measurements}},
  \href{http://dx.doi.org/10.1103/PhysRevD.89.063539}{\emph{Phys. Rev.} {\bf
  D89} (2014) 063539}, [\href{http://arxiv.org/abs/1309.2570}{{\tt
  1309.2570}}].

\bibitem{Aguilar:2014fea}
{\scshape AMS} collaboration, M.~Aguilar et~al., \emph{{Precision Measurement
  of the ($e^+ + e^?$) Flux in Primary Cosmic Rays from 0.5 GeV to 1 TeV with
  the Alpha Magnetic Spectrometer on the International Space Station}},
  \href{http://dx.doi.org/10.1103/PhysRevLett.113.221102}{\emph{Phys. Rev.
  Lett.} {\bf 113} (2014) 221102}.

\bibitem{Aguilar:2014mma}
{\scshape AMS} collaboration, M.~Aguilar et~al., \emph{{Electron and Positron
  Fluxes in Primary Cosmic Rays Measured with the Alpha Magnetic Spectrometer
  on the International Space Station}},
  \href{http://dx.doi.org/10.1103/PhysRevLett.113.121102}{\emph{Phys. Rev.
  Lett.} {\bf 113} (2014) 121102}.

\bibitem{Achterberg:2007qp}
{\scshape IceCube} collaboration, A.~Achterberg et~al., \emph{{Multi-year
  search for a diffuse flux of muon neutrinos with AMANDA-II}},
  \href{http://dx.doi.org/10.1103/PhysRevD.76.042008,
  10.1103/PhysRevD.77.089904}{\emph{Phys. Rev.} {\bf D76} (2007) 042008},
  [\href{http://arxiv.org/abs/0705.1315}{{\tt 0705.1315}}].

\bibitem{Abbasi:2012cu}
{\scshape IceCube} collaboration, R.~Abbasi et~al., \emph{{A Search for UHE Tau
  Neutrinos with IceCube}},
  \href{http://dx.doi.org/10.1103/PhysRevD.86.022005}{\emph{Phys. Rev.} {\bf
  D86} (2012) 022005}, [\href{http://arxiv.org/abs/1202.4564}{{\tt
  1202.4564}}].

\bibitem{Abbasi:2011ji}
{\scshape IceCube} collaboration, R.~Abbasi et~al., \emph{{Constraints on the
  Extremely-high Energy Cosmic Neutrino Flux with the IceCube 2008-2009 Data}},
  \href{http://dx.doi.org/10.1103/PhysRevD.84.079902,
  10.1103/PhysRevD.83.092003}{\emph{Phys. Rev.} {\bf D83} (2011) 092003},
  [\href{http://arxiv.org/abs/1103.4250}{{\tt 1103.4250}}].

\bibitem{Abreu:2011zze}
{\scshape Pierre Auger} collaboration, P.~Abreu et~al., \emph{{A Search for
  Ultra-High Energy Neutrinos in Highly Inclined Events at the Pierre Auger
  Observatory}}, \href{http://dx.doi.org/10.1103/PhysRevD.85.029902,
  10.1103/PhysRevD.84.122005}{\emph{Phys. Rev.} {\bf D84} (2011) 122005},
  [\href{http://arxiv.org/abs/1202.1493}{{\tt 1202.1493}}].

\bibitem{Gorham:2010kv}
{\scshape ANITA} collaboration, P.~W. Gorham et~al., \emph{{Observational
  Constraints on the Ultra-high Energy Cosmic Neutrino Flux from the Second
  Flight of the ANITA Experiment}},
  \href{http://dx.doi.org/10.1103/PhysRevD.82.022004,
  10.1103/PhysRevD.85.049901}{\emph{Phys. Rev.} {\bf D82} (2010) 022004},
  [\href{http://arxiv.org/abs/1011.5004}{{\tt 1011.5004}}].

\bibitem{Esmaili:2012us}
A.~Esmaili, A.~Ibarra and O.~L.~G. Peres, \emph{{Probing the stability of
  superheavy dark matter particles with high-energy neutrinos}},
  \href{http://dx.doi.org/10.1088/1475-7516/2012/11/034}{\emph{JCAP} {\bf 1211}
  (2012) 034}, [\href{http://arxiv.org/abs/1205.5281}{{\tt 1205.5281}}].

\bibitem{Barbieri:1979ag}
R.~Barbieri, D.~V. Nanopoulos, G.~Morchio and F.~Strocchi, \emph{{Neutrino
  Masses in Grand Unified Theories}},
  \href{http://dx.doi.org/10.1016/0370-2693(80)90058-1}{\emph{Phys. Lett.} {\bf
  B90} (1980) 91}.

\bibitem{Davidson:2002qv}
S.~Davidson and A.~Ibarra, \emph{{A Lower bound on the right-handed neutrino
  massfrom leptogenesis}},
  \href{http://dx.doi.org/10.1016/S0370-2693(02)01735-5}{\emph{Phys. Lett.}
  {\bf B535} (2002) 25--32}, [\href{http://arxiv.org/abs/hep-ph/0202239}{{\tt
  hep-ph/0202239}}].

\bibitem{DiBari:2005st}
P.~Di~Bari, \emph{{Seesaw geometry and leptogenesis}},
  \href{http://dx.doi.org/10.1016/j.nuclphysb.2005.08.032}{\emph{Nucl. Phys.}
  {\bf B727} (2005) 318--354}, [\href{http://arxiv.org/abs/hep-ph/0502082}{{\tt
  hep-ph/0502082}}].

\bibitem{Vives:2005ra}
O.~Vives, \emph{{Flavor dependence of CP asymmetries and thermal leptogenesis
  with strong right-handed neutrino mass hierarchy}},
  \href{http://dx.doi.org/10.1103/PhysRevD.73.073006}{\emph{Phys. Rev.} {\bf
  D73} (2006) 073006}, [\href{http://arxiv.org/abs/hep-ph/0512160}{{\tt
  hep-ph/0512160}}].

\bibitem{Blanchet:2008pw}
S.~Blanchet and P.~Di~Bari, \emph{{New aspects of leptogenesis bounds}},
  \href{http://dx.doi.org/10.1016/j.nuclphysb.2008.08.026}{\emph{Nucl. Phys.}
  {\bf B807} (2009) 155--187}, [\href{http://arxiv.org/abs/0807.0743}{{\tt
  0807.0743}}].

\bibitem{Barbieri:1999ma}
R.~Barbieri, A.~Creminelli, Paolo~andStrumia and N.~Tetradis,
  \emph{{Baryogenesis through leptogenesis}},
  \href{http://dx.doi.org/10.1016/S0550-3213(00)00011-0}{\emph{Nucl. Phys.}
  {\bf B575} (2000) 61--77}, [\href{http://arxiv.org/abs/hep-ph/9911315}{{\tt
  hep-ph/9911315}}].

\bibitem{Engelhard:2006yg}
G.~Engelhard, Y.~Grossman, E.~Nardi and Y.~Nir, \emph{{The Importance of N2
  leptogenesis}},
  \href{http://dx.doi.org/10.1103/PhysRevLett.99.081802}{\emph{Phys. Rev.
  Lett.} {\bf 99} (2007) 081802},
  [\href{http://arxiv.org/abs/hep-ph/0612187}{{\tt hep-ph/0612187}}].

\bibitem{Bertuzzo:2009im}
E.~Bertuzzo, P.~Di~Bari and E.~Feruglio, F.~andNardi, \emph{{Flavor symmetries,
  leptogenesis and the absolute neutrino mass scale}},
  \href{http://dx.doi.org/10.1088/1126-6708/2009/11/036}{\emph{JHEP} {\bf 0911}
  (2009) 036}, [\href{http://arxiv.org/abs/0908.0161}{{\tt 0908.0161}}].

\bibitem{Antusch:2011nz}
S.~Antusch, P.~Di~Bari and S.~F. Jones, D. A.~andKing, \emph{{Leptogenesis in
  the Two Right-Handed NeutrinoModel Revisited}},
  \href{http://dx.doi.org/10.1103/PhysRevD.86.023516}{\emph{Phys. Rev.} {\bf
  D86} (2012) 023516}, [\href{http://arxiv.org/abs/1107.6002}{{\tt
  1107.6002}}].

\bibitem{Nardi:2006fx}
E.~Nardi, Y.~Nir and J.~Roulet, Estebanand~Racker, \emph{{The Importance of
  flavor in leptogenesis}},
  \href{http://dx.doi.org/10.1088/1126-6708/2006/01/164}{\emph{JHEP} {\bf 0601}
  (2006) 164}, [\href{http://arxiv.org/abs/hep-ph/0601084}{{\tt
  hep-ph/0601084}}].

\bibitem{Abada:2006fw}
A.~Abada, S.~Davidson, F.-X. Josse-Michaux, M.~Losada and A.~Riotto,
  \emph{{Flavor issues in leptogenesis}},
  \href{http://dx.doi.org/10.1088/1475-7516/2006/04/004}{\emph{JCAP} {\bf 0604}
  (2006) 004}, [\href{http://arxiv.org/abs/hep-ph/0601083}{{\tt
  hep-ph/0601083}}].

\bibitem{Blanchet:2006be}
S.~Blanchet and P.~Di~Bari, \emph{{Flavor effects on leptogenesis
  predictions}},
  \href{http://dx.doi.org/10.1088/1475-7516/2007/03/018}{\emph{JCAP} {\bf 0703}
  (2007) 018}, [\href{http://arxiv.org/abs/hep-ph/0607330}{{\tt
  hep-ph/0607330}}].

\bibitem{Buchmuller:2003gz}
W.~Buchmuller, P.~Di~Bari and M.~Plumacher, \emph{{The Neutrino mass window for
  baryogenesis}},
  \href{http://dx.doi.org/10.1016/S0550-3213(03)00449-8}{\emph{Nucl. Phys.}
  {\bf B665} (2003) 445--468}, [\href{http://arxiv.org/abs/hep-ph/0302092}{{\tt
  hep-ph/0302092}}].

\bibitem{Giudice:2003jh}
G.~F. Giudice, A.~Notari, A.~Raidal, M.~andRiotto and A.~Strumia,
  \emph{{Towards a complete theory of thermal leptogenesisin the SM and MSSM}},
  \href{http://dx.doi.org/10.1016/j.nuclphysb.2004.02.019}{\emph{Nucl. Phys.}
  {\bf B685} (2004) 89--149}, [\href{http://arxiv.org/abs/hep-ph/0310123}{{\tt
  hep-ph/0310123}}].

\bibitem{Buchmuller:2004nz}
W.~Buchmuller, P.~Di~Bari and M.~Plumacher, \emph{{Leptogenesis for
  pedestrians}},
  \href{http://dx.doi.org/10.1016/j.aop.2004.02.003}{\emph{Annals Phys.} {\bf
  315} (2005) 305--351}, [\href{http://arxiv.org/abs/hep-ph/0401240}{{\tt
  hep-ph/0401240}}].

\bibitem{Pilaftsis:1997jf}
A.~Pilaftsis, \emph{{CP violation and baryogenesis due to heavyMajorana
  neutrinos}}, \href{http://dx.doi.org/10.1103/PhysRevD.56.5431}{\emph{Phys.
  Rev.} {\bf D56} (1997) 5431--5451},
  [\href{http://arxiv.org/abs/hep-ph/9707235}{{\tt hep-ph/9707235}}].

\bibitem{Pilaftsis:2005rv}
A.~Pilaftsis and T.~E.~J. Underwood, \emph{{Electroweak-scale resonant
  leptogenesis}},
  \href{http://dx.doi.org/10.1103/PhysRevD.72.113001}{\emph{Phys. Rev.} {\bf
  D72} (2005) 113001}, [\href{http://arxiv.org/abs/hep-ph/0506107}{{\tt
  hep-ph/0506107}}].

\bibitem{Blanchet:2009kk}
S.~Blanchet and F.-X. Hambye, Thomas andJosse-Michaux, \emph{{Reconciling
  leptogenesis with observable mu --->e gamma rates}},
  \href{http://dx.doi.org/10.1007/JHEP04(2010)023}{\emph{JHEP} {\bf 1004}
  (2010) 023}, [\href{http://arxiv.org/abs/0912.3153}{{\tt 0912.3153}}].

\bibitem{Akhmedov:1998qx}
E.~K. Akhmedov, V.~A. Rubakov and A.~Y. Smirnov, \emph{{Baryogenesis via
  neutrino oscillations}},
  \href{http://dx.doi.org/10.1103/PhysRevLett.81.1359}{\emph{Phys. Rev. Lett.}
  {\bf 81} (1998) 1359--1362}, [\href{http://arxiv.org/abs/hep-ph/9803255}{{\tt
  hep-ph/9803255}}].

\bibitem{Dick:1999je}
K.~Dick, M.~Lindner, M.~Ratz and D.~Wright, \emph{{Leptogenesis with Dirac
  neutrinos}}, \href{http://dx.doi.org/10.1103/PhysRevLett.84.4039}{\emph{Phys.
  Rev. Lett.} {\bf 84} (2000) 4039--4042},
  [\href{http://arxiv.org/abs/hep-ph/9907562}{{\tt hep-ph/9907562}}].

\bibitem{Canetti:2012vf}
L.~Canetti, M.~Drewes and M.~Shaposhnikov, \emph{{Sterile Neutrinos as the
  Origin of Dark and Baryonic Matter}},
  \href{http://dx.doi.org/10.1103/PhysRevLett.110.061801}{\emph{Phys. Rev.
  Lett.} {\bf 110} (2013) 061801}, [\href{http://arxiv.org/abs/1204.3902}{{\tt
  1204.3902}}].

\bibitem{Canetti:2012kh}
L.~Canetti, M.~Drewes, T.~Frossard and M.~Shaposhnikov, \emph{{Dark Matter,
  Baryogenesis and Neutrino Oscillations from Right Handed Neutrinos}},
  \href{http://dx.doi.org/10.1103/PhysRevD.87.093006}{\emph{Phys. Rev.} {\bf
  D87} (2013) 093006}, [\href{http://arxiv.org/abs/1208.4607}{{\tt
  1208.4607}}].

\bibitem{Shaposhnikov:2008pf}
M.~Shaposhnikov, \emph{{The nuMSM, leptonic asymmetries, and properties of
  singlet fermions}},
  \href{http://dx.doi.org/10.1088/1126-6708/2008/08/008}{\emph{JHEP} {\bf 0808}
  (2008) 008}, [\href{http://arxiv.org/abs/0804.4542}{{\tt 0804.4542}}].

\bibitem{Garbrecht:2014bfa}
B.~Garbrecht, \emph{{More Viable Parameter Space for Leptogenesis}},
  \href{http://dx.doi.org/10.1103/PhysRevD.90.063522}{\emph{Phys. Rev.} {\bf
  D90} (2014) 063522}, [\href{http://arxiv.org/abs/1401.3278}{{\tt
  1401.3278}}].

\bibitem{Drewes:2012ma}
M.~Drewes and B.~Garbrecht, \emph{{Leptogenesis from a GeV Seesaw without Mass
  Degeneracy}}, \href{http://dx.doi.org/10.1007/JHEP03(2013)096}{\emph{JHEP}
  {\bf 03} (2013) 096}, [\href{http://arxiv.org/abs/1206.5537}{{\tt
  1206.5537}}].

\bibitem{Alekhin:2015byh}
S.~Alekhin et~al., \emph{{A facility to Search for Hidden Particles at the CERN
  SPS: the SHiP physics case}},  \href{http://arxiv.org/abs/1504.04855}{{\tt
  1504.04855}}.

\bibitem{Cvetic:2014nla}
G.~Cveti?, C.~S. Kim and J.~Zamora-Sa�, \emph{{CP violation in lepton number
  violating semihadronic decays of $K,D,D_s,B,B_c$}},
  \href{http://dx.doi.org/10.1103/PhysRevD.89.093012}{\emph{Phys. Rev.} {\bf
  D89} (2014) 093012}, [\href{http://arxiv.org/abs/1403.2555}{{\tt
  1403.2555}}].

\bibitem{Shuve:2014zua}
B.~Shuve and I.~Yavin, \emph{{Baryogenesis through Neutrino Oscillations:
  AUnified Perspective}},
  \href{http://dx.doi.org/10.1103/PhysRevD.89.075014}{\emph{Phys. Rev.} {\bf
  D89} (2014) 075014}, [\href{http://arxiv.org/abs/1401.2459}{{\tt
  1401.2459}}].

\bibitem{Anisimov:2006hv}
A.~Anisimov, \emph{{Majorana Dark Matter}},
  \href{http://arxiv.org/abs/hep-ph/0612024}{{\tt hep-ph/0612024}}.

\bibitem{Anisimov:2008gg}
A.~Anisimov and P.~Di~Bari, \emph{{Cold Dark Matter from heavy Right-Handed
  neutrino mixing}},
  \href{http://dx.doi.org/10.1103/PhysRevD.80.073017}{\emph{Phys. Rev.} {\bf
  D80} (2009) 073017}, [\href{http://arxiv.org/abs/0812.5085}{{\tt
  0812.5085}}].

\bibitem{Hambye:2012fh}
T.~Hambye, \emph{{Leptogenesis: beyond the minimal type I seesaw scenario}},
  \href{http://dx.doi.org/10.1088/1367-2630/14/12/125014}{\emph{New J. Phys.}
  {\bf 14} (2012) 125014}, [\href{http://arxiv.org/abs/1212.2888}{{\tt
  1212.2888}}].

\bibitem{Branco:2002kt}
G.~Branco, R.~Gonzalez~Felipe, F.~R. Joaquim and M.~N. Rebelo,
  \emph{{Leptogenesis, CP violation and neutrino data:What can we learn?}},
  \href{http://dx.doi.org/10.1016/S0550-3213(02)00478-9}{\emph{Nucl. Phys.}
  {\bf B640} (2002) 202--232}, [\href{http://arxiv.org/abs/hep-ph/0202030}{{\tt
  hep-ph/0202030}}].

\bibitem{Akhmedov:2003dg}
E.~K. Akhmedov and A.~Y. Frigerio, Michele~andSmirnov, \emph{{Probing the
  seesaw mechanism with neutrino dataand leptogenesis}},
  \href{http://dx.doi.org/10.1088/1126-6708/2003/09/021}{\emph{JHEP} {\bf 0309}
  (2003) 021}, [\href{http://arxiv.org/abs/hep-ph/0305322}{{\tt
  hep-ph/0305322}}].

\bibitem{DiBari:2008mp}
P.~Di~Bari and A.~Riotto, \emph{{Successful type I Leptogenesis
  withSO(10)-inspired mass relations}},
  \href{http://dx.doi.org/10.1016/j.physletb.2008.12.054}{\emph{Phys. Lett.}
  {\bf B671} (2009) 462--469}, [\href{http://arxiv.org/abs/0809.2285}{{\tt
  0809.2285}}].

\bibitem{DiBari:2010ux}
P.~Di~Bari and A.~Riotto, \emph{{Testing SO(10)-inspired leptogenesis with
  lowenergy neutrino experiments}},
  \href{http://dx.doi.org/10.1088/1475-7516/2011/04/037}{\emph{JCAP} {\bf 1104}
  (2011) 037}, [\href{http://arxiv.org/abs/1012.2343}{{\tt 1012.2343}}].

\bibitem{DiBari:2013qja}
P.~Di~Bari and L.~Marzola, \emph{{SO(10)-inspired solution to the problem of
  theinitial conditions in leptogenesis}},
  \href{http://dx.doi.org/10.1016/j.nuclphysb.2013.10.027}{\emph{Nucl. Phys.}
  {\bf B877} (2013) 719--751}, [\href{http://arxiv.org/abs/1308.1107}{{\tt
  1308.1107}}].

\bibitem{DiBari:2014eqa}
P.~Di~Bari, S.~King and M.~ReFiorentin, \emph{{Strong thermal leptogenesis and
  the absoluteneutrino mass scale}},
  \href{http://dx.doi.org/10.1088/1475-7516/2014/03/050}{\emph{JCAP} {\bf 1403}
  (2014) 050}, [\href{http://arxiv.org/abs/1401.6185}{{\tt 1401.6185}}].

\bibitem{DiBari:2014eya}
P.~Di~Bari, L.~Marzola and M.~ReFiorentin, \emph{{Decrypting $SO(10)$-inspired
  leptogenesis}},
  \href{http://dx.doi.org/10.1016/j.nuclphysb.2015.02.005}{\emph{Nucl. Phys.}
  {\bf B893} (2015) 122--157}, [\href{http://arxiv.org/abs/1411.5478}{{\tt
  1411.5478}}].

\bibitem{Hooper:2010mq}
D.~Hooper and L.~Goodenough, \emph{{Dark Matter Annihilation in The Galactic
  Center As Seen by the Fermi Gamma Ray Space Telescope}},
  \href{http://dx.doi.org/10.1016/j.physletb.2011.02.029}{\emph{Phys. Lett.}
  {\bf B697} (2011) 412--428}, [\href{http://arxiv.org/abs/1010.2752}{{\tt
  1010.2752}}].

\bibitem{Bullock:2000qf}
J.~S. Bullock, A.~V. Kravtsov and D.~H. Weinberg, \emph{{Hierarchical galaxy
  formation and substructure in the Galaxy's stellar halo}},
  \href{http://dx.doi.org/10.1086/318681}{\emph{Astrophys. J.} {\bf 548} (2001)
  33--46}, [\href{http://arxiv.org/abs/astro-ph/0007295}{{\tt
  astro-ph/0007295}}].

\bibitem{Benson:2001at}
A.~J. Benson, C.~S. Frenk, C.~G. Lacey, C.~M. Baugh and S.~Cole, \emph{{The
  effects of photoionization on galaxy formation. 2. Satellites in the local
  group}}, \href{http://dx.doi.org/10.1046/j.1365-8711.2002.05388.x}{\emph{Mon.
  Not. Roy. Astron. Soc.} {\bf 333} (2002) 177},
  [\href{http://arxiv.org/abs/astro-ph/0108218}{{\tt astro-ph/0108218}}].

\bibitem{Navarro:1996gj}
J.~F. Navarro, C.~S. Frenk and S.~D.~M. White, \emph{{A Universal density
  profile from hierarchical clustering}},
  \href{http://dx.doi.org/10.1086/304888}{\emph{Astrophys. J.} {\bf 490} (1997)
  493--508}, [\href{http://arxiv.org/abs/astro-ph/9611107}{{\tt
  astro-ph/9611107}}].

\bibitem{Horiuchi:2015qri}
S.~Horiuchi, B.~Bozek, K.~N. Abazajian, M.~Boylan-Kolchin, J.~S. Bullock,
  S.~Garrison-Kimmel et~al., \emph{{Properties of resonantly produced sterile
  neutrino dark matter subhaloes}},
  \href{http://dx.doi.org/10.1093/mnras/stv2922}{\emph{Mon. Not. Roy. Astron.
  Soc.} {\bf 456} (2016) 4346--4353},
  [\href{http://arxiv.org/abs/1512.04548}{{\tt 1512.04548}}].

\bibitem{Sawala:2014baa}
T.~Sawala et~al., \emph{{Bent by baryons: the low mass galaxy-halo relation}},
  \href{http://dx.doi.org/10.1093/mnras/stu2753}{\emph{Mon. Not. Roy. Astron.
  Soc.} {\bf 448} (2015) 2941--2947},
  [\href{http://arxiv.org/abs/1404.3724}{{\tt 1404.3724}}].

\bibitem{Kauffmann:1993gv}
G.~Kauffmann, S.~D.~M. White and B.~Guiderdoni, \emph{{The Formation and
  Evolution of Galaxies Within Merging Dark Matter Haloes}}, {\emph{Mon. Not.
  Roy. Astron. Soc.} {\bf 264} (1993) 201}.

\bibitem{Cole:1994ab}
S.~Cole, A.~Aragon-Salamanca, C.~S. Frenk, J.~F. Navarro and S.~E. Zepf,
  \emph{{A Recipe for galaxy formation}}, {\emph{Mon. Not. Roy. Astron. Soc.}
  {\bf 271} (1994) 781}, [\href{http://arxiv.org/abs/astro-ph/9402001}{{\tt
  astro-ph/9402001}}].

\bibitem{Monaco:1999cu}
P.~Monaco, P.~Salucci and L.~Danese, \emph{{Joint cosmological formation of
  qsos and bulge-dominated galaxies}},
  \href{http://dx.doi.org/10.1046/j.1365-8711.2000.03043.x}{\emph{Mon. Not.
  Roy. Astron. Soc.} {\bf 311} (2000) 279},
  [\href{http://arxiv.org/abs/astro-ph/9907095}{{\tt astro-ph/9907095}}].

\bibitem{Kauffmann:1999ce}
G.~Kauffmann and M.~Haehnelt, \emph{{A Unified model for the evolution of
  galaxies and quasars}},
  \href{http://dx.doi.org/10.1046/j.1365-8711.2000.03077.x}{\emph{Mon. Not.
  Roy. Astron. Soc.} {\bf 311} (2000) 576--588},
  [\href{http://arxiv.org/abs/astro-ph/9906493}{{\tt astro-ph/9906493}}].

\bibitem{Granato:2003ch}
G.~L. Granato, G.~De~Zotti, L.~Silva, A.~Bressan and L.~Danese, \emph{{A
  physical model for the co-evolution of qsos and of their spheroidal hosts}},
  \href{http://dx.doi.org/10.1086/379875}{\emph{Astrophys. J.} {\bf 600} (2004)
  580--594}, [\href{http://arxiv.org/abs/astro-ph/0307202}{{\tt
  astro-ph/0307202}}].

\bibitem{Menci:2006me}
N.~Menci, A.~Fontana, E.~Giallongo, A.~Grazian and S.~Salimbeni, \emph{{The
  Abundance of Distant and Extremely Red Galaxies: The Role of AGN Feedback in
  Hierarchical Models}},
  \href{http://dx.doi.org/10.1086/505528}{\emph{Astrophys. J.} {\bf 647} (2006)
  753--762}, [\href{http://arxiv.org/abs/astro-ph/0605123}{{\tt
  astro-ph/0605123}}].

\bibitem{Croton:2005fe}
D.~J. Croton, V.~Springel, S.~D.~M. White, G.~De~Lucia, C.~S. Frenk, L.~Gao
  et~al., \emph{{The Many lives of AGN: Cooling flows, black holes and the
  luminosities and colours of galaxies}},
  \href{http://dx.doi.org/10.1111/j.1365-2966.2006.09994.x,
  10.1111/j.1365-2966.2005.09675.x}{\emph{Mon. Not. Roy. Astron. Soc.} {\bf
  365} (2006) 11--28}, [\href{http://arxiv.org/abs/astro-ph/0602065}{{\tt
  astro-ph/0602065}}].

\bibitem{Bower:2005vb}
R.~G. Bower, A.~J. Benson, R.~Malbon, J.~C. Helly, C.~S. Frenk, C.~M. Baugh
  et~al., \emph{{The broken hierarchy of galaxy formation}},
  \href{http://dx.doi.org/10.1111/j.1365-2966.2006.10519.x}{\emph{Mon. Not.
  Roy. Astron. Soc.} {\bf 370} (2006) 645--655},
  [\href{http://arxiv.org/abs/astro-ph/0511338}{{\tt astro-ph/0511338}}].

\bibitem{Somerville:1998bb}
R.~S. Somerville and J.~R. Primack, \emph{{Semianalytic modeling of galaxy
  formation. The Local Universe}},
  \href{http://dx.doi.org/10.1046/j.1365-8711.1999.03032.x}{\emph{Mon. Not.
  Roy. Astron. Soc.} {\bf 310} (1999) 1087},
  [\href{http://arxiv.org/abs/astro-ph/9802268}{{\tt astro-ph/9802268}}].

\bibitem{Cole:2000ex}
S.~Cole, C.~G. Lacey, C.~M. Baugh and C.~S. Frenk, \emph{{Hierarchical galaxy
  formation}},
  \href{http://dx.doi.org/10.1046/j.1365-8711.2000.03879.x}{\emph{Mon. Not.
  Roy. Astron. Soc.} {\bf 319} (2000) 168},
  [\href{http://arxiv.org/abs/astro-ph/0007281}{{\tt astro-ph/0007281}}].

\bibitem{Poli:2001fq}
F.~Poli, N.~Menci, E.~Giallongo, A.~Fontana, S.~Cristiani and S.~D'Odorico,
  \emph{{The evolution of the luminosity function in deep fields: a comparison
  with cdm models}}, \href{http://dx.doi.org/10.1086/319840}{\emph{Astrophys.
  J.} {\bf 551} (2001) L45}, [\href{http://arxiv.org/abs/astro-ph/0103113}{{\tt
  astro-ph/0103113}}].

\bibitem{Cirasuolo:2008en}
M.~Cirasuolo, R.~J. McLure, J.~S. Dunlop, O.~Almaini, S.~Foucaud and
  C.~Simpson, \emph{{A new measurement of the evolving near-infrared galaxy
  luminosity function out to z~4: a continuing challenge to theoretical models
  of galaxy formation}},
  \href{http://dx.doi.org/10.1111/j.1365-2966.2009.15710.x}{\emph{Mon. Not.
  Roy. Astron. Soc.} {\bf 401} (2010) 1166},
  [\href{http://arxiv.org/abs/0804.3471}{{\tt 0804.3471}}].

\bibitem{Faro:2009si}
B.~L. Faro, P.~Monaco, E.~Vanzella, F.~Fontanot, L.~Silva and S.~Cristiani,
  \emph{{Faint Lyman-Break galaxies as a crucial test for galaxy formation
  models}},
  \href{http://dx.doi.org/10.1111/j.1365-2966.2009.15316.x}{\emph{Mon. Not.
  Roy. Astron. Soc.} {\bf 399} (2009) 827},
  [\href{http://arxiv.org/abs/0906.4998}{{\tt 0906.4998}}].

\bibitem{Salimbeni:2007dq}
S.~Salimbeni et~al., \emph{{The red and blue galaxy populations in the GOODS
  field: Evidence for an excess of red dwarfs}},
  \href{http://dx.doi.org/10.1051/0004-6361:20077959}{\emph{Astron. Astrophys.}
  (2007) }, [\href{http://arxiv.org/abs/0710.3704}{{\tt 0710.3704}}].

\bibitem{Fontana:2006xg}
A.~Fontana et~al., \emph{{The Galaxy Mass Function up to z=4 in the GOODS-MUSIC
  sample: iIto the epoch of formation of massive galaxies}},
  \href{http://dx.doi.org/10.1051/0004-6361:20065475}{\emph{Astron. Astrophys.}
  {\bf 459} (2006) 745--757},
  [\href{http://arxiv.org/abs/astro-ph/0609068}{{\tt astro-ph/0609068}}].

\bibitem{Fontanot:2009sy}
F.~Fontanot, G.~De~Lucia, P.~Monaco, R.~S. Somerville and P.~Santini,
  \emph{{The Many Manifestations of Downsizing: Hierarchical Galaxy Formation
  Models confront Observations}},
  \href{http://dx.doi.org/10.1111/j.1365-2966.2009.15058.x}{\emph{Mon. Not.
  Roy. Astron. Soc.} {\bf 397} (2009) 1776},
  [\href{http://arxiv.org/abs/0901.1130}{{\tt 0901.1130}}].

\bibitem{Marchesini:2008xk}
D.~Marchesini, P.~G. van Dokkum, N.~M.~F. Schreiber, M.~Franx, I.~Labbe' and
  S.~Wuyts, \emph{{The Evolution of the Stellar Mass Function of Galaxies from
  z=4.0 and the First Comprehensive Analysis of its Uncertainties: Evidence for
  Mass-dependent Evolution}},
  \href{http://dx.doi.org/10.1088/0004-637X/701/2/1765}{\emph{Astrophys. J.}
  {\bf 701} (2009) 1765--1796}, [\href{http://arxiv.org/abs/0811.1773}{{\tt
  0811.1773}}].

\bibitem{Guo:2010ap}
Q.~Guo, S.~White, M.~Boylan-Kolchin, G.~De~Lucia, G.~Kauffmann, G.~Lemson
  et~al., \emph{{From dwarf spheroidals to cDs: Simulating the galaxy
  population in a LCDM cosmology}},
  \href{http://dx.doi.org/10.1111/j.1365-2966.2010.18114.x}{\emph{Mon. Not.
  Roy. Astron. Soc.} {\bf 413} (2011) 101},
  [\href{http://arxiv.org/abs/1006.0106}{{\tt 1006.0106}}].

\bibitem{Santini:2011fr}
P.~Santini et~al., \emph{{The evolving slope of the stellar mass function at
  0.6 <= z < 4.5 from deep WFC3 data}},
  \href{http://dx.doi.org/10.1051/0004-6361/201117513}{\emph{Astron.
  Astrophys.} {\bf 538} (2012) A33},
  [\href{http://arxiv.org/abs/1111.5728}{{\tt 1111.5728}}].

\bibitem{Papastergis:2011xe}
E.~Papastergis, A.~M. Martin, R.~Giovanelli and M.~P. Haynes, \emph{{The
  velocity width function of galaxies from the 40\% ALFALFA survey: shedding
  light on the cold dark matter overabundance problem}},
  \href{http://dx.doi.org/10.1088/0004-637X/739/1/38}{\emph{Astrophys. J.} {\bf
  739} (2011) 38}, [\href{http://arxiv.org/abs/1106.0710}{{\tt 1106.0710}}].

\bibitem{Simon:2007dq}
J.~D. Simon and M.~Geha, \emph{{The Kinematics of the Ultra-Faint Milky Way
  Satellites: Solving the Missing Satellite Problem}},
  \href{http://dx.doi.org/10.1086/521816}{\emph{Astrophys. J.} {\bf 670} (2007)
  313--331}, [\href{http://arxiv.org/abs/0706.0516}{{\tt 0706.0516}}].

\bibitem{Bullock:2000wn}
J.~S. Bullock, A.~V. Kravtsov and D.~H. Weinberg, \emph{{Reionization and the
  abundance of galactic satellites}},
  \href{http://dx.doi.org/10.1086/309279}{\emph{Astrophys. J.} {\bf 539} (2000)
  517}, [\href{http://arxiv.org/abs/astro-ph/0002214}{{\tt astro-ph/0002214}}].

\bibitem{Somerville:2001km}
R.~S. Somerville, \emph{{Can photoionization squelching resolve the
  sub-structure crisis?}},
  \href{http://dx.doi.org/10.1086/341444}{\emph{Astrophys. J.} {\bf 572} (2002)
  L23--L26}, [\href{http://arxiv.org/abs/astro-ph/0107507}{{\tt
  astro-ph/0107507}}].

\bibitem{Governato:2006cq}
F.~Governato, B.~Willman, L.~Mayer, A.~Brooks, G.~Stinson, O.~Valenzuela
  et~al., \emph{{Forming disk galaxies in lambda-cdm simulations}},
  \href{http://dx.doi.org/10.1111/j.1365-2966.2006.11266.x}{\emph{Mon. Not.
  Roy. Astron. Soc.} {\bf 374} (2007) 1479--1494},
  [\href{http://arxiv.org/abs/astro-ph/0602351}{{\tt astro-ph/0602351}}].

\bibitem{Mashchenko:2007jp}
S.~Mashchenko, J.~Wadsley and H.~M.~P. Couchman, \emph{{Stellar Feedback in
  Dwarf Galaxy Formation}},
  \href{http://dx.doi.org/10.1126/science.1148666}{\emph{Science} {\bf 319}
  (2008) 174}, [\href{http://arxiv.org/abs/0711.4803}{{\tt 0711.4803}}].

\bibitem{Nierenberg:2013lqa}
A.~M. Nierenberg, T.~Treu, N.~Menci, Y.~Lu and W.~Wang, \emph{{The Cosmic
  Evolution of Faint Satellite Galaxies as a Test of Galaxy Formation and the
  Nature of Dark Matter}},
  \href{http://dx.doi.org/10.1088/0004-637X/772/2/146}{\emph{Astrophys. J.}
  {\bf 772} (2013) 146}, [\href{http://arxiv.org/abs/1302.3243}{{\tt
  1302.3243}}].

\bibitem{Papastergis:2014aba}
E.~Papastergis, R.~Giovanelli, M.~P. Haynes and F.~Shankar, \emph{Is there a
  "too big to fail?" problem in the field?},
  \href{http://dx.doi.org/10.1051/0004-6361/201424909}{\emph{Astron.
  Astrophys.} {\bf 574} (2015) A113},
  [\href{http://arxiv.org/abs/1407.4665}{{\tt 1407.4665}}].

\bibitem{Kimm:2008rp}
T.~Kimm et~al., \emph{{The Correlation of Star Formation Quenching with
  Internal Galaxy Properties and Environment}},
  \href{http://dx.doi.org/10.1111/j.1365-2966.2009.14414.x}{\emph{Mon. Not.
  Roy. Astron. Soc.} {\bf 394} (2009) 1131},
  [\href{http://arxiv.org/abs/0810.2794}{{\tt 0810.2794}}].

\bibitem{Geha:2012nq}
M.~Geha, M.~Blanton, R.~Yan and J.~Tinker, \emph{{A Stellar Mass Threshold for
  Quenching of Field Galaxies}},
  \href{http://dx.doi.org/10.1088/0004-637X/757/1/85}{\emph{Astrophys. J.} {\bf
  757} (2012) 85}, [\href{http://arxiv.org/abs/1206.3573}{{\tt 1206.3573}}].

\bibitem{Wetzel:2012nn}
A.~R. Wetzel, J.~L. Tinker, C.~Conroy and F.~C. v.~d. Bosch, \emph{{Galaxy
  evolution in groups and clusters: satellite star formation histories and
  quenching timescales in a hierarchical Universe}},
  \href{http://dx.doi.org/10.1093/mnras/stt469}{\emph{Mon. Not. Roy. Astron.
  Soc.} {\bf 432} (2013) 336}, [\href{http://arxiv.org/abs/1206.3571}{{\tt
  1206.3571}}].

\bibitem{Phillips:2013lca}
J.~I. Phillips, C.~Wheeler, M.~Boylan-Kolchin, J.~S. Bullock, M.~C. Cooper and
  E.~J. Tollerud, \emph{{A Dichotomy in Satellite Quenching Around L*
  Galaxies}}, \href{http://dx.doi.org/10.1093/mnras/stt2023}{\emph{Mon. Not.
  Roy. Astron. Soc.} {\bf 437} (2014) 1930--1941},
  [\href{http://arxiv.org/abs/1307.3552}{{\tt 1307.3552}}].

\bibitem{Slater:2014yca}
C.~T. Slater and E.~F. Bell, \emph{{The Mass Dependence of Dwarf Satellite
  Galaxy Quenching}},
  \href{http://dx.doi.org/10.1088/0004-637X/792/2/141}{\emph{Astrophys. J.}
  {\bf 792} (2014) 141}, [\href{http://arxiv.org/abs/1407.6006}{{\tt
  1407.6006}}].

\bibitem{Wheeler:2014ega}
C.~Wheeler, J.~I. Phillips, M.~C. Cooper, M.~Boylan-Kolchin and J.~S. Bullock,
  \emph{{The surprising inefficiency of dwarf satellite quenching}},
  \href{http://dx.doi.org/10.1093/mnras/stu965}{\emph{Mon. Not. Roy. Astron.
  Soc.} {\bf 442} (2014) 1396--1404},
  [\href{http://arxiv.org/abs/1402.1498}{{\tt 1402.1498}}].

\bibitem{Larson:1980mv}
R.~B. Larson, B.~M. Tinsley and C.~N. Caldwell, \emph{{The evolution of disk
  galaxies and the origin of S0 galaxies}},
  \href{http://dx.doi.org/10.1086/157917}{\emph{Astrophys. J.} {\bf 237} (1980)
  692--707}.

\bibitem{Balogh:2000sf}
M.~L. Balogh, J.~F. Navarro and S.~L. Morris, \emph{{The origin of star
  formation gradients in rich galaxy clusters}},
  \href{http://dx.doi.org/10.1086/309323}{\emph{Astrophys. J.} {\bf 540} (2000)
  113--121}, [\href{http://arxiv.org/abs/astro-ph/0004078}{{\tt
  astro-ph/0004078}}].

\bibitem{Kawata:2007gk}
D.~Kawata and J.~S. Mulchaey, \emph{{Strangulation in Galaxy Groups}},
  \href{http://dx.doi.org/10.1086/526544}{\emph{Astrophys. J.} {\bf 672} (2008)
  L103}, [\href{http://arxiv.org/abs/0707.3814}{{\tt 0707.3814}}].

\bibitem{McCarthy:2007ff}
I.~G. McCarthy, C.~S. Frenk, A.~S. Font, C.~G. Lacey, R.~G. Bower, N.~L.
  Mitchell et~al., \emph{{Ram pressure stripping the hot gaseous halos of
  galaxies in groups and clusters}},
  \href{http://dx.doi.org/10.1111/j.1365-2966.2007.12577.x}{\emph{Mon. Not.
  Roy. Astron. Soc.} {\bf 383} (2008) 593},
  [\href{http://arxiv.org/abs/0710.0964}{{\tt 0710.0964}}].

\bibitem{Hirschmann:2012xp}
M.~Hirschmann, R.~S. Somerville, T.~Naab and A.~Burkert, \emph{{Origin of the
  anti-hierarchical growth of black holes}},
  \href{http://dx.doi.org/10.1111/j.1365-2966.2012.21626.x}{\emph{Mon. Not.
  Roy. Astron. Soc.} {\bf 426} (2012) 237},
  [\href{http://arxiv.org/abs/1206.6112}{{\tt 1206.6112}}].

\bibitem{Bower:2011aa}
R.~G. Bower, A.~J. Benson and R.~A. Crain, \emph{{What Shapes the Galaxy Mass
  Function? Exploring the Roles of Supernova-Driven Winds and AGN}},
  \href{http://dx.doi.org/10.1111/j.1365-2966.2012.20516.x}{\emph{Mon. Not.
  Roy. Astron. Soc.} {\bf 422} (2012) 2816},
  [\href{http://arxiv.org/abs/1112.2712}{{\tt 1112.2712}}].

\bibitem{Weinmann:2012za}
S.~M. Weinmann, A.~Pasquali, B.~D. Oppenheimer, K.~Finlator, J.~T. Mendel,
  R.~A. Crain et~al., \emph{{A fundamental problem in our understanding of low
  mass galaxy evolution}},
  \href{http://dx.doi.org/10.1111/j.1365-2966.2012.21931.x}{\emph{Mon. Not.
  Roy. Astron. Soc.} {\bf 426} (2012) 2797},
  [\href{http://arxiv.org/abs/1204.4184}{{\tt 1204.4184}}].

\bibitem{Dave:2011pn}
R.~Dave, B.~D. Oppenheimer and K.~Finlator, \emph{{Galaxy Evolution in
  Cosmological Simulations With Outflows I: Stellar Masses and Star Formation
  Rates}}, \href{http://dx.doi.org/10.1111/j.1365-2966.2011.18680.x}{\emph{Mon.
  Not. Roy. Astron. Soc.} {\bf 415} (2011) 11},
  [\href{http://arxiv.org/abs/1103.3528}{{\tt 1103.3528}}].

\bibitem{Hirschmann:2013ivp}
M.~Hirschmann et~al., \emph{{The effect of metal enrichment and galactic winds
  on galaxy formation in cosmological zoom simulations}},
  \href{http://dx.doi.org/10.1093/mnras/stt1770}{\emph{Mon. Not. Roy. Astron.
  Soc.} {\bf 436} (2013) 2929}, [\href{http://arxiv.org/abs/1309.2946}{{\tt
  1309.2946}}].

\bibitem{Wang:2011gma}
L.~Wang, S.~M. Weinmann and E.~Neistein, \emph{{A modified star formation law
  as a solution to open problems in galaxy evolution}},
  \href{http://dx.doi.org/10.1111/j.1365-2966.2012.20569.x}{\emph{Mon. Not.
  Roy. Astron. Soc.} {\bf 421} (2012) 3450},
  [\href{http://arxiv.org/abs/1107.4419}{{\tt 1107.4419}}].

\bibitem{Henriques:2012ku}
B.~Henriques, S.~White, P.~Thomas, R.~Angulo, Q.~Guo, G.~Lemson et~al.,
  \emph{{Simulations of the galaxy population constrained by observations from
  z=3 to the present day: implications for galactic winds and the fate of their
  ejecta}},  \href{http://arxiv.org/abs/1212.1717}{{\tt 1212.1717}}.

\bibitem{Simha:2008hd}
V.~Simha, D.~H. Weinberg, R.~Dave, O.~Y. Gnedin, N.~Katz and D.~Keres,
  \emph{{The Growth of Central and Satellite Galaxies in Cosmological Smoothed
  Particle Hydrodynamics Simulations}},
  \href{http://dx.doi.org/10.1111/j.1365-2966.2009.15341.x}{\emph{Mon. Not.
  Roy. Astron. Soc.} {\bf 399} (2009) 650},
  [\href{http://arxiv.org/abs/0809.2999}{{\tt 0809.2999}}].

\bibitem{Font:2008pc}
A.~S. Font, R.~G. Bower, I.~G. McCarthy, A.~J. Benson, C.~S. Frenk, J.~C. Helly
  et~al., \emph{{The Colours of Satellite Galaxies in Groups and Clusters}},
  \href{http://dx.doi.org/10.1111/j.1365-2966.2008.13698.x}{\emph{Mon. Not.
  Roy. Astron. Soc.} {\bf 389} (2008) 1619--1629},
  [\href{http://arxiv.org/abs/0807.0001}{{\tt 0807.0001}}].

\bibitem{DeLucia:2011ac}
G.~De~Lucia, S.~Weinmann, B.~Poggianti, A.~Aragon-Salamanca and D.~Zaritsky,
  \emph{{The environmental history of group and cluster galaxies in a
  $\Lambda$CDM Universe}},
  \href{http://dx.doi.org/10.1111/j.1365-2966.2012.20983.x}{\emph{Mon. Not.
  Roy. Astron. Soc.} {\bf 423} (2012) 1277},
  [\href{http://arxiv.org/abs/1111.6590}{{\tt 1111.6590}}].

\bibitem{Coe:2012kj}
D.~Coe et~al., \emph{{CLASH: Precise New Constraints on the Mass Profile of
  Abell 2261}},
  \href{http://dx.doi.org/10.1088/0004-637X/757/1/22}{\emph{Astrophys. J.} {\bf
  757} (2012) 22}, [\href{http://arxiv.org/abs/1201.1616}{{\tt 1201.1616}}].

\bibitem{Hirschmann:2014jta}
M.~Hirschmann, G.~De~Lucia, D.~Wilman, S.~Weinmann, A.~Iovino, O.~Cucciati
  et~al., \emph{{The influence of the environmental history on quenching star
  formation in a $\Lambda$ cold dark matter universe}},
  \href{http://dx.doi.org/10.1093/mnras/stu1609}{\emph{Mon. Not. Roy. Astron.
  Soc.} {\bf 444} (2014) 2938--2959},
  [\href{http://arxiv.org/abs/1407.5621}{{\tt 1407.5621}}].

\bibitem{Fillingham:2015vya}
S.~P. Fillingham, M.~C. Cooper, C.~Wheeler, S.~Garrison-Kimmel,
  M.~Boylan-Kolchin and J.~S. Bullock, \emph{{Taking care of business in a
  flash: constraining the time-scale for low-mass satellite quenching with
  ELVIS}}, \href{http://dx.doi.org/10.1093/mnras/stv2058}{\emph{Mon. Not. Roy.
  Astron. Soc.} {\bf 454} (2015) 2039--2049},
  [\href{http://arxiv.org/abs/1503.06803}{{\tt 1503.06803}}].

\bibitem{Benson:2012su}
{\scshape Michigan Center for Theoretical Physics, The University of Michigan,
  Ann Arbor, MI, USA} collaboration, A.~J. Benson, A.~Farahi, S.~Cole, L.~A.
  Moustakas, A.~Jenkins, M.~Lovell et~al., \emph{{Dark Matter Halo Merger
  Histories Beyond Cold Dark Matter: I - Methods and Application to Warm Dark
  Matter}}, \href{http://dx.doi.org/10.1093/mnras/sts159}{\emph{Mon. Not. Roy.
  Astron. Soc.} {\bf 428} (2013) 1774},
  [\href{http://arxiv.org/abs/1209.3018}{{\tt 1209.3018}}].

\bibitem{Schneider:2014rda}
A.~Schneider, \emph{{Structure formation with suppressed small-scale
  perturbations}}, \href{http://dx.doi.org/10.1093/mnras/stv1169}{\emph{Mon.
  Not. Roy. Astron. Soc.} {\bf 451} (2015) 3117--3130},
  [\href{http://arxiv.org/abs/1412.2133}{{\tt 1412.2133}}].

\bibitem{Menci:2012kk}
N.~Menci, F.~Fiore and A.~Lamastra, \emph{{Galaxy Formation in WDM Cosmology}},
  \href{http://dx.doi.org/10.1111/j.1365-2966.2012.20470.x}{\emph{Mon. Not.
  Roy. Astron. Soc.} {\bf 421} (2012) 2384},
  [\href{http://arxiv.org/abs/1201.1617}{{\tt 1201.1617}}].

\bibitem{Calura:2014pla}
F.~Calura, N.~Menci and A.~Gallazzi, \emph{{The ages of stellar populations in
  a warm dark matter universe}},
  \href{http://dx.doi.org/10.1093/mnras/stu339}{\emph{Mon. Not. Roy. Astron.
  Soc.} {\bf 440} (2014) 2066--2076},
  [\href{http://arxiv.org/abs/1402.4828}{{\tt 1402.4828}}].

\bibitem{Drory:2004eh}
N.~Drory, R.~Bender, G.~Feulner, U.~Hopp, C.~Maraston, J.~Snigula et~al.,
  \emph{{The Munich Near-Infrared Cluster Survey (MUNICS) - 6. The Stellar
  masses of k-band selected field galaxies to Z ~ 1.2}},
  \href{http://dx.doi.org/10.1086/420781}{\emph{Astrophys. J.} {\bf 608} (2004)
  742--751}, [\href{http://arxiv.org/abs/astro-ph/0403041}{{\tt
  astro-ph/0403041}}].

\bibitem{Monaco:2006df}
P.~Monaco, F.~Fontanot and G.~Taffoni, \emph{{The MORGANA model for the rise of
  galaxies and active nuclei}},
  \href{http://dx.doi.org/10.1111/j.1365-2966.2006.11253.x}{\emph{Mon. Not.
  Roy. Astron. Soc.} {\bf 375} (2007) 1189--1219},
  [\href{http://arxiv.org/abs/astro-ph/0610805}{{\tt astro-ph/0610805}}].

\bibitem{DeLucia:2005yk}
G.~De~Lucia, V.~Springel, S.~D.~M. White, D.~Croton and G.~Kauffmann,
  \emph{{The formation history of elliptical galaxies}},
  \href{http://dx.doi.org/10.1111/j.1365-2966.2005.09879.x}{\emph{Mon. Not.
  Roy. Astron. Soc.} {\bf 366} (2006) 499--509},
  [\href{http://arxiv.org/abs/astro-ph/0509725}{{\tt astro-ph/0509725}}].

\bibitem{Somerville:2006wp}
R.~S. Somerville et~al., \emph{{An Explanation for the Observed Weak Size
  Evolution of Disk Galaxies}},
  \href{http://dx.doi.org/10.1086/523661}{\emph{Astrophys. J.} {\bf 672} (2008)
  776}, [\href{http://arxiv.org/abs/astro-ph/0612428}{{\tt astro-ph/0612428}}].

\bibitem{Dayal:2014nva}
P.~Dayal, A.~Mesinger and F.~Pacucci, \emph{{Early galaxy formation in warm
  dark matter cosmologies}},
  \href{http://dx.doi.org/10.1088/0004-637X/806/1/67}{\emph{Astrophys. J.} {\bf
  806} (2015) 67}, [\href{http://arxiv.org/abs/1408.1102}{{\tt 1408.1102}}].

\bibitem{Abazajian:2014gza}
K.~N. Abazajian, \emph{{Resonantly Produced 7 keV Sterile Neutrino Dark Matter
  Models and the Properties of Milky Way Satellites}},
  \href{http://dx.doi.org/10.1103/PhysRevLett.112.161303}{\emph{Phys. Rev.
  Lett.} {\bf 112} (2014) 161303}, [\href{http://arxiv.org/abs/1403.0954}{{\tt
  1403.0954}}].

\bibitem{Pacucci:2013jfa}
F.~Pacucci, A.~Mesinger and Z.~Haiman, \emph{{Focusing on Warm Dark Matter with
  Lensed High-redshift Galaxies}},  \href{http://arxiv.org/abs/1306.0009}{{\tt
  1306.0009}}.

\bibitem{Polisensky:2010rw}
E.~Polisensky and M.~Ricotti, \emph{{Constraints on the Dark Matter Particle
  Mass from the Number of Milky Way Satellites}},
  \href{http://dx.doi.org/10.1103/PhysRevD.83.043506}{\emph{Phys. Rev.} {\bf
  D83} (2011) 043506}, [\href{http://arxiv.org/abs/1004.1459}{{\tt
  1004.1459}}].

\bibitem{Belokurov:2010rf}
V.~Belokurov et~al., \emph{{Big fish, small fish: Two New Ultra-Faint
  Satellites of the Milky Way}},
  \href{http://dx.doi.org/10.1088/2041-8205/712/1/L103}{\emph{Astrophys. J.}
  {\bf 712} (2010) L103}, [\href{http://arxiv.org/abs/1002.0504}{{\tt
  1002.0504}}].

\bibitem{Abazajian:2011dt}
K.~N. Abazajian et~al., \emph{{Cosmological and Astrophysical Neutrino Mass
  Measurements}},
  \href{http://dx.doi.org/10.1016/j.astropartphys.2011.07.002}{\emph{Astropart.
  Phys.} {\bf 35} (2011) 177--184}, [\href{http://arxiv.org/abs/1103.5083}{{\tt
  1103.5083}}].

\bibitem{Watson:2011dw}
C.~R. Watson, Z.-Y. Li and N.~K. Polley, \emph{{Constraining Sterile Neutrino
  Warm Dark Matter with Chandra Observations of the Andromeda Galaxy}},
  \href{http://dx.doi.org/10.1088/1475-7516/2012/03/018}{\emph{JCAP} {\bf 1203}
  (2012) 018}, [\href{http://arxiv.org/abs/1111.4217}{{\tt 1111.4217}}].

\bibitem{Schultz:2014eia}
C.~Schultz, J.~O\~{n}orbe, K.~N. Abazajian and J.~S. Bullock, \emph{{The
  High-$z$ Universe Confronts Warm Dark Matter: Galaxy Counts, Reionization and
  the Nature of Dark Matter}},
  \href{http://dx.doi.org/10.1093/mnras/stu976}{\emph{Mon. Not. Roy. Astron.
  Soc.} {\bf 442} (2014) 1597--1609},
  [\href{http://arxiv.org/abs/1401.3769}{{\tt 1401.3769}}].

\bibitem{Diemand:2005wv}
J.~Diemand, M.~Zemp, B.~Moore, J.~Stadel and M.~Carollo, \emph{{Cusps in cold
  dark matter haloes}},
  \href{http://dx.doi.org/10.1111/j.1365-2966.2005.09601.x}{\emph{Mon. Not.
  Roy. Astron. Soc.} {\bf 364} (2005) 665},
  [\href{http://arxiv.org/abs/astro-ph/0504215}{{\tt astro-ph/0504215}}].

\bibitem{Binney:2001wu}
J.~J. Binney and N.~W. Evans, \emph{{Cuspy dark-matter haloes and the Galaxy}},
  \href{http://dx.doi.org/10.1046/j.1365-8711.2001.04968.x}{\emph{Mon. Not.
  Roy. Astron. Soc.} {\bf 327} (2001) L27},
  [\href{http://arxiv.org/abs/astro-ph/0108505}{{\tt astro-ph/0108505}}].

\bibitem{Bissantz:2001wx}
N.~Bissantz and O.~Gerhard, \emph{{Spiral arms, bar shape and bulge
  microlensing in the milky way}},
  \href{http://dx.doi.org/10.1046/j.1365-8711.2002.05116.x}{\emph{Mon. Not.
  Roy. Astron. Soc.} {\bf 330} (2002) 591},
  [\href{http://arxiv.org/abs/astro-ph/0110368}{{\tt astro-ph/0110368}}].

\bibitem{vandenBosch:2000rza}
F.~C. van~den Bosch and R.~A. Swaters, \emph{{Dwarf galaxy rotation curves and
  the core problem of dark matter halos}},
  \href{http://dx.doi.org/10.1046/j.1365-8711.2001.04456.x}{\emph{Mon. Not.
  Roy. Astron. Soc.} {\bf 325} (2001) 1017},
  [\href{http://arxiv.org/abs/astro-ph/0006048}{{\tt astro-ph/0006048}}].

\bibitem{Marchesini:2002vm}
D.~Marchesini, E.~D'Onghia, G.~Chincarini, C.~Firmani, P.~Conconi, E.~Molinari
  et~al., \emph{{Halpha rotation curves: the soft core question}},
  \href{http://dx.doi.org/10.1086/341475}{\emph{Astrophys. J.} {\bf 575} (2002)
  801--813}, [\href{http://arxiv.org/abs/astro-ph/0202075}{{\tt
  astro-ph/0202075}}].

\bibitem{Oh:2008ww}
S.-H. Oh, W.~J.~G. de~Blok, F.~Walter, E.~Brinks and R.~C. Kennicutt, Jr,
  \emph{{High-resolution dark matter density profiles of THINGS dwarf galaxies:
  Correcting for non-circular motions}},
  \href{http://dx.doi.org/10.1088/0004-6256/136/6/2761}{\emph{Astron. J.} {\bf
  136} (2008) 2761}, [\href{http://arxiv.org/abs/0810.2119}{{\tt 0810.2119}}].

\bibitem{Oh:2010ea}
S.-H. Oh, W.~J.~G. de~Blok, E.~Brinks, F.~Walter and R.~C. Kennicutt, Jr,
  \emph{{Dark and luminous matter in THINGS dwarf galaxies}},
  \href{http://dx.doi.org/10.1088/0004-6256/141/6/193}{\emph{Astron. J.} {\bf
  141} (2011) 193}, [\href{http://arxiv.org/abs/1011.0899}{{\tt 1011.0899}}].

\bibitem{deBlok:2001mf}
W.~J.~G. de~Blok, S.~S. McGaugh and V.~C. Rubin, \emph{{High-Resolution
  Rotation Curves of Low Surface Brightness Galaxies. II. Mass Models}},
  \href{http://dx.doi.org/10.1086/323450}{\emph{Astron. J.} {\bf 122} (2001)
  2396--2427}.

\bibitem{ma-an-de}
M.~A. Breddels, A.~Helmi, R.~C.~E. van~den Bosch, G.~van~de Ven and
  G.~Battaglia, \emph{{Orbit-based dynamical models of the Sculptor dSph
  galaxy}}, \href{http://dx.doi.org/10.1093/mnras/stt956}{\emph{Mon. Not. Roy.
  Astron. Soc.} {\bf 433} (2013) 3173--3189},
  [\href{http://arxiv.org/abs/1205.4712}{{\tt 1205.4712}}].

\bibitem{Kleyna:2001us}
J.~T. Kleyna, M.~I. Wilkinson, N.~W. Evans and G.~Gilmore, \emph{{Dark matter
  in dwarf spheroidals. 2. Observations and modeling of draco}},
  \href{http://dx.doi.org/10.1046/j.1365-8711.2002.05155.x}{\emph{Mon. Not.
  Roy. Astron. Soc.} {\bf 330} (2002) 792},
  [\href{http://arxiv.org/abs/astro-ph/0109450}{{\tt astro-ph/0109450}}].

\bibitem{Walker:2009zp}
M.~G. Walker, M.~Mateo, E.~W. Olszewski, J.~Penarrubia, N.~W. Evans and
  G.~Gilmore, \emph{{A Universal Mass Profile for Dwarf Spheroidal Galaxies}},
  \href{http://dx.doi.org/10.1088/0004-637X/704/2/1274,
  10.1088/0004-637X/710/1/886}{\emph{Astrophys. J.} {\bf 704} (2009)
  1274--1287}, [\href{http://arxiv.org/abs/0906.0341}{{\tt 0906.0341}}].

\bibitem{Wolf:2009tu}
J.~Wolf, G.~D. Martinez, J.~S. Bullock, M.~Kaplinghat, M.~Geha, R.~R. Munoz
  et~al., \emph{{Accurate Masses for Dispersion-supported Galaxies}},
  \href{http://dx.doi.org/10.1111/j.1365-2966.2010.16753.x}{\emph{Mon. Not.
  Roy. Astron. Soc.} {\bf 406} (2010) 1220},
  [\href{http://arxiv.org/abs/0908.2995}{{\tt 0908.2995}}].

\bibitem{Evans:2008ik}
N.~W. Evans, J.~An and M.~G. Walker, \emph{{Cores and Cusps in the Dwarf
  Spheroidals}},
  \href{http://dx.doi.org/10.1111/j.1745-3933.2008.00596.x}{\emph{Mon. Not.
  Roy. Astron. Soc.} {\bf 393} (2009) 50},
  [\href{http://arxiv.org/abs/0811.1488}{{\tt 0811.1488}}].

\bibitem{Walker:2011zu}
M.~G. Walker and J.~Penarrubia, \emph{{A Method for Measuring (Slopes of) the
  Mass Profiles of Dwarf Spheroidal Galaxies}},
  \href{http://dx.doi.org/10.1088/0004-637X/742/1/20}{\emph{Astrophys. J.} {\bf
  742} (2011) 20}, [\href{http://arxiv.org/abs/1108.2404}{{\tt 1108.2404}}].

\bibitem{Amorisco:2011hb}
N.~C. Amorisco and N.~W. Evans, \emph{{Dark Matter Cores and Cusps: The Case of
  Multiple Stellar Populations in Dwarf Spheroidals}},
  \href{http://dx.doi.org/10.1111/j.1365-2966.2011.19684.x}{\emph{Mon. Not.
  Roy. Astron. Soc.} {\bf 419} (2012) 184--196},
  [\href{http://arxiv.org/abs/1106.1062}{{\tt 1106.1062}}].

\bibitem{Agnello:2012uc}
A.~Agnello and N.~W. Evans, \emph{{A Virial Core in the Sculptor Dwarf
  Spheroidal Galaxy}},
  \href{http://dx.doi.org/10.1088/2041-8205/754/2/L39}{\emph{Astrophys. J.}
  {\bf 754} (2012) L39}, [\href{http://arxiv.org/abs/1205.6673}{{\tt
  1205.6673}}].

\bibitem{Amorisco:2012rd}
N.~C. Amorisco, A.~Agnello and N.~W. Evans, \emph{{The core size of the Fornax
  dwarf Spheroidal}}, \href{http://dx.doi.org/10.1093/mnrasl/sls031}{\emph{Mon.
  Not. Roy. Astron. Soc.} {\bf 429} (2013) 89},
  [\href{http://arxiv.org/abs/1210.3157}{{\tt 1210.3157}}].

\bibitem{Strigari:2014yea}
L.~E. Strigari, C.~S. Frenk and S.~D.~M. White, \emph{{Dynamical models for the
  Sculptor dwarf spheroidal in a Lambda CDM universe}},
  \href{http://arxiv.org/abs/1406.6079}{{\tt 1406.6079}}.

\bibitem{Gnedin:2001ec}
O.~Y. Gnedin and H.~Zhao, \emph{{Maximum feedback and dark matter profiles of
  dwarf galaxies}},
  \href{http://dx.doi.org/10.1046/j.1365-8711.2002.05361.x}{\emph{Mon. Not.
  Roy. Astron. Soc.} {\bf 333} (2002) 299},
  [\href{http://arxiv.org/abs/astro-ph/0108108}{{\tt astro-ph/0108108}}].

\bibitem{Read:2004xc}
J.~I. Read and G.~Gilmore, \emph{{Mass loss from dwarf spheroidal galaxies: The
  Origins of shallow dark matter cores and exponential surface brightness
  profiles}},
  \href{http://dx.doi.org/10.1111/j.1365-2966.2004.08424.x/abs/}{\emph{Mon.
  Not. Roy. Astron. Soc.} {\bf 356} (2005) 107--124},
  [\href{http://arxiv.org/abs/astro-ph/0409565}{{\tt astro-ph/0409565}}].

\bibitem{Governato:2012fa}
F.~Governato, A.~Zolotov, A.~Pontzen, C.~Christensen, S.~H. Oh, A.~M. Brooks
  et~al., \emph{{Cuspy No More: How Outflows Affect the Central Dark Matter and
  Baryon Distribution in Lambda CDM Galaxies}},
  \href{http://dx.doi.org/10.1111/j.1365-2966.2012.20696.x}{\emph{Mon. Not.
  Roy. Astron. Soc.} {\bf 422} (2012) 1231--1240},
  [\href{http://arxiv.org/abs/1202.0554}{{\tt 1202.0554}}].

\bibitem{Pontzen:2011ty}
A.~Pontzen and F.~Governato, \emph{{How supernova feedback turns dark matter
  cusps into cores}},
  \href{http://dx.doi.org/10.1111/j.1365-2966.2012.20571.x}{\emph{Mon. Not.
  Roy. Astron. Soc.} {\bf 421} (2012) 3464},
  [\href{http://arxiv.org/abs/1106.0499}{{\tt 1106.0499}}].

\bibitem{deNaray:2011hy}
R.~K. de~Naray and K.~Spekkens, \emph{{Do Baryons Alter the Halos of Low
  Surface Brightness Galaxies?}},
  \href{http://dx.doi.org/10.1088/2041-8205/741/2/L29}{\emph{Astrophys. J.}
  {\bf 741} (2011) L29}, [\href{http://arxiv.org/abs/1109.1288}{{\tt
  1109.1288}}].

\bibitem{VillaescusaNavarro:2010qy}
F.~Villaescusa-Navarro and N.~Dalal, \emph{{Cores and Cusps in Warm Dark Matter
  Halos}}, \href{http://dx.doi.org/10.1088/1475-7516/2011/03/024}{\emph{JCAP}
  {\bf 1103} (2011) 024}, [\href{http://arxiv.org/abs/1010.3008}{{\tt
  1010.3008}}].

\bibitem{Springel:2008cc}
V.~Springel, J.~Wang, M.~Vogelsberger, A.~Ludlow, A.~Jenkins et~al., \emph{{The
  Aquarius Project: the subhalos of galactic halos}},
  \href{http://dx.doi.org/10.1111/j.1365-2966.2008.14066.x}{\emph{Mon.Not.Roy.Astron.Soc.}
  {\bf 391} (Dec., 2008) 1685--1711},
  [\href{http://arxiv.org/abs/0809.0898}{{\tt 0809.0898}}].

\bibitem{Diemand:2007qr}
J.~Diemand, M.~Kuhlen and P.~Madau, \emph{{Formation and evolution of galaxy
  dark matter halos and their substructure}},
  \href{http://dx.doi.org/10.1086/520573}{\emph{Astrophys. J.} {\bf 667} (2007)
  859--877}, [\href{http://arxiv.org/abs/astro-ph/0703337}{{\tt
  astro-ph/0703337}}].

\bibitem{Wang:2012sv}
J.~Wang, C.~S. Frenk, J.~F. Navarro and L.~Gao, \emph{{The Missing Massive
  Satellites of the Milky Way}},
  \href{http://dx.doi.org/10.1111/j.1365-2966.2012.21357.x}{\emph{Mon. Not.
  Roy. Astron. Soc.} {\bf 424} (2012) 2715--2721},
  [\href{http://arxiv.org/abs/1203.4097}{{\tt 1203.4097}}].

\bibitem{VeraCiro:2012na}
C.~A. Vera-Ciro, A.~Helmi, E.~Starkenburg and M.~A. Breddels, \emph{{Not too
  big, not too small: the dark halos of the dwarf spheroidals in the Milky
  Way}}, \href{http://dx.doi.org/10.1093/mnras/sts148}{\emph{Mon. Not. Roy.
  Astron. Soc.} {\bf 428} (2013) 1696},
  [\href{http://arxiv.org/abs/1202.6061}{{\tt 1202.6061}}].

\bibitem{Purcell:2012kd}
C.~W. Purcell and A.~R. Zentner, \emph{{Bailing Out the Milky Way: Variation in
  the Properties of Massive Dwarfs Among Galaxy-Sized Systems}},
  \href{http://dx.doi.org/10.1088/1475-7516/2012/12/007}{\emph{JCAP} {\bf 1212}
  (2012) 007}, [\href{http://arxiv.org/abs/1208.4602}{{\tt 1208.4602}}].

\bibitem{Tollerud:2014zha}
E.~J. Tollerud, M.~Boylan-Kolchin and J.~S. Bullock, \emph{{M31 Satellite
  Masses Compared to LCDM Subhaloes}},
  \href{http://dx.doi.org/10.1093/mnras/stu474}{\emph{Mon. Not. Roy. Astron.
  Soc.} {\bf 440} (2014) 3511--3519},
  [\href{http://arxiv.org/abs/1403.6469}{{\tt 1403.6469}}].

\bibitem{Strigari:2011ps}
L.~E. Strigari and R.~H. Wechsler, \emph{{The Cosmic Abundance of Classical
  Milky Way Satellites}},
  \href{http://dx.doi.org/10.1088/0004-637X/749/1/75}{\emph{Astrophys. J.} {\bf
  749} (2012) 75}, [\href{http://arxiv.org/abs/1111.2611}{{\tt 1111.2611}}].

\bibitem{Rodriguez-Puebla:2013aza}
A.~Rodriguez-Puebla, V.~Avila-Reese and N.~Drory, \emph{{The Massive Satellite
  Population of Milky-Way Sized Galaxies}},
  \href{http://dx.doi.org/10.1088/0004-637X/773/2/172}{\emph{Astrophys. J.}
  {\bf 773} (2013) 172}, [\href{http://arxiv.org/abs/1306.4328}{{\tt
  1306.4328}}].

\bibitem{Zolotov:2012xd}
A.~Zolotov, A.~M. Brooks, B.~Willman, F.~Governato, A.~Pontzen, C.~Christensen
  et~al., \emph{{Baryons Matter: Why Luminous Satellite Galaxies Have Reduced
  Central Masses}},
  \href{http://dx.doi.org/10.1088/0004-637X/761/1/71}{\emph{Astrophys. J.} {\bf
  761} (2012) 71}, [\href{http://arxiv.org/abs/1207.0007}{{\tt 1207.0007}}].

\bibitem{Brooks:2012ah}
A.~M. Brooks, M.~Kuhlen, A.~Zolotov and D.~Hooper, \emph{{A Baryonic Solution
  to the Missing Satellites Problem}},
  \href{http://dx.doi.org/10.1088/0004-637X/765/1/22}{\emph{Astrophys. J.} {\bf
  765} (2013) 22}, [\href{http://arxiv.org/abs/1209.5394}{{\tt 1209.5394}}].

\bibitem{Kirby:2014sya}
E.~N. Kirby, J.~S. Bullock, M.~Boylan-Kolchin, M.~Kaplinghat and J.~G. Cohen,
  \emph{{The dynamics of isolated Local Group galaxies}},
  \href{http://dx.doi.org/10.1093/mnras/stu025}{\emph{Mon. Not. Roy. Astron.
  Soc.} {\bf 439} (2014) 1015--1027},
  [\href{http://arxiv.org/abs/1401.1208}{{\tt 1401.1208}}].

\bibitem{Garrison-Kimmel:2014vqa}
S.~Garrison-Kimmel, M.~Boylan-Kolchin, J.~S. Bullock and E.~N. Kirby,
  \emph{{Too Big to Fail in the Local Group}},
  \href{http://dx.doi.org/10.1093/mnras/stu1477}{\emph{Mon. Not. Roy. Astron.
  Soc.} {\bf 444} (2014) 222--236}, [\href{http://arxiv.org/abs/1404.5313}{{\tt
  1404.5313}}].

\bibitem{Garrison-Kimmel:2013eoa}
S.~Garrison-Kimmel, M.~Boylan-Kolchin, J.~Bullock and K.~Lee, \emph{{ELVIS:
  Exploring the Local Volume in Simulations}},
  \href{http://dx.doi.org/10.1093/mnras/stt2377}{\emph{Mon. Not. Roy. Astron.
  Soc.} {\bf 438} (2014) 2578--2596},
  [\href{http://arxiv.org/abs/1310.6746}{{\tt 1310.6746}}].

\bibitem{Ferrero:2011au}
I.~Ferrero, M.~G. Abadi, J.~F. Navarro, L.~V. Sales and S.~Gurovich, \emph{{The
  dark matter halos of dwarf galaxies: a challenge for the LCDM paradigm?}},
  \href{http://dx.doi.org/10.1111/j.1365-2966.2012.21623.x}{\emph{Mon. Not.
  Roy. Astron. Soc.} {\bf 425} (2012) 2817--2823},
  [\href{http://arxiv.org/abs/1111.6609}{{\tt 1111.6609}}].

\bibitem{Guo:2009fn}
Q.~Guo, S.~White, C.~Li and M.~Boylan-Kolchin, \emph{{How do galaxies populate
  Dark Matter halos?}}, {\emph{Mon. Not. Roy. Astron. Soc.} {\bf 404} (2010)
  1111}, [\href{http://arxiv.org/abs/0909.4305}{{\tt 0909.4305}}].

\bibitem{Haynes:2011hi}
M.~P. Haynes et~al., \emph{{The Arecibo Legacy Fast ALFA Survey: The alpha.40
  HI Source Catalog, its Characteristics and their Impact on the Derivation of
  the HI Mass Function}},
  \href{http://dx.doi.org/10.1088/0004-6256/142/5/170}{\emph{Astron. J.} {\bf
  142} (2011) 170}, [\href{http://arxiv.org/abs/1109.0027}{{\tt 1109.0027}}].

\bibitem{Sawala:2012cn}
T.~Sawala, C.~S. Frenk, R.~A. Crain, A.~Jenkins, J.~Schaye, T.~Theuns et~al.,
  \emph{{The abundance of (not just) dark matter haloes}},
  \href{http://dx.doi.org/10.1093/mnras/stt259}{\emph{Mon. Not. Roy. Astron.
  Soc.} {\bf 431} (2013) 1366--1382},
  [\href{http://arxiv.org/abs/1206.6495}{{\tt 1206.6495}}].

\bibitem{Governato:2009bg}
F.~Governato et~al., \emph{{At the heart of the matter: the origin of bulgeless
  dwarf galaxies and Dark Matter cores}},
  \href{http://dx.doi.org/10.1038/nature08640}{\emph{Nature} {\bf 463} (2010)
  203--206}, [\href{http://arxiv.org/abs/0911.2237}{{\tt 0911.2237}}].

\bibitem{DiCintio:2014xia}
A.~Di~Cintio, C.~B. Brook, A.~A. Dutton, A.~V. Macci??, G.~S. Stinson and
  A.~Knebe, \emph{{A mass-dependent density profile for dark matter haloes
  including the influence of galaxy formation}},
  \href{http://dx.doi.org/10.1093/mnras/stu729}{\emph{Mon. Not. Roy. Astron.
  Soc.} {\bf 441} (2014) 2986--2995},
  [\href{http://arxiv.org/abs/1404.5959}{{\tt 1404.5959}}].

\bibitem{DiCintio:2013qxa}
A.~Di~Cintio, C.~B. Brook, A.~V. Macci??, G.~S. Stinson, A.~Knebe, A.~A. Dutton
  et~al., \emph{{The dependence of dark matter profiles on the stellar-to-halo
  mass ratio: a prediction for cusps versus cores}},
  \href{http://dx.doi.org/10.1093/mnras/stt1891}{\emph{Mon. Not. Roy. Astron.
  Soc.} {\bf 437} (2014) 415--423}, [\href{http://arxiv.org/abs/1306.0898}{{\tt
  1306.0898}}].

\bibitem{Brook:2014hda}
C.~B. Brook and A.~Di~Cintio, \emph{{Expanded haloes, abundance matching and
  too-big-to-fail in the Local Group}},
  \href{http://dx.doi.org/10.1093/mnras/stv864}{\emph{Mon. Not. Roy. Astron.
  Soc.} {\bf 450} (2015) 3920--3934},
  [\href{http://arxiv.org/abs/1410.3825}{{\tt 1410.3825}}].

\bibitem{Christensen:2012zh}
C.~R. Christensen, F.~Governato, T.~Quinn, A.~M. Brooks, D.~B. Fisher, S.~Shen
  et~al., \emph{{The effect of models of the interstellar media on the central
  mass distribution of galaxies}},
  \href{http://dx.doi.org/10.1093/mnras/stu399}{\emph{Mon. Not. Roy. Astron.
  Soc.} {\bf 440} (2014) 2843--2859},
  [\href{http://arxiv.org/abs/1211.0326}{{\tt 1211.0326}}].

\bibitem{Zavala:2009ms}
J.~Zavala, Y.~P. Jing, A.~Faltenbacher, G.~Yepes, Y.~Hoffman, S.~Gottlober
  et~al., \emph{{The velocity function in the local environment from LCDM and
  LWDM constrained simulations}},
  \href{http://dx.doi.org/10.1088/0004-637X/700/2/1779}{\emph{Astrophys. J.}
  {\bf 700} (2009) 1779--1793}, [\href{http://arxiv.org/abs/0906.0585}{{\tt
  0906.0585}}].

\bibitem{Lovell:2015psz}
M.~R. Lovell, S.~Bose, A.~Boyarsky, S.~Cole, C.~S. Frenk, V.~Gonzalez-Perez
  et~al., \emph{{Satellite galaxies in semi-analytic models of galaxy formation
  with sterile neutrino dark matter}},
  \href{http://arxiv.org/abs/1511.04078}{{\tt 1511.04078}}.

\bibitem{Lovell_eagle}
M.~R. Lovell et~al., \emph{{Properties of Local Group galaxies in
  hydrodynamical simulations of sterile neutrino dark matter cosmologies}},
  \href{http://arxiv.org/abs/1611.00010}{{\tt 1611.00010}}.

\bibitem{Anderhalden:2012jc}
D.~Anderhalden, A.~Schneider, A.~V. Maccio, J.~Diemand and G.~Bertone,
  \emph{{Hints on the Nature of Dark Matter from the Properties of Milky Way
  Satellites}},
  \href{http://dx.doi.org/10.1088/1475-7516/2013/03/014}{\emph{JCAP} {\bf 1303}
  (2013) 014}, [\href{http://arxiv.org/abs/1212.2967}{{\tt 1212.2967}}].

\bibitem{Lovell_tbtf}
M.~R. Lovell, V.~Gonzalez-Perez, S.~Bose, A.~Boyarsky, S.~Cole, C.~S. Frenk
  et~al., \emph{{Addressing the too big to fail problem with baryon physics and
  sterile neutrino dark matter}},  \href{http://arxiv.org/abs/1611.00005}{{\tt
  1611.00005}}.

\bibitem{Garzilli:2015iwa}
A.~Garzilli, A.~Boyarsky and O.~Ruchayskiy, \emph{{Cutoff in the Lyman $\alpha$
  forest power spectrum: warm IGM or warm dark matter?}},
  \href{http://arxiv.org/abs/1510.07006}{{\tt 1510.07006}}.

\bibitem{Klypin:2014ira}
A.~Klypin, I.~Karachentsev, D.~Makarov and O.~Nasonova, \emph{{Abundance of
  Field Galaxies}},  \href{http://arxiv.org/abs/1405.4523}{{\tt 1405.4523}}.

\bibitem{Maio:2014qwa}
U.~Maio and M.~Viel, \emph{{The First Billion Years of a Warm Dark Matter
  Universe}}, \href{http://dx.doi.org/10.1093/mnras/stu2304}{\emph{Mon. Not.
  Roy. Astron. Soc.} {\bf 446} (2015) 2760--2775},
  [\href{http://arxiv.org/abs/1409.6718}{{\tt 1409.6718}}].

\bibitem{Governato:2014gja}
F.~Governato et~al., \emph{{Faint dwarfs as a test of DM models: WDM vs. CDM}},
  \href{http://dx.doi.org/10.1093/mnras/stu2720}{\emph{Mon. Not. Roy. Astron.
  Soc.} {\bf 448} (2015) 792}, [\href{http://arxiv.org/abs/1407.0022}{{\tt
  1407.0022}}].

\bibitem{Wang:2015fia}
M.-Y. Wang, L.~E. Strigari, M.~R. Lovell, C.~S. Frenk and A.~R. Zentner,
  \emph{{Mass assembly history and infall time of the Fornax dwarf spheroidal
  galaxy}},  \href{http://arxiv.org/abs/1509.04308}{{\tt 1509.04308}}.

\bibitem{Strigari:2010un}
L.~E. Strigari, C.~S. Frenk and S.~D.~M. White, \emph{{Kinematics of Milky Way
  Satellites in a Lambda Cold Dark Matter Universe}},
  \href{http://dx.doi.org/10.1111/j.1365-2966.2010.17287.x}{\emph{Mon. Not.
  Roy. Astron. Soc.} {\bf 408} (2010) 2364--2372},
  [\href{http://arxiv.org/abs/1003.4268}{{\tt 1003.4268}}].

\bibitem{Behroozi:2012iw}
P.~S. Behroozi, R.~H. Wechsler and C.~Conroy, \emph{{The Average Star Formation
  Histories of Galaxies in Dark Matter Halos from $z=$0-8}},
  \href{http://dx.doi.org/10.1088/0004-637X/770/1/57}{\emph{Astrophys. J.} {\bf
  770} (2013) 57}, [\href{http://arxiv.org/abs/1207.6105}{{\tt 1207.6105}}].

\bibitem{Behroozi:2011js}
P.~S. Behroozi, R.~H. Wechsler, H.-Y. Wu, M.~T. Busha, A.~A. Klypin and J.~R.
  Primack, \emph{{Gravitationally Consistent Halo Catalogs and Merger Trees for
  Precision Cosmology}},
  \href{http://dx.doi.org/10.1088/0004-637X/763/1/18}{\emph{Astrophys. J.} {\bf
  763} (2013) 18}, [\href{http://arxiv.org/abs/1110.4370}{{\tt 1110.4370}}].

\bibitem{Gorbunov:2011zz}
D.~S. Gorbunov and V.~A. Rubakov, \emph{Introduction to the theory of the early
  universe: Hot big bang theory, hackensack, usa: World scientific (2011) 489
  p}, .

\bibitem{Gorbunov:2008ui}
D.~Gorbunov, A.~Khmelnitsky and V.~Rubakov, \emph{{Is gravitino still a warm
  dark matter candidate?}},
  \href{http://dx.doi.org/10.1088/1126-6708/2008/12/055}{\emph{JHEP} {\bf 0812}
  (2008) 055}, [\href{http://arxiv.org/abs/0805.2836}{{\tt 0805.2836}}].

\bibitem{Meiksin:2007rz}
A.~A. Meiksin, \emph{{The Physics of the Intergalactic Medium}},
  \href{http://dx.doi.org/10.1103/RevModPhys.81.1405}{\emph{Rev. Mod. Phys.}
  {\bf 81} (2009) 1405--1469}, [\href{http://arxiv.org/abs/0711.3358}{{\tt
  0711.3358}}].

\bibitem{Bi:1996fh}
H.~Bi and A.~F. Davidsen, \emph{{Evolution of structure in the intergalactic
  medium and the nature of the ly-alpha forest}},
  \href{http://dx.doi.org/10.1086/303908}{\emph{Astrophys. J.} {\bf 479} (1997)
  523}, [\href{http://arxiv.org/abs/astro-ph/9611062}{{\tt astro-ph/9611062}}].

\bibitem{Viel:2001hd}
M.~Viel, S.~Matarrese, H.~J. Mo, M.~G. Haehnelt and T.~Theuns, \emph{{Probing
  the intergalactic medium with the lyman alpha forest along multiple lines of
  sight to distant qsos}},
  \href{http://dx.doi.org/10.1046/j.1365-8711.2002.05060.x}{\emph{Mon. Not.
  Roy. Astron. Soc.} {\bf 329} (2002) 848},
  [\href{http://arxiv.org/abs/astro-ph/0105233}{{\tt astro-ph/0105233}}].

\bibitem{Busca:2012bu}
N.~G. Busca, T.~Delubac, J.~Rich, S.~Bailey, A.~Font-Ribera et~al.,
  \emph{{Baryon Acoustic Oscillations in the Ly-$\alpha$ forest of BOSS
  quasars}}, \href{http://dx.doi.org/10.1051/0004-6361/201220724}{\emph{Astron.
  Astrophys.} {\bf 552} (2013) A96},
  [\href{http://arxiv.org/abs/1211.2616}{{\tt 1211.2616}}].

\bibitem{Slosar:2013fi}
A.~Slosar, V.~Irsic, D.~Kirkby, S.~Bailey, N.~G. Busca et~al.,
  \emph{{Measurement of Baryon Acoustic Oscillations in the Lyman-alpha Forest
  Fluctuations in BOSS Data Release 9}},
  \href{http://dx.doi.org/10.1088/1475-7516/2013/04/026}{\emph{JCAP} {\bf 1304}
  (2013) 026}, [\href{http://arxiv.org/abs/1301.3459}{{\tt 1301.3459}}].

\bibitem{Markovic:2013iza}
K.~Markovic and M.~Viel, \emph{{Lyman-${\alpha}$ Forest and Cosmic Weak Lensing
  in a Warm Dark Matter Universe}},  \href{http://arxiv.org/abs/1311.5223}{{\tt
  1311.5223}}.

\bibitem{Bond:1980ha}
J.~R. Bond, G.~Efstathiou and J.~Silk, \emph{{Massive Neutrinos and the Large
  Scale Structure of the Universe}},
  \href{http://dx.doi.org/10.1103/PhysRevLett.45.1980}{\emph{Phys. Rev. Lett.}
  {\bf 45} (1980) 1980--1984}.

\bibitem{Narayanan:2000tp}
V.~K. Narayanan, D.~N. Spergel, R.~Dave and C.-P. Ma, \emph{{Constraints on the
  mass of warm dark matter particles and the shape of the linear power spectrum
  from the Ly$\alpha$ forest}},
  \href{http://dx.doi.org/10.1086/317269}{\emph{Astrophys. J.} {\bf 543} (2000)
  L103--L106}, [\href{http://arxiv.org/abs/astro-ph/0005095}{{\tt
  astro-ph/0005095}}].

\bibitem{Croft:2000hs}
R.~A.~C. Croft, D.~H. Weinberg, M.~Bolte, S.~Burles, L.~Hernquist et~al.,
  \emph{{Towards a precise measurement of matter clustering: Lyman alpha forest
  data at redshifts 2-4}},
  \href{http://dx.doi.org/10.1086/344099}{\emph{Astrophys. J.} {\bf 581} (2002)
  20--52}, [\href{http://arxiv.org/abs/astro-ph/0012324}{{\tt
  astro-ph/0012324}}].

\bibitem{Klypin:1994nf}
A.~Klypin, S.~Borgani, J.~Holtzman and J.~Primack, \emph{{Damped Lyman alpha
  systems versus cold + hot dark matter}},
  \href{http://arxiv.org/abs/astro-ph/9410022}{{\tt astro-ph/9410022}}.

\bibitem{Viel:2004bf}
M.~Viel, M.~G. Haehnelt and V.~Springel, \emph{{Inferring the dark matter power
  spectrum from the Lyman-alpha forest in high-resolution QSO absorption
  spectra}},
  \href{http://dx.doi.org/10.1111/j.1365-2966.2004.08224.x}{\emph{Mon. Not.
  Roy. Astron. Soc.} {\bf 354} (2004) 684},
  [\href{http://arxiv.org/abs/astro-ph/0404600}{{\tt astro-ph/0404600}}].

\bibitem{mcdonald05}
J.~McDonald, \emph{{F-term inflation Q-balls}},
  \href{http://dx.doi.org/10.1103/PhysRevD.73.043501}{\emph{Phys. Rev.} {\bf
  D73} (2006) 043501}, [\href{http://arxiv.org/abs/hep-th/0509141}{{\tt
  hep-th/0509141}}].

\bibitem{Abazajian:2006yn}
K.~Abazajian and S.~M. Koushiappas, \emph{{Constraints on Sterile Neutrino Dark
  Matter}}, \href{http://dx.doi.org/10.1103/PhysRevD.74.023527}{\emph{Phys.
  Rev.} {\bf D74} (2006) 023527},
  [\href{http://arxiv.org/abs/astro-ph/0605271}{{\tt astro-ph/0605271}}].

\bibitem{Viel:2011bk}
M.~Viel, K.~Markovic, M.~Baldi and J.~Weller, \emph{{The Non-Linear Matter
  Power Spectrum in Warm Dark Matter Cosmologies}},
  \href{http://dx.doi.org/10.1111/j.1365-2966.2011.19910.x}{\emph{Mon. Not.
  Roy. Astron. Soc.} {\bf 421} (2012) 50--62},
  [\href{http://arxiv.org/abs/1107.4094}{{\tt 1107.4094}}].

\bibitem{Carucci:2015bra}
I.~P. Carucci, F.~Villaescusa-Navarro, M.~Viel and A.~Lapi, \emph{{Warm dark
  matter signatures on the 21cm power spectrum: Intensity mapping forecasts for
  SKA}},  \href{http://arxiv.org/abs/1502.06961}{{\tt 1502.06961}}.

\bibitem{Shrock:1974nd}
R.~Shrock, \emph{{Decay l0 ---> nu(lepton) gamma in gauge theories of weak and
  electromagnetic interactions}},
  \href{http://dx.doi.org/10.1103/PhysRevD.9.743}{\emph{Phys. Rev.} {\bf D9}
  (1974) 743--748}.

\bibitem{Marciano:1977wx}
W.~J. Marciano and A.~I. Sanda, \emph{{Exotic Decays of the Muon and Heavy
  Leptons in Gauge Theories}},
  \href{http://dx.doi.org/10.1016/0370-2693(77)90377-X}{\emph{Phys. Lett.} {\bf
  B67} (1977) 303--305}.

\bibitem{Petcov:1976ff}
S.~T. Petcov, \emph{{The Processes mu --> e Gamma, mu --> e e anti-e, Neutrino'
  --> Neutrino gamma in the Weinberg-Salam Model with Neutrino Mixing}},
  {\emph{Sov. J. Nucl. Phys.} {\bf 25} (1977) 340}.

\bibitem{Abazajian:2001vt}
K.~Abazajian, G.~M. Fuller and W.~H. Tucker, \emph{{Direct detection of warm
  dark matter in the X-ray}},
  \href{http://dx.doi.org/10.1086/323867}{\emph{Astrophys. J.} {\bf 562} (2001)
  593--604}, [\href{http://arxiv.org/abs/astro-ph/0106002}{{\tt
  astro-ph/0106002}}].

\bibitem{Herder:2009im}
J.~W. den Herder et~al., \emph{{The Search for decaying Dark Matter}},
  \href{http://arxiv.org/abs/0906.1788}{{\tt 0906.1788}}.

\bibitem{Mapelli:05}
M.~{Mapelli} and A.~{Ferrara}, \emph{{Background radiation from sterile
  neutrino decay and reionization}},
  \href{http://dx.doi.org/10.1111/j.1365-2966.2005.09507.x}{\emph{MNRAS} {\bf
  364} (Nov., 2005) 2--12}.

\bibitem{Boyarsky:2005us}
A.~Boyarsky, A.~Neronov, O.~Ruchayskiy and M.~Shaposhnikov, \emph{{Constraints
  on sterile neutrino as a dark matter candidate from the diffuse x-ray
  background}},
  \href{http://dx.doi.org/10.1111/j.1365-2966.2006.10458.x}{\emph{Mon. Not.
  Roy. Astron. Soc.} {\bf 370} (2006) 213--218},
  [\href{http://arxiv.org/abs/astro-ph/0512509}{{\tt astro-ph/0512509}}].

\bibitem{Gruber:99}
D.~E. {Gruber}, J.~L. {Matteson}, L.~E. {Peterson} and G.~V. {Jung}, \emph{{The
  Spectrum of Diffuse Cosmic Hard X-Rays Measured with HEAO 1}},
  \href{http://dx.doi.org/10.1086/307450}{\emph{Ap. J.} {\bf 520} (July, 1999)
  124--129}.

\bibitem{Churazov:07}
E.~{Churazov}, R.~{Sunyaev}, M.~{Revnivtsev}, S.~{Sazonov}, S.~{Molkov},
  S.~{Grebenev} et~al., \emph{{INTEGRAL observations of the cosmic X-ray
  background in the 5-100 keV range via occultation by the Earth}},
  \href{http://dx.doi.org/10.1051/0004-6361:20066230}{\emph{A\&A} {\bf 467}
  (May, 2007) 529--540}.

\bibitem{Abazajian:2006jc}
K.~N. Abazajian, M.~Markevitch, S.~M. Koushiappas and R.~C. Hickox,
  \emph{{Limits on the Radiative Decay of Sterile Neutrino Dark Matter from the
  Unresolved Cosmic and Soft X-ray Backgrounds}},
  \href{http://dx.doi.org/10.1103/PhysRevD.75.063511}{\emph{Phys. Rev.} {\bf
  D75} (2007) 063511}, [\href{http://arxiv.org/abs/astro-ph/0611144}{{\tt
  astro-ph/0611144}}].

\bibitem{Boyarsky:2009rb}
A.~Boyarsky, O.~Ruchayskiy, D.~Iakubovskyi, A.~V. Maccio' and D.~Malyshev,
  \emph{{New evidence for dark matter}},
  \href{http://arxiv.org/abs/0911.1774}{{\tt 0911.1774}}.

\bibitem{Boyarsky:2009af}
A.~Boyarsky, A.~Neronov, O.~Ruchayskiy and I.~Tkachev, \emph{{Universal
  properties of Dark Matter halos}},
  \href{http://dx.doi.org/10.1103/PhysRevLett.104.191301}{\emph{Phys. Rev.
  Lett.} {\bf 104} (2010) 191301}, [\href{http://arxiv.org/abs/0911.3396}{{\tt
  0911.3396}}].

\bibitem{Cuesta:10}
A.~J. {Cuesta}, T.~E. {Jeltema}, F.~{Zandanel}, S.~{Profumo}, F.~{Prada},
  G.~{Yepes} et~al., \emph{{Dark Matter Decay and Annihilation in the Local
  Universe: Clues from Fermi}},
  \href{http://dx.doi.org/10.1088/2041-8205/726/1/L6}{\emph{Ap. J. Lett.} {\bf
  726} (Jan., 2011) L6}.

\bibitem{Cavaliere:76}
A.~{Cavaliere} and R.~{Fusco-Femiano}, \emph{{X-rays from hot plasma in
  clusters of galaxies}}, {\emph{A\&A} {\bf 49} (May, 1976) 137--144}.

\bibitem{Mushotzky:84}
R.~F. {Mushotzky}, \emph{{X-ray emission from clusters of galaxies}},
  \href{http://dx.doi.org/10.1088/0031-8949/1984/T7/036}{\emph{Physica Scripta
  Volume T} {\bf 7} (1984) 157--162}.

\bibitem{Sarazin:86}
C.~L. {Sarazin}, \emph{{X-ray emission from clusters of galaxies}},
  \href{http://dx.doi.org/10.1103/RevModPhys.58.1}{\emph{Reviews of Modern
  Physics} {\bf 58} (Jan., 1986) 1--115}.

\bibitem{Briel:93}
U.~G. {Briel} and J.~P. {Henry}, \emph{{X-ray emission from a complete sample
  of Abell clusters of galaxies}}, {\emph{A\&A} {\bf 278} (Nov., 1993)
  379--390}.

\bibitem{Bonamente:02}
M.~{Bonamente}, R.~{Lieu}, M.~K. {Joy} and J.~H. {Nevalainen}, \emph{{The Soft
  X-Ray Emission in a Large Sample of Galaxy Clusters with the ROSAT Position
  Sensitive Proportional Counter}},
  \href{http://dx.doi.org/10.1086/341806}{\emph{Ap. J.} {\bf 576} (Sept., 2002)
  688--707}.

\bibitem{Boyarsky:2006zi}
A.~Boyarsky, A.~Neronov, O.~Ruchayskiy and M.~Shaposhnikov, \emph{{Restrictions
  on parameters of sterile neutrino dark matter from observations of galaxy
  clusters}}, \href{http://dx.doi.org/10.1103/PhysRevD.74.103506}{\emph{Phys.
  Rev.} {\bf D74} (2006) 103506},
  [\href{http://arxiv.org/abs/astro-ph/0603368}{{\tt astro-ph/0603368}}].

\bibitem{Pratt:08}
G.~W. {Pratt}, J.~H. {Croston}, M.~{Arnaud} and H.~{B{\"o}hringer},
  \emph{{Galaxy cluster X-ray luminosity scaling relations from a
  representative local sample (REXCESS)}},
  \href{http://dx.doi.org/10.1051/0004-6361/200810994}{\emph{A\&A} {\bf 498}
  (May, 2009) 361--378}.

\bibitem{Sarazin:77}
C.~L. {Sarazin} and J.~N. {Bahcall}, \emph{{X-ray line emission for clusters of
  galaxies. II - Numerical models}},
  \href{http://dx.doi.org/10.1086/190457}{\emph{Ap. J. Suppl.} {\bf 34} (Aug.,
  1977) 451--467}.

\bibitem{Bahcall:78}
J.~N. {Bahcall} and C.~L. {Sarazin}, \emph{{X-ray line spectroscopy for
  clusters of galaxies. I}}, \href{http://dx.doi.org/10.1086/155839}{\emph{Ap.
  J.} {\bf 219} (Feb., 1978) 781--794}.

\bibitem{Smith:01a}
R.~K. {Smith}, N.~S. {Brickhouse}, D.~A. {Liedahl} and J.~C. {Raymond},
  \emph{{Collisional Plasma Models with APEC/APED: Emission-Line Diagnostics of
  Hydrogen-like and Helium-like Ions}},
  \href{http://dx.doi.org/10.1086/322992}{\emph{Ap. J. Lett.} {\bf 556} (Aug.,
  2001) L91--L95}.

\bibitem{Mateo:98}
M.~L. {Mateo}, \emph{{Dwarf Galaxies of the Local Group}},
  \href{http://dx.doi.org/10.1146/annurev.astro.36.1.435}{\emph{Ann. Rev.
  Astron. Astrophys.} {\bf 36} (1998) 435--506}.

\bibitem{Boyarsky:2006ag}
A.~Boyarsky, J.~Nevalainen and O.~Ruchayskiy, \emph{{Constraints on the
  parameters of radiatively decaying dark matter from the dark matter halo of
  the Milky Way and Ursa Minor}},
  \href{http://dx.doi.org/10.1051/0004-6361:20066774}{\emph{Astron. Astrophys.}
  {\bf 471} (2007) 51--57}, [\href{http://arxiv.org/abs/astro-ph/0610961}{{\tt
  astro-ph/0610961}}].

\bibitem{Loewenstein:2008yi}
M.~Loewenstein, A.~Kusenko and P.~L. Biermann, \emph{{New Limits on Sterile
  Neutrinos from Suzaku Observations of the Ursa Minor Dwarf Spheroidal
  Galaxy}},
  \href{http://dx.doi.org/10.1088/0004-637X/700/1/426}{\emph{Astrophys. J.}
  {\bf 700} (2009) 426--435}, [\href{http://arxiv.org/abs/0812.2710}{{\tt
  0812.2710}}].

\bibitem{RiemerSorensen:2009jp}
S.~Riemer-Sorensen and S.~H. Hansen, \emph{{Decaying dark matter in Draco}},
  \href{http://dx.doi.org/10.1051/0004-6361/200912430}{\emph{Astron.
  Astrophys.} {\bf 500} (2009) L37--L40},
  [\href{http://arxiv.org/abs/0901.2569}{{\tt 0901.2569}}].

\bibitem{Loewenstein:2009cm}
M.~Loewenstein and A.~Kusenko, \emph{{Dark Matter Search Using Chandra
  Observations of Willman 1, and a Spectral Feature Consistent with a Decay
  Line of a 5 keV Sterile Neutrino}},
  \href{http://dx.doi.org/10.1088/0004-637X/714/1/652}{\emph{Astrophys. J.}
  {\bf 714} (2010) 652--662}, [\href{http://arxiv.org/abs/0912.0552}{{\tt
  0912.0552}}].

\bibitem{Boyarsky:2010ci}
A.~Boyarsky, O.~Ruchayskiy, D.~Iakubovskyi, M.~G. Walker, S.~Riemer-Sorensen
  et~al., \emph{{Searching for Dark Matter in X-Rays: How to Check the Dark
  Matter origin of a spectral feature}},
  \href{http://dx.doi.org/10.1111/j.1365-2966.2010.17004.x}{\emph{Mon.Not.Roy.Astron.Soc.}
  {\bf 407} (2010) 1188--1202}, [\href{http://arxiv.org/abs/1001.0644}{{\tt
  1001.0644}}].

\bibitem{Mirabal:2010jj}
N.~Mirabal and D.~Nieto, \emph{{Willman 1: An X-ray shot in the dark with
  Chandra}},  \href{http://arxiv.org/abs/1003.3745}{{\tt 1003.3745}}.

\bibitem{Mirabal:2010an}
N.~Mirabal, \emph{{Swift observation of Segue 1: constraints on sterile
  neutrino parameters in the darkest galaxy}},
  \href{http://dx.doi.org/10.1111/j.1745-3933.2010.00963.x}{\emph{Mon. Not.
  Roy. Astron. Soc.} {\bf 409} (2010) 128},
  [\href{http://arxiv.org/abs/1010.4706}{{\tt 1010.4706}}].

\bibitem{Loewenstein:2012px}
M.~Loewenstein and A.~Kusenko, \emph{{Dark Matter Search Using XMM-Newton
  Observations of Willman 1}},
  \href{http://dx.doi.org/10.1088/0004-637X/751/2/82}{\emph{Astrophys. J.} {\bf
  751} (2012) 82}, [\href{http://arxiv.org/abs/1203.5229}{{\tt 1203.5229}}].

\bibitem{Kusenko:2012ch}
A.~Kusenko, M.~Loewenstein and T.~T. Yanagida, \emph{{Moduli dark matter and
  the search for its decay line using Suzaku X-ray telescope}},
  \href{http://dx.doi.org/10.1103/PhysRevD.87.043508}{\emph{Phys. Rev.} {\bf
  D87} (2013) 043508}, [\href{http://arxiv.org/abs/1209.6403}{{\tt
  1209.6403}}].

\bibitem{Malyshev:2014xqa}
D.~Malyshev, A.~Neronov and D.~Eckert, \emph{{Constraints on 3.55 keV line
  emission from stacked observations of dwarf spheroidal galaxies}},
  \href{http://dx.doi.org/10.1103/PhysRevD.90.103506}{\emph{Phys. Rev.} {\bf
  D90} (Nov., 2014) 103506}, [\href{http://arxiv.org/abs/1408.3531}{{\tt
  1408.3531}}].

\bibitem{Sonbas:15}
E.~{Sonbas}, B.~{Rangelov}, O.~{Kargaltsev}, K.~S. {Dhuga} and J.~{Hare},
  \emph{{X-ray Sources in the Dwarf Spheroidal Galaxy Draco}}, {\emph{ArXiv
  e-prints} (May, 2015) }.

\bibitem{RiemerSorensen:2006pi}
S.~Riemer-Sorensen, K.~Pedersen, S.~H. Hansen and H.~Dahle, \emph{{Probing the
  nature of dark matter with Cosmic X-rays: Constraints from Dark blobs and
  grating spectra of galaxy clusters}},
  \href{http://dx.doi.org/10.1103/PhysRevD.76.043524}{\emph{Phys. Rev.} {\bf
  D76} (2007) 043524}, [\href{http://arxiv.org/abs/astro-ph/0610034}{{\tt
  astro-ph/0610034}}].

\bibitem{Boyarsky:2006kc}
A.~Boyarsky, O.~Ruchayskiy and M.~Markevitch, \emph{{Constraints on parameters
  of radiatively decaying dark matter from the galaxy cluster 1E0657-56}},
  \href{http://dx.doi.org/10.1086/524397}{\emph{Astrophys. J.} {\bf 673} (2008)
  752--757}, [\href{http://arxiv.org/abs/astro-ph/0611168}{{\tt
  astro-ph/0611168}}].

\bibitem{Watson:2006qb}
C.~R. Watson, J.~F. Beacom, H.~Yuksel and T.~P. Walker, \emph{{Direct X-ray
  Constraints on Sterile Neutrino Warm Dark Matter}},
  \href{http://dx.doi.org/10.1103/PhysRevD.74.033009}{\emph{Phys. Rev.} {\bf
  D74} (2006) 033009}, [\href{http://arxiv.org/abs/astro-ph/0605424}{{\tt
  astro-ph/0605424}}].

\bibitem{Yuksel:2007xh}
H.~Yuksel, J.~F. Beacom and C.~R. Watson, \emph{{Strong Upper Limits on Sterile
  Neutrino Warm Dark Matter}},
  \href{http://dx.doi.org/10.1103/PhysRevLett.101.121301}{\emph{Phys. Rev.
  Lett.} {\bf 101} (2008) 121301}, [\href{http://arxiv.org/abs/0706.4084}{{\tt
  0706.4084}}].

\bibitem{Boyarsky:2006hr}
A.~Boyarsky, J.~W. den Herder, A.~Neronov and O.~Ruchayskiy, \emph{{Search for
  the light dark matter with an X-ray spectrometer}},
  \href{http://dx.doi.org/10.1016/j.astropartphys.2007.06.003}{\emph{Astropart.
  Phys.} {\bf 28} (2007) 303--311},
  [\href{http://arxiv.org/abs/astro-ph/0612219}{{\tt astro-ph/0612219}}].

\bibitem{Horiuchi:2015pda}
S.~Horiuchi, K.~C.~Y. Ng, J.~M. Gaskins, M.~Smith and R.~Preece,
  \emph{{Improved limits on sterile neutrino dark matter from full-sky
  observations by the Fermi-GBM}},  in \emph{{Fifth International Fermi
  Symposium Nagoya, Japan, October 20-24, 2014}}, 2015.
\newblock \href{http://arxiv.org/abs/1502.03399}{{\tt 1502.03399}}.

\bibitem{Prokhorov:2010us}
D.~A. Prokhorov and J.~Silk, \emph{{Can the Excess in the FeXXVI Ly Gamma Line
  from the Galactic Center Provide Evidence for 17 keV Sterile Neutrinos?}},
  \href{http://dx.doi.org/10.1088/2041-8205/725/2/L131}{\emph{Astrophys. J.}
  {\bf 725} (2010) L131}, [\href{http://arxiv.org/abs/1001.0215}{{\tt
  1001.0215}}].

\bibitem{Koyama:06}
K.~{Koyama}, Y.~{Hyodo}, T.~{Inui}, H.~{Nakajima}, H.~{Matsumoto}, T.~G.
  {Tsuru} et~al., \emph{{Iron and Nickel Line Diagnostics for the Galactic
  Center Diffuse Emission}}, {\emph{PASJ} {\bf 59} (Jan., 2007) 245--255}.

\bibitem{Borriello:2011un}
E.~Borriello, M.~Paolillo, G.~Miele, G.~Longo and R.~Owen, \emph{{Constraints
  on sterile neutrino dark matter from XMM--Newton observation of M33}},
  \href{http://dx.doi.org/10.1111/j.1365-2966.2012.21498.x}{\emph{Mon. Not.
  Roy. Astron. Soc.} {\bf 425} (2012) 1628--1632},
  [\href{http://arxiv.org/abs/1109.5943}{{\tt 1109.5943}}].

\bibitem{Iakubovskyi:13}
D.~Iakubovskyi, \emph{{Constraining properties of dark matter particles using
  astrophysical data}}.
\newblock PhD thesis, Leiden University, 2013.

\bibitem{Boyarsky:2014ska}
A.~Boyarsky, J.~Franse, D.~Iakubovskyi and O.~Ruchayskiy, \emph{{Checking the
  dark matter origin of 3.53~keV line with the Milky Way center}},
  \href{http://arxiv.org/abs/1408.2503}{{\tt 1408.2503}}.

\bibitem{Riemer-Sorensen:2014yda}
S.~Riemer-Sorensen, \emph{{Questioning a 3.5 keV dark matter emission line}},
  \href{http://arxiv.org/abs/1405.7943}{{\tt 1405.7943}}.

\bibitem{Jeltema:2014qfa}
T.~E. Jeltema and S.~Profumo, \emph{{Discovery of a 3.5 keV line in the
  Galactic Centre and a critical look at the origin of the line across
  astronomical targets}},
  \href{http://dx.doi.org/10.1093/mnras/stv768}{\emph{Mon. Not. Roy. Astron.
  Soc.} {\bf 450} (2015) 2143--2152},
  [\href{http://arxiv.org/abs/1408.1699}{{\tt 1408.1699}}].

\bibitem{Anderson:2014tza}
M.~E. Anderson, E.~Churazov and J.~N. Bregman, \emph{{Non-Detection of X-Ray
  Emission From Sterile Neutrinos in Stacked Galaxy Spectra}},
  \href{http://dx.doi.org/10.1093/mnras/stv1559}{\emph{Mon. Not. Roy. Astron.
  Soc.} {\bf 452} (2015) 3905}, [\href{http://arxiv.org/abs/1408.4115}{{\tt
  1408.4115}}].

\bibitem{Urban:2014yda}
O.~Urban, N.~Werner, S.~W. Allen, A.~Simionescu, J.~S. Kaastra and L.~E.
  Strigari, \emph{{A Suzaku Search for Dark Matter Emission Lines in the X-ray
  Brightest Galaxy Clusters}},
  \href{http://dx.doi.org/10.1093/mnras/stv1142}{\emph{Mon. Not. Roy. Astron.
  Soc.} {\bf 451} (2015) 2447--2461},
  [\href{http://arxiv.org/abs/1411.0050}{{\tt 1411.0050}}].

\bibitem{Carlson:2014lla}
E.~Carlson, T.~Jeltema and S.~Profumo, \emph{{Where do the 3.5 keV photons come
  from? A morphological study of the Galactic Center and of Perseus}},
  \href{http://dx.doi.org/10.1088/1475-7516/2015/02/009}{\emph{JCAP} {\bf 1502}
  (2015) 009}, [\href{http://arxiv.org/abs/1411.1758}{{\tt 1411.1758}}].

\bibitem{Tamura:2014mta}
T.~Tamura, R.~Iizuka, Y.~Maeda, K.~Mitsuda and N.~Y. Yamasaki, \emph{{An X-ray
  Spectroscopic Search for Dark Matter in the Perseus Cluster with Suzaku}},
  \href{http://dx.doi.org/10.1093/pasj/psu156}{\emph{Publ. Astron. Soc. Jap.}
  {\bf 67} (2015) 23}, [\href{http://arxiv.org/abs/1412.1869}{{\tt
  1412.1869}}].

\bibitem{Koyama:2014zca}
{\scshape ASTRO-H Science Working Group} collaboration, K.~Koyama et~al.,
  \emph{{ASTRO-H White Paper - Plasma Diagnostic and Dynamics of the Galactic
  Center Region}},  \href{http://arxiv.org/abs/1412.1170}{{\tt 1412.1170}}.

\bibitem{Ng:2015gfa}
K.~C.~Y. Ng, S.~Horiuchi, J.~M. Gaskins, M.~Smith and R.~Preece,
  \emph{{Improved Limits on Sterile Neutrino Dark Matter using Full-Sky Fermi
  Gamma-Ray Burst Monitor Data}},
  \href{http://dx.doi.org/10.1103/PhysRevD.92.043503}{\emph{Phys. Rev.} {\bf
  D92} (2015) 043503}, [\href{http://arxiv.org/abs/1504.04027}{{\tt
  1504.04027}}].

\bibitem{Sekiya:2015jsa}
N.~Sekiya, N.~Y. Yamasaki and K.~Mitsuda, \emph{{A Search for a keV Signature
  of Radiatively Decaying Dark Matter with Suzaku XIS Observations of the X-ray
  Diffuse Background}},  \href{http://arxiv.org/abs/1504.02826}{{\tt
  1504.02826}}.

\bibitem{Jeltema:2015mee}
T.~E. Jeltema and S.~Profumo, \emph{{Deep XMM Observations of Draco rule out at
  the 99\% Confidence Level a Dark Matter Decay Origin for the 3.5 keV Line}},
  \href{http://dx.doi.org/10.1093/mnras/stw578}{\emph{Mon. Not. Roy. Astron.
  Soc.} {\bf 458} (2016) 3592--3596},
  [\href{http://arxiv.org/abs/1512.01239}{{\tt 1512.01239}}].

\bibitem{Ruchayskiy:2015onc}
O.~Ruchayskiy, A.~Boyarsky, D.~Iakubovskyi, E.~Bulbul, D.~Eckert, J.~Franse
  et~al., \emph{{Searching for decaying dark matter in deep XMM-Newton
  observation of the Draco dwarf spheroidal}},
  \href{http://arxiv.org/abs/1512.07217}{{\tt 1512.07217}}.

\bibitem{Neronov:2016wdd}
A.~Neronov, D.~Malyshev and D.~Eckert, \emph{{Decaying dark matter search with
  NuSTAR deep sky observations}},  \href{http://arxiv.org/abs/1607.07328}{{\tt
  1607.07328}}.

\bibitem{Perez:2016tcq}
K.~Perez, K.~C.~Y. Ng, J.~F. Beacom, C.~Hersh, S.~Horiuchi and R.~Krivonos,
  \emph{{(Almost) Closing the Sterile Neutrino Dark Matter Window with
  NuSTAR}},  \href{http://arxiv.org/abs/1609.00667}{{\tt 1609.00667}}.

\bibitem{Iakubovskyi:2015dna}
D.~Iakubovskyi, E.~Bulbul, A.~R. Foster, D.~Savchenko and V.~Sadova,
  \emph{{Testing the origin of ~3.55 keV line in individual galaxy clusters
  observed with XMM-Newton}},  \href{http://arxiv.org/abs/1508.05186}{{\tt
  1508.05186}}.

\bibitem{Iakubovskyi:2015wma}
D.~Iakubovskyi, \emph{{Observation of the new line at ~3.55 keV in X-ray
  spectra of galaxies and galaxy clusters}},
  \href{http://arxiv.org/abs/1510.00358}{{\tt 1510.00358}}.

\bibitem{Boyarsky:2014paa}
A.~Boyarsky, J.~Franse, D.~Iakubovskyi and O.~Ruchayskiy, \emph{{Comment on the
  paper "Dark matter searches going bananas: the contribution of Potassium (and
  Chlorine) to the 3.5 keV line" by T. Jeltema and S. Profumo}},
  \href{http://arxiv.org/abs/1408.4388}{{\tt 1408.4388}}.

\bibitem{Bulbul:2014ala}
E.~Bulbul, M.~Markevitch, A.~R. Foster, R.~K. Smith, M.~Loewenstein and S.~W.
  Randall, \emph{Comment on "dark matter searches going bananas: the
  contribution of potassium (and chlorine) to the 3.5 kev line"},
  \href{http://arxiv.org/abs/1409.4143}{{\tt 1409.4143}}.

\bibitem{Jeltema:2014mla}
T.~Jeltema and S.~Profumo, \emph{Reply to two comments on "dark matter searches
  going bananas the contribution of potassium (and chlorine) to the 3.5 kev
  line"},  \href{http://arxiv.org/abs/1411.1759}{{\tt 1411.1759}}.

\bibitem{Franse:2016dln}
J.~Franse et~al., \emph{{Radial Profile of the 3.55 keV line out to $R_{200}$
  in the Perseus Cluster}},  \href{http://arxiv.org/abs/1604.01759}{{\tt
  1604.01759}}.

\bibitem{Moster:10}
B.~P. {Moster}, R.~S. {Somerville}, C.~{Maulbetsch}, F.~C. {van den Bosch},
  A.~V. {Macci{\`o}}, T.~{Naab} et~al., \emph{{Constraints on the Relationship
  between Stellar Mass and Halo Mass at Low and High Redshift}},
  \href{http://dx.doi.org/10.1088/0004-637X/710/2/903}{\emph{ApJ} {\bf 710}
  (Feb., 2010) 903--923}.

\bibitem{Prada:2011jf}
F.~Prada, A.~A. Klypin, A.~J. Cuesta, J.~E. Betancort-Rijo and J.~Primack,
  \emph{{Halo concentrations in the standard LCDM cosmology}},
  \href{http://dx.doi.org/10.1111/j.1365-2966.2012.21007.x}{\emph{Mon. Not.
  Roy. Astron. Soc.} {\bf 428} (2012) 3018--3030}.

\bibitem{Hofmann:2016urz}
F.~Hofmann, J.~S. Sanders, K.~Nandra, N.~Clerc and M.~Gaspari, \emph{{7.1 keV
  sterile neutrino constraints from X-ray observations of 33 clusters of
  galaxies with Chandra ACIS}},
  \href{http://dx.doi.org/10.1051/0004-6361/201527977}{\emph{Astron.
  Astrophys.} {\bf 592} (2016) A112},
  [\href{http://arxiv.org/abs/1606.04091}{{\tt 1606.04091}}].

\bibitem{Aharonian:2016gzq}
{\scshape Hitomi} collaboration, F.~A. Aharonian et~al., \emph{{Hitomi
  constraints on the 3.5 keV line in the Perseus galaxy cluster}},
  \href{http://arxiv.org/abs/1607.07420}{{\tt 1607.07420}}.

\bibitem{Phillips:2015wla}
K.~J.~H. Phillips, B.~Sylwester and J.~Sylwester, \emph{{THE X-RAY LINE FEATURE
  AT 3.5 KeV IN GALAXY CLUSTER SPECTRA}},
  \href{http://dx.doi.org/10.1088/0004-637X/809/1/50}{\emph{Astrophys. J.} {\bf
  809} (2015) 50}.

\bibitem{Gu:2015gqm}
L.~Gu, J.~Kaastra, A.~J.~J. Raassen, P.~D. Mullen, R.~S. Cumbee, D.~Lyons
  et~al., \emph{{A novel scenario for the possible X-ray line feature at ~3.5
  keV: Charge exchange with bare sulfur ions}},
  \href{http://dx.doi.org/10.1051/0004-6361/201527634}{\emph{Astron.
  Astrophys.} {\bf 584} (2015) L11},
  [\href{http://arxiv.org/abs/1511.06557}{{\tt 1511.06557}}].

\bibitem{Drag15}
O.~Dragoun and D.~Venos, \emph{Searches for active and sterile neutrinos in
  beta-ray spectra},  \href{http://arxiv.org/abs/1504.07496}{{\tt 1504.07496}}.

\bibitem{Cook48}
C.~S. Cook, L.~M. Langer and H.~C. Price{\ }Jr., \emph{{The $\beta$-spectrum of
  ${}^{35}$S and the mass of the neutrino}}, {\emph{Phys. Rev.} {\bf 73} (1948)
  1395}.

\bibitem{Berg72}
K.-E. Bergkvist, \emph{A high-luminosity, high resolution study of the
  end-point behavior of the tritium $\beta$-spectrum (i). basic experimental
  procedure and analysis with regard to neutrino mass and neutrino degeneracy},
  {\emph{Nucl. Phys. B} {\bf 39} (1972) 317--70}.

\bibitem{Drex13}
G.~Drexlin, V.~Hannen, S.~Mertens and C.~Weinheimer, \emph{Current direct
  neutrino mass experiments}, {\emph{Adv. High Energy Phys.} {\bf 2013} (2013)
  36pp}, [\href{http://arxiv.org/abs/1307.0101}{{\tt 1307.0101}}].

\bibitem{Drag83}
O.~Dragoun, \emph{Internal conversion-electron spectroscopy}, {\emph{Advances
  in Electronics and Electron Physics} {\bf 60} (1983) 1--94}.

\bibitem{Schreckenbach:1983cg}
K.~Schreckenbach, G.~Colvin and F.~von Feilitzsch, \emph{Search for mixing of
  heavy neutrinos in the $\beta^+$ and $\beta^-$ spectra of the ${}^{64}${C}u
  decay}, \href{http://dx.doi.org/10.1016/0370-2693(83)90858-4}{\emph{Phys.
  Lett. B} {\bf 129} (1983) 265--8}.

\bibitem{Deutsch:1990ut}
J.~Deutsch, M.~Lebrun and R.~Prieels, \emph{{Searches for admixture of massive
  neutrinos into the electron flavor}},
  \href{http://dx.doi.org/10.1016/0375-9474(90)90541-S}{\emph{Nucl. Phys.} {\bf
  A518} (1990) 149--155}.

\bibitem{Simp85}
J.~J. Simpson, \emph{Evidence of heavy-neutrino emission in beta decay},
  {\emph{Phys Rev. Lett.} {\bf 54} (1985) 1891--3}.

\bibitem{Holzschuh:1999vy}
E.~Holzschuh, W.~Kundig, L.~Palermo, H.~Stussi and P.~Wenk, \emph{{Search for
  heavy neutrinos in the beta spectrum of Ni-63}},
  \href{http://dx.doi.org/10.1016/S0370-2693(99)00200-2}{\emph{Phys. Lett.}
  {\bf B451} (1999) 247--255}.

\bibitem{OHI1985322}
T.~Ohi, M.~Nakajima, H.~Tamura, T.~Matsuzaki, T.~Yamazaki, O.~Hashimoto et~al.,
  \emph{Search for heavy neutrinos in the beta decay of 35s. evidence against
  the 17 kev heavy neutrino},
  \href{http://dx.doi.org/http://dx.doi.org/10.1016/0370-2693(85)91336-X}{\emph{Physics
  Letters B} {\bf 160} (1985) 322 -- 324}.

\bibitem{PhysRevC.32.2215}
J.~Markey and F.~Boehm, \emph{Search for admixture of heavy neutrinos with
  masses between 5 and 55 kev},
  \href{http://dx.doi.org/10.1103/PhysRevC.32.2215}{\emph{Phys. Rev. C} {\bf
  32} (Dec, 1985) 2215--2216}.

\bibitem{PhysRevLett.55.799}
T.~Altzitzoglou, F.~Calaprice, M.~Dewey, M.~Lowry, L.~Piilonen, J.~Brorson
  et~al., \emph{Experimental search for a heavy neutrino in the beta spectrum
  of $^{35}\mathrm{S}$},
  \href{http://dx.doi.org/10.1103/PhysRevLett.55.799}{\emph{Phys. Rev. Lett.}
  {\bf 55} (Aug, 1985) 799--802}.

\bibitem{PhysRevC.36.1504}
D.~W. Hetherington, R.~L. Graham, M.~A. Lone, J.~S. Geiger and G.~E.
  Lee-Whiting, \emph{Upper limits on the mixing of heavy neutrinos in the beta
  decay of $^{63}\mathrm{Ni}$},
  \href{http://dx.doi.org/10.1103/PhysRevC.36.1504}{\emph{Phys. Rev. C} {\bf
  36} (Oct, 1987) 1504--1513}.

\bibitem{1992pnap.conf..217R}
T.~J. {Radcliffe}, M.~{Chen}, D.~A. {Imel}, H.~{Henrikson} and F.~{Boehm},
  \emph{{New limits on the 17 keV neutrino.}},  in \emph{Progress in Atomic
  Physics, Neutrinos and Gravitation} (G.~{Chardin}, O.~{Fackler} and
  J.~{Tr{\^a}n Thanh V{\^a}n}, eds.), 1992.

\bibitem{Wiet96}
F.~E. Wietfeldt and E.~B. Norman, \emph{The 17 kev neutrino}, {\emph{Phys.
  Rep.} {\bf 273} (1996) 149--97}.

\bibitem{Muel94}
S.~Muller et~al., \emph{Search for an admixture of a 17 kev neutrino in the
  $\beta$ decay of ${}^{35}${S}}, {\emph{Z Naturforsch.} {\bf 49a} (1994)
  874--84}.

\bibitem{Kraus:2012he}
C.~Kraus, A.~Singer, K.~Valerius and C.~Weinheimer, \emph{Limit on sterile
  neutrino contribution from the {M}ainz neutrino mass experiment},
  \href{http://dx.doi.org/10.1140/epjc/s10052-013-2323-z}{\emph{Eur. Phys. J.}
  {\bf C73} (2013) 2323}, [\href{http://arxiv.org/abs/1210.4194}{{\tt
  1210.4194}}].

\bibitem{Belesev:2013cba}
A.~I. Belesev, A.~I. Berlev, E.~V. Geraskin, A.~A. Golubev, N.~A. Likhovid
  et~al., \emph{{The search for an additional neutrino mass eigenstate in the
  2-100 eV region from Troitsk nu-mass data: a detailed analysis.}},
  \href{http://dx.doi.org/10.1088/0954-3899/41/1/015001}{\emph{J. Phys.} {\bf
  G41} (2014) 015001}, [\href{http://arxiv.org/abs/1307.5687}{{\tt
  1307.5687}}].

\bibitem{Galeazzi:2001py}
M.~Galeazzi, F.~Fontanelli, F.~Gatti and S.~Vitale, \emph{{Limits on the
  existence of heavy neutrinos in the range 50-1000 eV from the study of the
  ${}^{187}$Re beta spectrum}},
  \href{http://dx.doi.org/10.1103/PhysRevLett.86.1978}{\emph{Phys. Rev. Lett.}
  {\bf 86} (2001) 1978--1981}.

\bibitem{Hiddemann:1995ce}
K.~H. Hiddemann, H.~Daniel and O.~Schwentker, \emph{{Limits on neutrino masses
  from the tritium beta spectrum}},
  \href{http://dx.doi.org/10.1088/0954-3899/21/5/008}{\emph{J. Phys.} {\bf G21}
  (1995) 639--650}.

\bibitem{Simp81}
J.~J. Simpson, \emph{Measurement of the $\beta$-energy spectrum of ${}^3${H} to
  determine the antineutrino mass}, {\emph{Phys. Rev. D} {\bf 23} (1981)
  649--62}.

\bibitem{Ohsh93}
T.~Ohshima et~al., \emph{No 17 kev neutrino: Admixture < 0.073\% (95\%
  {C}.{L}.).}, {\emph{Phys. Rev. D} {\bf 47} (1993) 4840--56}.

\bibitem{Mort93}
J.~L. Mortara et~al., \emph{Evidence against a 17 kev neutrino from
  ${}^{35}${S} beta decay}, {\emph{Phys. Rev. Lett.} {\bf 70} (1993) 394--7}.

\bibitem{Drag99}
O.~Dragoun et~al., \emph{Search for an admixture of heavy neutrinos in the
  $\beta$-decay of ${}^{241}${P}u}, {\emph{J. Phys G} {\bf 25} (1999)
  1839--58}.

\bibitem{Drag00}
O.~Dragoun et~al., \emph{Improved methods of measurement and analysis of
  conversion electron and $\beta$-particle spectra}, {\emph{Appl. Radiat.
  Isot.} {\bf 52} (2000) 387--91}.

\bibitem{Chen:2015dka}
M.-C. Chen, M.~Ratz and A.~Trautner, \emph{{Nonthermal cosmic neutrino
  background}}, \href{http://dx.doi.org/10.1103/PhysRevD.92.123006}{\emph{Phys.
  Rev.} {\bf D92} (2015) 123006}, [\href{http://arxiv.org/abs/1509.00481}{{\tt
  1509.00481}}].

\bibitem{Abazajian:2002bh}
K.~N. {Abazajian} and G.~M. {Fuller}, \emph{{Bulk QCD thermodynamics and
  sterile neutrino dark matter}},
  \href{http://dx.doi.org/10.1103/PhysRevD.66.023526}{\emph{\prd} {\bf 66}
  (July, 2002) 023526--+},
  [\href{http://arxiv.org/abs/arXiv:astro-ph/0204293}{{\tt
  arXiv:astro-ph/0204293}}].

\bibitem{Dolgov:2003sg}
A.~D. Dolgov and F.~L. Villante, \emph{{BBN bounds on active sterile neutrino
  mixing}},
  \href{http://dx.doi.org/10.1016/j.nuclphysb.2003.11.031}{\emph{Nucl. Phys.}
  {\bf B679} (2004) 261--298}, [\href{http://arxiv.org/abs/hep-ph/0308083}{{\tt
  hep-ph/0308083}}].

\bibitem{Asaka:2006ek}
T.~Asaka, M.~Shaposhnikov and A.~Kusenko, \emph{{Opening a new window for warm
  dark matter}},
  \href{http://dx.doi.org/10.1016/j.physletb.2006.05.067}{\emph{Phys. Lett.}
  {\bf B638} (2006) 401--406}, [\href{http://arxiv.org/abs/hep-ph/0602150}{{\tt
  hep-ph/0602150}}].

\bibitem{Kusenko:2006rh}
A.~Kusenko, \emph{{Sterile neutrinos, dark matter, and the pulsar velocities in
  models with a Higgs singlet}},
  \href{http://dx.doi.org/10.1103/PhysRevLett.97.241301}{\emph{Phys. Rev.
  Lett.} {\bf 97} (2006) 241301},
  [\href{http://arxiv.org/abs/hep-ph/0609081}{{\tt hep-ph/0609081}}].

\bibitem{Shaposhnikov:2006xi}
M.~Shaposhnikov and I.~Tkachev, \emph{{The nuMSM, inflation, and dark matter}},
  \href{http://dx.doi.org/10.1016/j.physletb.2006.06.063}{\emph{Phys. Lett.}
  {\bf B639} (2006) 414--417}, [\href{http://arxiv.org/abs/hep-ph/0604236}{{\tt
  hep-ph/0604236}}].

\bibitem{Boyanovsky:2006it}
D.~Boyanovsky and C.~M. Ho, \emph{{Sterile neutrino production via
  active-sterile oscillations: The Quantum Zeno effect}},
  \href{http://dx.doi.org/10.1088/1126-6708/2007/07/030}{\emph{JHEP} {\bf 07}
  (2007) 030}, [\href{http://arxiv.org/abs/hep-ph/0612092}{{\tt
  hep-ph/0612092}}].

\bibitem{Boyanovsky:2007as}
D.~Boyanovsky, \emph{{Production of a sterile species via active-sterile
  mixing: An Exactly solvable model}},
  \href{http://dx.doi.org/10.1103/PhysRevD.76.103514}{\emph{Phys. Rev.} {\bf
  D76} (2007) 103514}, [\href{http://arxiv.org/abs/0706.3167}{{\tt
  0706.3167}}].

\bibitem{Gorbunov:2007ak}
D.~Gorbunov and M.~Shaposhnikov, \emph{{How to find neutral leptons of the
  $\nu$MSM?}}, \href{http://dx.doi.org/10.1007/JHEP11(2013)101,
  10.1088/1126-6708/2007/10/015}{\emph{JHEP} {\bf 10} (2007) 015},
  [\href{http://arxiv.org/abs/0705.1729}{{\tt 0705.1729}}].

\bibitem{Kishimoto:2008ic}
C.~T. Kishimoto and G.~M. Fuller, \emph{{Lepton Number-Driven Sterile Neutrino
  Production in the Early Universe}},
  \href{http://dx.doi.org/10.1103/PhysRevD.78.023524}{\emph{Phys. Rev.} {\bf
  D78} (2008) 023524}, [\href{http://arxiv.org/abs/0802.3377}{{\tt
  0802.3377}}].

\bibitem{Petraki:2008ef}
K.~Petraki, \emph{{Small-scale structure formation properties of chilled
  sterile neutrinos as dark matter}},
  \href{http://dx.doi.org/10.1103/PhysRevD.77.105004}{\emph{Phys. Rev.} {\bf
  D77} (2008) 105004}, [\href{http://arxiv.org/abs/0801.3470}{{\tt
  0801.3470}}].

\bibitem{Petraki:2007gq}
K.~Petraki and A.~Kusenko, \emph{{Dark-matter sterile neutrinos in models with
  a gauge singlet in the Higgs sector}},
  \href{http://dx.doi.org/10.1103/PhysRevD.77.065014}{\emph{Phys. Rev.} {\bf
  D77} (2008) 065014}, [\href{http://arxiv.org/abs/0711.4646}{{\tt
  0711.4646}}].

\bibitem{Bezrukov:2009th}
F.~Bezrukov, H.~Hettmansperger and M.~Lindner, \emph{{keV sterile neutrino Dark
  Matter in gauge extensions of the Standard Model}},
  \href{http://dx.doi.org/10.1103/PhysRevD.81.085032}{\emph{Phys. Rev.} {\bf
  D81} (2010) 085032}, [\href{http://arxiv.org/abs/0912.4415}{{\tt
  0912.4415}}].

\bibitem{Kusenko:2010ik}
A.~Kusenko, F.~Takahashi and T.~T. Yanagida, \emph{{Dark Matter from Split
  Seesaw}}, \href{http://dx.doi.org/10.1016/j.physletb.2010.08.031}{\emph{Phys.
  Lett.} {\bf B693} (2010) 144--148},
  [\href{http://arxiv.org/abs/1006.1731}{{\tt 1006.1731}}].

\bibitem{Nemevsek:2012cd}
M.~Nemevsek, G.~Senjanovic and Y.~Zhang, \emph{{Warm Dark Matter in Low Scale
  Left-Right Theory}},
  \href{http://dx.doi.org/10.1088/1475-7516/2012/07/006}{\emph{JCAP} {\bf 1207}
  (2012) 006}, [\href{http://arxiv.org/abs/1205.0844}{{\tt 1205.0844}}].

\bibitem{Bezrukov:2012as}
F.~Bezrukov, A.~Kartavtsev and M.~Lindner, \emph{{Leptogenesis in models with
  keV sterile neutrino dark matter}},
  \href{http://dx.doi.org/10.1088/0954-3899/40/9/095202}{\emph{J. Phys.} {\bf
  G40} (2013) 095202}, [\href{http://arxiv.org/abs/1204.5477}{{\tt
  1204.5477}}].

\bibitem{Tsuyuki:2014aia}
T.~Tsuyuki, \emph{{Neutrino masses, leptogenesis, and sterile neutrino dark
  matter}}, \href{http://dx.doi.org/10.1103/PhysRevD.90.013007}{\emph{Phys.
  Rev.} {\bf D90} (2014) 013007}, [\href{http://arxiv.org/abs/1403.5053}{{\tt
  1403.5053}}].

\bibitem{Merle:2013wta}
A.~Merle, V.~Niro and D.~Schmidt, \emph{{New Production Mechanism for keV
  Sterile Neutrino Dark Matter by Decays of Frozen-In Scalars}},
  \href{http://dx.doi.org/10.1088/1475-7516/2014/03/028}{\emph{JCAP} {\bf 1403}
  (2014) 028}, [\href{http://arxiv.org/abs/1306.3996}{{\tt 1306.3996}}].

\bibitem{Bezrukov:2014nza}
F.~Bezrukov and D.~Gorbunov, \emph{{Relic Gravity Waves and 7 keV Dark Matter
  from a GeV scale inflaton}},
  \href{http://dx.doi.org/10.1016/j.physletb.2014.07.060}{\emph{Phys. Lett.}
  {\bf B736} (2014) 494--498}, [\href{http://arxiv.org/abs/1403.4638}{{\tt
  1403.4638}}].

\bibitem{Roland:2014vba}
S.~B. Roland, B.~Shakya and J.~D. Wells, \emph{{Neutrino Masses and Sterile
  Neutrino Dark Matter from the PeV Scale}},
  \href{http://arxiv.org/abs/1412.4791}{{\tt 1412.4791}}.

\bibitem{Abada:2014zra}
A.~Abada, G.~Arcadi and M.~Lucente, \emph{{Dark Matter in the minimal Inverse
  Seesaw mechanism}},
  \href{http://dx.doi.org/10.1088/1475-7516/2014/10/001}{\emph{JCAP} {\bf 1410}
  (2014) 001}, [\href{http://arxiv.org/abs/1406.6556}{{\tt 1406.6556}}].

\bibitem{Lello:2014yha}
L.~Lello and D.~Boyanovsky, \emph{{Cosmological Implications of Light Sterile
  Neutrinos produced after the QCD Phase Transition}},
  \href{http://dx.doi.org/10.1103/PhysRevD.91.063502}{\emph{Phys. Rev.} {\bf
  D91} (2015) 063502}, [\href{http://arxiv.org/abs/1411.2690}{{\tt
  1411.2690}}].

\bibitem{Humbert:2015epa}
P.~Humbert, M.~Lindner and J.~Smirnov, \emph{{The Inverse Seesaw in Conformal
  Electro-Weak Symmetry Breaking and Phenomenological Consequences}},
  \href{http://dx.doi.org/10.1007/JHEP06(2015)035}{\emph{JHEP} {\bf 06} (2015)
  035}, [\href{http://arxiv.org/abs/1503.03066}{{\tt 1503.03066}}].

\bibitem{Humbert:2015yva}
P.~Humbert, M.~Lindner, S.~Patra and J.~Smirnov, \emph{{Lepton Number Violation
  within the Conformal Inverse Seesaw}},
  \href{http://dx.doi.org/10.1007/JHEP09(2015)064}{\emph{JHEP} {\bf 09} (2015)
  064}, [\href{http://arxiv.org/abs/1505.07453}{{\tt 1505.07453}}].

\bibitem{Adulpravitchai:2014xna}
A.~Adulpravitchai and M.~A. Schmidt, \emph{{A Fresh Look at keV Sterile
  Neutrino Dark Matter from Frozen-In Scalars}},
  \href{http://dx.doi.org/10.1007/JHEP01(2015)006}{\emph{JHEP} {\bf 1501}
  (2015) 006}, [\href{http://arxiv.org/abs/1409.4330}{{\tt 1409.4330}}].

\bibitem{Adulpravitchai:2015mna}
A.~Adulpravitchai and M.~A. Schmidt, \emph{{Sterile Neutrino Dark Matter
  Production in the Neutrino-phillic Two Higgs Doublet Model}},
  \href{http://arxiv.org/abs/1507.05694}{{\tt 1507.05694}}.

\bibitem{Drewes:2015eoa}
M.~Drewes and J.~U. Kang, \emph{{Sterile neutrino Dark Matter production from
  scalar decay in a thermal bath}},
  \href{http://arxiv.org/abs/1510.05646}{{\tt 1510.05646}}.

\bibitem{Merle:2015vzu}
A.~Merle, A.~Schneider and M.~Totzauer, \emph{{Dodelson-Widrow Production of
  Sterile Neutrino Dark Matter with Non-Trivial Initial Abundance}},
  \href{http://dx.doi.org/10.1088/1475-7516/2016/04/003}{\emph{JCAP} {\bf 1604}
  (2016) 003}, [\href{http://arxiv.org/abs/1512.05369}{{\tt 1512.05369}}].

\bibitem{Abazajian:2009hx}
K.~N. Abazajian, \emph{{Detection of Dark Matter Decay in the X-ray}},
  \href{http://arxiv.org/abs/0903.2040}{{\tt 0903.2040}}.

\bibitem{Merle:2014xpa}
A.~Merle and A.~Schneider, \emph{{Production of Sterile Neutrino Dark Matter
  and the 3.5 keV line}},
  \href{http://dx.doi.org/10.1016/j.physletb.2015.07.080}{\emph{Phys. Lett.}
  {\bf B749} (2015) 283--288}, [\href{http://arxiv.org/abs/1409.6311}{{\tt
  1409.6311}}].

\bibitem{Barbieri:1989ti}
R.~Barbieri and A.~Dolgov, \emph{{Bounds on Sterile-neutrinos from
  Nucleosynthesis}},
  \href{http://dx.doi.org/10.1016/0370-2693(90)91203-N}{\emph{Phys. Lett.} {\bf
  B237} (1990) 440}.

\bibitem{Kainulainen:1990ds}
K.~Kainulainen, \emph{{Light Singlet Neutrinos and the Primordial
  Nucleosynthesis}},
  \href{http://dx.doi.org/10.1016/0370-2693(90)90054-A}{\emph{Phys. Lett.} {\bf
  B244} (1990) 191--195}.

\bibitem{Ghiglieri:2015jua}
J.~Ghiglieri and M.~Laine, \emph{{Improved determination of sterile neutrino
  dark matter spectrum}},  \href{http://arxiv.org/abs/1506.06752}{{\tt
  1506.06752}}.

\bibitem{Asaka:2006rw}
T.~Asaka, M.~Laine and M.~Shaposhnikov, \emph{{On the hadronic contribution to
  sterile neutrino production}},
  \href{http://dx.doi.org/10.1088/1126-6708/2006/06/053}{\emph{JHEP} {\bf 06}
  (2006) 053}, [\href{http://arxiv.org/abs/hep-ph/0605209}{{\tt
  hep-ph/0605209}}].

\bibitem{Bodeker:2014hqa}
D.~Bodeker and M.~Laine, \emph{{Kubo relations and radiative corrections for
  lepton number washout}},
  \href{http://dx.doi.org/10.1088/1475-7516/2014/05/041}{\emph{JCAP} {\bf 1405}
  (2014) 041}, [\href{http://arxiv.org/abs/1403.2755}{{\tt 1403.2755}}].

\bibitem{Schwinger:1960qe}
J.~S. Schwinger, \emph{{Brownian motion of a quantum oscillator}},
  {\emph{J.Math.Phys.} {\bf 2} (1961) 407--432}.

\bibitem{Bakshi:1962dv}
P.~M. Bakshi and K.~T. Mahanthappa, \emph{{Expectation value formalism in
  quantum field theory. 1.}}, {\emph{J.Math.Phys.} {\bf 4} (1963) 1--11}.

\bibitem{Bakshi:1963bn}
P.~M. Bakshi and K.~T. Mahanthappa, \emph{{Expectation value formalism in
  quantum field theory. 2.}}, {\emph{J.Math.Phys.} {\bf 4} (1963) 12--16}.

\bibitem{Keldysh:1964ud}
L.~Keldysh, \emph{{Diagram technique for nonequilibrium processes}},
  {\emph{Zh.Eksp.Teor.Fiz.} {\bf 47} (1964) 1515--1527}.

\bibitem{Buchmuller:2000nd}
W.~Buchmuller and S.~Fredenhagen, \emph{{Quantum mechanics of baryogenesis}},
  \href{http://dx.doi.org/10.1016/S0370-2693(00)00573-6}{\emph{Phys. Lett.}
  {\bf B483} (2000) 217--224}, [\href{http://arxiv.org/abs/hep-ph/0004145}{{\tt
  hep-ph/0004145}}].

\bibitem{Anisimov:2008dz}
A.~Anisimov, W.~Buchmuller, M.~Drewes and S.~Mendizabal, \emph{{Nonequilibrium
  Dynamics of Scalar Fields in a Thermal Bath}},
  \href{http://dx.doi.org/10.1016/j.aop.2009.01.001}{\emph{Annals Phys.} {\bf
  324} (2009) 1234--1260}, [\href{http://arxiv.org/abs/0812.1934}{{\tt
  0812.1934}}].

\bibitem{Anisimov:2010aq}
A.~Anisimov, W.~Buchm�ller, M.~Drewes and S.~Mendizabal, \emph{{Leptogenesis
  from Quantum Interference in a Thermal Bath}},
  \href{http://dx.doi.org/10.1103/PhysRevLett.104.121102}{\emph{Phys. Rev.
  Lett.} {\bf 104} (2010) 121102}, [\href{http://arxiv.org/abs/1001.3856}{{\tt
  1001.3856}}].

\bibitem{Anisimov:2010dk}
A.~Anisimov, W.~Buchmueller, M.~Drewes and S.~Mendizabal, \emph{{Quantum
  Leptogenesis I}}, \href{http://dx.doi.org/10.1016/j.aop.2011.02.002,
  10.1016/j.aop.2013.05.00}{\emph{Annals Phys.} {\bf 326} (2011) 1998--2038},
  [\href{http://arxiv.org/abs/1012.5821}{{\tt 1012.5821}}].

\bibitem{Fidler:2011yq}
C.~Fidler, M.~Herranen, K.~Kainulainen and P.~M. Rahkila, \emph{{Flavoured
  quantum Boltzmann equations from cQPA}},
  \href{http://dx.doi.org/10.1007/JHEP02(2012)065}{\emph{JHEP} {\bf 02} (2012)
  065}, [\href{http://arxiv.org/abs/1108.2309}{{\tt 1108.2309}}].

\bibitem{Beneke:2010dz}
M.~Beneke, B.~Garbrecht, C.~Fidler, M.~Herranen and P.~Schwaller,
  \emph{{Flavoured Leptogenesis in the CTP Formalism}},
  \href{http://dx.doi.org/10.1016/j.nuclphysb.2010.10.001}{\emph{Nucl.Phys.}
  {\bf B843} (2011) 177--212}, [\href{http://arxiv.org/abs/1007.4783}{{\tt
  1007.4783}}].

\bibitem{Beneke:2010wd}
M.~Beneke, B.~Garbrecht, M.~Herranen and P.~Schwaller, \emph{{Finite Number
  Density Corrections to Leptogenesis}},
  \href{http://dx.doi.org/10.1016/j.nuclphysb.2010.05.003}{\emph{Nucl.Phys.}
  {\bf B838} (2010) 1--27}, [\href{http://arxiv.org/abs/1002.1326}{{\tt
  1002.1326}}].

\bibitem{Drewes:2010pf}
M.~Drewes, \emph{{On the Role of Quasiparticles and thermal Masses in
  Nonequilibrium Processes in a Plasma}},
  \href{http://arxiv.org/abs/1012.5380}{{\tt 1012.5380}}.

\bibitem{Drewes:2012qw}
M.~Drewes, S.~Mendizabal and C.~Weniger, \emph{{The Boltzmann Equation from
  Quantum Field Theory}},
  \href{http://dx.doi.org/10.1016/j.physletb.2012.11.046}{\emph{Phys. Lett.}
  {\bf B718} (2013) 1119--1124}, [\href{http://arxiv.org/abs/1202.1301}{{\tt
  1202.1301}}].

\bibitem{Cirigliano:2009yt}
V.~Cirigliano, C.~Lee, M.~J. Ramsey-Musolf and S.~Tulin, \emph{{Flavored
  Quantum Boltzmann Equations}},
  \href{http://dx.doi.org/10.1103/PhysRevD.81.103503}{\emph{Phys. Rev.} {\bf
  D81} (2010) 103503}, [\href{http://arxiv.org/abs/0912.3523}{{\tt
  0912.3523}}].

\bibitem{Garny:2011hg}
M.~Garny, A.~Kartavtsev and A.~Hohenegger, \emph{{Leptogenesis from first
  principles in the resonant regime}},
  \href{http://dx.doi.org/10.1016/j.aop.2012.10.007}{\emph{Annals Phys.} {\bf
  328} (2013) 26--63}, [\href{http://arxiv.org/abs/1112.6428}{{\tt
  1112.6428}}].

\bibitem{Millington:2012pf}
P.~Millington and A.~Pilaftsis, \emph{{Perturbative nonequilibrium thermal
  field theory}},
  \href{http://dx.doi.org/10.1103/PhysRevD.88.085009}{\emph{Phys. Rev.} {\bf
  D88} (2013) 085009}, [\href{http://arxiv.org/abs/1211.3152}{{\tt
  1211.3152}}].

\bibitem{Dev:2014wsa}
P.~S. Bhupal~Dev, P.~Millington, A.~Pilaftsis and D.~Teresi,
  \emph{{Kadanoff-��Baym approach to flavour mixing and oscillations in
  resonant leptogenesis}},
  \href{http://dx.doi.org/10.1016/j.nuclphysb.2014.12.003}{\emph{Nucl. Phys.}
  {\bf B891} (2015) 128--158}, [\href{http://arxiv.org/abs/1410.6434}{{\tt
  1410.6434}}].

\bibitem{Frossard:2012pc}
T.~Frossard, M.~Garny, A.~Hohenegger, A.~Kartavtsev and D.~Mitrouskas,
  \emph{{Systematic approach to thermal leptogenesis}},
  \href{http://dx.doi.org/10.1103/PhysRevD.87.085009}{\emph{Phys. Rev.} {\bf
  D87} (2013) 085009}, [\href{http://arxiv.org/abs/1211.2140}{{\tt
  1211.2140}}].

\bibitem{Hohenegger:2014cpa}
A.~Hohenegger and A.~Kartavtsev, \emph{{Leptogenesis in crossing and runaway
  regimes}}, \href{http://dx.doi.org/10.1007/JHEP07(2014)130}{\emph{JHEP} {\bf
  07} (2014) 130}, [\href{http://arxiv.org/abs/1404.5309}{{\tt 1404.5309}}].

\bibitem{Dev:2015dka}
P.~S. Bhupal~Dev, P.~Millington, A.~Pilaftsis and D.~Teresi, \emph{{Flavour
  effects in Resonant Leptogenesis from semi-classical and Kadanoff-Baym
  approaches}},
  \href{http://dx.doi.org/10.1088/1742-6596/631/1/012087}{\emph{J. Phys. Conf.
  Ser.} {\bf 631} (2015) 012087}, [\href{http://arxiv.org/abs/1502.07987}{{\tt
  1502.07987}}].

\bibitem{Kartavtsev:2015vto}
A.~Kartavtsev, P.~Millington and H.~Vogel, \emph{{Lepton asymmetry from mixing
  and oscillations}},  \href{http://arxiv.org/abs/1601.03086}{{\tt
  1601.03086}}.

\bibitem{Notzold:1987ik}
D.~Notzold and G.~Raffelt, \emph{{Neutrino Dispersion at Finite Temperature and
  Density}}, \href{http://dx.doi.org/10.1016/0550-3213(88)90113-7}{\emph{Nucl.
  Phys.} {\bf B307} (1988) 924}.

\bibitem{Enqvist:1990ad}
K.~Enqvist, K.~Kainulainen and J.~Maalampi, \emph{{Refraction and Oscillations
  of Neutrinos in the Early Universe}},
  \href{http://dx.doi.org/10.1016/0550-3213(91)90397-G}{\emph{Nucl. Phys.} {\bf
  B349} (1991) 754--790}.

\bibitem{D'Olivo:1992vm}
J.~C. D'Olivo, J.~F. Nieves and M.~Torres, \emph{{Finite temperature
  corrections to the effective potential of neutrinos in a medium}},
  \href{http://dx.doi.org/10.1103/PhysRevD.46.1172}{\emph{Phys. Rev.} {\bf D46}
  (1992) 1172--1179}.

\bibitem{Quimbay:1995jn}
C.~Quimbay and S.~Vargas-Castrillon, \emph{{Fermionic dispersion relations in
  the standard model at finite temperature}},
  \href{http://dx.doi.org/10.1016/0550-3213(95)00298-7}{\emph{Nucl. Phys.} {\bf
  B451} (1995) 265--304}, [\href{http://arxiv.org/abs/hep-ph/9504410}{{\tt
  hep-ph/9504410}}].

\bibitem{Weldon:1982bn}
H.~A. Weldon, \emph{{Effective Fermion Masses of Order gT in High Temperature
  Gauge Theories with Exact Chiral Invariance}},
  \href{http://dx.doi.org/10.1103/PhysRevD.26.2789}{\emph{Phys. Rev.} {\bf D26}
  (1982) 2789}.

\bibitem{Tututi:2002gz}
E.~S. Tututi, M.~Torres and J.~C. D'Olivo, \emph{{Neutrino damping rate at
  finite temperature and density}},
  \href{http://dx.doi.org/10.1103/PhysRevD.66.043001}{\emph{Phys. Rev.} {\bf
  D66} (2002) 043001}, [\href{http://arxiv.org/abs/hep-th/0209006}{{\tt
  hep-th/0209006}}].

\bibitem{Enqvist:1991qj}
K.~Enqvist, K.~Kainulainen and M.~J. Thomson, \emph{{Stringent cosmological
  bounds on inert neutrino mixing}},
  \href{http://dx.doi.org/10.1016/0550-3213(92)90442-E}{\emph{Nucl. Phys.} {\bf
  B373} (1992) 498--528}.

\bibitem{Langacker:1992xk}
P.~Langacker and J.~Liu, \emph{{Standard Model contributions to the neutrino
  index of refraction in the early universe}},
  \href{http://dx.doi.org/10.1103/PhysRevD.46.4140}{\emph{Phys. Rev.} {\bf D46}
  (1992) 4140--4160}, [\href{http://arxiv.org/abs/hep-ph/9206209}{{\tt
  hep-ph/9206209}}].

\bibitem{Weldon:1983jn}
H.~A. Weldon, \emph{{Simple Rules for Discontinuities in Finite Temperature
  Field Theory}}, \href{http://dx.doi.org/10.1103/PhysRevD.28.2007}{\emph{Phys.
  Rev.} {\bf D28} (1983) 2007}.

\bibitem{Bodeker:2015exa}
D.~Bodeker, M.~Sangel and M.~Wormann, \emph{{Equilibration, particle
  production, and self-energy}},  \href{http://arxiv.org/abs/1510.06742}{{\tt
  1510.06742}}.

\bibitem{Laine:2006cp}
M.~Laine and Y.~Schroder, \emph{{Quark mass thresholds in QCD thermodynamics}},
  \href{http://dx.doi.org/10.1103/PhysRevD.73.085009}{\emph{Phys. Rev.} {\bf
  D73} (2006) 085009}, [\href{http://arxiv.org/abs/hep-ph/0603048}{{\tt
  hep-ph/0603048}}].

\bibitem{Lello:2015uma}
L.~Lello and D.~Boyanovsky, \emph{{The case for mixed dark matter from sterile
  neutrinos}},  \href{http://arxiv.org/abs/1508.04077}{{\tt 1508.04077}}.

\bibitem{Weldon:1989ys}
H.~A. Weldon, \emph{{Dynamical Holes in the Quark - Gluon Plasma}},
  \href{http://dx.doi.org/10.1103/PhysRevD.40.2410}{\emph{Phys. Rev.} {\bf D40}
  (1989) 2410}.

\bibitem{Anisimov:2010gy}
A.~Anisimov, D.~Besak and D.~Bodeker, \emph{{Thermal production of relativistic
  Majorana neutrinos: Strong enhancement by multiple soft scattering}},
  \href{http://dx.doi.org/10.1088/1475-7516/2011/03/042}{\emph{JCAP} {\bf 1103}
  (2011) 042}, [\href{http://arxiv.org/abs/1012.3784}{{\tt 1012.3784}}].

\bibitem{Garbrecht:2013urw}
B.~Garbrecht, F.~Glowna and P.~Schwaller, \emph{{Scattering Rates For
  Leptogenesis: Damping of Lepton Flavour Coherence and Production of Singlet
  Neutrinos}},
  \href{http://dx.doi.org/10.1016/j.nuclphysb.2013.08.020}{\emph{Nucl. Phys.}
  {\bf B877} (2013) 1--35}, [\href{http://arxiv.org/abs/1303.5498}{{\tt
  1303.5498}}].

\bibitem{Merle:2015oja}
A.~Merle and M.~Totzauer, \emph{{keV Sterile Neutrino Dark Matter from Singlet
  Scalar Decays: Basic Concepts and Subtle Features}},
  \href{http://dx.doi.org/10.1088/1475-7516/2015/06/011}{\emph{JCAP} {\bf 1506}
  (2015) 011}, [\href{http://arxiv.org/abs/1502.01011}{{\tt 1502.01011}}].

\bibitem{Bezrukov:2009yw}
F.~Bezrukov and D.~Gorbunov, \emph{{Light inflaton Hunter's Guide}},
  \href{http://dx.doi.org/10.1007/JHEP05(2010)010}{\emph{JHEP} {\bf 1005}
  (2010) 010}, [\href{http://arxiv.org/abs/0912.0390}{{\tt 0912.0390}}].

\bibitem{Boyanovsky:2008nc}
D.~Boyanovsky, \emph{{Clustering properties of a sterile neutrino dark matter
  candidate}}, \href{http://dx.doi.org/10.1103/PhysRevD.78.103505}{\emph{Phys.
  Rev.} {\bf D78} (2008) 103505}, [\href{http://arxiv.org/abs/0807.0646}{{\tt
  0807.0646}}].

\bibitem{Shuve:2014doa}
B.~Shuve and I.~Yavin, \emph{{Dark matter progenitor: Light vector boson decay
  into sterile neutrinos}},
  \href{http://dx.doi.org/10.1103/PhysRevD.89.113004}{\emph{Phys. Rev.} {\bf
  D89} (2014) 113004}, [\href{http://arxiv.org/abs/1403.2727}{{\tt
  1403.2727}}].

\bibitem{Schechter:1981cv}
J.~Schechter and J.~W.~F. Valle, \emph{{Neutrino Decay and Spontaneous
  Violation of Lepton Number}},
  \href{http://dx.doi.org/10.1103/PhysRevD.25.774}{\emph{Phys. Rev.} {\bf D25}
  (1982) 774}.

\bibitem{Frigerio:2011in}
M.~Frigerio, T.~Hambye and E.~Masso, \emph{{Sub-GeV dark matter as
  pseudo-Goldstone from the seesaw scale}},
  \href{http://dx.doi.org/10.1103/PhysRevX.1.021026}{\emph{Phys. Rev.} {\bf X1}
  (2011) 021026}, [\href{http://arxiv.org/abs/1107.4564}{{\tt 1107.4564}}].

\bibitem{Queiroz:2014yna}
F.~S. Queiroz and K.~Sinha, \emph{{The Poker Face of the Majoron Dark Matter
  Model: LUX to keV Line}},
  \href{http://dx.doi.org/10.1016/j.physletb.2014.06.016}{\emph{Phys. Lett.}
  {\bf B735} (2014) 69--74}, [\href{http://arxiv.org/abs/1404.1400}{{\tt
  1404.1400}}].

\bibitem{Cheung:2015iqa}
Y.-K.~E. Cheung, M.~Drewes, J.~U. Kang and J.~C. Kim, \emph{{Effective Action
  for Cosmological Scalar Fields at Finite Temperature}},
  \href{http://dx.doi.org/10.1007/JHEP08(2015)059}{\emph{JHEP} {\bf 08} (2015)
  059}, [\href{http://arxiv.org/abs/1504.04444}{{\tt 1504.04444}}].

\bibitem{Drewes:2013iaa}
M.~Drewes and J.~U. Kang, \emph{{The Kinematics of Cosmic Reheating}},
  \href{http://dx.doi.org/10.1016/j.nuclphysb.2013.07.009,
  10.1016/j.nuclphysb.2014.09.008}{\emph{Nucl. Phys.} {\bf B875} (2013)
  315--350}, [\href{http://arxiv.org/abs/1305.0267}{{\tt 1305.0267}}].

\bibitem{Drewes:2014pfa}
M.~Drewes, \emph{{On finite density effects on cosmic reheating and moduli
  decay and implications for Dark Matter production}},
  \href{http://dx.doi.org/10.1088/1475-7516/2014/11/020}{\emph{JCAP} {\bf 1411}
  (2014) 020}, [\href{http://arxiv.org/abs/1406.6243}{{\tt 1406.6243}}].

\bibitem{Schneider:2016uqi}
A.~Schneider, \emph{{Astrophysical constraints on resonantly produced sterile
  neutrino dark matter}},
  \href{http://dx.doi.org/10.1088/1475-7516/2016/04/059}{\emph{JCAP} {\bf 1604}
  (2016) 059}, [\href{http://arxiv.org/abs/1601.07553}{{\tt 1601.07553}}].

\bibitem{Giudice:1999fb}
G.~F. Giudice, M.~Peloso, A.~Riotto and I.~Tkachev, \emph{{Production of
  massive fermions at preheating and leptogenesis}},
  \href{http://dx.doi.org/10.1088/1126-6708/1999/08/014}{\emph{JHEP} {\bf 9908}
  (1999) 014}, [\href{http://arxiv.org/abs/hep-ph/9905242}{{\tt
  hep-ph/9905242}}].

\bibitem{Chung:1999ve}
D.~J.~H. Chung, E.~W. Kolb, A.~Riotto and I.~I. Tkachev, \emph{{Probing
  Planckian physics: Resonant production of particles during inflation and
  features in the primordial power spectrum}},
  \href{http://dx.doi.org/10.1103/PhysRevD.62.043508}{\emph{Phys. Rev.} {\bf
  D62} (2000) 043508}, [\href{http://arxiv.org/abs/hep-ph/9910437}{{\tt
  hep-ph/9910437}}].

\bibitem{Bezrukov:2013fca}
F.~Bezrukov and D.~Gorbunov, \emph{{Light inflaton after LHC8 and WMAP9
  results}}, \href{http://dx.doi.org/10.1007/JHEP07(2013)140}{\emph{JHEP} {\bf
  1307} (2013) 140}, [\href{http://arxiv.org/abs/1303.4395}{{\tt 1303.4395}}].

\bibitem{Bezrukov:2007ep}
F.~L. Bezrukov and M.~Shaposhnikov, \emph{{The Standard Model Higgs boson as
  the inflaton}},
  \href{http://dx.doi.org/10.1016/j.physletb.2007.11.072}{\emph{Phys. Lett.}
  {\bf B659} (2008) 703--706}, [\href{http://arxiv.org/abs/0710.3755}{{\tt
  0710.3755}}].

\bibitem{Shakya:2015xnx}
B.~Shakya, \emph{{Sterile Neutrino Dark Matter from Freeze-In}},
  \href{http://dx.doi.org/10.1142/S0217732316300056}{\emph{Mod. Phys. Lett.}
  {\bf A31} (2016) 1630005}, [\href{http://arxiv.org/abs/1512.02751}{{\tt
  1512.02751}}].

\bibitem{Frigerio:2014ifa}
M.~Frigerio and C.~E. Yaguna, \emph{{Sterile Neutrino Dark Matter and Low Scale
  Leptogenesis from a Charged Scalar}},
  \href{http://dx.doi.org/10.1140/epjc/s10052-014-3252-1}{\emph{Eur. Phys. J.}
  {\bf C75} (2015) 31}, [\href{http://arxiv.org/abs/1409.0659}{{\tt
  1409.0659}}].

\bibitem{Konig:2016dzg}
J.~König, A.~Merle and M.~Totzauer, \emph{{keV Sterile Neutrino Dark Matter
  from Singlet Scalar Decays: The Most General Case}},
  \href{http://arxiv.org/abs/1609.01289}{{\tt 1609.01289}}.

\bibitem{Patwardhan:2015kga}
A.~V. Patwardhan, G.~M. Fuller, C.~T. Kishimoto and A.~Kusenko, \emph{{Diluted
  equilibrium sterile neutrino dark matter}},
  \href{http://dx.doi.org/10.1103/PhysRevD.92.103509}{\emph{Phys. Rev.} {\bf
  D92} (2015) 103509}, [\href{http://arxiv.org/abs/1507.01977}{{\tt
  1507.01977}}].

\bibitem{Martin:2014nya}
J.~Martin, C.~Ringeval and V.~Vennin, \emph{{Observing Inflationary
  Reheating}},
  \href{http://dx.doi.org/10.1103/PhysRevLett.114.081303}{\emph{Phys. Rev.
  Lett.} {\bf 114} (2015) 081303}, [\href{http://arxiv.org/abs/1410.7958}{{\tt
  1410.7958}}].

\bibitem{Drewes:2015coa}
M.~Drewes, \emph{{What can the CMB tell about the microphysics of cosmic
  reheating?}},
  \href{http://dx.doi.org/10.1088/1475-7516/2016/03/013}{\emph{JCAP} (2016)
  013}, [\href{http://arxiv.org/abs/1511.03280}{{\tt 1511.03280}}].

\bibitem{Scherrer:1984fd}
R.~J. Scherrer and M.~S. Turner, \emph{{Decaying Particles Do Not Heat Up the
  Universe}}, \href{http://dx.doi.org/10.1103/PhysRevD.31.681}{\emph{Phys.
  Rev.} {\bf D31} (1985) 681}.

\bibitem{Ma:2012if}
E.~Ma, \emph{{Radiative Scaling Neutrino Mass and Warm Dark Matter}},
  \href{http://dx.doi.org/10.1016/j.physletb.2012.09.046}{\emph{Phys. Lett.}
  {\bf B717} (2012) 235--237}, [\href{http://arxiv.org/abs/1206.1812}{{\tt
  1206.1812}}].

\bibitem{Hu:2012az}
P.-K. Hu, \emph{{Radiative Seesaw Model with Non-zero $\theta_{13}$ and Warm
  Dark Matter Scenario}},  \href{http://arxiv.org/abs/1208.2613}{{\tt
  1208.2613}}.

\bibitem{Robinson:2012wu}
D.~J. Robinson and Y.~Tsai, \emph{{KeV Warm Dark Matter and Composite
  Neutrinos}}, \href{http://dx.doi.org/10.1007/JHEP08(2012)161}{\emph{JHEP}
  {\bf 1208} (2012) 161}, [\href{http://arxiv.org/abs/1205.0569}{{\tt
  1205.0569}}].

\bibitem{King:2012wg}
S.~F. King and A.~Merle, \emph{{Warm Dark Matter from keVins}},
  \href{http://dx.doi.org/10.1088/1475-7516/2012/08/016}{\emph{JCAP} {\bf 1208}
  (2012) 016}, [\href{http://arxiv.org/abs/1205.0551}{{\tt 1205.0551}}].

\bibitem{Babu:2014uoa}
K.~S. Babu, S.~Chakdar and R.~N. Mohapatra, \emph{{Warm Dark Matter in Two
  Higgs Doublet Models}},
  \href{http://dx.doi.org/10.1103/PhysRevD.91.075020}{\emph{Phys. Rev.} {\bf
  D91} (2015) 075020}, [\href{http://arxiv.org/abs/1412.7745}{{\tt
  1412.7745}}].

\bibitem{Takahashi:2013eva}
R.~Takahashi, \emph{{Separate seesaw and its applications to dark matter and
  baryogenesis}}, \href{http://dx.doi.org/10.1093/ptep/ptt042}{\emph{PTEP} {\bf
  2013} (2013) 063B04}, [\href{http://arxiv.org/abs/1303.0108}{{\tt
  1303.0108}}].

\bibitem{Merle:2011yv}
A.~Merle and V.~Niro, \emph{{Deriving Models for keV sterile Neutrino Dark
  Matter with the Froggatt-Nielsen mechanism}},
  \href{http://dx.doi.org/10.1088/1475-7516/2011/07/023}{\emph{JCAP} {\bf 1107}
  (2011) 023}, [\href{http://arxiv.org/abs/1105.5136}{{\tt 1105.5136}}].

\bibitem{Datta:2005ci}
A.~Datta, L.~Everett and P.~Ramond, \emph{{Cabibbo haze in lepton mixing}},
  \href{http://dx.doi.org/10.1016/j.physletb.2005.05.075}{\emph{Phys. Lett.}
  {\bf B620} (2005) 42--51}, [\href{http://arxiv.org/abs/hep-ph/0503222}{{\tt
  hep-ph/0503222}}].

\bibitem{Kamikado:2008jx}
H.~Kamikado, T.~Shindou and E.~Takasugi, \emph{{Froggatt-Nielsen hierarchy and
  the neutrino mass matrix}},  \href{http://arxiv.org/abs/0805.1338}{{\tt
  0805.1338}}.

\bibitem{Kanemura:2007yy}
S.~Kanemura, K.~Matsuda, T.~Ota, S.~Petcov, T.~Shindou et~al., \emph{{CP
  violation due to multi Froggatt-Nielsen fields}},
  \href{http://dx.doi.org/10.1140/epjc/s10052-007-0343-2}{\emph{Eur. Phys. J.}
  {\bf C51} (2007) 927--931}, [\href{http://arxiv.org/abs/0704.0697}{{\tt
  0704.0697}}].

\bibitem{Choi:2001rm}
K.~Choi, E.~J. Chun, K.~Hwang and W.~Y. Song, \emph{{Bimaximal neutrino mixing
  and small $U_{e3}$ from Abelian flavor symmetry}},
  \href{http://dx.doi.org/10.1103/PhysRevD.64.113013}{\emph{Phys. Rev.} {\bf
  D64} (2001) 113013}, [\href{http://arxiv.org/abs/hep-ph/0107083}{{\tt
  hep-ph/0107083}}].

\bibitem{Wolfenstein:1981kw}
L.~Wolfenstein, \emph{{Different Varieties of Massive Dirac Neutrinos}},
  \href{http://dx.doi.org/10.1016/0550-3213(81)90096-1}{\emph{Nucl. Phys.} {\bf
  B186} (1981) 147}.

\bibitem{Asaka:2003fp}
T.~Asaka, \emph{{Lopsided mass matrices and leptogenesis in SO(10) GUT}},
  \href{http://dx.doi.org/10.1016/S0370-2693(03)00611-7}{\emph{Phys. Lett.}
  {\bf B562} (2003) 291--298}, [\href{http://arxiv.org/abs/hep-ph/0304124}{{\tt
  hep-ph/0304124}}].

\bibitem{Sato:2000kj}
J.~Sato and T.~Yanagida, \emph{{Low-energy predictions of lopsided family
  charges}}, \href{http://dx.doi.org/10.1016/S0370-2693(00)01153-9}{\emph{Phys.
  Lett.} {\bf B493} (2000) 356--365},
  [\href{http://arxiv.org/abs/hep-ph/0009205}{{\tt hep-ph/0009205}}].

\bibitem{Mohapatra:1986aw}
R.~N. Mohapatra, \emph{{Mechanism for Understanding Small Neutrino Mass in
  Superstring Theories}},
  \href{http://dx.doi.org/10.1103/PhysRevLett.56.561}{\emph{Phys. Rev. Lett.}
  {\bf 56} (1986) 561--563}.

\bibitem{Nandi:1985uh}
S.~Nandi and U.~Sarkar, \emph{{A Solution to the Neutrino Mass Problem in
  Superstring E6 Theory}},
  \href{http://dx.doi.org/10.1103/PhysRevLett.56.564}{\emph{Phys. Rev. Lett.}
  {\bf 56} (1986) 564}.

\bibitem{Akhmedov:1995vm}
E.~K. Akhmedov, M.~Lindner, E.~Schnapka and J.~W.~F. Valle, \emph{{Dynamical
  left-right symmetry breaking}},
  \href{http://dx.doi.org/10.1103/PhysRevD.53.2752}{\emph{Phys. Rev.} {\bf D53}
  (1996) 2752--2780}, [\href{http://arxiv.org/abs/hep-ph/9509255}{{\tt
  hep-ph/9509255}}].

\bibitem{Ma:2009du}
E.~Ma, \emph{{Deciphering the Seesaw Nature of Neutrino Mass from Unitarity
  Violation}}, \href{http://dx.doi.org/10.1142/S0217732309031776}{\emph{Mod.
  Phys. Lett.} {\bf A24} (2009) 2161--2165},
  [\href{http://arxiv.org/abs/0904.1580}{{\tt 0904.1580}}].

\bibitem{Mohapatra:2005wk}
R.~N. Mohapatra, S.~Nasri and H.-B. Yu, \emph{{Seesaw right handed neutrino as
  the sterile neutrino for LSND}},
  \href{http://dx.doi.org/10.1103/PhysRevD.72.033007}{\emph{Phys. Rev.} {\bf
  D72} (2005) 033007}, [\href{http://arxiv.org/abs/hep-ph/0505021}{{\tt
  hep-ph/0505021}}].

\bibitem{Fong:2011xh}
C.~S. Fong, R.~N. Mohapatra and I.~Sung, \emph{{Majorana Neutrinos from Inverse
  Seesaw in Warped Extra Dimension}},
  \href{http://dx.doi.org/10.1016/j.physletb.2011.08.069}{\emph{Phys. Lett.}
  {\bf B704} (2011) 171--178}, [\href{http://arxiv.org/abs/1107.4086}{{\tt
  1107.4086}}].

\bibitem{Dev:2012sg}
P.~S.~B. Dev and A.~Pilaftsis, \emph{{Minimal Radiative Neutrino Mass Mechanism
  for Inverse Seesaw Models}},
  \href{http://dx.doi.org/10.1103/PhysRevD.86.113001}{\emph{Phys. Rev.} {\bf
  D86} (2012) 113001}, [\href{http://arxiv.org/abs/1209.4051}{{\tt
  1209.4051}}].

\bibitem{Chun:1995bb}
E.~J. Chun, A.~S. Joshipura and A.~Y. Smirnov, \emph{{QuasiGoldstone fermion as
  a sterile neutrino}},
  \href{http://dx.doi.org/10.1103/PhysRevD.54.4654}{\emph{Phys. Rev.} {\bf D54}
  (1996) 4654--4661}, [\href{http://arxiv.org/abs/hep-ph/9507371}{{\tt
  hep-ph/9507371}}].

\bibitem{Zhang:2011vh}
H.~Zhang, \emph{{Light Sterile Neutrino in the Minimal Extended Seesaw}},
  \href{http://dx.doi.org/10.1016/j.physletb.2012.06.074}{\emph{Phys. Lett.}
  {\bf B714} (2012) 262--266}, [\href{http://arxiv.org/abs/1110.6838}{{\tt
  1110.6838}}].

\bibitem{Heeck:2012bz}
J.~Heeck and H.~Zhang, \emph{{Exotic Charges, Multicomponent Dark Matter and
  Light Sterile Neutrinos}},
  \href{http://dx.doi.org/10.1007/JHEP05(2013)164}{\emph{JHEP} {\bf 1305}
  (2013) 164}, [\href{http://arxiv.org/abs/1211.0538}{{\tt 1211.0538}}].

\bibitem{Pilaftsis:1991ug}
A.~Pilaftsis, \emph{{Radiatively induced neutrino masses and large Higgs
  neutrino couplings in the standard model with Majorana fields}},
  \href{http://dx.doi.org/10.1007/BF01482590}{\emph{Z. Phys.} {\bf C55} (1992)
  275--282}, [\href{http://arxiv.org/abs/hep-ph/9901206}{{\tt
  hep-ph/9901206}}].

\bibitem{Dev:2012bd}
P.~S.~B. Dev and A.~Pilaftsis, \emph{{Light and Superlight Sterile Neutrinos in
  the Minimal Radiative Inverse Seesaw Model}},
  \href{http://dx.doi.org/10.1103/PhysRevD.87.053007}{\emph{Phys. Rev.} {\bf
  D87} (2013) 053007}, [\href{http://arxiv.org/abs/1212.3808}{{\tt
  1212.3808}}].

\bibitem{deHolanda:2003tx}
P.~C. de~Holanda and A.~Y. Smirnov, \emph{{Homestake result, sterile neutrinos
  and low-energy solar neutrino experiments}},
  \href{http://dx.doi.org/10.1103/PhysRevD.69.113002}{\emph{Phys. Rev.} {\bf
  D69} (2004) 113002}, [\href{http://arxiv.org/abs/hep-ph/0307266}{{\tt
  hep-ph/0307266}}].

\bibitem{deHolanda:2010am}
P.~C. de~Holanda and A.~Y. Smirnov, \emph{{Solar neutrino spectrum, sterile
  neutrinos and additional radiation in the Universe}},
  \href{http://dx.doi.org/10.1103/PhysRevD.83.113011}{\emph{Phys. Rev.} {\bf
  D83} (2011) 113011}, [\href{http://arxiv.org/abs/1012.5627}{{\tt
  1012.5627}}].

\bibitem{Bakhti:2013ora}
P.~Bakhti and Y.~Farzan, \emph{{Constraining Super-light Sterile Neutrino
  Scenario by JUNO and RENO-50}},
  \href{http://dx.doi.org/10.1007/JHEP10(2013)200}{\emph{JHEP} {\bf 1310}
  (2013) 200}, [\href{http://arxiv.org/abs/1308.2823}{{\tt 1308.2823}}].

\bibitem{Pilaftsis:2003gt}
A.~Pilaftsis and T.~E.~J. Underwood, \emph{{Resonant leptogenesis}},
  \href{http://dx.doi.org/10.1016/j.nuclphysb.2004.05.029}{\emph{Nucl. Phys.}
  {\bf B692} (2004) 303--345}, [\href{http://arxiv.org/abs/hep-ph/0309342}{{\tt
  hep-ph/0309342}}].

\bibitem{Dev:2014laa}
P.~S. Bhupal~Dev, P.~Millington, A.~Pilaftsis and D.~Teresi, \emph{{Flavour
  Covariant Transport Equations: an Application to Resonant Leptogenesis}},
  \href{http://dx.doi.org/10.1016/j.nuclphysb.2014.06.020}{\emph{Nucl. Phys.}
  {\bf B886} (2014) 569--664}, [\href{http://arxiv.org/abs/1404.1003}{{\tt
  1404.1003}}].

\bibitem{Sierra:2008wj}
D.~Aristizabal~Sierra, J.~Kubo, D.~Restrepo, D.~Suematsu and O.~Zapata,
  \emph{{Radiative seesaw: Warm dark matter, collider and lepton flavour
  violating signals}},
  \href{http://dx.doi.org/10.1103/PhysRevD.79.013011}{\emph{Phys. Rev.} {\bf
  D79} (2009) 013011}, [\href{http://arxiv.org/abs/0808.3340}{{\tt
  0808.3340}}].

\bibitem{Suematsu:2009ww}
D.~Suematsu, T.~Toma and T.~Yoshida, \emph{{Reconciliation of CDM abundance and
  mu ---> e gamma in a radiative seesaw model}},
  \href{http://dx.doi.org/10.1103/PhysRevD.79.093004}{\emph{Phys. Rev.} {\bf
  D79} (2009) 093004}, [\href{http://arxiv.org/abs/0903.0287}{{\tt
  0903.0287}}].

\bibitem{Ahn:2012cga}
Y.~H. Ahn and H.~Okada, \emph{{Non-zero $\theta_{13}$ linking to Dark Matter
  from Non-Abelian Discrete Flavor Model in Radiative Seesaw}},
  \href{http://dx.doi.org/10.1103/PhysRevD.85.073010}{\emph{Phys. Rev.} {\bf
  D85} (2012) 073010}, [\href{http://arxiv.org/abs/1201.4436}{{\tt
  1201.4436}}].

\bibitem{Gelmini:2009xd}
G.~B. Gelmini, E.~Osoba and S.~Palomares-Ruiz, \emph{{Inert-Sterile Neutrino:
  Cold or Warm Dark Matter Candidate}},
  \href{http://dx.doi.org/10.1103/PhysRevD.81.063529}{\emph{Phys. Rev.} {\bf
  D81} (2010) 063529}, [\href{http://arxiv.org/abs/0912.2478}{{\tt
  0912.2478}}].

\bibitem{Kubo:2006yx}
J.~Kubo, E.~Ma and D.~Suematsu, \emph{{Cold Dark Matter, Radiative Neutrino
  Mass, $\mu \to e\gamma$, and Neutrinoless Double Beta Decay}},
  \href{http://dx.doi.org/10.1016/j.physletb.2006.08.085}{\emph{Phys. Lett.}
  {\bf B642} (2006) 18--23}, [\href{http://arxiv.org/abs/hep-ph/0604114}{{\tt
  hep-ph/0604114}}].

\bibitem{Adulpravitchai:2009gi}
A.~Adulpravitchai, M.~Lindner and A.~Merle, \emph{{Confronting Flavour
  Symmetries and extended Scalar Sectors with Lepton Flavour Violation
  Bounds}}, \href{http://dx.doi.org/10.1103/PhysRevD.80.055031}{\emph{Phys.
  Rev.} {\bf D80} (2009) 055031}, [\href{http://arxiv.org/abs/0907.2147}{{\tt
  0907.2147}}].

\bibitem{Adulpravitchai:2009re}
A.~Adulpravitchai, M.~Lindner, A.~Merle and R.~N. Mohapatra, \emph{{Radiative
  Transmission of Lepton Flavor Hierarchies}},
  \href{http://dx.doi.org/10.1016/j.physletb.2009.09.042}{\emph{Phys. Lett.}
  {\bf B680} (2009) 476--479}, [\href{http://arxiv.org/abs/0908.0470}{{\tt
  0908.0470}}].

\bibitem{Bouchand:2012dx}
R.~Bouchand and A.~Merle, \emph{{Running of Radiative Neutrino Masses: The
  Scotogenic Model}},
  \href{http://dx.doi.org/10.1007/JHEP07(2012)084}{\emph{JHEP} {\bf 07} (2012)
  084}, [\href{http://arxiv.org/abs/1205.0008}{{\tt 1205.0008}}].

\bibitem{Toma:2013zsa}
T.~Toma and A.~Vicente, \emph{{Lepton Flavor Violation in the Scotogenic
  Model}}, \href{http://dx.doi.org/10.1007/JHEP01(2014)160}{\emph{JHEP} {\bf
  01} (2014) 160}, [\href{http://arxiv.org/abs/1312.2840}{{\tt 1312.2840}}].

\bibitem{Vicente:2014wga}
A.~Vicente and C.~E. Yaguna, \emph{{Probing the scotogenic model with lepton
  flavor violating processes}},
  \href{http://dx.doi.org/10.1007/JHEP02(2015)144}{\emph{JHEP} {\bf 02} (2015)
  144}, [\href{http://arxiv.org/abs/1412.2545}{{\tt 1412.2545}}].

\bibitem{Merle:2015gea}
A.~Merle and M.~Platscher, \emph{{Parity Problem of the Scotogenic Neutrino
  Model}}, \href{http://dx.doi.org/10.1103/PhysRevD.92.095002}{\emph{Phys.
  Rev.} {\bf D92} (2015) 095002}, [\href{http://arxiv.org/abs/1502.03098}{{\tt
  1502.03098}}].

\bibitem{Merle:2015ica}
A.~Merle and M.~Platscher, \emph{{Running of radiative neutrino masses: the
  scotogenic model -- revisited}},
  \href{http://dx.doi.org/10.1007/JHEP11(2015)148}{\emph{JHEP} {\bf 11} (2015)
  148}, [\href{http://arxiv.org/abs/1507.06314}{{\tt 1507.06314}}].

\bibitem{Adhikari:2008uc}
R.~Adhikari, J.~Erler and E.~Ma, \emph{{Seesaw Neutrino Mass and New U(1) Gauge
  Symmetry}},
  \href{http://dx.doi.org/10.1016/j.physletb.2009.01.017}{\emph{Phys. Lett.}
  {\bf B672} (2009) 136--140}, [\href{http://arxiv.org/abs/0810.5547}{{\tt
  0810.5547}}].

\bibitem{Borah:2013waa}
D.~Borah and R.~Adhikari, \emph{{Common Radiative Origin of Active and Sterile
  Neutrino Masses}},
  \href{http://dx.doi.org/10.1016/j.physletb.2014.01.018}{\emph{Phys. Lett.}
  {\bf B729} (2014) 143--148}, [\href{http://arxiv.org/abs/1310.5419}{{\tt
  1310.5419}}].

\bibitem{Adhikari:2014nea}
R.~Adhikari, D.~Borah and E.~Ma, \emph{{Common Origin of Active and Sterile
  Neutrino Masses with Dark Matter}},
  \href{http://arxiv.org/abs/1411.4602}{{\tt 1411.4602}}.

\bibitem{Adhikari:2015woo}
R.~Adhikari, D.~Borah and E.~Ma, \emph{{New U(1) Gauge Model of Radiative
  Lepton Masses with Sterile Neutrino and Dark Matter}},
  \href{http://arxiv.org/abs/1512.05491}{{\tt 1512.05491}}.

\bibitem{Borah:2012qr}
D.~Borah and R.~Adhikari, \emph{{Abelian Gauge Extension of Standard Model:
  Dark Matter and Radiative Neutrino Mass}},
  \href{http://dx.doi.org/10.1103/PhysRevD.85.095002}{\emph{Phys. Rev.} {\bf
  D85} (2012) 095002}, [\href{http://arxiv.org/abs/1202.2718}{{\tt
  1202.2718}}].

\bibitem{Lindner:2010wr}
M.~Lindner, A.~Merle and V.~Niro, \emph{{Soft $L_e - L_\mu - L_\tau$ flavour
  symmetry breaking and sterile neutrino keV Dark Matter}},
  \href{http://dx.doi.org/10.1088/1475-7516/2011/01/034}{\emph{JCAP} {\bf 1101}
  (2011) 034}, [\href{http://arxiv.org/abs/1011.4950}{{\tt 1011.4950}}].

\bibitem{Lindner:2010wr_Erratum}
M.~Lindner, A.~Merle and V.~Niro, \emph{{\emph{Erratum}: Soft $L_e - L_\mu -
  L_\tau$ flavour symmetry breaking and sterile neutrino keV Dark Matter}},
  \href{http://dx.doi.org/10.1088/1475-7516/2014/07/E01}{\emph{JCAP} {\bf 1407}
  (2014) E01}, [\href{http://arxiv.org/abs/1011.4950}{{\tt 1011.4950}}].

\bibitem{Merle:2012ya}
A.~Merle, \emph{{keV sterile Neutrino Dark Matter and Neutrino Model
  Building}}, \href{http://dx.doi.org/10.1088/1742-6596/375/1/012047}{\emph{J.
  Phys. Conf. Ser.} {\bf 375} (2012) 012047},
  [\href{http://arxiv.org/abs/1201.0881}{{\tt 1201.0881}}].

\bibitem{Lavoura:2000ci}
L.~Lavoura and W.~Grimus, \emph{{Seesaw model with softly broken $L_e - L_\mu -
  L_\tau$}}, \href{http://dx.doi.org/10.1088/1126-6708/2000/09/007}{\emph{JHEP}
  {\bf 0009} (2000) 007}, [\href{http://arxiv.org/abs/hep-ph/0008020}{{\tt
  hep-ph/0008020}}].

\bibitem{Frampton:2004ud}
P.~H. Frampton, S.~T. Petcov and W.~Rodejohann, \emph{{On deviations from
  bimaximal neutrino mixing}},
  \href{http://dx.doi.org/10.1016/j.nuclphysb.2004.03.014}{\emph{Nucl. Phys.}
  {\bf B687} (2004) 31--54}, [\href{http://arxiv.org/abs/hep-ph/0401206}{{\tt
  hep-ph/0401206}}].

\bibitem{Mohapatra:2001ns}
R.~N. Mohapatra, \emph{{Connecting bimaximal neutrino mixing to a light sterile
  neutrino}}, \href{http://dx.doi.org/10.1103/PhysRevD.64.091301}{\emph{Phys.
  Rev.} {\bf D64} (2001) 091301},
  [\href{http://arxiv.org/abs/hep-ph/0107264}{{\tt hep-ph/0107264}}].

\bibitem{Gehrlein:2015ena}
J.~Gehrlein, A.~Merle and M.~Spinrath, \emph{{Renormalisation Group Corrections
  to Neutrino Mass Sum Rules}},
  \href{http://dx.doi.org/10.1007/JHEP09(2015)066}{\emph{JHEP} {\bf 09} (2015)
  066}, [\href{http://arxiv.org/abs/1506.06139}{{\tt 1506.06139}}].

\bibitem{Asaka:2010kk}
T.~Asaka and H.~Ishida, \emph{{Flavour Mixing of Neutrinos and Baryon Asymmetry
  of the Universe}},
  \href{http://dx.doi.org/10.1016/j.physletb.2010.07.016}{\emph{Phys. Lett.}
  {\bf B692} (2010) 105--113}, [\href{http://arxiv.org/abs/1004.5491}{{\tt
  1004.5491}}].

\bibitem{Barry:2011fp}
J.~Barry, W.~Rodejohann and H.~Zhang, \emph{{Sterile Neutrinos for Warm Dark
  Matter and the Reactor Anomaly in Flavor Symmetry Models}},
  \href{http://dx.doi.org/10.1088/1475-7516/2012/01/052}{\emph{JCAP} {\bf 1201}
  (2012) 052}, [\href{http://arxiv.org/abs/1110.6382}{{\tt 1110.6382}}].

\bibitem{Merle:2011vy}
A.~Merle and R.~Zwicky, \emph{{Explicit and spontaneous breaking of SU(3) into
  its finite subgroups}},
  \href{http://dx.doi.org/10.1007/JHEP02(2012)128}{\emph{JHEP} {\bf 02} (2012)
  128}, [\href{http://arxiv.org/abs/1110.4891}{{\tt 1110.4891}}].

\bibitem{Merle:2014eja}
A.~Merle, S.~Morisi and W.~Winter, \emph{{Common origin of reactor and sterile
  neutrino mixing}},
  \href{http://dx.doi.org/10.1007/JHEP07(2014)039}{\emph{JHEP} {\bf 07} (2014)
  039}, [\href{http://arxiv.org/abs/1402.6332}{{\tt 1402.6332}}].

\bibitem{Ma:1995gf}
E.~Ma and P.~Roy, \emph{{Model of four light neutrinos in the light of all
  present data}},
  \href{http://dx.doi.org/10.1103/PhysRevD.52.R4780}{\emph{Phys. Rev.} {\bf
  D52} (1995) 4780--4783}, [\href{http://arxiv.org/abs/hep-ph/9504342}{{\tt
  hep-ph/9504342}}].

\bibitem{Chun:1995js}
E.~J. Chun, A.~S. Joshipura and A.~Y. Smirnov, \emph{{Models of light singlet
  fermion and neutrino phenomenology}},
  \href{http://dx.doi.org/10.1016/0370-2693(95)00967-P}{\emph{Phys. Lett.} {\bf
  B357} (1995) 608--615}, [\href{http://arxiv.org/abs/hep-ph/9505275}{{\tt
  hep-ph/9505275}}].

\bibitem{Banks:2010zn}
T.~Banks and N.~Seiberg, \emph{{Symmetries and Strings in Field Theory and
  Gravity}}, \href{http://dx.doi.org/10.1103/PhysRevD.83.084019}{\emph{Phys.
  Rev.} {\bf D83} (2011) 084019}, [\href{http://arxiv.org/abs/1011.5120}{{\tt
  1011.5120}}].

\bibitem{Zeldovich:1974uw}
Y.~B. Zeldovich, I.~Y. Kobzarev and L.~B. Okun, \emph{{Cosmological
  Consequences of the Spontaneous Breakdown of Discrete Symmetry}}, {\emph{Zh.
  Eksp. Teor. Fiz.} {\bf 67} (1974) 3--11}.

\bibitem{Babu:2003is}
K.~S. Babu and G.~Seidl, \emph{{Simple model for (3+2) neutrino oscillations}},
  \href{http://dx.doi.org/10.1016/j.physletb.2004.03.086}{\emph{Phys. Lett.}
  {\bf B591} (2004) 127--136}, [\href{http://arxiv.org/abs/hep-ph/0312285}{{\tt
  hep-ph/0312285}}].

\bibitem{Babu:2004mj}
K.~S. Babu and G.~Seidl, \emph{{Chiral gauge models for light sterile
  neutrinos}}, \href{http://dx.doi.org/10.1103/PhysRevD.70.113014}{\emph{Phys.
  Rev.} {\bf D70} (2004) 113014},
  [\href{http://arxiv.org/abs/hep-ph/0405197}{{\tt hep-ph/0405197}}].

\bibitem{Batra:2005rh}
P.~Batra, B.~A. Dobrescu and D.~Spivak, \emph{{Anomaly-free sets of fermions}},
  \href{http://dx.doi.org/10.1063/1.2222081}{\emph{J. Math. Phys.} {\bf 47}
  (2006) 082301}, [\href{http://arxiv.org/abs/hep-ph/0510181}{{\tt
  hep-ph/0510181}}].

\bibitem{Batell:2010bp}
B.~Batell, \emph{{Dark Discrete Gauge Symmetries}},
  \href{http://dx.doi.org/10.1103/PhysRevD.83.035006}{\emph{Phys. Rev.} {\bf
  D83} (2011) 035006}, [\href{http://arxiv.org/abs/1007.0045}{{\tt
  1007.0045}}].

\bibitem{Robinson:2014bma}
D.~J. Robinson and Y.~Tsai, \emph{{Dynamical framework for keV Dirac neutrino
  warm dark matter}},
  \href{http://dx.doi.org/10.1103/PhysRevD.90.045030}{\emph{Phys. Rev.} {\bf
  D90} (2014) 045030}, [\href{http://arxiv.org/abs/1404.7118}{{\tt
  1404.7118}}].

\bibitem{ArkaniHamed:1998pf}
N.~Arkani-Hamed and Y.~Grossman, \emph{{Light active and sterile neutrinos from
  compositeness}},
  \href{http://dx.doi.org/10.1016/S0370-2693(99)00672-3}{\emph{Phys. Lett.}
  {\bf B459} (1999) 179--182}, [\href{http://arxiv.org/abs/hep-ph/9806223}{{\tt
  hep-ph/9806223}}].

\bibitem{Okui:2004xn}
T.~Okui, \emph{{Searching for composite neutrinos in the cosmic microwave
  background}},
  \href{http://dx.doi.org/10.1088/1126-6708/2005/09/017}{\emph{JHEP} {\bf 0509}
  (2005) 017}, [\href{http://arxiv.org/abs/hep-ph/0405083}{{\tt
  hep-ph/0405083}}].

\bibitem{Grossman:2008xb}
Y.~Grossman and Y.~Tsai, \emph{{Leptogenesis with Composite Neutrinos}},
  \href{http://dx.doi.org/10.1088/1126-6708/2008/12/016}{\emph{JHEP} {\bf 0812}
  (2008) 016}, [\href{http://arxiv.org/abs/0811.0871}{{\tt 0811.0871}}].

\bibitem{Grossman:2010iq}
Y.~Grossman and D.~J. Robinson, \emph{{Composite Dirac Neutrinos}},
  \href{http://dx.doi.org/10.1007/JHEP01(2011)132}{\emph{JHEP} {\bf 1101}
  (2011) 132}, [\href{http://arxiv.org/abs/1009.2781}{{\tt 1009.2781}}].

\bibitem{McDonald:2010jm}
K.~L. McDonald, \emph{{Light Neutrinos from a Mini-Seesaw Mechanism in Warped
  Space}}, \href{http://dx.doi.org/10.1016/j.physletb.2010.12.059}{\emph{Phys.
  Lett.} {\bf B696} (2011) 266--272},
  [\href{http://arxiv.org/abs/1010.2659}{{\tt 1010.2659}}].

\bibitem{Duerr:2011ks}
M.~Duerr, D.~P. George and K.~L. McDonald, \emph{{Neutrino Mass and $\mu
  \rightarrow e + \gamma $ from a Mini-Seesaw}},
  \href{http://dx.doi.org/10.1007/JHEP07(2011)103}{\emph{JHEP} {\bf 1107}
  (2011) 103}, [\href{http://arxiv.org/abs/1105.0593}{{\tt 1105.0593}}].

\bibitem{Singer:1980sw}
M.~Singer, J.~W.~F. Valle and J.~Schechter, \emph{{Canonical Neutral Current
  Predictions From the Weak Electromagnetic Gauge Group $SU(3) \times U(1)$}},
  \href{http://dx.doi.org/10.1103/PhysRevD.22.738}{\emph{Phys. Rev.} {\bf D22}
  (1980) 738}.

\bibitem{Valle:1983dk}
J.~W.~F. Valle and M.~Singer, \emph{{Lepton Number Violation With Quasi Dirac
  Neutrinos}}, \href{http://dx.doi.org/10.1103/PhysRevD.28.540}{\emph{Phys.
  Rev.} {\bf D28} (1983) 540}.

\bibitem{Montero:1992jk}
J.~C. Montero, F.~Pisano and V.~Pleitez, \emph{{Neutral currents and GIM
  mechanism in $SU(3)_L \times U(1)_N$ models for electroweak interactions}},
  \href{http://dx.doi.org/10.1103/PhysRevD.47.2918}{\emph{Phys. Rev.} {\bf D47}
  (1993) 2918--2929}, [\href{http://arxiv.org/abs/hep-ph/9212271}{{\tt
  hep-ph/9212271}}].

\bibitem{Foot:1994ym}
R.~Foot, H.~N. Long and T.~A. Tran, \emph{{$SU(3)_L \times U(1)_N$ and $SU(4)_L
  \times U(1)_N$ gauge models with right-handed neutrinos}},
  \href{http://dx.doi.org/10.1103/PhysRevD.50.R34}{\emph{Phys. Rev.} {\bf D50}
  (1994) 34--38}, [\href{http://arxiv.org/abs/hep-ph/9402243}{{\tt
  hep-ph/9402243}}].

\bibitem{Dias:2005yh}
A.~G. Dias, C.~A. de~S.~Pires and P.~S. Rodrigues~da Silva, \emph{{Naturally
  light right-handed neutrinos in a 3-3-1 model}},
  \href{http://dx.doi.org/10.1016/j.physletb.2005.09.028}{\emph{Phys. Lett.}
  {\bf B628} (2005) 85--92}, [\href{http://arxiv.org/abs/hep-ph/0508186}{{\tt
  hep-ph/0508186}}].

\bibitem{Dinh:2006ia}
D.~N. Dinh, N.~A. Ky, N.~T. Van and P.~Q. Van, \emph{{Model of neutrino
  effective masses}},
  \href{http://dx.doi.org/10.1103/PhysRevD.74.077701}{\emph{Phys. Rev.} {\bf
  D74} (2006) 077701}.

\bibitem{Cogollo:2009yi}
D.~Cogollo, H.~Diniz and C.~A. de~S.~Pires, \emph{{KeV right-handed neutrinos
  from type II seesaw mechanism in a 3-3-1 model}},
  \href{http://arxiv.org/abs/0903.0370}{{\tt 0903.0370}}.

\bibitem{Ky:2005yq}
N.~A. Ky and N.~T.~H. Van, \emph{{Scalar sextet in the 331 model with
  right-handed neutrinos}},
  \href{http://dx.doi.org/10.1103/PhysRevD.72.115017}{\emph{Phys. Rev.} {\bf
  D72} (2005) 115017}, [\href{http://arxiv.org/abs/hep-ph/0512096}{{\tt
  hep-ph/0512096}}].

\bibitem{Pisano:1991ee}
F.~Pisano and V.~Pleitez, \emph{{An $SU(3) \times U(1)$ model for electroweak
  interactions}}, \href{http://dx.doi.org/10.1103/PhysRevD.46.410}{\emph{Phys.
  Rev.} {\bf D46} (1992) 410--417},
  [\href{http://arxiv.org/abs/hep-ph/9206242}{{\tt hep-ph/9206242}}].

\bibitem{Frampton:1992wt}
P.~H. Frampton, \emph{{Chiral dilepton model and the flavor question}},
  \href{http://dx.doi.org/10.1103/PhysRevLett.69.2889}{\emph{Phys. Rev. Lett.}
  {\bf 69} (1992) 2889--2891}.

\bibitem{Pleitez:1992xh}
V.~Pleitez and M.~D. Tonasse, \emph{{Heavy charged leptons in an $SU(3)_L
  \times U(1)_N$ model}},
  \href{http://dx.doi.org/10.1103/PhysRevD.48.2353}{\emph{Phys. Rev.} {\bf D48}
  (1993) 2353--2355}, [\href{http://arxiv.org/abs/hep-ph/9301232}{{\tt
  hep-ph/9301232}}].

\bibitem{Dong:2013wca}
P.~V. Dong, H.~T. Hung and T.~D. Tham, \emph{{3-3-1-1 model for dark matter}},
  \href{http://dx.doi.org/10.1103/PhysRevD.87.115003}{\emph{Phys. Rev.} {\bf
  D87} (2013) 115003}, [\href{http://arxiv.org/abs/1305.0369}{{\tt
  1305.0369}}].

\bibitem{Kelso:2013nwa}
C.~Kelso, C.~A. de~S.~Pires, S.~Profumo, F.~S. Queiroz and P.~S. Rodrigues~da
  Silva, \emph{{A 331 WIMPy Dark Radiation Model}},
  \href{http://dx.doi.org/10.1140/epjc/s10052-014-2797-3}{\emph{Eur. Phys. J.}
  {\bf C74} (2014) 2797}, [\href{http://arxiv.org/abs/1308.6630}{{\tt
  1308.6630}}].

\bibitem{Ma:1998dx}
E.~Ma and U.~Sarkar, \emph{{Neutrino masses and leptogenesis with heavy Higgs
  triplets}}, \href{http://dx.doi.org/10.1103/PhysRevLett.80.5716}{\emph{Phys.
  Rev. Lett.} {\bf 80} (1998) 5716--5719},
  [\href{http://arxiv.org/abs/hep-ph/9802445}{{\tt hep-ph/9802445}}].

\bibitem{Palcu:2006ti}
A.~Palcu, \emph{{Implementing canonical seesaw mechanism in the exact solution
  of a 3-3-1 gauge model without exotic electric charges}},
  \href{http://dx.doi.org/10.1142/S0217732306021566}{\emph{Mod. Phys. Lett.}
  {\bf A21} (2006) 2591--2598},
  [\href{http://arxiv.org/abs/hep-ph/0605155}{{\tt hep-ph/0605155}}].

\bibitem{Dong:2008sw}
P.~V. Dong and H.~N. Long, \emph{{Neutrino masses and lepton flavor violation
  in the 3-3-1 model with right-handed neutrinos}},
  \href{http://dx.doi.org/10.1103/PhysRevD.77.057302}{\emph{Phys. Rev.} {\bf
  D77} (2008) 057302}, [\href{http://arxiv.org/abs/0801.4196}{{\tt
  0801.4196}}].

\bibitem{Cogollo:2008zc}
D.~Cogollo, H.~Diniz, C.~A. de~S.~Pires and P.~S. Rodrigues~da Silva,
  \emph{{The Seesaw mechanism at TeV scale in the 3-3-1 model with right-handed
  neutrinos}},
  \href{http://dx.doi.org/10.1140/epjc/s10052-008-0749-5}{\emph{Eur. Phys. J.}
  {\bf C58} (2008) 455--461}, [\href{http://arxiv.org/abs/0806.3087}{{\tt
  0806.3087}}].

\bibitem{Dias:2010vt}
A.~G. Dias, C.~A. de~S.~Pires and P.~S. Rodrigues~da Silva, \emph{{The
  Left-Right $SU(3)_L \times SU(3)_R \times U(1)_X$ Model with Light, keV and
  Heavy Neutrinos}},
  \href{http://dx.doi.org/10.1103/PhysRevD.82.035013}{\emph{Phys. Rev.} {\bf
  D82} (2010) 035013}, [\href{http://arxiv.org/abs/1003.3260}{{\tt
  1003.3260}}].

\bibitem{Pati:1974yy}
J.~C. Pati and A.~Salam, \emph{{Lepton Number as the Fourth Color}},
  \href{http://dx.doi.org/10.1103/PhysRevD.10.275,
  10.1103/PhysRevD.11.703.2}{\emph{Phys. Rev.} {\bf D10} (1974) 275--289}.

\bibitem{Senjanovic:1975rk}
G.~Senjanovic and R.~N. Mohapatra, \emph{{Exact Left-Right Symmetry and
  Spontaneous Violation of Parity}},
  \href{http://dx.doi.org/10.1103/PhysRevD.12.1502}{\emph{Phys. Rev.} {\bf D12}
  (1975) 1502}.

\bibitem{Rosner:2014cha}
J.~L. Rosner, \emph{{Three sterile neutrinos in E6}},
  \href{http://dx.doi.org/10.1103/PhysRevD.90.035005}{\emph{Phys. Rev.} {\bf
  D90} (2014) 035005}, [\href{http://arxiv.org/abs/1404.5198}{{\tt
  1404.5198}}].

\bibitem{Pilaftsis:2012hq}
A.~Pilaftsis, \emph{{Anomalous Fermion Mass Generation at Three Loops}},
  \href{http://dx.doi.org/10.1142/S0217732313500831}{\emph{Mod. Phys. Lett.}
  {\bf A28} (2013) 1350083}, [\href{http://arxiv.org/abs/1207.0544}{{\tt
  1207.0544}}].

\bibitem{Mavromatos:2012cc}
N.~E. Mavromatos and A.~Pilaftsis, \emph{{Anomalous Majorana Neutrino Masses
  from Torsionful Quantum Gravity}},
  \href{http://dx.doi.org/10.1103/PhysRevD.86.124038}{\emph{Phys. Rev.} {\bf
  D86} (2012) 124038}, [\href{http://arxiv.org/abs/1209.6387}{{\tt
  1209.6387}}].

\bibitem{Hehl:1976kj}
F.~W. Hehl, P.~Von Der~Heyde, G.~D. Kerlick and J.~M. Nester, \emph{{General
  Relativity with Spin and Torsion: Foundations and Prospects}},
  \href{http://dx.doi.org/10.1103/RevModPhys.48.393}{\emph{Rev. Mod. Phys.}
  {\bf 48} (1976) 393--416}.

\bibitem{Shapiro:2001rz}
I.~L. Shapiro, \emph{{Physical aspects of the space-time torsion}},
  \href{http://dx.doi.org/10.1016/S0370-1573(01)00030-8}{\emph{Phys. Rept.}
  {\bf 357} (2002) 113}, [\href{http://arxiv.org/abs/hep-th/0103093}{{\tt
  hep-th/0103093}}].

\bibitem{Kibble:1961ba}
T.~W.~B. Kibble, \emph{{Lorentz invariance and the gravitational field}},
  \href{http://dx.doi.org/10.1063/1.1703702}{\emph{J. Math. Phys.} {\bf 2}
  (1961) 212--221}.

\bibitem{Sciama:1964wt}
D.~W. Sciama, \emph{{The Physical structure of general relativity}},
  \href{http://dx.doi.org/10.1103/RevModPhys.36.1103}{\emph{Rev. Mod. Phys.}
  {\bf 36} (1964) 463--469}.

\bibitem{Duncan:1992vz}
M.~J. Duncan, N.~Kaloper and K.~A. Olive, \emph{{Axion hair and dynamical
  torsion from anomalies}},
  \href{http://dx.doi.org/10.1016/0550-3213(92)90052-D}{\emph{Nucl. Phys.} {\bf
  B387} (1992) 215--238}.

\bibitem{Weinberg:1996kr}
S.~Weinberg, \emph{{The quantum theory of fields. Vol. 2: Modern
  applications}}.
\newblock Cambridge, UK: Univ. Pr., 1996.

\bibitem{Metsaev:1987zx}
R.~R. Metsaev and A.~A. Tseytlin, \emph{{Order alpha-prime (Two Loop)
  Equivalence of the String Equations of Motion and the Sigma Model Weyl
  Invariance Conditions: Dependence on the Dilaton and the Antisymmetric
  Tensor}}, \href{http://dx.doi.org/10.1016/0550-3213(87)90077-0}{\emph{Nucl.
  Phys.} {\bf B293} (1987) 385--419}.

\bibitem{Gross:1986mw}
D.~J. Gross and J.~H. Sloan, \emph{{The Quartic Effective Action for the
  Heterotic String}},
  \href{http://dx.doi.org/10.1016/0550-3213(87)90465-2}{\emph{Nucl. Phys.} {\bf
  B291} (1987) 41--89}.

\bibitem{Kalb:1974yc}
M.~Kalb and P.~Ramond, \emph{{Classical direct interstring action}},
  \href{http://dx.doi.org/10.1103/PhysRevD.9.2273}{\emph{Phys. Rev.} {\bf D9}
  (1974) 2273--2284}.

\bibitem{Donoghue:1994dn}
J.~F. Donoghue, \emph{{General relativity as an effective field theory: The
  leading quantum corrections}},
  \href{http://dx.doi.org/10.1103/PhysRevD.50.3874}{\emph{Phys. Rev.} {\bf D50}
  (1994) 3874--3888}, [\href{http://arxiv.org/abs/gr-qc/9405057}{{\tt
  gr-qc/9405057}}].

\bibitem{Arvanitaki:2009fg}
A.~Arvanitaki, S.~Dimopoulos, S.~Dubovsky, N.~Kaloper and J.~March-Russell,
  \emph{{String Axiverse}},
  \href{http://dx.doi.org/10.1103/PhysRevD.81.123530}{\emph{Phys. Rev.} {\bf
  D81} (2010) 123530}, [\href{http://arxiv.org/abs/0905.4720}{{\tt
  0905.4720}}].

\bibitem{Cicoli:2012sz}
M.~Cicoli, M.~Goodsell and A.~Ringwald, \emph{{The type IIB string axiverse and
  its low-energy phenomenology}},
  \href{http://dx.doi.org/10.1007/JHEP10(2012)146}{\emph{JHEP} {\bf 1210}
  (2012) 146}, [\href{http://arxiv.org/abs/1206.0819}{{\tt 1206.0819}}].

\bibitem{RiemerSorensen:2006fh}
S.~Riemer-Sorensen, S.~H. Hansen and K.~Pedersen, \emph{{Sterile neutrinos in
  the Milky Way: Observational constraints}},
  \href{http://dx.doi.org/10.1086/505330}{\emph{Astrophys. J.} {\bf 644} (2006)
  L33--L36}, [\href{http://arxiv.org/abs/astro-ph/0603661}{{\tt
  astro-ph/0603661}}].

\bibitem{Weber:2009pt}
M.~Weber and W.~de~Boer, \emph{{Determination of the Local Dark Matter Density
  in our Galaxy}},
  \href{http://dx.doi.org/10.1051/0004-6361/200913381}{\emph{Astron.
  Astrophys.} {\bf 509} (Jan., 2010) A25},
  [\href{http://arxiv.org/abs/0910.4272}{{\tt 0910.4272}}].

\bibitem{Lovell:2014lea}
M.~R. Lovell, G.~Bertone, A.~Boyarsky, A.~Jenkins and O.~Ruchayskiy,
  \emph{{Decaying dark matter: the case for a deep X-ray observation of
  Draco}}, \href{http://dx.doi.org/10.1093/mnras/stv963}{\emph{Mon. Not. Roy.
  Astron. Soc.} {\bf 451} (2015) 1573},
  [\href{http://arxiv.org/abs/1411.0311}{{\tt 1411.0311}}].

\bibitem{Takahashi:2012jn}
T.~Takahashi et~al., \emph{{The ASTRO-H X-ray Observatory}},
  \href{http://dx.doi.org/10.1117/12.926190}{\emph{Proc. SPIE Int. Soc. Opt.
  Eng.} {\bf 8443} (2012) 1Z}, [\href{http://arxiv.org/abs/1210.4378}{{\tt
  1210.4378}}].

\bibitem{Figueroa-Feliciano:2015gwa}
{\scshape XQC} collaboration, E.~Figueroa-Feliciano et~al., \emph{{Searching
  for keV Sterile Neutrino Dark Matter with X-ray Microcalorimeter Sounding
  Rockets}},
  \href{http://dx.doi.org/10.1088/0004-637X/814/1/82}{\emph{Astrophys. J.} {\bf
  814} (2015) 82}, [\href{http://arxiv.org/abs/1506.05519}{{\tt 1506.05519}}].

\bibitem{2012arXiv1209.3114M}
A.~{Merloni}, P.~{Predehl}, W.~{Becker}, H.~{B{\"o}hringer}, T.~{Boller},
  H.~{Brunner} et~al., \emph{{eROSITA Science Book: Mapping the Structure of
  the Energetic Universe}}, {\emph{ArXiv e-prints} (Sept., 2012) },
  [\href{http://arxiv.org/abs/1209.3114}{{\tt 1209.3114}}].

\bibitem{Zandanel:2015xca}
F.~Zandanel, C.~Weniger and S.~Ando, \emph{{The role of the eROSITA all-sky
  survey in searches for sterile neutrino dark matter}},
  \href{http://arxiv.org/abs/1505.07829}{{\tt 1505.07829}}.

\bibitem{Garzilli:2015bha}
A.~Garzilli, T.~Theuns and J.~Schaye, \emph{{The broadening of Lyman-$\alpha$
  forest absorption lines}},
  \href{http://dx.doi.org/10.1093/mnras/stv394}{\emph{Mon. Not. Roy. Astron.
  Soc.} {\bf 450} (2015) 1465--1476},
  [\href{http://arxiv.org/abs/1502.05715}{{\tt 1502.05715}}].

\bibitem{Arzoumanian:2001dv}
Z.~Arzoumanian, D.~F. Chernoffs and J.~M. Cordes, \emph{{The Velocity
  distribution of isolated radio pulsars}},
  \href{http://dx.doi.org/10.1086/338805}{\emph{Astrophys. J.} {\bf 568} (2002)
  289--301}, [\href{http://arxiv.org/abs/astro-ph/0106159}{{\tt
  astro-ph/0106159}}].

\bibitem{Hobbs:2005yx}
G.~Hobbs, D.~R. Lorimer, A.~G. Lyne and M.~Kramer, \emph{{A Statistical study
  of 233 pulsar proper motions}},
  \href{http://dx.doi.org/10.1111/j.1365-2966.2005.09087.x}{\emph{Mon. Not.
  Roy. Astron. Soc.} {\bf 360} (2005) 974--992},
  [\href{http://arxiv.org/abs/astro-ph/0504584}{{\tt astro-ph/0504584}}].

\bibitem{Janka:2012wk}
H.-T. Janka, \emph{{Explosion Mechanisms of Core-Collapse Supernovae}},
  \href{http://dx.doi.org/10.1146/annurev-nucl-102711-094901}{\emph{Ann. Rev.
  Nucl. Part. Sci.} {\bf 62} (2012) 407--451},
  [\href{http://arxiv.org/abs/1206.2503}{{\tt 1206.2503}}].

\bibitem{Wang:2005jg}
C.~Wang, D.~Lai and J.~Han, \emph{{Neutron star kicks in isolated and binary
  pulsars: observational constraints and implications for kick mechanisms}},
  \href{http://dx.doi.org/10.1086/499397}{\emph{Astrophys. J.} {\bf 639} (2006)
  1007--1017}, [\href{http://arxiv.org/abs/astro-ph/0509484}{{\tt
  astro-ph/0509484}}].

\bibitem{Ng:2007aw}
C.-Y. Ng and R.~W. Romani, \emph{{Birth Kick Distributions and the Spin-Kick
  Correlation of Young Pulsars}},
  \href{http://dx.doi.org/10.1086/513597}{\emph{Astrophys. J.} {\bf 660} (2007)
  1357--1374}, [\href{http://arxiv.org/abs/astro-ph/0702180}{{\tt
  astro-ph/0702180}}].

\bibitem{Kusenko:1997sp}
A.~Kusenko and G.~Segre, \emph{{Neutral current induced neutrino oscillations
  in a supernova}},
  \href{http://dx.doi.org/10.1016/S0370-2693(97)00121-4}{\emph{Phys. Lett.}
  {\bf B396} (1997) 197--200}, [\href{http://arxiv.org/abs/hep-ph/9701311}{{\tt
  hep-ph/9701311}}].

\bibitem{Kusenko:2008gh}
A.~Kusenko, B.~P. Mandal and A.~Mukherjee, \emph{{Delayed pulsar kicks from the
  emission of sterile neutrinos}},
  \href{http://dx.doi.org/10.1103/PhysRevD.77.123009}{\emph{Phys. Rev.} {\bf
  D77} (2008) 123009}, [\href{http://arxiv.org/abs/0801.4734}{{\tt
  0801.4734}}].

\bibitem{Kishimoto:2011mw}
C.~T. Kishimoto, \emph{{Pulsar Kicks from Active-Sterile Neutrino
  Transformation in Supernovae}},  \href{http://arxiv.org/abs/1101.1304}{{\tt
  1101.1304}}.

\bibitem{Fuller:2003gy}
G.~M. Fuller, A.~Kusenko, I.~Mocioiu and S.~Pascoli, \emph{{Pulsar kicks from a
  dark-matter sterile neutrino}},
  \href{http://dx.doi.org/10.1103/PhysRevD.68.103002}{\emph{Phys. Rev.} {\bf
  D68} (2003) 103002}, [\href{http://arxiv.org/abs/astro-ph/0307267}{{\tt
  astro-ph/0307267}}].

\bibitem{Raffelt:1999tx}
G.~G. Raffelt, \emph{{Particle physics from stars}},
  \href{http://dx.doi.org/10.1146/annurev.nucl.49.1.163}{\emph{Ann. Rev. Nucl.
  Part. Sci.} {\bf 49} (1999) 163--216},
  [\href{http://arxiv.org/abs/hep-ph/9903472}{{\tt hep-ph/9903472}}].

\bibitem{Raffelt:2011nc}
G.~G. Raffelt and S.~Zhou, \emph{{Supernova bound on keV-mass sterile neutrinos
  reexamined}}, \href{http://dx.doi.org/10.1103/PhysRevD.83.093014}{\emph{Phys.
  Rev.} {\bf D83} (2011) 093014}, [\href{http://arxiv.org/abs/1102.5124}{{\tt
  1102.5124}}].

\bibitem{Kainulainen:1990bn}
K.~Kainulainen, J.~Maalampi and J.~T. Peltoniemi, \emph{{Inert neutrinos in
  supernovae}},
  \href{http://dx.doi.org/10.1016/0550-3213(91)90354-Z}{\emph{Nucl. Phys.} {\bf
  B358} (1991) 435--446}.

\bibitem{Raffelt:1992bs}
G.~Raffelt and G.~Sigl, \emph{{Neutrino flavor conversion in a supernova
  core}},
  \href{http://dx.doi.org/10.1016/0927-6505(93)90020-E}{\emph{Astropart. Phys.}
  {\bf 1} (1993) 165--184}, [\href{http://arxiv.org/abs/astro-ph/9209005}{{\tt
  astro-ph/9209005}}].

\bibitem{Zhou:2015jha}
S.~Zhou, \emph{Supernova bounds on kev-mass sterile neutrinos},
  {\emph{Int.J.Mod.Phys.} {\bf A30} (2015) 0033},
  [\href{http://arxiv.org/abs/1504.02729}{{\tt 1504.02729}}].

\bibitem{Hidaka:2007se}
J.~Hidaka and G.~M. Fuller, \emph{{Sterile Neutrino-Enhanced Supernova
  Explosions}}, \href{http://dx.doi.org/10.1103/PhysRevD.76.083516}{\emph{Phys.
  Rev.} {\bf D76} (2007) 083516}, [\href{http://arxiv.org/abs/0706.3886}{{\tt
  0706.3886}}].

\bibitem{Fryer:2005sz}
C.~L. Fryer and A.~Kusenko, \emph{{Effects of neutrino-driven kicks on the
  supernova explosion mechanism}},
  \href{http://dx.doi.org/10.1086/500933}{\emph{Astrophys. J. Suppl.} {\bf 163}
  (2006) 335}, [\href{http://arxiv.org/abs/astro-ph/0512033}{{\tt
  astro-ph/0512033}}].

\bibitem{Hansen:2003yj}
S.~H. Hansen and Z.~Haiman, \emph{{Do we need stars to reionize the universe at
  high redshifts? Early reionization by decaying heavy sterile neutrinos}},
  \href{http://dx.doi.org/10.1086/379636}{\emph{Astrophys. J.} {\bf 600} (2004)
  26--31}, [\href{http://arxiv.org/abs/astro-ph/0305126}{{\tt
  astro-ph/0305126}}].

\bibitem{Werner}
W.~Rodejohann. personal communication.

\bibitem{de2013role}
H.~de~Vega, O.~Moreno, E.~Moya~de Guerra, M.~Ram{\'o}n~Medrano and
  N.~S{\'a}nchez, \emph{{Role of sterile neutrino warm dark matter in rhenium
  and tritium beta decays}}, {\emph{Nucl. Phys. B} {\bf 866} (2013) 177--195}.

\bibitem{Rod14}
W.~Rodejohann and H.~Zhang, \emph{{Signatures of extra dimensional sterile
  neutrinos}}, {\emph{Phys. Lett} {\bf B 737} (2014) 81--89}.

\bibitem{Rod14b}
J.~Barry, J.~Heeck and W.~Rodejohann, \emph{{Sterile neutrinos and right-handed
  currents in KATRIN}}, {\emph{JHEP} {\bf 81} (2014) 7}.

\bibitem{Monreal:PhysRevD80051301:2009}
B.~Monreal and J.~A. Formaggio, \emph{Relativistic cyclotron radiation
  detection of tritium decay electrons as a new technique for measuring the
  neutrino mass},
  \href{http://dx.doi.org/10.1103/PhysRevD.80.051301}{\emph{Phys.Rev.D} {\bf
  80} (4, 2009) 051301}.

\bibitem{PhysRevLett.114.162501}
{\scshape Project 8 Collaboration} collaboration, D.~M. Asner, R.~F. Bradley,
  L.~de~Viveiros, P.~J. Doe, J.~L. Fernandes, M.~Fertl et~al.,
  \emph{Single-electron detection and spectroscopy via relativistic cyclotron
  radiation},
  \href{http://dx.doi.org/10.1103/PhysRevLett.114.162501}{\emph{Phys. Rev.
  Lett.} {\bf 114} (Apr, 2015) 162501}.

\bibitem{betts2013development}
S.~Betts, W.~Blanchard, R.~Carnevale, C.~Chang, C.~Chen, S.~Chidzik et~al.,
  \emph{Development of a relic neutrino detection experiment at ptolemy:
  Princeton tritium observatory for light, early-universe, massive-neutrino
  yield},  \href{http://arxiv.org/abs/1307.4738}{{\tt 1307.4738}}.

\bibitem{Bezrukov:2006cy}
F.~Bezrukov and M.~Shaposhnikov, \emph{Searching for dark matter sterile
  neutrinos in the laboratory},
  \href{http://dx.doi.org/10.1103/PhysRevD.75.053005}{\emph{Phys. Rev. D} {\bf
  75} (Mar, 2007) 053005}.

\bibitem{Belesev:2012hx}
A.~I. Belesev, A.~I. Berlev, E.~V. Geraskin, A.~A. Golubev, N.~A. Likhovid,
  A.~A. Nozik et~al., \emph{{An upper limit on additional neutrino mass
  eigenstate in 2 to 100 eV region from 'Troitsk nu-mass' data}},
  \href{http://dx.doi.org/10.1134/S0021364013020033}{\emph{JETP Lett.} {\bf 97}
  (2013) 67--69}, [\href{http://arxiv.org/abs/1211.7193}{{\tt 1211.7193}}].

\bibitem{Abdurashitov:2015jha}
D.~Abdurashitov, A.~Belesev, A.~Berlev, V.~Chernov, E.~Geraskin et~al.,
  \emph{The current status of "troitsk nu-mass" experiment in search for
  sterile neutrino},  \href{http://arxiv.org/abs/1504.00544}{{\tt 1504.00544}}.

\bibitem{Abdurashitov:2014vqa}
D.~N. Abdurashitov, A.~I. Berlev, N.~A. Likhovid, A.~V. Lokhov, I.~I. Tkachev
  and V.~E. Yants, \emph{{Searches for a Sterile-Neutrino Admixture in
  Detecting Tritium Decays in a Proportional Counter: New Possibilities}},
  \href{http://dx.doi.org/10.1134/S1063778815020027}{\emph{Phys. Atom. Nucl.}
  {\bf 78} (2015) 268--280}, [\href{http://arxiv.org/abs/1403.2935}{{\tt
  1403.2935}}].

\bibitem{Osipowicz:2001sq}
{\scshape KATRIN} collaboration, A.~Osipowicz et~al., \emph{{KATRIN: A Next
  generation tritium beta decay experiment with sub-eV sensitivity for the
  electron neutrino mass. Letter of intent}},
  \href{http://arxiv.org/abs/hep-ex/0109033}{{\tt hep-ex/0109033}}.

\bibitem{Stu10}
M.~Sturm et~al., \emph{Monitoring of all hydrogen isotopologues at tritium
  laboratory karlsruhe using raman spectroscopy}, {\emph{Laser Phys.} {\bf 20}
  (2010) 493--507}.

\bibitem{Sch11}
M.~Schl{\"osser} et~al., \emph{Design implications for laser raman measurement
  systems for tritium sample-analysis, accountancy or process-control
  applications}, {\emph{Fus. Sci. and Techn.} {\bf (60) 3} (2011) 976--981}.

\bibitem{Fis11}
M.~Sturm et~al., \emph{Monitoring of tritium purity during long-term
  circulation in the katrin test experiment loopino using laser raman
  spectroscopy}, {\emph{Fus. Sci. and Techn.} {\bf (60) 3} (2011) 925--930}.

\bibitem{Babutzka:2012xd}
M.~Babutzka et~al., \emph{{Monitoring of the operating parameterss of the
  KATRIN Windowless Gaseous Tritium Source}},
  \href{http://dx.doi.org/10.1088/1367-2630/14/10/103046}{\emph{New J. Phys.}
  {\bf 14} (2012) 103046}, [\href{http://arxiv.org/abs/1205.5421}{{\tt
  1205.5421}}].

\bibitem{Grohmann:2011zz}
S.~Grohmann, T.~Bode, H.~Schon and M.~Susser, \emph{{Precise temperature
  measurement at 30-K in the KATRIN source cryostat}},
  \href{http://dx.doi.org/10.1016/j.cryogenics.2011.05.001}{\emph{Cryogenics}
  {\bf 51} (2011) 438--445}.

\bibitem{Aseev:2000}
V.~Aseev, A.~Belesev, A.~Berlev, E.~Geraskin, O.~Kazachenko, Y.~Kuznetsov
  et~al., \emph{Energy loss of 18 kev electrons in gaseous t and quench
  condensed d films}, \href{http://dx.doi.org/10.1007/s100530050525}{\emph{The
  European Physical Journal D - Atomic, Molecular, Optical and Plasma Physics}
  {\bf 10} (2000) 39--52}.

\bibitem{Lukic:2011fw}
S.~Lukic, B.~Bornschein, L.~Bornschein, G.~Drexlin, A.~Kosmider, K.~Schloesser
  et~al., \emph{{Measurement of the gas-flow reduction factor of the KATRIN
  DPS2-F differential pumping section}},
  \href{http://dx.doi.org/10.1016/j.vacuum.2011.10.017}{\emph{Vacuum} {\bf 86}
  (2012) 1126--1133}, [\href{http://arxiv.org/abs/1107.0220}{{\tt 1107.0220}}].

\bibitem{Eichelhardt:2011zz}
{\scshape KATRIN} collaboration, F.~Eichelhardt, \emph{{The cryogenic pumping
  section of KATRIN and the test experiment TRAP}},
  \href{http://dx.doi.org/10.1016/j.nuclphysbps.2011.09.042}{\emph{Nucl. Phys.
  Proc. Suppl.} {\bf 221} (2011) 342}.

\bibitem{Luo08}
X.~Luo and C.~Day, \emph{Test particle monte carlo study of the cryogenic
  pumping system of the karlsruhe tritium neutrino experiment}, {\emph{J. of
  Vac. Sci. and Techn. A} {\bf 26} (2008) 1319--1325}.

\bibitem{Pra12}
M.~Prall and Others, \emph{The katrin pre-spectrometer at reduced filter
  energy}, {\emph{New Journal of Physics} {\bf 14} (2012) 073054}.

\bibitem{Amsbaugh:2014uca}
J.~F. Amsbaugh et~al., \emph{{Focal-plane detector system for the KATRIN
  experiment}}, \href{http://dx.doi.org/10.1016/j.nima.2014.12.116}{\emph{Nucl.
  Instrum. Meth.} {\bf A778} (2015) 40--60},
  [\href{http://arxiv.org/abs/1404.2925}{{\tt 1404.2925}}].

\bibitem{Mertens:2014nha}
S.~Mertens, T.~Lasserre, S.~Groh, G.~Drexlin, F.~Glueck, A.~Huber et~al.,
  \emph{{Sensitivity of Next-Generation Tritium Beta-Decay Experiments for
  keV-Scale Sterile Neutrinos}},
  \href{http://dx.doi.org/10.1088/1475-7516/2015/02/020}{\emph{JCAP} {\bf 1502}
  (2015) 020}, [\href{http://arxiv.org/abs/1409.0920}{{\tt 1409.0920}}].

\bibitem{Mertens:2014osa}
S.~Mertens, K.~Dolde, M.~Korzeczek, F.~Glueck, S.~Groh, R.~D. Martin et~al.,
  \emph{{Wavelet approach to search for sterile neutrinos in tritium
  $\beta$-decay spectra}},
  \href{http://dx.doi.org/10.1103/PhysRevD.91.042005}{\emph{Phys. Rev.} {\bf
  D91} (2015) 042005}, [\href{http://arxiv.org/abs/1410.7684}{{\tt
  1410.7684}}].

\bibitem{merle}
A.~Merle, A.~Schneider and Totzauer. personal communication.

\bibitem{Steinbrink:2013ska}
N.~Steinbrink, V.~Hannen, E.~L. Martin, R.~H. Robertson, M.~Zacher et~al.,
  \emph{{Neutrino mass sensitivity by MAC-E-Filter based time-of-flight
  spectroscopy with the example of KATRIN}},
  \href{http://dx.doi.org/10.1088/1367-2630/15/11/113020}{\emph{New J.Phys.}
  {\bf 15} (2013) 113020}, [\href{http://arxiv.org/abs/1308.0532}{{\tt
  1308.0532}}].

\bibitem{Bonn1999256}
J.~Bonn, L.~Bornschein, B.~Degen, E.~Otten and C.~Weinheimer, \emph{A high
  resolution electrostatic time-of-flight spectrometer with adiabatic magnetic
  collimation},
  \href{http://dx.doi.org/http://dx.doi.org/10.1016/S0168-9002(98)01263-7}{\emph{Nuclear
  Instruments and Methods in Physics Research Section A: Accelerators,
  Spectrometers, Detectors and Associated Equipment} {\bf 421} (1999) 256 --
  265}.

\bibitem{Steinbrink:prep}
N.~Steinbrink, J.~Behrens, V.~Hannen, S.~Mertens and C.~Weinheimer,
  ``{Sensitivity on keV scale sterile neutrinos by Time-Of-Flight spectroscopy
  with KATRIN}.''.

\bibitem{Behrens:prep}
J.~Behrens, M.~Fedkevych, V.~Hannen, P.~C.-O. Ranitzsch, O.~Rest, N.~Steinbrink
  et~al., ``{Time-focussing time of flight spectroscopy for neutrino mass
  measurements}.''.

\bibitem{takahashi2012astro}
T.~Takahashi, K.~Mitsuda, R.~Kelley, H.~Aarts, F.~Aharonian, H.~Akamatsu
  et~al., \emph{The astro-h x-ray observatory},  in \emph{SPIE Astronomical
  Telescopes+ Instrumentation}, pp.~84431Z--84431Z, International Society for
  Optics and Photonics, 2012.

\bibitem{Pontecorvo:47}
B.~Pontecorvo, \emph{The neutrino and the recoil of nuclei in beta
  disintegrations}, {\emph{Reports on Progress in Physics} {\bf 11} (1947) 32}.

\bibitem{PhysRevD.46.R888}
G.~Finocchiaro and R.~E. Shrock, \emph{Suggestion for an experiment to search
  for a massive admixed neutrino in nuclear beta decay by complete kinematic
  reconstruction of the final state},
  \href{http://dx.doi.org/10.1103/PhysRevD.46.R888}{\emph{Phys. Rev. D} {\bf
  46} (Aug, 1992) R888--R891}.

\bibitem{PhysRevD.46.R6}
S.~Cook, M.~Fink, S.~Thomas and H.~Wellenstein, \emph{Proposal to isolate the
  origin of the 17-kev kink in the $\ensuremath{\beta}$ spectrum},
  \href{http://dx.doi.org/10.1103/PhysRevD.46.R6}{\emph{Phys. Rev. D} {\bf 46}
  (Jul, 1992) R6--R8}.

\bibitem{PhysRevLett.90.012501}
M.~Trinczek, A.~Gorelov, D.~Melconian, W.~P. Alford, D.~Asgeirsson, D.~Ashery
  et~al., \emph{Novel search for heavy $\ensuremath{\nu}$ mixing from the
  ${\ensuremath{\beta}}^{+}$ decay of $^{\mathrm{38}\mathrm{m}}\mathrm{K}$
  confined in an atom trap},
  \href{http://dx.doi.org/10.1103/PhysRevLett.90.012501}{\emph{Phys. Rev.
  Lett.} {\bf 90} (Jan, 2003) 012501}.

\bibitem{doi:10.1088/0953-4075/30/13/006}
J.~Ullrich, R.~Moshammer, R.~D\"orner, O.~Jagutzki, V.~Mergel,
  H.~Schmidt-B�cking et~al., \emph{Recoil-ion momentum spectroscopy},
  {\emph{Journal of Physics B: Atomic, Molecular and Optical Physics} {\bf 30}
  (1997) 2917}.

\bibitem{Doerner200095}
R.~D{\"o}rner, V.~Mergel, O.~Jagutzki, L.~Spielberger, J.~Ullrich, R.~Moshammer
  et~al., \emph{{Cold Target Recoil Ion Momentum Spectroscopy: a 'momentum
  microscope' to view atomic collision dynamics}},
  \href{http://dx.doi.org/http://dx.doi.org/10.1016/S0370-1573(99)00109-X}{\emph{Physics
  Reports} {\bf 330} (2000) 95 -- 192}.

\bibitem{Gastaldo:2013wha}
L.~Gastaldo, K.~Blaum, A.~Doerr, C.~Duellmann, K.~Eberhardt et~al., \emph{{The
  Electron Capture $^{163}$Ho Experiment ECHo: an overview}},
  \href{http://dx.doi.org/10.1007/s10909-014-1187-4}{\emph{J.Low Temp.Phys.}
  {\bf 176} (2014) 876--884}, [\href{http://arxiv.org/abs/1309.5214}{{\tt
  1309.5214}}].

\bibitem{Kem13b}
S.~Kempf, M.~Wegner, L.~Gastaldo, A.~Fleischmann and C.~Enss, \emph{Multiplexed
  readout of mmc detector arrays using non-hysteretic rf-squids},
  \href{http://dx.doi.org/10.1007/s10909-013-1041-0}{\emph{J. Low. Temp. Phys.}
  {\bf 176} (2014) 426--432}.

\bibitem{http://dx.doi.org/10.1103/PhysRevA.73.012710}
A.~S. Kadyrov, I.~Bray and A.~T. Stelbovics, \emph{On-shell coupled-channel
  approach to proton-hydrogen collisions without partial-wave expansion},
  \href{http://dx.doi.org/10.1103/PhysRevA.73.012710}{\emph{Phys. Rev. A} {\bf
  73} (Jan, 2006) 012710}.

\bibitem{PhysRevLett.115.062501}
S.~Eliseev, K.~Blaum, M.~Block, S.~Chenmarev, H.~Dorrer, C.~E. D\"ullmann
  et~al., \emph{Direct measurement of the mass difference of
  $^{163}\mathrm{Ho}$ and $^{163}\mathrm{Dy}$ solves the $q$-value puzzle for
  the neutrino mass determination},
  \href{http://dx.doi.org/10.1103/PhysRevLett.115.062501}{\emph{Phys. Rev.
  Lett.} {\bf 115} (Aug, 2015) 062501}.

\bibitem{Kopp:2009yp}
J.~Kopp and A.~Merle, \emph{{Ultra-low Q values for neutrino mass
  measurements}},
  \href{http://dx.doi.org/10.1103/PhysRevC.81.045501}{\emph{Phys. Rev.} {\bf
  C81} (2010) 045501}, [\href{http://arxiv.org/abs/0911.3329}{{\tt
  0911.3329}}].

\bibitem{De_Rujula_1982}
A.~De~Rujula and M.~Lusignoli, \emph{{Calorimetric Measurements of $^{163}$Ho
  Decay as Tools to Determine the Electron Neutrino Mass}},
  \href{http://dx.doi.org/10.1016/0370-2693(82)90218-0}{\emph{Phys. Lett.} {\bf
  B118} (1982) 429}.

\bibitem{ECHo_web}
``{ECHo Website}.'' \url{http://www.kip.uni-heidelberg.de/echo/}.

\bibitem{Ranitzsch:2014kma}
P.~C.~O. Ranitzsch, C.~Hassel, M.~Wegner, S.~Kempf, A.~Fleischmann, C.~Enss
  et~al., \emph{{First Calorimetric Measurement of OI-line in the Electron
  Capture Spectrum of $^{163}$Ho}},  \href{http://arxiv.org/abs/1409.0071}{{\tt
  1409.0071}}.

\bibitem{HOLMES}
B.~Alpert et~al., \emph{The electron capture decay of 163ho to measure the
  electron neutrino mass with sub-ev sensitivity},
  \href{http://dx.doi.org/10.1140/epjc/s10052-015-3329-5}{\emph{Eu. Phys. J.}
  {\bf C 7} (2015) 112}.

\bibitem{NuMECS}
``{NuMecs Position Paper}.'' \url{http://p25ext.lanl.gov/~kunde/NuMECS/}.

\bibitem{lusignoli_2012}
M.~Lusignoli and M.~Vignati, \emph{{Relic Antineutrino Capture on 163-Ho
  decaying Nuclei}}, \href{http://dx.doi.org/10.1016/j.physletb.2011.01.047,
  10.1016/j.physletb.2011.06.008}{\emph{Phys. Lett.} {\bf B697} (2011) 11--14},
  [\href{http://arxiv.org/abs/1012.0760}{{\tt 1012.0760}}].

\bibitem{Deslattes2003}
R.~D. Deslattes, E.~G. Kessler, P.~Indelicato, L.~de~Billy, E.~Lindroth and
  J.~Anton, \emph{{X-ray transition energies: new approach to a comprehensive
  evaluation}}, \href{http://dx.doi.org/10.1103/RevModPhys.75.35}{\emph{Rev.
  Mod. Phys.} {\bf 75} (2003) 35--99}.

\bibitem{XRay2009}
A.~Thompson et~al., \emph{X-ray data booklet}, {\emph{X-ray data booklet}
  (2009) }.

\bibitem{Campbell2001}
J.~Campbell and T.~Papp, \emph{Widths of the atomic k-n7 levels}, {\emph{At.
  Data Nucl. Data Table} {\bf 77} (2001) 1}.

\bibitem{Cohen1972}
R.~L. Cohen, G.~K. Wertheim, A.~Rosencwaig and H.~J. Guggenheim,
  \emph{Multiplet splitting of the 4s and 5s electrons of the rare earths},
  {\emph{Phys. Rev.} {\bf B 5} (1972) 1037}.

\bibitem{Enss:2005md}
C.~Enss, \emph{Cryogenic particle detection}.
\newblock Springer: Topics in applied physics, Berlin, Germany, 2005.

\bibitem{Fleischmann_LTD13}
A.~Fleischmann, L.~Gastaldo, S.~Kempf, A.~Kirsch, A.~Pabinger, C.~Pies et~al.,
  \emph{Metallic magnetic calorimeters}, {\emph{AIP Conf. Proc.} {\bf 1185}
  (2009) }.

\bibitem{Fleischmann_tbs}
A.~Fleischmann et~al.

\bibitem{Pies_LTD14}
C.~Pies et~al., \emph{Metallic magnetic calorimeters}, {\emph{Journal of Low
  Temperature Physics} {\bf 167} (2012) }.

\bibitem{Mates}
J.~A.~B. Mates et~al., \emph{Demonstration of a multiplexer of dissipationless
  superconducting quantum interference devices}, {\emph{Applied Physics
  Letters} {\bf 92} (2008) }.

\bibitem{Kempf_LTD15}
S.~Kempf et~al., \emph{Microcalorimeters with inductively read out paramagnetic
  and superconducting temperature sensors}, {\emph{Journal of Low Temperature
  Physics} {\bf 176} (2014) 426}.

\bibitem{Gastaldo:2012nv}
L.~Gastaldo, P.-O. Ranitzsch, F.~von Seggern, J.-P. Porst, S.~Sch�fer et~al.,
  \emph{{Characterization of low temperature metallic magnetic calorimeters
  having gold absorbers with implanted $^{163}$Ho ions}},
  \href{http://dx.doi.org/10.1016/j.nima.2013.01.027}{\emph{Nucl.Instrum.Meth.}
  {\bf A711} (2013) 150--159}, [\href{http://arxiv.org/abs/1206.5647}{{\tt
  1206.5647}}].

\bibitem{Meunier:1996ge}
{\scshape Cryogenic Detector Group of Genoa} collaboration, P.~Meunier and
  C.~Salvo, \emph{{A Calorimetric measurement of Ho-163 spectrum by means of a
  cryogenic detector}},
  \href{http://dx.doi.org/10.1016/S0370-2693(97)00239-6}{\emph{Phys.Lett.} {\bf
  B398} (1997) 415--419}.

\bibitem{Q_EC}
G.~Audi et~al., \emph{The ame2012 atomic mass evaluation}, {\emph{Chinese
  Physics C} {\bf 36} (2012) }.

\bibitem{PT-MS}
K.~Blaum, {\relax Yu}.~N. Novikov and G.~Werth, \emph{{Penning traps as a
  versatile tool for precise experiments in fundamental physics}},
  \href{http://dx.doi.org/10.1080/00107510903387652}{\emph{Contemp. Phys.} {\bf
  51} (2010) 149--175}, [\href{http://arxiv.org/abs/0909.1095}{{\tt
  0909.1095}}].

\bibitem{SHIPTRAP}
M.~Block et~al., \emph{Towards direct mass measurements of nobelium at
  shiptrap}, {\emph{Eur. Phys. J. D} {\bf 45} (2007) }.

\bibitem{ICRT}
S.~Eliseev et~al., \emph{A phase-imaging technique for cyclotron-frequency
  measurements}, {\emph{Appl. Phys. B} {\bf 114} (2014) }.

\bibitem{Ketelaer:2008by}
J.~Ketelaer, J.~Kramer, D.~Beck, K.~Blaum, M.~Block et~al., \emph{{TRIGA-SPEC:
  A Setup for mass spectrometry and laser spectroscopy at the research reactor
  TRIGA Mainz}},
  \href{http://dx.doi.org/10.1016/j.nima.2008.06.023}{\emph{Nucl.Instrum.Meth.}
  {\bf A594} (2008) 162--177}, [\href{http://arxiv.org/abs/0805.4475}{{\tt
  0805.4475}}].

\bibitem{Fabian_2015}
F.~Schneider et~al., \emph{Preparatory studies for a high-precision penning
  trap measurement of the 163ho electron capture q-value}, {\emph{submitted to
  EPJ} (2015) }.

\bibitem{Repp:2011hm}
J.~Repp, C.~Boehm, J.~R.~C. Lopez-Urrutia, A.~Doerr, S.~Eliseev, S.~George
  et~al., \emph{{PENTATRAP}: a novel cryogenic multi-penning-trap experiment
  for high-precision mass measurements on highly charged ions}, {\emph{Appl.
  Phys. B} {\bf 107} (2012) }, [\href{http://arxiv.org/abs/1110.2919}{{\tt
  1110.2919}}].

\bibitem{Roux:2011hn}
C.~Roux, C.~Boehm, A.~Doerr, S.~Eliseev, S.~G.~M. Goncharov, Y.~Novikov et~al.,
  \emph{The trap design of {PENTATRAP}}, {\emph{Appl. Phys.} {\bf B 107} (2012)
  }, [\href{http://arxiv.org/abs/1110.2920}{{\tt 1110.2920}}].

\bibitem{Robertson:2014fka}
R.~Robertson, \emph{{Examination of the calorimetric spectrum to determine the
  neutrino mass in low-energy electron capture decay}},
  \href{http://dx.doi.org/10.1103/PhysRevC.91.035504}{\emph{Phys.Rev.} {\bf
  C91} (2015) 035504}, [\href{http://arxiv.org/abs/1411.2906}{{\tt
  1411.2906}}].

\bibitem{Faessler:2015pka}
A.~Faessler and F.~Simkovic, \emph{{Improved Description of One- and Two-Hole
  States after Electron Capture in 163 Holmium and the Determination of the
  Neutrino Mass}},
  \href{http://dx.doi.org/10.1103/PhysRevC.91.045505}{\emph{Phys.Rev.} {\bf
  C91} (2015) 045505}, [\href{http://arxiv.org/abs/1501.04338}{{\tt
  1501.04338}}].

\bibitem{Faessler:2015txa}
A.~Faessler, C.~Enss, L.~Gastaldo and F.~Simkovic, \emph{{Determination of the
  neutrino mass by electron capture in 163 Holmium and the role of the
  three-hole states in 163 Dysprosium}},
  \href{http://dx.doi.org/10.1103/PhysRevC.91.064302}{\emph{Phys.Rev.} {\bf
  C91} (2015) 064302}, [\href{http://arxiv.org/abs/1503.02282}{{\tt
  1503.02282}}].

\bibitem{ADR_ML_2015}
A.~D. Rujula and M.~Lusignoli, \emph{The calorimetric spectrum of the
  electron-capture decay of 163ho. a preliminary analysis of the preliminary
  data},  \href{http://arxiv.org/abs/1510.05462}{{\tt 1510.05462}}.

\bibitem{Springer_1987}
P.~T. Springer, C.~L. Bennett and P.~A. Baisden, \emph{{Measurement of the
  Neutrino Mass Using the Inner Bremsstrahlung Emitted in the Electron-Capture
  Decay og 163Ho}},
  \href{http://dx.doi.org/10.1103/PhysRevA.35.679}{\emph{Phys. Rev.} {\bf A35}
  (1987) 679--689}.

\bibitem{Ranitzsch_2015}
P.~C.-O. Ranitzsch et~al.

\bibitem{Filianin_2014}
P.~E. Filianin, K.~Blaum, S.~Eliseev, L.~Gastaldo, Y.~N. Novikov, V.~M. Shabaev
  et~al., \emph{On the kev sterile neutrino search in electron capture},
  {\emph{J.Phys. G} {\bf 41} (2014) }.

\bibitem{halflife}
``\url{http://nucleardata.nuclear.lu.se/toi/}.''

\bibitem{Q_EC-values}
G.~Audi, M.~Wang, A.~H. Wapstra, F.~G. Kondev, M.~MacCormick, X.~Xu et~al.,
  \emph{The ame2012 atomic mass evaluation}, {\emph{Chinese Phys. C} {\bf 36}
  (2012) }.

\bibitem{binding_E}
J.~A. Bearden and A.~F. Burr, \emph{{Reevaluation of X-Ray Atomic Energy
  Levels}}, \href{http://dx.doi.org/10.1103/RevModPhys.39.125}{\emph{Rev. Mod.
  Phys.} {\bf 39} (1967) 125--142}.

\bibitem{amplitudes}
W.~Bambynek, H.~Behrens, M.~H. Chen, B.~Crasemann, M.~L. Fitzpatrick, K.~W.~D.
  Ledingham et~al., \emph{Orbital electron capture by the nucleus}, {\emph{Rev.
  Mod. Phys.} {\bf 49} (1977) }.

\bibitem{Liao:2010yx}
W.~Liao, \emph{{keV scale $\nu_{R}$ dark matter and its detection in $\beta$
  decay experiment}}, {\emph{Phys.\ Rev.\ D} {\bf 82} (2010) 073001}.

\bibitem{Li:2010vy}
Y.~F. Li and Z.~z.~Xing, \emph{{Possible Capture of keV Sterile Neutrino Dark
  Matter on Radioactive $\beta$-decaying Nuclei}}, {\emph{Phys.\ Lett.\ B} {\bf
  695} (2011) 205}.

\bibitem{Cocco:2007za}
A.~G. Cocco, G.~Mangano and M.~Messina, \emph{{Probing low energy neutrino
  backgrounds with neutrino capture on beta decaying nuclei}}, {\emph{JCAP}
  {\bf 0706} (2007) 015}.

\bibitem{Lazauskas:2007da}
R.~Lazauskas, P.~Vogel and C.~Volpe, \emph{{Charged current cross section for
  massive cosmological neutrinos impinging on radioactive nuclei}}, {\emph{J.\
  Phys.\ G} {\bf 35} (2008) 025001}.

\bibitem{Li:2010sn}
Y.~F. Li, Z.~z.~Xing and S.~Luo, \emph{{Direct Detection of the Cosmic Neutrino
  Background Including Light Sterile Neutrinos}}, {\emph{Phys.\ Lett.\ B} {\bf
  692} (2010) 261}.

\bibitem{Li:2011px}
Y.~F. Li and Z.~z.~Xing, \emph{{Neutrinos as Hot or Warm Dark Matter}},
  {\emph{Acta Phys.\ Polon.\ B} {\bf 42} (2011) 2193}.

\bibitem{Li:2015koa}
Y.~F. Li, \emph{{Detection Prospects of the Cosmic Neutrino Background}},
  {\emph{Int.\ J.\ Mod.\ Phys.\ A} {\bf 30} (2015) 1530031}.

\bibitem{Liao:2013jwa}
W.~Liao, X.~H. Wu and H.~Zhou, \emph{{Electron events from the scattering with
  solar neutrinos in the search of keV scale sterile neutrino dark matter}},
  {\emph{Phys.\ Rev.\ D} {\bf 89} (2014) 093017}.

\bibitem{Ando:2010ye}
S.~Ando and A.~Kusenko, \emph{{Interactions of keV sterile neutrinos with
  matter}}, {\emph{Phys.\ Rev.\ D} {\bf 81} (2010) 113006}.

\bibitem{Campos:2016gjh}
M.~D. Campos and W.~Rodejohann, \emph{{Testing keV sterile neutrino dark matter
  in future direct detection experiments}},
  \href{http://arxiv.org/abs/1605.02918}{{\tt 1605.02918}}.

\bibitem{Li:2011mw}
Y.~F. Li and Z.~z.~Xing, \emph{{Captures of Hot and Warm Sterile Antineutrino
  Dark Matter on EC-decaying Ho-163 Nuclei}}, {\emph{JCAP} {\bf 1108} (2011)
  006}.

\bibitem{Bovy:2012tw}
J.~Bovy and S.~Tremaine, \emph{{On the local dark matter density}},
  \href{http://dx.doi.org/10.1088/0004-637X/756/1/89}{\emph{Astrophys. J.} {\bf
  756} (2012) 89}, [\href{http://arxiv.org/abs/1205.4033}{{\tt 1205.4033}}].

\bibitem{Weinberg:1962zza}
S.~Weinberg, \emph{{Universal Neutrino Degeneracy}},
  \href{http://dx.doi.org/10.1103/PhysRev.128.1457}{\emph{Phys. Rev.} {\bf 128}
  (1962) 1457--1473}.

\bibitem{Irvine:1983nr}
J.~M. Irvine and R.~Humphreys, \emph{{Neutrino Masses and the Cosmic Neutrino
  Background}}, \href{http://dx.doi.org/10.1088/0305-4616/9/7/017}{\emph{J.
  Phys.} {\bf G9} (1983) 847--852}.

\bibitem{Long:2014zva}
A.~J. Long, C.~Lunardini and E.~Sabancilar, \emph{{Detecting non-relativistic
  cosmic neutrinos by capture on tritium: phenomenology and physics
  potential}},
  \href{http://dx.doi.org/10.1088/1475-7516/2014/08/038}{\emph{JCAP} {\bf 1408}
  (2014) 038}, [\href{http://arxiv.org/abs/1405.7654}{{\tt 1405.7654}}].

\bibitem{Morrissey:2009ur}
D.~E. Morrissey, D.~Poland and K.~M. Zurek, \emph{{Abelian Hidden Sectors at a
  GeV}}, \href{http://dx.doi.org/10.1088/1126-6708/2009/07/050}{\emph{JHEP}
  {\bf 07} (2009) 050}, [\href{http://arxiv.org/abs/0904.2567}{{\tt
  0904.2567}}].

\bibitem{Chun:2008by}
E.~J. Chun and J.-C. Park, \emph{{Dark matter and sub-GeV hidden U(1) in GMSB
  models}}, \href{http://dx.doi.org/10.1088/1475-7516/2009/02/026}{\emph{JCAP}
  {\bf 0902} (2009) 026}, [\href{http://arxiv.org/abs/0812.0308}{{\tt
  0812.0308}}].

\bibitem{Anelli:2015pba}
{\scshape SHiP} collaboration, M.~Anelli et~al., \emph{{A facility to Search
  for Hidden Particles (SHiP) at the CERN SPS}},
  \href{http://arxiv.org/abs/1504.04956}{{\tt 1504.04956}}.

\bibitem{Drewes:2016jae}
M.~Drewes, B.~Garbrecht, D.~Gueter and J.~Klaric, \emph{{Testing the low scale
  seesaw and leptogenesis}},  \href{http://arxiv.org/abs/1609.09069}{{\tt
  1609.09069}}.

\bibitem{DiBari:2013dna}
P.~Di~Bari, S.~F. King and A.~Merle, \emph{{Dark Radiation or Warm Dark Matter
  from long lived particle decays in the light of Planck}},
  \href{http://dx.doi.org/10.1016/j.physletb.2013.06.003}{\emph{Phys. Lett.}
  {\bf B724} (2013) 77--83}, [\href{http://arxiv.org/abs/1303.6267}{{\tt
  1303.6267}}].

\bibitem{Lasserre:2016eot}
T.~Lasserre, K.~Altenmueller, M.~Cribier, A.~Merle, S.~Mertens and M.~Vivier,
  \emph{{Direct Search for keV Sterile Neutrino Dark Matter with a Stable
  Dysprosium Target}},  \href{http://arxiv.org/abs/1609.04671}{{\tt
  1609.04671}}.

\bibitem{Castorina:2012md}
E.~Castorina, U.~Franca, M.~Lattanzi, J.~Lesgourgues, G.~Mangano, A.~Melchiorri
  et~al., \emph{{Cosmological lepton asymmetry with a nonzero mixing angle
  $\theta_{13}$}},
  \href{http://dx.doi.org/10.1103/PhysRevD.86.023517}{\emph{Phys. Rev.} {\bf
  D86} (2012) 023517}, [\href{http://arxiv.org/abs/1204.2510}{{\tt
  1204.2510}}].

\bibitem{deVega:2013hpa}
H.~J. de~Vega, M.~C. Falvella and N.~G. Sanchez, \emph{{Towards the Chalonge
  17th Paris Cosmology Colloquium 2013: highlights and conclusions of the
  Chalonge 16th Paris Cosmology Colloquium 2012}},
  \href{http://arxiv.org/abs/1307.1847}{{\tt 1307.1847}}.

\bibitem{Biermann:2013nxa}
P.~L. Biermann, H.~J. de~Vega and N.~G. Sanchez, \emph{{Towards the Chalonge
  Meudon Workshop 2013. Highlights and Conclusions of the Chalonge Meudon
  workshop 2012: warm dark matter galaxy formation in agreement with
  observations}},  \href{http://arxiv.org/abs/1305.7452}{{\tt 1305.7452}}.

\bibitem{deVega:2012jm}
H.~J. de~Vega, M.~C. Falvella and N.~G. Sanchez, \emph{{Towards the Chalonge
  16th Paris Cosmology Colloquium 2012: Highlights and Conclusions of the
  Chalonge 15th Paris Cosmology Colloquium 2011}},
  \href{http://arxiv.org/abs/1203.3562}{{\tt 1203.3562}}.

\bibitem{deVega:2011si}
H.~J. de~Vega and N.~G. Sanchez, \emph{{Warm dark matter in the
  galaxies:theoretical and observational progresses. Highlights and conclusions
  of the chalonge meudon workshop 2011}},
  \href{http://arxiv.org/abs/1109.3187}{{\tt 1109.3187}}.

\bibitem{deVega:2010zk}
H.~J. de~Vega and N.~G. Sanchez, \emph{{Highlights and Conclusions of the
  Chalonge Meudon Workshop Dark Matter in the Universe}},
  \href{http://arxiv.org/abs/1007.2411}{{\tt 1007.2411}}.

\bibitem{deVega:2010wj}
H.~J. de~Vega, M.~C. Falvella and N.~G. Sanchez, \emph{{Highlights and
  Conclusions of the Chalonge 14th Paris Cosmology Colloquium 2010: `The
  Standard Model of the Universe: Theory and Observations'}},
  \href{http://arxiv.org/abs/1009.3494}{{\tt 1009.3494}}.

\end{thebibliography}\endgroup
